\documentclass[twoside]{report}
\usepackage{a4}
\usepackage{symbols}
\usepackage{graphicx}
\usepackage{amsmath}
\usepackage{amssymb}
\begin{document}

\thispagestyle{empty}

\begin{center}
{{\Huge{E\textsc{lastic} C\textsc{ompton} S\textsc{cattering}}}}

\vspace{.5cm}
{{\Huge{\textsc{from the}}}}

\vspace{.5cm}
{{\Huge{N\textsc{ucleon and} D\textsc{euteron}}}}
\end{center}

\vspace{2cm}

\begin{center}

\includegraphics[width=.5\linewidth]{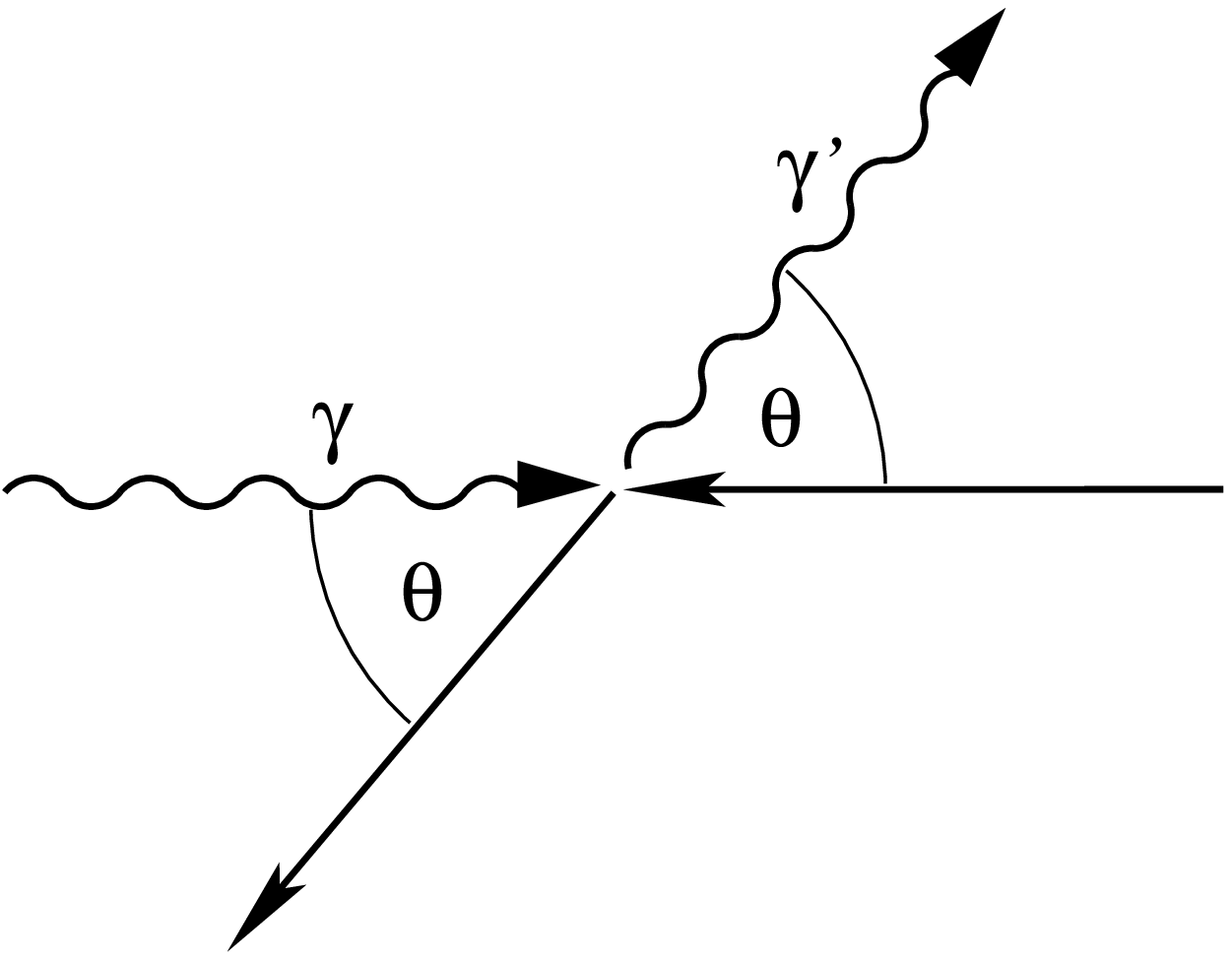}

\vspace{2cm}

{\Large{R\textsc{obert} P. H\textsc{ildebrandt}}}

\vfill

\includegraphics[width=.25\linewidth]{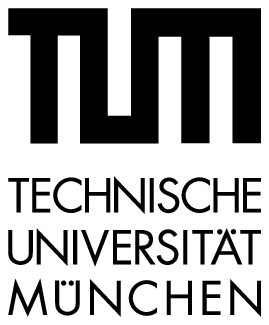}

\vspace{.5cm}

{\Large{P\textsc{hysik}-D\textsc{epartment}}}
\end{center}

\newpage
\thispagestyle{empty}
\phantom{hello}

\newpage
\thispagestyle{empty}
\setcounter{page}{1}
\begin{minipage}[t][\height][b]{0.48\linewidth}
\begin{center}
\textbf{Technische Universit\"at M\"unchen}\\
Physik-Department\\
Institut f\"ur Theoretische Physik T39\\
Univ.-Prof. Dr. W. Weise
\end{center}
\end{minipage}
\hfill
\begin{minipage}[t][\height][b]{0.48\linewidth}
\hfill
\includegraphics[width=.5\linewidth]{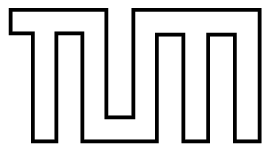}
\end{minipage}

\vspace{2cm}

\begin{center}
{\LARGE{\textbf{Elastic Compton Scattering\\from the Nucleon and Deuteron}}}

\vspace{.4cm}

{\large{Dipl.-Phys. Univ. Robert P. Hildebrandt}}
\end{center}

\vspace{4cm}

\noindent
Vollst\"andiger Abdruck  
der von der Fakult\"at f\"ur Physik der Technischen Universit\"at M\"unchen
zur Erlangung des akademischen Grades eines

\begin{center}
\textit{Doktors der Naturwissenschaften (Dr. rer. nat.)}
\end{center}
\noindent
genehmigten Dissertation.

\vspace{2cm}

\begin{tabular}{lll}
Vorsitzender:&&Univ.-Prof. Dr. Reiner Kr\"ucken \\ 
&&\\
Pr\"ufer der Dissertation:&&\\
&1.&Univ.-Prof. Dr. Wolfram Weise\\
&&\\
&2.&Univ.-Prof. Dr. Harald Friedrich
\end{tabular}

\vspace{1cm}

\noindent
Die Dissertation wurde am 28.10.2005 bei der Technischen 
Universit\"at M\"unchen eingereicht und durch die Fakult\"at f\"ur Physik am 
23.11.2005 angenommen.

\newpage
\thispagestyle{empty}
\phantom{hello}

\newpage
\noindent
{\Large{\textbf{Summary}}}

\vspace{.2cm}

\noindent
Goal of this work is the consistent description of elastic Compton 
scattering from the single nucleon and the deuteron. The theoretical 
framework chosen is
Chiral Perturbation Theory, which is the low-energy formulation of 
Quantum Chromodynamics, where we extend the spectrum of active degrees of 
freedom from only pions and nucleons to also the $\Delta(1232)$ resonance. 
In the deuteron sector, we treat
the nucleon-nucleon interaction non-perturbatively. 
Besides the Compton cross sections, our main concern is with  
the nucleon polarizabilities, which are a useful tool to 
describe the reaction of the internal nucleonic degrees of freeedom to the 
external electromagnetic field. Experimentally, only a few nucleon 
polarizabilities are known. Especially our present knowledge of the neutron 
polarizabilities is not satisfying. The reason 
why it is so difficult to determine these quantities experimentally
is the finite lifetime of the free neutron. Therefore, we want
to contribute to the ongoing discussion of the neutron polarizabilities 
by fitting the average over proton and neutron polarizabilities to the 
elastic deuteron Compton-scattering data. Similarly, we determine the 
proton polarizabilities from fits to the 
single-proton Compton data. We finally combine 
both results in order to identify the neutron polarizabilities. 

\vfill
\noindent
{\Large{\textbf{Zusammenfassung}}}

\vspace{.2cm}

\noindent
Ziel dieser Arbeit ist die konsistente Beschreibung von elastischer
Compton-Streu-ung am einzelnen Nukleon und am Deuteron. Als
theoretischen Rahmen w\"ahlen wir chirale St\"orungsrechnung, die 
Niederenergieformulierung der Quantenchromodynamik, wobei wir das
Spektrum der aktiven Freiheitsgrade von Pionen und Nu-kleonen um die 
$\Delta(1232)$-Resonanz erweitern. Im Deuteron-Sektor behandeln wir die 
Nukleon-Nukleon-Wechselwirkung nichtperturbativ. Neben den 
Compton-Streu-querschnitten$\,$ gilt$\,$ unser$\,$ Hauptinteresse$\,$ den$\,$ 
Nukleon-Polarisierbarkeiten,$\,$ einem n\"utzlichen Instrument, um die 
Reaktion der inneren Freiheitsgrade des Nukleons auf das \"au{\ss}ere 
elektromagnetische Feld 
zu beschreiben. Experimentell sind nur wenige der Nukleon-Polarisierbarkeiten
bekannt. Insbesondere unsere momentane Kenntnis der Neutron-Polarisierbarkeiten
ist nicht zufriedenstellend. Der Grund, warum es so schwierig ist, diese 
Gr\"o{\ss}en experimentell zu bestimmen, ist die endliche Lebensdauer des 
freien Neutrons. Deshalb
wollen wir zu der aktuellen Diskussion der Neutron-Polarisierbarkeiten 
beitragen, indem wir das Mittel von Proton- und Neutron-Polarisierbarkeiten 
an die  elastischen Deuteron-Compton-Streudaten anfitten. Ebenso bestimmen
wir die Proton-Polarisierbarkeiten aus Fits an die Proton-Compton-Streudaten.
Schlie{\ss}lich kombinieren wir beide Ergebnisse, um die 
Neutron-Polarisierbarkeiten zu ermitteln.

\newpage
\thispagestyle{empty}
\phantom{hello}

\newpage
\tableofcontents
\contentsline{chapter}{Bibliography}{\pageref{biblio}}
\listoffigures
\listoftables

\pagestyle{headings}
\chapter{Introduction \label{chap:intro} }
\section[Compton Scattering and the Nucleon Polarizabilities]
{Compton Scattering and the Nucleon\\Polarizabilities 
\label{sec:intro}}
\markboth{CHAPTER \ref{chap:intro}. INTRODUCTION}{\ref{sec:intro}. COMPTON
SCATTERING AND THE NUCLEON POLARIZABILITIES}

The structure of protons and neutrons as analyzed with electromagnetic probes 
has been under experimental and theoretical investigation for a number of 
decades. While for photon energies below 25 MeV the experimental cross
section is well described with the assumption of a point-like spin-1/2 nucleon
with an additional anomalous magnetic moment $\kappa$ \cite{Powell}, the
internal structure of the nucleon starts to play a role at higher energies.
In Compton scattering, the external electromagnetic field of the photon
attempts to deform the nucleon. The electromagnetic polarizabilities provide 
a measure of the global resistance of the
nucleon's internal degrees of freedom against displacement in an external 
electric or magnetic field, which makes 
them an excellent tool to study the structure of the nucleon. 
The most prominent two nucleon polarizabilities, the electric and magnetic 
dipole polarizabilities\footnote{We denote the static polarizabilities, i.e. 
the polarizabilities in the limit of vanishing photon energy, by a bar 
throughout this work.} 
$\bar{\alpha}_{E1}$ and $\bar{\beta}_{M1}$, are connected via the famous
Baldin sum rule with the total photon-nucleon cross section~\cite{Baldin}.
This sum rule is derived via the Kramers-Kronig relation from the 
low-energy theorems, an expansion of the forward-scattering amplitude
in the photon energy~\cite{Low}. It reads
\be
\bar{\alpha}_{E1}+\bar{\beta}_{M1}=
\frac{1}{2\,\pi^2}\int_{\w_\mathrm{th}}^\infty d\w \,
\frac{\sigma_\mathrm{tot}(\w)}{\w^2}
\label{eq:Baldinsumrule}
\ee
with $\w$ the photon energy 
and $\w_\mathrm{th}$ the threshold energy for pion photo-production.

The leading-order effective Hamiltonian, which arises due to the nucleon 
being polarized by the photon, is (see e.g.~\cite{Babusci})
\be
H_\mathrm{eff}=-\frac{1}{2}\,4\pi\left[\bar{\alpha}_{E1}\,\vec{E}^2+
                                       \bar{\beta} _{M1}\,\vec{B}^2\right].
\label{eq:Heff}
\ee
This Hamiltonian, however, includes only terms of second order  in $\w$ and 
can be extended, as long as symmetries like gauge-, Lorentz- and 
isospin-invariance are respected. One of the $\calO(\w^4)$-terms is 
proportional to the square of the time derivative of the electric field: 
\be
\delta H_\mathrm{eff}=
-\frac{1}{2}\,4\pi\left[\alpha_{E1}^\nu\,\dot{\vec{E}}^2\right]
\label{eq:deltaHeff}
\ee
Such corrections may be resummed to a dynamical, i.e. energy-dependent 
polarizability
\be
\bar{\alpha}_{E1}+\alpha_{E1}^\nu\,\omega^2+\dots\rightarrow
\alpha_{E1}(\omega),
\label{eq:alphaseries}
\ee
which in turn provides information on the energy dependence of the internal 
nucleonic degrees of freedom. 
In this work we deal with dynamical polarizabilities in terms of multipole
expansions.

Apart from $\alpha_{E1}(\w)$ and $\beta_{M1}(\w)$ there exist further
polarizabilities, such as the spin-dependent dipole polarizabilities and 
quadrupole polarizabilities. In this work we are mainly 
concerned with the dipole polarizabilities because we will see that in the 
energy range where our calculation is supposed to be valid, polarizabilities 
of higher multipole order are negligible in describing nucleon Compton cross 
sections. Stated differently, we observe a fast convergence of the Compton 
multipole expansion.
On the other hand, the spin polarizabilities do give important 
contributions. Therefore, we rewrite Eq.~(\ref{eq:Heff}), extended to the 
(energy-dependent) spin dipole polarizabilities:
\ba
H_\mathrm{eff}&=-2\pi\,\left[\alpha_{E1}(\w)\,\vec{E}^2+
\beta_{M1}(\w)\,\vec{B}^2+
\gamma_{E1E1}(\w)\,\vec{\sigma}\cdot\vec{E}\times\dot{\vec{E}}\right.\\
&+\left.\gamma_{M1M1}(\w)\,\vec{\sigma}\cdot\vec{B}\times\dot{\vec{B}}-
2\,\gamma_{M1E2}(\w)\,\sigma_i\,E_{ij}\,B_j+
2\,\gamma_{E1M2}(\w)\,\sigma_i\,B_{ij}\,E_j\right]\nonumber
\end{align}
This Hamiltonian can be found e.g. in \cite{Babusci}, with $\vec{\sigma}$ 
denoting the nucleon spin, $T_{ij}=\frac{1}{2}(\partial_i T_j+\partial_j T_i)$
and the indices of the spin polarizabilities chosen such that 
$\gamma_{TlT'l'}$ parameterizes a $Tl$ ($T'l'$) transition at the first 
(second) photon vertex.

The experimentally best-known nucleon polarizabilities are the static 
electric and magnetic dipole polarizabilities of the proton, 
$\bar{\alpha}_{E1}^p$ and $\bar{\beta}_{M1}^p$. 
They are extracted from proton Compton-scattering data, and a large number of 
such experiments has been performed over the past decades, see e.g. 
Refs.~[\ref{OLMOS}-\ref{GALLER}]. 
In~\cite{Olmos}, a global, Baldin-sum-rule-constrained fit to the wealth 
of Compton-scattering data on the proton yielded 
\begin{align}
\bar{\alpha}_{E1}^p&=(12.1\pm0.3\,(\mathrm{stat})\mp0.4\,(\mathrm{syst})
            \pm0.3\,(\mathrm{model}))\cdot 10^{-4}\;\mathrm{fm}^3,\nonumber\\
\bar{\beta}_{M1}^p &=( 1.6\pm0.4\,(\mathrm{stat})\pm0.4\,(\mathrm{syst})
            \pm0.4\,(\mathrm{model}))\cdot 10^{-4}\;\mathrm{fm}^3,
\label{eq:expp}
\end{align}
where the errors include statistical and systematic uncertainties as well as 
estimates of model dependence in the analysis.
These numbers are very close to those recommended in a recent 
review~\cite{Schumacher}, which were obtained as the weighted average over 
several experiments:
\begin{align}
\bar{\alpha}_{E1}^p&=(12.0\pm0.6)\cdot 10^{-4}\;\mathrm{fm}^3\nonumber\\
\bar{\beta}_{M1}^p &=( 1.9\mp0.6)\cdot 10^{-4}\;\mathrm{fm}^3
\label{eq:reviewp}
\end{align}
Comparing these values to the typical size of a proton ($\sim1\,\fm^3$) we 
find that nucleons are hard to deform, i.e. they are rather stiff objects. 
Note that the experiments from which the numbers of Eq.~(\ref{eq:expp}) are 
derived, have not been performed for near-static
photons but in an energy range of about 55~-~800~MeV. Therefore theoretical 
input, e.g. from a Dispersion-Relation Analysis, is unavoidable in the 
extraction of $\bar{\alpha}_{E1}$ and $\bar{\beta}_{M1}$.

Stable single-neutron targets do not exist. It is therefore 
much harder to access the neutron polarizabilities experimentally. 
An experiment on quasi-free Compton scattering from the proton 
and neutron bound in the deuteron~\cite{Kossert} gives results for the neutron 
polarizabilities which suggest 
very small isovector components\footnote{The isovector polarizabilities are 
defined as $\alpha_{E1}^v\equiv\frac{1}{2}\,(\alpha_{E1}^p-\alpha_{E1}^n)$, 
           $\beta_{M1}^v\equiv\frac{1}{2}\,(\beta _{M1}^p-\beta _{M1}^n)$.} 
when compared to Eqs.~(\ref{eq:expp},~\ref{eq:reviewp}):
\begin{align}
\bar{\alpha}_{E1}^n&=
(12.5\pm1.8\,(\mathrm{stat})\,^{+1.1}_{-0.6}\,(\mathrm{syst})
            \pm1.1\,(\mathrm{model}))\cdot 10^{-4}\;\mathrm{fm}^3\nonumber\\
\bar{\beta}_{M1}^n &=
( 2.7\mp1.8\,(\mathrm{stat})\,^{+0.6}_{-1.1}\,(\mathrm{syst})
            \mp1.1\,(\mathrm{model}))\cdot 10^{-4}\;\mathrm{fm}^3
\label{eq:expn}
\end{align}
The central values of Eq.~(\ref{eq:expn}) are identical to those suggested 
in~\cite{Schumacher}, which were obtained~-- like the proton values given in 
Eq.~(\ref{eq:reviewp})~-- as the weighted average over several experiments:
\begin{align}
\bar{\alpha}_{E1}^n&=(12.5\pm1.7)\cdot 10^{-4}\;\mathrm{fm}^3\nonumber\\
\bar{\beta}_{M1}^n &=( 2.7\mp1.8)\cdot 10^{-4}\;\mathrm{fm}^3
\label{eq:reviewn}
\end{align}
The two procedures entering
Eq.~(\ref{eq:reviewn}) 
are quasi-free Compton scattering from the neutron and electromagnetic 
scattering of neutrons on lead.
A similar observation for $\bar{\alpha}_{E1}^n$ has been made 
in~\cite{Schmied}, where the latter process was investigated:
\begin{equation}
\bar{\alpha}_{E1}^n=(12.6\pm2.5)\cdot 10^{-4}\;\mathrm{fm}^3
\label{eq:expSchmied}
\end{equation}
However, the precision of this result has been questioned by the authors of 
\cite{Enik}. Their estimate of the correct range for the result 
from~\cite{Schmied} is $7\leq\bar{\alpha}_{E1}^n\leq19$.
On the other hand, another experiment~\cite{Koester}, using the same 
technique, gives a completely  different result:
\be
\bar{\alpha}_{E1}^n=(0.6\pm5.0)\cdot 10^{-4}\;\mathrm{fm}^3
\ee

On the theory side, non-relativistic Chiral Perturbation Theory predicts that 
the proton and neutron 
polarizabilities are equal at leading-one-loop order \cite{BKKM}, 
since the pion loops that generate these contributions are isoscalar in 
nature. The absence of large isovector pieces in $\alpha_{E1}$ and 
$\beta_{M1}$ is therefore in accord with this picture.
The isoscalar $\mathcal{O}(p^4)$-HB$\chi$PT estimate~\cite{BKMS},
$\bar{\alpha}_{E1}^s=(11.95\pm2.5)\cdot10^{-4}\;\mathrm{fm}^3$,
$\bar{\beta}_{M1}^s = (5.65\pm5.1)\cdot10^{-4}\;\mathrm{fm}^3$, 
is consistent with vanishing isovector polarizabilities, when compared to the 
numbers given in Eq.~(\ref{eq:reviewn}), but no 
meaningful conclusion can be drawn due to the large error bars in the 
$\mathcal{O}(p^4)$ estimate, which in addition is cutoff dependent. 
The reason for the huge error bars is  
sensitivity to short-distance contributions that were estimated using the 
resonance-saturation hypothesis.

Another possible way to determine the neutron polarizabilities is elastic 
low-energy Compton scattering from light nuclei, e.g. from the deuteron. 
Several experiments have already been performed~\cite{Lucas,Lund,Hornidge} 
and further proposals exist -- e.g. Compton scattering on the  
deuteron or $^3\mathrm{He}$ at TUNL/HI$\gamma$S 
\cite{Gao}\footnote{Experiments on $^3\mathrm{He}$ are especially well suited 
to investigate the spin structure of the neutron, which mainly carries the 
spin of the $^3$He-nucleus.} and on 
deuteron targets at MAXlab~\cite{Schroeder}.  The latter proposal has already
been accepted and promises an extensive study of elastic deuteron Compton 
scattering below the pion-production threshold. 
From a theorist's point of view, extracting the neutron 
polarizabilities from elastic $\gamma d$ scattering 
requires an accurate description of the nucleon structure \textit{and} of  the 
dynamics of the low-energy degrees of freedom within the deuteron, as one has 
to account for the proton polarizabilities as well as for nuclear binding 
effects. A first attempt to fit the isoscalar polarizabilities 
$\bar{\alpha}_{E1}^s\equiv
\frac{1}{2}\,(\bar{\alpha}_{E1}^p+\bar{\alpha}_{E1}^n)$, 
$\bar{\beta}_{M1}^s\equiv
\frac{1}{2}\,(\bar{\beta} _{M1}^p+\bar{\beta} _{M1}^n)$ to 
the elastic deuteron Compton-scattering data from~\cite{Lucas, Hornidge} has 
been made in~\cite{Lvov}. The extracted neutron polarizabilities 
$\bar{\alpha}_{E1}^n=( 9.0\pm3.0)\cdot 10^{-4}\;\mathrm{fm}^3$,
$\bar{\beta}_{M1}^n=(11.0\pm3.0)\cdot 10^{-4}\;\mathrm{fm}^3$
indicate the possibility of a rather \textit{large} isovector part, in 
contrast to the quasi-elastic results from \cite{Kossert}. 
They also disagree with the isoscalar Baldin sum rule, 
$\bar{\alpha}_{E1}^s+\bar{\beta}_{M1}^s=
(14.5\pm0.6)\cdot10^{-4}\;\mathrm{fm}^3$, cf. Chapter~\ref{chap:perturbative}
for details.
The same pattern is observed, albeit less pronounced, in the fit of 
Ref.~\cite{Ji} to the data measured at 49~MeV~\cite{Lucas} and 
55~MeV~\cite{Lund} within an Effective Field Theory with pions integrated out:
$\bar{\alpha}_{E1}^s=(12.3\pm1.4)\cdot 10^{-4}\;\mathrm{fm}^3$,
$\bar{\beta}_{M1}^s =( 5.0\pm1.6)\cdot 10^{-4}\;\mathrm{fm}^3$.
On the other hand, comparing
the elastic deuteron Compton calculation of Ref.~\cite{Karakowski} with the 
data from~\cite{Lucas} is in good agreement with nearly vanishing isovector 
polarizabilities: 
$\bar{\alpha}_{E1}^n=(12.0\pm4.0)\cdot 10^{-4}\;\mathrm{fm}^3$,
$\bar{\beta}_{M1}^n=( 2.0\pm4.0)\cdot 10^{-4}\;\mathrm{fm}^3$, albeit within 
rather large error bars. 

It is obvious from these partly 
contradictory results that there is still a lot of work to be 
done in order to have reliable values for $\bar{\alpha}_{E1}^n$ and 
$\bar{\beta}_{M1}^n$. Therefore, in this thesis we contribute to the 
ongoing discussion of the neutron polarizabilities, investigating 
single-nucleon Compton scattering and elastic Compton scattering from the 
deuteron. The framework is Chiral Effective Field Theory ($\chi$EFT),
allowing for non-perturbative methods in the deuteron section. In general, 
$\chi$EFT provides a consistent, controlled framework for elastic $\gamma p$ 
and $\gamma d$ scattering within 
which nucleon-structure effects can be disentangled from meson-exchange 
currents, deuteron binding, etc.  It also allows for an estimate of the 
uncertainties arising from higher-order corrections.

We perform a multipole expansion of the Compton amplitude in the 
single-nucleon Compton calculation, mainly for two reasons:
Firstly, we are interested in the Compton multipoles themselves, which we 
combine to dynamical polarizabilities, cf. Eq.~(\ref{eq:alphaseries}). These 
multipole amplitudes exhibit the response of the various nucleonic degrees of 
freedom to the external electromagnetic field of the scattered photon 
and therefore 
provide valuable information on the structure of the
nucleon. The second reason is that it is often sufficient to include only
the leading terms of the multipole expansion in order to achieve a good 
approximation of the full calculation. We demonstrate that this scenario also 
holds in nucleon Compton scattering and conclude that the few contributing  
parameters, i.e. the six dipole polarizabilities, may be extracted from 
a combination of spin-averaged and polarized Compton experiments. 
The two parameters $\bar{\alpha}_{E1}^p$ and 
$\bar{\beta}_{M1}^p$, which are a priori free in our calculation, are 
determined via fits to proton Compton data.

The central goal of this work is to extract not only the proton but also  
the neutron values for $\bar{\alpha}_{E1}$ and $\bar{\beta}_{M1}$ from 
data. Therefore in the second main part we are 
concerned with elastic deuteron Compton scattering, which was introduced 
before as one of the possible ways to determine the isoscalar 
polarizabilities. We present two partly different calculations of deuteron 
Compton scattering. The first one agrees well with the high-energy data but it
fails to describe the low-energy regime, say the region $\w\ll 50$~MeV, 
correctly, a feature that we have in 
common with other calculations such as~\cite{McGPhil}. On the other hand, all
calculations on elastic deuteron Compton scattering existing so far, which 
reach the correct low-energy limit, have 
problems to describe the high-energy data from \cite{Hornidge}, see e.g.
\cite{Lvov,Karakowski}. Therefore, we consider it as one of the central points
of this work that in our second approach to $\gamma d$ scattering 
we obtain a consistent and novel description of
all existing deuteron Compton data, which also fulfills the low-energy theorem,
i.e. we obtain the well-known Thomson-limit of elastic deuteron Compton 
scattering. The good description of the data enables us to perform a global 
fit of the isoscalar polarizabilities to all data points. These numbers are 
combined with our fit results for the proton polarizabilities, yielding values 
which prove that the data basis on elastic $\gamma d$ scattering is in good
agreement with even vanishing isovector polarizabilities.
 
Parts of this work have been published in our papers 
Refs.~\cite{HGHP},~\cite{polarizedpaper} and~\cite{deuteronpaper}. 
Some elements are based on my diploma thesis~\cite{DA}.

\section{Outline
\label{sec:outline}}
This thesis is structured in the following way:

In Chapter~\ref{chap:theory} we give a brief introduction to the two versions
of Chiral Effective Field Theory applied in this work, starting with
``Heavy Baryon Chiral Perturbation Theory'' (HB$\chi$PT, 
Section~\ref{sec:HBchiPT}), which is the 
low-energy formulation of Quantum Chromodynamics (QCD) with pions and nucleons
as active degrees of freedom. In Section~\ref{sec:SSE} we review how this 
theory is modified in the so-called ``Small Scale Expansion'', an extension 
of HB$\chi$PT including the $\Delta(1232)$-resonance field as an additional
explicit degree of freedom. In both sections we write down the respective 
Lagrangeans relevant for this work.

Chapter~\ref{chap:spinaveraged} is dedicated to spin-averaged single-nucleon 
Compton scattering. We derive a multipole expansion for this process
and compare the Compton cross sections, calculated 
in leading-one-loop order HB$\chi$PT and SSE, respectively, to a 
Dispersion-Relation Analysis and experiments. In the SSE description we 
allow for free polarizabilities $\bar{\alpha}_{E1}$ and $\bar{\beta}_{M1}$, 
which we fit to proton Compton data. Furthermore, we use the various 
multipole amplitudes to define dynamical, i.e. energy-dependent 
nucleon polarizabilities, cf. Section~\ref{sec:intro}, and we compare the
results achieved for these quantities within the three theoretical frameworks.

After the chapter on unpolarized cross sections we turn to polarized 
Compton scattering in Chapter~\ref{chap:spinpolarized}, where we calculate 
several asymmetries, using circularly and linearly polarized photons. We 
demonstrate that determining the six dipole polarizabilities directly from 
experiment is possible, due to the strong sensitivity of selected observables 
to the spin polarizabilities.

Our most prominent aim is to extract the proton and the neutron values
for $\bar{\alpha}_{E1}$ and $\bar{\beta}_{M1}$ from data within 
one consistent framework. As there are no data on Compton scattering from 
the neutron, we turn in Chapters~\ref{chap:perturbative} 
and~\ref{chap:nonperturbative} to elastic deuteron Compton scattering 
in order to determine the isoscalar polarizabilities.
The main difference between the two chapters is that in 
Chapter~\ref{chap:perturbative} we restrict ourselves to photon energies of 
the order of 100~MeV which enables us to
calculate the $\gamma d$ kernel strictly 
according to the power-counting rules of the Small Scale Expansion. In 
Chapter~\ref{chap:nonperturbative}, we allow for non-perturbative aspects
via the inclusion of the two-nucleon $T$-matrix in the intermediate
state, which turns out to be a necessary modification in order 
to describe low-energy deuteron Compton scattering correctly. In both 
approaches we fit the isoscalar polarizabilities  $\bar{\alpha}_{E1}^s$, 
$\bar{\beta}_{M1}^s$ and combine the respective numbers with our results for 
the proton polarizabilities in order to determine those of the neutron. Both 
extractions agree well with each other within their (small) error bars
and with the results of 
Ref.~\cite{Kossert}, however only in Chapter~\ref{chap:nonperturbative} we are
able to fit to all existing elastic deuteron Compton data.

We conclude in Chapter~\ref{chap:conclusion}, having shifted the most
technical parts of this thesis to the appendices.

\chapter{Chiral Effective Field Theories \label{chap:theory}}
In this chapter we want to give a brief survey of the two Effective Field 
Theories applied in this work, starting with ``Heavy 
Baryon Chiral Perturbation Theory'' (HB$\chi$PT), which is the low-energy 
formulation of Quantum Chromodynamics including nucleons and pions as active 
degrees of freedom. In the succeeding section we shortly 
introduce the so-called ``Small Scale Expansion'' (SSE), an extension of the 
former theory that, in addition, also includes the $\Delta(1232)$ resonance as
an explicit degree of freedom. 

\section{Heavy Baryon Chiral Perturbation Theory \label{sec:HBchiPT}}

The theory of strong interactions, QCD, describes point-like fermions of 
spin~$\frac{1}{2}$, the so-called quarks, which interact with each other via 
the exchange of gauge-bosons, the gluons. These particles couple to the 
``color'' of the quarks, an additional degree of freedom which was introduced
in order 
to maintain the demand of totally antisymmetric fermion wave functions, see 
e.g.~\cite{Muta}. At low momentum transfer, however, the coupling constant of 
QCD becomes rather strong and quarks and gluons are no longer the relevant
degrees of freedom. In fact they are confined in color-neutral objects, the 
hadrons, which therefore are the active degrees of freedom of low-energy QCD.

When we compare the masses of the six different
quark flavors, we find three flavors~-- ``up'', ``down'' and ``strange''~--  
which are considerably lighter than the nucleon. Their masses (at a 
renormalization scale of about 1 GeV) are~\cite{Weise}
\be
m_u\approx   4\;\text{MeV},\qquad
m_d\approx   8\;\text{MeV},\qquad
m_s\approx 164\;\text{MeV}.
\label{eq:quarkmasses}
\ee
The other quark flavors (``charm'', ``bottom'' and ``top'') are heavy, 
i.e. their masses exceed the nucleon mass. Therefore, it is sufficient 
for many applications to only calculate with the three light flavors, or even 
with only up and down quarks.

Due to the low masses of the light quark flavors, it is instructive to 
investigate the so-called ``chiral limit'', i.e. the limit of vanishing 
quark masses. In this limit, the QCD-Lagrangean exhibits another symmetry, 
the ``chiral symmetry'', which forbids the coupling of right-handed to 
left-handed quarks. In order to demonstrate this symmetry we write the 
Lagrangean of a free, mass-less fermion,
\be
\mathcal{L}=i\bar{\psi}\gamma_\mu\partial^\mu\psi
\ee
in terms of right- and left-handed particles:
\be
\mathcal{L}=i\bar{\psi}_R\gamma_\mu\partial^\mu\psi_R
           +i\bar{\psi}_L\gamma_\mu\partial^\mu\psi_L
\label{eq:LQCDRL}
\ee
with
\be
\psi_R=\frac{1}{2}\left(1+\gamma_5\right)\psi,\qquad 
\psi_L=\frac{1}{2}\left(1-\gamma_5\right)\psi.
\ee
Obviously, the two kinds of fermions in Eq.~(\ref{eq:LQCDRL}) do not interact 
with each other, i.e. in the chiral limit the Lagrangean of QCD is invariant
under the global, unitary transformations in flavor space
\be
\psi_R\rightarrow U_R\,\psi_R=
\exp\left[i\theta_R^a\frac{\tau_a}{2}\right]\psi_R,\qquad
\psi_L\rightarrow U_L\,\psi_L=
\exp\left[i\theta_L^a\frac{\tau_a}{2}\right]\psi_L
\label{eq:Trafos}
\ee
with $\tau_a$ the Pauli isospin matrices. Here we restrict ourselves to 
the lightest two  quark flavors, i.e. $U_R,\;U_L\in SU(2)$. The 
corresponding conserved Noether currents are 
\be
J_{R,a}^\mu=\bar{\psi}_R\gamma^\mu\frac{\tau_a}{2}\psi_R,\qquad
J_{L,a}^\mu=\bar{\psi}_L\gamma^\mu\frac{\tau_a}{2}\psi_L,
\ee
which can be rewritten by addition or subtraction as the often used isospin
(vector) and axial-vector currents
\be
V_a^\mu=\bar{\psi}\gamma^\mu        \frac{\tau_a}{2}\psi, \qquad 
A_a^\mu=\bar{\psi}\gamma^\mu\gamma_5\frac{\tau_a}{2}\psi.
\ee

Including the $SU(2)$-quark-mass matrix $\hat{m}$ in the Lagrangean explicitly
breaks chiral symmetry, as such a term couples left- to right-handed quarks:
\begin{equation}
\mathcal{L}_m=\bar{\psi}\,\hat{m}\,\psi=
              \bar{\psi}_L\,\hat{m}\,\psi_R+\bar{\psi}_R\,\hat{m}\,\psi_L
\label{eq:Lmass}
\end{equation}
Nevertheless, due to the small size of the quark masses, 
chiral symmetry might still be fulfilled to a good approximation.
In this case, all hadrons
would appear as doublets of nearly equal mass but opposite parity.  
However, investigating the hadronic spectrum we find that chiral symmetry has
to be spontaneously broken, as e.g. the pseudoscalar ($J^P=0^-$) 
mesons are considerably lighter than their pendants with positive parity.  
On the other hand, $m_u\approx m_d$, cf. Eq.~(\ref{eq:quarkmasses}), i.e. 
isospin symmetry is nearly exactly preserved. Therefore, the symmetry 
$SU(2)_L\times SU(2)_R$ is obviously broken to $SU(2)_V$. 

Whenever a global, continuous 
symmetry of the Lagrangean is spontaneously broken, 
Goldstone's theorem applies and massless Goldstone-bosons occur. In the case
of chiral symmetry with two active quark flavors there are three generators
broken, i.e. we expect the observation of three Goldstone-bosons. These are the
three pions, which are by far the lightest of the hadrons. Their quantum 
numbers coincide with breaking the axial-vector symmetry: Like the 
$0$-component of the axial current they are of pseudoscalar 
nature~\cite{Weise}.  Although they are the Goldstone-bosons of chiral 
symmetry breaking, the pions do have a small mass due to the non-vanishing 
quark masses, which explicitly break chiral symmetry. Nevertheless, 
the mass gap between the pions and all other hadrons suggests that
the pions are the relevant low-energy degrees of freedom of QCD. 

It is at the heart of perturbation theory to have a small expansion parameter.
In Chiral Perturbation Theory ($\chi$PT), the low-energy formulation of 
QCD with only pions as active degrees of freedom~\cite{Gasser}, this 
parameter is given by the pion mass or a small momentum, divided by the 
characteristic scale, which is $\Lambda_\chi=4\pi f_\pi\approx1161$~MeV 
with the pion-decay constant $f_\pi$. Obviously, the expansion only 
converges for momenta of the external probe~-- in our case the scattered 
photon~-- which are much smaller than the breakdown scale $\Lambda_\chi$. 

In this work we are 
interested in processes involving one or two nucleons. Therefore we use 
Heavy Baryon Chiral Perturbation Theory, an extension of $\chi$PT which 
explicitly includes the nucleon as an additional degree of 
freedom~\cite{Manohar}. Weinberg showed~\cite{Weinberg79}, how one can 
systematically include contributions from pion loops to the tree-level 
diagrams. These corrections are often referred to as contributions from the 
``pion-cloud'' around the nucleon.

In HB$\chi$PT one assumes small 
momenta of the heavy nucleons and therefore includes their kinetic energy 
only perturbatively. As long as the nucleons are on-shell, this is a valid 
procedure for low energies, however we will encounter a severe problem in our 
deuteron calculation, where even for external sources of vanishing energy the 
nucleons are non-static, due to their slight off-shellness.

For completeness, in the following we list all HB$\chi$PT Lagrangeans relevant
for real Compton scattering up to $\calO(p^3)$, the order to which we are 
working. However, except for the leading-order Lagrangeans we restrict 
ourselves to those parts which actually contribute to Compton scattering. 
We choose the Weyl-gauge, i.e. $v\cdot A=0$ with the four-velocity 
$v_\mu$ of the nucleon and the photon field $A_\mu$. Further details can be 
found in Ref.~\cite{BKM}, albeit our convention differs from~\cite{BKM} in 
the sense, that we split the photon field into isoscalar and isovector parts.

The Lagrangean is composed of baryonic and purely pionic parts,
\be
\mathcal{L}_{CS}^{(3)}=\mathcal{L}_{\pi N}^{(1)}+\mathcal{L}_{\pi N}^{(2)}+
\mathcal{L}_{\pi N}^{(3)}+\mathcal{L}_{\pi\pi}^{(2)}+\mathcal{L}_{\pi\pi}^{(4)}
\ee
with 
\begin{align}
\mathcal{L}_{\pi N}^{(1)}&=\bar{N}_v\left(iv\cdot D+g_A\,S\cdot u\right)N_v,\\
\mathcal{L}_{\pi N}^{(2)}&=\frac{1}{2\,m_N}\bar{N}_v\biggl\{(v\cdot D)^2-D^2
\nonumber\\
&-\frac{i}{2}\left[S^\mu,S^\nu\right]
\left[(1+\kappa_v)\,f_{\mu\nu}^+ +2\,(1+\kappa_s)\,v_{\mu\nu}^{(s)}
\right]+\dots\biggl\}N_v,\\
\mathcal{L}_{\pi N}^{(3)}&=\frac{-1}{8\,m_N^2}\bar{N}_v\biggl\{(1+2\kappa_v)\,
\left[S_\mu,S_\nu\right]\,f_+^{\mu\sigma}\,v_\sigma\,D^\nu\nonumber\\
&+2\,(\kappa_s-\kappa_v)\,\left[S_\mu,S_\nu\right]\,v^{\mu\sigma}_{(s)}\,
v_\sigma\,D^\nu+h.\,c.+\dots\biggl\}N_v.
\end{align}
$g_A$ is the axial pion-nucleon coupling constant at leading order, $m_N$ the
nucleon mass and $\kappa_s=\kappa_p+\kappa_n$ ($\kappa_v=\kappa_p-\kappa_n$) 
the anomalous isoscalar (isovector) magnetic moment of the nucleon.
The velocity projection operator
\begin{equation}
P_v^+=\frac{1}{2}(1+v\!\!\!/)
\end{equation}
projects from the relativistic nucleon Dirac field $\Psi_N$ to the 
velocity-dependent nucleon field via
\begin{equation}
N_v=\mathrm{exp}\left[im_Nv\cdot x\right]P_v^+\Psi_N.
\end{equation}
$S_\mu$ is the Pauli-Lubanski vector (see e.g. \cite{BKM}) and 
\begin{equation}
D_\mu=\partial_\mu+\Gamma_\mu-iv_\mu^{(s)}
\end{equation}
denotes the covariant derivative of the nucleon.
The chiral tensors contained in the above equations are
\begin{align}
U&=u^2=\mathrm{exp}\left[i\vec{\tau}\cdot\vec{\pi}/f_\pi\right],\\
\Gamma_\mu&=\frac{1}{2}\left\{
 u^\dagger\left(\partial_\mu-i\,e\,\frac{\tau^3}{2}\,A_\mu\right)u
+u        \left(\partial_\mu-i\,e\,\frac{\tau^3}{2}\,A_\mu\right)u^\dagger
\right\}
\end{align}
and
\begin{align}
u_\mu&=i\left\{
 u^\dagger\left(\partial_\mu-i\,e\,\frac{\tau^3}{2}\,A_\mu\right)u
-u        \left(\partial_\mu-i\,e\,\frac{\tau^3}{2}\,A_\mu\right)u^\dagger
\right\}\;\,
\end{align}
with  $\vec{\pi}$ representing the  pion field. 
$v_\mu^{(s)}=\frac{e}{2}\,A_\mu$ denotes an isoscalar photon field and 
the corresponding field-strength tensors are 
\begin{align}
v_{\mu\nu}^{(s)}&=\partial_\mu v_\nu^{(s)}-\partial_\nu v_\mu^{(s)},\\
f^{\mu\nu}_{+}  &=u        \,e\,\frac{\tau^3}{2}
\left(\partial_\mu A_\nu-\partial_\nu A_\mu\right)\,u^\dagger
                 +u^\dagger\,e\,\frac{\tau^3}{2}
\left(\partial_\mu A_\nu-\partial_\nu A_\mu\right)\,u.
\label{eq:fmunu+}
\end{align}

As far as the purely mesonic sector is concerned, we need
\begin{equation}
\mathcal{L}_{\pi\pi}^{(2)}=\frac{f_\pi^2}{4}\,\mathrm{tr}
\left[\left(\nabla_\mu\,U\right)^\dagger\,\nabla^\mu\,U
     +\chi^\dagger\,U+\chi\,U^\dagger\right]
\label{eq:Lpi2}
\end{equation}
and
\begin{equation}
\mathcal{L}_{\pi\pi}^{(4)}=-\frac{e^2}{32\,\pi^2\,f_\pi}\,
\epsilon^{\mu\nu\alpha\beta}\,f_{\mu\nu}\,f_{\alpha\beta}\,\pi^0+\dots,
\label{eq:Lpi4}
\end{equation}
where $\epsilon_{0123}=1$. Eq.~(\ref{eq:Lpi4}) is responsible for the decay of
an uncharged pion into two photons, i.e. for the pion-pole diagram, 
Fig.~\ref{fig:app:pole}(d). For further details on this decay we refer the 
reader to Refs.~\cite{Itzykson,Peskin}. 
The chiral tensors in Eqs.~(\ref{eq:Lpi2}) and~(\ref{eq:Lpi4}) are
\begin{align}
\nabla_\mu U&=\partial_\mu U-i\,\frac{e}{2}\,A_\mu\left[\tau_3,U\right],\\
\chi&=2\,\mathcal{B}\,\hat{m}
\end{align}
with the quark-condensate parameter $\mathcal{B}$ and the  $SU(2)$-quark-mass 
matrix $\hat{m}$ in the isospin limit $m_u=m_d$, which has  already been 
introduced in Eq.~(\ref{eq:Lmass}). 
$f_{\mu\nu}$ is the well-known field-strength tensor 
\be
f_{\mu\nu}=\partial_\mu A_\nu-\partial_\nu A_\mu.
\ee

Now we have prepared all tools in order to calculate real Compton scattering 
up to leading-one-loop order in HB$\chi$PT. However, it has been known for 
many decades that the $\Delta(1232)$ resonance plays a crucial role in 
this process. Therefore, in the next section we extend the spectrum of 
explicit degrees of freedom for this first nucleonic resonance.

\section{Small Scale Expansion \label{sec:SSE}}

Concerning the mass difference $\Delta_0$ between the $\Delta(1232)$ and the 
nucleon, there exist two complementary points of view: 
Some authors consider $\Delta_0$ to be much larger than $m_\pi$ and therefore
argue that the contributions from the $\Delta$ resonance may be absorbed into 
higher-order contact terms~\cite{BKM}. 
In other publications the $\Delta$ resonance is included as an explicit degree
of freedom because one might expect that due to the strong $N\Delta$-coupling
the $\Delta(1232)$ gives as important contributions as the pion cloud.
The idea to extend HB$\chi$PT in such a way has 
its origin in the early 1990's~\cite{Jenkins}.
Whether or not it is advantageous to include the 
$\Delta(1232)$ explicitly of course depends on the process under 
investigation, e.g. the pion-mass dependence of the nucleon mass is  rather 
weakly influenced by the $\Delta(1232)$~\cite{Massi}, whereas the anomalous 
magnetic moment of the nucleon depends strongly on the 
$\Delta$~resonance~\cite{HW}.

In nucleon Compton scattering, the $\Delta(1232)$ resonance is well-known to 
play a crucial role, see e.g.~\cite{Galler}. Therefore, in this work we 
include the $\Delta$ resonance explicitly, 
i.e. we need to specify how the $\Delta N$ mass splitting $\Delta_0$ is 
treated in the power counting. Here we use the so-called Small Scale Expansion
(SSE)~\cite{HHKLett,HHK}\footnote{We note that there also exist alternative 
approaches for Chiral Effective Field Theories with explicit $\pi$, $N$ and 
$\Delta$ degrees of freedom, e.g. the $\delta$-expansion~\cite{deltaexp}, 
which was recently shown to describe $\gamma p$ cross-section 
data well in an energy range from $\omega=0$~MeV to $\omega=300$~MeV.}, 
where the expansion parameter is called $\epsilon$, denoting either a small 
momentum, the pion mass or the mass difference $\Delta_0$ between the 
real part of the $\Delta$ mass and the nucleon mass: 
\be
\Delta_0=\mathrm{Re}[m_\Delta]-m_N
\ee
We note that the SSE
power counting is constructed such that up to a certain order in 
$\epsilon$ all HB$\chi$PT diagrams to the same order are included as well.
In this work, real Compton scattering on the nucleon and on the deuteron is 
investigated up to third order ($\calO(\epsilon^3)$) in the Small Scale 
Expansion.

In the following we summarize all parts of the 
SSE Lagrangean, which are relevant for our calculation and contain the 
explicit $\Delta$ field. The notation is adapted from~\cite{HHK} 
and~\cite{HGHP}. 
\begin{align}
\mathcal{L}_  \Delta ^{(1)}&=
-\bar{T}_i^\mu\,g_{\mu\nu}\,\left[iv\cdot D^{ij}-\Delta\,\delta^{ij}
+\dots\right]T_j^\nu
\label{eq:LD1}\\
\mathcal{L}_{N\Delta}^{(1)}&=g_{\pi N\Delta}\,\bar{T}_i^\mu\,w^i_\mu\,N+h.c.
\label{eq:LND1}\\
\mathcal{L}_{N\Delta}^{(2)}&=
\bar{T}_i^\mu\,\left[\frac{i\,b_1}{M}\,S^\nu\,f^i_{+\mu\nu}
+\dots\right]\,N+h.c.
\label{eq:LND2}
\end{align}  
Here
\begin{equation}
T_\mu^i(x)=
P_v^+\,P^{3/2}_{(33)\mu\nu}\,\psi_i^\nu(x)\,\mathrm{exp}(iM\,v\cdot x)
\end{equation}
denotes the ``light'' part of the spin-3/2 baryon field $\psi_\mu^i$, 
which is projected from the relativistic Rarita-Schwinger field via the 
spin-3/2 projection operator for fields of constant ``velocity'',
\begin{equation}
P^{3/2}_{(33)\mu\nu}=
g_{\mu\nu}-\frac{1}{3}\gamma_\mu\,\gamma_\nu-\frac{1}{3}\,
\left(v\!\!\!/\,\gamma_\mu\,v_\nu
     +v_\mu\,\gamma_\nu\,v\!\!\!/\right),
\end{equation}
whereas the remaining ``heavy'' parts are integrated out. For further details
cf. Ref.~\cite{HHK}. The chiral tensors in Eqs.~(\ref{eq:LD1}-\ref{eq:LND2}) 
are
\begin{align}
D_\mu^{ij}   &=
\partial_\mu\,\delta^{ij}-i\,\frac{e}{2}\,(1+\tau_3)\,A_\mu\,\delta^{ij}
+e\,\epsilon^{i3j}\,A_\mu+\dots\,,\\
w_\mu^i      &=
-\frac{1}{f_\pi}\,\partial_\mu\,\pi^i
-\frac{e}{f_\pi}\,A_\mu\,\epsilon^{i3j}\,\pi^j+\dots\,,\\
f^i_{+\mu\nu}&=
e\,\delta^{i3}\,\left(\partial_\mu\,A_\nu-\partial_\nu\,A_\mu\right)+\dots\;.
\end{align}
The two coupling constants in Eqs.~(\ref{eq:LND1}) and~(\ref{eq:LND2}) are 
$g_{\pi N\Delta}$, which parameterizes the leading-order $\pi N\Delta$ 
coupling, and $b_1$, the leading-order constant of the 
$\Delta\rightarrow N\gamma$ transition.

In the original formulation of SSE, the only contributions
to a leading-one-loop order calculation of nucleon Compton scattering come from
the Lagrangeans given in Section~\ref{sec:HBchiPT} and 
Eqs.~(\ref{eq:LD1}-\ref{eq:LND2}). However, we found that 
including only these tools we are missing strong diamagnetic contributions,
which are necessary to render the magnetic dipole polarizability a small 
number, as found in experiments such as~\cite{Olmos, Kossert}.
In fact, using only the above Lagrangeans, 
$\bar{\beta}_{M1}$ turns out to be of the same order of magnitude as
the electric dipole polarizability $\bar{\alpha}_{E1}$, in clear contradiction
to experiment. Therefore, we include two additional  
$\gamma\gamma NN$ couplings $g_{1},\,g_{2}$ 
\cite{Fettes}\footnote{The coupling constants 
$g_1$ and $g_2$ correspond to $g_{117}$ and $g_{118}$ in \cite{Fettes}.}
in the leading-one-loop SSE analysis,
which are formally of higher order, but will turn out to be anomalously large.
The corresponding short-distance 
Lagrangeans are
\begin{align}
\mathcal{L}_1^{sd}&=
\frac{2\,g_{1}}{(4\pi \,f_\pi)^2\,m_N}\,\bar{N}\,v^\mu \,v^\nu 
\left<\tilde{f}_{\lambda\mu}\,\tilde{f}^{\lambda}_\nu\right>N,
\label{eq:LCT1} \\
\mathcal{L}_2^{sd}&=
\frac{2\,g_{2}}{(4\pi \,f_\pi)^2\,m_N}\,\bar{N} 
\left<\tilde{f}_{\mu\nu}\,\tilde{f}^{\mu\nu}\right>N
\label{eq:LCT2}
\end{align}
with the electromagnetic field-strength tensor 
$\tilde{f}^{\mu\nu}=
\frac{e}{2}\,\tau_3\,\left(\partial^\mu A^\nu-\partial^\nu A^\mu\right)$
\cite{BKM}. 
To promote these two
structures to leading-one-loop order modifies the power counting, as they are
formally part of the next-to-leading one-loop order chiral
Lagrangean \cite{Fettes}. However, we will see in the next chapter that due to
their unnaturally strong contributions to nucleon Compton scattering 
we cannot avoid this modification.

Now that we have fixed~-- albeit very briefly~-- the theoretical framework
of our calculation, we turn to the first main part of this work: Elastic, 
spin-averaged Compton scattering from the single nucleon.

\chapter{Unpolarized Compton Scattering and Nucleon Polarizabilities 
\label{chap:spinaveraged} }
\markboth{CHAPTER \ref{chap:spinaveraged}. 
COMPTON SCATTERING AND POLARIZABILITIES}
{}
We turn now to the first main part of this work: Compton scattering on the 
single nucleon. While we discuss both proton and neutron Compton scattering, 
we are aware of the fact that experiments using single-neutron targets are 
not feasible due to the finite life-time of the free neutron. Therefore 
the main focus will be on proton Compton cross sections, where a large amount 
of data exists (Section~\ref{sec:spin-averaged}). We also 
discuss in detail the response of the internal degrees of freedom of the 
nucleon to an external electromagnetic field, represented e.g. by 
a photon that is scattered on the nucleon. This reaction is described in 
Section~\ref{sec:dynpolas} in terms of  dynamical nucleon polarizabilities, 
which show a characteristic dependence on the photon energy.
These polarizabilities are defined via a multipole expansion 
in Compton scattering, derived in Section~\ref{sec:multipoleexpansion}. 
One of the advantages 
is that the inclusion of only a few multipoles is often sufficient to obtain 
a good approximation of the full calculation. 
Only when such a reduction of the parameters can be achieved, there is hope 
that the unknown structure can be fitted to experimental data. We show
that our full calculation is nearly indistinguishable from an 
approximation which includes only the six dynamical dipole polarizabilities.
Therefore we conclude that determining the dipole polarizabilities of the 
nucleon is possible, combining spin-averaged cross sections
(Chapter~\ref{chap:spinaveraged}) and polarized ones
(Chapter~\ref{chap:spinpolarized}).

The results of this chapter are published in our
Ref.~\cite{HGHP} and partly in my diploma thesis, Ref.~\cite{DA}.
Throughout the chapter we will indicate new results with respect to \cite{DA}.

\newpage

\section[Multipole Expansion for Nucleon Compton Scattering]
{Multipole Expansion for Nucleon Compton\\Scattering 
\label{sec:multipoleexpansion}}
\markboth{CHAPTER \ref{chap:spinaveraged}. COMPTON SCATTERING AND 
POLARIZABILITIES}{\ref{sec:multipoleexpansion}. 
MULTIPOLE EXPANSION FOR NUCLEON COMPTON SCATTERING}

\subsection{From Amplitudes to Multipoles \label{sec:amplitudestomultipoles}}

The $T$-matrix for real Compton scattering off the nucleon is written in terms
of six structure amplitudes $A_i(\w,z),\;i=1,\dots,6$:
\begin{align}
T_{fi}(\omega,z)&=   
 A_1(\omega,z)\,\vec{\epsilon}\,'\cdot         \vec{\epsilon}
+A_2(\omega,z)\,\vec{\epsilon}\,'\cdot\hat{k}\,\vec{\epsilon}\cdot\hat{k}'
\nonumber\\
&+i\,A_3(\omega,z)\,\vec{\sigma}\cdot(\vec{\epsilon}\,'\times\vec{\epsilon}\,)
 +i\,A_4(\omega,z)\,\vec{\sigma}\cdot(\hat{k}'\times\hat{k})\,
\vec{\epsilon}\,'\cdot\vec{\epsilon}\nonumber\\ 
&+i\,A_5(\omega,z)\,\vec{\sigma}\cdot
  \left[(\vec{\epsilon}\,'\times\hat{k} )\,\vec{\epsilon}   \cdot\hat{k}'
       -(\vec{\epsilon}   \times\hat{k}')\,\vec{\epsilon}\,'\cdot\hat{k}\right]
\nonumber\\
&+i\,A_6(\omega,z)\,\vec{\sigma}\cdot
  \left[(\vec{\epsilon}\,'\times\hat{k}')\,\vec{\epsilon}   \cdot\hat{k}'
       -(\vec{\epsilon}   \times\hat{k} )\,\vec{\epsilon}\,'\cdot\hat{k}\right]
\label{eq:Tfinucleon}
\end{align}
For the Compton multipole expansion, we follow Ritus et al.~\cite{Ritus} 
and work in the center-of-mass (cm) frame, i.e.
$\w$ denotes the cm energy of a real photon scattering under the cm angle 
$\theta$ ($z=\cos\theta$) off a nucleon. 
$\vec{\sigma}$ is the vector of the Pauli spin matrices, 
 $\hat{\vec{k}}_i=\vec{k}_i/\w$
($\hat{\vec{k}}_f=\vec{k}_f/\w$) denotes the unit vector in the momentum 
direction of the incoming (outgoing) photon with polarization
$\epsvec$ ($\epspvec$).

Expanding the Compton-scattering amplitude into multipoles has a long 
tradition~\cite{Ritus}, albeit then it was common to use a slightly different 
basis. The connection between the amplitudes used in~\cite{Ritus} and written 
in Eq.~(\ref{eq:ritus}), and those of Eq.~(\ref{eq:Tfinucleon}), reads
\begin{align}
A_1&= \frac{4\pi\,W}{m_N}( R_1+z R_2)            ,\nonumber\\
A_2&= \frac{4\pi\,W}{m_N}(-R_2)                  ,\nonumber\\
A_3&= \frac{4\pi\,W}{m_N}( R_3+z R_4+2z R_5+2R_6),\nonumber\\
A_4&= \frac{4\pi\,W}{m_N}( R_4)                  ,\nonumber\\
A_5&= \frac{4\pi\,W}{m_N}(-R_4-  R_5)            ,\nonumber\\
A_6&= \frac{4\pi\,W}{m_N}(-R_6)                  
\label{eq:Ritus}
\end{align}
with $W=\sqrt{s}=\w+\sqrt{m_N^2+\w^2}$ denoting the total cm energy and $m_N$  
the nucleon mass.

\begin{figure}[!htb]
\begin{center} 
\includegraphics*[width=.65\textwidth]{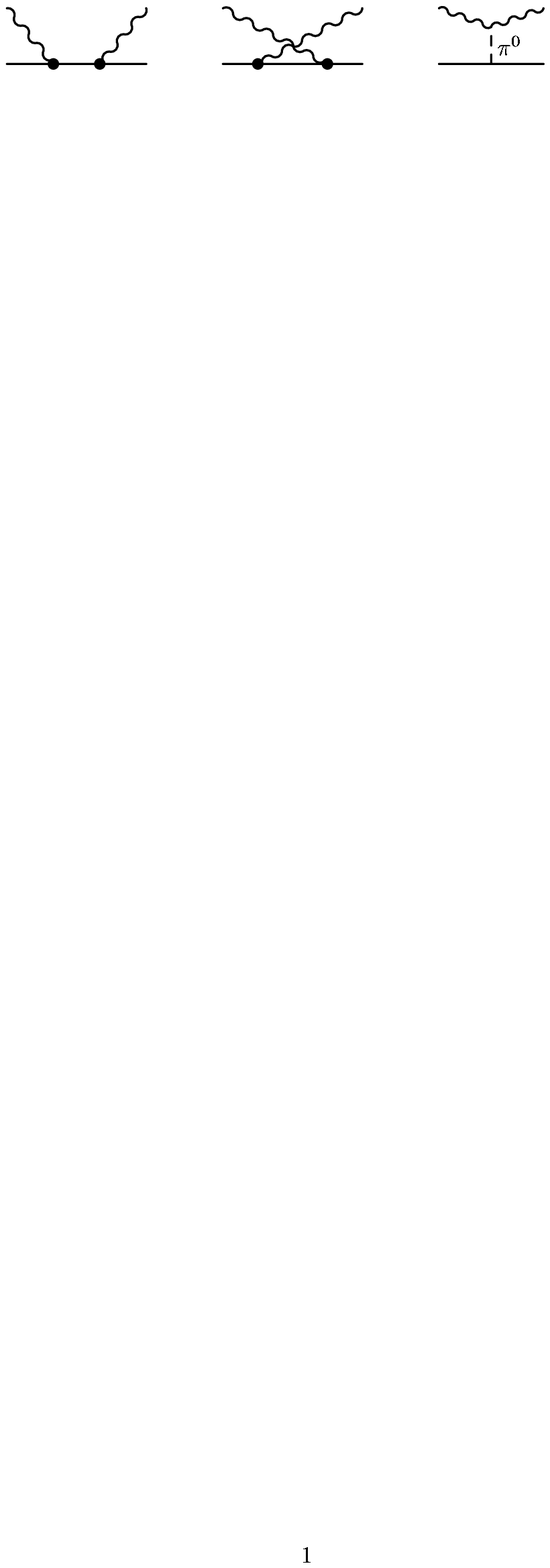}
\caption[Schematic representation of the three types of pole contributions]
{Schematic representation of the three types of pole contributions to 
nucleon Compton scattering in the $s$-, $u$- and $t$-channel (left to right).}
\label{fig:pole}
\end{center}
\end{figure}

The multipole expansion 
is defined for the complete
Compton amplitude. Nucleon structure effects as for example expressed in
$\bar{\alpha}_{E1}$ and $\bar{\beta}_{M1}$, cf. Section~\ref{sec:intro}, 
involve processes for which the particle content between the interactions of 
the in- and outgoing photon goes beyond the single nucleon.
This corresponds to subtracting from the full amplitudes the Powell amplitudes
\cite{Powell} of Compton scattering on a point-like nucleon of spin
$\frac{1}{2}$ and anomalous magnetic moment $\kappa$. Therefore, we separate
the six amplitudes into structure-independent (pole) and structure-dependent
(non-pole) parts,
\be
R_i(\w,z)=R_i^{\mathrm{pole}}(\w,z)+\bar{R}_i(\w,z). 
\ee
We specify the pole contributions as those terms which have a
nucleon pole in the $s$- or $u$-channel and {\em in addition} as terms
with a pion pole in the $t$-channel. Schematically, we show these three
contributions in Fig.~\ref{fig:pole}
and note that any theoretical framework utilized to calculate Compton
scattering has to clearly state the separation of these pole contributions 
before any information on static or dynamical polarizabilities can be
obtained. Obviously, any Compton observable 
is independent of the choice of separating the amplitudes.

As the pole contributions to nucleon Compton scattering and the resulting 
low-energy theorems have been known for 
many decades \cite{Low}, the main interest in Compton studies over the past few
years focused on the non-pole contributions $\bar{R}_i$. 
Ref.~\cite{GH1} suggested that the Compton multipole expansion should
be applied only to these structure-dependent terms. In analogy to
Ref.~\cite{Ritus}, one obtains \cite{Babusci}\footnote{
We correct here the factors of 2 in front of $f_{ME}^{l+}$ ($f_{EM}^{l+}$) 
in $R_5$ ($R_6$), which appear in Ref.~\cite{Babusci}.}
\ba\label{eq:ritus}
\bar{R}_1(\w,z)&=\sum_{l=1}^\infty\bigg\{
  \left[(l+1)\,f_{EE}^{l+}(\w)+l\,f_{EE}^{l-}(\w)\right]\,
  \left(l\,P_l'(z)+ P_{l-1}''(z)\right)\nonumber\\
&-\left[(l+1)\,f_{MM}^{l+}(\w)+l\,f_{MM}^{l-}(\w)\right]\,P_l''(z)\bigg\},
\nonumber\\
\bar{R}_2(\w,z)&=\sum_{l=1}^\infty\bigg\{
  \left[(l+1)\,f_{MM}^{l+}(\w)+l\,f_{MM}^{l-}(\w)\right]\,
  \left(l\,P_l'(z)+P_{l-1}''(z)\right)\nonumber\\
&-\left[(l+1)\,f_{EE}^{l+}(\w)+l\,f_{EE}^{l-}(\w)\right]\,P_l''(z)\bigg\},
\nonumber\\
\bar{R}_3(\w,z)&=\sum_{l=1}^\infty\bigg\{
  \left[       f_{EE}^{l+}(\w)-   f_{EE}^{l-}(\w)\right]\,
  \left(P_{l-1}''(z)-l^2\,P_l'(z)\right)\nonumber\\
&-\left[       f_{MM}^{l+}(\w)-   f_{MM}^{l-}(\w)\right]\,P_l''(z)
  +2f_{EM}^{l+}(\w)\,P_{l+1}''(z)-2f_{ME}^{l+}(\w)\,P_l''(z)\bigg\},\nonumber\\
\bar{R}_4(\w,z)&=\sum_{l=1}^\infty\bigg\{
  \left[       f_{MM}^{l+}(\w)-   f_{MM}^{l-}(\w)\right]\,
  \left(P_{l-1}''(z)-l^2\,P_l'(z)\right)\nonumber\\
&-\left[       f_{EE}^{l+}(\w)-   f_{EE}^{l-}(\w)\right]\,P_l''(z)
  +2f_{ME}^{l+}(\w)\,P_{l+1}''(z)-2f_{EM}^{l+}(\w)\,P_l''(z)\bigg\},\nonumber\\
\bar{R}_5(\w,z)&=\sum_{l=1}^\infty\bigg\{
  \left[       f_{EE}^{l+}(\w)-   f_{EE}^{l-}(\w)\right]\,
  \left(l\,P_l''(z)+P_{l-1}'''(z)\right)
 -\left[       f_{MM}^{l+}(\w)-   f_{MM}^{l-}(\w)\right]\,P_l'''(z)\nonumber\\ 
& +f_{EM}^{l+}(\w)\,\left[(3l+1)\,P_l    ''(z)+2P_{l-1}'''(z)\right]
  -f_{ME}^{l+}(\w)\,\left[( l+1)\,P_{l+1}''(z)+2P_l    '''(z)\right]\bigg\},
\nonumber\\
\bar{R}_6(\w,z)&=\sum_{l=1}^\infty\bigg\{
  \left[       f_{MM}^{l+}(\w)-   f_{MM}^{l-}(\w)\right]\,
  \left(l\,P_l''(z)+P_{l-1}'''(z)\right)
 -\left[       f_{EE}^{l+}(\w)-   f_{EE}^{l-}(\w)\right]\,P_l'''(z)\nonumber\\
& +f_{ME}^{l+}(\w)\,\left[(3l+1)\,P_l    ''(z)+2P_{l-1}'''(z)\right] 
  -f_{EM}^{l+}(\w)\,\left[( l+1)\,P_{l+1}''(z)+2P_l    '''(z)\right]\bigg\}.
\end{align}
The prime denotes differentiation with respect to $z=\cos\theta$ in the cm
system, and $P_l(z)$ is the $l$th Legendre polynomial. The functions
$f_{TT'}^{l\pm}(\w)$ are the Compton multipoles and correspond to
transitions $Tl\rightarrow T'l'$, where $T,T'=E,M$ labels
the coupling of the incoming or outgoing photon as electric or magnetic. Here
$l$ ($l'=l\pm\{1,\,0\}$) denotes the angular momentum of the initial
(final) photon, whereas the total angular momentum is 
$j_\pm=l\pm\frac{1}{2}$, abbreviated by $l\pm$.  
We note that mixed multipole amplitudes $T\not=T'$ 
only occur in the spin-dependent amplitudes $\bar{R}_i,\,i=3,\ldots, 6$.

Having defined the structure-dependent Compton multipoles in the cm 
frame, we now move on to connect them to polarizabilities.

\subsection{Dynamical and Static Polarizabilities 
\label{sec:polarizabilities1}}

In order to derive a consistent connection between the Compton multipoles
$f_{TT'}$
and the polarizabilities of definite spin
structure and multipolarity at a certain energy, we recall the low-energy
behavior of the multipoles in the cm frame \cite{Ritus}:
\ba
f_{TT'}^{l\pm}(\w)&\sim\w^{2l  },\qquad\;\;\;\, T=    T',\nonumber\\
f_{TT'}^{l+  }(\w)&\sim\w^{2l+1},\qquad T\neq T'.&
\end{align}
With this information, dynamical spin-independent electric or magnetic dipole
and quadrupole polarizabilities were defined as linear combinations of Compton
multipoles in \cite{GH1}:
\ba
\alpha _{E1}(\w)&=  \left[2f_{EE}^{1+}(\w)+ f_{EE}^{1-}(\w)\right]/\w^2
\nonumber\\
\beta  _{M1}(\w)&=  \left[2f_{MM}^{1+}(\w)+ f_{MM}^{1-}(\w)\right]/\w^2
\nonumber\\
\alpha _{E2}(\w)&=36\left[3f_{EE}^{2+}(\w)+2f_{EE}^{2-}(\w)\right]/\w^4
\nonumber\\
\beta  _{M2}(\w)&=36\left[3f_{MM}^{2+}(\w)+2f_{MM}^{2-}(\w)\right]/\w^4
\label{eq:spinindiepolasdef}
\end{align}
We note that the normalization of these linear superpositions has been chosen
in such a way that the usual (static) electric and magnetic polarizabilities
of the nucleon 
can be recovered from the dynamical dipole polarizabilities via
\be
\bar{\alpha}_{E1}=\lim_{\w\rightarrow 0}\alpha_{E1}(\w),\qquad\qquad
\bar{\beta}_{M1} =\lim_{\w\rightarrow 0}\beta_ {M1}(\w).
\ee
Likewise, the static electric and magnetic quadrupole polarizabilities
$\bar{\alpha}_{E2}$ and $\bar{\beta}_{M2}$ discussed in Refs.~\cite{Babusci} 
and \cite{Holst} can be obtained as the zero-energy
limit of the corresponding dynamical quadrupole polarizabilities.

Extending Ref.~\cite{GH1}, we also introduce dynamical {\em spin-dependent}
dipole polarizabilities\footnote{The definition of the dynamical spin 
quadrupole polarizabilities can be found in Ref.~\cite{DA}.} via
\begin{align}
\gamma _{E1E1}(\w)&=\left[f_{EE}^{1+}(\w)-f_{EE}^{1-}(\w)\right]/\w^3,
\nonumber\\
\gamma _{M1M1}(\w)&=\left[f_{MM}^{1+}(\w)-f_{MM}^{1-}(\w)\right]/\w^3,
\nonumber\\
\gamma _{E1M2}(\w)&=6\,   f_{EM}^{1+}(\w)                       /\w^3,
\nonumber\\
\gamma _{M1E2}(\w)&=6\,   f_{ME}^{1+}(\w)                       /\w^3.
\label{eq:spinpolasdef}
\end{align}
The notation is such that $\gamma_{TlT'l'}$ parameterizes a $Tl$ ($T'l'$) 
transition at the first (second) photon vertex.
In the limit of zero photon energy, one again recovers
the four static spin polarizabilities $\bar{\gamma}_{E1E1}$,
$\bar{\gamma}_{M1M1}$, $\bar{\gamma}_{E1M2}$, $\bar{\gamma}_{M1E2}$ of the
nucleon:
\be
\bar{\gamma}_{TlT'l'}=
\lim_{\w\rightarrow 0}\gamma_{TlT'l'}(\w),\qquad T,T'=E,M.
\ee
Here, these four static spin polarizabilities are written in the so called
multipole-basis \cite{Babusci}. The connection to the Ragusa-basis
$\gamma_i,\,i=1,\ldots, 4$ \cite{Ragusa}, is discussed in Ref.~\cite{Hem01}. We
note that at present there exists little information on the spin-dependent
nucleon polarizabilities. Only two linear combinations~-- typically denoted 
as the forward $\bar{\gamma}_0$ and the backward $\bar{\gamma}_\pi$ spin 
polarizabilities of the nucleon~-- are constrained from
experiments, e.g. see the discussion in Ref.~\cite{review}. 
Both quantities involve all four (dipole) spin polarizabilities:
\ba
\bar{\gamma}_0  &=-\bar{\gamma}_{E1E1}-\bar{\gamma}_{E1M2}
                  -\bar{\gamma}_{M1M1}-\bar{\gamma}_{M1E2}\nonumber\\
\bar{\gamma}_\pi&=-\bar{\gamma}_{E1E1}-\bar{\gamma}_{E1M2}
                  +\bar{\gamma}_{M1M1}+\bar{\gamma}_{M1E2}
\label{eq:gpi}
\end{align}

While the static polarizabilities of the nucleon are real, we note that the
dynamical polarizabilities become complex once the energy in the intermediate
state is high enough to create an on-shell intermediate state, the first being
the physical $\pi N$ intermediate state. Below the two-pion-production 
threshold, the imaginary parts of the dynamical polarizabilities can be 
understood very easily. They are simply given by the well-known multipoles of 
single-pion photoproduction (see e.g.~\cite{HDT}). One obtains~\cite{HGHP}
\be
\label{eq:im_dyn_pol} 
\parbox{4.cm}{
\ba
&\mathrm{Im}[\alpha_{E1}]  =    \frac{k_{\pi}}{\w^2}
\sum_{\sss{C}}(2|E^{\sss{(C)}}_{2-}|^2+ |E^{\sss{(C)}}_{0+}|^2),\nonumber\\
&\mathrm{Im}[\alpha_{E2}]  =36\,\frac{k_{\pi}}{\w^4} 
\sum_{\sss{C}}(3|E^{\sss{(C)}}_{3-}|^2+2|E^{\sss{(C)}}_{1+}|^2),\nonumber\\
&\mathrm{Im}[\gamma_{E1E1}]=    \frac{k_{\pi}}{\w^3} 
\sum_{\sss{C}}(|E^{\sss{(C)}}_{2-}|^2-|E^{\sss{(C)}}_{0+}|^2),\nonumber\\
&\mathrm{Im}[\gamma_{E1M2}]= 6\,\frac{k_{\pi}}{\w^3} 
\sum_{\sss{C}}\mathrm{Re}[E^{\sss{(C)}}_{2-}(M^{\sss{(C)}}_{2-})^*],\nonumber
\end{align}}
\quad
\parbox{4cm}{
\ba
&\mathrm{Im}[\beta_ {M1}]  =    \frac{k_{\pi}}{\w^2}
\sum_{\sss{C}}(2|M^{\sss{(C)}}_{1+}|^2+ |M^{\sss{(C)}}_{1-}|^2),\nonumber\\
&\mathrm{Im}[\beta_ {M2}]  =36\,\frac{k_{\pi}}{\w^4} 
\sum_{\sss{C}}(3|M^{\sss{(C)}}_{2+}|^2+2|M^{\sss{(C)}}_{2-}|^2),\nonumber\\
&\mathrm{Im}[\gamma_{M1M1}]=    \frac{k_{\pi}}{\w^3} 
\sum_{\sss{C}}(|M^{\sss{(C)}}_{1+}|^2-|M^{\sss{(C)}}_{1-}|^2),\nonumber\\
&\mathrm{Im}[\gamma_{M1E2}]=-6\,\frac{k_{\pi}}{\w^3} 
\sum_{\sss{C}}\mathrm{Re}[E^{\sss{(C)}}_{1+}(M^{\sss{(C)}}_{1+})^*],\nonumber
\end{align}}
\ee
where $k_{\pi}$ is the magnitude of the pion momentum and 
$E_{l\pm}^{{\sss{(C)}}}$ and $M_{l\pm}^{{\sss{(C})}}$ are the 
pion-photoproduction multipoles which are summed over the
different isotopic or charge channels $C$. In the following, we therefore
focus only on the real parts of the dynamical polarizabilities. 
The imaginary parts of our calculation can be found in Ref.~\cite{DA}.

This concludes our section pertaining to the definitions of the dynamical
polarizabilities and their connection to static polarizabilities as well as to
single-pion photoproduction. Before we discuss the numerical values of the
(static) polarizabilities in the upcoming section, we first provide some
background on the theoretical machinery employed to analyze nucleon Compton
scattering.

\section{Theoretical Framework \label{sec:spinaveragedtheory}}

Many calculations of nucleon Compton scattering~-- some even up to
next-to-leading one-loop order~-- have been performed using Chiral Effective 
Field Theory ($\chi$EFT) during the
past decade \cite{BKM,deltaexp,CD97,CD00,McGovern}. 
Here, we extract information on the dynamical polarizabilities of the 
nucleon both from the leading-one-loop
Heavy Baryon Chiral Perturbation
Theory (HB$\chi$PT) calculation of Ref.~\cite{BKM} as well as from the
leading-one-loop ``Small Scale Expansion" (SSE) calculations of
Refs.~\cite{HHK97,HHKK}. We remind the reader of Chapter~\ref{chap:theory}, 
where we introduced the HB$\chi$PT as well as the SSE formalism. The first 
one only involves explicit
$\pi N$ degrees of freedom, whereas the latter is one possibility to
also systematically include explicit spin-3/2 nucleon-resonance degrees of
freedom, i.e. the $\Delta$(1232), in $\chi$EFT. 

\begin{figure}[!htb]
\begin{center}
\includegraphics*[width=.8\textwidth]{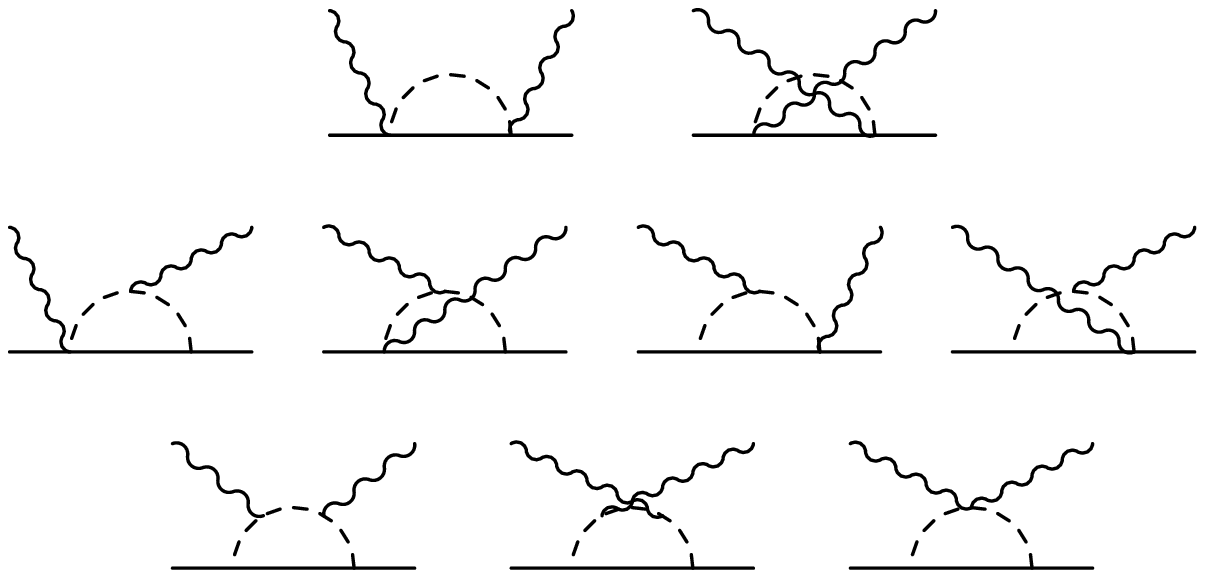}
\caption[$N\pi$-continuum contributions to nucleon 
polarizabilities]
{Leading-one-loop $N\pi$-continuum contributions to nucleon polarizabilities.}
\label{fig:Npicontinuum}
\end{center}
\end{figure}

\begin{figure}[!htb]
\begin{center}
\includegraphics*[width=.8\textwidth]{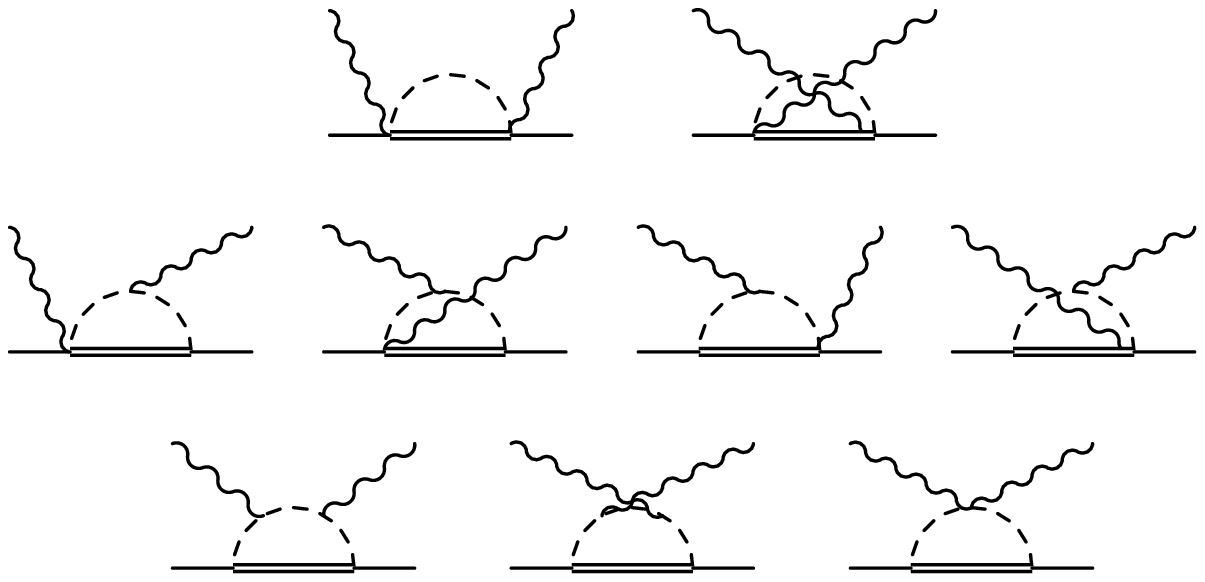}
\caption[$\Delta\pi$-continuum contributions to nucleon 
polarizabilities]{Leading-one-loop $\Delta\pi$-continuum contributions to 
nucleon polarizabilities.}
\label{fig:Deltapicontinuum}  
\end{center}
\end{figure}

\begin{figure}[!htb]
\begin{center}
\includegraphics*[width=.6\textwidth]{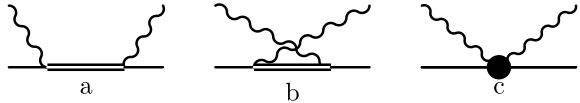}
\caption[$\Delta$-pole and short-distance contributions to nucleon 
polarizabilities]
{$\Delta$-pole and short-distance contributions to nucleon polarizabilities.} 
\label{fig:Deltapolediagrams}
\end{center}
\end{figure}

The pole contributions to nucleon Compton scattering at
leading-one-loop order in $\chi$EFT are given in Appendix~\ref{app:poleterms}.
As discussed in Section~\ref{sec:multipoleexpansion}, it is the non-pole 
contribution to Compton scattering which determines the
polarizabilities.  In HB$\chi$PT, these structure-dependent contributions are
solely given by $\pi N$ intermediate states (Fig.~\ref{fig:Npicontinuum}), 
whereas SSE takes into account in addition $\pi\Delta$ diagrams
(Fig.~\ref{fig:Deltapicontinuum}) as well as the $\Delta$(1232) $s$- and 
$u$-channel pole terms (Fig.~\ref{fig:Deltapolediagrams} (a),(b)).

We note that we go beyond the existing leading-one-loop HB$\chi$PT/SSE
calculations \cite{BKM,HHK97,HHKK} of nucleon Compton scattering in four
aspects:
\begin{itemize}
\item[1)] Both HB$\chi$PT and SSE are non-relativistic frameworks leading to a
$1/m_N$ expansion of the amplitudes, where $m_N$ is the mass of the
target nucleon. In the leading-one-loop structure amplitudes $\bar{R}_i$,
the one-pion-production threshold
\be
\w_\pi=\frac{m_\pi^2+2\,m_\pi\,m_N}{2\,(m_\pi+m_N)}\approx 131\;\MeV
\ee
is therefore not at the correct location. We correct for this purely
kinematical effect by replacing the photon energy $\w$ with the Mandelstam
variable $s$ via
\begin{equation}
\w\rightarrow\sqrt{s(\w)}-m_N.
\label{eq:substitut}
\end{equation}
There are various possibilities to perform such a threshold 
correction (cf. Ref.~\cite{BKMS}), but it is clear that all these 
choices agree within the strict truncation of the $\chi$EFT employed.
Obviously this replacement should only be applied in those
places where an imaginary part arises above threshold. In contrast to 
Ref.~\cite{DA}, the $u$-channel diagrams
are left unchanged. We are aware that this procedure violates crossing 
symmetry, but the crossing-violating effects in the $u$-channel are 
small. Formally, the terms correcting for the exact location of the pion 
threshold start to appear at $\mathcal{O}(p^4)$.
We implement this kinematical correction in the leading-one-loop 
$\pi N$-continuum contribution to the $\chi$EFT amplitudes throughout all 
chapters of this work on single-nucleon Compton scattering. 
Such kinematical corrections should be employed in non-relativistic 
$\chi$EFTs at all particle thresholds. However, as our calculation is
valid only below the $\Delta$(1232) resonance, the one-pion-production 
threshold is the only one to be taken care of.
The formulae for the SSE amplitudes 
are given in Refs.~\cite{HGHP} and~\cite{DA}, albeit in Ref.~\cite{DA} we also
corrected the $u$-channel, in analogy to Eq.~(\ref{eq:substitut}). 

\item[2)] Another kinematical effect concerns the exact location of the
$\Delta$(1232) pole. In Ref.~\cite{HDT}, it was determined as a $T$-matrix
pole in the complex $W=\sqrt{s}$ plane at the location
$m_\Delta=(1210-\,i\,50)$~MeV. We therefore employ the same substitution 
prescription $\w\rightarrow \sqrt{s}-m_N$ as in Eq.~(\ref{eq:substitut}) in 
$s$-channel pole contributions of the $\Delta$(1232) resonance, generating a 
pole at $\sqrt{s}=m_N+\Delta_0=1210$~MeV. Given that 
$\Delta$(1232)-pole contributions in the $u$-channel can also affect higher 
multipoles, we make an analogous replacement $\w\rightarrow m_N-\sqrt{u}$ in 
the $\Delta$(1232) $u$-channel-pole contributions and note that the 
corrections of the $\Delta$ pole are new with respect to Ref.~\cite{DA}. 
While these kinematical
details are of minor importance when one only discusses static
polarizabilities (with the exception of $\bar{\beta}_{M2}$, see
Section~\ref{sec:staticpolas}), they do become important in dynamical
polarizabilities once the photon energy is higher than 100 MeV. 
We note again that via these modifications, we
have not introduced any additional physics content into the $\chi$EFT
calculations, as in the $m_N\rightarrow\infty$ limit all these purely
kinematical modifications reduce to the strict ${\cal O}(\epsilon^3)$
truncation of SSE \cite{HHK97,HHKK}.  The detailed form of the modified
amplitudes can be found in Appendix~B of our Ref.~\cite{HGHP}.
  
\item[3)] The parameters required for the leading-one-loop HB$\chi$PT
calculation are well known. For completeness, we list them in 
Table~\ref{tab:const}. 
Also shown are the two parameters $\Delta_0$ and 
$g_{\pi N\Delta}$ utilized in the leading-one-loop SSE Compton-scattering
calculation of Refs.~\cite{HHK97,HHKK}. The numbers given here differ
slightly from Ref.~\cite{HHK97}, as we determine them now from the exact
kinematical location of the $\Delta$(1232)-pole in the complex $W$-plane,
discussed in the previous paragraph.
  
To leading-one-loop order, the HB$\chi$PT calculation for nucleon Compton
scattering is therefore parameter-free (in the sense that all parameters
shown in Table~\ref{tab:const} can be determined outside Compton
scattering). On the other hand, in the corresponding SSE calculation we are
left with one free parameter $b_1$~-- which in $\chi$EFT corresponds to the
leading $\gamma N\Delta$ coupling \cite{HHKLett,HHK97,review}. In 
Ref.~\cite{HHKK}, $b_1$ was estimated from the measured 
$\Delta\rightarrow N\gamma$ decay
width to be $|b_1|\approx 3.9$. As this determination is very sensitive to
the numerical value of the parameter $\Delta_0$ (for the value
$\Delta_0=271$ MeV shown in Table~\ref{tab:const}, we would obtain
$|b_1|\approx 4.4$), we choose a different strategy here and determine $b_1$
directly from a fit to Compton cross-section data, whereas in Ref.~\cite{DA}
we used the Dispersion-Relation result for 
$\bar{\gamma}_{M1M1}$~\cite{HGHP} to fix this coupling. 

\item[4)] With the $\gamma N\Delta$ coupling constant $b_1$ as a fit parameter
in the SSE analysis, we can constrain the crucial paramagnetic contribution
from the $\Delta$ directly from data. In~\cite{Weise} an estimate of its 
contribution to $\bar{\beta}_{M1}$, based on the $N\rightarrow\Delta$ 
transition matrix element from~\cite{Ericson} gives 
$\beta_\Delta\approx12\cdot10^{-4}\;\fm^3$. This number agrees with the 
result $\beta_\Delta=(13\pm3)\cdot10^{-4}\;\fm^3$ from~\cite{Mukho}.
However, it has been known for a long
time that there must also be substantial diamagnetism in the
nucleon~-- otherwise the small numbers for the static magnetic polarizability
of the proton cannot be understood, see e.g. Ref.~\cite{review} for
details. At leading-one-loop order neither HB$\chi$PT nor SSE in their
respective counting schemes, based on (na\"ive) dimensional analysis, allow 
for such a contribution \cite{HHK97}. Both
schemes assume that this is a ``small'' higher-order effect, which can be
accounted for at the next-to-leading one-loop order. As a side remark we
remind the reader that in Ref.~\cite{BKMS} it was shown in a next-to-leading
one-loop HB$\chi$PT calculation that for ``reasonable'' values of the
regularization scale $\lambda$, a large part of this diamagnetism could be
accounted for by $\pi N$ loop effects. 
Working only to leading-one-loop
order, we cannot contribute to the discussion of the physical nature of this
diamagnetism in the nucleon. However, from our combined analysis of proton and
deuteron Compton data we conclude that dynamics beyond our 
leading-one-loop order calculation indeed strongly contributes to 
$\bar{\alpha}_{E1}$ and $\bar{\beta}_{M1}$. 
We can constrain these contributions to be of isoscalar nature and, 
reminiscent of short-distance dynamics, they are largely energy independent.

As we determine the paramagnetic response of the nucleon from data and as 
there is a well-known delicate interplay between para- and diamagnetic 
contributions at small photon energy, we introduce two additional 
$\calO(p^4)$ $\gamma\gamma NN$ 
couplings $g_1$ and $g_2$, cf. Eqs.~(\ref{eq:LCT1}, \ref{eq:LCT2}). 
If they turned out to give only small corrections, we could safely
neglect them as a higher-order effect in accordance with the counting
assumptions of SSE. However, as will be demonstrated in 
Section~\ref{sec:protonfits}, this is not the case and these two couplings 
have to be included already at leading-one-loop order, modifying the 
na\"ive power counting due to their unnaturally large sizes.
Two independent structures are needed to separate magnetic and 
electric contributions via different
linear combinations of $g_{1}$ and $g_{2}$. Promoting these two
structures from $\calO(\epsilon^4)$ to leading-one-loop order obviously 
modifies the power counting.
Nevertheless, in light of the reasoning given above, we must include them as 
free parameters in our SSE fit to Compton cross sections.  We find that the 
two couplings in Eqs.~(\ref{eq:LCT1}, \ref{eq:LCT2}) are sufficient to 
parameterize any quark-mass independent 
unknown magnetic and electric short-distance physics in 
nucleon Compton scattering (cf. Fig.~\ref{fig:spinindependentpolas}).  
The contributions of $g_{1},\,g_{2}$ to the Compton structure-amplitudes are 
shown explicitly in Appendix~B of our Ref.~\cite{HGHP}\footnote{We correct for
a missing factor of -2 in Ref.~\cite{DA}.}.
\end{itemize}
The leading-one-loop structure-dependent Compton amplitudes 
of Ref.~\cite{HGHP} include the four modifications discussed above.
In order to extract from them the dynamical polarizabilities of the nucleon in
$\chi$EFT frameworks, one first projects out the Compton multipoles
$f_{TT'}(\w)$ of Section~\ref{sec:amplitudestomultipoles}, using the formulae 
in Appendix~\ref{app:projection}. The dynamical polarizabilities at definite 
multipolarity as a function of the photon energy follow then from
Eqs.~(\ref{eq:spinindiepolasdef}, \ref{eq:spinpolasdef}).

This concludes our brief summary of leading-one-loop $\chi$EFT calculations
for nucleon Compton scattering. We now move on to a determination of the three
free parameters $b_1,\,g_{1}$ and $g_{2}$ from cross-section data.

\section{Spin-Averaged Compton Cross Sections \label{sec:spin-averaged}}

\subsection{General Remarks \label{sec:crosssectionsgeneral}}

In the previous section, we have briefly introduced the theoretical framework
of our nucleon Compton-scattering calculation, which we now confront with 
actual proton Compton data.  This will also serve as a check for the parameters
employed (in the case of HB$\chi$PT), respectively allow us to constrain some 
parameters (in the case of
SSE). To be precise, we compare the experimental differential cross sections
with predictions from leading-one-loop HB$\chi$PT, which does not contain any 
additional free parameters to be determined from Compton scattering, and with 
the Dispersion-Relation Analysis from~\cite{HGHP}. This method makes use of 
the optical theorem to deduce the imaginary part of the transition amplitude 
from measured break-up cross sections, in our case $\gamma N\rightarrow X$. 
The real part of the amplitude is derived from the imaginary part via the 
Kramers-Kronig dispersion relations. Although we do not show any error bands 
for the 
Dispersion-Analysis curves in our figures, we must stress that there are 
non-negligible uncertainties also in this framework, which arise e.g. due
to error bars in the input parameters $\bar{\alpha}_{E1}$, $\bar{\beta}_{M1}$
and $\bar{\gamma}_\pi$, cf. Ref.~\cite{HGHP}. For further details on the 
Dispersion-Theory formalism we refer the reader to the literature, e.g. to 
the review given in Ref.~\cite{review}.

In the case of leading-one-loop SSE calculations, we perform a fit of the 
three free 
parameters $b_1,\,g_{1},\,g_{2}$ discussed in the previous section to proton 
Compton data. In this section, we can therefore only check whether the two 
curves from Field Theory are consistent with data and Dispersion Theory. A 
detailed discussion of the electromagnetic structure of the proton 
will be given in Section~\ref{sec:dynpolas}.

So far, only spin-averaged cross sections on the proton have been measured. We
draw from a large set of data \cite{Olmos,Hallin,Fed91,Mac95}, covering
proton Compton scattering from low energies to above  pion-production
threshold. We present the low-energy data as functions of the differential
cross section in the cm system versus the photon energy (in the cm system) at
different angles $\theta_\mathrm{lab}$. Note that in the plots we work
in the cm system when comparing with the SAL data, and in the lab system for 
all other cases.

In the differential Compton-scattering cross sections, the artificial
separation between pole and non-pole contributions is absent and both terms 
have to be added. The differences between lab and cm system are expressed 
via the flux factors
\begin{equation}
\Phi_\mathrm{cm} =\frac{m_N}             {4\pi\,\sqrt{s(\w)}},\qquad
\Phi_\mathrm{lab}=\frac{\w_f}{4\pi\,\w_i}, 
\label{eq:phasespace}
\end{equation}
with
\ba
\w_f&=\frac{m_N\,\w_i}{m_N+\w_i\,(1-\cos\theta_\mathrm{lab})}.
\label{eq:sandwf}
\end{align} 
$\w_f$ ($\w_i$) denote the energy of the
outgoing (incoming) photon in the lab frame, $\w_i$ being related to the 
photon energy in the $\gamma N$ cm frame by
\be
\w=\frac{\w_i}{\sqrt{1+2\w_i/m_N}}.
\label{eq:wirelation}
\ee  
In the spin-averaged case, the differential cross section is then given by
\begin{equation}
\left.\frac{d\sigma}{d\Omega}\right|_\mathrm{frame}=
\Phi_\mathrm{frame}^2\;|T|^2.
\label{eq:diffcrosssection}
\end{equation}
The absolute square of the Compton amplitude, averaged over the initial and 
summed over the final nucleon and photon polarizations, is~\cite{BKM}
\ba
|T|^2&=\frac{1}{2}\left|A_1\right|^2\,\left(1+z^2\right)
      +\frac{1}{2}\left|A_3\right|^2\,\left(3-z^2\right)\nonumber\\
&+\left(1-z^2\right)\,\left[4\,\mathrm{Re}[A^{\ast}_3\,A_6]+\mathrm{Re}
[A_3^{\ast}\,A_4+2A_3^{\ast}\,A_5-A_1^{\ast}\,A_2]\,z
\right]\nonumber\\
&+\left(1-z^2\right)\,
\bigg[ \frac{1}{2}\left|A_2\right|^2\,\left(1-z^2\right)
+\frac{1}{2}\left|A_4\right|^2\,\left(1+z^2\right)
\nonumber\\
&+\left|A_5\right|^2\,\left(1+2z^2\right)+3\left|A_6\right|^2+
2\mathrm{Re}[A_6^{\ast}\,\left(A_4+3A_5\right)]\,z+
2\mathrm{Re}[A_4^{\ast}\,A_5]\,z^2\bigg].
\label{eq:Tmatrix}
\end{align} 
After these general remarks, we now move on to the comparison with experiment.

\subsection[Comparison to Experiments on Proton Compton Scattering]
{Comparison to Experiments on Proton Compton\\Scattering 
\label{sec:protoncrosssections}}

\begin{figure}[!htb]
\begin{center}
\includegraphics*[width=0.48\textwidth]{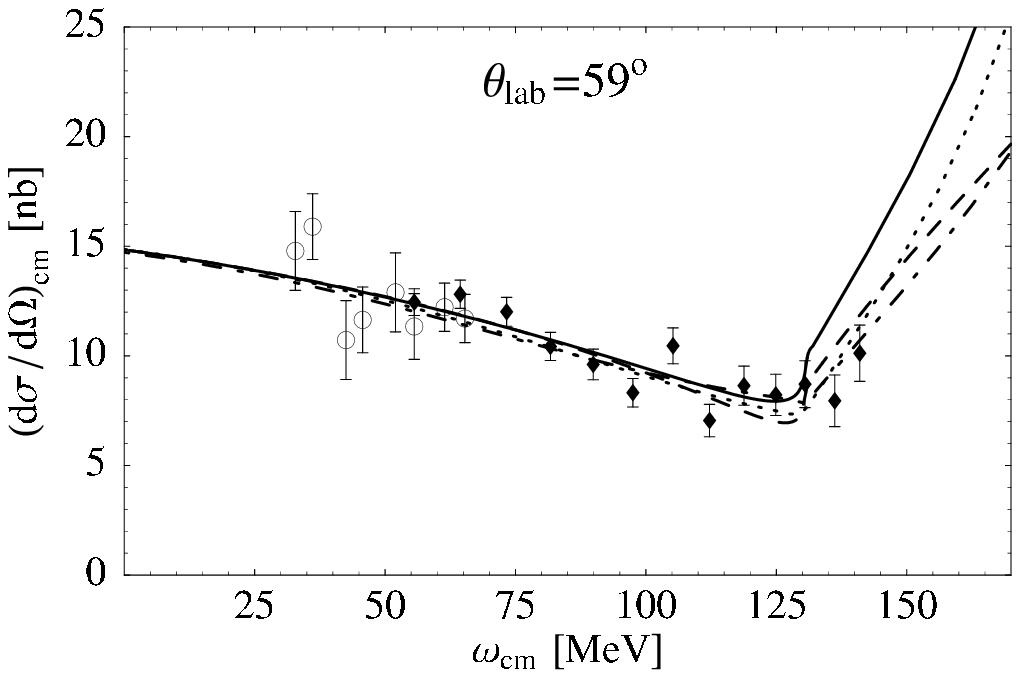}
\hfill
\includegraphics*[width=0.48\textwidth]{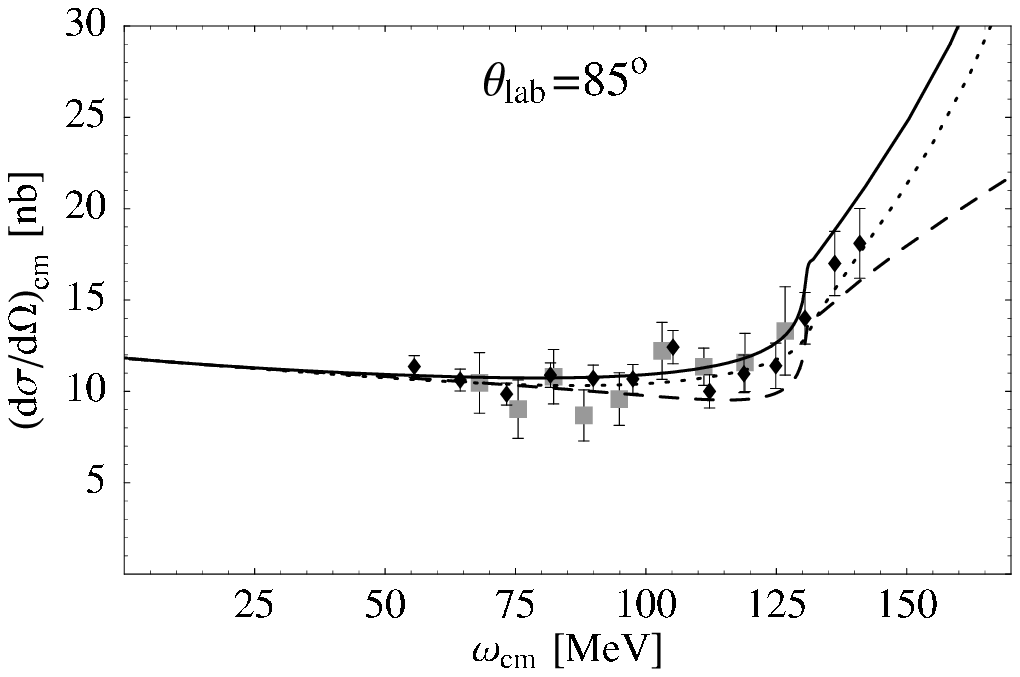}
\includegraphics*[width=0.48\textwidth]{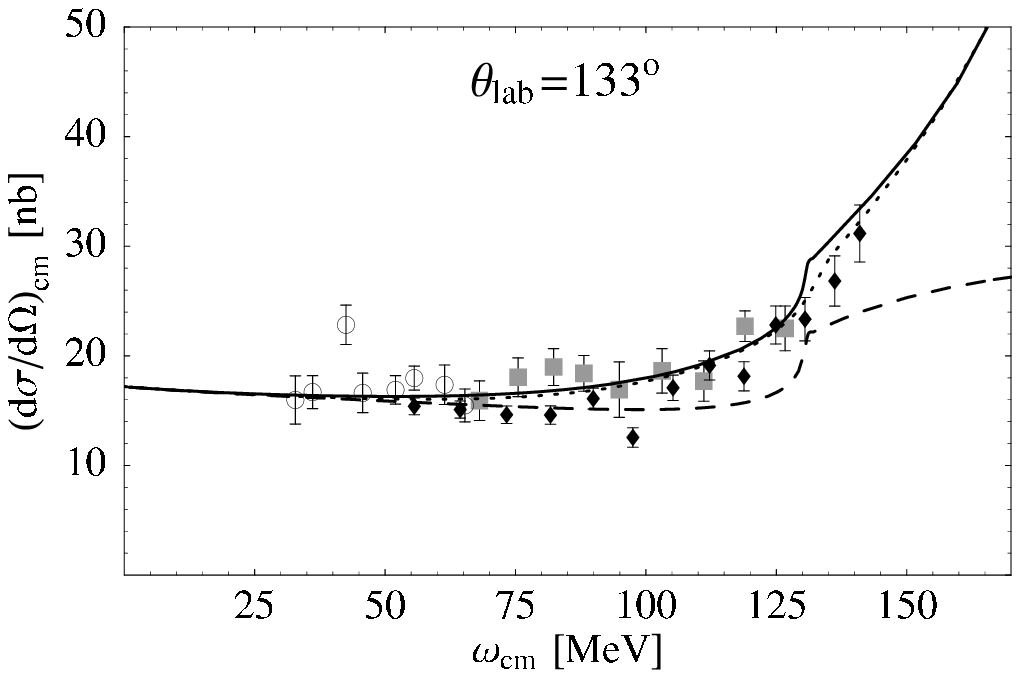}
\hfill
\includegraphics*[width=0.48\textwidth]{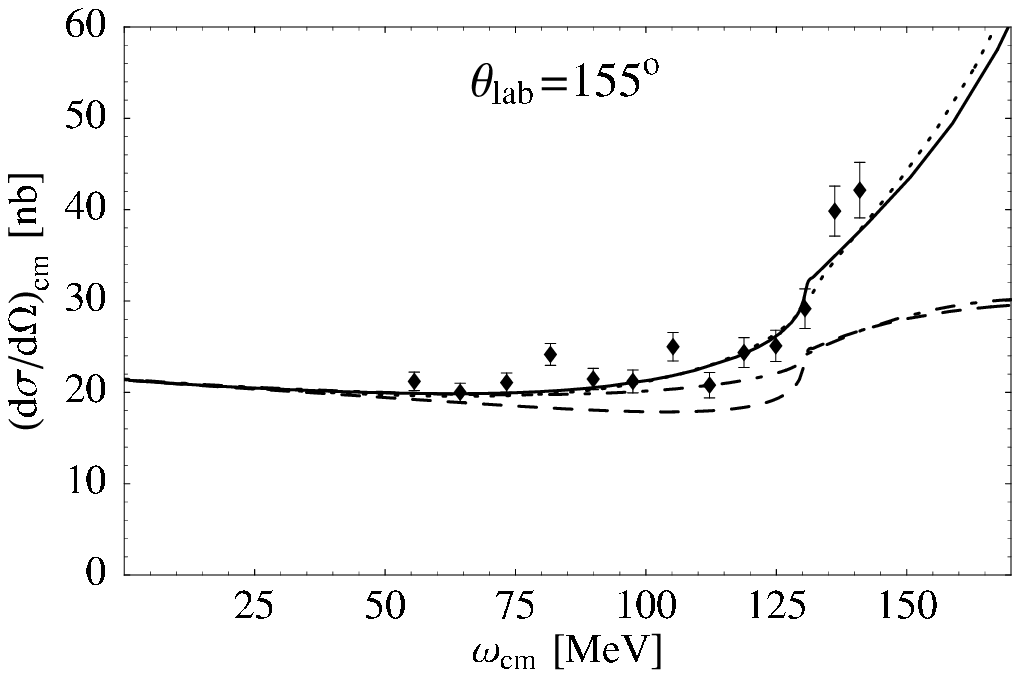}
\caption[Comparison of theoretical results and experimental data 
for spin-averaged Compton scattering off the proton below 150~MeV]
{Comparison of the differential cross-section data for 
Compton scattering off the proton (diamonds: Olmos de Leon et
al.~\cite{Olmos}, circles: Federspiel et al.~\cite{Fed91}, boxes:
MacGibbon et al.~\cite{Mac95}) with Dispersion Theory and leading-one-loop
order HB$\chi$PT respectively SSE at fixed lab angle. At $59^\circ$ and
$155^\circ$ we also compare to the $\calO(p^4)$~HB$\chi$PT calculation of 
\cite{McGovern}. Solid line: SSE results; 
dashed line: $\calO(p^3)$~HB$\chi$PT; dotted line: DR; dotdashed line: 
$\calO(p^4)$~HB$\chi$PT. Note that
the data of~\cite{Fed91} are not given at $59^\circ$ and $133^\circ$ but at
$60^\circ$ and $135^\circ$; the data of~\cite{Mac95} are not given at
$85^\circ$ and $133^\circ$ but at $90^\circ$ and $135^\circ$.}
\label{fig:Olmos}
\end{center}
\end{figure}

\begin{figure}[!htb]
\begin{center}
\includegraphics*[width=0.48\textwidth]{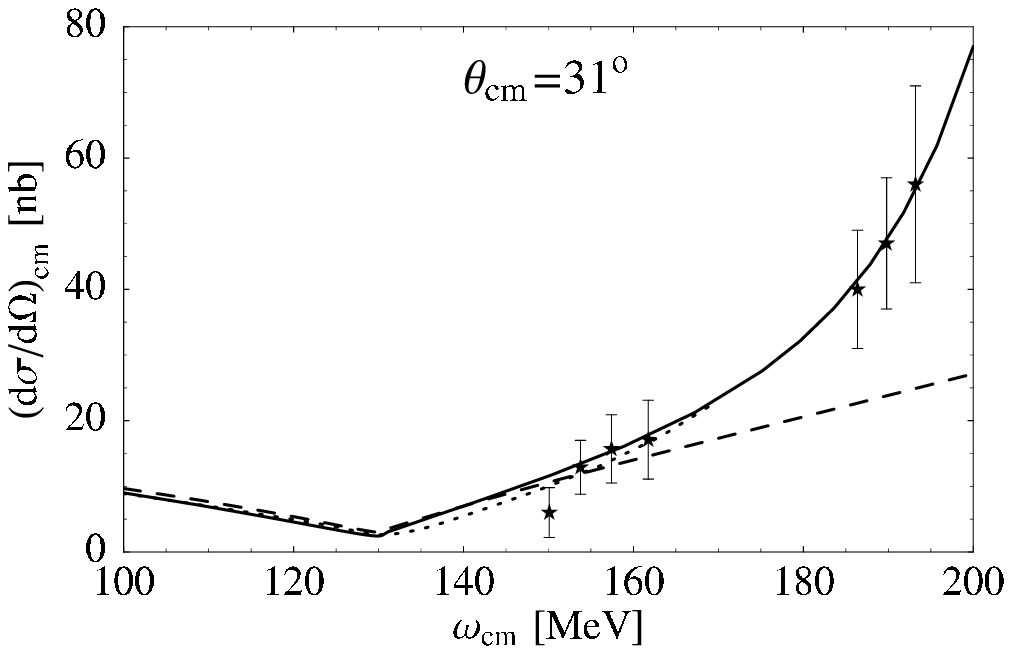}
\hfill
\includegraphics*[width=0.48\textwidth]{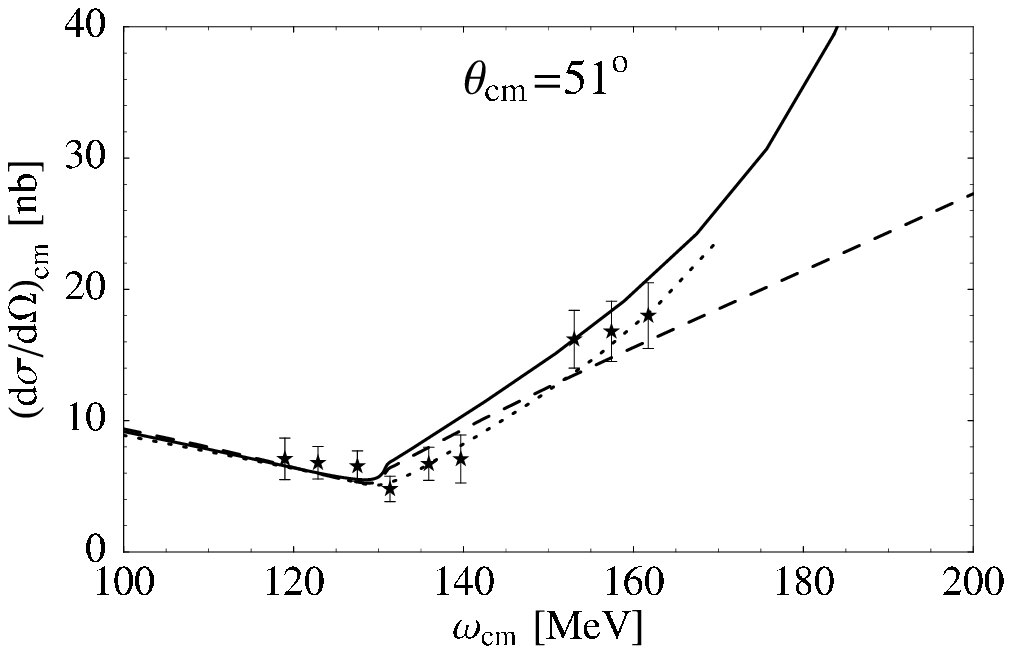}
\includegraphics*[width=0.48\textwidth]{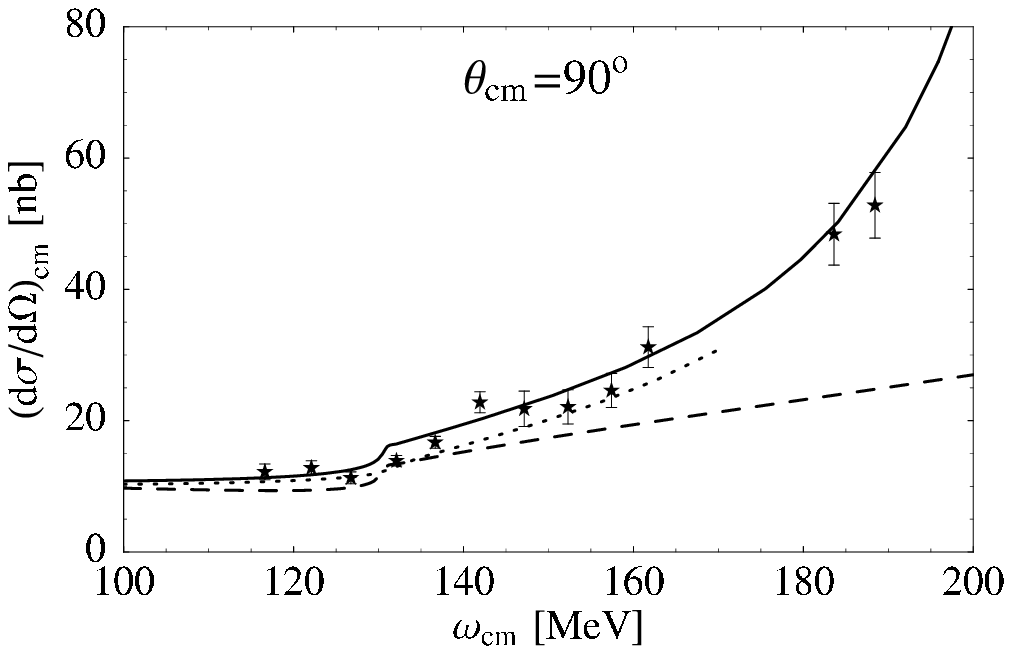}
\hfill
\includegraphics*[width=0.48\textwidth]{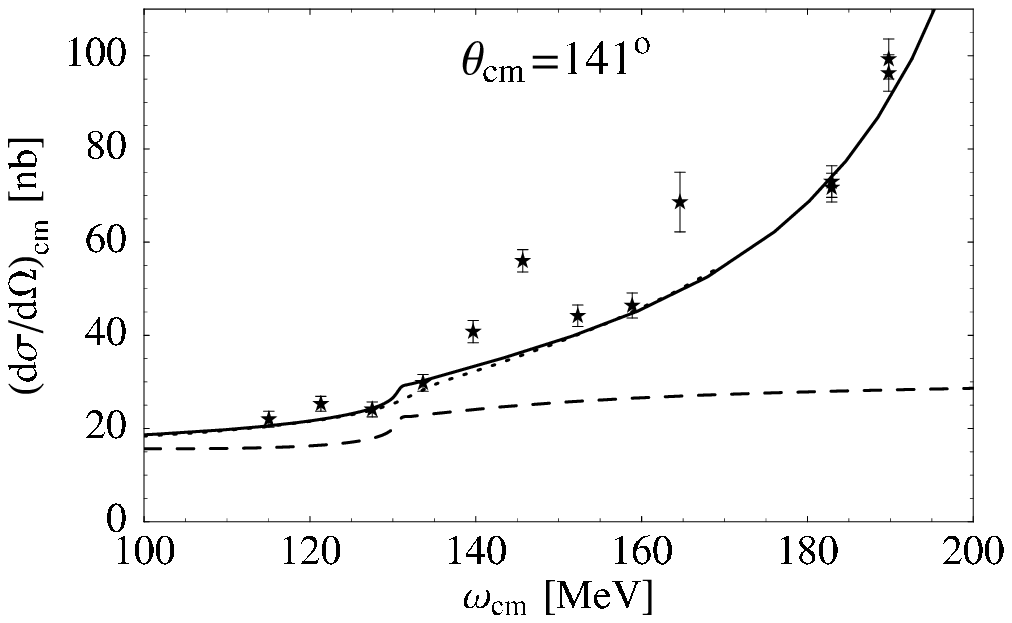}
\caption[Comparison of theoretical results and experimental data 
for spin-averaged Compton scattering off the proton above 100~MeV]
{Comparison of the differential cross-section data for
Compton scattering off the proton from Hallin et al. \cite{Hallin} with
leading-one-loop order HB$\chi$PT respectively SSE and DR at fixed cm angle.
Solid: SSE; dashed: HB$\chi$PT; dotted: DR.}
\label{fig:Hallin}
\end{center}
\end{figure}

Figs.~\ref{fig:Olmos} and \ref{fig:Hallin} compare several different 
cross-section data at selected angles with the third-order HB$\chi$PT 
prediction, the DR prediction from Ref.~\cite{HGHP} and with the result of our
SSE fit (details of the fit will be discussed in the next section). 
Similar pictures can already be found in Ref.~\cite{DA}.
In Fig.~\ref{fig:Olmos} we also give the comparison to the 
$\calO(p^4)$-HB$\chi$PT result of \cite{McGovern} for two angles: $59^\circ$ 
and $155^\circ$. The data of 
Hallin et al. \cite{Hallin} (Fig.~\ref{fig:Hallin}) provide
important constraints for the fit above pion threshold. However, we must 
caution that close to 200~MeV there may already be a sizeable error in the 
SSE calculation due to our treatment of the $\Delta$(1232), which behaves
like a stable particle in the SSE. 
Its width is only included perturbatively and it is zero at 
leading-one-loop order. We are aware that such procedure 
violates unitarity. 
However, sufficiently far below the resonance we can expand in powers of 
$\w/\Delta_0$, so that unitarity and the width are built up order by order in
this expansion parameter. Nevertheless, violating unitarity
may cause a non-negligible uncertainty for $\w\approx200$~MeV. 
The DR calculation is only shown up to 170~MeV in Fig.~\ref{fig:Hallin}, 
which is the upper energy limit of the plots in \cite{HGHP}. 

All four theoretical curves in Fig.~\ref{fig:Olmos} describe the available 
data quite well in the 
forward direction.  The upwards trend in the data above 130 MeV related to the
opening of the $\pi N$ channel is also present in all three frameworks. 
However, while the SSE and DR results are rather similar at all angles, the 
HB$\chi$PT curve deviates from the data significantly in the backward 
direction, starting from photon energies around 80 MeV. This observation 
holds for both the $\calO(p^3)$ \textit{and} the $\calO(p^4)$ calculation. 
The detailed analysis of the dynamical polarizabilities in the next section 
will show that this different energy dependence is due to the lack of explicit 
$\Delta$(1232)-resonance degrees of freedom in HB$\chi$PT.
Even a broad variation of the two 
$\calO(p^4)$-HB$\chi$PT counter terms, such that 
$\bar{\alpha}_{E1}=8\cdot10^{-4}\;\fm^3$, 
$\bar{\beta }_{M1}=6\cdot10^{-4}\;\fm^3$, is not sufficient to 
properly account for the $\Delta$ resonance, cf. Ref.~\cite{McGovern}.  
We find the well-known fact that cross-section calculations in 
leading-one-loop order $\chi$EFT discarding the $\Delta$ as explicit degree 
of freedom fail for large-angle scattering $\theta>90^\circ$, even at 
energies well below pion threshold. The reason is that the $\Delta$ resonance 
dominates in spin-flip $M1$ processes, which are primarily observed in 
back-angle scattering, see e.g. \cite{Huen97,Galler,review}.

\begin{figure}[!htb]
\begin{center}
\includegraphics*[width=0.48\textwidth]{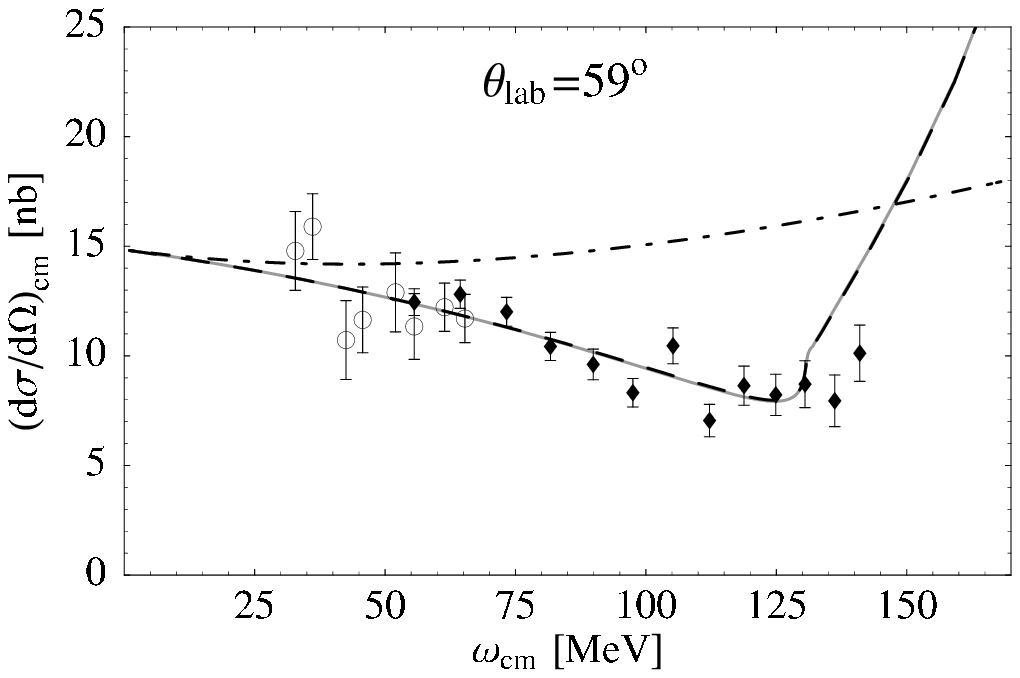}
\hfill
\includegraphics*[width=0.48\textwidth]{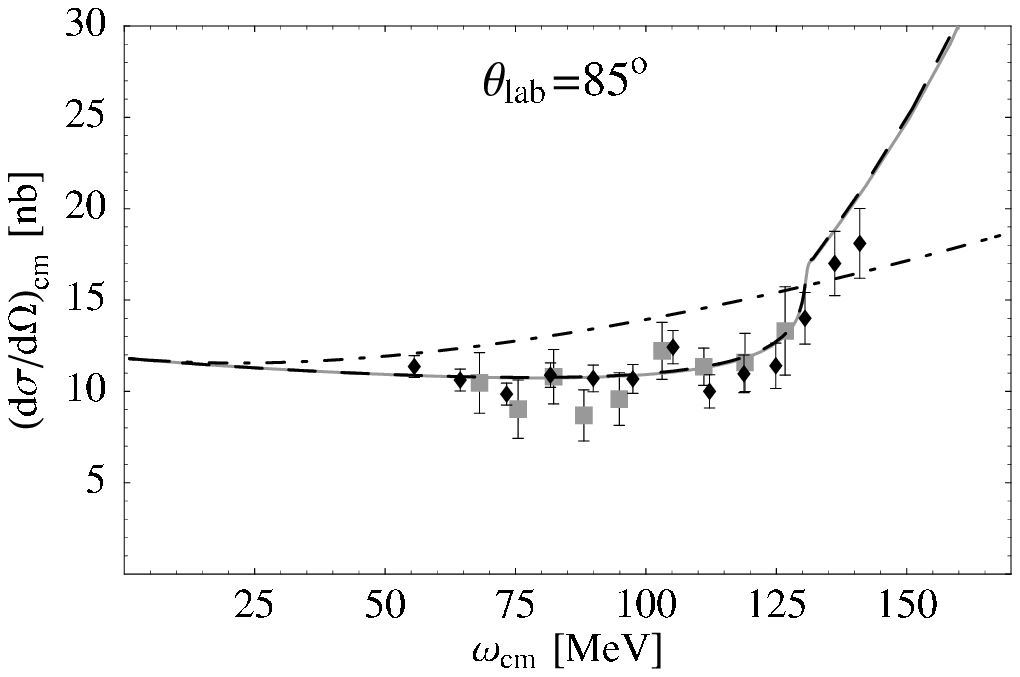}
\includegraphics*[width=0.48\textwidth]{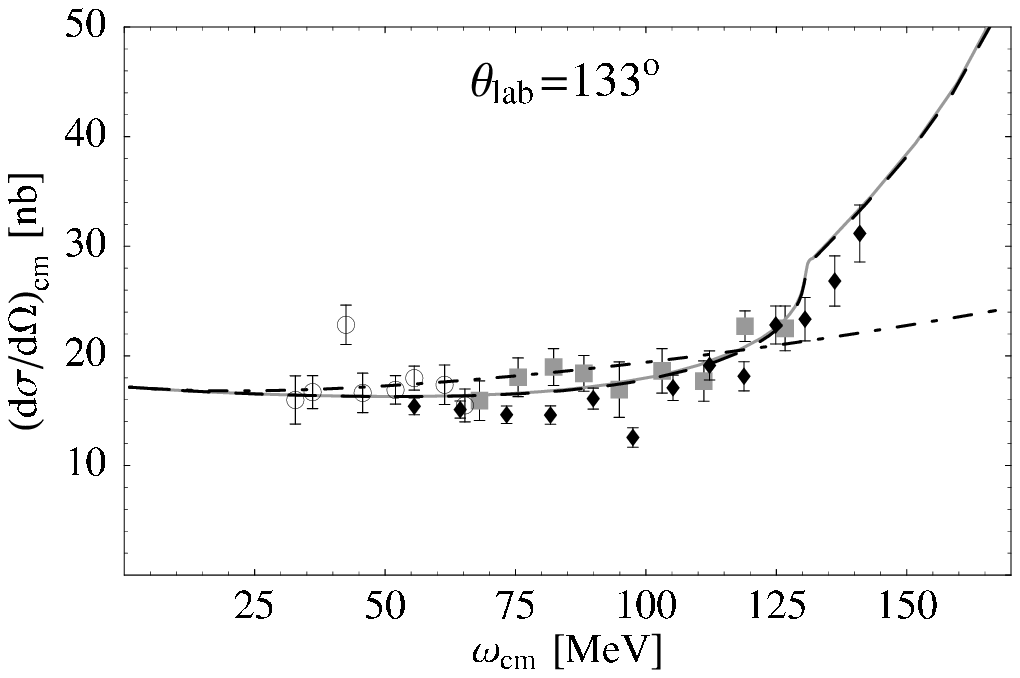}
\hfill
\includegraphics*[width=0.48\textwidth]{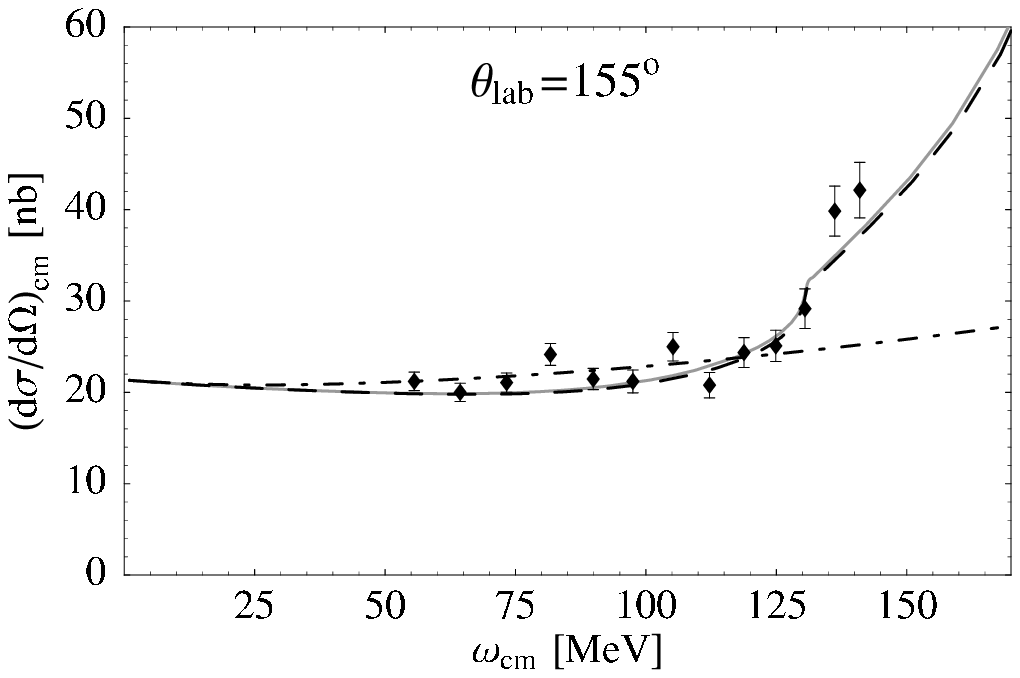}
\caption[Comparison of the SSE multipole expansion to the differential 
cross-section data for Compton scattering off the proton]
{Comparison of the SSE multipole expansion to the differential cross-section 
data for Compton scattering off the proton. Note that the $l=1$ and 
$l=2$ truncations are indistinguishable in the energy region shown here. ($l=0$
truncation: dash-dotted curve; $l=1$ truncation: dashed curve; $l=2$
truncation: solid grey curve)}
\label{fig:fig9}
\end{center}
\end{figure}

Having shown that the full Compton amplitudes $A_1,\ldots, A_6$ of 
leading-one-loop SSE provide a good description of the available Compton data
up to energies above pion threshold, we now examine what kind of physics
dominates in the kinematic regime considered here. A well-established
procedure to answer this question is of course a systematic multipole
expansion of the Compton amplitudes $A_i(\w,z)$ as discussed in
Section~\ref{sec:amplitudestomultipoles}. In Fig.~\ref{fig:fig9}, we compare 
the contributions of the first three terms of the Compton multipole expansion 
to the same data as shown in Fig.~\ref{fig:Olmos}. 
The $l=0$ truncation only contains the pole
contributions to nucleon Compton scattering as shown by the diagrams in
Fig.~\ref{fig:pole}, see also Appendix~\ref{app:poleterms}. 
Truncating the multipole expansion at $l=1$, the curve includes in
addition all dynamical dipole polarizabilities.  All dynamical
quadrupole polarizabilities are contained in the $l=2$ truncation. As has been
known for a long time, a theoretical framework which only incorporates the pole
contributions for nucleon Compton scattering gives a rather poor description
of the cross sections, especially at small angles. The discrepancy between the
$l=0$ result and the data therefore is a clear indication of internal nucleon 
structure not contained in the standard pole terms. According to $\chi$EFT 
calculations, this structure can be interpreted as chiral dynamics in the 
nucleon: It is largely the contributions from the pions as the Goldstone 
Bosons of low-energy QCD (cf. Section~\ref{sec:HBchiPT})~-- 
or in other words the contribution from the pion cloud of the
nucleon~-- which closes the gap between the pole contributions and the Compton
data, at least for energies below the pion threshold. While this is
after many years of $\chi$EFT calculations in nucleon Compton scattering  
a well-known fact, the surprising find from our multipole analysis 
is that up to
energies of $\w\approx 200\;\MeV$, \emph{there is no visible difference}
between the $l=1$ and the $l=2$ truncation.  Therefore, the multipole
expansion for the nucleon Compton cross section
converges very fast in the entire energy region considered. 
In Chapter~\ref{chap:spinpolarized}, this suppression of higher-order
multipoles is also demonstrated in a variety of asymmetries in Compton
scattering. Furthermore, we see that aside from the well-known standard pole 
terms of Fig.~\ref{fig:pole}, all one needs to know for a
good description of nucleon Compton scattering are the six dynamical dipole
polarizabilities $\alpha_{E1}(\w)$, $\beta_{M1}(\w)$, $\gamma_{E1E1}(\w)$,
$\gamma_{M1M1}(\w)$, $\gamma_{E1M2}(\w)$
and $\gamma_{M1E2}(\w)$ 
representing the complete
$l=1$ information. In Section~\ref{sec:extracting} we present a proposal how 
to directly determine the spin-dependent polarizabilities $\gamma_i$ from
experiment, and in Section~\ref{sec:spincontributions} we demonstrate their 
non-negligible contribution even to spin-averaged Compton cross sections. 
While $\chi$EFT calculations for nucleon Compton scattering
in the past have either focused on the static values of the polarizabilities
or on the (rather complicated) full Compton amplitudes, one
can now dissect the role of chiral dynamics (and of explicit resonance
contributions) in this process by looking at the individual multipole
channels.

Before we move on to a detailed comparison of the results for these six 
dynamical polarizabilities derived in leading-one-loop order
HB$\chi$PT, SSE and DR, respectively,  
we first give details regarding the three 
free parameters of SSE fitted to the Compton data.

\section[Fit to Proton Compton Data and Static Polarizabilities]
{Fit to Proton Compton Data and Static\\Polarizabilities 
\label{sec:protonfitsandpolas}}
\markboth{CHAPTER \ref{chap:spinaveraged}. COMPTON SCATTERING AND 
POLARIZABILITIES}{\ref{sec:protonfitsandpolas}. FIT TO PROTON COMPTON DATA 
AND STATIC POLARIZABILITIES}

\subsection{Small Scale Expansion Fit \label{sec:protonfits}}

The two energy-independent short-distance terms with couplings 
$g_{1}$ and $g_{2}$ of Eqs. (\ref{eq:LCT1}, \ref{eq:LCT2}) give contributions 
only to the electric and magnetic dipole polarizabilities. The three free
parameters of the leading-one-loop SSE analysis therefore correspond to a fit
which determines $\bar{\alpha}_{E1},\,\bar{\beta}_{M1}$ plus the leading 
$\gamma N\Delta$ coupling $b_1$, cf. Section~\ref{sec:staticpolas}. 
Note that we go beyond Ref.~\cite{DA}, where we only fitted the two  
polarizabilities. We are able to fit the
two static spin-independent dipole polarizabilities because the fourth-order 
Lagrangeans Eqs.~(\ref{eq:LCT1}, \ref{eq:LCT2}) are promoted to
leading-one-loop order.  For the fit, we use the data from \cite{Olmos}, which
cover the low-energy region ($\w\leq m_\pi$) very well with extremely 
small error bars, and those from \cite{Hallin}, 
which is the only data set available in the energy regime 
$m_\pi\leq\w\leq200$~MeV. 
The results are displayed in Table~\ref{tab:protonfit}, together with 
their corresponding $\chi^2/d.o.f.$-values, which we calculate using the 
standard definition of $\chi^2$, i.e.
\be
\chi^2=\sum
\left(\frac{\sigma_\mathrm{exp}-\sigma_\mathrm{theo}}{\Delta\sigma}\right)^2
\label{eq:chisquared}
\ee
with $\sigma_\mathrm{exp}$ the experimental, $\sigma_\mathrm{theo}$ the
calculated cross sections and $\Delta\sigma$ the experimental error bars.
In a first step, the number of degrees of freedom ($d.o.f.$) is the
number of data points (115) minus the number of free parameters (3). Note that
the value of $\bar{\alpha}_{E1}+\bar{\beta}_{M1}$ from the three-parameter fit
is consistent within error bars with the Baldin sum rule for the proton,
$\bar{\alpha}_{E1}+\bar{\beta}_{M1}=(13.8\pm0.4)\cdot10^{-4}\;
\mathrm{fm}^3$~\cite{Olmos}, cf. Section~\ref{sec:intro}.\footnote{We note 
that an alternative extraction yielded 
$\bar{\alpha}_{E1}+\bar{\beta}_{M1}=(14.0\pm0.3)\cdot10^{-4}\;
\mathrm{fm}^3$~\cite{Lvov}, which was combined with~\cite{Olmos} to 
$(13.9\pm0.3)\cdot10^{-4}\;\mathrm{fm}^3$ in Ref.~\cite{Schumacher}. The error
bars of the value adopted in this work cover the whole range of this result.} 
One can therefore in a second step use the value of the Baldin sum rule as
additional fit constraint and thus reduce the number of free parameters to 
two. If not stated differently, we use these Baldin-sum-rule constrained 
values  in all plots throughout this work. The resulting
static spin-independent dipole polarizabilities, given in 
Table~\ref{tab:protonfit}, compare very well with
state-of-the-art Dispersion Analyses \cite{review} and the values recommended 
in the recent review~\cite{Schumacher}, cf. Eq.~(\ref{eq:reviewp}). 
Nevertheless, the
$\chi^2/d.o.f.$-values of our fits are relatively large, but they are more an
indication of the spread in the Compton data, which we have not allowed to
float with a free normalization constant. We note that at $\theta=133^\circ$
our calculation yields values which are systematically larger than 
the data from~\cite{Olmos}, but agree very well with the data from 
\cite{Fed91,Mac95}, cf. Fig.~\ref{fig:Olmos}.
The encouraging results of Table~\ref{tab:protonfit}
therefore prove that by utilizing the SSE amplitudes of \cite{HGHP},
one has an alternative technique to extract the static polarizabilities
$\bar{\alpha}_{E1},\;\bar{\beta}_{M1}$ from low-energy Compton data below the
$\Delta$-resonance. We note that a determination of
$\bar{\alpha}_{E1},\;\bar{\beta}_{M1}$ from Compton data using next-to-leading
one-loop order HB$\chi$PT was presented in \cite{McGPhil}. The results
obtained there are comparable to ours, although the authors had to restrict
their fit to the lower-energy data to exclude effects from the
$\Delta(1232)$, due to the known inadequate description of the Compton cross
sections in the backward direction in HB$\chi$PT, cf. 
Section~\ref{sec:protoncrosssections}.

The values we obtain in the two fits for the leading $\gamma N\Delta$
coupling $b_1$, cf. Table~\ref{tab:protonfit}, agree with the previous 
analysis~\cite{HHKK} from the radiative $\Delta$-width as
discussed in Section~\ref{sec:spinaveragedtheory}. Note that
we could also employ the strategy to rely on the DR-results for 
$\bar{\alpha}_{E1}$, $\bar{\beta}_{M1}$ and $\bar{\gamma}_{M1M1}$ to determine 
the three unknowns. In this case, the whole energy-dependence is predicted. 
The values thus obtained are identical with the fit-results within the error 
bars, see~\cite{DA}.

As our leading-one-loop order SSE calculation only describes an isoscalar
nucleon, we cannot contribute to the ongoing controversies over the size of
the neutron polarizabilities \cite{review,McGPhil,Rupak} at this point. 
These quantities will be discussed in length in 
Chapters~\ref{chap:perturbative} and~\ref{chap:nonperturbative}, where we fit 
the \textit{isoscalar} polarizabilities, i.e. the average between proton and 
neutron, to elastic deuteron Compton data. From these values and the known
proton numbers we are then able to deduce the elusive neutron polarizabilities.

\begin{table}[!htb]
\begin{center}
\begin{tabular}{|c||r|r|r|}
\hline
Quantity&3-parameter fit&2-parameter fit&\cite{Olmos}\\
\hline
$\chi^2/d.o.f.$&2.87&2.83&1.14 \\
\hline
$\bar{\alpha}_{E1}$&$11.52\pm2.43$&$11.04\pm1.36$
     &$12.4\pm0.6(\mathrm{stat})\mp0.5(\mathrm{syst})\pm0.1(\mathrm{mod})$\\
$\bar{\beta} _{M1}$&$ 3.42\pm1.70$&$ 2.76\mp1.36$
     & $1.4\pm0.7(\mathrm{stat})\pm0.4(\mathrm{syst})\pm0.1(\mathrm{mod})$\\
$b_1           $&$ 4.66\pm0.14$&$ 4.67\pm0.14$&\\
\hline
\end{tabular}
\caption[SSE results from fitting $\bar{\alpha}_{E1}$, $\bar{\beta}_{M1}$ and 
$b_1$ to spin-averaged proton Compton cross sections]
{Values for $\bar{\alpha}_{E1}$, $\bar{\beta}_{M1}$ (in 
$10^{-4}\,\mathrm{fm}^3$) and $b_1$ from a fit to MAMI-~\cite{Olmos} and 
SAL-data~\cite{Hallin}, compared
to the results from \cite{Olmos}. Note that the definition of $\chi^2/d.o.f.$ 
used in~\cite{Olmos} is different from Eq.~(\ref{eq:chisquared}). The error
bars in our fits are only statistical, i.e. the error $\pm0.4$ due to the 
Baldin sum rule in the 2-parameter fit and uncertainties from higher orders 
are not included.}
\label{tab:protonfit}
\end{center}
\end{table}

\subsection{Static Spin-Independent Polarizabilities \label{sec:staticpolas}}

The spin-independent static dipole polarizabilities to leading-one-loop order
in SSE consist of the following individual contributions:
\begin{align}
\label{eq:alpharesult}
\bar{\alpha}_{E1}&=\frac{5\,\alpha\,g_A^2}{ 96\,f_\pi^2\,m_\pi\,\pi}\,
\left(1-\frac{m_\pi}{m_N}\,\frac{1}{\pi}\right)
-\frac{2\,\alpha\,\left(g_{1}+2g_{2}\right)}{(4\pi\,f_\pi)^2\,m_N}
\nonumber\\
&+\frac{\alpha\,g_{\pi N\Delta_0}^2}{54\,\left(f_\pi\,\pi\right)^2}\,
\left[\frac{9\,\Delta_0}{\Delta_0^2-m_\pi^2}+
\frac{\Delta_0^2-10\,m_\pi^2}{(\Delta_0^2-m_\pi^2)^{3/2}}\,\ln R\right]
\nonumber\\
&=\left[11.87\,(N\pi)-( 5.92\pm1.36)\,(\text{c.t.})+0.0\, 
(\Delta\text{-pole})
+5.09\,(\Delta\pi)\right]\times 10^{-4}\;\mathrm{fm}^3\nonumber\\
&=(11.04\pm1.36)         \times 10^{-4}\;\mathrm{fm}^3
\end{align}
\begin{align}
\label{eq:betaresult}
\bar{\beta}_{M1} &=\frac{   \alpha\,g_A^2}{192\,f_\pi^2\,m_\pi\,\pi}
+\frac{4\,\alpha\,      g_{2}                }{(4\pi\,f_\pi)^2\,m_N} 
+\frac{2\,\alpha\,b_{1}^2}{9\,\Delta_0\,m_N^2} 
 +\frac{\alpha\,g_{\pi N\Delta_0}^2}{54\,\left(f_\pi\,\pi\right)^2}\,
\frac{1}{\sqrt{\Delta_0^2-m_\pi^2}}\,\ln R\quad\nonumber\\
&=[ 1.25\,(N\pi)-(10.68\pm1.17)\,(\mathrm{c.t.})+(11.33\pm0.70)\,
(\Delta\text{-pole})\nonumber\\
&+0.86\,(\Delta\pi)]\times 10^{-4}\;\mathrm{fm}^3
=( 2.76\mp1.36)     \times 10^{-4}\;\mathrm{fm}^3
\end{align}
where $ R=\left(\Delta_0+\sqrt{\Delta_0^2-m_\pi^2}\right)/{m_\pi}$ is a
dimensionless parameter \cite{HHK97} and the results from the 
Baldin-constrained fit have been used, cf.~Table~\ref{tab:protonfit}.

In the case of $\bar{\alpha}_{E1}$, one notices a strong cancellation between 
the $\pi\Delta$ contributions and the short-distance physics contained in
$g_{1},\,g_{2}$. In Section~\ref{sec:dynpolas}, we will demonstrate that
this mutual cancellation holds throughout the low-energy region also in the
case of the dynamical electric dipole polarizability, forcing us to the not
surprising conclusion that for photon energies far below on-shell $\Delta\pi$
intermediate states 
such contributions are indistinguishable from
counterterms parameterizing the short-distance physics. 
However, the physics content of the short-distance operators is not yet 
completely understood.
Assuming that corrections from next-to-leading order are suppressed 
by $m_\pi/m_N$ with respect to the leading-order result, comparing the 
leading-order pion-cloud contribution to $\bar{\alpha}_{E1}$ and 
$\delta\bar{\alpha}_{E1}^{sd}$ demonstrates that the ``natural'' size of  
$g_1,\,g_2$ is unity. Therefore, na\"ive dimensional analysis for 
next-to-leading order contributions to $\bar{\alpha}_{E1}$ and 
$\bar{\beta}_{M1}$ predicts that their size is  
\begin{equation}
   \label{eq:higherorder}
  |\bar{\alpha}_\mathrm{NLO}|\sim|\bar{\beta}_\mathrm{NLO}|\sim
        \frac{\alpha}{\Lambda_\chi^2\,m_N} \sim 1 \cdot 10^{-4}\; \fm^3 
\end{equation}
with $\Lambda_\chi\sim 1$~GeV the breakdown scale of HB$\chi$PT, cf. 
Section~\ref{sec:HBchiPT},
but apparently fails to give a correct estimate of their magnitudes. 
The numerical values of Eqs.~(\ref{eq:alpharesult}, \ref{eq:betaresult}) tell 
us post factum that they must already be included in the analysis at leading 
order. We note that the
extra, quark-mass-independent term in the $\pi N$ contribution arises from our
pion-threshold correction discussed in Section~\ref{sec:spinaveragedtheory}.
This term does not appear in Ref.~\cite{DA} due to the analogous correction of
the $u$-channel applied there.

In $\bar{\beta}_{M1}$, we encounter the well-known cancellation between a large
paramagnetism from the $\Delta$(1232)-pole contributions and the
nucleon diamagnetism, arising from short-distance effects  
parameterized by the coupling $g_{2}$.
Several proposals to explain this effect 
were put forward in the literature. 
One attributes it to an interplay between short-distance 
physics and the pion-cloud occurring from the next-to-leading order chiral 
Lagrangean \cite{BKMS}, another one to the $t$-channel exchange of a meson or 
correlated two-pion exchange~\cite{GuichonWeise, Schumacher}. 
An alternative explanation for the smallness of $\bar{\beta}_{M1}$ 
due to off-shell effects in the $\gamma N \Delta$-transition form factors has 
been presented in \cite{BSS}. Whether either of these explanations gives a 
convincing quantitative description of the short-distance coefficients is not 
clear yet.

In contrast to the cancellation in $\bar{\alpha}_{E1}$ discussed above, the sum
of dia- and paramagnetic effects is strongly energy dependent and therefore
leads to a clear signal in the dynamical magnetic dipole polarizability
$\beta_{M1}(\w)$, see Section~\ref{sec:dynpolas}. Apart from the contribution 
proportional to $g_{2}$, Eq.~(\ref{eq:betaresult}) agrees with the
result found in Ref.~\cite{HHK97} (modulo the different convention for the
coupling $b_1$), where it was already noted that the $\Delta\pi$ contributions
to $\bar{\beta}_{M1}$ are considerably smaller than in the case of
$\bar{\alpha}_{E1}$.

Already  from this discussion, one can see that the two extra terms
$g_{1},\,g_{2}$ are not just small higher-order effects. For a consistent
description both of the data and of the static polarizabilities, they are in
contrast required in a leading-one-loop SSE analysis. Translating the fit 
results of Table~\ref{tab:protonfit} back into these two unknown couplings, 
one obtains
\be
g_{1}=17.44\pm2.11\,(\text{stat}),\;\;g_{2}=-5.64\pm0.88\,(\text{stat})
\ee
for the 3-parameter fit and 
\be
g_{1}=18.82\pm0.79\,(\text{stat})\pm0.4\,(\text{Baldin}),\;\;
g_{2}=-6.05\mp0.66\,(\text{stat})\pm0.4\,(\text{Baldin}) 
\label{eq:g1g2Baldin}
\ee
for the Baldin-constrained fit.
Therefore, these two couplings are significantly larger than their ``natural''
values, which in the Lagrangean employed (Eqs.~(\ref{eq:LCT1}, \ref{eq:LCT2}))
would be expected to be unity. These couplings~-- though 
formally being part of the next-to-leading one-loop order Lagrangean~-- 
therefore break the na\"ive power counting underlying SSE and have to be taken 
into account already at leading-one-loop order. 
There are indications that this feature is not 
specific to SSE but occurs in all chiral calculations of $\alpha_{E1}(\w)$
and $\beta_{M1}(\w)$, if high-energy modes in the pion-loop graphs are to be 
properly accounted for, cf. \cite{Holstein}. 
Having determined $g_{1},\;g_{2}$ from fits to
Compton-scattering data, we now have fixed all our unknown parameters and have
full predictive power in the determination of the energy dependence of the 
polarizabilities discussed in Section~\ref{sec:dynpolas}. 

Finally, we note again that not only the energy dependence of the dynamical
polarizabilities is independent of the two extra couplings
$g_{1},\,g_{2}$, but also the values of the four spin-dependent static
dipole polarizabilities $\bar{\gamma}_{E1E1}$, $\bar{\gamma}_{M1M1}$,
$\bar{\gamma}_{E1M2}$, $\bar{\gamma}_{M1E2}$. The results obtained in
Ref.~\cite{HHKK} are therefore reproduced\footnote{We note that in the case of
$\bar{\gamma}_{E1E1}$ we obtain a small extra term
$\delta\,\bar{\gamma}_{E1E1}=-\frac{\alpha g_A^2}{96 f_\pi^2\pi m_N m_\pi}$
due to our correction of the pion threshold discussed in point 1 of
Section~\ref{sec:spinaveragedtheory}. This term is part of the next-to-leading
one-loop order contributions to this polarizability discussed in 
Ref.~\cite{CD00}.}, as expected. For better comparison with Dispersion Theory 
and experiment, we present in Table~\ref{tab:gamma0gammapi} the numbers for 
the linear combinations $\bar{\gamma}_0,\,\bar{\gamma}_\pi$ of 
Eq.~(\ref{eq:gpi}). For more detail, we refer the interested reader
to the extensive literature on these two elusive structures \cite{review}.

\begin{table}[!htb]
\begin{center}
\begin{tabular}{|c||c|c|c|c|}
\hline
Quantity&SSE&experiment&DR&\cite{Schumacher}\\
\hline
$\bar{\gamma}_0  ^{p}$&$ 0.62\mp0.25$&
 $-1.01\pm0.08(\text{stat})\pm0.10(\text{syst})$&$-0.7$&---\\
$\bar{\gamma}_\pi^{p}$&$ 8.86\pm0.25$
&$10.6 \pm2.1 (\text{stat})\mp0.4 (\text{syst})\pm0.8(\text{mod})$
& $\;\;\;9.3$&$8\pm1.8$\\
\hline
$\bar{\gamma}_0  ^{n}$&$0.62\mp0.25$&---&$\;\;-0.07$&---\\
$\bar{\gamma}_\pi^{n}$&$8.86\pm0.25$&---& $\;\,13.7$&$11.9\pm4.0$\\
\hline
\end{tabular}
\caption[Comparison of proton and neutron spin polarizabilities]
{Comparison of proton $\bar{\gamma}_0^{p}$, $\bar{\gamma}_\pi^{p}$ and neutron
$\bar{\gamma}_0^{n}$, $\bar{\gamma}_\pi^{n}$ spin polarizabilities (in 
$10^{-4}\,\mathrm{fm}^4$) between leading-one-loop SSE, experiment and 
fixed-$t$ Hyperbolic Dispersion Theory \cite{review}; the experimental values 
for $\bar{\gamma}_0^{p}$, $\bar{\gamma}_\pi^{p}$ are taken from \cite{GDH} and 
\cite{Olmos}, respectively. The last column corresponds to the averaged values 
recommended in Ref.~\cite{Schumacher}.
Note that the $\mathcal{O}(\epsilon^3)$-SSE results are purely isoscalar.}
\label{tab:gamma0gammapi} 
\end{center}
\end{table}

The spin-independent static quadrupole polarizabilities $\bar{\alpha}_{E2}$, 
$\bar{\beta}_{M2}$ have been analyzed in
Ref.~\cite{Holst} to leading-one-loop order in SSE. Here we present the
details of our results for these $l=2$ polarizabilities, as they 
include some additional features with respect to Refs.~\cite{Holst} 
and~\cite{DA}.
\begin{align}
\bar{\alpha}_{E2}&= \frac{   \alpha\,g_A^2}{ 32\,f_\pi^2\,m_\pi^3\,\pi}
\left(\frac{7}{5}+\frac{9}{10}\,\frac{m_\pi}{m_N}\,\frac{1}{\pi}\right)
\nonumber\\
&+\frac{\alpha\,g_{\pi N\Delta_0}^2}{135\,(f_\pi\,\pi)^2\,m_\pi^2}
\left[\frac{ \Delta_0\,\left(11\,\Delta_0^2-41\,m_\pi^2\right)}
{\left(\Delta_0^2-m_\pi^2\right)^2}+       \frac{3\,m_\pi^2\,
\left(3\,\Delta_0^2+7\,m_\pi^2\right)}{\left(\Delta_0^2-m_\pi^2\right)^{5/2}}\,
\ln R\right]\qquad\qquad\nonumber\\
&=\left[21.48\,(N\pi)+0.0\,(\text{c.t.})+0.0\,(\Delta\text{-pole})
+4.99\,(\Delta\pi)\right]
\times 10^{-4}\;\mathrm{fm}^5\nonumber\\
&= 26.47\times 10^{-4}\;\mathrm{fm}^5
\end{align}
\begin{align}
\bar{\beta}_{M2} &=-\frac{3\,\alpha\,g_A^2}{160\,f_\pi^2\,m_\pi^3\,\pi}
-\frac{2\,\alpha\,b_1^2}{3\,\Delta_0^2\,m_N^3}
 +\frac{\alpha\,g_{\pi N\Delta_0}^2}{ 15\,(f_\pi\,\pi)^2\,m_\pi^2}
\left[\frac{-\Delta_0}{\Delta_0^2-m_\pi^2}+\frac{   m_\pi^2}
                                      {\left(\Delta_0^2-m_\pi^2\right)^{3/2}}\,
\ln R\right]\nonumber\\
&=\left[-8.93\,(N\pi)+0.0\,(\text{c.t.})-(5.18\pm0.32)\,(\Delta\text{-pole})
-3.37\,(\Delta\pi)\right]\times 10^{-4}\;\mathrm{fm}^5\nonumber\\
&=\left(-17.48\pm0.32\right)\times 10^{-4}\;\mathrm{fm}^5
\end{align}
In particular we note the extra piece $\sim m_\pi^{-2}$ from the 
one-pion-threshold correction in $\bar{\alpha}_{E2}$, as well as the 
kinematically induced $u$-channel $\Delta$(1232) contribution to 
$\bar{\beta}_{M2}$, which was not yet included in Ref.~\cite{DA}. 
For details on the origin of these terms we refer to 
Section~\ref{sec:spinaveragedtheory}, items 1 and 2. 
Each of these effects seems to improve the agreement between SSE and 
Dispersion Theory, as we will see in the plots of the corresponding dynamical 
quadrupole polarizabilities, given in Section~\ref{sec:dynpolasspinindie}. 
However, as discussed
in Section~\ref{sec:protoncrosssections}, we remind the reader that the $l=2$
polarizabilities are in effect so small that they cannot be determined 
directly from state-of-the-art nucleon Compton-scattering experiments.

\section[Chiral Dynamics and Dynamical Polarizabilities]
{Chiral Dynamics and\\Dynamical Polarizabilities \label{sec:dynpolas}}
\markboth{CHAPTER \ref{chap:spinaveraged}. COMPTON SCATTERING AND 
POLARIZABILITIES}{\ref{sec:dynpolas}. CHIRAL DYNAMICS AND DYNAMICAL 
POLARIZABILITIES}

Dynamical polarizabilities are a concept complementary to \emph{generalized}
polarizabilities of the nucleon \cite{Guichon, genpoleff, genpolDR}. The
latter probe the nucleon in virtual Compton scattering, i.e. with an incoming
photon of non-zero virtuality and an outgoing, soft real photon. Therefore,
they provide information about the spatial distribution of charges and
magnetism inside the nucleon at zero energy. Dynamical polarizabilities 
on the other hand test the global low-energy
excitation spectrum of the nucleon at {\em non-zero} energy and answer the
question, which internal degrees of freedom govern the structure of the
nucleon at low energies.

In the following detailed discussion of the dynamical
polarizabilities, the error bars for the input parameters as discussed in
Section~\ref{sec:protoncrosssections} induce uncertainties in the static and
dynamical polarizabilities. The grey bands in the figures around the SSE
curves arise from the (statistical) uncertainty in the fit parameters 
determined with the help of the Baldin sum rule in 
Section~\ref{sec:protonfits}. Albeit only a full
higher-order calculation will give a good estimate of the higher-order
effects in EFT, this permits already a rough guess of their size, at least 
in $\alpha_{E1}^{s}(\w)$ and $\beta_{M1}^{s}(\w)$, given that the statistical 
uncertainties in Table~\ref{tab:protonfit} are at least as large as the 
corrections one expects from higher orders, cf. Eq.~(\ref{eq:higherorder}).
Similar pictures of the dynamical polarizabilities have already been 
presented in \cite{DA}, however without error bars.

\subsection{Isoscalar Spin-Independent Polarizabilities 
\label{sec:dynpolasspinindie}}

\begin{figure}[!htb]
\begin{center}

\includegraphics*[width=0.48\textwidth]{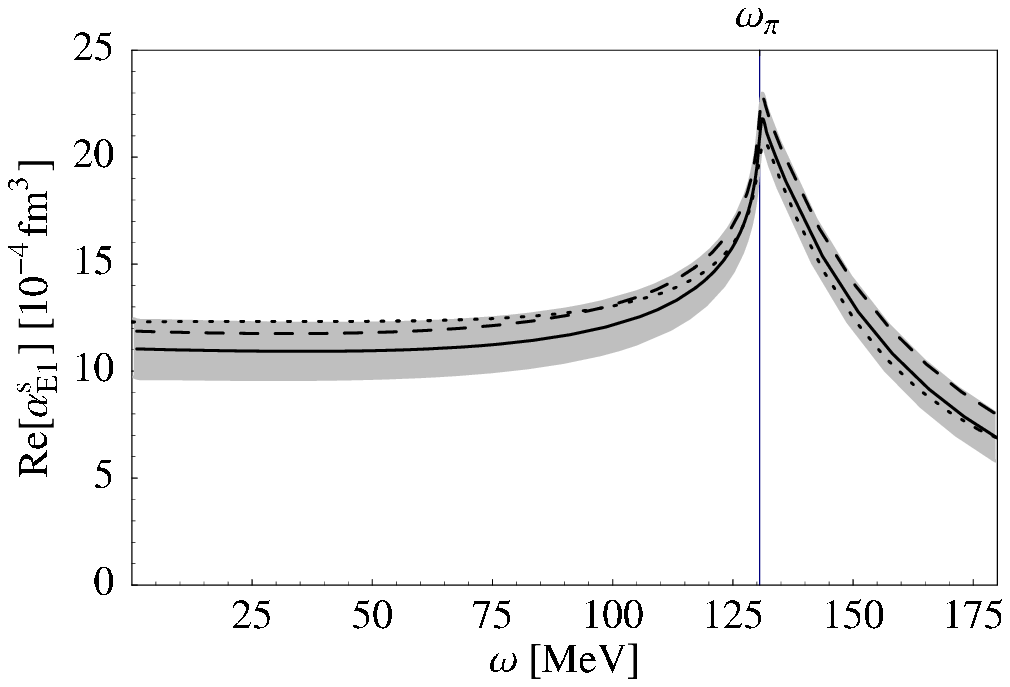}
\hfill
\includegraphics*[width=0.48\textwidth]{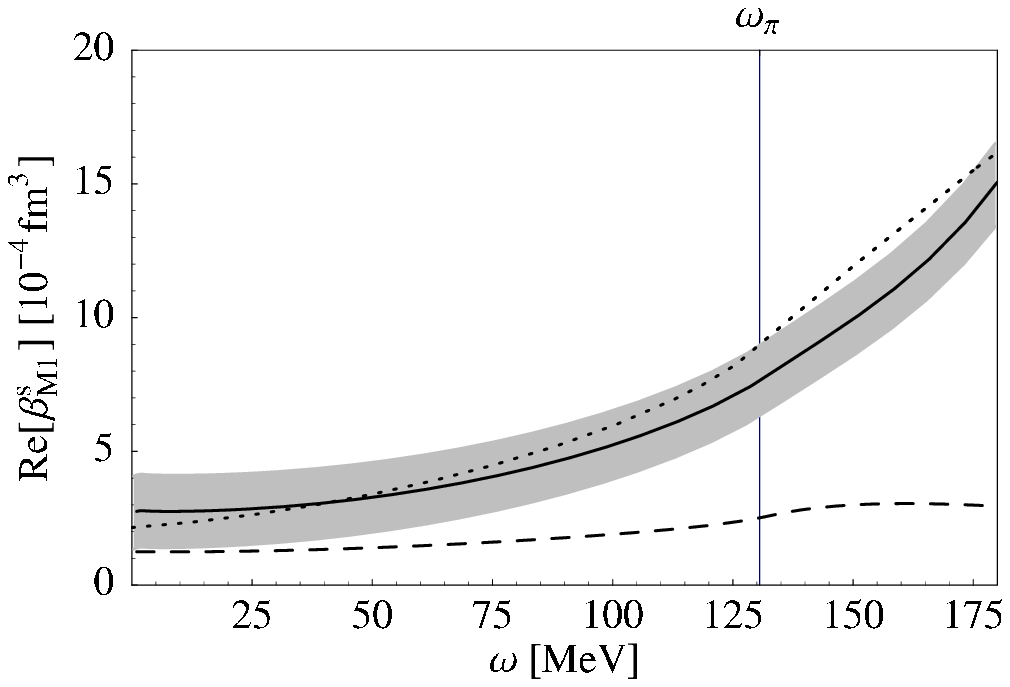}
\includegraphics*[width=0.48\textwidth]{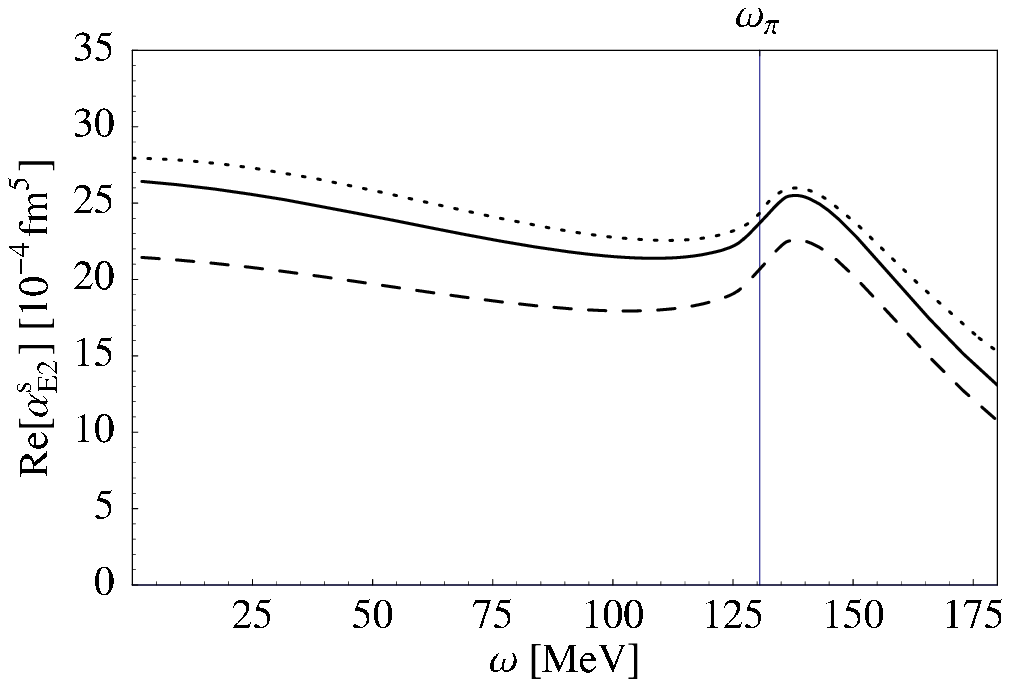}
\hfill
\includegraphics*[width=0.48\textwidth]{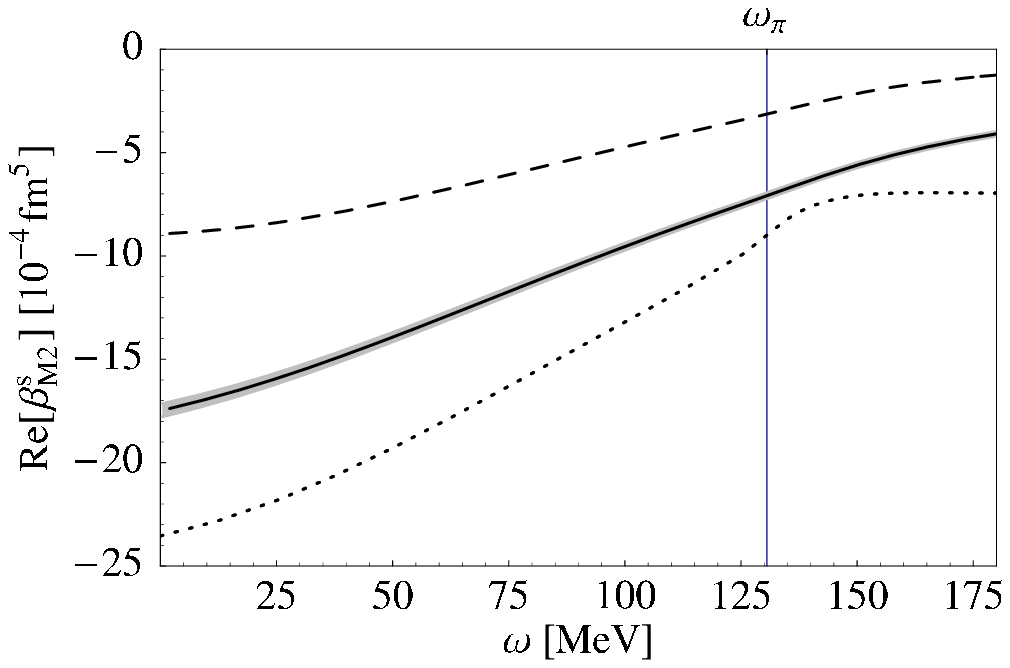}
\caption[Theoretical results for the isoscalar, spin-independent 
dynamical dipole polarizabilities]
{Comparison of the leading-one-loop order SSE (solid) and HB$\chi$PT 
(dashed) results for the real parts of the isoscalar, spin-independent 
dynamical electric and magnetic dipole (top) and quadrupole (bottom) 
polarizabilities with Dispersion Theory (dotted).}
\label{fig:spinindependentpolas}
\end{center}
\end{figure}

Turning first to $\alpha_{E1}^{s}(\w)$ as shown in 
Fig.~\ref{fig:spinindependentpolas}, it
is obvious that its energy dependence in the low-energy region is entirely
controlled by chiral dynamics arising from single-$\pi N$ intermediate states.
All three theoretical analyses agree rather well within the statistical
uncertainty band of the SSE calculation. As already discussed for the static
electric polarizability $\bar{\alpha}_{E1}$ in the previous section, no effects
from any inherent $\pi\Delta$ intermediate states can be detected, pointing to
the fact that these rather heavy degrees of freedom are effectively frozen out
at low energies. This makes them~-- as far as the energy dependence of
the dynamical polarizabilities is concerned~-- indistinguishable from 
short-distance contributions represented by the couplings $g_{1},\, g_{2}$. 
We also note that the strength and shape of the cusp associated with the 
one-pion-production threshold is reproduced extremely well by the 
leading-one-loop chiral calculations. It will serve as an interesting check 
for the convergence properties of the chiral theories to see whether the 
rather good agreement is maintained, once the higher-order corrections are 
included.

The other spin-independent $l=1$ dynamical polarizability,
$\beta_{M1}^{s}(\w)$, shows quite a different picture. We note that the
three theoretical frameworks only agree (within the uncertainty of the SSE
parameters) for the value of the static magnetic polarizability
$\bar{\beta}_{M1}$. For increasing values of the photon energy, it becomes
obvious from the agreement between SSE and Dispersion Theory that explicit
$\Delta$(1232) contributions via $s$-channel pole graphs lead to a
paramagnetic behaviour quickly rising with energy. Any $\Delta\pi$
contributions remain small and are effectively frozen out.  The near
cancellation between para- and diamagnetic contributions for the static value
discussed in the previous section is completely taken over by
$\Delta$(1232)-induced paramagnetism when the photon energy goes up. We
explicitly point to the scale on the $y$-axis of this plot, indicating a rise
by a factor of four at photon energies near the one-pion-production threshold.
While the leading-one-loop HB$\chi$PT calculation \cite{BKKM} provides a
good prediction for $\bar{\beta}_{M1}$, it fails to describe the energy
dependence of $\beta_{M1}^{s}(\w)$, as shown in 
Fig.~\ref{fig:spinindependentpolas}. In
contrast to $\alpha_{E1}^{s}(\w)$, hardly any cusp is visible in
$\beta_{M1}^{s}(\w)$. Beyond the static limit, the chiral $\pi N$
contributions play a minor role in this polarizability. We note that while the
fine details of the rising paramagnetism in $\beta_{M1}^{s}(\w)$ differ
between SSE and Dispersion Theory, they are consistent within the
uncertainties of the SSE curve. The discrepancy between the two schemes above
the one-pion-production threshold is likely to be connected to a detailed
treatment of the width of the $\Delta$ resonance, which is neglected in
leading-one-loop SSE.

We further note that the good agreement between SSE and Dispersion Theory for
the $l=1$ spin-independent dynamical polarizabilities provides a non-trivial
check regarding the physics parameterized in the couplings
$g_{1},\,g_{2}$. Given that these two structures are energy independent,
cf.  Eqs.~(\ref{eq:LCT1}, \ref{eq:LCT2}), the fact that only the $\pi N$ and
$\Delta$ degrees of freedom suffice to describe the energy dependence in the
low-energy region quite well supports our idea that the physics underlying
$g_{1},\, g_{2}$ is ``short-distance'' from the point of view of
$\chi$EFTs.

It is also interesting to look at the spin-independent $l=2$ dynamical
polarizabilities, even if in actual analyses of Compton data they only play a
minor role. In $\alpha_{E2}^{s}(\w)$ we observe a visible contribution
from $\Delta\pi$ intermediate states. It hardly modifies the shape of the
energy dependence, but does affect the overall normalization of this
polarizability, as can be seen from the difference between the SSE and the
HB$\chi$PT curve. The agreement between SSE and Dispersion Theory is
surprisingly good throughout the entire low-energy region. Another interesting
higher-order dynamical polarizability is $\beta_{M2}^{s}(\w)$. The
chiral $\pi N$ contribution seems to play only a minor role in the energy
dependence of this polarizability. $\Delta\pi$ and a surprisingly large
$\Delta$(1232) $u$-channel pole contribution can close a significant part of
the gap between the HB$\chi$PT and the Dispersion Theory result. The remaining
gap between SSE and Dispersion Theory might well be accounted for by
next-to-leading one-loop chiral $\pi N$ corrections, given that the slope of
the energy dependence below pion threshold seems to agree between the two
frameworks. Nevertheless, the energy dependence of this polarizability is
quite peculiar. The magnetic quadrupole strength has decreased rather fast by
more than a factor of two when the photon energy reaches the 
one-pion-production threshold. This shape is reminiscent of a relaxation effect
typically discussed in textbook examples for dispersive effects \cite{GH1}.
While both in HB$\chi$PT and SSE the strengths for $\beta_{M2}(\w)$ tend
to zero for large photon energies, the DR-curve seems to point to additional
physics contributions above the pion threshold.

We now move on to a discussion of the $l=1$ spin-dependent dynamical
polarizabilities.

\subsection{Isoscalar Spin-Dependent Polarizabilities  
\label{sec:dynpolasspin}}

\begin{figure}[!htb]
\begin{center}
\includegraphics*[width=0.48\textwidth]{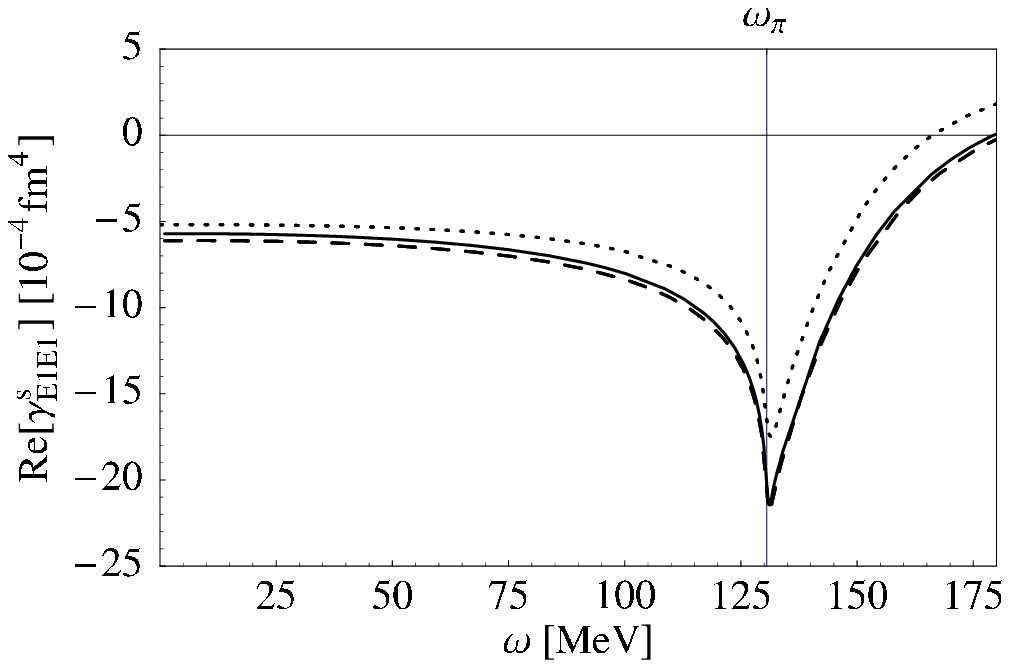}
\hfill
\includegraphics*[width=0.48\textwidth]{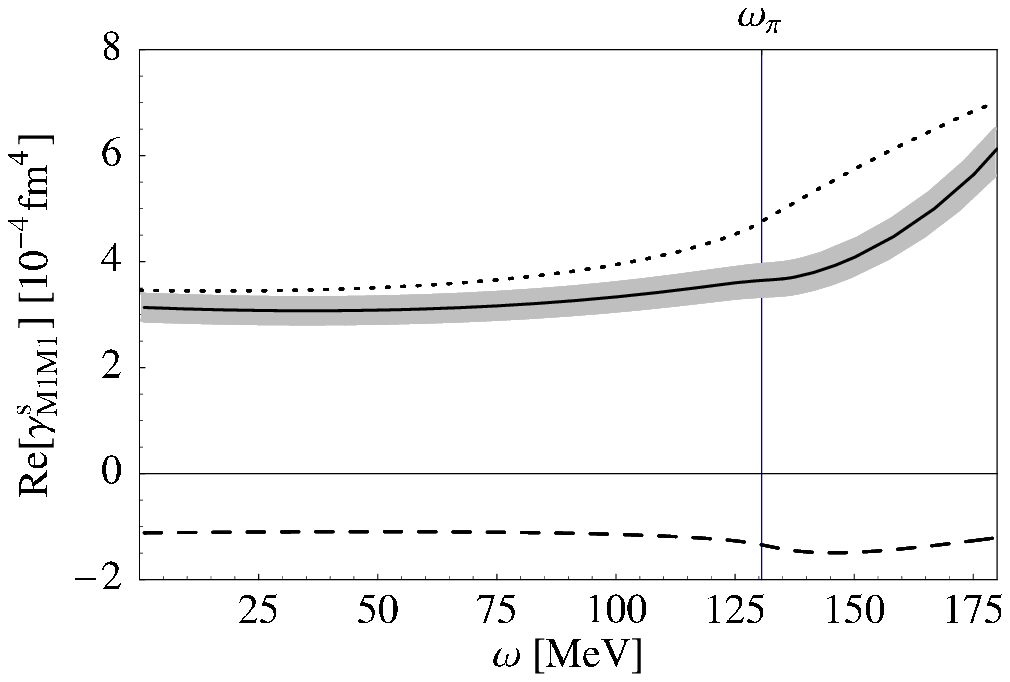}
\includegraphics*[width=0.48\textwidth]{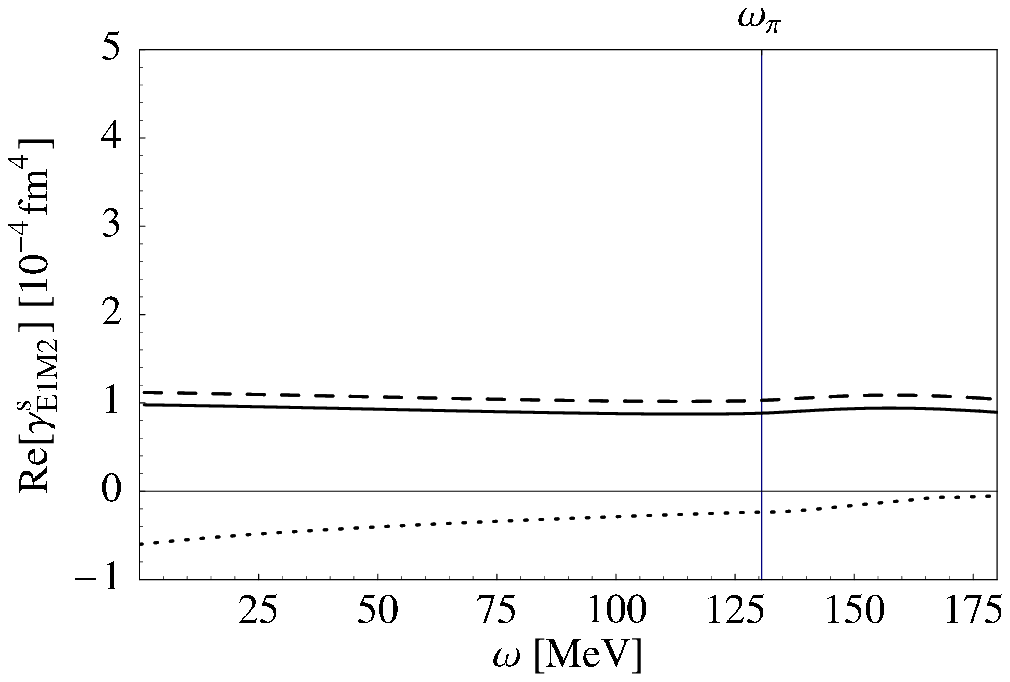}
\hfill
\includegraphics*[width=0.48\textwidth]{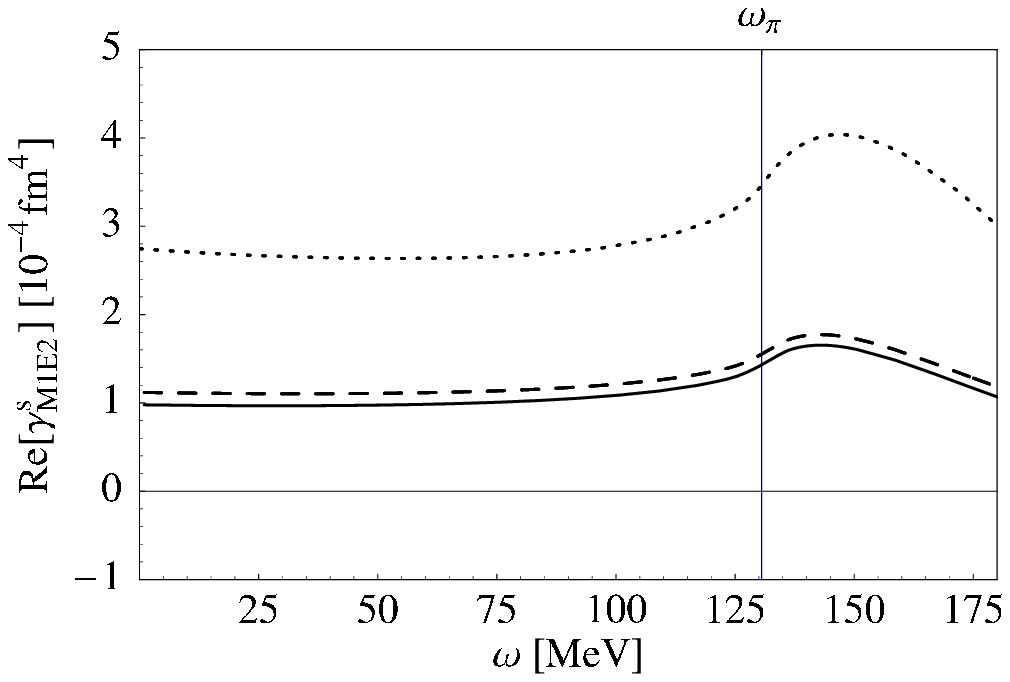}
\caption[Theoretical results for the isoscalar, spin-dependent 
dynamical dipole polarizabilities]
{Comparison of the leading-one-loop order SSE (solid) and HB$\chi$PT 
(dashed) results for the real parts of the isoscalar, spin-dependent 
dynamical dipole polarizabilities with Dispersion Theory (dotted).}
\label{fig:spindependentpolas}
\end{center}
\end{figure}

We again remind the reader that no fit parameters analogously to $g_{1}$ and
$g_{2}$ are present in the leading-one-loop SSE results for the
spin-dependent polarizabilities. The only free parameter entering the
dynamical spin polarizabilities is $b_1$, which we have determined from the
fit to Compton cross sections in Section~\ref{sec:protonfits}; however, it 
influences only $\gamma_{M1M1}^{s}(\w)$. As Fig.~\ref{fig:spindependentpolas}
demonstrates, the contributions of the $\Delta\pi$ continuum to the spin
polarizabilities are small throughout the low-energy region. The energy
dependence in $\gamma_{E1E1}^{s}(\w)$ is completely governed by chiral
dynamics and agrees well among the three frameworks, quite analogously to the
situation in $\alpha_{E1}^{s}(\w)$. The $\Delta$(1232)-pole
contribution~-- rising with energy~-- is visible in
$\gamma_{M1M1}^{s}(\w)$, but it does not rise as dramatically as in the
case of $\beta_{M1}^{s}(\w)$ (cf. Fig.~\ref{fig:spinindependentpolas}). The 
HB$\chi$PT calculation for $\gamma_{M1M1}^{s}(\w)$ deviates strongly from both
the SSE and DR result, signaling again the need for explicit $\Delta$(1232)
degrees of freedom in resonant multipoles. The slight disagreement between SSE
and DR for photon energies above pion threshold in
$\gamma_{M1M1}^{s}(\w)$ might be connected to a detailed treatment of
the width of the $\Delta$ resonance, which is not included in
leading-one-loop SSE, see Refs.~\cite{DA, HGHP}\footnote{We do not include a 
chapter on polarizabilities in the resonance region in this work, as we did 
not investigate this issue further than reported in Ref.~\cite{DA}.}. Both 
HB$\chi$PT and SSE predictions for the mixed spin polarizabilities are rather 
similar, disagreeing with the DR result. While $\gamma_{E1M2}^{s}(\w)$
constitutes a rather tiny structure effect which will be hard to pin down
precisely, the ``large'' gap between SSE and the DR result in
$\gamma_{M1E2}^{s}(\w)$ could possibly arise from the missing $E2$
excitation of the $\Delta$ resonance in a leading-one-loop SSE calculation.
This effect can be accounted for at next-to-leading one-loop order. On the
other hand, the overall shape of the energy dependence in
$\gamma_{M1E2}^{s}(\w)$ is rather similar between the chiral and the DR
results, indicating that a $\pi N$ loop contribution at the next-higher chiral
order might also suffice to close the gap.

In conclusion, among the four isoscalar \hspace{.01cm} spin-dependent dipole
polarizabilities, only $\gamma_{E1E1}^{s}(\w)$ seems to be dominated by
$\pi N$ chiral dynamics, which can be accounted for rather well already at
leading-one-loop order throughout the low-energy region. A detailed
understanding of the dynamical spin dipole polarizabilities requires explicit
$\Delta$(1232)-resonance degrees of freedom.

We now close our discussion of the nucleon polarizabilities, which will, 
however, show up several times in this work, as polarizability is one 
of the fundamental properties of the nucleon when tested by an 
external electromagnetic field. In the next chapter we turn to polarized 
nucleon Compton cross sections and present our predictions for various
asymmetries. The aim of this investigation is to demonstrate the possibility
to extract spin polarizabilities from Compton experiments. In this context, we 
will have another look at spin-averaged Compton cross sections, as these
observables may also give important constraints on the spin-dependent 
polarizabilities.

\chapter{Polarized Nucleon Compton Scattering 
\label{chap:spinpolarized} }
In Chapter~\ref{chap:spinaveraged}, we showed results for unpolarized proton 
Compton cross sections, which are derived by averaging over the initial and 
summing over the final spin states. Now, we concentrate on spin-polarized 
cross sections for proton and neutron, albeit we will briefly return to 
spin-averaged ones in Section~\ref{sec:spincontributions}. These 
investigations aim to demonstrate that determining spin polarizabilities 
directly from experiment is possible. They 
serve as a guideline for forthcoming experiments on spin-polarized 
Compton cross sections.

\section{Asymmetries~-- Formalism \label{sec:asymmetries}}

In this section we define the various quantities under investigation. These are
asymmetries with the incoming photon polarized circularly, 
Section~\ref{sec:asymmetriescircularly}, and linearly,  
Section~\ref{sec:asymmetrieslinearly}. Our predictions for the asymmetries 
using circularly polarized photons are published in Ref.~\cite{polarizedpaper}.

\subsection{Asymmetries with Circularly Polarized Photons}
\label{sec:asymmetriescircularly}

\begin{figure}[!htb]
\begin{center}
\includegraphics*[width=.8\textwidth]
{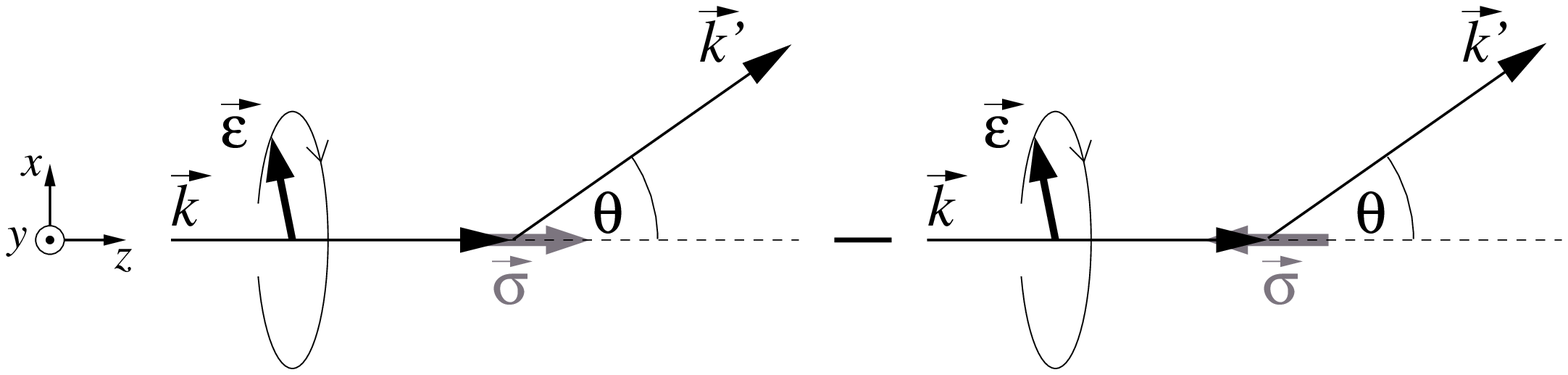}\\\vspace{.5cm}
\includegraphics*[width=.8\textwidth]
{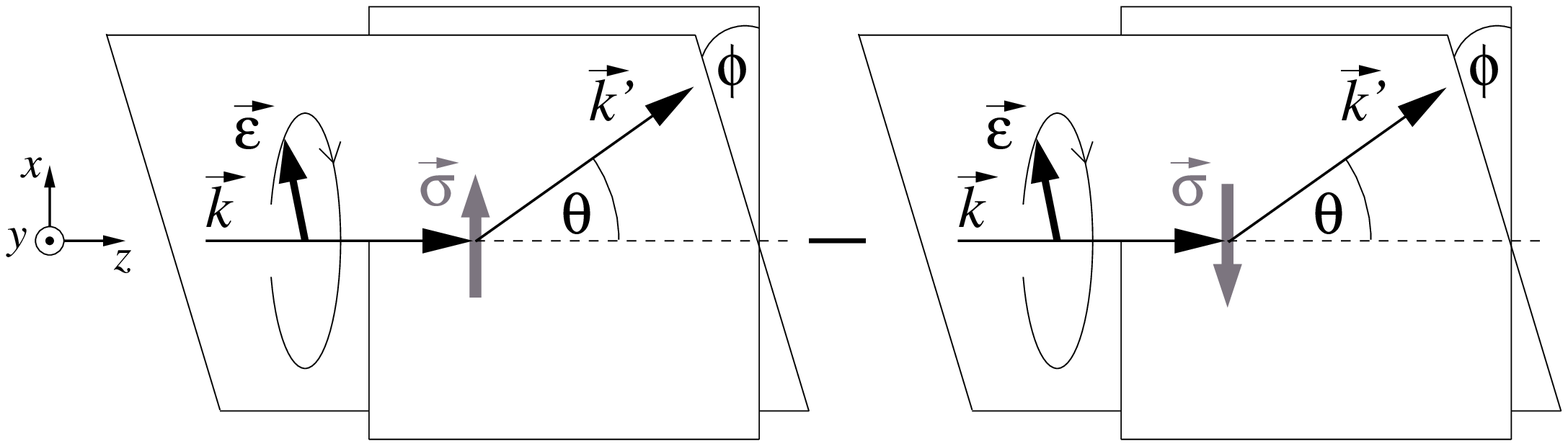}
\caption[The two configurations investigated using circularly polarized 
photons]
{The two configurations investigated using circularly polarized photons; the 
nucleon spin is aligned parallel (upper panel) and perpendicular (lower panel) 
to the direction of the photon momentum.}
\label{fig:circular}
\end{center}
\end{figure}

Triggered by a forthcoming proposal on polarized Compton scattering off
$^3\!\mathrm{He}$ at the HI$\gamma$S lab of TUNL \cite{Gao}, we choose the
incoming photon to be right-circularly polarized,
\be
\vec{\epsilon}=\frac{1}{\sqrt{2}}
\begin{pmatrix}
1\\
i\\
0
\end{pmatrix}
\ee
and to move along the positive $z$-direction, while the final polarization and
nucleon spin remain undetected. The two nucleon spin configurations we
investigate are 
\begin{itemize}
\item[1)] the difference between the target-nucleon spin pointing parallel or
antiparallel to the incident photon momentum
\be
\frac{d\sigma_{\uparrow  \Uparrow}}{d\Omega_{\mathrm{cm}}} -
\frac{d\sigma_{\downarrow\Uparrow}}{d\Omega_{\mathrm{cm}}};
\ee
\item[2)] the difference between the target-nucleon spin aligned in positive
or negative $x$-direction:
\be
\frac{d\sigma_{\rightarrow\Uparrow}}{d\Omega_{\mathrm{cm}}} -
\frac{d\sigma_{\leftarrow \Uparrow}}{d\Omega_{\mathrm{cm}}}.
\ee
\end{itemize}
The first arrow in our notation denotes the direction of the nucleon spin, the
second one the direction of the incoming, right-circularly polarized photon. 
Both configurations are sketched in Fig.~\ref{fig:circular}.

The corresponding formulae for $|T|^2$ have already been derived in \cite{BKM},
albeit there they are given only for real amplitudes $A_1-A_6$. However, 
these amplitudes become complex for photon energies above the 
pion-production threshold $\w_\pi$, cf. Section~\ref{sec:polarizabilities1}. 
Including the imaginary part of the amplitudes, the formulae read
\ba
\frac{1}{2}\,\left(|T|^2_{\uparrow\Uparrow}\right.
&\left.-|T|^2_{\downarrow\Uparrow}\right)
=-\mathrm{Re}[A_1\,A_3^*]\,(1+\cos^2\theta)
-\biggl[|A_3|^2+2\,|A_6|^2+2\,|A_5|^2\,\cos^2\theta\nonumber\\
&+\mathrm{Re}[A_6\,(A_1^*+3\,A_3^*)]
+\biggl(\mathrm{Re}[A_3\,(3\,A_5^*+A_4^*-A_2^*)]
+\mathrm{Re}[A_5\,(4\,A_6^*-A_1^*)]\biggl)\,\cos\theta\nonumber\\
&+\mathrm{Re}[A_5\,(A_2^*-A_4^*)]\,\sin^2\theta \biggl]\,\sin^2\theta
\label{eq:paranti}
\end{align}
and
\ba
\frac{1}{2}\,\left(|T|^2_{\rightarrow\Uparrow}\right.&
\left.-|T|^2_{\leftarrow\Uparrow}\right)
=\biggl[ \mathrm{Im}[A_1\,(A_3^*+2\,A_6^*+2\,A_5^*\,\cos\theta)]\,\cos\theta
+\mathrm{Im}[A_1\,A_4^*]\,(1+\cos^2\theta)\nonumber\\
&-\mathrm{Im}[A_2\,(A_3^*+2\,A_6^*)]\,\sin^2\theta
-\mathrm{Im}[A_2\,(A_4^*+2\,A_5^*)]\,\cos\theta\,\sin^2\theta\biggl]
\sin\theta\,\sin\phi\nonumber\\
&+\biggl[\mathrm{Re}[A_3\,(A_3^*-A_1^*+2\,A_6^*)]\,\cos\theta
+\mathrm{Re}[A_3\,A_5^*]\,(3\,\cos^2\theta-1)\nonumber\\
&+\biggl(\mathrm{Re}[A_1\,A_5^*]+\mathrm{Re}[A_2\,A_3^*]
+\mathrm{Re}[A_6\,(A_2^*+A_4^*-2\,A_5^*)]\biggl)\,\sin^2\theta\nonumber\\
&+\mathrm{Re}[A_3\,A_4^*]\,(\cos^2\theta+1)
+\mathrm{Re}[A_5\,(A_2^*-A_4^*-2\,A_5^*)]\,\cos\theta\,\sin^2\theta
\biggl]\sin\theta\,\cos\phi.
\label{eq:rightleft}
\end{align}
Here, $\phi$ is the angle between the reaction plane and the plane spanned by
the momentum of the incoming photon and the target-nucleon spin, cf. 
Fig.~\ref{fig:circular}. Obviously,
the difference Eq.~(\ref{eq:rightleft}) takes on the largest values~-- at
least below the pion-production threshold~-- for $\phi=0$. Therefore, we
choose the nucleon spin in the reaction plane, which simplifies
Eq.~(\ref{eq:rightleft}) considerably. Using left- instead of
right-circularly polarized photons changes the overall sign in
Eqs.~(\ref{eq:paranti}) and~(\ref{eq:rightleft}).
The spin-averaged cross section can be derived by taking the sum 
instead of the difference in Eq.~(\ref{eq:paranti}) 
(or as well in Eq.~(\ref{eq:rightleft})). The corresponding $|T|^2$ is given 
in Eq.~(\ref{eq:Tmatrix}).

The asymmetries we consider\footnote{$\Sigma_{z}^\mathrm{circ}$ corresponds 
to $\Sigma_{2z}$ in the notation of \cite{Babusci}, $\Sigma_x^\mathrm{circ}$ 
to $\Sigma_{2x}$.} are
\begin{align}
\Sigma_z^\mathrm{circ}
&=\frac{|T|^2_{\uparrow\Uparrow}-|T|^2_{\downarrow\Uparrow}}
       {|T|^2_{\uparrow\Uparrow}+|T|^2_{\downarrow\Uparrow}},
\label{eq:Sigmazcirc}\\
\Sigma_x^\mathrm{circ}
&=\frac{|T|^2_{\rightarrow\Uparrow}-|T|^2_{\leftarrow\Uparrow}}
       {|T|^2_{\rightarrow\Uparrow}+|T|^2_{\leftarrow\Uparrow}}.
\label{eq:Sigmaxcirc}
\end{align}
$\Sigma$ is a frame-independent quantity, as the frame-dependent flux factor
cancels in the ratio between difference and sum of the cross section, while
$|T|^2$ can be written in terms of the frame-independent Mandelstam variables.

From the experimentalist's point of view, it is more convenient to measure the
\textit{asymmetry}~-- i.e.~the difference divided by the sum~-- instead of the
differences Eqs.~(\ref{eq:paranti}) and~(\ref{eq:rightleft}), as the former
is more tolerant to systematic errors in experiments. 
However, division
by a small quantity, say a small spin-averaged cross section, may enhance
theoretical uncertainties. Sensitivity on the nucleon structure, e.g. the
spin polarizabilities, may be lost by dividing the difference by the sum. 
Whenever this happens in Sections \ref{sec:protonasymmetries} and 
\ref{sec:neutronasymmetries}, we will give hints in the text, but we refrain 
from showing our results for the differences~(\ref{eq:paranti}, 
\ref{eq:rightleft}) for reasons of compactification, as it is questionable 
whether one can compare with experimental data for absolute values of 
polarized Compton cross sections within the next few years. 

\subsection{Asymmetries with Linearly Polarized Photons}
\label{sec:asymmetrieslinearly}

\begin{figure}[!htb]
\begin{center}
\includegraphics*[width=.8\textwidth]
{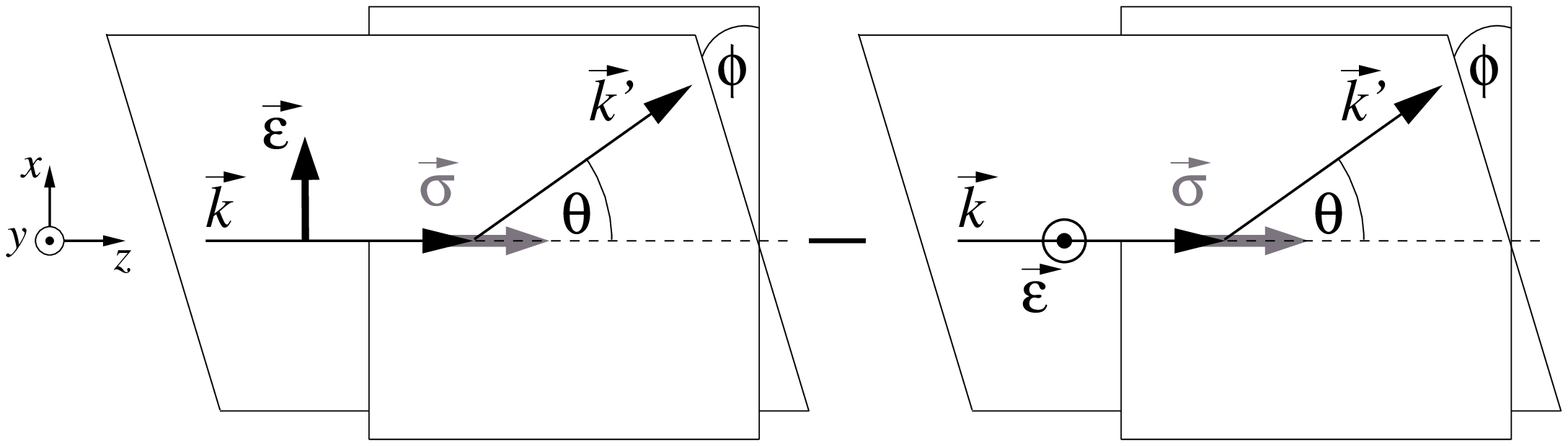}\\\vspace{.5cm}
\includegraphics*[width=.8\textwidth]
{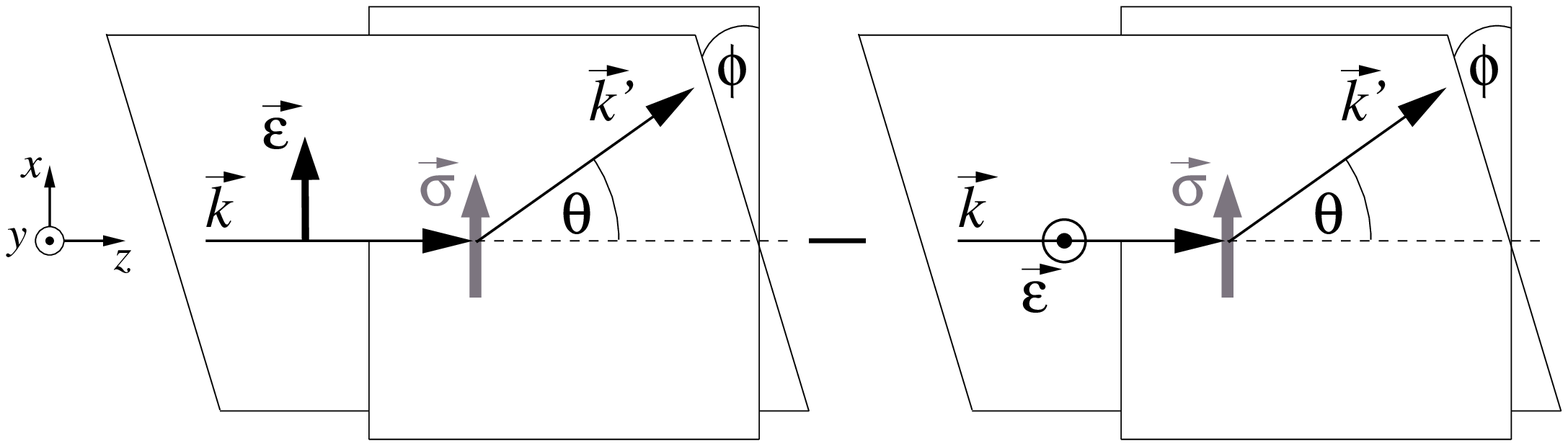}
\caption[The two configurations investigated using linearly polarized 
photons]
{The two configurations investigated using linearly polarized photons; the 
nucleon spin is aligned parallel (upper panel) and perpendicular (lower panel) 
to the direction of the photon momentum. Note that in both systems the 
photon polarization is flipped between the two measurements.}
\label{fig:linear}
\end{center}
\end{figure}

Besides the asymmetries with circularly polarized photons in the initial state,
defined in Eqs.~(\ref{eq:Sigmazcirc}, \ref{eq:Sigmaxcirc}), 
we also investigate asymmetries with linearly polarized photons. 
Again there are two configurations to be considered. The only difference 
between these asymmetries is the direction of the nucleon spin, 
which is aligned in the positive $z$- or $x$-direction, respectively. In both 
cases we define $\phi$ as the angle between the plane 
perpendicular to the $y$-direction and the reaction plane. Our 
calculation shows that for fixed angle $\phi$ it is necessary to change 
the photon polarization between the two measurements, in order to obtain a 
non-vanishing asymmetry below the pion-production threshold.
The two asymmetries read
\begin{align}
\Sigma_z^\mathrm{lin}
&=\frac{|T|^2_{\uparrow\rightarrow}-|T|^2_{\uparrow\odot}}
       {|T|^2_{\uparrow\rightarrow}+|T|^2_{\uparrow\odot}},
\label{eq:Sigmazlin}\\
\Sigma_x^\mathrm{lin}
&=\frac{|T|^2_{\rightarrow\rightarrow}-|T|^2_{\rightarrow\odot}}
       {|T|^2_{\rightarrow\rightarrow}+|T|^2_{\rightarrow\odot}}.
\label{eq:Sigmaxlin}
\end{align}
Again the first arrow corresponds to the direction of the nucleon spin, 
whereas here the 
second one denotes the photon polarization, with the positive $y$-axis 
symbolised by $\odot$. 
Obviously, $\Sigma_x^\mathrm{lin}$ is equivalent to having the nucleon spin 
aligned in the positive $y$-direction~-- both cases emerge from each other 
under the rotation $\phi\rightarrow \phi+90^\circ$.
The two kinematical systems are illustrated in Fig.~\ref{fig:linear}.

The numerators of Eqs.~(\ref{eq:Sigmazlin}) and (\ref{eq:Sigmaxlin}) are 
\begin{align}
\frac{1}{2}\,\left(|T|^2_{\uparrow\rightarrow}\right.
&-\left.|T|^2_{\uparrow\odot}\right)
 =\frac{1}{2}\,\biggl\{\biggl[\R[A_3\,(A_3^*+4\,A_6^*)]-|A_1|^2+4\,|A_6|^2
\nonumber\\
&+2\,\left(\R[A_3\,A_4^*]-\R[A_1\,A_2^*]+2\,\R[A_3\,A_5^*]
+2\,\R[(A_4+2\,A_5)\,A_6^*]\right)\,\cos\theta\nonumber\\
&+4\,\R[(A_4+A_5)\,A_5^*]\,\cos^2\theta+(|A_2|^2-|A_4|^2)\,\sin^2\theta
\biggl]\,\left(2\,\cos^2\phi-1\right)\nonumber\\
&+4\,\biggl[\I[(A_1+A_3)\,A_6^*]+\I[A_1\,A_3^*]+\left(\I[A_1\,A_5^*]+
\I[A_2\,A_3^*]+\I[A_3\,(A_4^*+A_5^*)]\right.\nonumber\\
&\left.-2\,\I[A_4\,A_6^*]\right)\,\cos\theta
-\I[A_2\,A_5^*]\,\sin^2\theta-\I[A_4\,A_5^*]\,(\cos^2\theta+1)
\biggl]\,\sin\phi\,\cos\phi\biggl\}\,\sin^2\theta
\label{eq:diffzlin}
\end{align}
and
\begin{align}
\frac{1}{2}\,\left(|T|^2_{\rightarrow\rightarrow}\right.
&-\left.|T|^2_{\rightarrow\odot}\right)
 =\frac{1}{2}\,\biggl\{\biggl[\R[A_3\,(A_3^*+4\,A_6^*)]-|A_1|^2+4\,|A_6|^2
\nonumber\\
&+2\,\left(\R[A_3\,A_4^*]-\R[A_1\,A_2^*]+2\,\R[A_3\,A_5^*]
+2\,\R[(A_4+2\,A_5)\,A_6^*]\right)\,\cos\theta\nonumber\\
&+4\,\R[(A_4+A_5)\,A_5^*]\,\cos^2\theta+(|A_2|^2-|A_4|^2)\,\sin^2\theta
\biggl]\,\left(2\,\cos^2\phi-1\right)\,\sin\theta\nonumber\\
&+\biggl[2\,\biggl(\left(\I[A_2\,A_3^*]+2\,\I[A_5\,A_6^*]+\I[A_1\,A_4^*]\,
(1-2\,\cos^2\phi)+2\,\left(\I[A_3\,A_4^*]\right.\right.\nonumber\\
&\left.\left.-\I[A_4\,A_6^*]\right)\,\cos^2\phi
+2\,\I[A_2\,A_6^*]\,\sin^2\phi\right)\,\sin^2\theta-2\,\I[A_1\,A_5^*]\,
(\cos^2\theta+\cos^2\phi\,\sin^2\theta)\nonumber\\
&-2\,\I[A_3\,A_5^*]\,(1-\cos^2\phi\,\sin^2\theta)\biggl)
+2\,\biggl(\left(\I[A_2\,A_4^*]\,(1-2\,\cos^2\phi)
-2\,\I[A_4\,A_5^*]\,\cos^2\phi\right.\nonumber\\
&\left.+2\,\I[A_2\,A_5^*]\,\sin^2\phi\right)\,
\sin^2\theta-\I[A_1\,A_3^*]-2\,\I[(A_1+A_3)\,A_6^*]\biggl)\,
\cos\theta\biggl]\,\sin\phi\biggl\}\,\sin\theta.
\label{eq:diffxlin}
\end{align}
In the first configuration the sum gives the spin-averaged cross 
section, however in the latter this is not the case. Here an additional
term arises above threshold:
\begin{align}
\frac{1}{2}\, \left(|T|^2_{\rightarrow\rightarrow}\right.
&+\left.|T|^2_{\rightarrow\odot}\right)
=\frac{1}{2}\,(|T|^2_{\uparrow\Uparrow} +|T|^2_{\downarrow\Uparrow})\nonumber\\
&+\biggl[
\I[A_1\,(A_3^*+2\,A_6^*)]\,\cos\theta+\I[A_1\,A_4^*]\,(1+\cos^2\theta)
+2\,\I[A_1\,A_5^*]\,\cos^2\theta\nonumber\\
&-\left(\I[A_2\,(A_3^*+2\,A_6^*)]
 +\I[A_2\,(A_4^*+2\,A_5^*)]\,\cos\theta\right)\,\sin^2\theta
\biggl]\,\sin\phi\,\sin\theta
\end{align}

Comparing Eqs.~(\ref{eq:diffzlin}) and (\ref{eq:diffxlin}), we find that the 
real parts of the various products of amplitudes 
contributing to the two differences are identical. Therefore there is no 
difference between the two configurations below the pion-production threshold. 
Furthermore the factor $(2\,\cos^2\phi-1)$ is largest for $\phi=0$. For this 
choice of $\phi$, however, the real parts are the only contributions in both 
cases. Therefore
we set $\phi\equiv 0$ and only investigate one asymmetry with linearly 
polarized photons in Sections~\ref{sec:protonasymmetries} 
and~\ref{sec:neutronasymmetries}, denoted by $\Sigma^\mathrm{lin}$.

Another interesting observation in Eqs.~(\ref{eq:diffzlin}, \ref{eq:diffxlin}) 
is the fact that there exist no interference terms in the real parts between 
spin-independent ($A_1,\;A_2$) and spin-dependent ($A_3$-$A_6$) amplitudes.
We are interested in the structure of the nucleon, especially in its spin 
structure. Therefore, we suspect that in the proton case this configuration is
less suited than the other two, where such effects are amplified by 
interference with the strong proton pole amplitudes $A_1^\mathrm{pole}$ and 
$A_2^\mathrm{pole}$, cf. Appendix~\ref{app:poleterms}.

\section[Extracting Spin Polarizabilities from Experiment]
{Extracting Spin Polarizabilities from\\Experiment 
\label{sec:extracting}}
\markboth{CHAPTER \ref{chap:spinpolarized}. POLARIZED NUCLEON COMPTON 
SCATTERING}{\ref{sec:extracting}. EXTRACTING SPIN POLARIZABILITIES FROM 
EXPERIMENT}
One of our aims in this chapter is to prove the possibility of a direct 
determination of dynamical spin polarizabilities from experiment.
We start from our findings for the spin-independent dipole
polarizabilities $\alpha_{E1}(\w)$ and $\beta_{M1}(\w)$, which show
very good agreement with Dispersion-Relation Analysis up to about $170$~MeV, 
cf.~Fig.~\ref{fig:spinindependentpolas}. 
Truncating the multipole decomposition 
at $l=1$, this leaves no unknowns in $A_1$ and $A_2$. As higher 
polarizabilities are negligible 
(cf. Sects.~\ref{sec:spincontributions}-\ref{sec:neutronasymmetries}), the 
spin-dependent dipole polarizabilities could then be fitted to
data sets which combine polarized and spin-averaged experimental results,
taken at a fixed energy and varying the scattering angle. As starting values 
for the fit
one may use our $\chi$EFT results (see Fig.~\ref{fig:spindependentpolas}),
as indicated in Eq.~(\ref{eq:strucamp}), where we show the structure 
amplitudes up to $l=1$ with the spin polarizabilities $\gamma_i(\w)$ 
replaced by $\gamma_i(\w)+\delta_i$, introducing the fit parameters 
$\delta_i$. The expansion~(\ref{eq:strucamp}) has been derived using 
Eqs.~(\ref{eq:Ritus},~\ref{eq:ritus},~\ref{eq:spinindiepolasdef}) and 
(\ref{eq:spinpolasdef}).  
Small fit parameters mean correct prediction of the dynamical 
spin-dipole polarizabilities within the Small Scale Expansion.
\begin{align}
\bar{A}_1(\w,z) &=\frac{4\pi\,W}{M}\,\left[\alpha_{E1}(\w)
+z\,\beta_ {M1}(\w)\right]\,\w^2\nonumber\\
\bar{A}_2(\w,z) &=-\frac{4\pi\,W}{M}\,\beta_{M1}(\w)\,\w^2
\nonumber\\
\bar{A}_3^{\mathrm{fit}}(\w,z)
&=-\frac{4\pi\,W}{M}\,[(\gamma_{E1E1}(\w)+\delta_{E1E1})
+z\,(\gamma_{M1M1}(\w)+\delta_{M1M1})\nonumber\\
&+(\gamma_{E1M2}(\w)+\delta_{E1M2})
+z\,(\gamma_{M1E2}(\w)+\delta_{M1E2})]\,\w^3\nonumber\\
\bar{A}_4^{\mathrm{fit}}(\w,z) &=\frac{4\pi\,
W}{M}\,\left[-(\gamma_{M1M1}(\w)+\delta_{M1M1})
+(\gamma_{M1E2}(\w)+\delta_{M1E2})\right]\,\w^3\nonumber\\
\bar{A}_5^{\mathrm{fit}}(\w,z) &=\frac{4\pi\, W}{M}\,
(\gamma_{M1M1}(\w)+\delta_{M1M1})\,\w^3\nonumber\\
\bar{A}_6^{\mathrm{fit}}(\w,z) &=\frac{4\pi\,
W}{M}\,(\gamma_{E1M2}(\w)+\delta_{E1M2})\,\w^3
\label{eq:strucamp}
\end{align}
Thus, one obtains the spin-dipole polarizabilities at a definite energy.
Repeating this procedure for various energies gives the energy dependence,
i.e. the dynamics of the $l=1$ spin polarizabilities. Therefore, the
amplitudes Eq.~(\ref{eq:strucamp}) provide one possible way to extract 
dynamical spin polarizabilities directly from the asymmetry observables of 
the previous section, using $\chi$EFT.
Note that the $\delta_i$ may show a weak energy dependence.
At first trial, they can be taken 
as energy-independent quantities. This corresponds to a free 
normalization of the spin-dipole polarizabilities, assuming the energy 
dependence derived from $\chi$EFT to be correct. This assumption is well 
justified, as at low energies only $\Delta(1232)$ and pion degrees of freedom 
are supposed to give dispersive contributions to the polarizabilities, and 
usually the pion-cloud dispersion is well captured in Chiral Perturbation 
Theory.

\begin{figure}[!htb]
\begin{center}
\includegraphics*[width=.25\textwidth]
{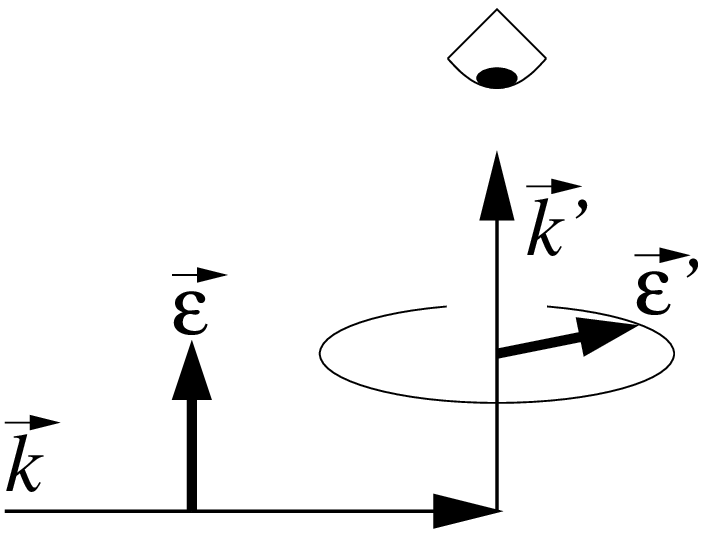}\hspace{.1\textwidth}
\includegraphics*[width=.25\textwidth]
{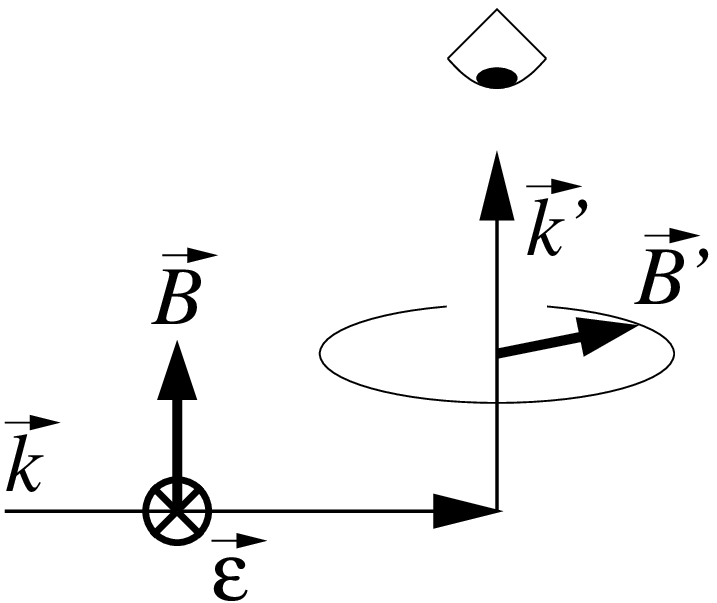}
\caption[Two configurations of special interest with linearly polarized 
photons]
{Two experimental configurations with linearly polarized photons, which 
exclude contributions from $\alpha_{E1}(\w)$ (left panel) and $\beta_{M1}(\w)$ 
(right panel).}
\label{fig:configpol}
\end{center}
\end{figure}

For certain scattering angles, the theoretical error due to the dynamical 
spin-independent dipole polarizabilities $\alpha_{E1}(\w)$ and 
$\beta_{M1}(\w)$ is even further reduced. In order to prove this claim we 
rewrite the effective Hamiltonian up to dipole order,
\ba
H_\mathrm{eff}&=-2\pi\,\left[\alpha_{E1}(\w)\,\vec{E}^2+
\beta_{M1}(\w)\,\vec{B}^2+
\gamma_{E1E1}(\w)\,\vec{\sigma}\cdot\vec{E}\times\dot{\vec{E}}\right.\\
&+\left.\gamma_{M1M1}(\w)\,\vec{\sigma}\cdot\vec{B}\times\dot{\vec{B}}-
2\,\gamma_{M1E2}(\w)\,\sigma_i\,E_{ij}\,B_j+
2\,\gamma_{E1M2}(\w)\,\sigma_i\,B_{ij}\,E_j\right]\nonumber
\end{align}
with $T_{ij}=\frac{1}{2}(\partial_i T_j+\partial_j T_i)$, cf. 
Section~\ref{sec:intro}.
Obviously, when the polarizations of the incoming and the outgoing photon 
are perpendicular to each other, i.e. $\vec{\epsilon}\perp\vec{\epsilon}\,'$, 
the scalar product $\vec{E}^2\propto\vec{\epsilon}\cdot\vec{\epsilon}\,'$ 
vanishes and therefore $\alpha_{E1}(\w)$ cannot
contribute. This scenario is depicted in the left panel of 
Fig.~\ref{fig:configpol}, where the incoming photon is polarized along the 
$x$-direction and the outgoing photon is detected under this very direction, 
i.e. under the scattering angles $\theta=\frac{\pi}{2}$, $\phi=0$.
In the right panel, the incoming photon is polarized along the $y$-direction,
and therefore the magnetic fields of initial and final photon are 
perpendicular to each other. In this case $\beta_{M1}(\w)$ gives no 
contribution. One might expect that there exist similar exclusion principles
for the spin polarizabilities $\gamma_{E1E1}(\w)$ and $\gamma_{M1M1}(\w)$, 
when the nucleon spin is aligned along a 
well-defined direction in the initial state. 
However, this would only be the case, if we would calculate asymmetries 
with polarized nucleons also in the final state.
Nevertheless, for $\theta=\frac{\pi}{2}$, $\phi=0$ and the initial nucleon
polarized parallel to the $x$-direction, one spin polarizability vanishes,
namely $\gamma_{M1E2}(\w)$ for the incoming photon  polarized 
parallel to the nucleon spin, $\gamma_{E1M2}(\w)$ for 
$\vec{\epsilon}\parallel\vec{e}_y$. 
Unfortunately, these mixed spin polarizabilities are highly non-intuitive. 
Therefore we cannot give a heuristic explanation for this behaviour.

In the next section we confirm our claim that determining spin polarizabilities
from experiment is possible by  
demonstrating that also spin-averaged Compton cross 
sections may contribute to this demanding task.

\section[Spin Contributions to Spin-Averaged Cross Sections]
{Spin Contributions to Spin-Averaged Cross\\Sections
\label{sec:spincontributions}}
\markboth{CHAPTER \ref{chap:spinpolarized}. POLARIZED NUCLEON COMPTON 
SCATTERING}{\ref{sec:spincontributions}. SPIN CONTRIBUTIONS TO SPIN-AVERAGED 
CROSS SECTIONS}

\begin{figure}[!htb]
\begin{center}
\includegraphics*[width=.31\textwidth]{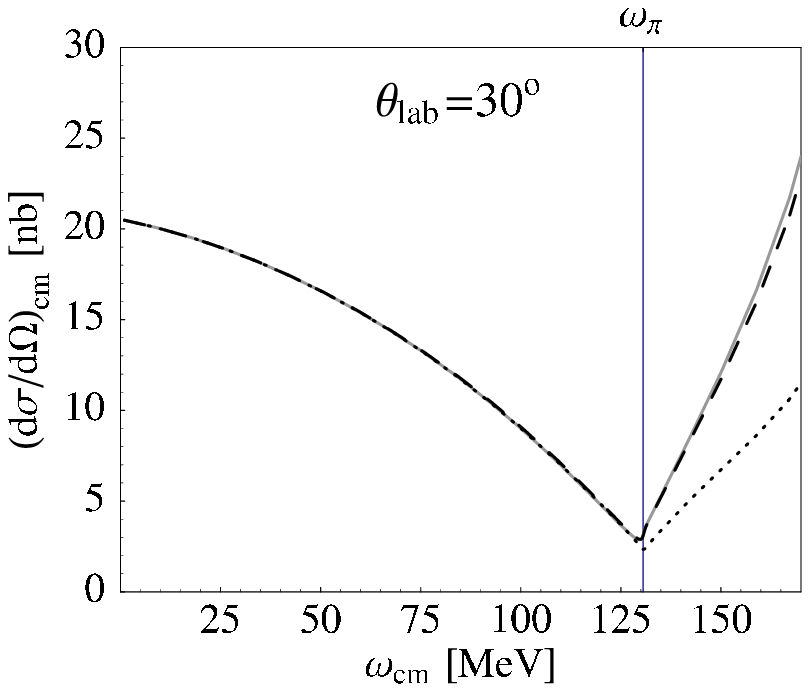}
\hspace{.01\textwidth}
\includegraphics*[width=.31\textwidth]{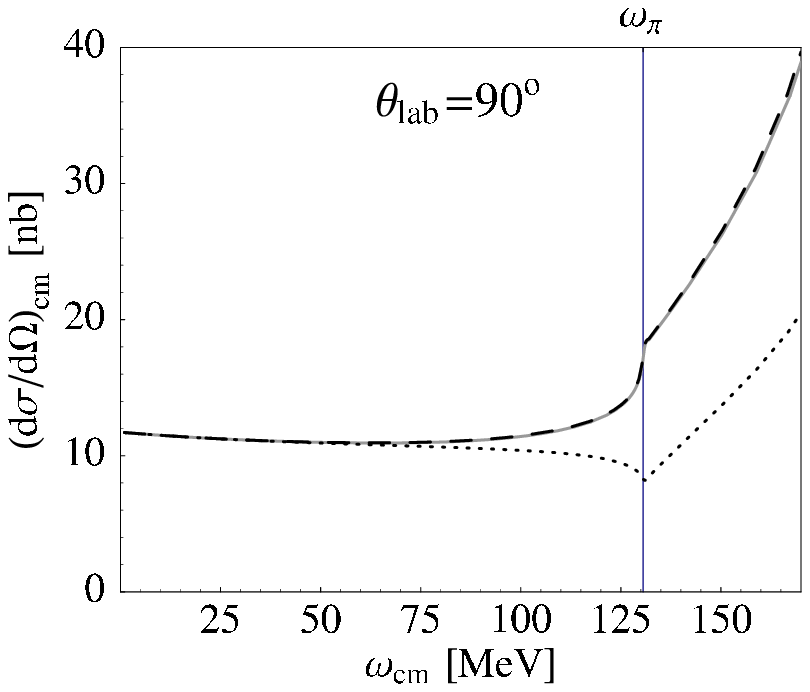}
\hspace{.01\textwidth}
\includegraphics*[width=.31\textwidth]{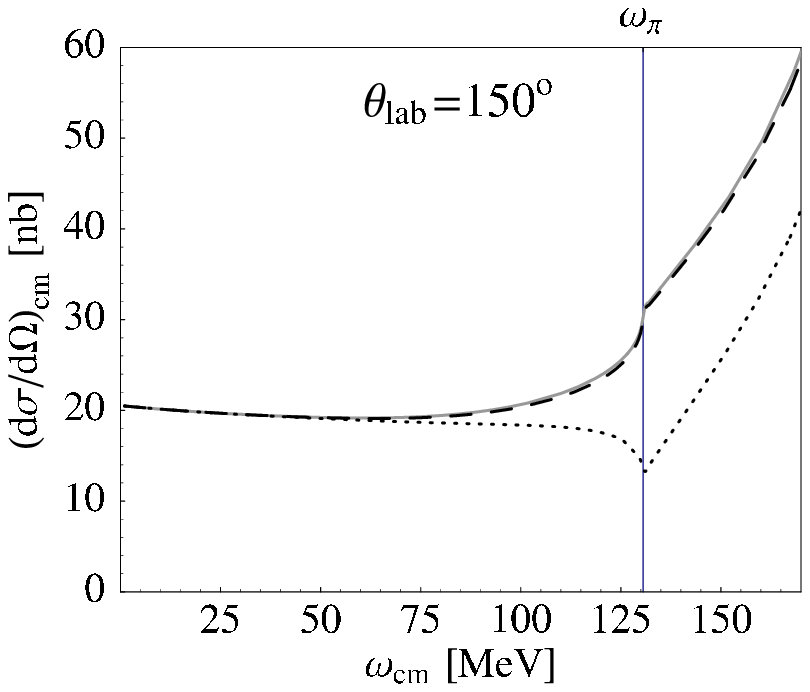}
\caption[SSE predictions for spin contributions to spin-averaged proton 
Compton cross sections]
{Complete $\mathcal{O}(\epsilon^3)$-SSE predictions (grey) for the 
spin-averaged proton cross section; dotted: spin polarizabilities not 
included, dashed: quadrupole polarizabilities not included.}
\label{fig:SSEindiesump}
\end{center}
\end{figure}

\begin{figure}[!htb]
\begin{center}
\includegraphics*[width=.31\textwidth]{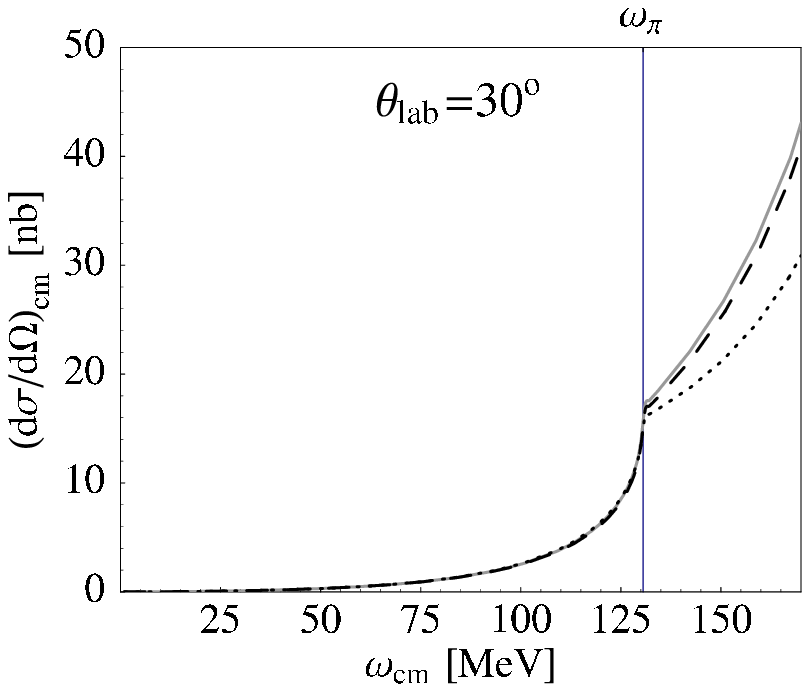}
\hspace{.01\textwidth}
\includegraphics*[width=.31\textwidth]{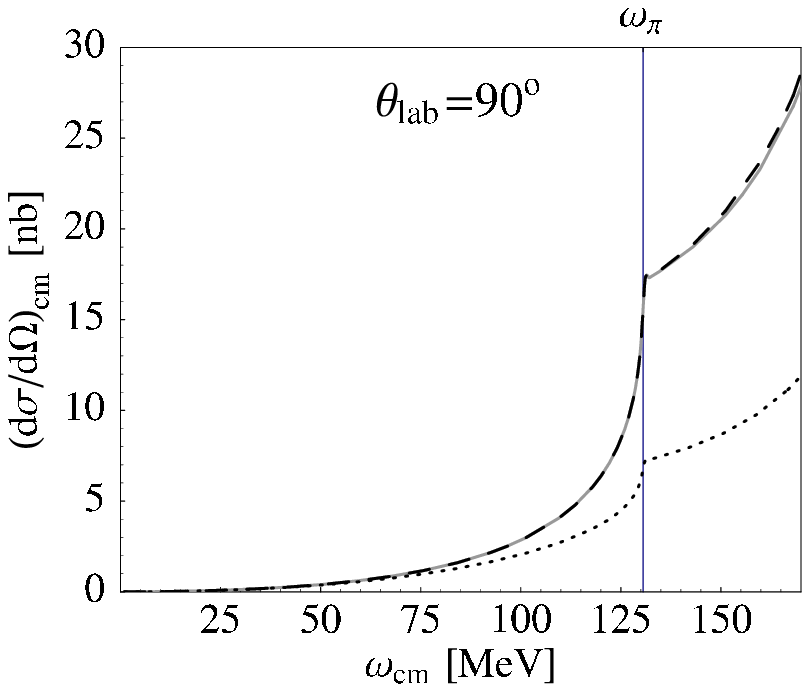}
\hspace{.01\textwidth}
\includegraphics*[width=.31\textwidth]{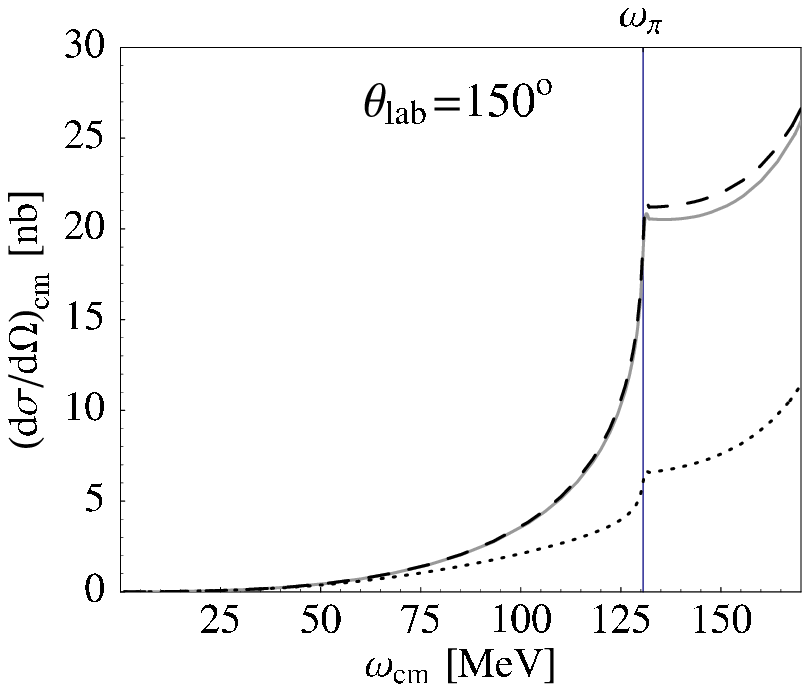}
\caption[SSE predictions for spin contributions to spin-averaged neutron 
Compton cross sections]
{Spin-averaged neutron cross section; for notation see 
Fig.~\ref{fig:SSEindiesump}.}
\label{fig:SSEindiesumn}
\end{center}
\end{figure}

Before discussing the asymmetries in detail, we briefly turn to the question
which polarizabilities are seen in unpolarized Compton cross sections.
To this end we have another look at the $\mathcal{O}(\epsilon^3)$ 
SSE results already discussed in Section~\ref{sec:spin-averaged}, but here we 
focus on the influence of the spin polarizabilities.
As shown in Fig.~\ref{fig:SSEindiesump}, we find a large contribution of the 
dynamical spin polarizabilities to
spin-averaged proton Compton cross sections above $\w\sim
100~\mathrm{MeV}$. We also show our results for the neutron
(Fig.~\ref{fig:SSEindiesumn}), exhibiting a huge sensitivity on the spin
polarizabilities in the backward direction. This can be well understood, as
the right hand side of Eq.~(\ref{eq:Tmatrix}) simplifies to 
$|A_1|^2+|A_3|^2$ for
$\theta=0^\circ$ and $\theta=180^\circ$. In the forward direction, the
spin-independent amplitude $|A_1|^2$ dominates, in the backward direction the
spin-dependent amplitude $|A_3|^2$, as can be seen in Appendix~A of 
Ref.~\cite{polarizedpaper}. 

Recall that we found in Section~\ref{sec:spin-averaged} that any effects of 
quadrupole polarizabilities are invisible at the level of the unpolarized 
proton cross sections. According to Fig.~\ref{fig:SSEindiesumn}, the same 
observation holds for the neutron. It suffices therefore
to terminate the multipole expansion at the dipole level, which leaves 
the six dipole polarizabilities at given energy $\w$ as parameters.

\begin{figure}[!htb]
\begin{center}
\includegraphics*[width=.31\textwidth]{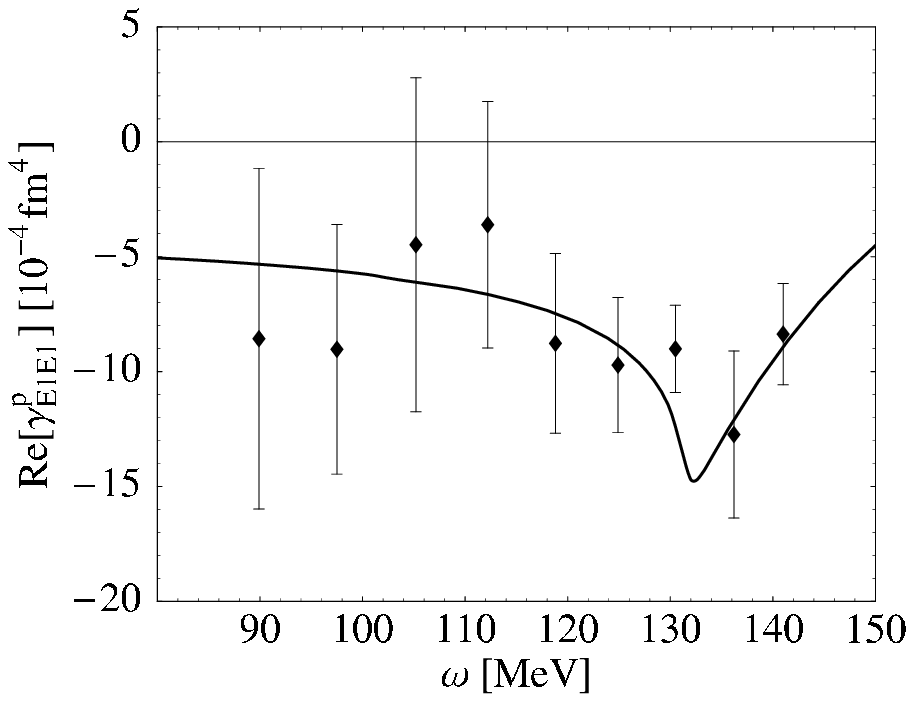}
\hspace{.01\textwidth}
\includegraphics*[width=.31\textwidth]{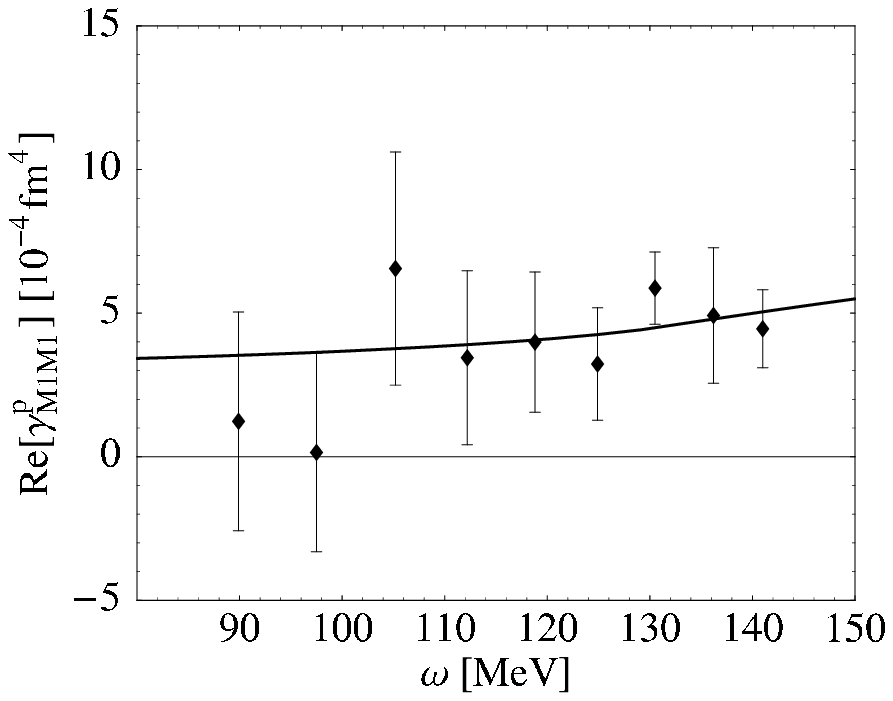}
\hspace{.01\textwidth}
\includegraphics*[width=.31\textwidth]{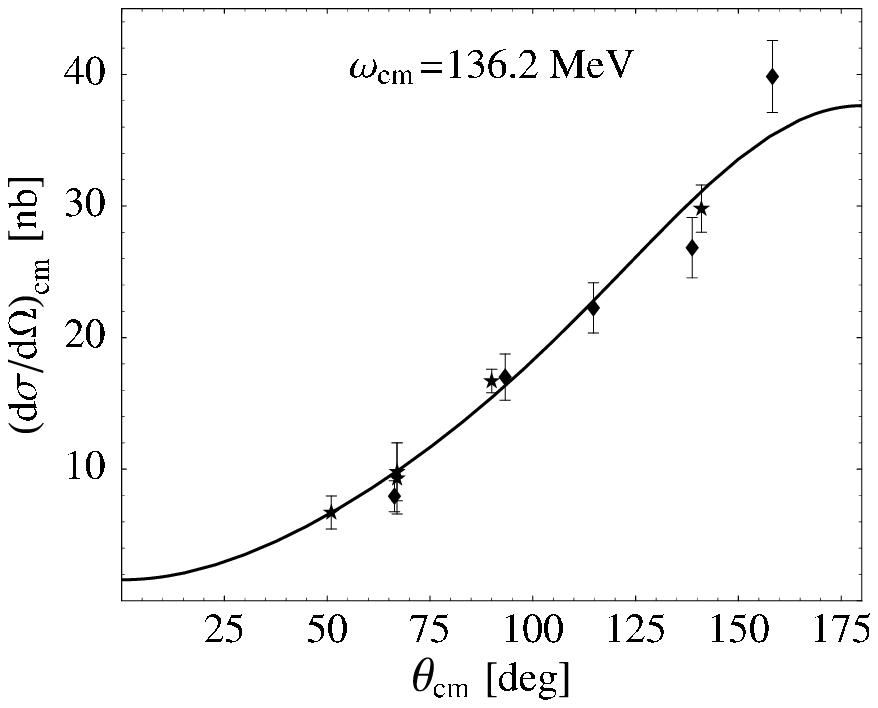}
\caption[Fits of dynamical spin polarizabilities to experimental data]
{Fits of the dynamical spin polarizabilities $\gamma_{E1E1}(\w)$ and 
$\gamma_{M1M1}(\w)$  to experimental proton Compton data from 
\cite{Olmos, Hallin, Fed91, Mac95} at fixed energy and varying angle. The 
solid line in the left two panels is the prediction from the 
Dispersion-Relation Analysis of \cite{HGHP}. The right panel shows one example
for the fitting procedure~-- for the experimental data cf. 
Figs.~\ref{fig:Olmos} and \ref{fig:Hallin}.}
\label{fig:spinpolasfit}
\end{center}
\end{figure}

While effects from the spin polarizabilities are non-negligible in unpolarized
experiments, to extract all four of them from such data is clearly impossible.
Thus, double-polarized experiments as discussed in the rest of this chapter
are necessary additional ingredients in a combined multipole analysis.
Nevertheless, spin-averaged Compton data may give valuable constraints on 
such fits. As proof of principle, we show in 
Fig.~\ref{fig:spinpolasfit} results from fitting the 
dynamical spin polarizabilities $\gamma_{E1E1}(\w)$ and $\gamma_{M1M1}(\w)$ to
 data from \cite{Olmos, Hallin, Fed91, Mac95} at fixed energy and varying 
angle. Note that here we extend the data base with respect to our global 
proton fits of Section~\ref{sec:protonfits}, as we try to increase the number 
of data points per energy value as far as possible. The results are compared  
to the prediction from the 
Dispersion-Relation Analysis of \cite{HGHP}. However,
these fits are rather unstable, as the sensitivity to $\gamma_{E1E1}(\w)$ and 
$\gamma_{M1M1}(\w)$ does not suffice to compensate for the spread in the 
data, cf. right panel of Fig.~\ref{fig:spinpolasfit}, which are oriented 
around a straight line, whereas the cross section is actually of cosine 
shape. In order to improve the convergence by minimizing the 
uncertainty in the fit amplitudes, we therefore use the multipoles from 
DR~\cite{HGHP} for their construction, rather than the SSE amplitudes. Thus,
Fig.~\ref{fig:spinpolasfit} is not supposed to be considered quantitatively, 
but we believe to have established that spin-averaged Compton cross sections
do help to pin down the dynamical spin polarizabilities.

\section{Proton Asymmetries \label{sec:protonasymmetries}}
\subsection[Proton Asymmetries from Circularly Polarized Photons]
{Proton Asymmetries from Circularly Polarized\\Photons 
\label{sec:protoncircularly}}

We turn now to the results for the asymmetries of the proton, using circularly
polarized photons in the initial state.
Analogously to the previous section, we confirm for each observable that
the quadrupole polarizabilities are negligible. Thus, the multipole expansion
of the amplitude can always be truncated at the dipole level, leaving at most
six parameters. However, it will turn out that not all asymmetries are equally
sensitive to the spin polarizabilities. As expected, most asymmetries are 
indeed governed by the pole part of the amplitudes.

In order to determine which asymmetries are most sensitive to the
structure parts of the Compton amplitudes, and which of the internal 
low-energy degrees of freedom in the nucleon dominate the structure-dependent 
part of the cross section, we will first compare three scenarios for each
asymmetry: (i) the result when only the pole terms of the amplitudes are kept;
(ii) the same when the effects from the pion cloud around the nucleon are
added, as described by the leading-one-loop order HB$\chi$PT result; and 
finally (iii) a leading-one-loop order calculation in SSE, including also the 
$\Delta(1232)$ as dynamical degree of freedom.

An ideal asymmetry should thus fulfill three criteria: It should be large to
give a good experimental signal, it should show sensitivity on the structure
amplitudes, and it should allow a differentiation between the pion cloud and
$\Delta$ resonance contributions in resonant channels, revealing as much as 
possible about the role
of at least these low-energy degrees of freedom in the nucleon. In
Section~\ref{sec:neutronasymmetries}, we will repeat this presentation for the
neutron asymmetries. To simplify connection to experiment, we give the 
scattering angle in the following plots in the lab frame.

Similar plots for the nucleon asymmetries are already shown in \cite{Babusci},
using Dispersion Theory techniques. Direct comparison to those plots is
however not possible, as in \cite{Babusci}  the asymmetries are plotted 
against $\w_\mathrm{lab}$ and because of a different choice of 
angles~-- the authors of
\cite{Babusci} concentrated on the extreme angles $0^\circ$, $90^\circ$,
$180^\circ$, whereas we show our results for $30^\circ$, $90^\circ$ and 
$150^\circ$, as the extreme forward and backward direction is nearly 
impossible to access experimentally.\footnote{Note that for reasons of 
compactification we reduce the number of angles with respect to 
Ref.~\cite{polarizedpaper}, where we present our results for $70^\circ$ and 
$110^\circ$ instead of $90^\circ$.} 
Nevertheless, qualitative agreement between our $\chi$EFT results
and \cite{Babusci} can be deduced. 

We emphasize also that our predictions are parameter free, as all constants
are determined from unpolarized Compton scattering, see 
Section~\ref{sec:protonfits}. 
In the following, the fit parameters 
$\delta_i$ introduced in Eq.~(\ref{eq:strucamp}) are
all set to zero, as no measured asymmetries exist at this point.

\subsubsection{Proton Spin Parallel to Photon Momentum}
\label{sec:pparanti}

\begin{figure}[!htb]
\begin{center}
\includegraphics*[width=.31\textwidth]
{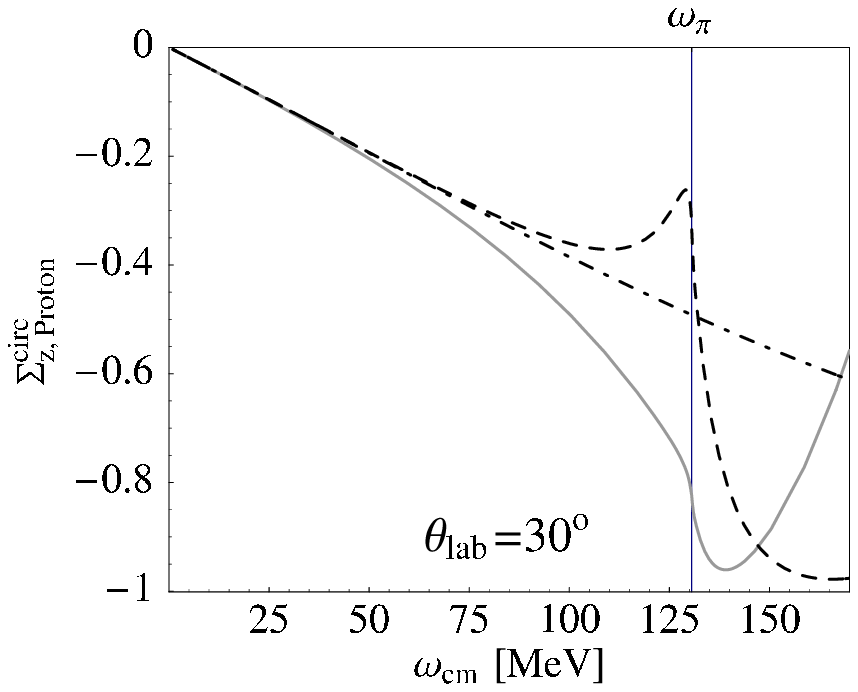}
\hspace{.01\textwidth}
\includegraphics*[width=.31\textwidth]
{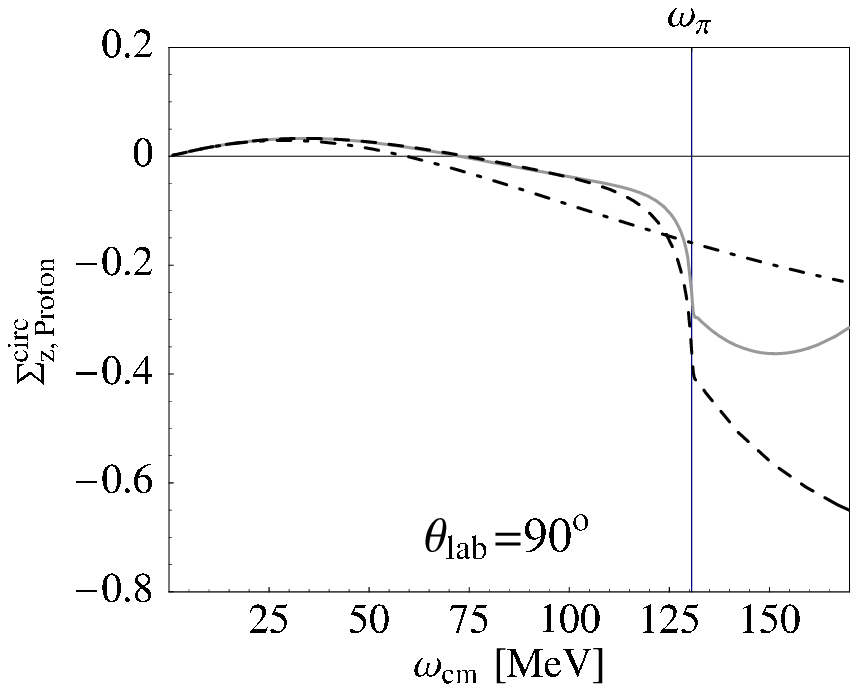}
\hspace{.01\textwidth}
\includegraphics*[width=.31\textwidth]
{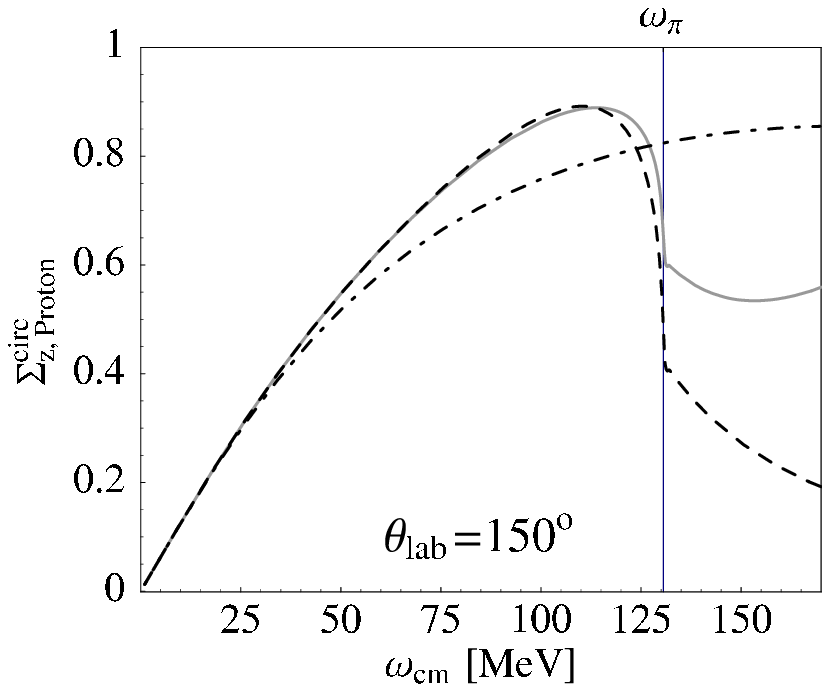}
\caption[Predictions for the proton asymmetry 
$\Sigma_{z,p}^\mathrm{circ}$]
{$\mathcal{O}(p^3)$-HB$\chi$PT (dashed) and
$\mathcal{O}(\epsilon^3)$-SSE predictions (grey) for the proton 
asymmetry $\Sigma_{z,p}^\mathrm{circ}$; the dotdashed line describes the 
third-order pole contributions.}
\label{fig:bornasyp}
\end{center}
\end{figure}

As one can see in Fig.~\ref{fig:bornasyp}, the proton asymmetry 
$\Sigma_{z,p}^\mathrm{circ}$ reaches values of $\mathcal{O}(1)$ and is 
therefore quite large, although for $\w=0$ it vanishes independently of the 
scattering angle, due to the vanishing 
difference and the finite static spin-averaged cross section, given by the 
familiar Thomson limit.

Comparing the three curves in Fig.~\ref{fig:bornasyp}~-- third-order pole,
$\mathcal{O}(p^3)$-HB$\chi$PT and $\mathcal{O}(\epsilon^3)$-SSE~-- one
recognizes the strong influence of the pole amplitudes, given by
Eq.~(\ref{eq:poleterms}). This is exactly what one expects for the charged 
proton, and can also be deduced from Eqs.~(\ref{eq:paranti}, 
\ref{eq:strucamp}, \ref{eq:poleterms}): The asymmetry starts linearly in 
$\w$, while the leading term of the structure part of 
$\Sigma_{z,p}^\mathrm{circ}$ is proportional to $\w^3$, as there is no term in
Eq.~(\ref{eq:paranti}) that contains only spin-independent amplitudes. As we 
are interested in information about the structure of the nucleon, i.e. in the 
deviation of the dashed and grey lines from the dotdashed (pole contributions 
only) line in Fig.~\ref{fig:bornasyp}, and as this deviation is not as strong 
as later in $\Sigma_{x,p}^\mathrm{circ}$ and in the 
neutron asymmetries, $\Sigma_{z,p}^\mathrm{circ}$ does not seem to be an 
ideal choice. 
  
Sizeable contributions from the explicit $\Delta$ degrees of freedom exist 
only above $\w_\pi$. The only exception is noticed in the
extreme forward direction, but this is an artifact of the asymmetry, which is
extremely sensitive at small angles due to the small spin-averaged cross
section at $\w_\pi$ (Fig.~\ref{fig:SSEindiesump}), and neither visible in 
the difference, described by Eq.~(\ref{eq:paranti}),
nor in the spin-averaged cross section.

At $\w_\pi$, the cusp at the pion-production threshold is clearly visible 
for all angles. 
Polarized cross sections are much more sensitive on the fine structure of the 
nucleon than their unpolarized pendants.
Therefore, our results might considerably deviate from experiment above
threshold, as there are sizeable uncertainties in our imaginary parts, due to 
the vanishing $\Delta$ width in our leading-one-loop order SSE calculation, 
cf.~\cite{DA}. Nevertheless,
qualitative agreement should be fulfilled, so we use the same plot 
range as for the unpolarized results in Section~\ref{sec:spin-averaged}, with 
a maximum photon energy of $170$~MeV. In \cite{Babusci}, the plots end below 
threshold since the Dispersion-Relation Analysis is compared to a low-energy 
expansion of the polarizabilities, which  cannot reproduce signatures like the 
steep rise in $\alpha_{E1}(\w)$, connected with the non-analyticity of the 
pion-production threshold, see Fig.~\ref{fig:spinindependentpolas}. 

\begin{figure}[!htb]
\begin{center}
\includegraphics*[width=.31\textwidth]{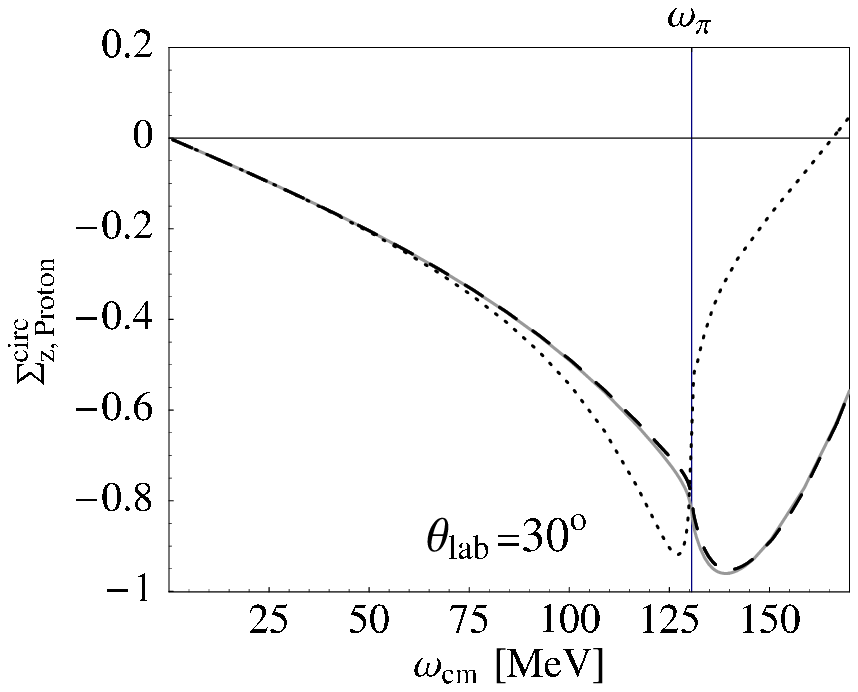}
\hspace{.01\textwidth}
\includegraphics*[width=.31\textwidth]{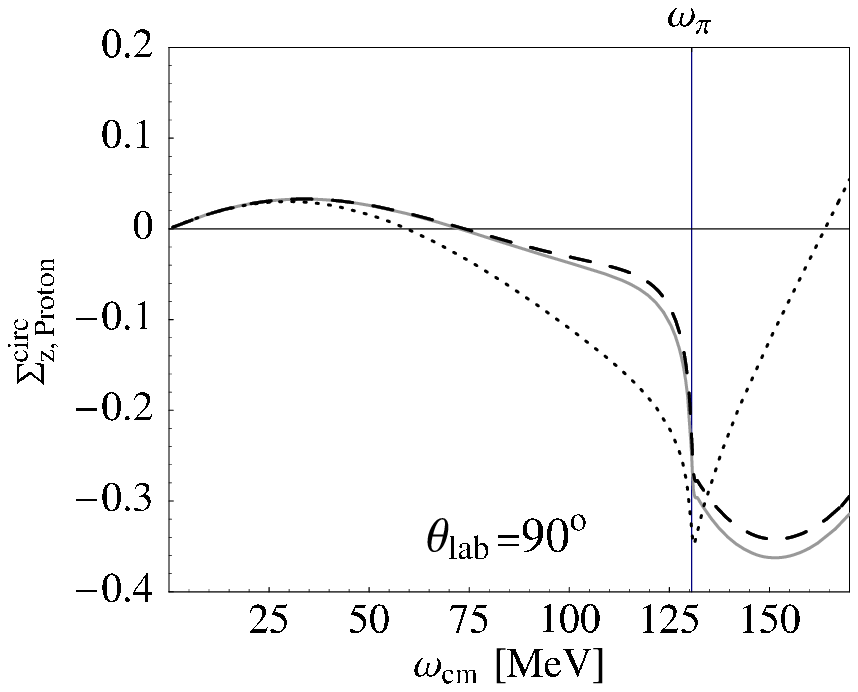}
\hspace{.01\textwidth}
\includegraphics*[width=.31\textwidth]{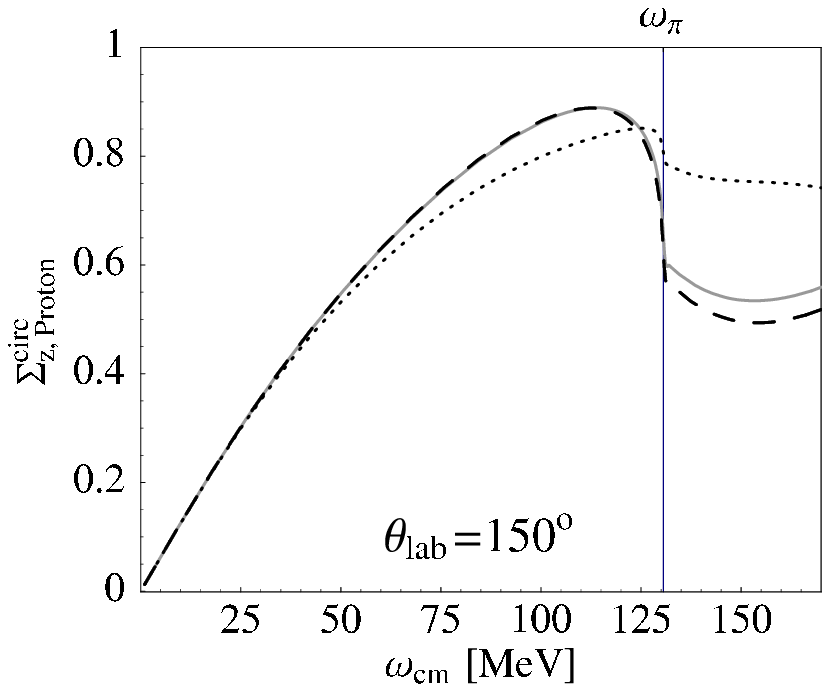}
\caption[Spin contributions to $\Sigma_{z,p}^\mathrm{circ}$]
{Dependence of the proton asymmetry $\Sigma_{z,p}^\mathrm{circ}$ on spin and 
quadrupole polarizabilities; for notation see Fig.~\ref{fig:SSEindiesump}.}
\label{fig:SSEindieasyp}
\end{center}
\end{figure}

The asymmetry $\Sigma_{z,p}^\mathrm{circ}$ exhibits only a weak dependence on
the spin polarizabilities in the forward direction below $\w_\pi$ 
(Fig.~\ref{fig:SSEindieasyp}).
The largest sensitivity is noted around $90^\circ$, 
whereas in the extreme backward direction the dependence on the spin 
polarizabilities partly cancels in the division of the difference by the sum.
The sharp rise of the result without spin polarizabilities in
Fig.~\ref{fig:SSEindieasyp} above the pion-production threshold in the forward
direction is due to a sharply rising difference and the small spin-averaged
cross section which enters the denominator presented in
Fig.~\ref{fig:SSEindiesump}.

In the literature, e.g.~in \cite{Sandorfi}, the pion pole
(Fig.~\ref{fig:app:pole}(d)) is often considered as one of the structure
diagrams, giving the dominant contribution to the static backward spin 
polarizability $\bar{\gamma}_\pi$. We 
treat the term as pole, as it contains a pion pole in the $t$-channel and we 
assume that its contribution to nucleon Compton scattering is well understood.
So the
question arises why we are sensitive to the spin polarizabilities, despite of
having removed this supposedly dominant part from them. The reason is that
the pion pole dominates over the structure part of $\gamma_\pi(\w)$ only for 
low energies. The pion-pole contribution to $\gamma_\pi(\w)$ looks like a
Lorentzian (Eq.~(\ref{eq:poleterms})) and becomes smaller than the structure
contribution above 100~MeV, as the latter rises due to the increasing
values of $\gamma_{E1E1}(\w)$ and $\gamma_{M1M1}(\w)$, cf. 
Section~\ref{sec:dynpolas}.

It is crucial to notice that the quadrupole polarizabilities ($l=2$) play
again a negligibly small role, see
Fig.~\ref{fig:SSEindieasyp}. The most important
quadrupole contribution is observed at $90^\circ$ and $150^\circ$ above 
threshold, but the
relative size is still $<0.1$ and therefore presumably within the experimental
error bars. As repeatedly stated, that these contributions are small is
mandatory if one wants to determine spin polarizabilities via polarized cross
section data, because only then can the multipole expansion be truncated at 
$l=1$ as in Eq.~(\ref{eq:strucamp}).

\subsubsection{Proton Spin Perpendicular to Photon Momentum}

\begin{figure}[!htb]
\begin{center}
\includegraphics*[width=.31\textwidth]
{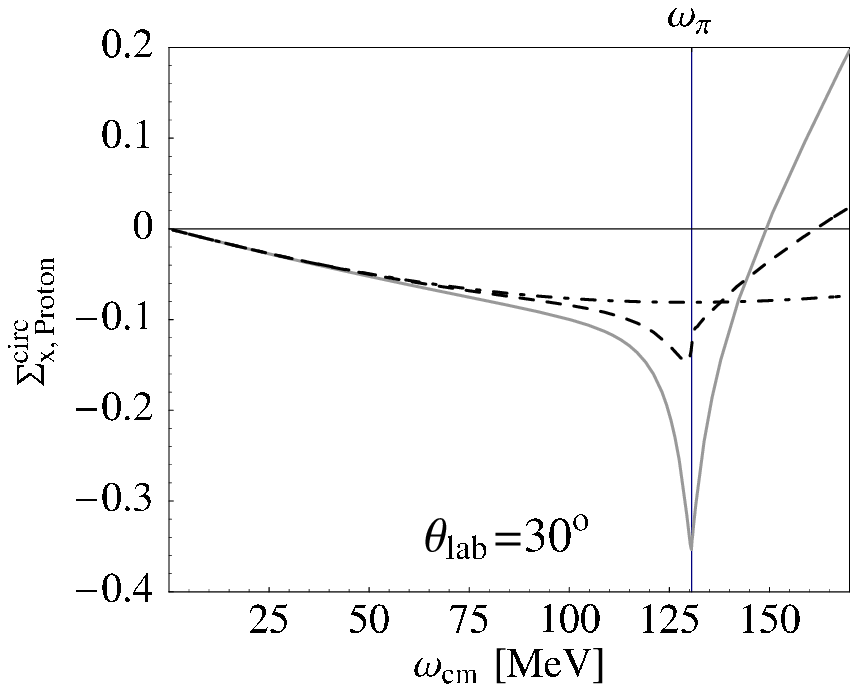}
\hspace{.01\textwidth}
\includegraphics*[width=.31\textwidth]
{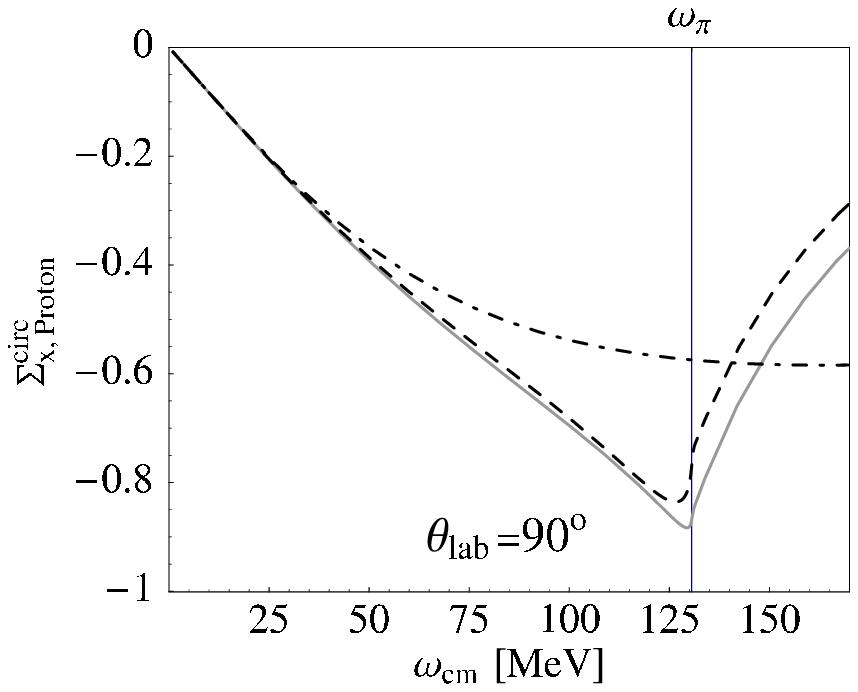}
\hspace{.01\textwidth}
\includegraphics*[width=.31\textwidth]
{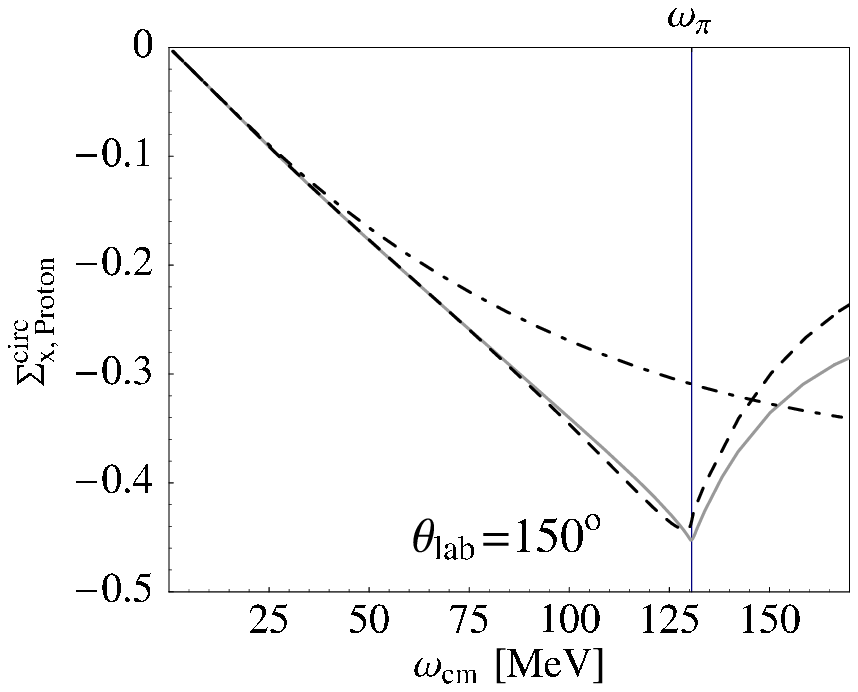}
\caption[Predictions for the proton asymmetry 
$\Sigma_{x,p}^\mathrm{circ}$]
{Dependence of the proton asymmetry $\Sigma_{x,p}^\mathrm{circ}$ on pion and 
$\Delta$ physics; for notation see Fig.~\ref{fig:bornasyp}.}
\label{fig:bornvertasyp}
\end{center}
\end{figure}

The asymmetry $\Sigma_{x,p}^\mathrm{circ}$
in Fig.~\ref{fig:bornvertasyp} looks quite similar for the different angles: It
always starts with a falling slope and exhibits a sharp minimum at the pion
cusp, therefore staying negative in a wide energy range. 

\begin{figure}[!htb]
\begin{center}
\includegraphics*[width=.31\textwidth]
{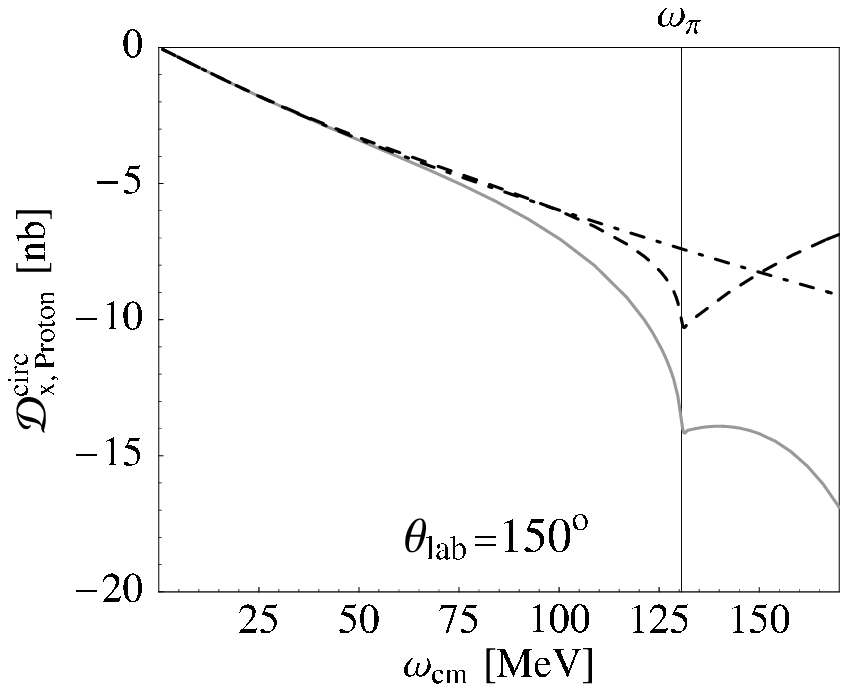}
\hspace{.01\textwidth}
\includegraphics*[width=.31\textwidth]
{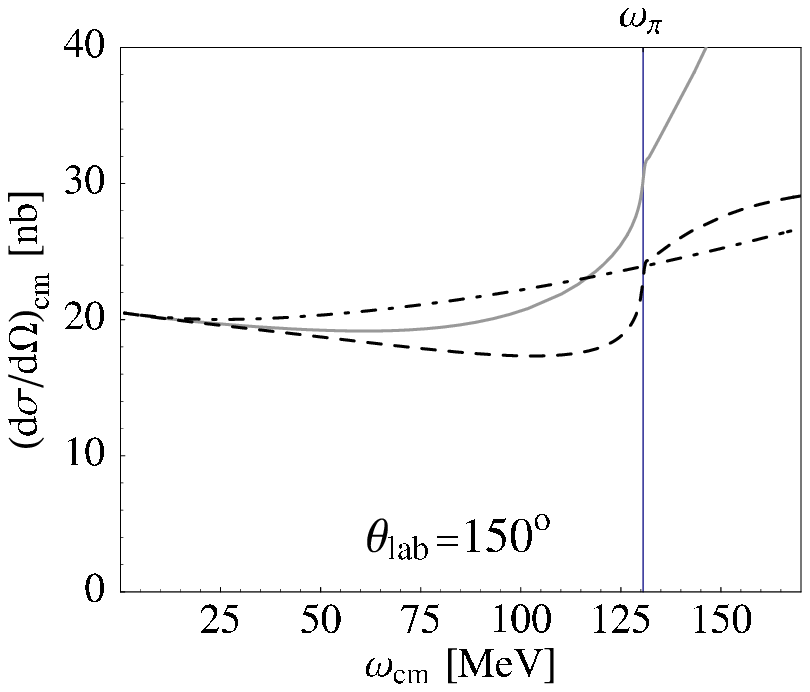}
\hspace{.01\textwidth}
\includegraphics*[width=.31\textwidth]
{labangle150bornChPTSSEpx.eps}
\caption[Cancellation of $\Delta$-contributions to 
$\Sigma_{x,p}^\mathrm{circ}$]
{Cancellation of contributions from the $\Delta$ resonance to the asymmetry 
$\Sigma_{x,p}^\mathrm{circ}$ (right) in the division of the difference (left) 
by the spin-averaged cross section (middle); for notation see 
Fig.~\ref{fig:bornasyp}.}
\label{fig:diff}
\end{center}
\end{figure}

Even more striking than for $\Sigma_{z,p}^\mathrm{circ}$ is the weak 
sensitivity of the asymmetry $\Sigma_{x,p}^\mathrm{circ}$ on explicit 
$\Delta$ degrees of freedom. Once again,
the only exception to this rule is the extreme forward direction around 
$\w_\pi$ because of
the small spin-averaged cross section which enhances the small deviations
between the HB$\chi$PT and the SSE calculation of the difference
Eq.~(\ref{eq:rightleft}) and makes $\Sigma_{x,p}^\mathrm{circ}$ extremely 
sensitive to theoretical uncertainties. Therefore, we consider the forward 
angle regime as inconvenient for measuring proton asymmetries. In the other 
panels of Fig.~\ref{fig:bornvertasyp}, the
$\Delta$ dependence cancels in the asymmetry, whereas we found the
$\Delta(1232)$ resonance to give sizeable contributions to both the difference
and the sum. This is demonstrated for $\theta_\text{lab}=150^\circ$ in 
Fig.~\ref{fig:diff}, where we show the difference 
$\mathcal{D}_{x,p}^\mathrm{circ}$, corresponding to Eq.~(\ref{eq:rightleft}), 
the spin-averaged cross section and the resulting asymmetry.
This is one example that an asymmetry actually hides interesting
physical information.

The dominance of the pole amplitudes is~-- as in 
$\Sigma_{z,p}^\mathrm{circ}$~-- clearly visible. 
The argument is the same as before.
Nonetheless, we find a stronger dependence of $\Sigma_{x,p}^\mathrm{circ}$ 
on the nucleon structure than for $\Sigma_{z,p}^\mathrm{circ}$, especially 
around $\w_\pi$.

\begin{figure}[!htb]
\begin{center}
\includegraphics*[width=.31\textwidth]{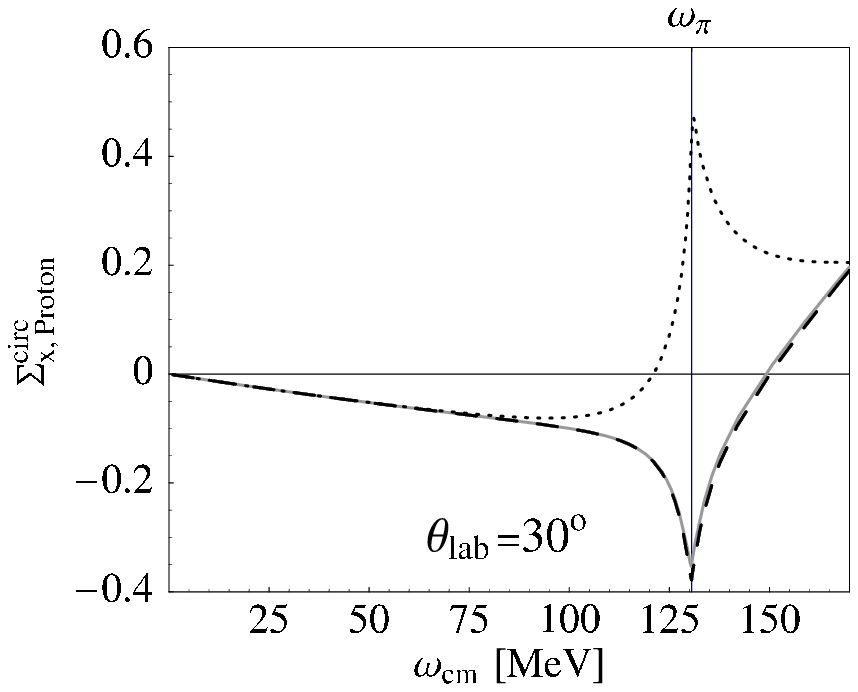}
\hspace{.01\textwidth}
\includegraphics*[width=.31\textwidth]{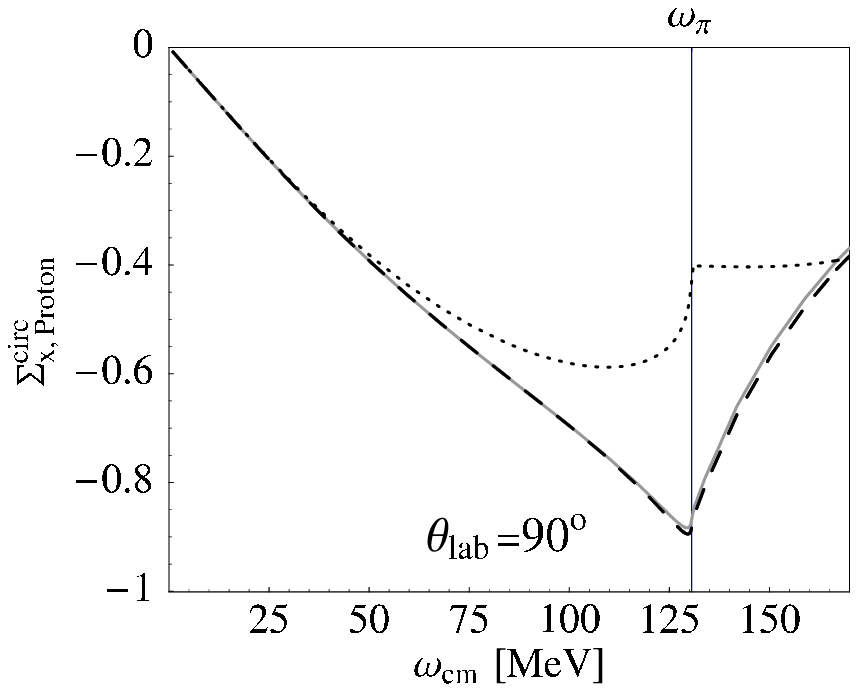}
\hspace{.01\textwidth}
\includegraphics*[width=.31\textwidth]{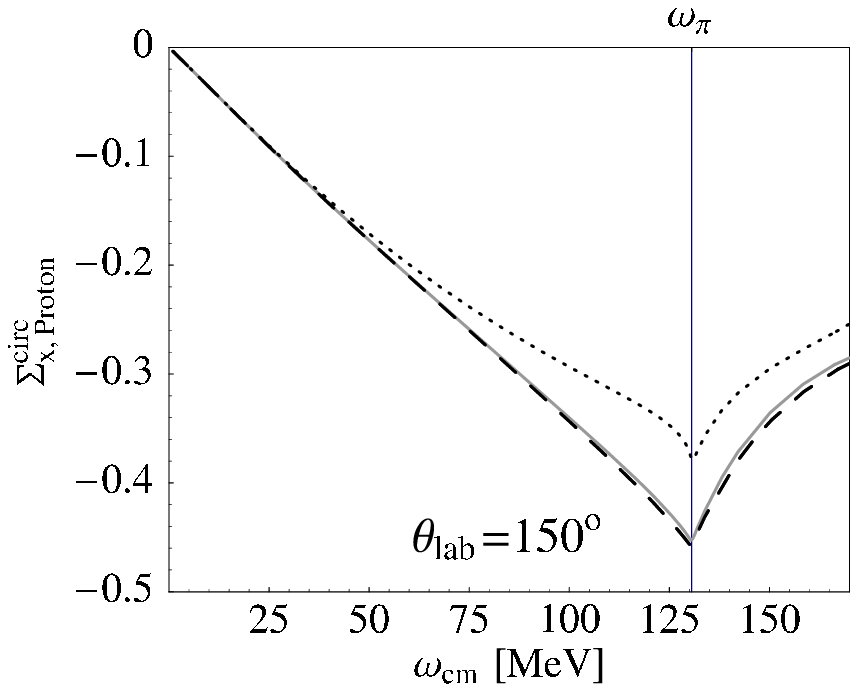}
\caption[Spin contributions to $\Sigma_{x,p}^\mathrm{circ}$]
{Dependence of the proton asymmetry $\Sigma_{x,p}^\mathrm{circ}$ on spin 
and quadrupole polarizabilities; for notation see Fig.~\ref{fig:SSEindiesump}.}
\label{fig:SSEindievertasyp}
\end{center}
\end{figure}

As one can see in Fig.~\ref{fig:SSEindievertasyp}, 
$\Sigma_{x,p}^\mathrm{circ}$ is very sensitive
to the spin polarizabilities for all angles.
Therefore~-- and because of our
findings in the previous subsection~-- choosing the proton spin perpendicular 
to the photon momentum seems to be more convenient than parallel
to examine the spin structure of the proton. In the
backward direction, the spin dependence of the asymmetry is less pronounced
than in the forward direction.

The quadrupole contributions are extremely small.

\subsection{Proton Asymmetries from Linearly Polarized Photons
\label{sec:protonlinearly}}

\begin{figure}[!htb]
\begin{center} 
\includegraphics*[width=.31\textwidth]
{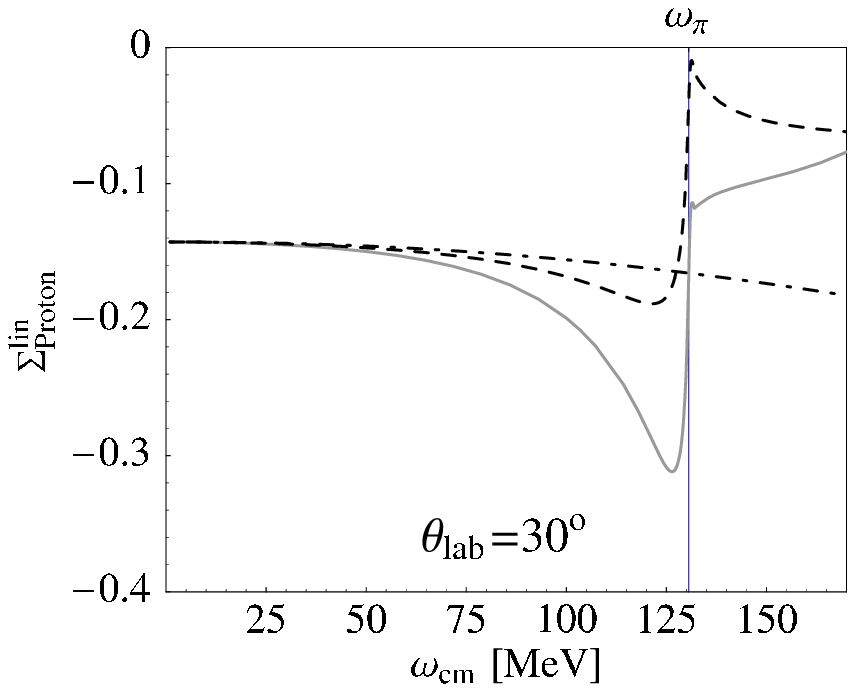}\hspace{.01\textwidth}
\includegraphics*[width=.31\textwidth]
{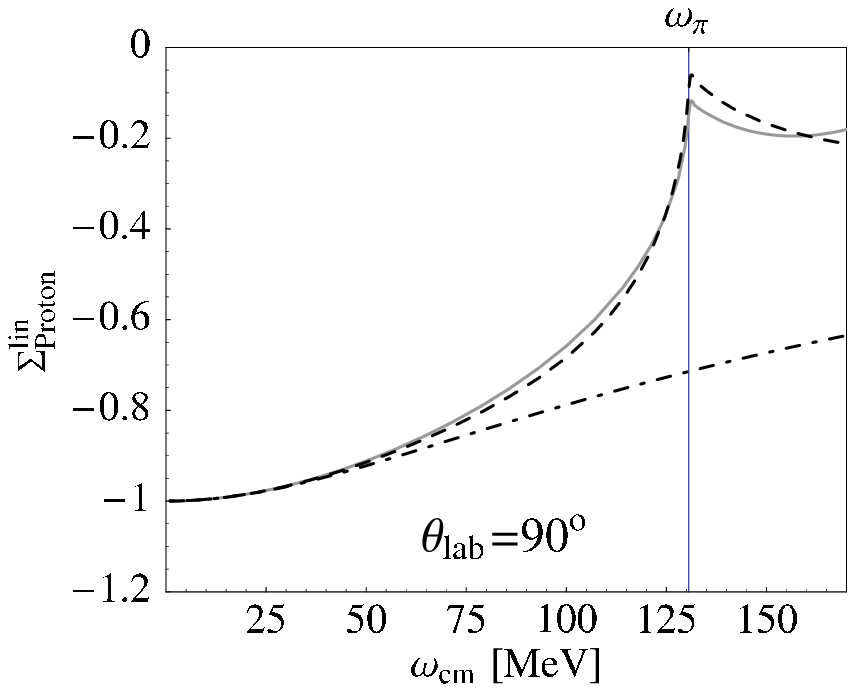}\hspace{.01\textwidth}
\includegraphics*[width=.31\textwidth]
{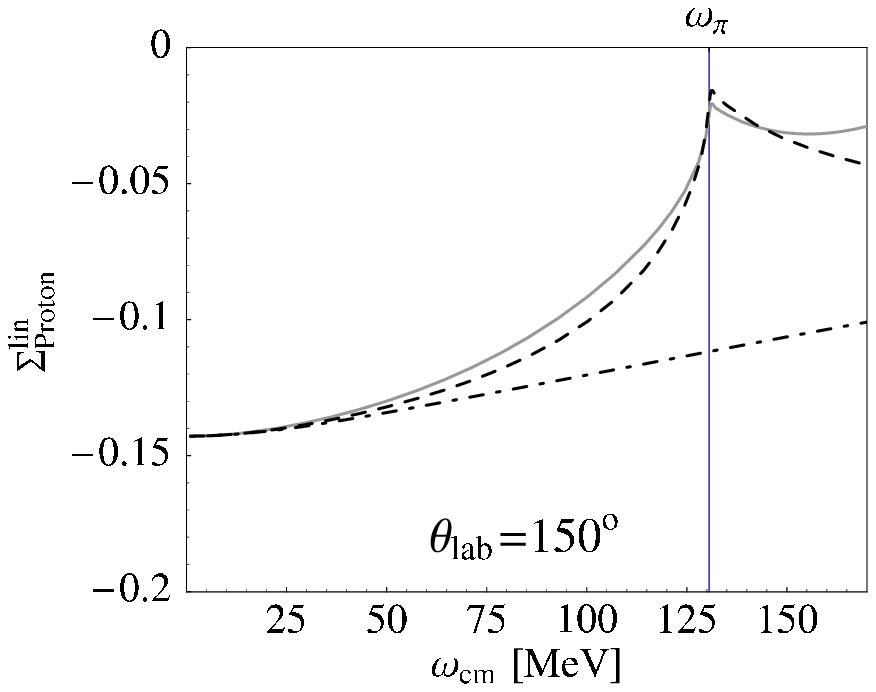}
\caption[Predictions for the proton asymmetry 
$\Sigma_{p}^\mathrm{lin}$]
{Dependence of the proton asymmetry $\Sigma_p^\mathrm{lin}$ on pion and 
$\Delta$ physics; for notation see Fig.~\ref{fig:bornasyp}.}
\label{fig:Sigmaplinborn}
\end{center}
\end{figure}

The most striking difference between Fig.~\ref{fig:Sigmaplinborn} 
and the asymmetries using circularly polarized incoming photons is
the fact that the proton asymmetry $\Sigma_{p}^\mathrm{lin}$ is non-vanishing 
even for zero photon energy. The reason is the term $|A_1|^2$ in 
Eq.~(\ref{eq:diffzlin}), which in leading order does not depend on the photon 
energy. The resulting formula for $\Sigma_p^\mathrm{lin}$ in the static limit 
is 
\be
\Sigma_{p}^\mathrm{lin}(\w,\theta)=-\frac{\sin^2\theta}{1+\cos^2\theta}, 
\ee
which can 
be derived from Eqs.~(\ref{eq:Tmatrix}, \ref{eq:paranti}, \ref{eq:poleterms}).
A strong dependence on $\Delta$ physics is only observed in the forward 
direction, which is, however, a most delicate region with respect to 
theoretical errors, as already discussed. 

\begin{figure}[!htb]
\begin{center} 
\includegraphics*[width=.31\textwidth]{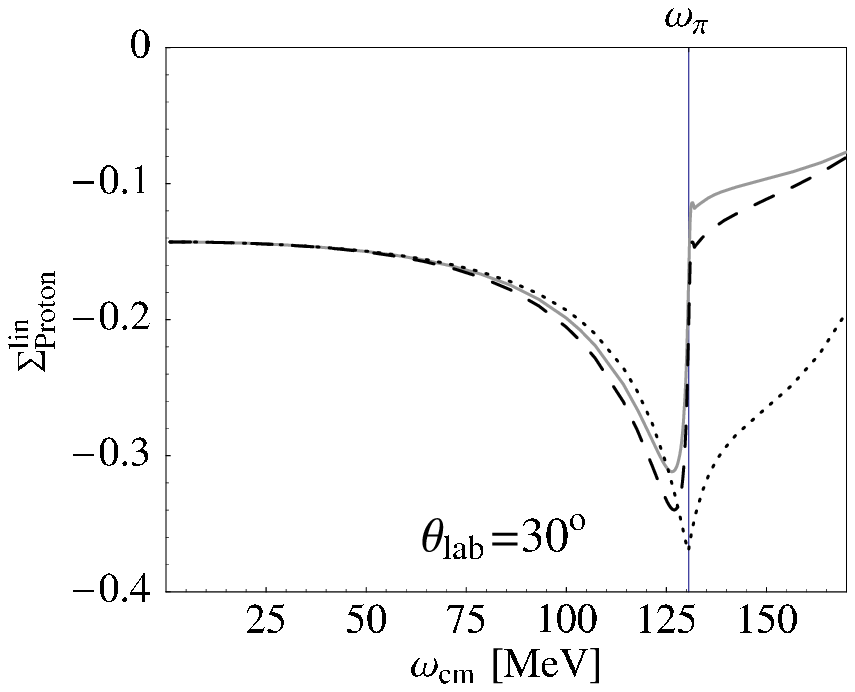}
\hspace{.01\textwidth}
\includegraphics*[width=.31\textwidth]{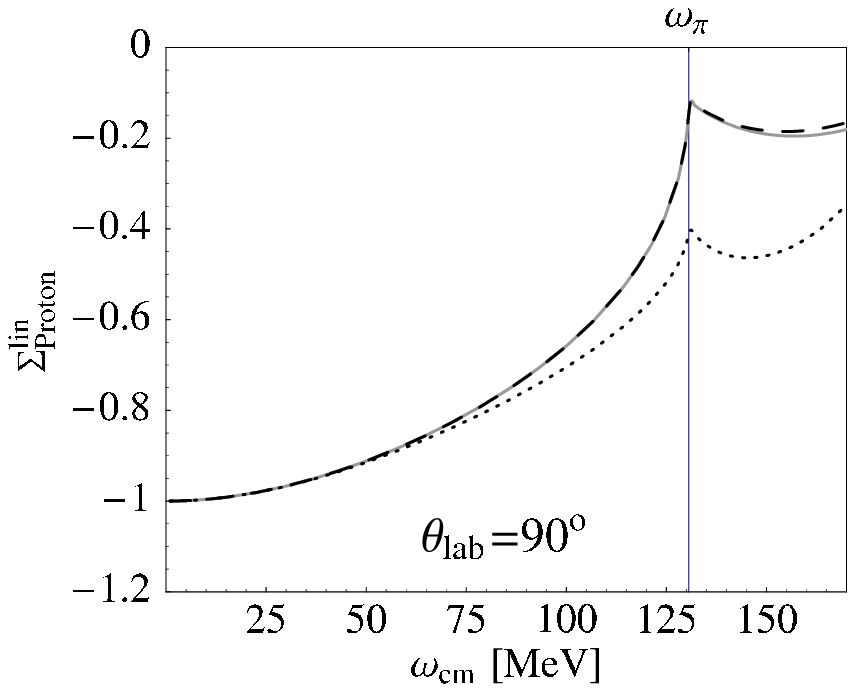}
\hspace{.01\textwidth}
\includegraphics*[width=.31\textwidth]{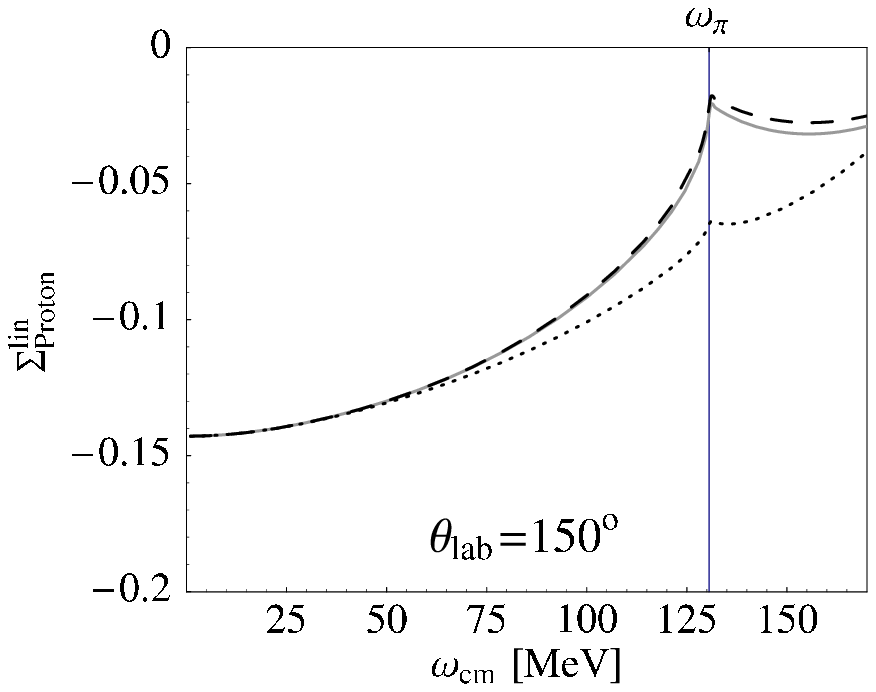}
\caption[Spin contributions to $\Sigma_{p}^\mathrm{lin}$]
{Dependence of the proton asymmetry $\Sigma_p^\mathrm{lin}$ on spin and 
quadrupole polarizabilities; for notation see Fig.~\ref{fig:SSEindiesump}.}
\label{fig:Sigmaplinindie}
\end{center}
\end{figure}

A similar picture arises for the spin-polarizability dependence, 
Fig.~\ref{fig:Sigmaplinindie}. Except for the energy region above threshold in
the extreme forward direction, these degrees of freedom are not as dominant 
as they are in 
Fig.~\ref{fig:SSEindievertasyp}. In the forward direction, however, we observe
a relatively strong influence of the quadrupole polarizabilities. As these
quantities were nearly invisible so far, this confirms our hypothesis that 
the forward direction is not the ideal choice to determine 
spin polarizabilities from asymmetries. Another disadvantage of 
$\Sigma_p^\mathrm{lin}$ is the small size of this asymmetry, except 
for $\theta\approx90^\circ$, due to the overall factor 
$\sin^2\theta$ in Eq.~(\ref{eq:diffzlin}).

As a short r\'esum\'e of the proton asymmetries, we consider the second 
configuration, i.e. $\Sigma_{x,p}^\mathrm{circ}$, as the best choice to 
determine spin polarizabilities from experiments. The reasons are 
the clear signal of the spin polarizabilities and the nearly invisible 
contributions of multipole order $l>1$. However, we have to caution that due
to the very small spin-averaged proton cross section around $\w_\pi$, 
there may be considerable theoretical errors in our calculation in the forward
direction. Therefore we consider experiments performed above $60^\circ$ most 
promising, see also Ref.~\cite{polarizedpaper} for a different set of angles.

\section{Neutron Asymmetries \label{sec:neutronasymmetries}}
\subsection[Neutron Asymmetries from Circularly Polarized Photons]
{Neutron Asymmetries from Circularly Polarized\\Photons 
\label{sec:neutroncircularly}}

In the absence of stable
single-neutron targets, the following results for the neutron have to be
corrected for binding and meson-exchange effects inside light nuclei~-- 
analogously to our spin-averaged deuteron Compton calculation, described in 
Chapters~\ref{chap:perturbative} and~\ref{chap:nonperturbative}.
Here, we present the neutron results
to guide considerations on future experiments using polarized deuterium or
$^3\!\mathrm{He}$~\cite{Gao}.

As in the proton case, the neutron asymmetries reach quite large values of
$\mathcal{O}(1)$ as the photon energy increases. In the neutron, pole
contributions might be expected to be small, because it is uncharged and thus
only the pion pole and the anomalous magnetic moment contribute. On the other
hand, spin polarizabilities are then not enhanced by interference with large
pole amplitudes. Therefore, whether and which neutron asymmetries are
sensitive to the structure parts, and hence to the spin polarizabilities, must 
be investigated carefully.

We follow the same line of presentation as outlined at the beginning of
Section~\ref{sec:protonasymmetries} for the proton asymmetries: First, we 
investigate which internal degrees of freedom are seen in a specific 
asymmetry, and then show that quadrupole polarizabilities give negligible 
contributions. Thus, the
asymmetries most sensitive to spin polarizabilities are identified.
Note that we may use the SSE parameters~-- $\bar{\alpha}_{E1}$, 
$\bar{\beta}_{M1}$ and $b_1$~-- which we derived from proton cross sections
in Section~\ref{sec:protonfits}, because 
non-relativistic Chiral Perturbation Theory predicts that the proton and 
neutron polarizabilities are equal at leading-one-loop order \cite{BKKM}.

\subsubsection{Neutron Spin Parallel to Photon Momentum}

\begin{figure}[!htb]
\begin{center}
\includegraphics*[width=.31\textwidth]
{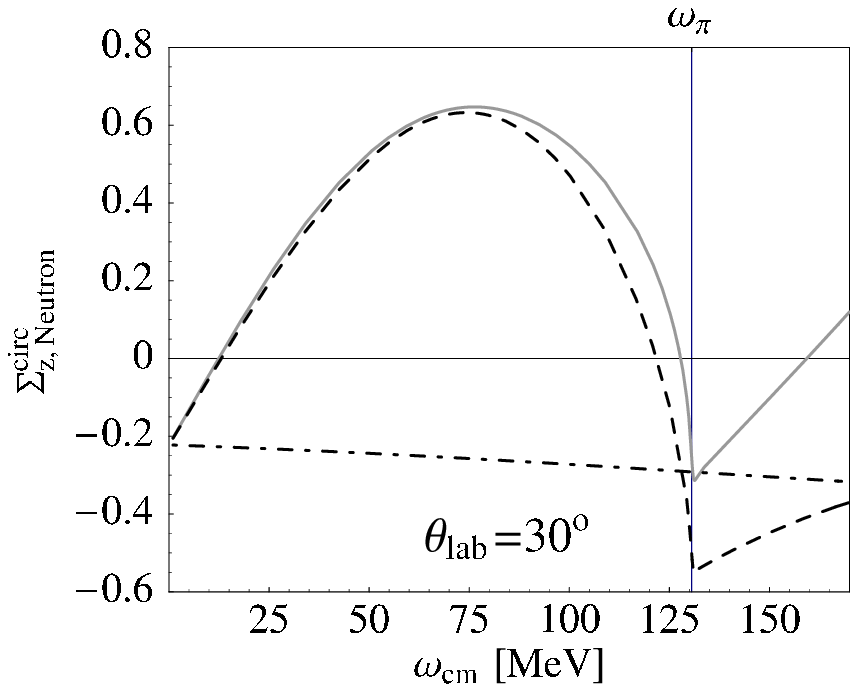}
\hspace{.01\textwidth}
\includegraphics*[width=.31\textwidth]
{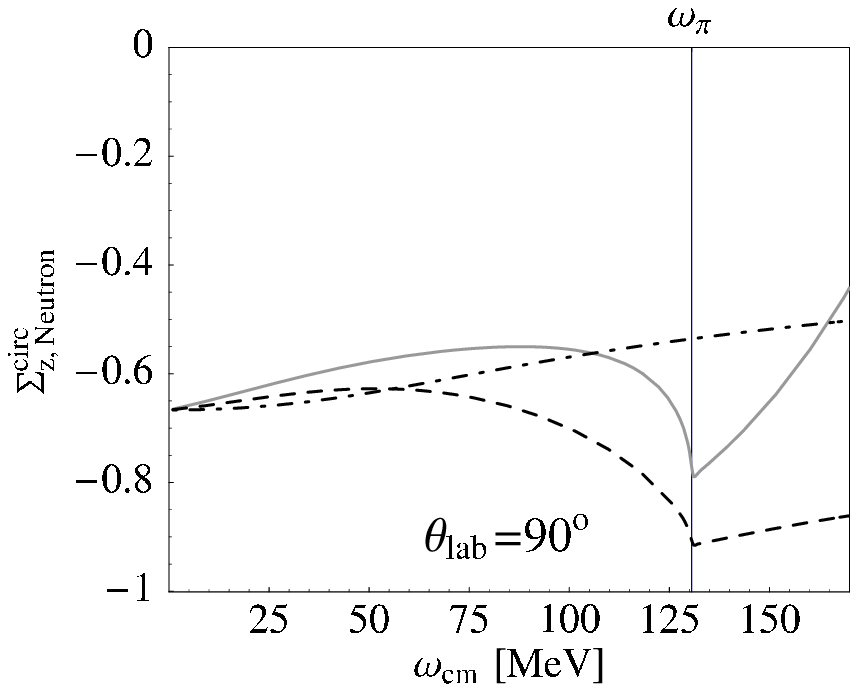}
\hspace{.01\textwidth}
\includegraphics*[width=.31\textwidth]
{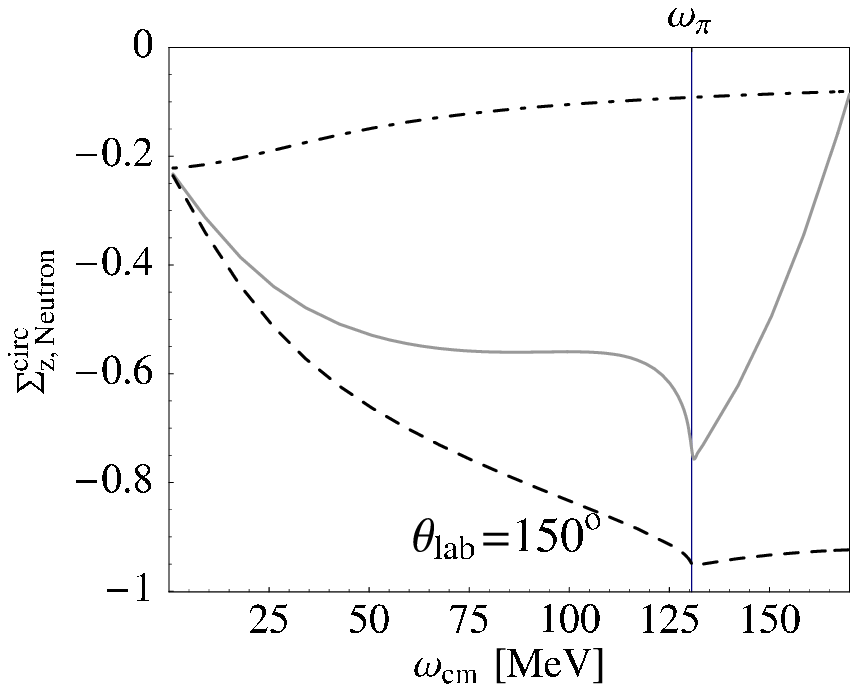}
\caption[Predictions for the neutron asymmetry 
$\Sigma_{z,n}^\mathrm{circ}$]
{Dependence of the neutron asymmetry $\Sigma_{z,n}^\mathrm{circ}$ on pion and 
$\Delta$ physics; for notation see Fig.~\ref{fig:bornasyp}.}
\label{fig:bornasyn}
\end{center}
\end{figure}

Comparing Fig.~\ref{fig:bornasyn} to the proton analogs, Figs. 
\ref{fig:bornasyp} and \ref{fig:bornvertasyp}, we notice that the neutron 
is much more sensitive to the
structure amplitudes. The pole curves show only a weak energy dependence,
so that nearly the whole dynamics is given by the neutron polarizabilities.
This minor influence of the pole amplitudes is due to the vanishing 
third-order pole contributions to $A_1$ and $A_2$, which make the difference
Eq.~(\ref{eq:paranti}) start with a term proportional to $\w^2$, whereas
the leading structure part is $\mathcal{O}(\w^3)$. The lowest order in $\w$ of
the spin-averaged cross section is $\w^2$, rendering finite static values of
$\Sigma_{z,n}^\mathrm{circ}$. The angular dependence of this static value can 
be derived from Eqs.~(\ref{eq:Tmatrix}, \ref{eq:paranti}, \ref{eq:poleterms}) 
as
\be
\Sigma_{z,n}^\mathrm{circ}(\w=0,\theta)=
\frac{4\,\sin^2\theta}{-5+\cos(2\,\theta)}.
\ee
The structure dependence of the neutron is also visible in the huge
sensitivity of $\Sigma_{z,n}^\mathrm{circ}$ to the $\Delta$ resonance, which 
influences the polarized cross sections considerably even for very low 
energies. As is well known, the influence of the $\Delta(1232)$ increases 
with increasing angle.

Concerning the shape of the asymmetry, one recognizes a similar behaviour for
the whole angular spectrum. It always reaches a local minimum at the pion
cusp. A precise interpretation of the shape of $\Sigma_{z,n}^\mathrm{circ}$ is
hard to give, as the denominator has the leading power $\w^2$, while it was 
$\w^0$ in the proton case. 

\begin{figure}[!htb]
\begin{center}
\includegraphics*[width=.31\textwidth]{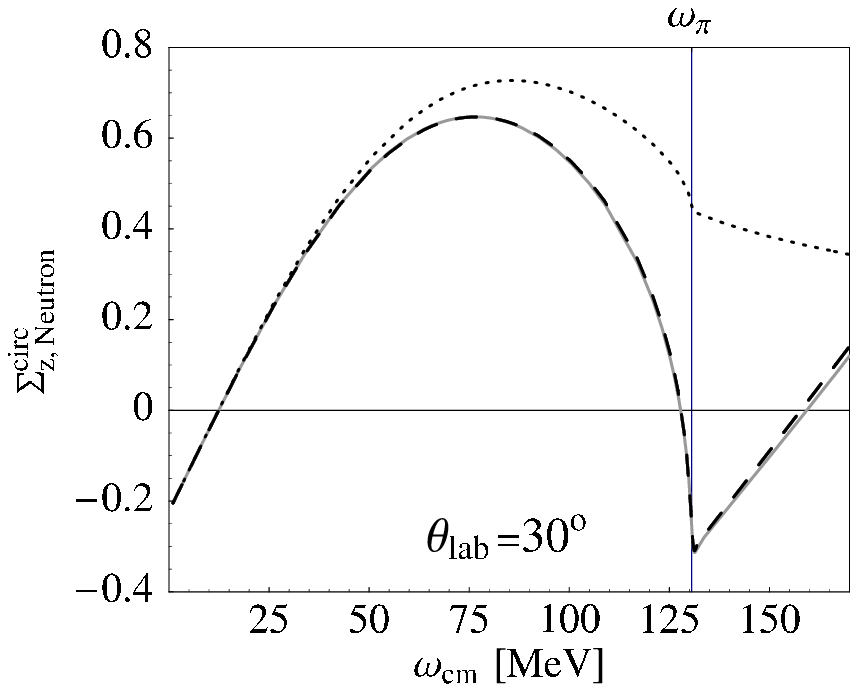}
\hspace{.01\textwidth}
\includegraphics*[width=.31\textwidth]{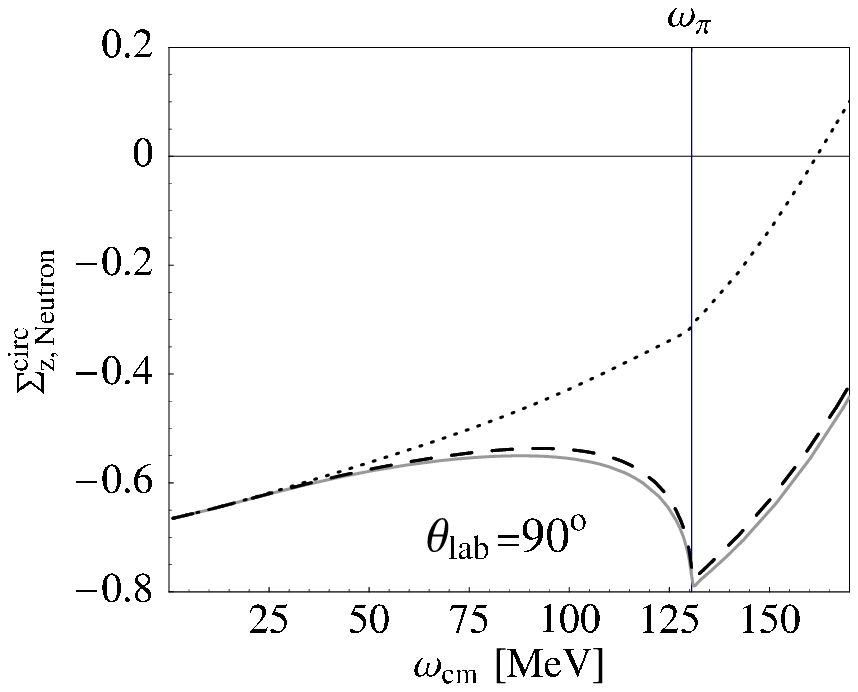}
\hspace{.01\textwidth}
\includegraphics*[width=.31\textwidth]{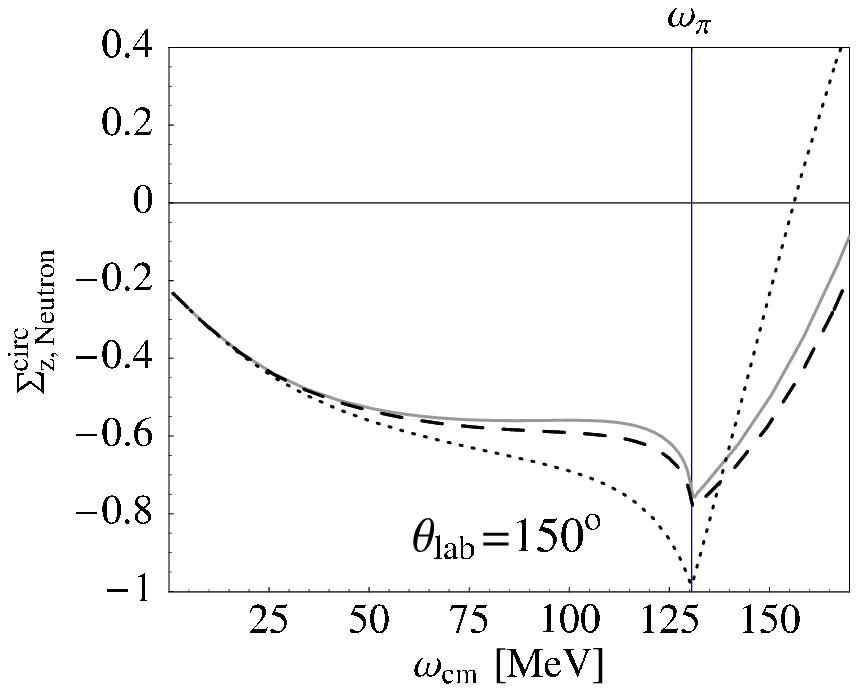}
\caption[Spin contributions to $\Sigma_{z,n}^\mathrm{circ}$]
{Dependence of the neutron asymmetry $\Sigma_{z,n}^\mathrm{circ}$ on spin and 
quadrupole polarizabilities; for notation see Fig.~\ref{fig:SSEindiesump}.}
\label{fig:SSEindieasyn}
\end{center}
\end{figure}

Fig.~\ref{fig:SSEindieasyn} exhibits that there are sizeable spin 
contributions to the asymmetry $\Sigma_{z,n}^\mathrm{circ}$ for each angle.
Nevertheless, one recognizes a decreasing spin 
dependence with increasing angle. 

As in the proton case we find the quadrupole part to be negligibly small
within the accuracy of this analysis (Fig.~\ref{fig:SSEindieasyn}).

\subsubsection{Neutron Spin Perpendicular to Photon Momentum}

\begin{figure}[!htb]
\begin{center}
\includegraphics*[width=.31\textwidth]
{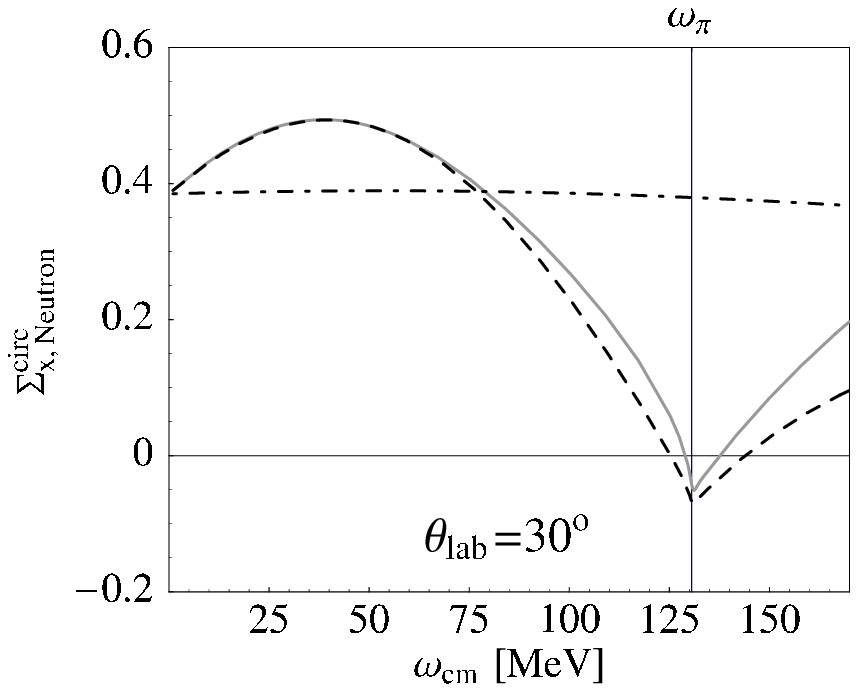}
\hspace{.01\textwidth}
\includegraphics*[width=.31\textwidth]
{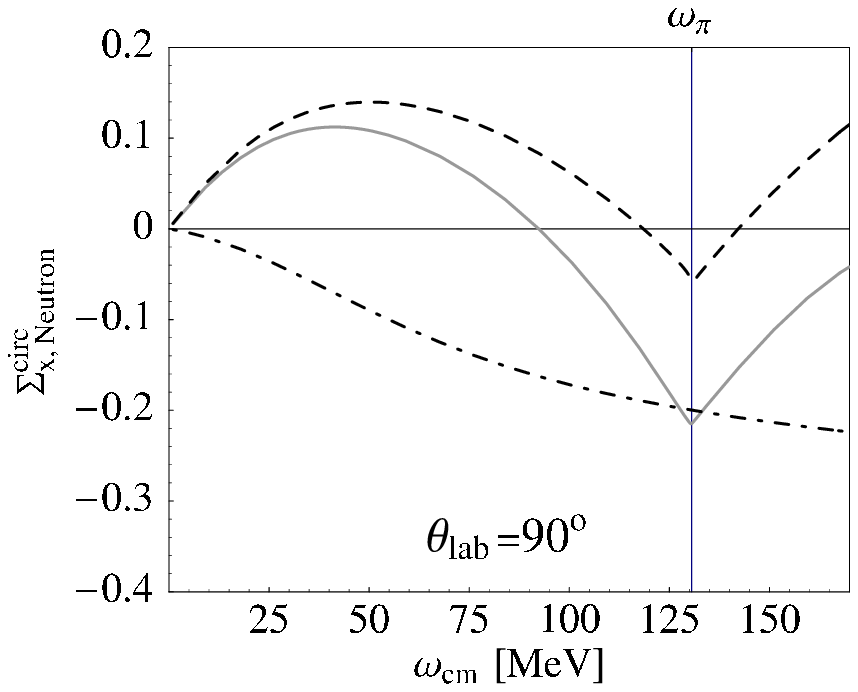}
\hspace{.01\textwidth}
\includegraphics*[width=.31\textwidth]
{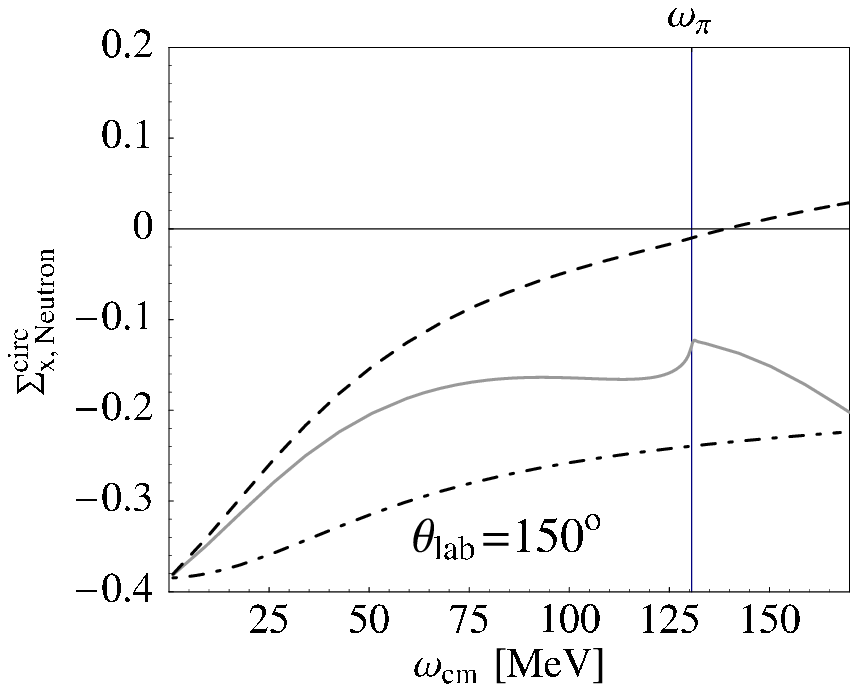}
\caption[Predictions for the neutron asymmetry 
$\Sigma_{x,n}^\mathrm{circ}$]
{Dependence of the neutron asymmetry $\Sigma_{x,n}^\mathrm{circ}$ on pion and 
$\Delta$ physics; for notation see Fig.~\ref{fig:bornasyp}.}
\label{fig:bornvertasyn}
\end{center}
\end{figure}

The shape of the asymmetry $\Sigma_{x,n}^\mathrm{circ}$
in Fig.~\ref{fig:bornvertasyn}
with the minimum at $\w_\pi$ is similar to $\Sigma_{z,n}^\mathrm{circ}$
(Fig.~\ref{fig:bornasyn}), especially in the forward direction. The curve is 
shifted downward with increasing angle $\theta$. 
The angular dependence of
the static value is determined by the pole contributions. It is
\begin{equation}
\Sigma_{x,n}^\mathrm{circ}(\w=0,\theta)=
\frac{4\,\sin\theta\,\cos\theta}{5-\cos(2\,\theta)},
\end{equation}
but as for $\Sigma_{z,n}^\mathrm{circ}$, the dynamics of 
$\Sigma_{x,n}^\mathrm{circ}$ is completely dominated by the neutron 
polarizabilities.

Another interesting feature in Fig.~\ref{fig:bornvertasyn} is the fact, that 
the explicit $\Delta$ degrees of freedom only play a minor role in the forward
direction but dominate for large angles.

\begin{figure}[!htb]
\begin{center}
\includegraphics*[width=.31\textwidth]{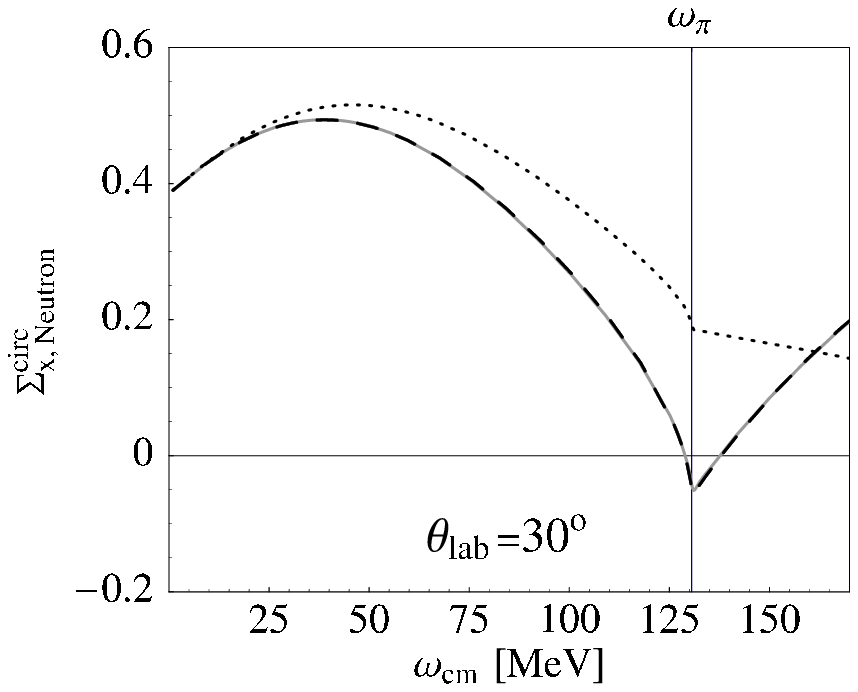}
\hspace{.01\textwidth}
\includegraphics*[width=.31\textwidth]{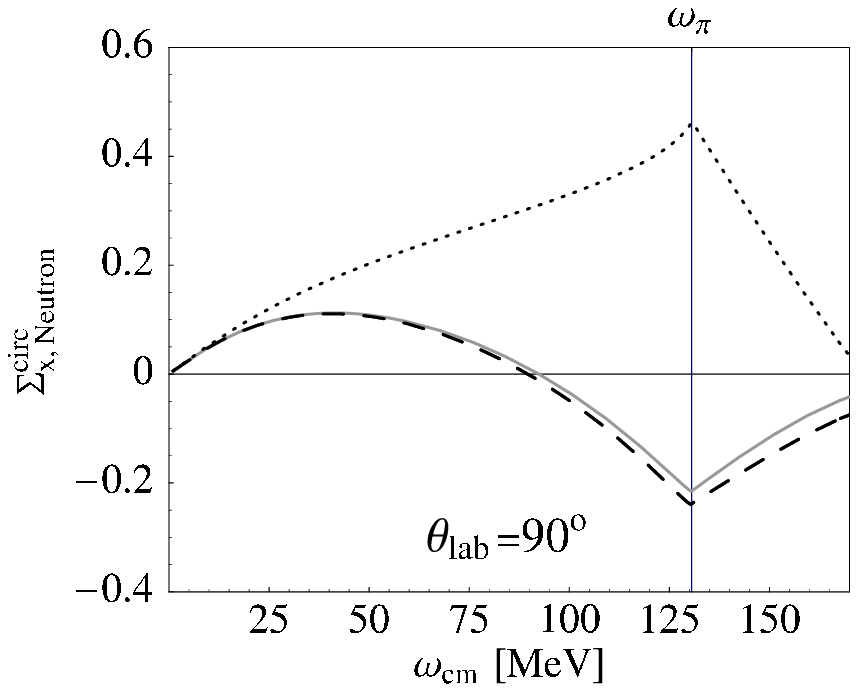}
\hspace{.01\textwidth}
\includegraphics*[width=.31\textwidth]{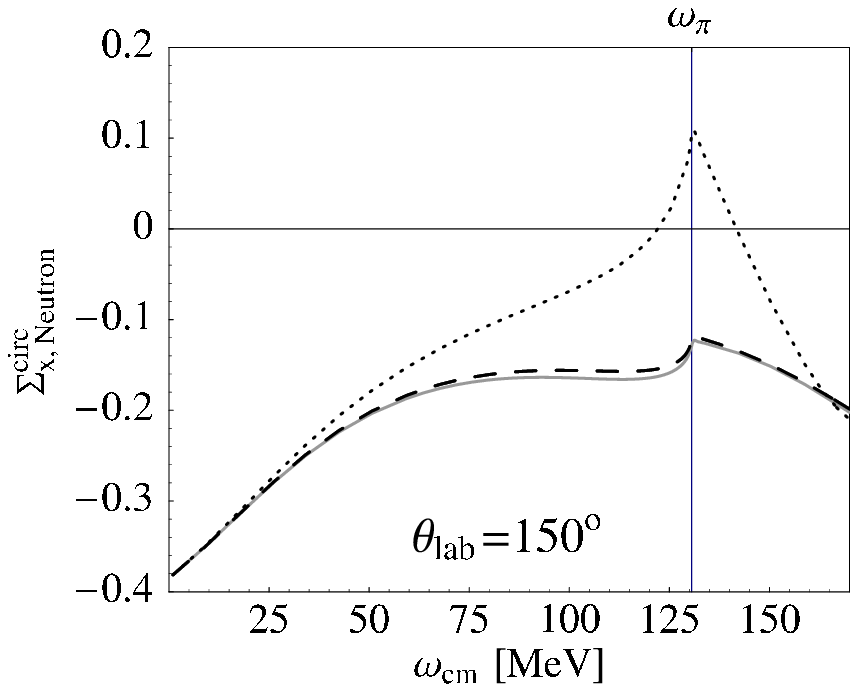}
\caption[Spin contributions to $\Sigma_{x,n}^\mathrm{circ}$]
{Dependence of the neutron asymmetry $\Sigma_{x,n}^\mathrm{circ}$ on spin and 
quadrupole polarizabilities; for notation see Fig.~\ref{fig:SSEindiesump}.}
\label{fig:SSEindievertasyn}
\end{center}
\end{figure}

Turning to Fig.~\ref{fig:SSEindievertasyn}, $\Sigma_{x,n}^\mathrm{circ}$ 
exhibits of all asymmmetries by far the largest
sensitivity to the spin polarizabilities. Therefore, and due to our 
observations for the proton asymmetries, an
experiment with the nucleon spin aligned perpendicularly to the photon momentum
seems from the theorist's point of view to be the most promising of the
considered configurations to extract the spin polarizabilities. 

As in $\Sigma_{z,n}^\mathrm{circ}$, the quadrupole polarizabilities 
are negligibly small in $\Sigma_{x,n}^\mathrm{circ}$
(Fig.~\ref{fig:SSEindievertasyn}). 

\subsection[Neutron Asymmetries from Linearly Polarized Photons]
{Neutron Asymmetries from Linearly Polarized\\Photons
\label{sec:neutronlinearly}}

\begin{figure}[!htb]
\begin{center} 
\includegraphics*[width=.31\textwidth]
{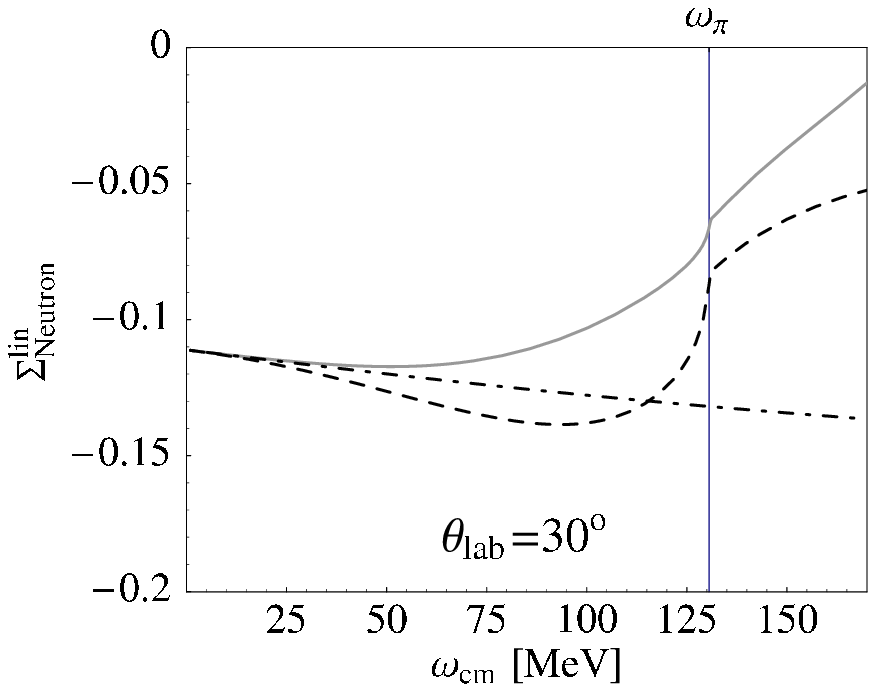}\hspace{.01\textwidth}
\includegraphics*[width=.31\textwidth]
{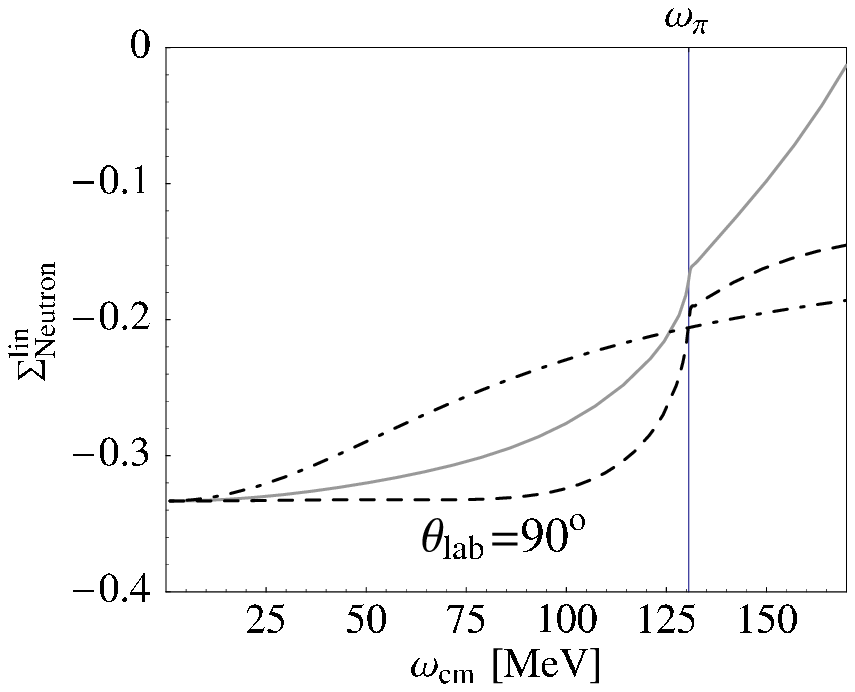}\hspace{.01\textwidth}
\includegraphics*[width=.31\textwidth]
{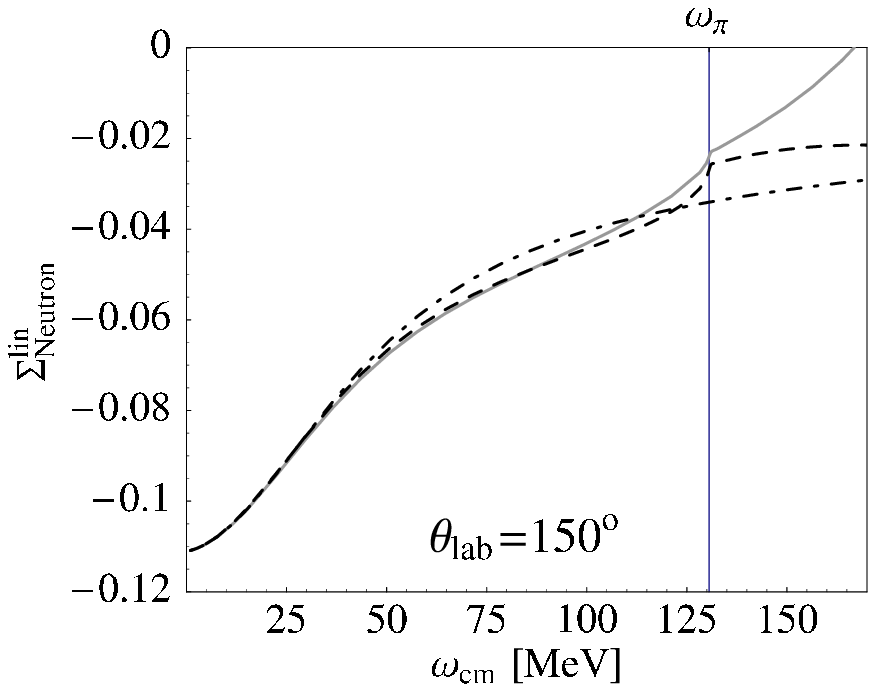}
\caption[Predictions for the neutron asymmetry 
$\Sigma_{n}^\mathrm{lin}$]
{Dependence of the neutron asymmetry $\Sigma_n^\mathrm{lin}$ on pion and 
$\Delta$ physics; for notation see Fig.~\ref{fig:bornasyp}.}
\label{eq:Sigmanlinborn}
\end{center}
\end{figure}

In the neutron asymmetries with circularly polarized photons only the
leading (static) terms are identical, when we expand the pole contributions 
and our full $\calO(\epsilon^3)$-calculation in the photon energy; in 
$\Sigma_n^\mathrm{lin}$ we find that also the slope at $\w=0$ is the same.
The reason for this behaviour is that in the real parts of the various
products of amplitudes in Eq.~(\ref{eq:diffzlin}) there is no interference
term between spin-independent ($A_1$,~$A_2$) and spin-dependent ($A_3$-$A_6$) 
amplitudes, as already mentioned at the end of 
Section~\ref{sec:asymmetrieslinearly}. 
Therefore there is no structure contribution at $\calO(\w^3)$.
This missing interference term already suggests that the asymmetry 
$\Sigma_n^\mathrm{lin}$ is not as sensitive to the neutron structure as are 
$\Sigma_{x,n}^\mathrm{circ}$ and $\Sigma_{z,n}^\mathrm{circ}$. In fact the 
three  approximations in Fig.~\ref{eq:Sigmanlinborn} are quite close to 
each other. The static limit in this configuration is described by
\be
\Sigma_{n}^\mathrm{lin}(\w=0,\theta)=\frac{2\,\sin^2\theta}{-5+\cos2\theta}.
\ee

\begin{figure}[!htb]
\begin{center} 
\includegraphics*[width=.31\textwidth]{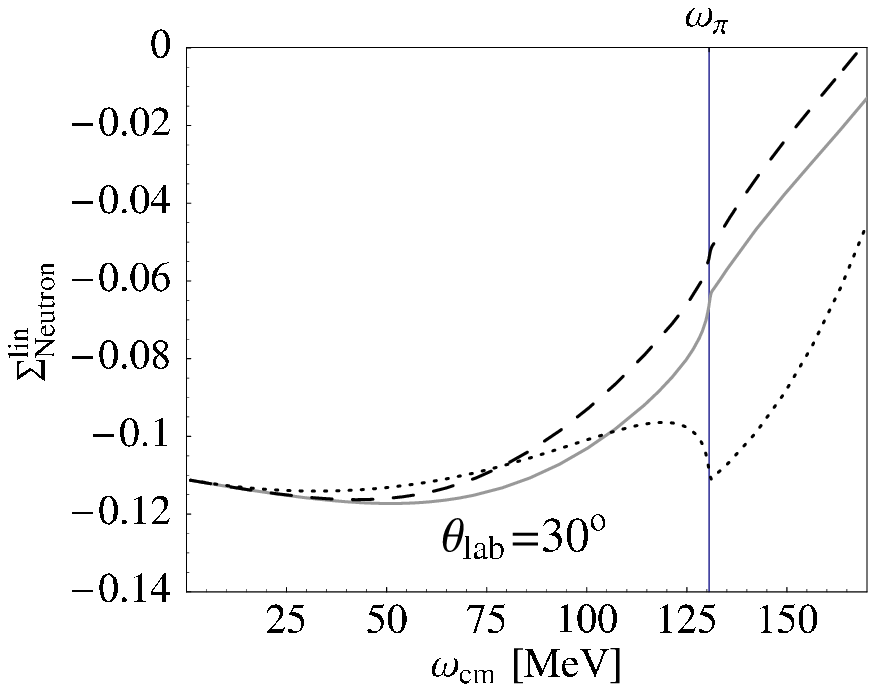}
\hspace{.01\textwidth}
\includegraphics*[width=.31\textwidth]{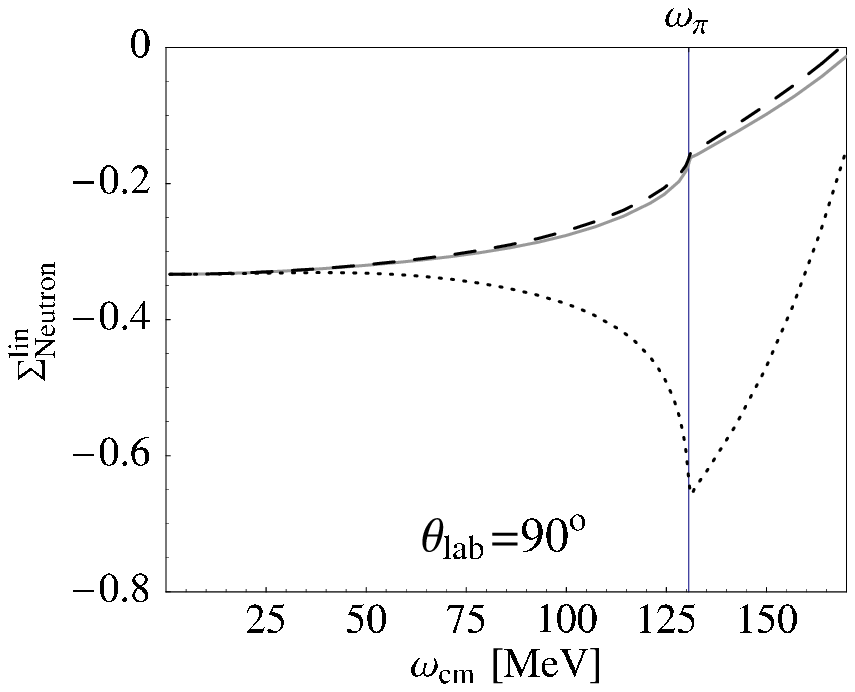}
\hspace{.01\textwidth}
\includegraphics*[width=.31\textwidth]{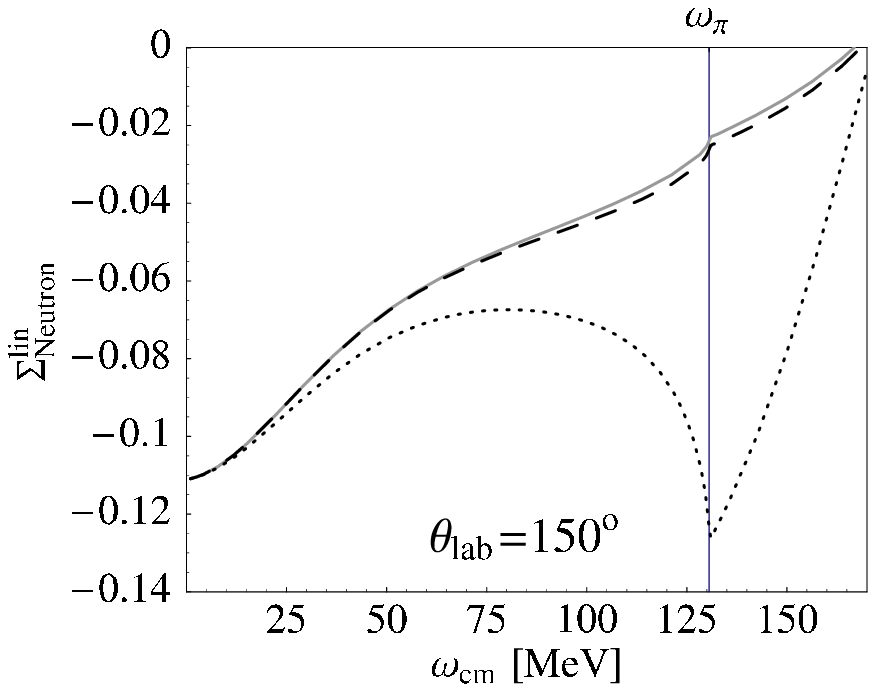}
\caption[Spin contributions to $\Sigma_{n}^\mathrm{lin}$]
{Dependence of the neutron asymmetry $\Sigma_n^\mathrm{lin}$ on spin and 
quadrupole polarizabilities; for notation see Fig.~\ref{fig:SSEindiesump}.}
\label{fig:Sigmanlinindie}
\end{center}
\end{figure}

On the other hand, Fig.~\ref{fig:Sigmanlinindie} suggests that the spin 
sensitivity of the configuration is huge, except for the forward direction. 
However, we have to caution that the absolute scale of the asymmetry in the 
backward (and also in the forward) direction is rather small~-- 
recall the factor 
$\sin^2\theta$ in Eq.~(\ref{eq:diffzlin}). Nevertheless we believe that this 
asymmetry is able to give valuable contributions to the determination of the 
spin polarizabilities, at least for scattering angles close to $90^\circ$. 
Another disadvantage appears in $\Sigma_n^\mathrm{lin}$ in the forward 
direction, namely the surprisingly strong contributions of the
dynamical quadrupole polarizabilities.

As a short conclusion of Sections~\ref{sec:protonasymmetries} and 
\ref{sec:neutronasymmetries} we find a much stronger sensitivity of the 
neutron asymmetries on the nucleon structure, while the
proton asymmetries are dominated by pole terms up to at least $50$~MeV. 
Contributions from the $\Delta(1232)$ resonance are crucial only for certain 
asymmetries and angles. For both nucleons,
the spin configuration $\Sigma_x^\mathrm{circ}$ turned out as the one which is
most sensitive to the nucleon spin structure, comparing to  
$\Sigma_z^\mathrm{circ}$ and 
$\Sigma^\mathrm{lin}$. However, this last configuration may give valuable 
contributions in the neutron case around $\theta=90^\circ$. Dynamical 
quadrupole contributions are negligible in each of the considered 
cases\footnote{The only exception to this rule is $\Sigma_n^\mathrm{lin}$ in 
the forward direction, but this configuration is also for other reasons  
unfavorable, e.g. due to the small absolute size.}.

\begin{figure}[!htb]
\begin{center} 
\includegraphics*[width=.48\textwidth]{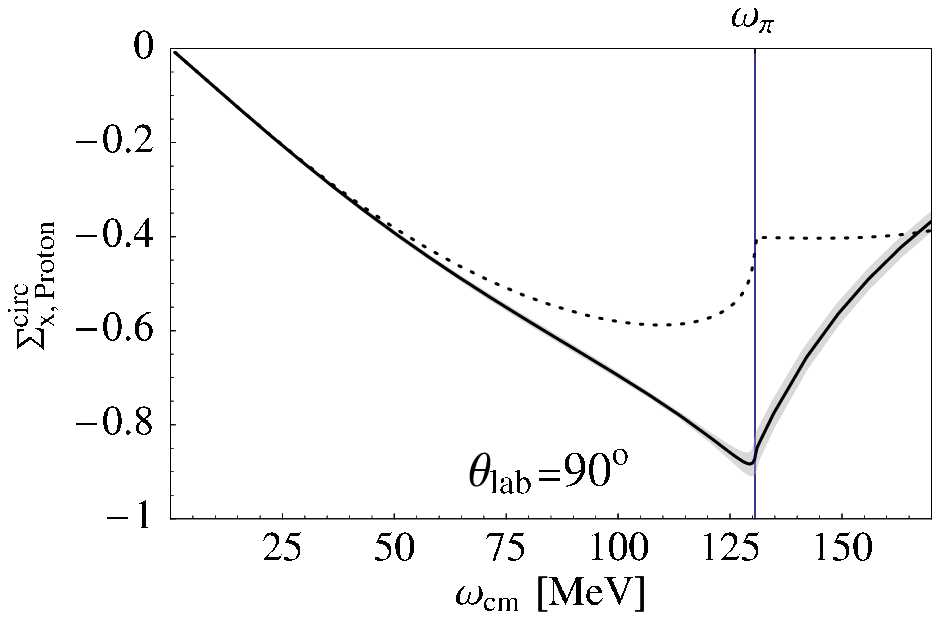}
\hfill
\includegraphics*[width=.48\textwidth]{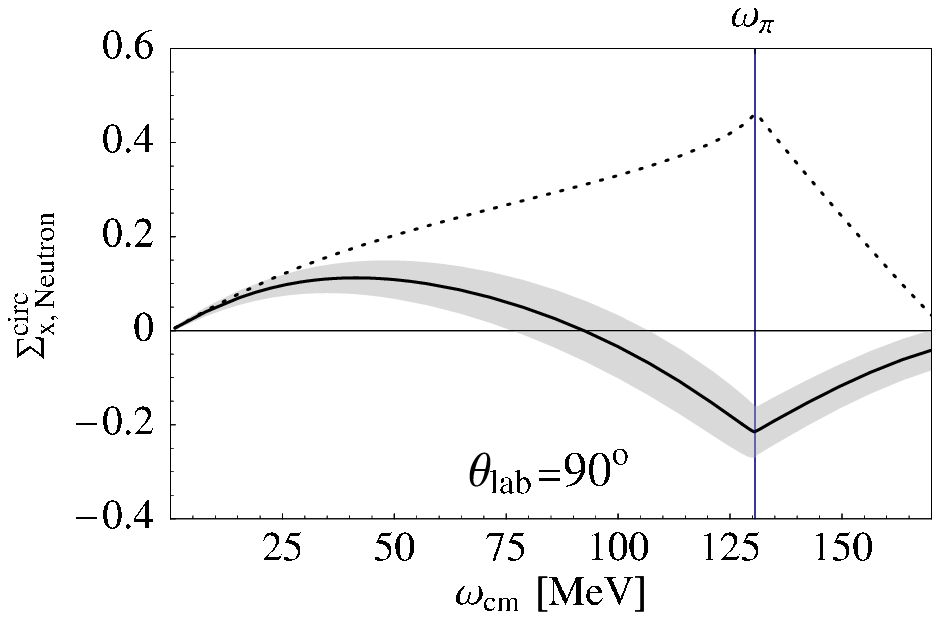}
\caption[Dependence of $\Sigma_x^\mathrm{circ}$ on the statistical errors 
of our input parameters]
{Dependence of the proton (left) and neutron (right) asymmetry 
$\Sigma_x^\mathrm{circ}$ on the statistical errors in $g_1,\;g_2,\;b_1$, 
denoted by the grey band around our full $\calO(\epsilon^3)$ SSE calculation 
(solid); in the dotted line the spin polarizabilities are switched off.}
\label{fig:asyerror}
\end{center}
\end{figure}

As we consider $\Sigma_x^\mathrm{circ}$ the most promising configuration,
we investigate two aspects of this quantity in more detail: Its 
sensitivity to the (statistical) errors of the three SSE fit parameters, cf. 
Table~\ref{tab:protonfit} and Eq.~(\ref{eq:g1g2Baldin}), and to the four 
spin-dependent dipole polarizabilities. Fig.~\ref{fig:asyerror} tells us that 
the statistical errors play only a minor role in the proton asymmetry, whereas 
for the neutron they are considerably larger, due to the weaker constraint by 
the pole terms, which renders the neutron more sensitive to its structure than 
the proton. This behaviour is observed independently of the scattering 
angle $\theta$. 

\begin{figure}[!htb]
\begin{center} 
\includegraphics*[width=.48\textwidth]{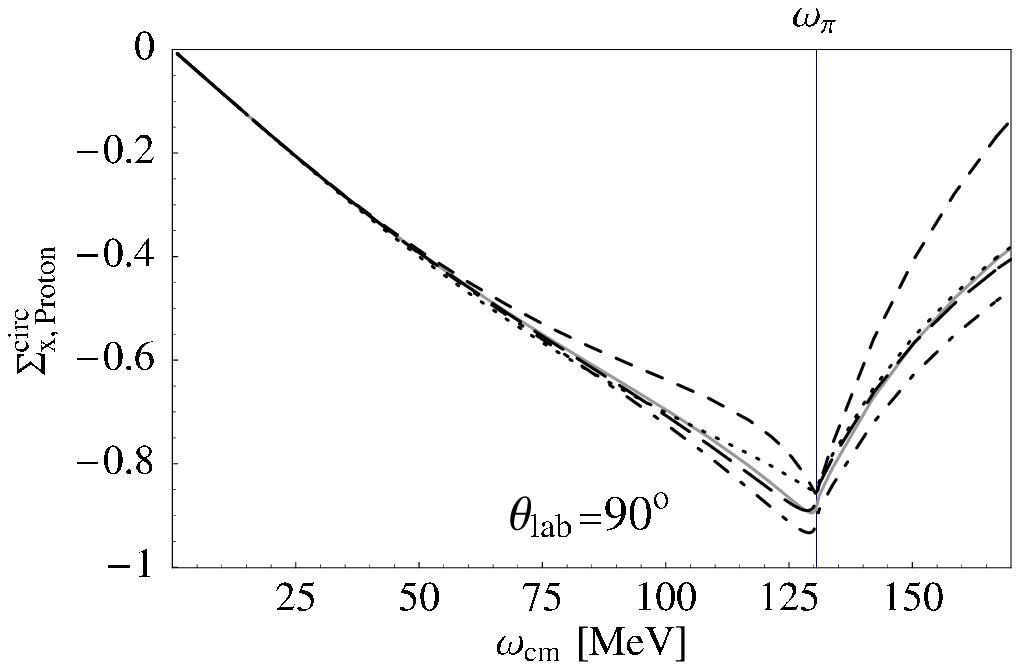}
\hfill
\includegraphics*[width=.48\textwidth]{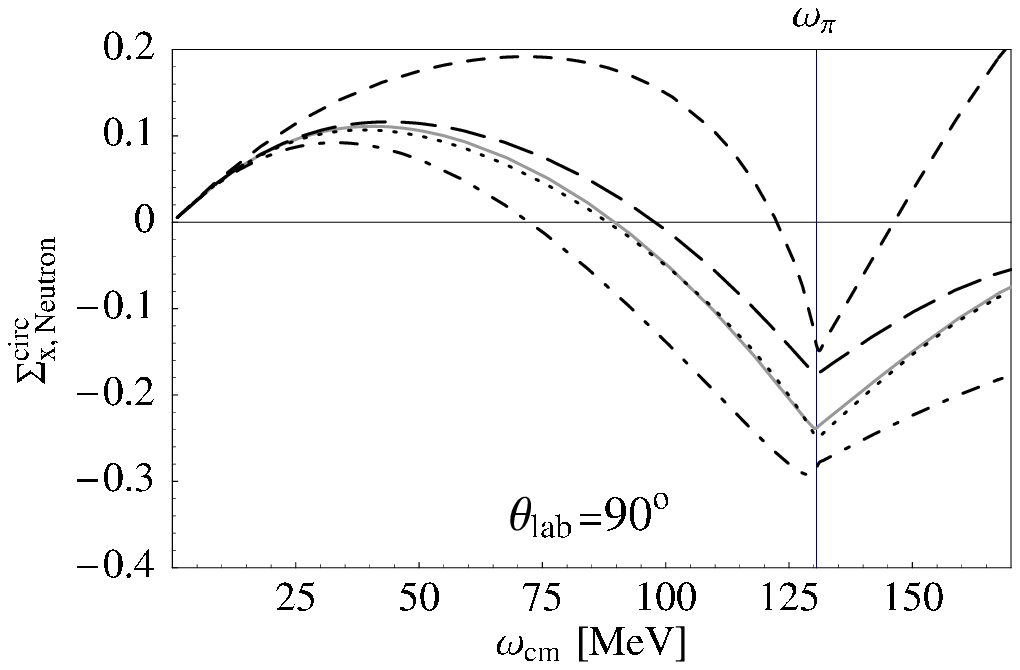}
\includegraphics*[width=.48\textwidth]{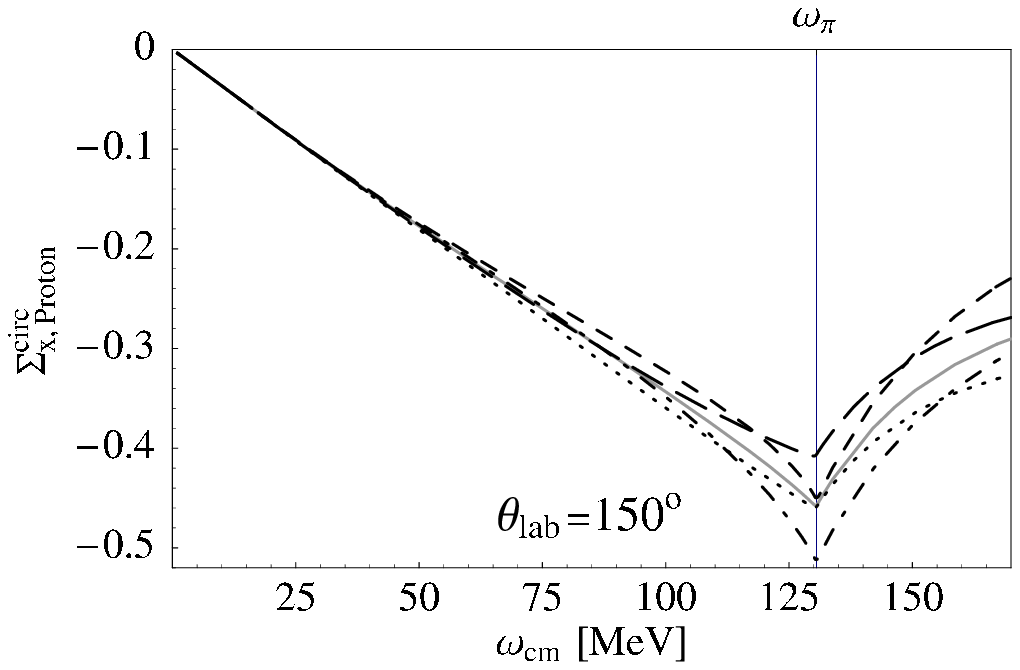}
\hfill
\includegraphics*[width=.48\textwidth]{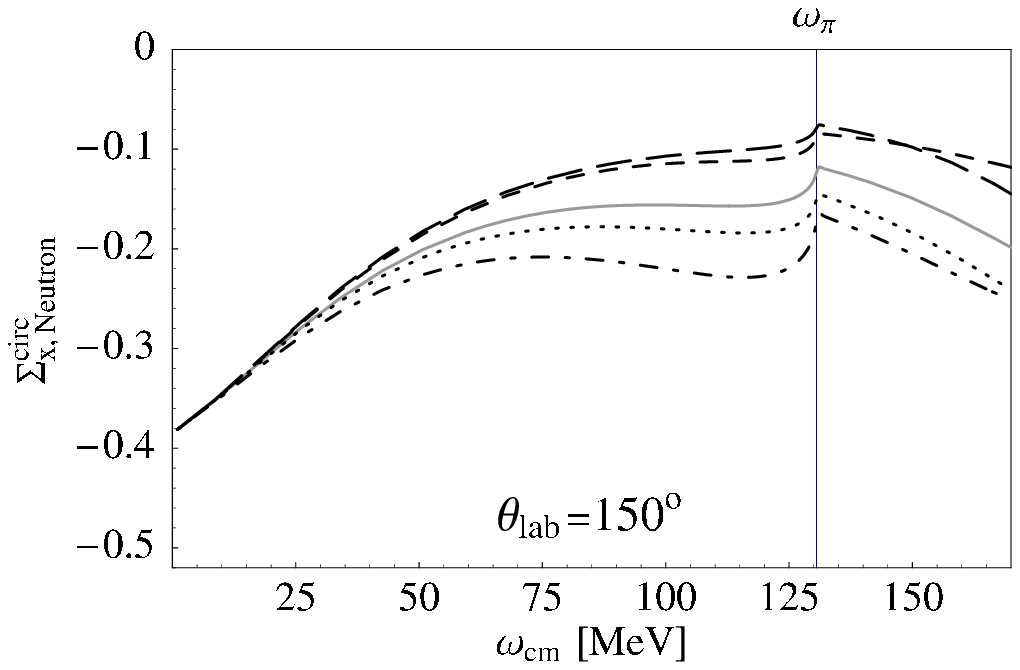}
\caption[Dependence of $\Sigma_x^\mathrm{circ}$ on the various spin dipole 
polarizabilities]
{Dependence of the proton (left) and neutron (right) asymmetry 
$\Sigma_x^\mathrm{circ}$ on the various spin dipole polarizabilities. 
We compare our full $\calO(\epsilon^3)$ SSE calculation (grey, solid) to the 
same result with a
variation by $+5\cdot10^{-4}\,\fm^4$ in $\gamma_{E1E1}(\w)$ (short-dashed), 
$\gamma_{M1M1}(\w)$ (dotted), $\gamma_{E1M2}(\w)$ (dotdashed) and 
$\gamma_{M1E2}(\w)$ (long-dashed).}
\label{fig:gammadep}
\end{center}
\end{figure}

In$\,$ Fig.~\ref{fig:gammadep},$\,$ we$\,$ vary$\,$ the$\,$ four$\,$ spin$\,$ 
dipole$\,$ polarizabilities$\,$ 
successively$\,$ by $+5\cdot10^{-4}\,\fm^4$, in order to investigate the 
dependence of $\Sigma_x^\mathrm{circ}$ on these quantities. 
For $\theta_\text{lab}=150^\circ$ the asymmetry is about equally sensitive 
to all four of them, whereas around $90^\circ$ we find that the spin 
dependence is dominated by $\gamma_{E1E1}(\w)$.
We also note that although 
the asymmetries are sensitive to all four spin polarizabilities, the dominant 
contribution comes from $\gamma_{E1E1}(\w)$, which is the largest in size.

Before we conclude the one-nucleon part of this work, let us
resume the main aspects of  Chapters~\ref{chap:spinaveraged} and 
\ref{chap:spinpolarized}. In Chapter~\ref{chap:spinaveraged} we derived
a multipole expansion for single-nucleon Compton scattering, which we use
to define dynamical, i.e. energy-dependent polarizabilities. These quantities
have been calculated in the framework of Chiral Effective Field Theory and are
compared to predictions from a Dispersion-Relation Analysis~\cite{HGHP}~-- 
the agreement between both approaches in most of the multipole channels 
was found to be very good. Before, we fixed three parameters, including the 
static dipole 
polarizabilities $\bar{\alpha}_{E1}$ and $\bar{\beta}_{M1}$, via fits to 
spin-averaged proton Compton cross sections, yielding results close to 
alternative extractions. Comparison of the resulting, fitted cross 
sections with experiment and Dispersion Relation looks very promising.
Contributions from quadrupole polarizabilities turned out nearly invisible 
below 170~MeV. 
Nevertheless it is clear that not all of the six dipole 
polarizabilities can be determined via fits to spin-averaged cross sections 
alone. Therefore we investigated in Chapter~\ref{chap:spinpolarized} several
single-nucleon asymmetries and demonstrated that 
neglecting quadrupole polarizabilities is a valid approximation also for 
spin-polarized observables. We conclude that determining the elusive 
spin polarizabilities from a combination of spin-polarized and spin-averaged
experiments is feasible. However, such experiments are nearly impossible for 
the unstable neutron. So in order to investigate the neutron 
polarizabilities, one has to rely on light nuclei like the 
deuteron or $^3\!$He. 
With this in mind we turn now to the second main part of this work, the 
calculation of Compton scattering from the deuteron.

\chapter{Deuteron Compton Scattering in Effective Field Theory
\label{chap:perturbative} }
\markboth{CHAPTER \ref{chap:perturbative}. DEUTERON COMPTON SCATTERING IN EFT}
{}
So far we were only concerned with Compton scattering from the single nucleon.
We found that describing
proton and neutron Compton scattering theoretically is quite similar 
in the framework and up to the order of the chiral expansion chosen in this 
work. As we saw in Section~\ref{sec:protoncrosssections}, our calculation 
gives a good description of the experimentally measured proton Compton cross 
sections below 200~MeV. Therefore, in Section~\ref{sec:protonfits}, we were 
able to fit the (static) proton polarizabilities $\bar{\alpha}_{E1}^p$ and 
$\bar{\beta}_{M1}^p$ reliably to the existing data. 

However, as there is no stable single-neutron target, 
the direct experimental investigation of 
neutron Compton scattering is nearly impossible. Therefore,
in order to access the neutron polarizabilities, one has to rely on other 
experimental methods, e.g. quasi-free Compton scattering from neutrons bound 
in a light nucleus, scattering neutrons on heavy nuclei or elastic Compton 
scattering from light nuclei.
In the Introduction (Section~\ref{sec:intro}), we reviewed several attempts 
to extract the neutron polarizabilities from  data and we found a 
rather broad range quoted for $\bar{\alpha}_{E1}^n$ and $\bar{\beta}_{M1}^n$, 
e.g. $\bar{\alpha}_{E1}^n\in[-4;\,19]$. Therefore, in order to contribute to 
the ongoing discussion of these quantities, we are in this and the next 
chapter concerned with elastic deuteron Compton scattering. The cross
sections resulting from our calculation are fitted to experimental data, with 
the isoscalar polarizabilities as fit parameters. These numbers are then 
combined with our fit results for $\bar{\alpha}_{E1}^p$ and 
$\bar{\beta}_{M1}^p$ from Section~\ref{sec:protonfits} in 
order to deduce the elusive neutron polarizabilities.

In this first chapter on deuteron Compton scattering we aim for an improved 
description of the elastic deuteron Compton data 
at $\omega\sim50$-100~MeV, 
compared to the calculations presented in \cite{Lvov,Karakowski}, which cannot 
describe the data from~\cite{Hornidge}, measured at 
$\w_\mathrm{lab}\sim95$~MeV. 
In the next chapter 
we show how to extend the region of validity of our calculation down to 
the limit of vanishing photon energy.

This chapter is based on the calculations of 
Refs.~\cite{Phillips,McGPhil}, where Compton 
scattering off the deuteron was examined for photon energies $\omega$ ranging 
from 50~MeV to 100~MeV. 
The central values for the isoscalar polarizabilities, derived in the 
recent $\calO(p^4)$-HB$\chi$PT analysis~\cite{McGPhil} of the data 
from~\cite{Lucas,Lund,Hornidge} are
\begin{align}
\bar{\alpha}_{E1}^s&=
(13.0\pm1.9)^{+3.9}_{-1.5}\cdot10^{-4}\;\mathrm{fm}^3,\nonumber\\
\bar{\beta}_ {M1}^s&=
(-1.8\pm1.9)^{+2.1}_{-0.9}\cdot10^{-4}\;\mathrm{fm}^3.
\label{eq:Oq4bestresults}
\end{align}
Comparing with Table~\ref{tab:protonfit}, these results indicate
a small isovector electric polarizability, but allow for 
values of $\bar{\beta}_{M1}^v$ which are considerably larger than 
$\bar{\beta}_{M1}^p$.
However, the range for  
$\bar{\alpha}_{E1}^s$ and $\bar{\beta}_{M1}^s$ quoted in~\cite{McGPhil} is 
large: $\bar{\alpha}_{E1}^s=( 9.6\dots 18.8)\cdot 10^{-4}\;\mathrm{fm}^3$, 
$\bar{\beta}_{M1}^s=(-4.6\dots  2.2)\cdot 10^{-4}\;\mathrm{fm}^3$.
The authors of Refs.~\cite{Phillips,McGPhil} followed Weinberg's 
proposal~\cite{Weinberg} 
to calculate the irreducible kernel for the 
$\gamma N N\rightarrow \gamma N N$ process in Heavy Baryon Chiral 
Perturbation Theory, cf. Section~\ref{sec:HBchiPT}. Proceeding in 
this fashion means working within an Effective Field Theory in which only 
nucleons and pions are active degrees of freedom. The kernel is then folded
with external deuteron wave functions, derived from high-precision 
$NN$-potentials such as Nijm93~\cite{Nijm}, 
CD-Bonn~\cite{Bonn} or AV18~\cite{AV18}. This combination of various
elements, calculated within different theoretical frameworks, is called 
``hybrid'' approach and has proven quite successful in describing e.g. 
radiative $np$ capture~\cite{Rho} and  $\pi d$~\cite{Beane}, 
$e^- d$~\cite{Danieled} and also $\gamma d$~\cite{Phillips,McGPhil} 
scattering. As we have seen in Chapter~\ref{chap:spinaveraged} that the 
$\Delta(1232)$ is an important degree of freedom in single-nucleon Compton 
scattering, we extend in this chapter the calculation of Ref.~\cite{Phillips} 
to the framework of the Small Scale Expansion, 
cf. Section~\ref{sec:SSE}.
The advantage of our approach with respect 
to the NNLO calculation of Ref.~\cite{McGPhil} is that we 
have a more realistic energy dependence of the Compton multipoles, which in 
Ref.~\cite{McGPhil} is only partially contained in the two short-distance 
parameters, contributing to $\bar{\alpha}_{E1}$ and $\bar{\beta}_{M1}$. 
As we already saw in Chapter~\ref{chap:spinaveraged}, the strong energy 
dependence induced by the $\Delta(1232)$ 
plays an important role in quantities such as the magnetic dipole 
polarizability $\beta_{M1}(\omega)$ and in the spin-averaged 
Compton cross sections.
It is therefore interesting to also investigate the role of these degrees of 
freedom in elastic $\gamma d$ scattering, which is the main focus in this 
chapter.

In Section~\ref{sec:results}, we discuss our predictions for the deuteron 
Compton cross sections for four different energies between 50~MeV and 100~MeV,
comparing to data and to the $\mathcal{O}(p^3)$-HB$\chi$PT 
calculation~\cite{Phillips}. Before that, we give a brief survey of
the theoretical formalism in Section~\ref{sec:theory} and show that 
combining Weinberg's counting ideas with the SSE power-counting scheme leads 
to no additional diagrams in the two-body part of the kernel with respect 
to~\cite{Phillips}. 
In Section~\ref{sec:fits1}, we present our results for the isoscalar
polarizabilities, derived from a fit to 
elastic deuteron Compton-scattering data, which turn out to be in good 
agreement with the theoretical expectation that the isovector 
components are small. The corresponding curves are compared to the plots 
resulting from one of the $\calO(p^4)$-HB$\chi$PT fits from 
Ref.~\cite{McGPhil}.  All the results 
reported in this chapter have been published in our paper 
Ref.~\cite{deuteronpaper}.

\section{Theory of Deuteron Compton Scattering}
\label{sec:theory}

\begin{figure}[!htb]
\begin{center} 
\includegraphics*[width=.18\textwidth]{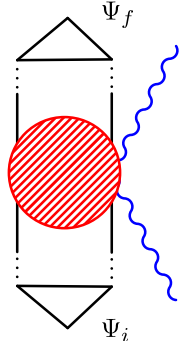}
\parbox{1.\textwidth}{
\caption[Anatomy of the deuteron Compton calculation]
{Anatomy of the deuteron Compton calculation. The blob symbolises the 
$\gamma\gamma d$ interaction kernel, $\Psi_{i,f}$ denote the deuteron wave 
function in the initial and final state, respectively.}
\label{fig:anatomy}}
\end{center}
\end{figure}

We  calculate Compton scattering off the deuteron in the framework of the
Small Scale Expansion~\cite{HHKLett}, cf. Section~\ref{sec:SSE}.
The power-counting scheme that we use for Compton scattering off light nuclei 
is motivated by Weinberg's idea to count powers only in the interaction 
kernel. 
While the kernel is 
power counted according to the rules of the Effective Field Theory, the 
deuteron wave functions we use are obtained from 
state-of-the-art $NN$ potentials: Nijm93~\cite{Nijm}, the 
CD-Bonn potential~\cite{Bonn}, the AV18 potential~\cite{AV18} (see also 
Appendix~\ref{app:AV18}) and the NNLO 
chiral potential~\cite{Epelbaum} with the cutoff chosen as 
$\Lambda=650$~MeV\footnote{We note that it has been questioned in  
Ref.~\cite{Nogga}, whether there is a unique 'NNLO' chiral potential due to 
the observed limit-cycle-like cutoff dependence of the leading-order chiral 
potential.}.
This last potential is derived by applying the HB$\chi$PT power counting, 
proposed by Weinberg for the two-nucleon sector, to the $N N$ potential $V$. 
The latter two wave functions are sketched in 
Appendix~\ref{app:wavefunction}. 
The anatomy of our deuteron Compton calculation is
illustrated in Fig.~\ref{fig:anatomy}.

\begin{figure}[!htb]
\begin{center} 
\includegraphics*[width=.25\textwidth]{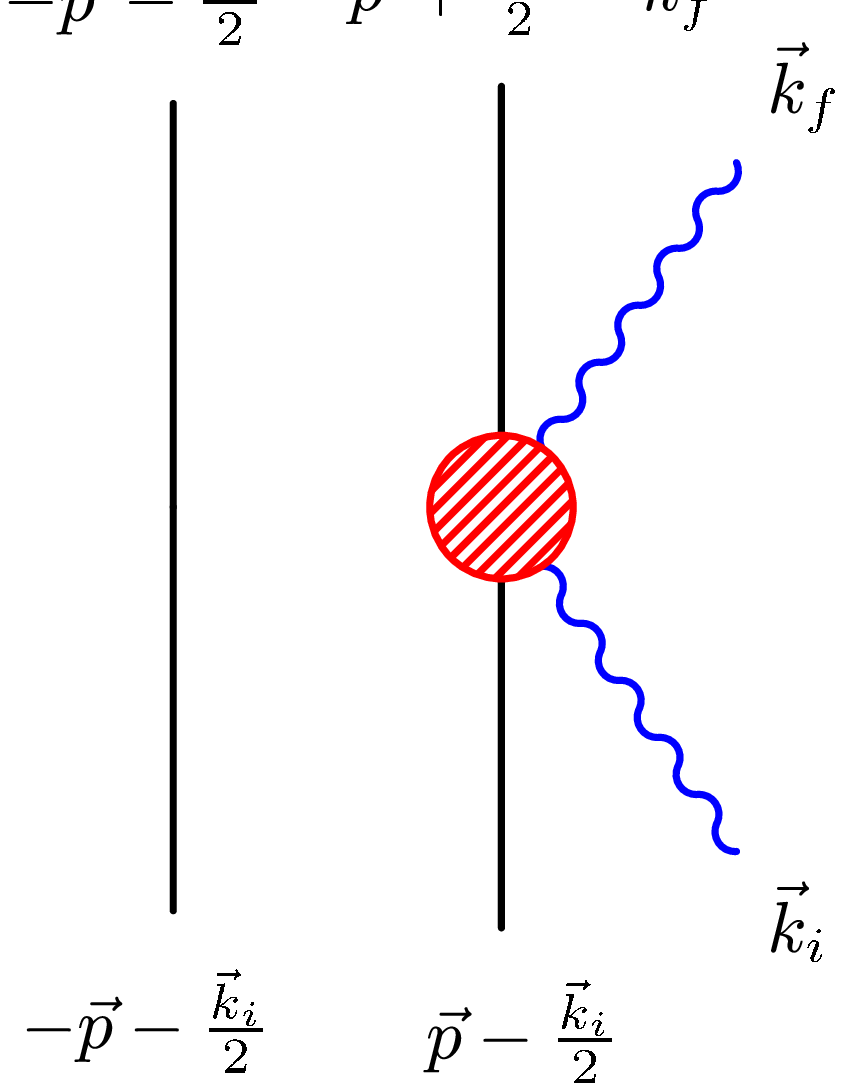}
\hspace{.5cm}
\includegraphics*[width=.25\textwidth]{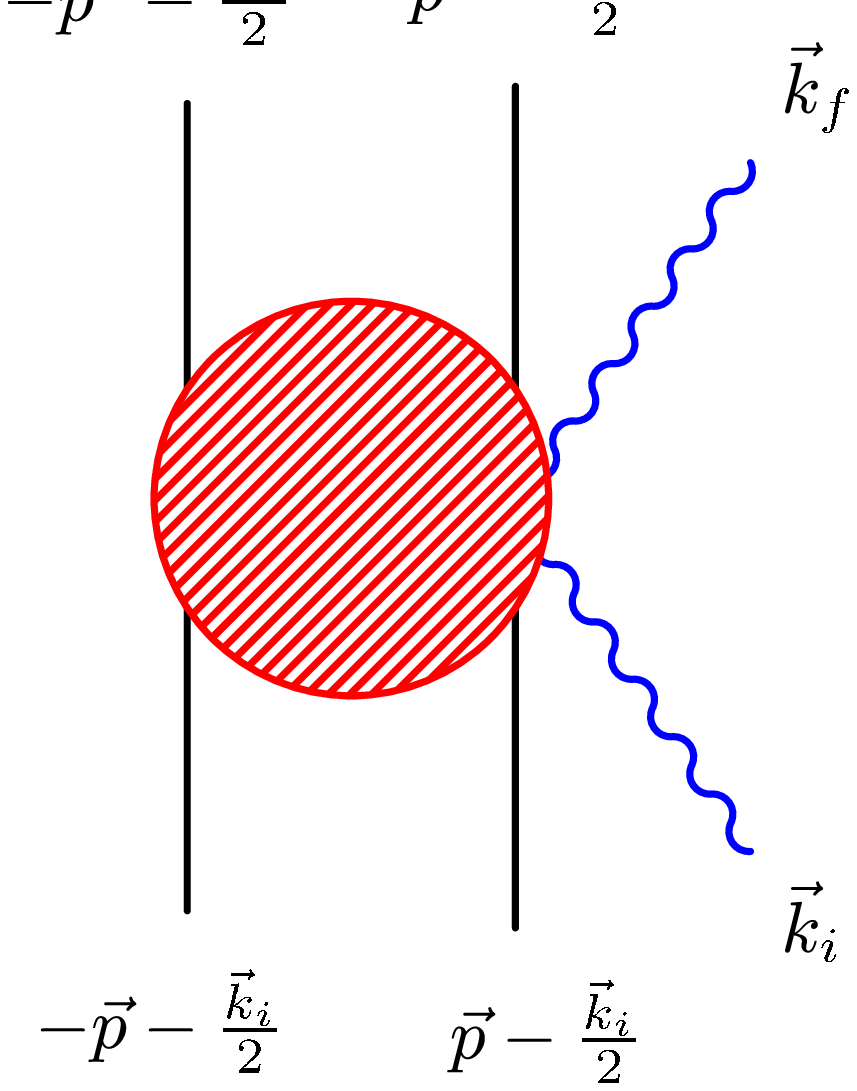}
\parbox{1.\textwidth}{
\caption[Nucleon momenta in deuteron Compton scattering]
{Momenta in single- and two-nucleon contributions to deuteron Compton 
scattering.}
\label{fig:deuteronmomenta}}
\end{center}
\end{figure}

The diagrams contributing to 
the scattering process considered can be classified
into ``one-body'' pieces, in which all photon interactions take place on a
single nucleon, and ``two-body'' pieces, in which both nucleons are
involved in the Compton-scattering process,
see Fig.~\ref{fig:deuteronmomenta}. As we calculate in the 
$\gamma d$ cm frame, the two nucleon momenta in the initial state must add up 
to $-\ki$. The relative momentum 
$\vec{p}=\frac{\vec{p}_1-\vec{p}_2}{2}$ of the two nucleons is non-vanishing. 
The momenta of the outgoing nucleons depend on whether only one or both 
nucleons are involved in the scattering process. In the first case, they are 
completely determined by the momenta of the incoming nucleons and the photons,
whereas momentum is transferred in the latter from one nucleon to the other, 
e.g. via the exchange of a pion. Of course, in both cases the 
momenta of the outgoing nucleons have to add up to $-\kf$
cf. Fig.~\ref{fig:deuteronmomenta}.

This, though, is not the only way to classify diagrams.
They can also be divided like
\be
G_{\gamma\gamma}=G\,K_\gamma\,G\,K_\gamma\,G+G\,K_{\gamma\gamma}\,G,
\label{eq:Ggammagamma}
\ee
where we defined the Green's function for Compton scattering from 
the $NN$ system as $G_{\gamma \gamma}$. It is the sum of all
Feynman graphs which contribute to $\gamma NN \rightarrow \gamma NN$ in
which the photon interacts with the $NN$ system. 
$G$ is the two-particle Green's function, constructed from the 
two-nucleon irreducible interaction $V$ and the free two-nucleon Green's 
function. $K_{\gamma}$ denotes the coupling of one photon to the two-nucleon
system, $K_{\gamma \gamma}$ is the two-nucleon irreducible kernel for the 
coupling of incoming and outgoing photon.

\begin{figure}[!htb]
\begin{center} 
\includegraphics*[width=.52\textwidth]
{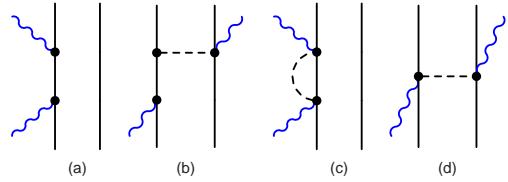}
\parbox{1.\textwidth}{
\caption[Examples of two-nucleon reducible and two-nucleon irreducible 
diagrams]
{Examples of two-nucleon reducible (a, b) and two-nucleon 
irreducible diagrams (c, d). Except for diagram (b), which is of 
$\calO(\epsilon^4)$, all graphs displayed are of third order in the Small 
Scale Expansion.}
\label{fig:reducibleirreducible}}
\end{center}
\end{figure}

The first piece in Eq.~(\ref{eq:Ggammagamma}) is called the 
``two-nucleon reducible'' part 
and the second is the ``two-nucleon irreducible'' part.  
Two-nucleon reducible diagrams are those which 
contain an intermediate state with only the two nucleons as particle content.
Note that, according to this classification,
$K_{\gamma} G K_{\gamma}$ and $K_{\gamma \gamma}$ each contain one-body
{\textit{and}} two-body pieces. An example of an $\mathcal{O}(\epsilon^3)$ 
one-body contribution to the
two-nucleon reducible part is given in Fig.~\ref{fig:reducibleirreducible}(a).
Two-nucleon reducible two-body contributions to 
$G_{\gamma \gamma}$ begin at $\mathcal{O}(\epsilon^4)$. 
One such diagram is given in Fig.~\ref{fig:reducibleirreducible}(b), whereas 
Figs.~\ref{fig:reducibleirreducible}(c) and (d) are two examples of 
two-nucleon irreducible diagrams in the one-nucleon and the two-nucleon 
sector, respectively.

The amplitude for Compton scattering off the deuteron is 
derived as the matrix element of the interaction kernel, consisting of the 
two-nucleon irreducible part $K_{\gamma\gamma}$ and the two-nucleon reducible 
part $K_\gamma\,G\,K_\gamma$, evaluated between an 
initial- and final-state deuteron wave function, cf. \cite{Phillips}:
\be
\mathcal{M}=\left<\Psi_f|K_{\gamma\gamma}+K_\gamma\,G\,K_\gamma|\Psi_i\right>
\label{eq:deuteronComptonamplitude},
\ee
see also Eq.~(\ref{eq:Mfi}).

In this chapter, we apply the same power-counting rules to both 
$K_{\gamma\gamma}$ and $K_\gamma G K_\gamma$, calculating all contributions 
to $\calO(\epsilon^3)$. Therefore, also the 
interaction between the two nucleons in the intermediate state of e.g. 
Fig.~\ref{fig:reducibleirreducible}(a) is built up only perturbatively. 
Up to the order to which we work, $G$ turns out as the Green's function of 
two free nucleons, while it represents the full $NN$-scattering Green's 
function in Chapter~\ref{chap:nonperturbative}.
We refer therefore to the approach to deuteron Compton scattering used in 
this chapter as the \textit{approach without rescattering}, whereas 
in Chapter~\ref{chap:nonperturbative}, we insert
the full two-particle Green's function in the intermediate state of 
Fig.~\ref{fig:reducibleirreducible}(a).

In order to determine which diagrams contribute to our leading-one-loop
order calculation, we first remind the reader that a diagram appearing 
at a certain order in $q$ in HB$\chi$PT contributes at the same order 
$\epsilon$ in SSE, cf. Section~\ref{sec:SSE}. 
In HB$\chi$PT, the leading-order 
propagator of a nucleon with the energy $\omega$ of the external probe 
flowing through it is $\frac{i}{\omega}$~\cite{BKM}. 
Corrections from the kinetic energy of the nucleon are
treated perturbatively. In the deuteron, such a perturbative treatment is not 
applicable for low photon energies, due to the relative momentum $\vec{p}$ 
between the two 
equally heavy
nucleons. Therefore, one has to use the full non-relativistic nucleon 
propagator $\frac{i}{\omega-p^2/2m_N}$ in the low-energy regime. Nonetheless, 
the approximation $\frac{i}{\omega}$ is useful for $\omega$ much larger than 
the expectation value $\left<p^2/m_N\right>$ of the kinetic energy inside the 
deuteron, which we found to be of the order of 20~MeV for the wave functions 
we are using. 
These considerations demonstrate that for $\omega\gg \left<p^2/m_N\right>$, 
the nucleon propagator may be counted as $\mathcal{O}(\epsilon^{-1})$  
like in standard HB$\chi$PT, whereas in the ``nuclear'' regime, i.e. 
$\omega\sim \mathcal{O}\left(\left<p^2/m_N\right>\right)$, 
it has to be counted as $\mathcal{O}(\epsilon^{-2})$, since $p\sim\epsilon$.
Therefore, from the point of view of $\chi$EFT, 
one has to strictly differentiate between two energy regimes: 
the nuclear regime $\omega\sim \mathcal{O}\left(\left<p^2/m_N\right>\right)$  
and the regime $\omega\sim \mathcal{O}(m_\pi)$.
In this chapter we restrict ourselves to the latter one, as we are mainly 
concerned with photon energies $\omega\geq50$~MeV, which is the energy region 
where one starts to be sensitive to the nucleon polarizabilities, cf. 
Ref.~\cite{Rupak}. 
Therefore, our calculation is only valid above some lower energy 
limit, which will turn out to be of the order of 50-60~MeV 
(cf. Section~\ref{sec:results}).
In Chapter~\ref{chap:nonperturbative}, we present an approach to deuteron 
Compton scattering which does not suffer from such a lower energy limit.
Nevertheless, the strict perturbative calculation of this chapter has some 
advantages with respect to the one presented in 
Chapter~\ref{chap:nonperturbative}: The most important point is the 
computational effort which as we shall see is considerably  smaller. 
The second advantage is that it is much easier to employ a 
systematic expansion scheme 
for the interaction kernel.
This systematicity leads to a high degree of transparency,
which makes it easy to disentangle the various nuclear 
degrees of freedom from each other. 

In the regime $\omega \sim \mathcal{O}(\mpi)$, the contributions from
$K_\gamma G K_\gamma$ can be treated using a perturbative chiral expansion.
Heuristically, this can be easily understood, because the absorption of a 
high-energy photon immediately separates the nucleons from each other, so 
the deuteron would be destroyed if the second photon
was not emitted near-instantaneously. 
Such a perturbative treatment is \textit{not} valid in the nuclear 
regime, $\omega\sim\mathcal{O}\left(\left<p^2/m_N\right>\right)$, 
as described in detail in~\cite{Phillips,McGPhil}, and applying it there 
leads~-- not surprisingly~-- to violations of the low-energy theorems which 
govern the limit $\omega\rightarrow 0$.
For example, the Thomson limit for Compton scattering from a nucleus of charge 
$Q\,e$ and mass $A\,m_N$,
\be
A^\mathrm{Thomson}=A(\omega=0)=
       -\frac{Q^2\,e^2}{A\,m_N}\,\vec{\epsilon}\cdot\vec{\epsilon}\,',
\label{eq:Thomson}
\ee
is a direct consequence of gauge invariance \cite{Friar} and cannot be 
recovered without the full 
two-nucleon Green's function in the intermediate state of 
diagram~\ref{fig:reducibleirreducible}(a). It is one of the central 
results of Chapter~\ref{chap:nonperturbative} that the correct low-energy 
limit will be obtained (see Section~\ref{sec:Thomson2}).  Some attempts to
reach this limit at least approximately by the inclusion of 
$\calO(p^4)$-pion-exchange diagrams
are reported in Section~\ref{sec:Thomson1}.

Due to the inapplicability of our calculation at small photon energies, we 
strictly constrain ourselves in this chapter to 
$\omega\sim \mathcal{O}(m_\pi)$, where a 
perturbative expansion of the kernel in the standard HB$\chi$PT counting 
scheme, i.e. counting the nucleon propagator as $\mathcal{O}(\epsilon^{-1})$, 
is possible\footnote{The only exception to this rule is 
Section~\ref{sec:Thomson1}, where we investigate the Thomson limit.}. 
The lower limit of this power counting  turns out to be 
$\omega\approx 50$~MeV, so we have to caution the reader that the 
calculation is  not supposed to work in the region  $\omega\ll 50$~MeV.
In this energy regime, pions may be treated as ``heavy'' compared to the 
photon energy, and it 
can therefore be more convenient to use an Effective Field Theory approach to 
deuteron Compton scattering where pions are integrated out, 
see Refs.~\cite{Rupak, Ji}. These calculations describe the 
very-low-energy region well and also reach the exact Thomson limit.

\begin{figure}[!htb]
\begin{center} 
\includegraphics*[width=.28\textwidth]{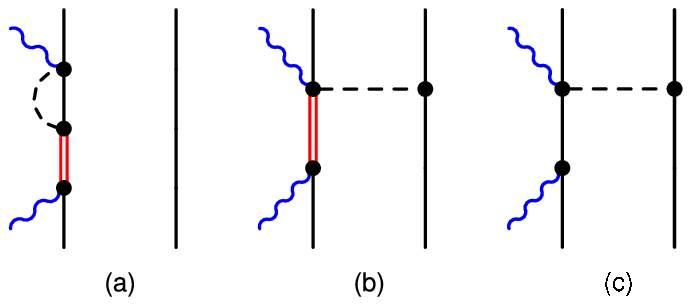}
\parbox{1.\textwidth}{
\caption[Two examples of $\mathcal{O}(\epsilon^4)$ contributions to deuteron 
Compton scattering]
{Two examples of $\mathcal{O}(\epsilon^4)$ contributions to deuteron Compton 
scattering with explicit $\Delta(1232)$ degrees of 
freedom in the  one- and two-body sector.}
\label{fig:SSEOQ4}}
\end{center}
\end{figure}

As we calculate $\gamma d$ scattering in the Small Scale Expansion, we 
have to fix our counting rules also for diagrams including $\Delta(1232)$ 
propagators. For the one-body contributions this is straightforward, as we 
apply the SSE counting scheme, cf. Refs.~\cite{HGHP, HHK}. 
As far as the two-body physics is concerned,
we combine the SSE counting rules, e.g. counting the $\Delta$-propagator as 
$\epsilon^{-1}$, with Weinberg's prescription of counting only within the 
interaction kernel. To $\mathcal{O}(\epsilon^3)$, the order up to which we are
working, this leads to identical meson-exchange diagrams as in the 
$\mathcal{O}(p^3)$-HB$\chi$PT calculation. All additional diagrams are at 
least one order higher, an example is given in Fig.~\ref{fig:SSEOQ4}(b) (an 
example of an $\mathcal{O}(\epsilon^4)$-one-body diagram is sketched 
in Fig.~\ref{fig:SSEOQ4}(a)). Note that only  nucleon propagators 
with the energy of the scattered photon flowing through them are 
supposed to be counted as $\epsilon^{-1}$. Therefore, up to 
$\mathcal{O}(\epsilon^3)$ 
there are no diagrams with $\Delta$ excitations in outgoing lines,
since the $\Delta$ propagator always has to be counted as $-1$.

As a side remark we note that a modified counting 
scheme in the two-body sector has been suggested in~\cite{Beane}, as certain
pion-exchange diagrams may be enhanced when the photon energy comes close 
enough to the pion mass that the pions in the two-body diagrams are almost on 
mass shell. We do not consider such a modification necessary for 
our calculation, as we restrict ourselves to photon energies 
$\omega\leq100$~MeV.
Therefore, the diagrams contributing to deuteron Compton scattering up to 
$\mathcal{O}(\epsilon^3)$ are:

\begin{itemize}
\item One-body contributions without explicit $\Delta(1232)$ degrees of 
freedom. These are the single-nucleon seagull with the two-photon vertices from
$\mathcal{L}_{N\pi}^{(2)}$ 
and $\mathcal{L}_{N\pi}^{(3)}$~(see Fig.~\ref{fig:chiPTsingle}(a) and 
Appendix~\ref{app:poleterms}). 
The former gives the only contribution at $\mathcal{O}(\epsilon^2)$, 
the latter enters at $\mathcal{O}(\epsilon^3)$. 
Also at $\mathcal{O}(\epsilon^3)$, the nucleon-pole terms 
(Fig.~\ref{fig:chiPTsingle}(b) and Appendix~\ref{app:poleterms})
and the contributions from the leading chiral dynamics of the pion cloud 
around the nucleon enter (Figs.~\ref{fig:chiPTsingle}(d)-(g), see also 
Fig.~\ref{fig:Npicontinuum}). We note that at $\mathcal{O}(\epsilon^3)$ 
the nucleon $s$-channel pole term is the 
only contribution from $K_\gamma\,G\,K_\gamma$, cf. Eq.~(\ref{eq:Ggammagamma}).
The pion pole (Fig.~\ref{fig:chiPTsingle}(c)), i.e. the $\pi^0$-exchange in 
the $t$-channel, does not contribute to  deuteron Compton scattering 
at this order, as it is an isovector and we neglect isospin-breaking effects.
\begin{figure}[!htb]
\begin{center} 
\includegraphics*[width=.121\linewidth]{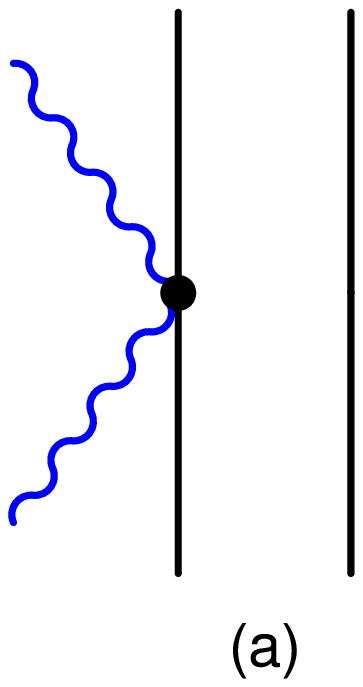}
\includegraphics*[width=.121\linewidth]{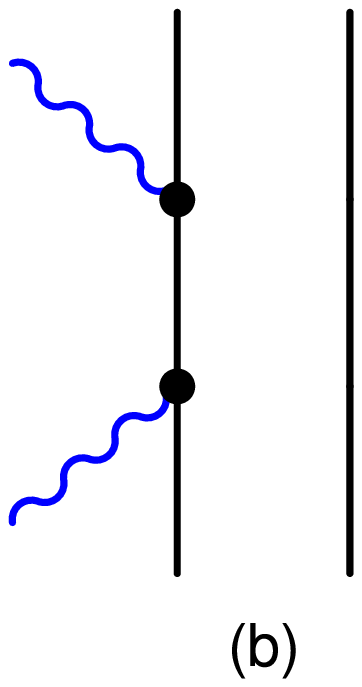}
\includegraphics*[width=.121\linewidth]{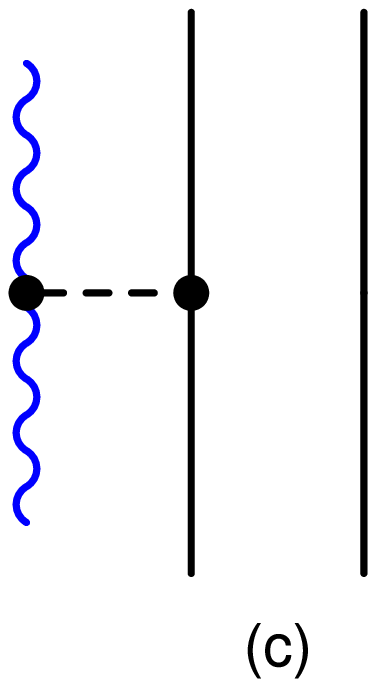}
\includegraphics*[width=.121\linewidth]{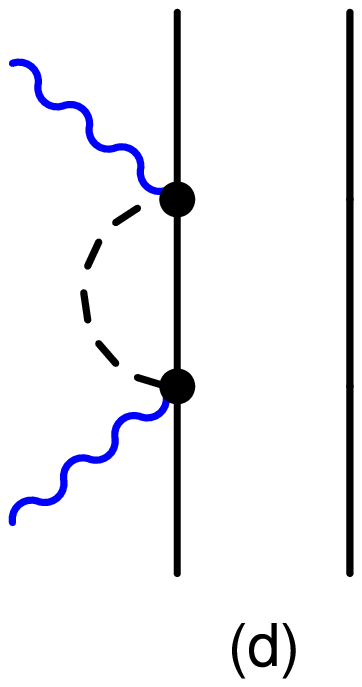}
\includegraphics*[width=.121\linewidth]{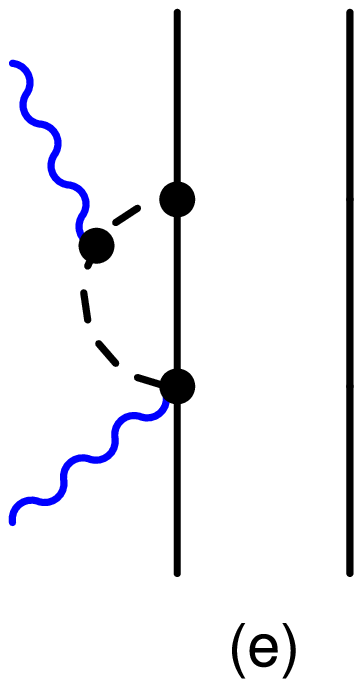}
\includegraphics*[width=.121\linewidth]{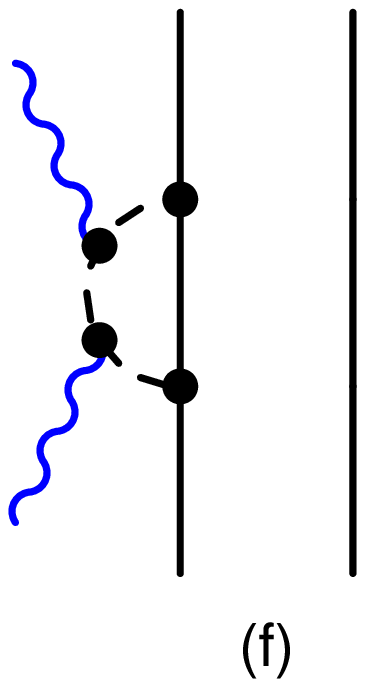}
\includegraphics*[width=.121\linewidth]{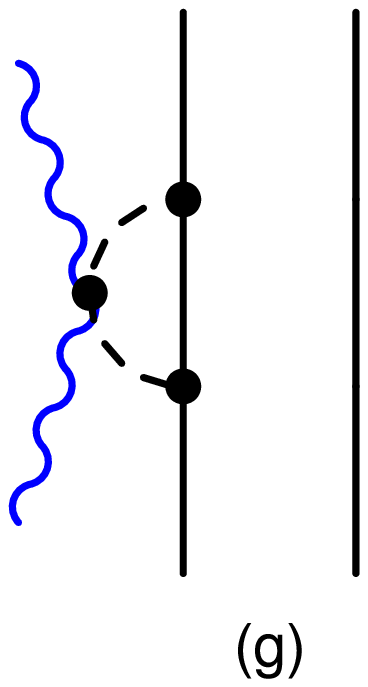}
\parbox{1.\textwidth}{
\caption[One-body interactions without $\Delta(1232)$ propagator 
contributing to $\gamma d$ scattering up to $\mathcal{O}(\epsilon^3)$ in SSE]
{One-body interactions without a $\Delta(1232)$ propagator 
contributing to deuteron Compton scattering up to $\mathcal{O}(\epsilon^3)$ 
in SSE. Permutations and crossed graphs are not shown.}
\label{fig:chiPTsingle}}
\end{center}
\end{figure}
\item One-body diagrams with explicit $\Delta$ degrees of freedom, as 
shown in Fig.~\ref{fig:SSEsingle}: The $\Delta$-pole diagrams 
(Fig.~\ref{fig:SSEsingle}(a)) and the contributions from the pion cloud around 
the $\Delta(1232)$ (Figs.~\ref{fig:SSEsingle}(b)-(e)). 
\item The two isoscalar short-distance one-body operators 
(Fig.~\ref{fig:SSEsingle}(f)), introduced in Section~\ref{sec:spin-averaged}. 
We note that except for these two contact operators, the 
$\delta$-expansion~\cite{deltaexp} 
(cf. Section~\ref{sec:SSE}) up to NNLO is equivalent to 
$\calO(\epsilon^3)$ SSE in the energy range $\omega\sim m_\pi$ considered. 
\begin{figure}[!htb]
\begin{center} 
\includegraphics*[width=.121\linewidth]{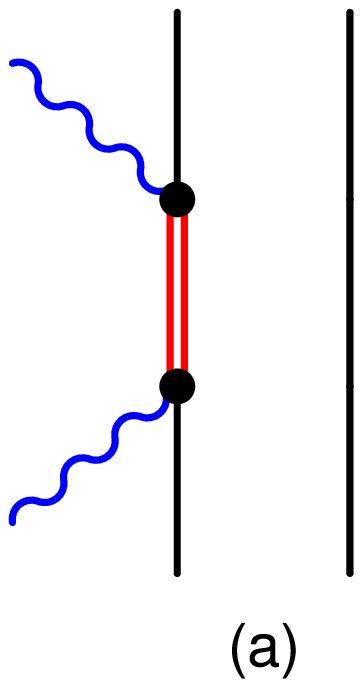}
\includegraphics*[width=.121\linewidth]{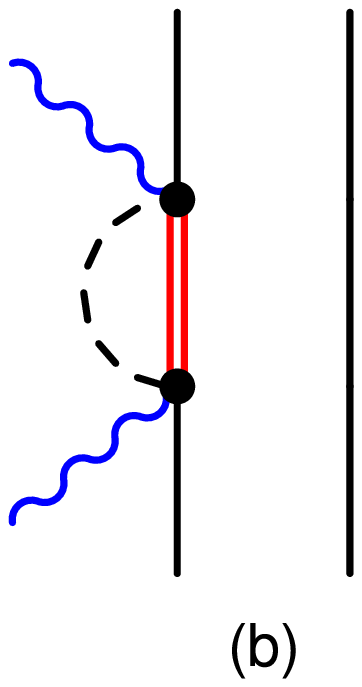}
\includegraphics*[width=.121\linewidth]{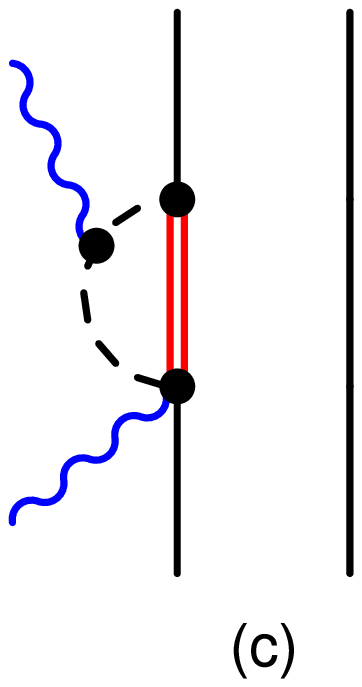}
\includegraphics*[width=.121\linewidth]{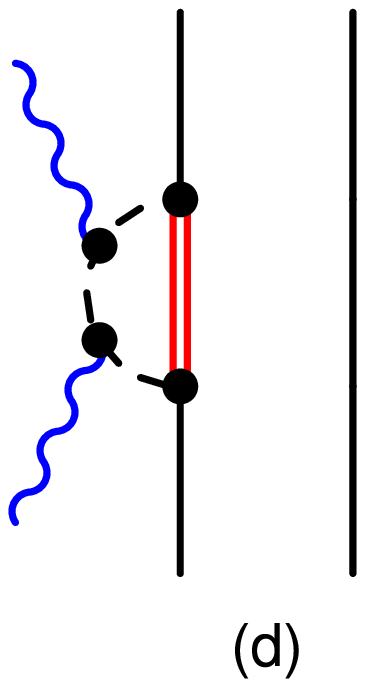}
\includegraphics*[width=.121\linewidth]{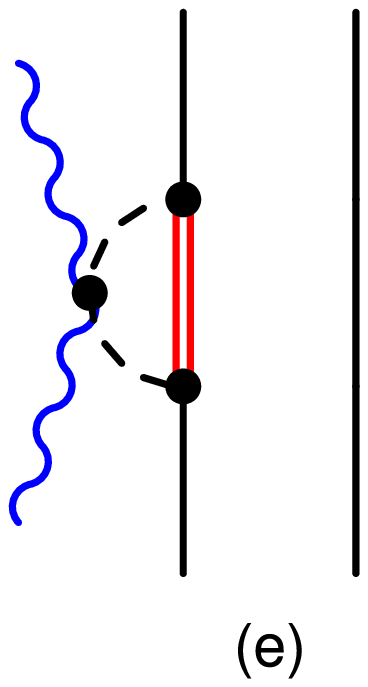}
\includegraphics*[width=.121\linewidth]{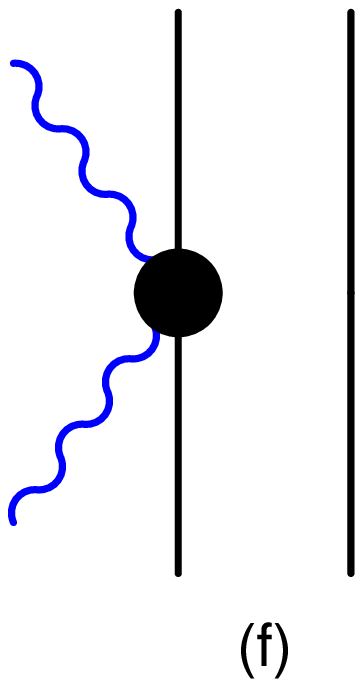}
\parbox{1.\textwidth}{
\caption[Additional one-body interactions contributing to $\gamma d$
scattering at $\mathcal{O}(\epsilon^3)$ in SSE  
compared to third-order HB$\chi$PT.]
{Additional one-body interactions which contribute to deuteron Compton 
scattering at $\mathcal{O}(\epsilon^3)$ in SSE  
compared to third-order HB$\chi$PT. Permutations and crossed graphs 
are not shown.}
\label{fig:SSEsingle}}
\end{center}
\end{figure}
\item Two-body contributions with one pion exchanged between the two 
nucleons (Fig.~\ref{fig:chiPTdouble}). 
In total there 
are nine two-body diagrams at $\mathcal{O}(\epsilon^3)$. As discussed 
before, the meson-exchange diagrams are identical in third-order HB$\chi$PT 
and SSE.
\begin{figure}[!htb]
\begin{center} 
\includegraphics*[width=.75\textwidth]{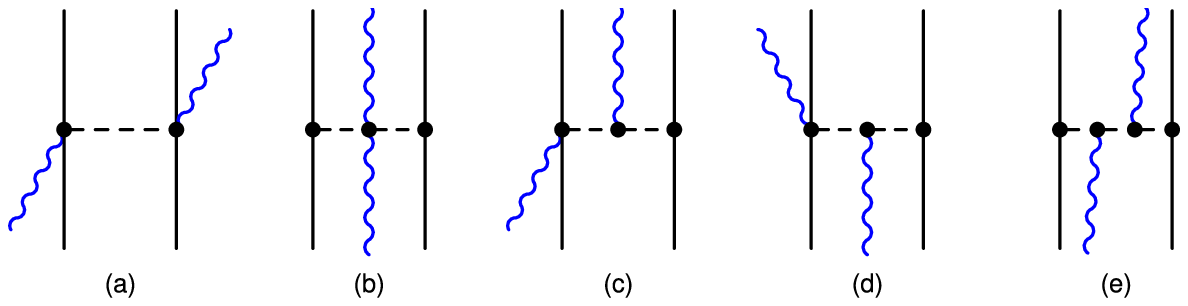}
\parbox{1.\textwidth}{
\caption[Two-body interactions contributing 
to deuteron Compton scattering at $\mathcal{O}(\epsilon^3)$ in SSE]
{Two-body interactions 
contributing to the kernel for deuteron Compton scattering 
at $\mathcal{O}(\epsilon^3)$ in SSE. Diagrams which differ only by 
nucleon interchange are not shown.}
\label{fig:chiPTdouble}}
\end{center}
\end{figure}
\end{itemize}

\begin{figure}[!htb]
\begin{center} 
\parbox[m]{.25\linewidth}{
\includegraphics*[width=.25\textwidth]{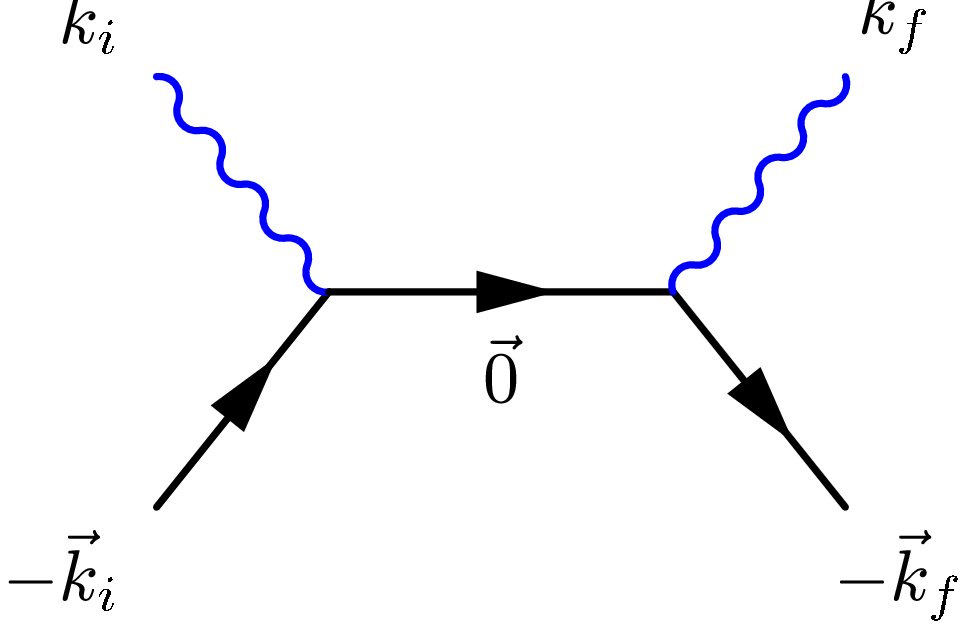}}
$\longrightarrow$
\parbox[m]{.25\linewidth}{
\includegraphics*[width=.28\textwidth]{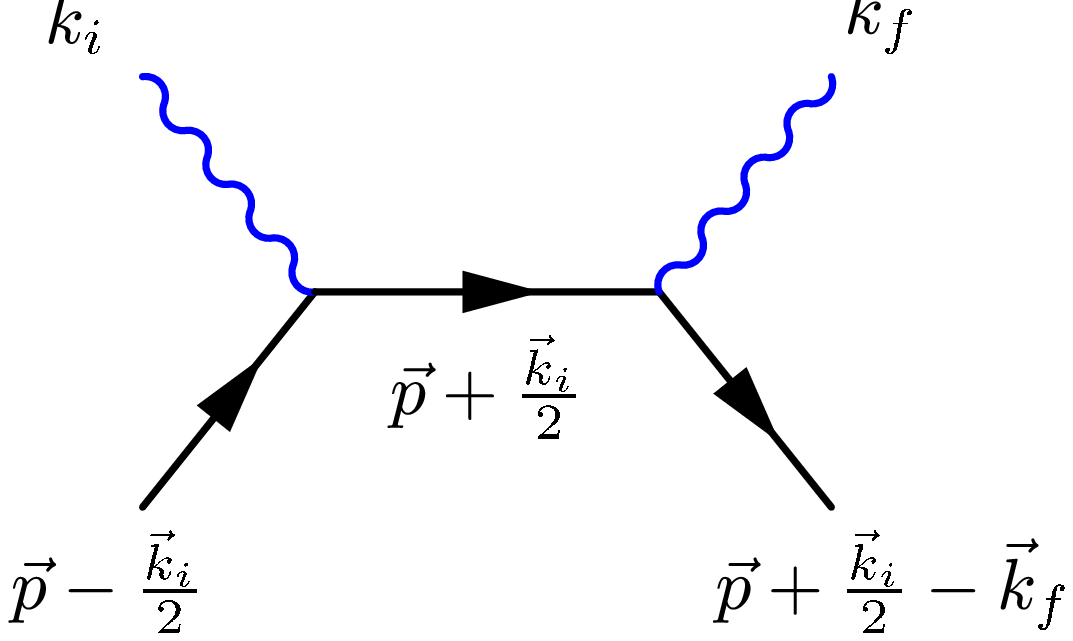}}
\parbox{1.\textwidth}{
\caption[Momenta in the $s$-channel nucleon-pole diagram in the $\gamma d$-cm 
frame]
{Momenta in the $s$-channel nucleon-pole diagram comparing the $\gamma d$-cm 
frame (right) to the  $\gamma N$-cm frame (left). The analogous replacement 
in the $u$-channel is not shown.}
\label{fig:boost}}
\end{center}
\end{figure}

All these diagrams (Figs.~\ref{fig:chiPTsingle}-\ref{fig:chiPTdouble}) make 
up our interaction kernel. The SSE single-nucleon amplitudes can 
be found in~\cite{HGHP}, however there is one difference  
compared to the $T$-matrix for Compton scattering off the single nucleon 
(Eq.~(\ref{eq:Tmatrix})):  The nucleon-pole amplitudes, which are given in the
$\gamma N$ center-of-mass frame in Appendix~\ref{app:poleterms}, have to be 
boosted to the $\gamma N N$ center-of-mass system, as our calculation is 
performed in the $\gamma d$ cm frame. This is easily accomplished by 
evaluating  the pole diagrams (Fig.~\ref{fig:chiPTsingle}(b) plus crossed) in 
a frame with non-zero total $\gamma N$ momentum, see Fig.~\ref{fig:boost}.
We verified the resulting formula for the boost, given in 
Ref.~\cite{Phillips}:
\be
T_\mathrm{boost}=-\frac{Q^2\,e^2}{2\,m_N^2\,\w}\,
\left\{(\vec{\epsilon}\cdot\kf    )\,(\vec{\epsilon}\,'\cdot\ki)+2\,
 \left[(\vec{\epsilon}\cdot\vec{p})\,(\vec{\epsilon}\,'\cdot\ki)+
       (\vec{\epsilon}\cdot\kf    )\,(\vec{\epsilon}\,'\cdot\vec{p})\right]
\right\}
\ee
We note that we have simplified the expressions for the single-nucleon 
amplitudes given in~\cite{HGHP} with respect
to the exact position of the pion threshold, 
cf. Section~\ref{sec:spinaveragedtheory}, as we are only analysing Compton 
scattering for photon energies $\leq 100$~MeV. For such low energies, 
shifting the pion threshold has only a minor effect~\cite{DA}.
An estimate of the (small) size of this simplification is given in 
Section~\ref{sec:threshold}. 

The two-nucleon amplitude corresponding to Fig.~\ref{fig:chiPTdouble} has 
been calculated in \cite{Phillips} and verified by us. 
It reads (with $\vec{\tau}^t=(\tau^x,\tau^y,\tau^z)$) 
\be
T^{\gamma NN}=
-\frac{e^2\,g_A^2}{2\,f_\pi^2}\,\left(\vec{\tau}_1\cdot\vec{\tau}_2-
\tau_1^z\,\tau_2^z\right)\,\left(t^{(a)}+t^{(b)}+t^{(c)}+t^{(d)}+t^{(e)}\right)
\label{eq:twonucleonamplitude}
\ee
with
\begin{align}
t^{(a)}&=\frac{(\epsvec\cdot\sigone)\,(\epspvec\cdot\sigtwo)}
{2\,[\w^2-m_\pi^2-(\pvec-\ppvec+\frac{1}{2}\,(\ki+\kf))^2]}+
(1\leftrightarrow 2),\nonumber\\
t^{(b)}&=\frac{(\epsvec\cdot\epspvec)\,
(\sigone\cdot(\pvec-\ppvec-\frac{1}{2}\,(\ki-\kf)))\,
(\sigtwo\cdot(\pvec-\ppvec+\frac{1}{2}\,(\ki-\kf)))}
{[(\pvec-\ppvec-\frac{1}{2}\,(\ki-\kf))^2+\mpi^2]\,
 [(\pvec-\ppvec+\frac{1}{2}\,(\ki-\kf))^2+\mpi^2]},\nonumber\\
t^{(c)}&=-\frac{(\epspvec\cdot(\pvec-\ppvec+\frac{1}{2}\,\ki))\,
(\sigone\cdot\epsvec)\,
(\sigtwo\cdot(\pvec-\ppvec+\frac{1}{2}\,(\ki-\kf)))}
{[\w^2-\mpi^2-(\pvec-\ppvec+\frac{1}{2}\,(\ki+\kf))^2]\,
 [(\pvec-\ppvec+\frac{1}{2}\,(\ki-\kf))^2+\mpi^2]}+
(1\leftrightarrow 2),\nonumber\\
t^{(d)}&=-\frac{(\epsvec\cdot(\pvec-\ppvec+\frac{1}{2}\,\kf))\,
(\sigone\cdot(\pvec-\ppvec-\frac{1}{2}\,(\ki-\kf)))\,
(\sigtwo\cdot\epspvec)}
{[\w^2-\mpi^2-(\pvec-\ppvec+\frac{1}{2}\,(\ki+\kf))^2]\,
 [(\pvec-\ppvec+\frac{1}{2}\,(\ki-\kf))^2+\mpi^2]}+
(1\leftrightarrow 2),\nonumber\\
t^{(e)}&=\frac{2\,(\epsvec \cdot(\pvec-\ppvec+\frac{1}{2}\,\kf))\,
                   (\epspvec\cdot(\pvec-\ppvec+\frac{1}{2}\,\ki))\,
(\sigone\cdot(\pvec-\ppvec-\frac{1}{2}\,(\ki-\kf)))}
{[\w^2-\mpi^2-(\pvec-\ppvec+\frac{1}{2}\,(\ki+\kf))^2]\,
 [(\pvec-\ppvec-\frac{1}{2}\,(\ki-\kf))^2+\mpi^2]}\nonumber\\
&\times\frac{(\sigtwo\cdot(\pvec-\ppvec+\frac{1}{2}\,(\ki-\kf)))}
{[(\pvec-\ppvec+\frac{1}{2}\,(\ki-\kf))^2+\mpi^2]}+(1\leftrightarrow 2).
\end{align}
$(1\leftrightarrow 2)$ is a shortcut for the corresponding amplitude with the 
two nucleons exchanged, as e.g. in Fig.~\ref{fig:chiPTdouble}(a) the incoming 
(outgoing) photon can couple to nucleon 1 (2) or vice versa. Such an exchange 
term obviously does not exist for diagram \ref{fig:chiPTdouble}(b), which 
is invariant under the exchange of nucleon~1 and nucleon~2.

In order to calculate the amplitude for deuteron Compton scattering we now
have to evaluate the kernel between an initial- and final-state deuteron wave 
function, cf. Eq.~(\ref{eq:deuteronComptonamplitude}). The isospin operator
$(\vec{\tau}_1\cdot\vec{\tau}_2-\tau_1^z\,\tau_2^z)$ in 
Eq.~(\ref{eq:twonucleonamplitude}) reflects the fact that the photon 
couples only to charged pions. 
Its evaluation yields
\be
\mx{d}{\vec{\tau}_1\cdot\vec{\tau}_2-\tau_1^z\,\tau_2^z}{d}
=-2,
\ee
with the isospin wave function of the isospin-0 deuteron  
\be
\bra{d}\,=\frac{1}{\sqrt{2}}\,\bra{p\,n-n\,p}.
\label{eq:isospinwavefunction}
\ee
Now we write down the amplitude 
according to \cite{Phillips}, separating the single-nucleon amplitude with 
its kernel $T^{\gamma N}$ from the two-nucleon contributions:
\begin{align}
\Mfi{}(\ki,\kf)&=\int\frac{d^3p}{(2\pi)^3}\,\Psi_f^\ast(\pvec+(\ki-\kf)/2)\,
T^{\gamma N}(\ki,\kf;\pvec)\,\Psi_i(\pvec)\nonumber\\
&+\int\frac{d^3p\,d^3p'}{(2\pi)^6}\,\Psi_f^\ast(\ppvec)\,
T^{\gamma NN}(\ki,\kf;\pvec,\ppvec)\,\Psi_i(\pvec)
\label{eq:Mfi}
\end{align}
The deuteron wave functions~-- explicit expressions can be found in 
Appendix~\ref{app:wavefunction}~-- depend on 
half of the difference between the momenta of the two nucleons,
see Fig.~\ref{fig:deuteronmomenta}. Due to the momentum transfer by 
the pion we have to integrate over momentum space twice in the two-nucleon 
amplitude, whereas there is only one loop integral to perform in the 
single-nucleon part. The indices $i,f$ denote the dependence of the 
amplitude on the initial and final photon polarization and also on the 
projections of the total angular momentum of the deuteron $M_f$ and $M_i$.
Note that we have to sum over all possible combinations of spins of the two 
nucleons in the initial and final state, cf. Eq.~(\ref{eq:Psi}). 
These sums are not explicitly shown in Eq.~(\ref{eq:Mfi}).

The spin-averaged differential cross section for deuteron Compton 
scattering is calculated like the single-nucleon cross section, cf. 
Section~\ref{sec:crosssectionsgeneral}: We take the absolute square of 
Eq.~(\ref{eq:Mfi}), average over the initial and sum over the final photon and 
deuteron states 
and multiply the result by the square of the appropriate phase-space factor, 
depending on the frame, in which we want to evaluate the cross section. 
Therefore we find, using $\sum_{i,f}$ as a shortcut for 
$\sum_{M_i,M_f,\li,\lf}$
\be
\left.\frac{d\sigma}{d\Omega}\right|_{\gamma d}=\Phi^2\cdot\frac{1}{6}\,
\sum_{i,f}|\Mfi{}|^2,
\label{eq:deuteroncrosssection}
\ee
as there are six possible initial states: three deuteron polarizations times 
the two polarizations of the photon. The phase-space factors $\Phi$ read 
\ba
\Phi_\mathrm{cm} &=\frac{m_d }{4\pi\,\sqrt{s_{\gamma d}}},\nonumber\\
\Phi_\mathrm{lab}&=\frac{\w_f}{4\pi\,\w_i},
\end{align}
with
\ba
\sqrt{s_{\gamma d}}&=\w+\sqrt{\w^2+m_d^2},\nonumber\\
\w_f&=\frac{m_d\,\w_i}{m_d+\w_i\,(1-\cos\theta_\mathrm{lab})},
\end{align} 
cf. Eqs.~(\ref{eq:phasespace},~\ref{eq:sandwf}). The relation between the 
initial photon energy in the lab frame $\w_i$ and the photon energy in the 
$\gamma d$-cm frame $\w$ is
\be
\w=\frac{\w_i}{\sqrt{1+2\w_i/m_d}},
\ee
in analogy to Eq.~(\ref{eq:wirelation}).

Now that we have defined the theoretical framework for $\gamma d$ scattering, 
we compare in the next section our $\mathcal{O}(\epsilon^3)$-SSE
results for the deuteron Compton cross 
sections to the $\mathcal{O}(p^3)$-HB$\chi$PT calculation performed 
in~\cite{Phillips} and to the available data. 
The comparison to the $\mathcal{O}(p^4)$-HB$\chi$PT fit from 
Ref.~\cite{McGPhil} is postponed to 
Section~\ref{sec:fits1}, where we fit the isoscalar polarizabilities 
$\bar{\alpha}_{E1}^s$, $\bar{\beta}_{M1}^s$ to these data.
Special emphasis is put on the energy and 
wave-function dependence of the cross sections.

\section[Predictions for Deuteron Compton Cross Sections]
{Predictions for Deuteron Compton Cross\\Sections\label{sec:results}}

\begin{figure}[!htb]
\begin{center} 
\includegraphics*[width=.48\textwidth]{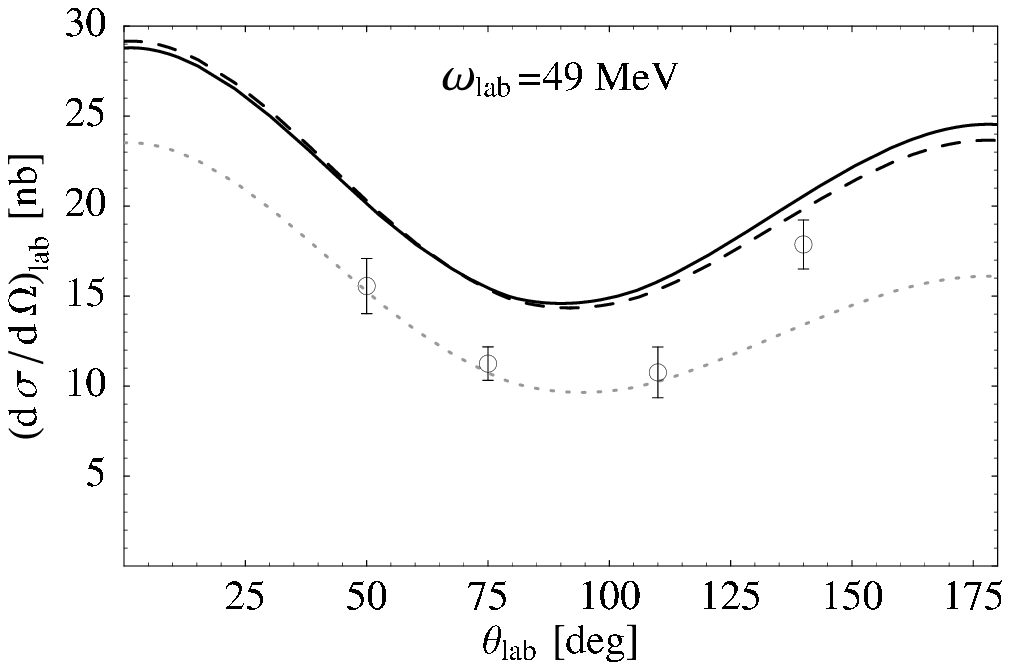}
\hfill
\includegraphics*[width=.48\textwidth]{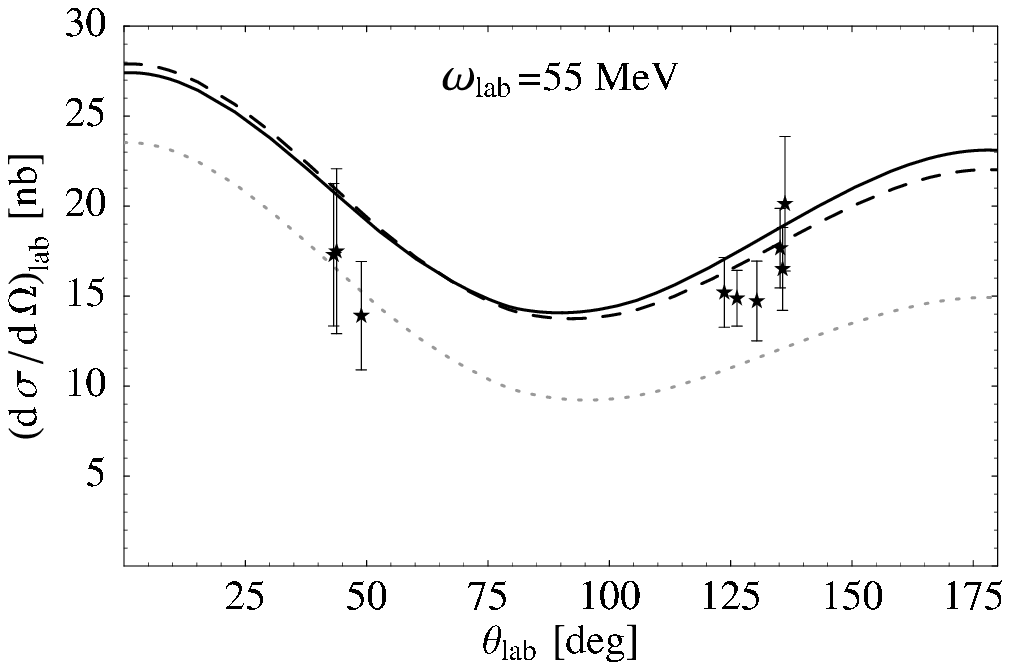}
\includegraphics*[width=.48\textwidth]{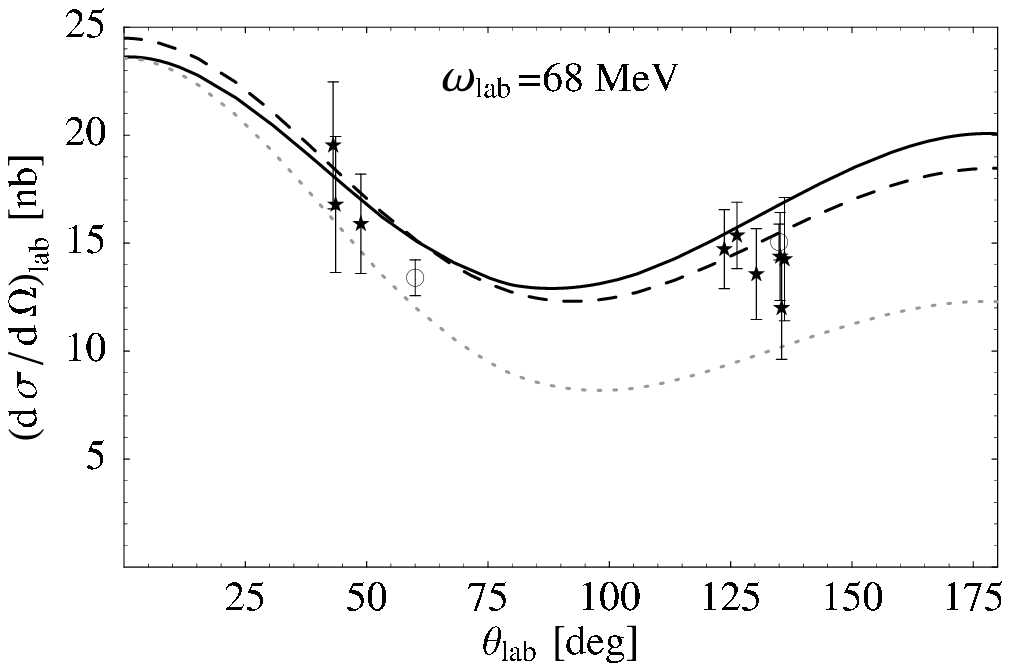}
\hfill
\includegraphics*[width=.48\textwidth]{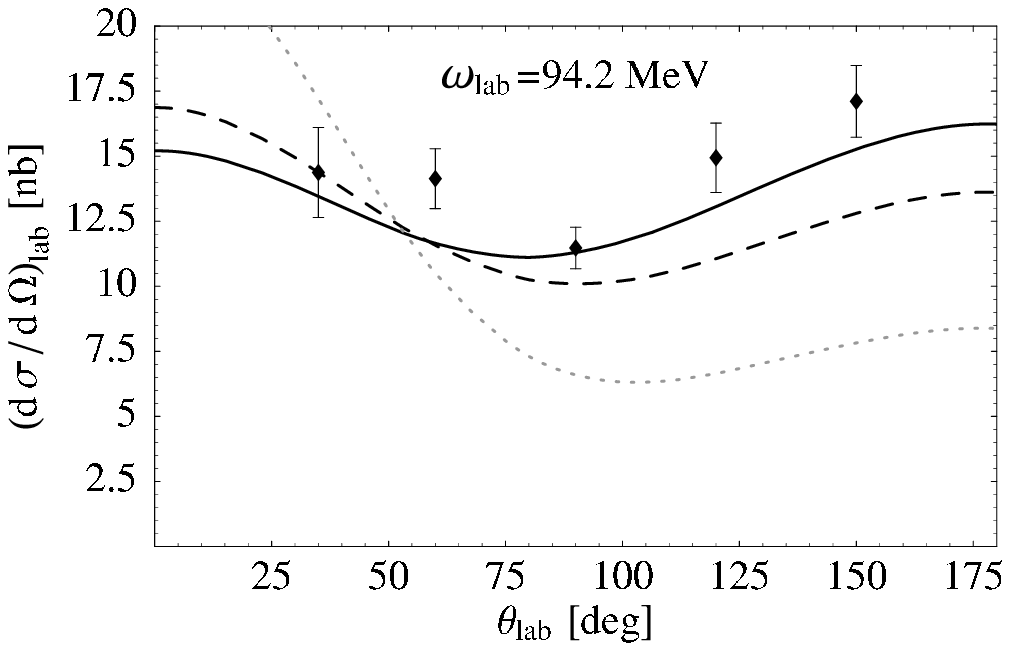}
\parbox{1.\textwidth}{
\caption[Comparison of the $\mathcal{O}(p^3)$-HB$\chi$PT and the 
$\mathcal{O}(\epsilon^3)$-SSE prediction for deuteron Compton scattering]
{Comparison of the $\mathcal{O}(p^3)$-HB$\chi$PT (dashed) and the 
$\mathcal{O}(\epsilon^3)$-SSE (solid) prediction at 
$\omega_\mathrm{lab}=49~\mathrm{MeV}$, $\omega_\mathrm{lab}=55~\mathrm{MeV}$, 
$\omega_\mathrm{lab}=68~\mathrm{MeV}$
and $\omega_\mathrm{lab}=94.2~\mathrm{MeV}$ using the chiral NNLO wave 
function~\cite{Epelbaum}. The data are from Illinois~\cite{Lucas} (circle), 
Lund~\cite{Lund} (star) and SAL~\cite{Hornidge} (diamond). The dotted line is 
the $\mathcal{O}(p^2)$ result.}
\label{fig:SSEHBplots}}
\end{center}
\end{figure}

In Fig.~\ref{fig:SSEHBplots}, we compare the $\mathcal{O}(\epsilon^3)$-SSE 
predictions to the $\mathcal{O}(p^3)$-HB$\chi$PT calculation of 
Ref.~\cite{Phillips}, using the wave function derived from the NNLO chiral 
potential with spectral-function regularization and cutoff 
$\Lambda=650$~MeV~\cite{Epelbaum}; this is the 
wave function that we always use if not stated differently. 
We also show the $\mathcal{O}(p^2)$ 
result, which consists only of the single-nucleon seagull 
(Fig.~\ref{fig:chiPTsingle}(a)). 
As we discuss only \textit{predictions} for deuteron Compton scattering in 
this section, we postpone the comparison to the $\mathcal{O}(p^4)$-HB$\chi$PT 
fits of Ref.~\cite{McGPhil} to Section~\ref{sec:fits1}.
The experiments shown have been performed at a lab energy of 
49~MeV~\cite{Lucas}, 55~MeV~\cite{Lund}, $\sim$67~MeV~\cite{Lund}, 
69~MeV~\cite{Lucas} and $\sim$94.2~MeV~\cite{Hornidge}. (The last experiment
used photons in an energy range from $84.2-104.5$~MeV; the deviation from the
central value has been corrected for~\cite{Hornidge}.) 

The numerical values for the various input parameters are given in 
Table~\ref{tab:const}.
We use for the coupling constants $g_{1}$ and $g_{2}$, 
connected with the two short-distance $\gamma N$-operators 
(cf.~Sections~\ref{sec:spin-averaged} and \ref{sec:theory}), 
and the $\gamma N\Delta$ coupling $b_1$ 
the results of the Baldin-sum-rule-constrained fit to the spin-averaged 
proton Compton-scattering data from Section~\ref{sec:protonfits}. 
Determining the isoscalar parameters $g_{1}$, $g_{2}$~-- or, 
eqivalently, the polarizabilities $\bar{\alpha}_{E1}^s$, 
$\bar{\beta}_{M1}^s$~-- from proton data alone is possible, because the 
$\mathcal{O}(\epsilon^3)$-SSE calculation predicts 
$\alpha_{E1}^n\equiv\alpha_{E1}^p$, $\beta _{M1}^n\equiv\beta _{M1}^p$, 
as the isovector contributions only come in at 
$\mathcal{O}(\epsilon^4)$. This is in agreement with small isovector 
polarizabilities found in Dispersion-Theoretical Analyses based on 
pion-photoproduction multipoles (see e.g. \cite{review}).
We use the central values of the fit, which are 
$\bar{\alpha}_{E1}^p=11.04\cdot10^{-4}\;\mathrm{fm}^3$, 
$\bar{\beta} _{M1}^p= 2.76\cdot10^{-4}\;\mathrm{fm}^3$, cf. 
Section~\ref{sec:protonfits}. Therefore, like in third-order HB$\chi$PT, 
there are no free parameters in our deuteron Compton calculation.

From the 49~MeV, 55~MeV and 68~MeV curves shown in Fig.~\ref{fig:SSEHBplots} 
it is obvious that explicit $\Delta$
degrees of freedom may well be neglected for these low energies. The two 
calculations~-- HB$\chi$PT and SSE~-- yield results which differ only within 
the uncertainties one expects from higher-order contributions. 
This is an important check, as it demonstrates the correct decoupling of the 
resonance, leading to the same low-energy limit in both 
theories. The 49~MeV data are best described by the 
$\mathcal{O}(p^2)$ calculation but we regard this as a coincidence, as 
the low-energy theorems are violated at this order too.

By comparing our results to data, we can now quantify the region of 
applicability of our calculation. The counting scheme described in 
Section~\ref{sec:theory} seems to break down for energies somewhere between 
50 and 60~MeV, as both theoretical descriptions miss the 49~MeV 
data points, whereas the 68~MeV data are well described within both 
theories\footnote{We neglect the minor corrections due to 
the data of~\cite{Lund} (\cite{Lucas}) being measured around 67~MeV (69~MeV), 
and account for this deviation by calculating at the averaged energy 68~MeV.}. 
The 55~MeV curves are still in agreement with the data 
within the (large) error bars but lie systematically above the central values.
As discussed before, we assume $\omega\sim m_\pi$ in this chapter. For 
$\omega\approx 20$~MeV, this counting is known to break down, see 
Section~\ref{sec:theory}.
In the remaining part of this chapter 
we are therefore only concerned with the data published for energies 
higher than 60~MeV. 

In the  high-energy regime of our calculation~-- i.e. for describing the 
94.2~MeV data correctly~-- the inclusion of the explicit $\Delta$ field seems 
to be advantageous in a third-order calculation, as can be seen in 
Fig.~\ref{fig:SSEHBplots}.
Here, $\mathcal{O}(p^3)$ HB$\chi$PT misses the data in the backward direction.
It also fails to reproduce the shape of 
the data points, which shows a slight tendency towards higher cross sections 
in the backward than in the forward direction. This shape is very well 
reproduced in SSE, demonstrating once again the importance of the $\Delta$ 
resonance in Compton backscattering, due to the strong $M1\rightarrow M1$ 
transition. This feature 
can be clearly seen in the dynamical magnetic dipole polarizability 
$\beta_{M1}(\omega)$, even for photon energies below the pion-production 
threshold, cf. Fig.~\ref{fig:spinindependentpolas}. 
We believe that this is the main reason why calculations 
like the ones presented in Refs.~\cite{Lvov,Karakowski}, which 
truncate the Compton amplitudes after 
the leading~\cite{Lvov,Karakowski} and subleading terms~\cite{Lvov} of a 
Taylor expansion in $\w$, fail to describe the data around 95~MeV, at least 
without introducing an unexpectedly large magnetic dipole polarizability
$\beta_{M1}$~\cite{Lvov}.

\subsection{Energy Dependence of the $\gamma d$ Cross Sections}

\begin{figure}[!htb]
\begin{center} 
\includegraphics*[width=.48\textwidth]{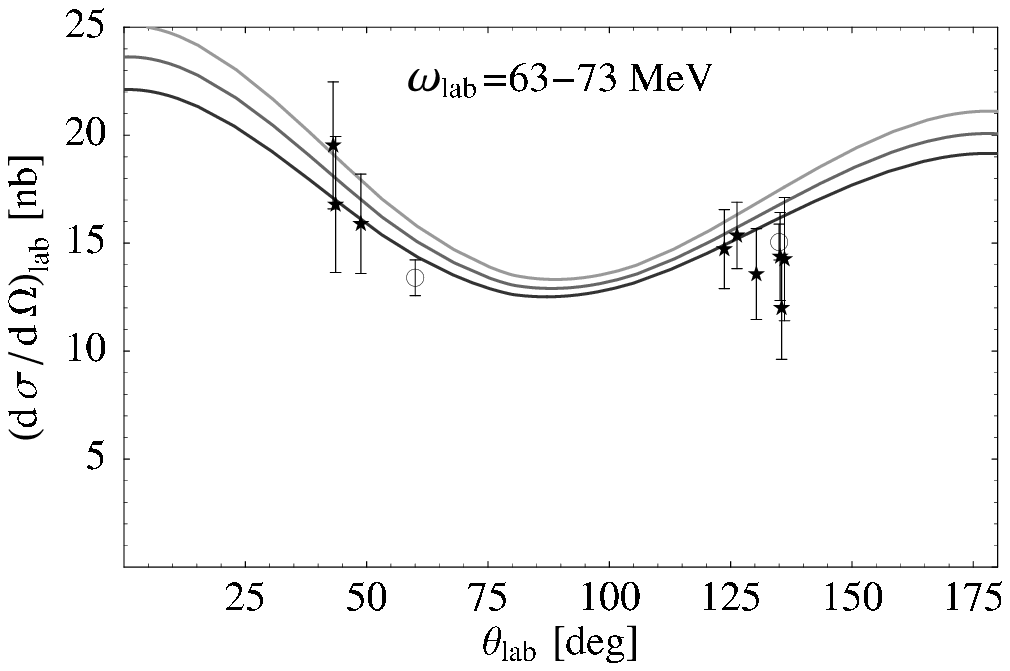}
\hfill
\includegraphics*[width=.48\textwidth]{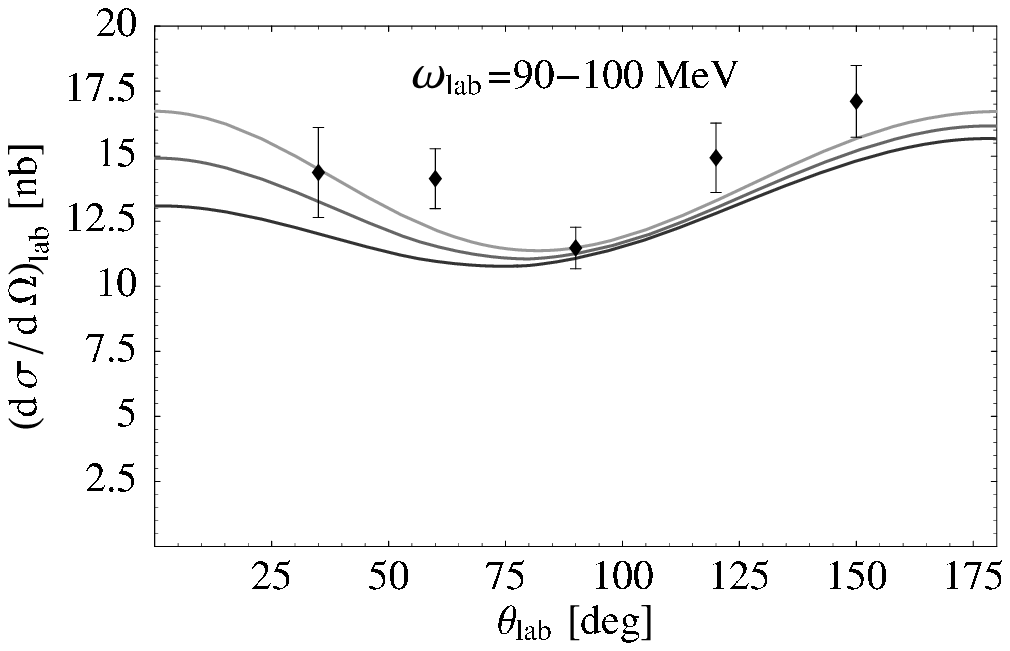}
\parbox{1.\textwidth}{
\caption[Energy dependence of the $\mathcal{O}(\epsilon^3)$-SSE result for 
deuteron Compton scattering]
{$\mathcal{O}(\epsilon^3)$-SSE results for 63~MeV, 68~MeV, 73~MeV and,
respectively, 90~MeV, 95~MeV, 100~MeV (from upper to lower curve 
in each panel), using the $\chi$PT wave function~\cite{Epelbaum}. }
\label{fig:omegadep}}
\end{center}
\end{figure}

In order to decrease the statistical uncertainties, the 
experiment~\cite{Hornidge} had to accept 
scattering events in an energy range of 20~MeV. Therefore we think it 
worthwhile to examine the sensitivity of our results to the photon energy.
In fact, our calculations suggest that the forward-angle cross section, in 
particular, has a sizeable energy dependence, which is, however, nearly 
linear. In 
Fig.~\ref{fig:omegadep} we show our results for three different photon energies
around 68~MeV and 95~MeV, respectively,
in steps of 5~MeV. This emphasizes the importance of having a well-known 
spectrum of the photon flux, especially in the forward direction, when one 
wants to examine the effects of $\alpha_{E1}$ and $\beta_{M1}$ experimentally.

\subsection{Correction due to the Pion-Production Threshold}
\label{sec:threshold}

\begin{figure}[!htb]
\begin{center} 
\includegraphics*[width=.5\textwidth]{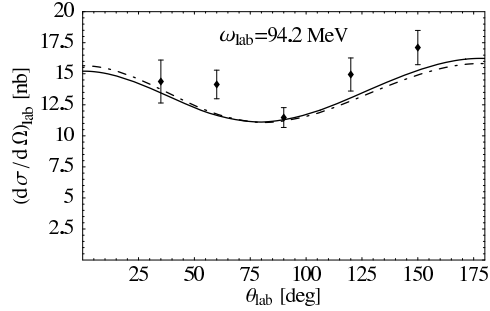}
\parbox{1.\textwidth}{
\caption[Estimate of the effect of a threshold correction]
{Estimate of the effect of a threshold correction (dotdashed) on the 
$\mathcal{O}(\epsilon^3)$-SSE results 
(solid), using the chiral NNLO wave function~\cite{Epelbaum}.}
\label{fig:SSEnsuns}}
\end{center}
\end{figure}

In low-order (non-relativistic) HB$\chi$PT/SSE calculations, the 
$\gamma d\rightarrow \pi N N$ threshold $\w_\pi$ is not at the 
correct position as demanded by relativistic kinematics. 
For a similar problem, regarding the correct position of the pion-production 
threshold in the single-nucleon sector, see 
Section~\ref{sec:polarizabilities1}. Thus far we have refrained from 
an analogous correction for $\gamma d$ scattering.
However, in Fig.~\ref{fig:SSEnsuns} we investigate what deviations one would 
expect from our present results, as indicated by an estimate which  
uses the single-nucleon SSE amplitudes~\cite{HGHP} with the exact expression 
for $\sqrt{s}-m_N$. Obviously, even at the highest photon energies considered 
here, 94.2~MeV, the corrections are negligible, given 
the sizeable error bars of the experimental data and the theoretical 
uncertainties of a leading-one-loop order calculation. This is no surprise, as
we found in Ref.~\cite{DA} that the threshold correction is only mandatory 
close to $\w_\pi$, i.e.
above 100~MeV. There, however, it should not be neglected anymore, see also 
discussion in Ref.~\cite{Deepshikha}.

\subsection{Wave-Function Dependence of the $\gamma d$ Cross Sections}
\label{sec:wfdep}

\begin{figure}[!htb]
\begin{center} 
\includegraphics*[width=.48\textwidth]
{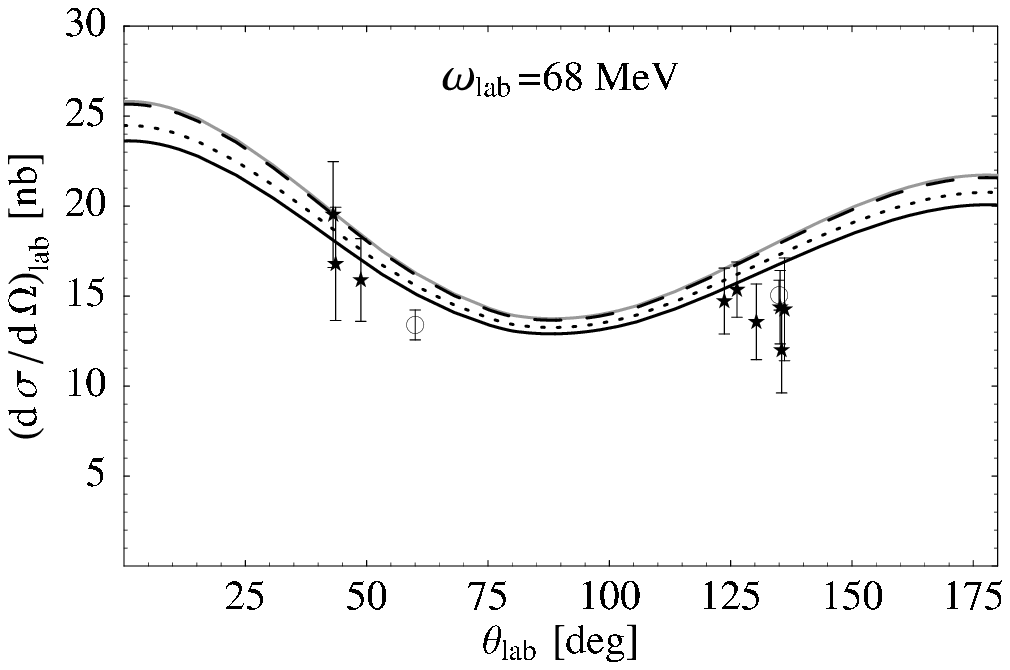}
\hfill
\includegraphics*[width=.48\textwidth]
{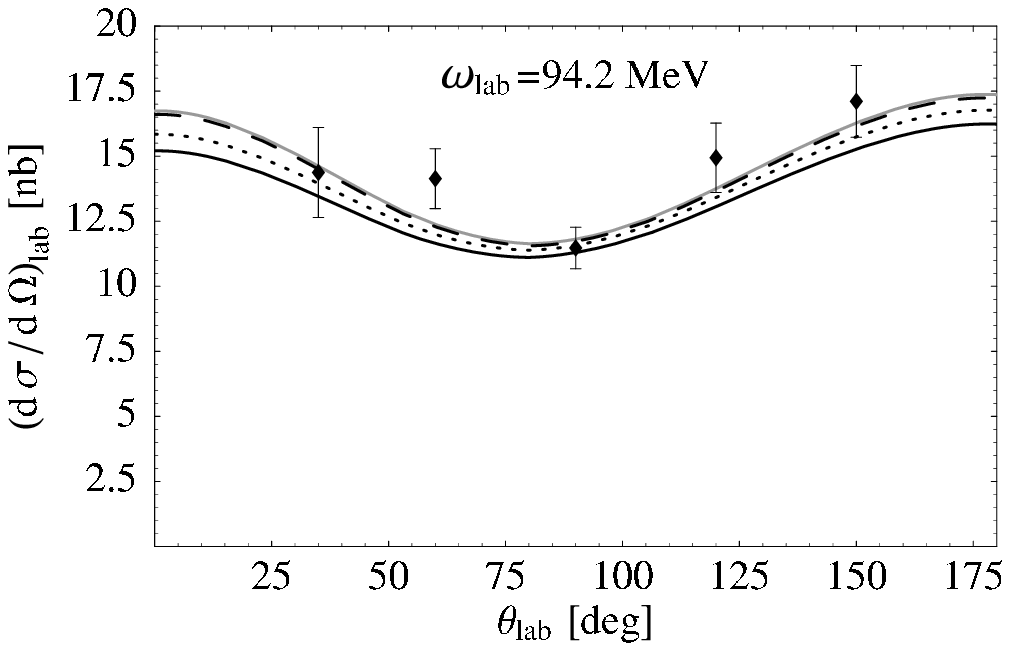}
\includegraphics*[width=.31\textwidth]{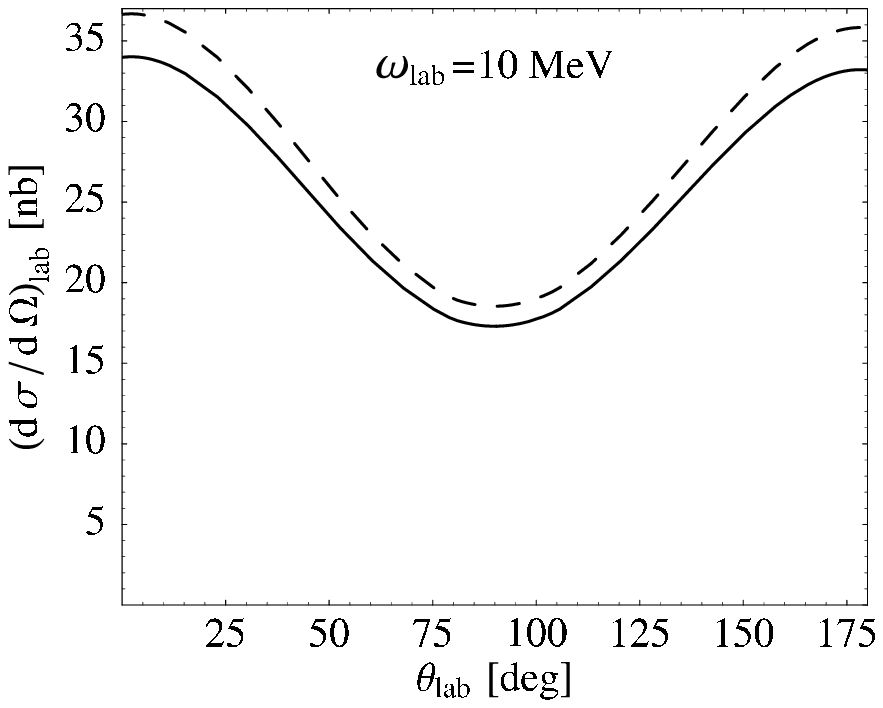}
\hspace{.01\textwidth}
\includegraphics*[width=.31\textwidth]{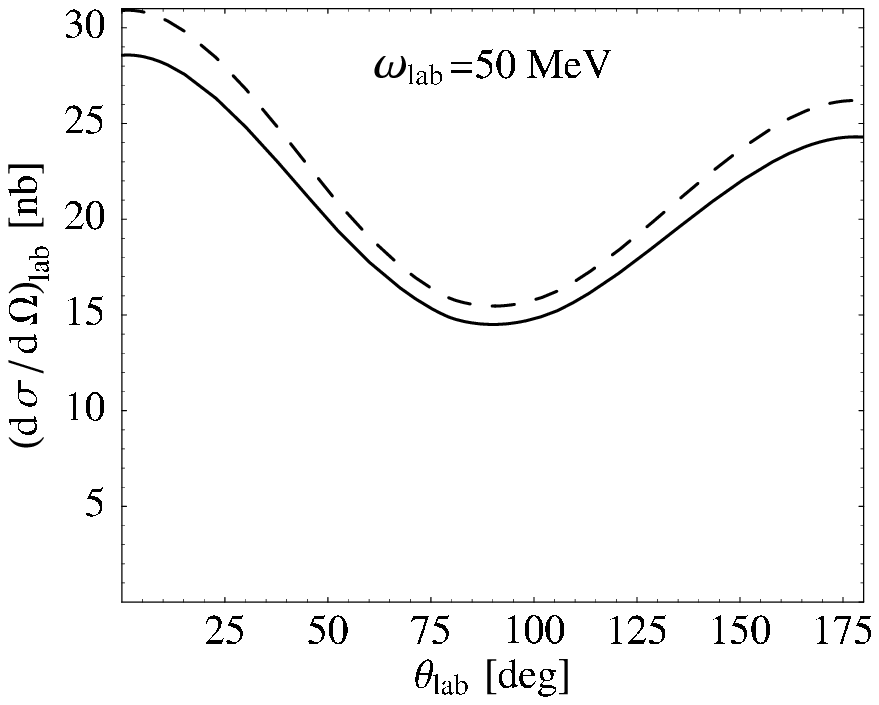}
\hspace{.01\textwidth}
\includegraphics*[width=.31\textwidth]{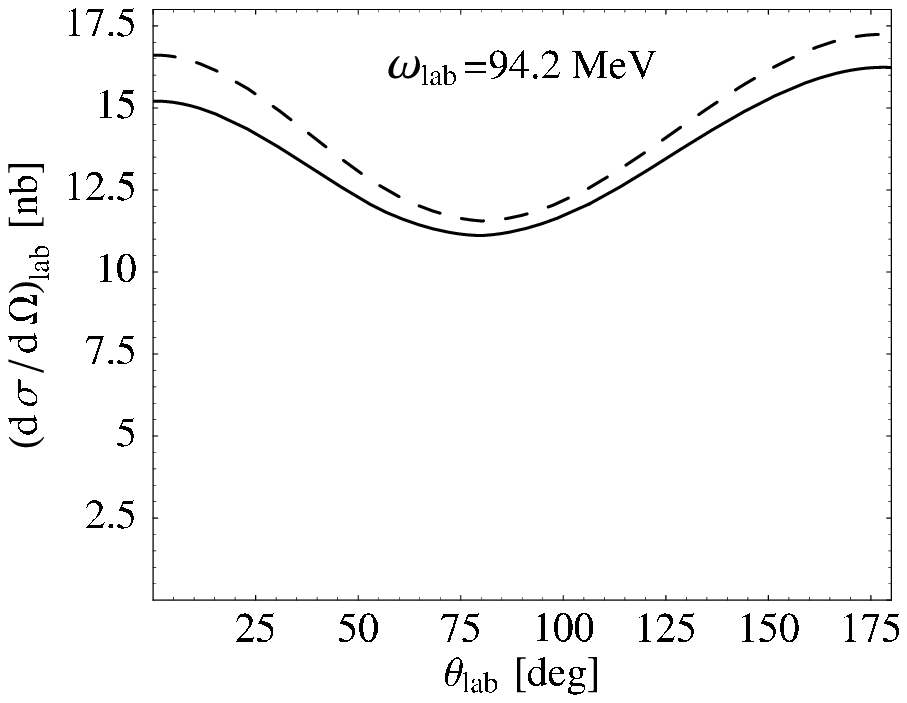}
\parbox{1.\textwidth}{
\caption[$\mathcal{O}(\epsilon^3)$-SSE results using different wave functions]
{Upper panels: 
$\mathcal{O}(\epsilon^3)$-SSE results for 68 and 94.2~MeV, using four
different wave functions: NNLO $\chi$PT with $\Lambda=650$~MeV~\cite{Epelbaum}
(solid), Nijm93~\cite{Nijm} (grey), 
CD-Bonn~\cite{Bonn} (dotted), AV18~\cite{AV18} (dashed).
Lower panels: 
$\mathcal{O}(\epsilon^3)$-SSE results for 10, 50 and 94.2~MeV, using 
the NNLO $\chi$PT wave function~\cite{Epelbaum} (solid) and the 
AV18 wave function~\cite{AV18} (dashed).} 
\label{fig:wavefdep}}
\end{center}
\end{figure}
Another interesting issue is the wave-function dependence of our results.
Fig.~\ref{fig:wavefdep} investigates the sensitivity to the 
wave function chosen, showing sizeable deviations between the NNLO $\chi$PT
wave function with cutoff $\Lambda=650$~MeV~\cite{Epelbaum} 
on one hand and the  wave function derived from 
the AV18 potential~\cite{AV18} on the other, which are both given in 
Appendix~\ref{app:wavefunction}. The latter yields results which are 
nearly indistinguishable from those obtained with the Nijm93 wave 
function~\cite{Nijm}, but are considerably higher than the cross sections 
found with the wave function from the chiral potential~\cite{Epelbaum}. 
We note that our results using the chiral wave function from 
Ref.~\cite{Epelbaum} with the cutoff chosen to be 450~MeV are even smaller 
by about 5\% than those achieved with $\Lambda=650$~MeV.
With the CD-Bonn wave function~\cite{Bonn} we obtain results in between 
NNLO $\chi$PT and Nijm93/AV18. This pattern is identical for both energies 
under investigation, 68~MeV and 94.2~MeV. 
In the lower three panels of Fig.~\ref{fig:wavefdep}, where we compare
our results obtained with the NNLO chiral wave 
function and the AV18 wave function at 10~MeV, 50~MeV and 94.2~MeV,
we demonstrate that the wave-function dependence is largely energy independent.
Note that the main difference between the two curves in each panel is an
angle-independent off-set, reminiscent of a systematic error. 

\begin{figure}[!htb]
\begin{center} 
\includegraphics*[width=.31\textwidth]{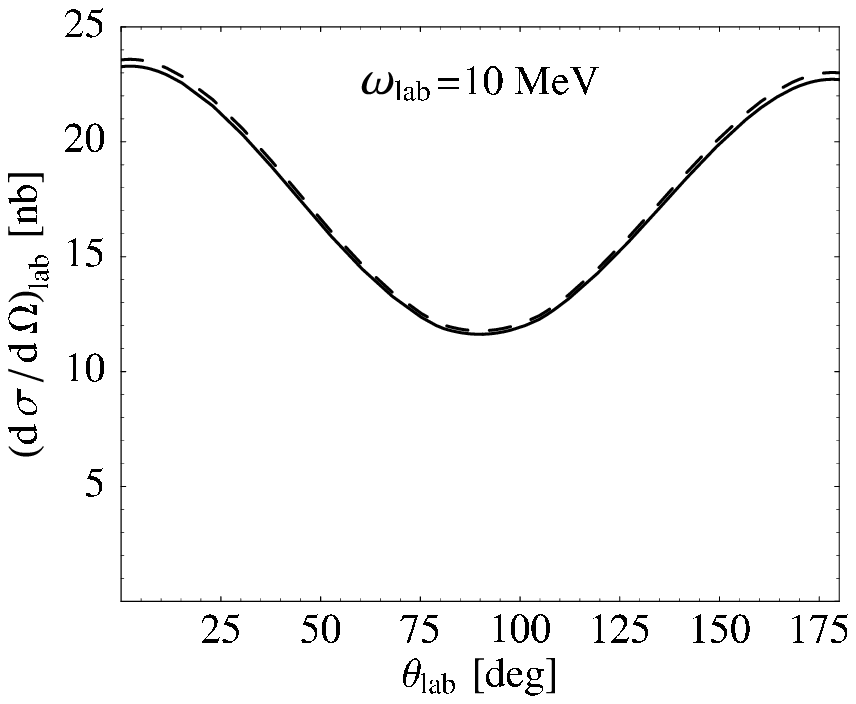}
\hspace{.01\textwidth}
\includegraphics*[width=.31\textwidth]{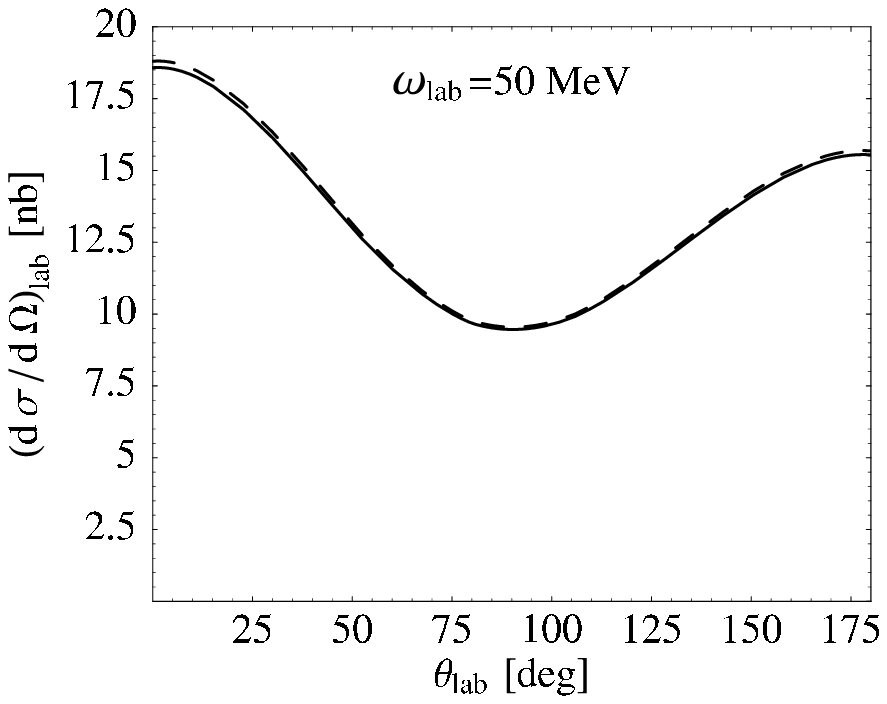}
\hspace{.01\textwidth}
\includegraphics*[width=.31\textwidth]{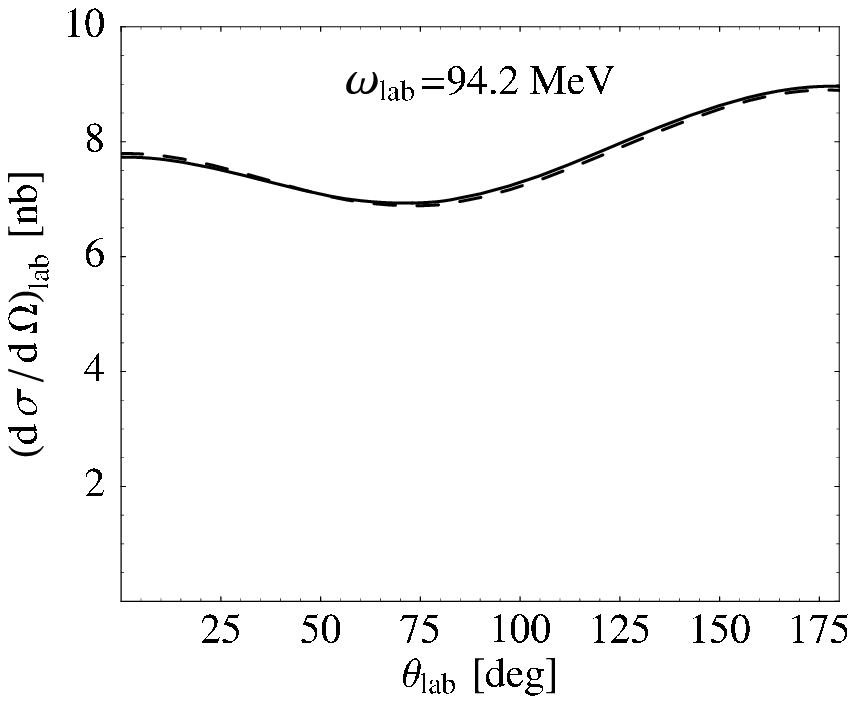}
\parbox{1.\textwidth}{
\caption[Wave-function sensitivity of the one-body diagrams]
{Results for 10, 50 and 94.2~MeV, including only the third-order SSE 
one-body diagrams, calculated with 
the NNLO $\chi$PT wave function~\cite{Epelbaum} (solid) and the 
AV18 wave function~\cite{AV18} (dashed).} 
\label{fig:wavefdepIA}}
\end{center}
\end{figure}

The main contribution to the sensitivity on the deuteron wave function comes 
from the two-body diagrams (Fig.~\ref{fig:chiPTdouble}), 
as is shown in Fig.~\ref{fig:wavefdepIA}, where we 
calculate deuteron Compton cross sections at 
various energies, including only the $\calO(\epsilon^3)$-SSE one-body 
diagrams. The curves in Fig.~\ref{fig:wavefdepIA}  corresponding to the 
NNLO $\chi$PT wave function~\cite{Epelbaum} are nearly indistinguishable from 
those obtained with the AV18 wave function~\cite{AV18}.
The same observation was made in Ref.~\cite{McGPhil}.

Given that our calculation is based on a low-energy Effective Field Theory of 
QCD, the dependence on the wave function is somewhat worrisome. 
Our calculation describes
deuteron Compton scattering up to next-to-leading one-loop order in the 
Small Scale Expansion, therefore instead of the 10\%-discrepancy observed in 
Fig.~\ref{fig:wavefdep} one would rather expect such higher-order corrections 
to be of the order of $(m_\pi/m_N)^2\sim2$\%.
We interpret this feature, which will be discussed further in 
Section~\ref{sec:wavefunctiondep2}, as an unwanted sensitivity to 
short-distance physics, because the 
long-range part of all wave functions, described by  one-pion exchange, is 
identical. 
However, one must caution that the 
NNLO $\chi$PT potential reproduces the Nijmegen partial-wave analysis 
with less precision than the CD-Bonn, AV18 or Nijm93 potentials. 
It will be one of the major successes of our deuteron
Compton calculation described in Chapter~\ref{chap:nonperturbative}, that it is
largely insensitive to the deuteron wave function chosen, cf. 
Section~\ref{sec:wavefunctiondep2}.

\subsection{Dependence on the Upper Integration Limit}
\label{sec:upperlimit}

\begin{figure}[!htb]
\begin{center} 
\includegraphics*[width=.48\textwidth]{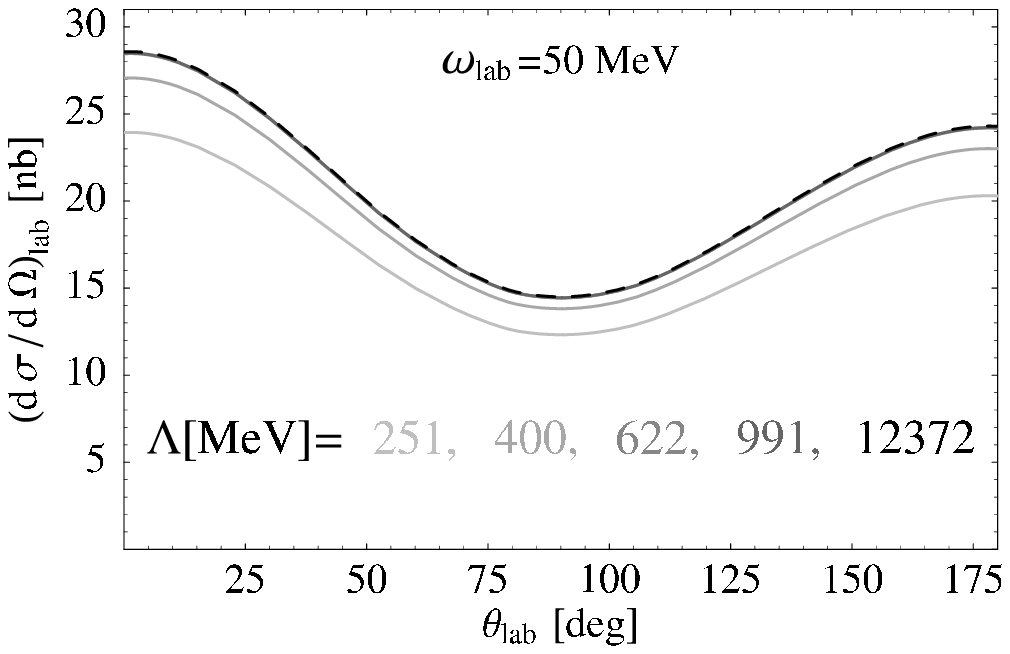}
\hfill
\includegraphics*[width=.48\textwidth]{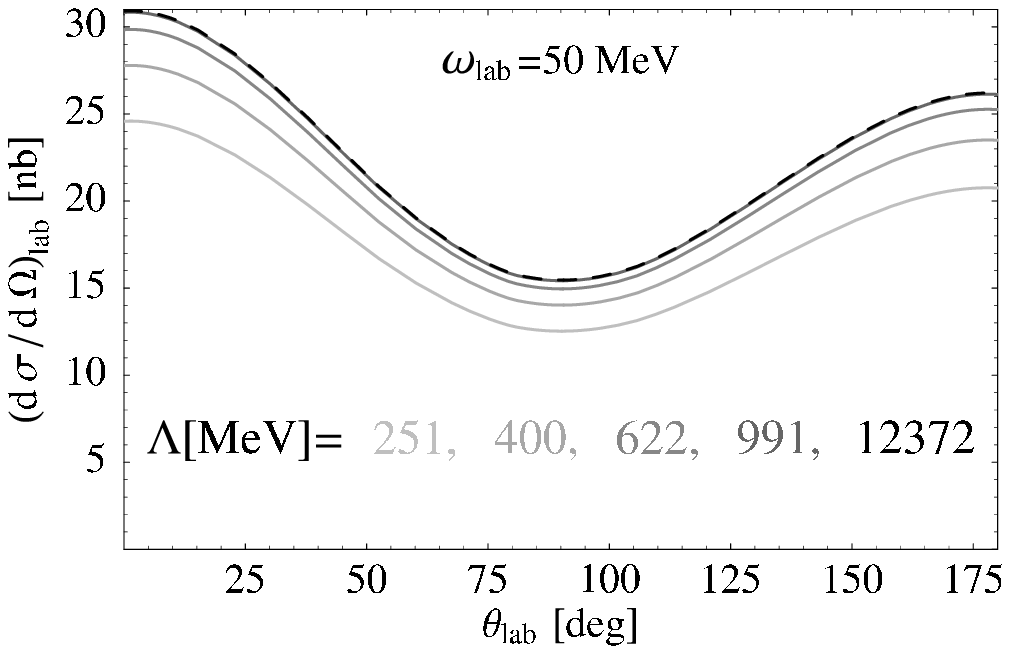}

\vspace{.5cm}
\includegraphics*[width=.48\textwidth]{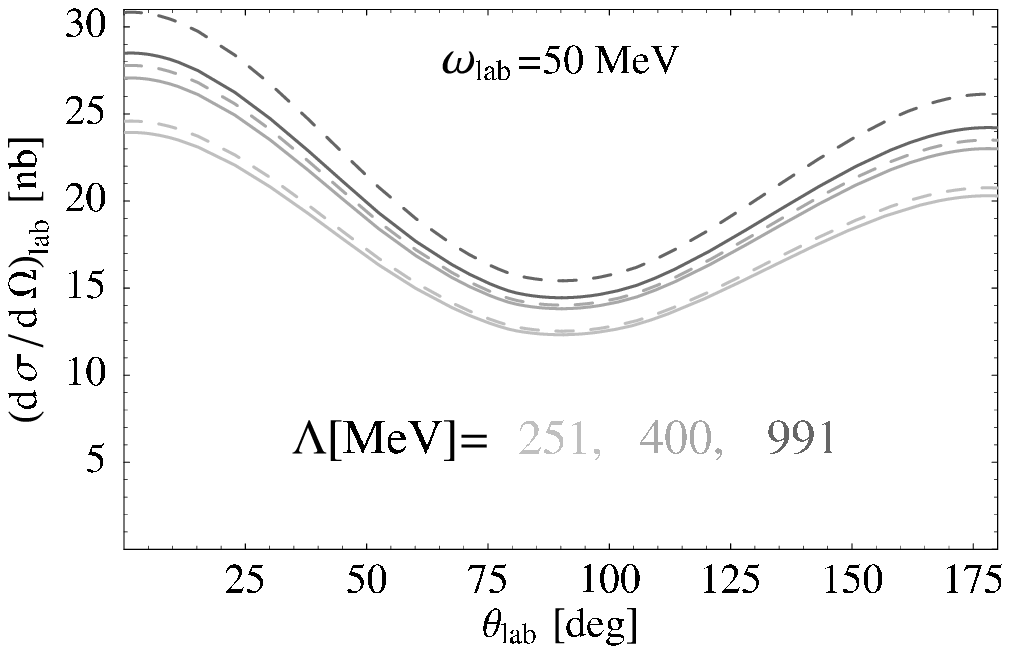}
\parbox{1.\textwidth}{
\caption[Dependence of the deuteron Compton cross section on the upper 
integration limit]
{$\calO(\epsilon^3)$-SSE results with various upper integration limits 
$\Lambda$. The curves in the upper left panel have been calculated with the 
NNLO $\chi$PT wave function~\cite{Epelbaum}, those in the upper right panel 
with the AV18 wave function~\cite{AV18}. In the lower panel the solid lines 
correspond to the chiral, the dashed lines to the AV18 wave function.} 
\label{fig:Lambdadep}}
\end{center}
\end{figure}
Here we want to investigate how fast our integrals converge, and in which 
energy regime the main part of the wave-function dependence enters, cf. 
Section~\ref{sec:wfdep}. This question is addressed in Fig.~\ref{fig:Lambdadep}
for an exemplary photon energy $\w=50$~MeV, with the respective upper 
integration limits $\Lambda$ given in the figures. The upper left panel 
corresponds to the NNLO $\chi$PT wave function, the upper right one to the 
wave function derived from the AV18 potential. Obviously, the integrals are 
well saturated below 1~GeV, the 
one using the chiral wave function already below 620~MeV, due to the cutoff 
$\Lambda=650$~MeV chosen. This observation 
agrees with Fig~\ref{fig:wavefunctions}, where we show the momentum-space 
representation of both wave functions. The only significant deviations between 
the two wave functions are observed in $w(p)$, 
corresponding to the orbital angular momentum $l=2$. Here we notice that the 
$\chi$PT wave function converges faster to zero for $p\rightarrow\infty$ 
than the AV18 wave 
function, resulting in a faster saturation of the integrals entering the 
cross section. In the lower panel of Fig.~\ref{fig:Lambdadep} we give a direct 
comparison of our results, derived from the two wave 
functions for three different upper integration limits. We note that the 
largest part of the wave-function dependence enters between 400~MeV and 
1~GeV, which is no surprise, because in this energy range the $l=2$ wave 
functions differ most, cf. Fig.~\ref{fig:wavefunctions}. 

In this section we presented our predictions for $\gamma d$ differential 
cross sections. These are parameter-free, as we fixed the isoscalar
nucleon polarizabilities via proton Compton data. 
The good agreement of the SSE results with experiment at 68~MeV and 94.2~MeV 
leaves little room for large isovector polarizabilities, since 
this formalism treats the proton and neutron polarizabilities on the same 
footing to the order we are working. The good description of the high-energy 
data further 
encourages us to determine the isoscalar dipole polarizabilities 
$\bar{\alpha}_{E1}^s$ and $\bar{\beta}_{M1}^s$ directly from the deuteron 
Compton cross sections. The results are given in the next section, 
together with the results one obtains from analogous 
fits using the $\mathcal{O}(p^3)$ and the 
$\mathcal{O}(p^4)$-HB$\chi$PT~\cite{McGPhil} amplitudes, respectively.

\section{Determining $\bar{\alpha}_{E1}^s$ and $\bar{\beta}_{M1}^s$ from 
$\gamma d$ Scattering}
\label{sec:fits1}
\markboth{CHAPTER \ref{chap:perturbative}. DEUTERON COMPTON SCATTERING IN EFT}
{\ref{sec:fits1}. DETERMINING $\bar{\alpha}_{E1}^s$ AND $\bar{\beta}_{M1}^s$ 
FROM $\gamma d$ SCATTERING}

An accurate and systematically-improvable description of Compton scattering 
on deuterium offers the 
possibility to extract the isoscalar polarizabilities directly 
from deuteron Compton-scattering experiments in a well-defined way. The 
results can then be 
combined with the known numbers for the proton to draw conclusions about 
isovector pieces $\bar{\alpha}_{E1}^v$ and $\bar{\beta}_{M1}^v$, 
or, equivalently, the elusive neutron polarizabilities. As our SSE 
calculation provides a reasonable description of the 68~MeV and the 94.2~MeV 
data (see Section~\ref{sec:results}), we present in the following our results 
from a least-$\chi^2$ fit of the isoscalar polarizabilities to these 
two data sets.
This corresponds to fitting the coupling strengths of the two short-distance 
isoscalar $\gamma N$-operators (Fig.~\ref{fig:SSEsingle}(f)), which we now fit
to $\gamma d$ rather than to $\gamma p$ data. In this way we can check 
the SSE claim that the short-distance operators are isoscalar at leading 
order. If their coefficients as extracted from $\gamma d$ data are 
consistent with those obtained
from $\gamma p$ data, that argues in favour of short-distance mechanisms which 
are predominantly isoscalar. The value for the $\gamma N\Delta$-coupling $b_1$
is adopted from our Baldin-constrained 2-parameter fit, cf. 
Table~\ref{tab:protonfit}, as there is no isovector contribution to this 
coupling up to third-order SSE.

Our SSE results are compared to the fit
results for $\bar{\alpha}_{E1}^s$ and $\bar{\beta}_{M1}^s$ when we
use modified $\mathcal{O}(p^3)$-HB$\chi$PT amplitudes. This modification 
consists of including in our calculation isoscalar short-distance $\gamma N$ 
operators which change 
both the electric and magnetic polarizability from their $\mathcal{O}(p^3)$ 
values. In other words, we write
\begin{align}
\bar{\alpha}_{E1}^s&=\frac{5\,\alpha\,g_A^2}{96\,f_\pi^2\,m_\pi\,\pi}
           +\delta_\alpha,\nonumber\\
\bar{\beta}_{M1}^s &=\frac{\alpha\,g_A^2}{192\,f_\pi^2\,m_\pi\,\pi}
           +\delta_\beta,
\label{eq:deltaalpha}
\end{align}
where $\delta_\alpha$, $\delta_\beta$ are energy-independent quantities, 
connected to $g_{1}$ and $g_{2}$, cf. Ref.~\ref{sec:SSE}. Therefore, the 
energy dependence of the 
polarizabilities is still given solely by the leading-order pion cloud.
Eq.~(\ref{eq:deltaalpha}) promotes the short-distance contributions to 
$\alpha$ and $\beta$ from $\mathcal{O}(p^4)$ to $\mathcal{O}(p^3)$. 
As the loops are isoscalar, we associate purely
isoscalar counter-terms which renormalize the loop integrals.
In order to avoid confusion we denote the fits done with this procedure
as HB$\chi$PT $\mathcal{O}(\bar{p}^3)$.

Fits similar to our $\mathcal{O}(\epsilon^3)$ and $\mathcal{O}(\bar{p}^3)$ 
ones have already been performed in~\cite{McGPhil}, calculating in HB$\chi$PT 
up to $\mathcal{O}(p^4)$. The authors of~\cite{McGPhil} used all available 
data sets but had to exclude the two 94.2~MeV data points measured in the 
backward direction, due to the inadequate description of back-angle 
Compton scattering in fourth-order HB$\chi$PT, cf. Fig.~\ref{fig:Olmos} and 
Ref.~\cite{McGovern}.
As \cite{Phillips,McGPhil} and the $\mathcal{O}(\epsilon^3)$-SSE calculation 
are not designed to work below 60~MeV, we decided to only include the data 
around 68~MeV~\cite{Lucas,Lund} and 94.2~MeV~\cite{Hornidge} in the fit. 
We do not make any cuts on the angles and,  
in contrast to~\cite{McGPhil}, 
we do not allow the normalizations in the various experiments to float 
in the fit within their quoted systematic errors.

\subsection{Wave-Function Dependence of the Fits}
\label{sec:wfdepfits}

To have an estimate on the systematic error in the deuteron fits 
due to the wave-function dependence, we show our results when we use two 
different wave functions for the fit: the NNLO 
chiral wave function~\cite{Epelbaum} and the wave function from the Nijm93 
potential~\cite{Nijm}. These wave functions mark the two extremes in 
our cross sections with the CD-Bonn wave function in between (cf. 
Fig.~\ref{fig:wavefdep}). Therefore, we consider the difference between the 
two fits using Nijm93 and NNLO $\chi$PT as a measure of our wave-function 
induced error, which is typically of the order of 10$\,$\% for 
$\bar{\alpha}_{E1}$. Higher-order effects are expected to contribute to
the systematic error by about $\pm1 \cdot 10^{-4}\; \fm^3$ from na\"ive
dimensional analysis, cf. Eq.~(\ref{eq:higherorder}),
but we refrain for the moment from including this error explicitly into our 
findings. However, 
we are aware that it is comparable in size with our statistical error.

We fit the 16 data points using 2 free parameters ($\bar{\alpha}_{E1}^s$ 
and $\bar{\beta}_{M1}^s$), leaving us with 14 degrees of freedom. 
Averaging over the results of our 2-parameter SSE fits with the NNLO 
$\chi$PT~\cite{Epelbaum}  and the Nijm93 wave function~\cite{Nijm}, 
respectively,  
given in Table~\ref{tab:fulldata}, results in the isoscalar polarizabilities
\begin{align}
\bar{\alpha}_{E1}^s&=(13.2\pm1.3\,(\mathrm{stat})\pm1.1\,(\mathrm{wf}))
            \cdot 10^{-4}\;\mathrm{fm}^3\,,\nonumber\\
\bar{\beta}_{M1}^s &=( 1.7\pm1.6\,(\mathrm{stat})\pm0.2\,(\mathrm{wf}))
            \cdot 10^{-4}\;\mathrm{fm}^3\,,
\label{eq:final}
\end{align}
where we assume the same statistical errors as in Table~\ref{tab:fulldata}. 
The systematic error 
due to the differing results when we use different wave functions (wf) is 
estimated to be half of the difference between the results obtained with the 
extreme wave functions, i.e. Nijm93 and NNLO $\chi$PT. 
Fitting the $\gamma d$ cross sections 
using the Nijm93 wave 
function yields larger results for $\bar{\alpha}_{E1}$ and smaller ones for 
$\bar{\beta}_{M1}$ with respect to the chiral wave function, but 
both extractions are in reasonable agreement with the values given in 
Eq.~(\ref{eq:expn})~\cite{Kossert}. 

The results for $\bar{\alpha}_{E1}^s$ and $\bar{\beta}_{M1}^s$, given in 
Eq.~(\ref{eq:final}), correspond to
the obviously important short-distance contributions
\ba
\bar{\alpha}_\mathrm{sd}^s&=(- 4.4\pm1.3\,(\mathrm{stat})\pm1.1\,(\mathrm{wf}))
\cdot10^{-4}\;\mathrm{fm}^3,\nonumber\\
\bar{\beta} _\mathrm{sd}^s&=(-11.7\pm1.6\,(\mathrm{stat})\pm0.2\,(\mathrm{wf}))
\cdot10^{-4}\;\mathrm{fm}^3.
\label{eq:sd}
\end{align}
In Section~\ref{sec:polarizabilities1}, we already saw that they are 
indeed comparable in size with the other leading-order contributions to the 
polarizabilities, demonstrating the necessity of including the couplings $g_1$ 
and $g_2$ at leading-one-loop order. 
The plots corresponding to the fits with the chiral wave 
function are displayed in Fig.~\ref{fig:errorfitsfull}, 
together with the results of our $\mathcal{O}(\bar{p}^3)$ fits. 

Using the experimental values for the proton polarizabilities 
from~\cite{Olmos} as input, 
one can derive the neutron polarizabilities from the isoscalar ones: 
\ba
\bar{\alpha}_{E1}^n&=(14.3\pm1.3\,(\mathrm{stat})\pm1.1\,(\mathrm{wf}))
\cdot10^{-4}\;\mathrm{fm}^3,\nonumber\\
\bar{\beta}_{M1}^n &=( 1.7\pm1.6\,(\mathrm{stat})\pm0.2\,(\mathrm{wf}))
\cdot10^{-4}\;\mathrm{fm}^3.
\end{align}
From these results we deduce that the 
isovector polarizabilities are 
rather small (see Table~\ref{tab:fulldata}), in good agreement with $\chi$PT
expectations, which predict the 
isovector part to be of higher than third order. Therefore we find no 
contradiction anymore between the results from quasi-free~\cite{Kossert} and 
elastic deuteron Compton scattering. Furthermore, our calculation 
demonstrates that the experiments performed at Illinois~\cite{Lucas} and 
Lund~\cite{Lund} are consistent with the SAL-data~\cite{Hornidge}.

Our results for $\bar{\alpha}_{E1}^s$ and $\bar{\beta}_{M1}^s$ in SSE (cf. 
Eq.~(\ref{eq:final})) are well consistent with the isoscalar Baldin sum rule 
\be
\left.\phantom{\PCsq}\bar{\alpha}_{E1}^s+\bar{\beta}_{M1}^s
\right|_{\mathrm{world}\;\, \mathrm{av.}}=
 (14.5\pm0.6)\cdot10^{-4}\;\mathrm{fm}^3,
\label{eq:Baldin}
\ee
which has been a serious problem in former 
extractions~\cite{Lvov,McGPhil}, cf. Eq.~(\ref{eq:Oq4bestresults}). 
The numerical value for the sum rule is derived from 
\begin{align}
\bar{\alpha}_{E1}^p+\bar{\beta}_{M1}^p&=
(13.8\pm0.4)\cdot10^{-4}\;\mathrm{fm}^3\;\;\;\;\;\cite{Olmos},\nonumber\\
\bar{\alpha}_{E1}^n+\bar{\beta}_{M1}^n&=
(15.2\pm0.5)\cdot10^{-4}\;\mathrm{fm}^3\;\;\;\;\;\cite{Lvov}.
\end{align}
Due to the consistency of our fit results with the sum-rule value from 
Eq.~(\ref{eq:Baldin}), one can in a second step use this number~-- we use the 
central value~-- as an additional fit constraint and thus reduce the number 
of free parameters to one. The resulting values of the 1-parameter SSE 
fit (see Table~\ref{tab:fulldata}),
\begin{align}
\bar{\alpha}_{E1}^s&=(13.1\pm0.7\,(\mathrm{stat})\pm0.8\,(\mathrm{wf})
                 \pm0.6\,(\mathrm{Baldin}))
            \cdot 10^{-4}\;\mathrm{fm}^3\,,\nonumber\\
\bar{\beta}_{M1}^s &=( 1.5\mp0.7\,(\mathrm{stat})\mp0.8\,(\mathrm{wf})
                 \pm0.6\,(\mathrm{Baldin}))
            \cdot 10^{-4}\;\mathrm{fm}^3\,,
\label{eq:finalBaldin}
\end{align}
are in good agreement with the average of the proton numbers from 
Table~\ref{tab:protonfit} and the neutron polarizabilities given in 
Eq.~(\ref{eq:expn}).

Comparing our fit results to 
the isoscalar $\mathcal{O}(p^4)$-HB$\chi$PT estimate~\cite{BKMS},
$\bar{\alpha}_{E1}^s=(11.95\pm2.5)\cdot10^{-4}\;\mathrm{fm}^3$,
$\bar{\beta}_{M1}^s = (5.65\pm5.1)\cdot10^{-4}\;\mathrm{fm}^3$, 
cf. Section~\ref{sec:intro}, 
we see only minor deviations in $\bar{\alpha}_{E1}^s$. 
Our values for $\bar{\beta}_{M1}^s$ are significantly smaller, but 
still consistent within the (large) error bars of the 
$\mathcal{O}(p^4)$ estimate.

\begin{table}[!htb]
\begin{center}
\begin{tabular}{|c||c||c|c|c|c|}
\hline
Amplitudes&Quantity&2-par. fit    &1-par. fit   &2-par. fit&1-par. fit\\
          &        &NNLO $\chi$PT &NNLO $\chi$PT&Nijm93    &Nijm93    \\    
\hline
\hline
$\mathcal{O}(\epsilon^3)$~SSE&$\chi^2/d.o.f.$ &1.78&1.67&2.45&2.35\\
\cline{2-6}
&$\bar{\alpha}_{E1}^s$&$12.1\pm1.3$
                 &$12.3\pm0.7$&$14.3\pm 1.3$&$13.8\pm0.7$\\
&$\bar{\beta} _{M1}^s$&$ 1.8\pm1.6$
                 &$ 2.2\mp0.7$&$ 1.5\pm 1.6$&$ 0.7\mp0.7$\\
\cline{2-6}
&$\bar{\alpha}_{E1}^s+\bar{\beta} _{M1}^s$
                       &$13.9\pm2.1$&$14.5\,(\mathrm{fit})$
                       &$15.8\pm2.1$&$14.5\,(\mathrm{fit})$\\
\cline{2-6}
&$\bar{\alpha}_{E1}^n$&$12.1\pm1.3$
                 &$12.5\pm0.8$&$16.5\pm 1.3$&$15.5\pm0.8$\\
&$\bar{\beta} _{M1}^n$&$ 2.0\pm1.6$
                 &$ 2.8\mp0.8$&$ 1.4\pm 1.6$&$ -0.2\mp0.8$\\
\hline
\hline
$\mathcal{O}(\bar{p}^3)$~HB$\chi$PT&$\chi^2/d.o.f.$ &2.14&2.01&2.87&2.75\\
\cline{2-6}
&$\bar{\alpha}_{E1}^s$&$11.0\pm1.3$
                 &$11.3\pm0.7$&$13.2\pm 1.2$&$12.7\pm0.7$\\
&$\bar{\beta} _{M1}^s$&$ 2.8\pm1.6$
                 &$ 3.2\mp0.7$&$ 2.5\pm 1.5$&$ 1.8\mp0.7$\\
\cline{2-6}
&$\bar{\alpha}_{E1}^s+\bar{\beta} _{M1}^s$
                       &$13.8\pm2.1$&$14.5\,(\mathrm{fit})$
                       &$15.7\pm1.9$&$14.5\,(\mathrm{fit})$\\
\cline{2-6}
&$\bar{\alpha}_{E1}^n$&$9.9\pm1.3$
                 &$10.5\pm0.8$&$14.3\pm 1.2$&$13.3\pm0.8$\\
&$\bar{\beta} _{M1}^n$&$ 4.0\pm1.6$
                 &$ 4.8\mp0.8$&$ 3.4\pm 1.5$&$ 2.0\mp0.8$\\
\hline
\end{tabular}
\caption[Values for polarizabilities from fit to deuteron Compton data, 
using $\mathcal{O}(\epsilon^3)$~SSE and  $\mathcal{O}(\bar{p}^3)$~HB$\chi$PT]
{Values for the isoscalar and neutron polarizabilities (in 
$10^{-4}\,\mathrm{fm}^3$) from a fit to the full 68~MeV and 
94.2~MeV data sets~\cite{Lucas, Lund, Hornidge}, using the 
$\mathcal{O}(\epsilon^3)$-SSE  and the 
$\mathcal{O}(\bar{p}^3)$-HB$\chi$PT amplitudes, respectively.
The neutron results are derived 
using the proton values from~\cite{Olmos} (Table~\ref{tab:protonfit}) as input.
All error bars displayed are only statistical (and added in quadrature).}
\label{tab:fulldata}
\end{center}
\end{table}

In the next section we demonstrate that even without an explicit $\Delta$ field
a good fit of the data can be achieved, however at the cost of an enhanced  
$\bar{\beta}_{M1}^s$.

\subsection{Comparison of $\mathcal{O}(\epsilon^3)$-SSE and 
                          $\mathcal{O}(\bar{p}^3)$-HB$\chi$PT Fits}
\label{sec:Comparison1}

When we compare the $\mathcal{O}(\epsilon^3)$-SSE fit results for 
$\bar{\alpha}_{E1}^s$ and $\bar{\beta}_{M1}^s$ with the corresponding 
$\mathcal{O}(\bar{p}^3)$-HB$\chi$PT results (Table~\ref{tab:fulldata}), 
we see that in the HB$\chi$PT fit (cf. Eq.~(\ref{eq:deltaalpha})) the electric 
dipole polarizability is smaller, whereas $\bar{\beta}_{M1}^s$ turns out to be 
larger. The reason for the systematic shift of the magnetic polarizability is 
that due to the missing $\Delta(1232)$ resonance in HB$\chi$PT the static 
value of $\beta_{M1}(\w)$ is inflated in order to compensate for the 
paramagnetic rise of the resonance, which can be clearly seen in 
Figs.~\ref{fig:spinindependentpolas} and~\ref{fig:SSEHBplots}. 

\begin{figure}[!htb]
\begin{center} 
\includegraphics*[width=.48\textwidth]
{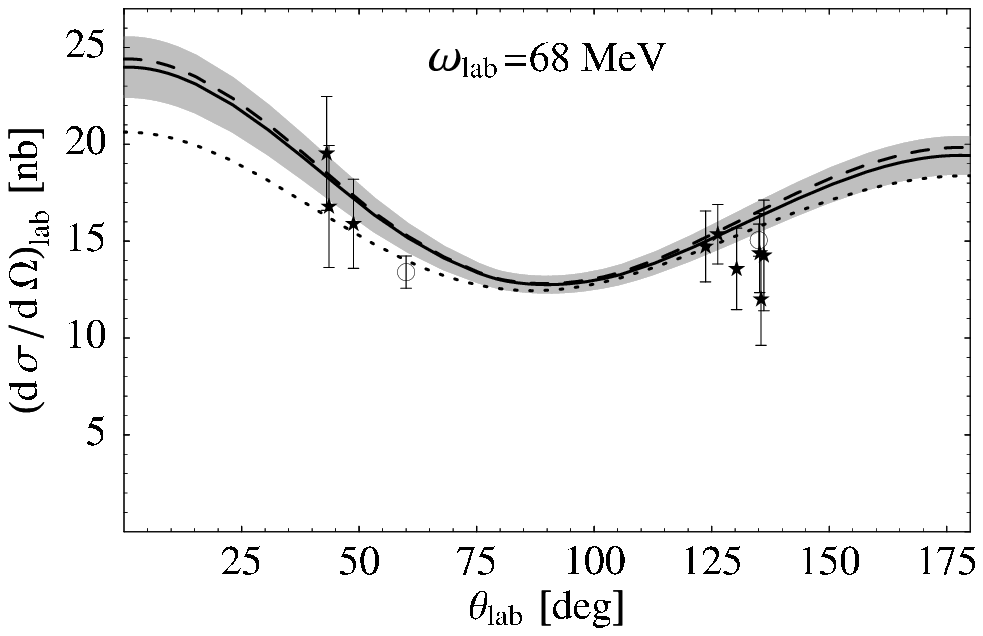}
\hfill
\includegraphics*[width=.48\textwidth]
{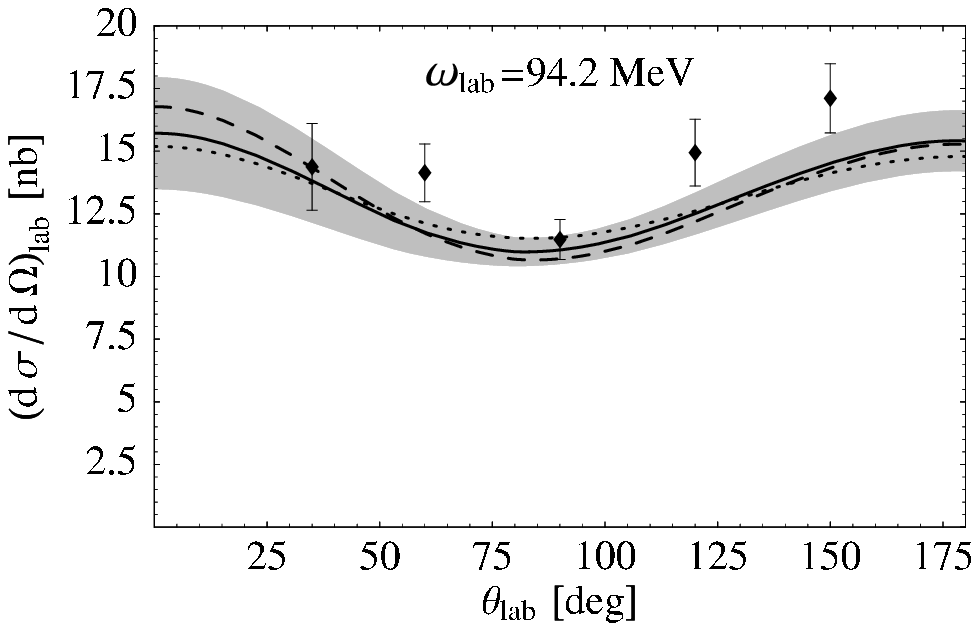}
\includegraphics*[width=.48\textwidth]
{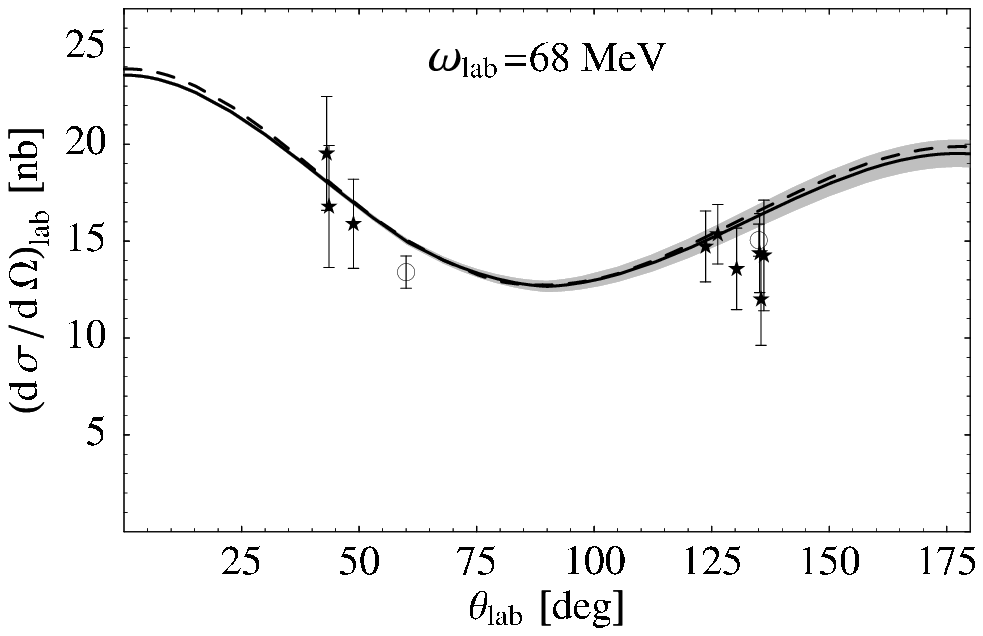}
\hfill
\includegraphics*[width=.48\textwidth]
{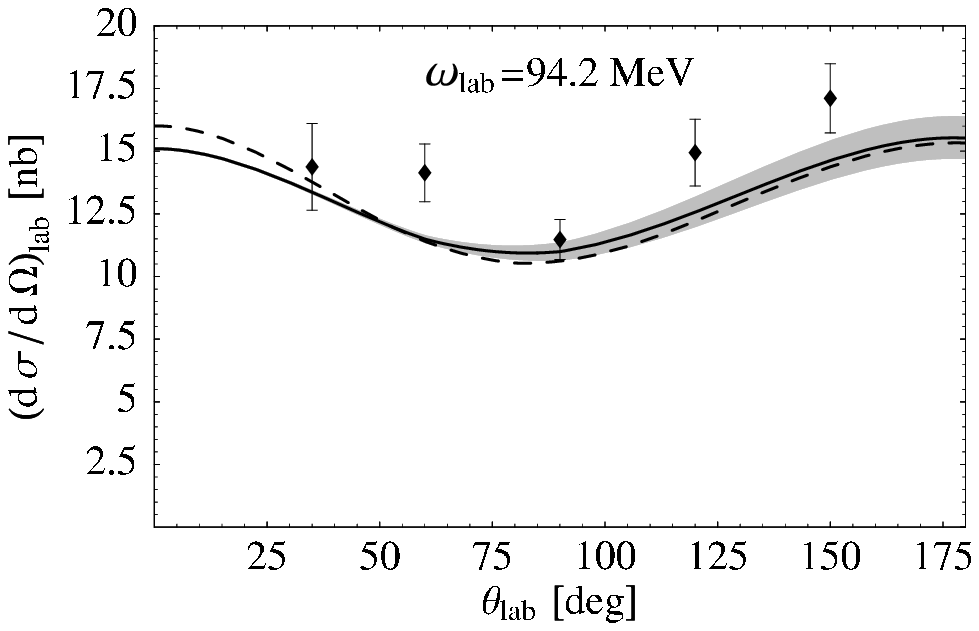}
\parbox{1.\textwidth}{
\caption[Results from fitting $\bar{\alpha}_{E1}^s$ and $\bar{\beta}_{M1}^s$ 
to deuteron Compton data, using 
$\mathcal{O}(\epsilon^3)$ SSE, $\mathcal{O}(\bar{p}^3)$ HB$\chi$PT and
$\mathcal{O}(p^4)$ HB$\chi$PT]
{$\mathcal{O}(\epsilon^3)$-SSE (solid) and 
$\mathcal{O}(\bar{p}^3)$-HB$\chi$PT (dashed) results with 
$\bar{\alpha}_{E1}^s$, $\bar{\beta}_{M1}^s$ from Table~\ref{tab:fulldata}, 
using the chiral NNLO wave function~\cite{Epelbaum}. The upper panels 
correspond to a fit of both polarizabilities, in the lower panels the Baldin 
sum rule (cf. Eq.~(\ref{eq:Baldin})) is 
used as additional fit constraint. The grey bands are derived from our 
(statistical) errors.
The dotted line in the upper panels represents ``fit IV'', one of the 
$\mathcal{O}(     q ^4)$-HB$\chi$PT fits from Ref.~\cite{McGPhil}, 
with central values $\bar{\alpha}_{E1}^s=11.5$, $\bar{\beta}_{M1}^s=0.3$. For 
the $\mathcal{O}(     q ^4)$ calculation the NLO chiral
wave function of Ref.~\cite{NLO} has been used.}
\label{fig:errorfitsfull}}
\end{center}
\end{figure}
As we see in Fig.~\ref{fig:errorfitsfull}, this compensation works very 
well in the $\gamma d$ cross sections: the curves, which correspond to the
$\mathcal{O}(\epsilon^3)$-SSE and to the $\mathcal{O}(\bar{p}^3)$-HB$\chi$PT 
fits are nearly indistinguishable, i.e.  the two fit results 
only differ in the associated pairs $\bar{\alpha}_{E1}$,~$\bar{\beta}_{M1}$. 
Therefore, from the available $\gamma d$ 
data alone one cannot draw any firm conclusion regarding the 
importance of explicit $\Delta(1232)$ degrees of freedom.
However, as we demonstrated in Section~\ref{sec:polarizabilities1},
it is clear from $\gamma p$ scattering experiments
that third-order HB$\chi$PT does not describe 
the dynamics in the $\gamma d$ process correctly. Given that the SSE 
calculation describes both the $\gamma p$ and the $\gamma d$ experiments we 
have shown that a Chiral Effective Field Theory which 
includes the explicit $\Delta$ field is an efficient framework 
to identify the relevant physics underlying low-energy Compton scattering.

\subsection{Comparison of $\mathcal{O}(\epsilon^3)$-SSE and 
                          $\mathcal{O}(p^4)$-HB$\chi$PT Fits}
\label{sec:Comparison2}

In Fig.~\ref{fig:errorfitsfull}, we also show the 
$\mathcal{O}(p^4)$-HB$\chi$PT fit of Ref.~\cite{McGPhil}.  
We consider the quality of our $\mathcal{O}(\bar{p}^3)$ fit to be comparable
to those curves. 
Note that the numbers used for $\bar{\alpha}_{E1}^s$ and $\bar{\beta}_{M1}^s$ 
in the $\mathcal{O}(p^4)$ curves 
differ from Eq.~(\ref{eq:Oq4bestresults}), as they were derived from fitting 
a different set of data. The values from Eq.~(\ref{eq:Oq4bestresults}) are the 
best fit results of Ref.~\cite{McGPhil} in the sense of the least $\chi^2$, 
which excluded, however, the two data points at 94.2~MeV in the backward 
direction. As here we are concerned with all five data points at this energy, 
we compare to ``Fit IV'' of Ref.~\cite{McGPhil} where all of the 
$\gamma d$ data were fitted.

\subsection{Why Equal Statistics at all Energies would be Useful}
\label{sec:unbiased}

\begin{table}[!htb]
\begin{center}
\begin{tabular}{|c||c|}
\hline
angle [deg]&$\mathrm{d}\sigma/\mathrm{d}\Omega$ [nb]\\
\hline
 45.6&$17.3\pm1.9$\\
130.5&$14.1\pm1.4$\\
\hline
\end{tabular}
\caption[Effective data points gained by rebinning the 67~MeV data 
from~\cite{Lund}]
{Effective data points gained by rebinning the 67~MeV data from~\cite{Lund}.}
\label{tab:eff}
\end{center}
\end{table}

\noindent
There are eleven data points at $\omega_\mathrm{lab}\approx 68$~MeV, centered 
around only two different angles, and five points at 
$\omega_\mathrm{lab}\approx 94.2$~MeV, distributed 
over the whole angular spectrum. 
Especially at 
$\theta_\mathrm{lab} \approx 130^\circ$, the experimental efforts at 
Lund and Illinois to get a large integrated number of counts around 65-69~MeV
mean that there is a wealth of data at these energies  
(six points from \cite{Lund} and one from \cite{Lucas}), which gives a strong 
constraint to our fit routines.

\begin{figure}[!htb]
\begin{center} 
\includegraphics*[width=.48\textwidth]
{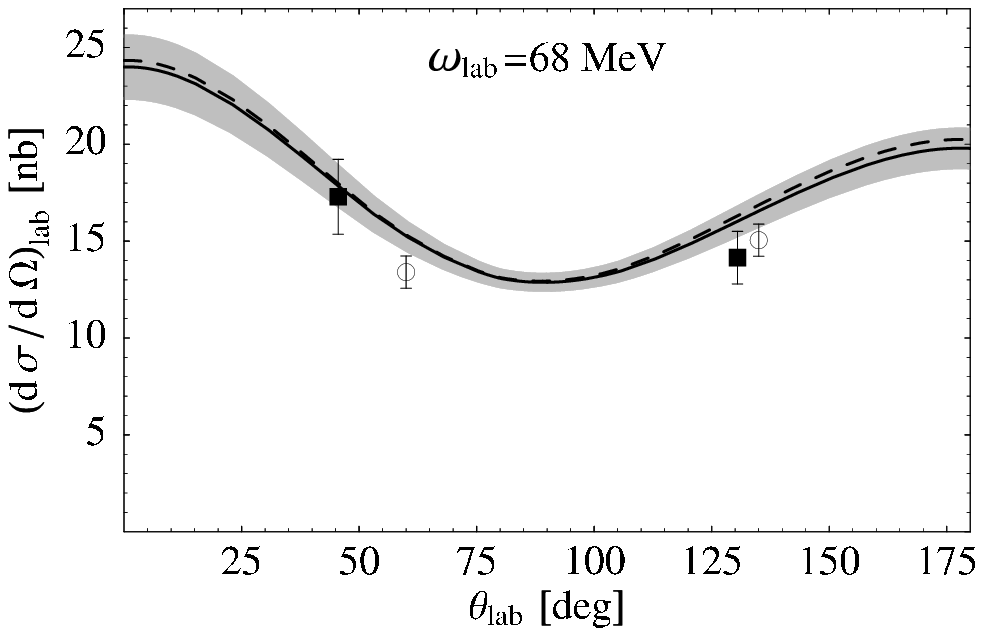}
\hfill
\includegraphics*[width=.48\textwidth]
{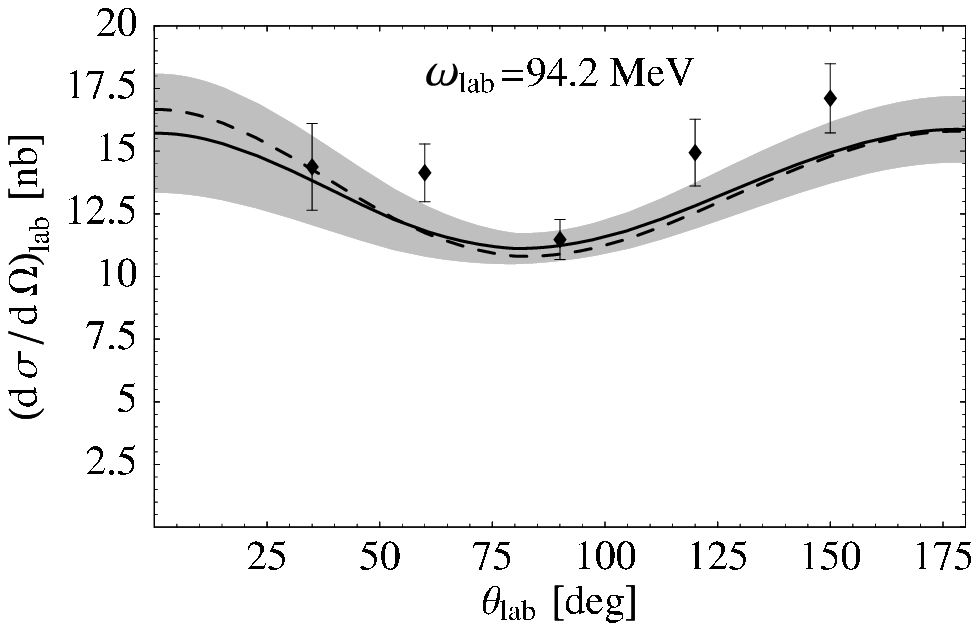}
\includegraphics*[width=.48\textwidth]
{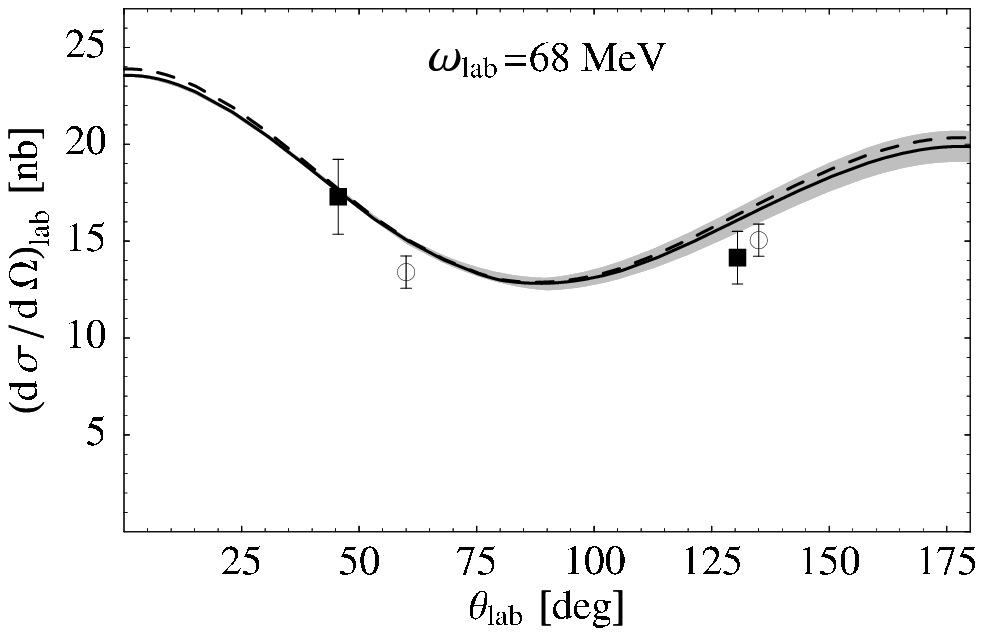}
\hfill
\includegraphics*[width=.48\textwidth]
{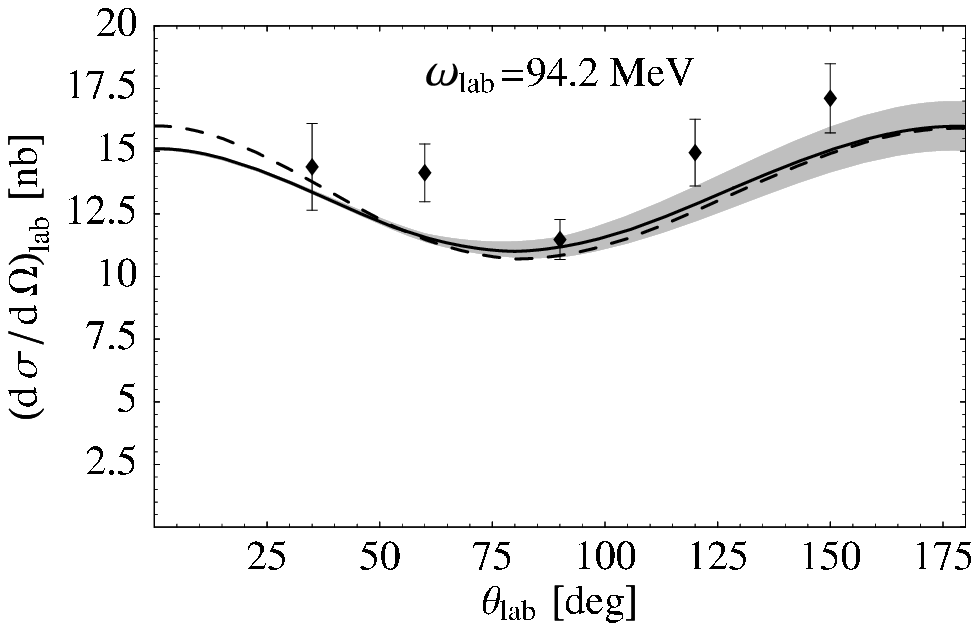}
\parbox{1.\textwidth}{
\caption[Results from fitting $\bar{\alpha}_{E1}^s$ and $\bar{\beta}_{M1}^s$ 
to a reduced set of deuteron Compton data, using $\mathcal{O}(\epsilon^3)$ SSE
and $\mathcal{O}(\bar{p}^3)$ HB$\chi$PT]
{$\mathcal{O}(\epsilon^3)$-SSE (solid) and 
$\mathcal{O}(\bar{p}^3)$-HB$\chi$PT (dashed) results with 
$\bar{\alpha}_{E1}^s$, $\bar{\beta}_{M1}^s$ from Table~\ref{tab:effdata}, using
the chiral NNLO wave function~\cite{Epelbaum}. The upper panels correspond 
to a fit of both polarizabilities, in the lower panels the Baldin sum rule 
(cf. Eq.~(\ref{eq:Baldin})) is used as additional fit constraint. The grey 
bands are derived from our (statistical) errors. The two 
data points plotted as boxes are the effective data that we use as 
representatives for the data from~\cite{Lund}, cf. Table~\ref{tab:eff}.} 
\label{fig:errorfitseff}}
\end{center}
\end{figure}

In the following we demonstrate what would happen if there 
were comparable statistics at 94.2~MeV and 68~MeV.
Of course we cannot generate data at higher energies.
Therefore, we replace in this subsection 
the Lund data~\cite{Lund} by two ``effective'' data points 
(cf. Table~\ref{tab:eff} and
Fig.~\ref{fig:errorfitseff}), which represent the data in the forward and 
backward direction, respectively. These data have been obtained by 
\textit{rebinning}\footnote{We are indebted to Bent Schr\"oder for 
valuable comments made on this point.} 
the data from~\cite{Lund}, 
and thus the statistical error of the effective data points is reduced with 
respect to the errors of the individual data points published in~\cite{Lund}.
In order to calculate the rebinned data we weight the angles and 
the differential cross-section values and also the systematic errors of the 
represented data points by the inverse of their errors, e.g. 
the cross sections are derived via
\be
\frac{\mathrm{d}\bar{\sigma}}{\mathrm{d}\Omega}=
\sum_i\left(\frac{\mathrm{d}\sigma}{\mathrm{d}\Omega}\right)_i\cdot
\frac{1}{\Delta \left(\frac{\mathrm{d}\sigma}{\mathrm{d}\Omega}\right)_i}\bigg/
\sum_i\frac{1}{\Delta \left(\frac{\mathrm{d}\sigma}{\mathrm{d}\Omega}\right)_i}
\ee
with $\Delta \left(\frac{\mathrm{d}\sigma}{\mathrm{d}\Omega}\right)_i$ denoting
the statistical errors of the rebinned data points.
The statistical error of the effective data is obtained as 
\be
\Delta\left(\frac{\mathrm{d}\bar{\sigma}}{\mathrm{d}\Omega}\right)=
\frac{\mathrm{d}\bar{\sigma}}{\mathrm{d}\Omega}\bigg/
\sqrt{\sum_i\left[\left(\frac{\mathrm{d}\sigma}{\mathrm{d}\Omega}\right)_i
\bigg/\Delta \left(\frac{\mathrm{d}\sigma}{\mathrm{d}\Omega}\right)_i\right]^2}
\;.
\ee
Therefore, the remaining data are the two data 
points from~\cite{Lucas} at 69~MeV, the two ``effective'' data at 
$\sim$67~MeV, shown in Table~\ref{tab:eff}, 
representing~\cite{Lund}, and the five data points 
from~\cite{Hornidge} around 94.2~MeV. With these 
nine data points we perform the same fits as we did before for the complete 
data sets.
The resulting values for $\bar{\alpha}_{E1}$ and $\bar{\beta}_{M1}$ are 
presented in Table~\ref{tab:effdata}. The plots (including the two effective 
data points), shown in Fig.~\ref{fig:errorfitseff}, exhibit better agreement 
with the 94.2~MeV data than the fits of Section~\ref{sec:Comparison1} 
(Fig.~\ref{fig:errorfitsfull}); that is exactly what we expected, because 
due to the reduced number of data points at 68~MeV, the sensitivity of the 
fits to the 94.2~MeV data is increased.

The results for the static isoscalar polarizabilities, averaged as before 
over the two wave functions are 
\begin{align}
\bar{\alpha}_{E1}^s&=(12.8\pm1.4\,(\mathrm{stat})\pm1.1\,(\mathrm{wf}))
            \cdot 10^{-4}\;\mathrm{fm}^3\,,\nonumber\\
\bar{\beta}_{M1}^s &=( 2.1\pm1.7\,(\mathrm{stat})\pm0.1\,(\mathrm{wf}))
            \cdot 10^{-4}\;\mathrm{fm}^3\,.
\label{eq:finalunbiased}
\end{align}
Including the Baldin constraint we get
\begin{align}
\bar{\alpha}_{E1}^s&=(12.6\pm0.8\,(\mathrm{stat})\pm0.7\,(\mathrm{wf})
                 \pm0.6\,(\mathrm{Baldin}))
            \cdot 10^{-4}\;\mathrm{fm}^3\,,\nonumber\\
\bar{\beta}_{M1}^s &=( 1.9\mp0.8\,(\mathrm{stat})\mp0.7\,(\mathrm{wf})
                 \pm0.6\,(\mathrm{Baldin}))
            \cdot 10^{-4}\;\mathrm{fm}^3\,.
\label{eq:finalunbiasedBaldin}
\end{align}
The results of all four extraction methods, given in Eqs.~(\ref{eq:final}, 
\ref{eq:finalBaldin}, \ref{eq:finalunbiased}, \ref{eq:finalunbiasedBaldin}),  
agree well with each other, albeit we note that the central values for 
$\bar{\alpha}_{E1}$ ($\bar{\beta}_{M1}$) given in 
Eqs.~(\ref{eq:finalunbiased}/\ref{eq:finalunbiasedBaldin}) are 
slightly smaller (slightly larger), comparing to the results  from our fits 
to all data points. However,
we consider the result from the rebinned data set as more 
reliable as the procedure weights the 69~MeV data and the 94.2~MeV data in a 
more equal fashion. 

We also see in Table~\ref{tab:effdata} once again that the theory 
without explicit $\Delta$ degrees of freedom leads to a 
systematically larger value for $\bar{\beta}_{M1}^s$ than obtained in the 
$\mathcal{O}(\epsilon^3)$ fit, supporting our hypothesis that 
this enhancement is due to insufficient dynamics in the multipoles
(cf. Table~\ref{tab:fulldata}).
In HB$\chi$PT, the static value is artificially enlarged by the fit constraint 
from the 94.2~MeV data in order to compensate for the missing dynamics. 
Both in Fig.~\ref{fig:errorfitsfull} and Fig.~\ref{fig:errorfitseff} the 
enhanced static magnetic dipole polarizability $\bar{\beta}_{M1}$ 
cures the $\gamma d$ cross 
sections and makes the resulting curves very similar to the plots from the 
SSE fits, albeit the resulting picture in the Compton multipoles is very 
different (see Fig.~\ref{fig:spinindependentpolas}). 

Therefore, for understanding the available $\gamma d$ data via fits of 
$\bar{\alpha}_{E1}^s$ and $\bar{\beta}_{M1}^s$,
it is essential to combine the pairs $\bar{\alpha}_{E1}$,~$\bar{\beta}_{M1}$ 
resulting from the $\gamma d$ analysis with an energy-dependent multipole 
analysis of $\gamma p$ scattering. 
From the information available on Compton multipoles from $\gamma p$ 
scattering experiments, it is clear that third-order HB$\chi$PT is too 
simplistic a picture for the dynamics of the $\gamma d$ process at energies of 
$\mathcal{O}(100~\mathrm{MeV})$. It is therefore
crucial that deuteron and proton Compton experiments are available at 
comparable energies and that they are analyzed within the same framework.

Putting equal statistical weight on the 68 and the 94.2~MeV data
can be seen as a demonstration of the importance of obtaining 
comparable statistics at all energies.
We therefore urge for more experimental information at photon energies 
around 100~MeV.
With such information, deuteron Compton cross sections below the pion mass 
provide an excellent window to investigate which internal nucleonic degrees 
of freedom contribute in both processes, $\gamma p\rightarrow \gamma p$ and 
$\gamma d\rightarrow \gamma d$.  

\begin{table}[!htb]
\begin{center}
\begin{tabular}{|c||c||c|c|c|c|}
\hline
Amplitudes&Quantity&2-par. fit    &1-par. fit   &2-par. fit&1-par. fit\\
          &        &NNLO $\chi$PT &NNLO $\chi$PT&Nijm93    &Nijm93    \\    
\hline
\hline
$\mathcal{O}(\epsilon^3)$~SSE&$\chi^2/d.o.f.$ &2.79&2.47&3.97&3.59\\
\cline{2-6}
&$\bar{\alpha}_{E1}^s$&$11.7\pm1.4$
                 &$11.9\pm0.8$&$13.8\pm 1.3$&$13.3\pm0.7$\\
&$\bar{\beta} _{M1}^s$&$ 2.2\pm1.7$
                 &$ 2.6\mp0.8$&$ 2.0\pm 1.6$&$ 1.2\mp0.7$\\
\cline{2-6}
&$\bar{\alpha}_{E1}^s+\bar{\beta} _{M1}^s$ 
                       &$13.9\pm2.2$&$14.5\,(\mathrm{fit})$
                       &$15.8\pm2.1$&$14.5\,(\mathrm{fit})$\\
\cline{2-6}
&$\bar{\alpha}_{E1}^n$&$11.3\pm1.4$
                 &$11.7\pm0.9$&$15.5\pm 1.3$&$14.5\pm0.8$\\
&$\bar{\beta} _{M1}^n$&$ 2.8\pm1.7$
                 &$ 3.6\mp0.9$&$ 2.4\pm 1.6$&$ 0.8\mp0.8$\\
\hline
\hline
$\mathcal{O}(\bar{p}^3)$~HB$\chi$PT&$\chi^2/d.o.f.$ &3.38&2.98&4.69&4.24\\
\cline{2-6}
&$\bar{\alpha}_{E1}^s$&$10.6\pm1.3$
                 &$10.8\pm0.8$&$12.7\pm 1.3$&$12.2\pm0.8$\\
&$\bar{\beta} _{M1}^s$&$ 3.3\pm1.7$
                 &$ 3.7\mp0.8$&$ 3.1\pm 1.6$&$ 2.3\mp0.8$\\
\cline{2-6}
&$\bar{\alpha}_{E1}^s+\bar{\beta} _{M1}^s$
                       &$13.9\pm2.1$&$14.5\,(\mathrm{fit})$
                       &$15.8\pm2.1$&$14.5\,(\mathrm{fit})$\\
\cline{2-6}
&$\bar{\alpha}_{E1}^n$&$9.1\pm1.3$
                 &$9.5\pm0.9$&$13.3\pm 1.3$&$12.3\pm0.9$\\
&$\bar{\beta} _{M1}^n$&$ 5.0\pm1.7$
                 &$ 5.8\mp0.9$&$ 4.6\pm 1.6$&$ 3.0\mp0.9$\\
\hline
\end{tabular}
\caption[Values for polarizabilities from fit to a reduced set of deuteron 
Compton data, using $\mathcal{O}(\epsilon^3)$~SSE and  
$\mathcal{O}(\bar{p}^3)$~HB$\chi$PT]
{Values for the isoscalar and neutron polarizabilities 
(in $10^{-4}\,\mathrm{fm}^3$) from a fit to the 68~MeV and 
94.2~MeV data sets~\cite{Lucas, Hornidge}, using the 
$\mathcal{O}(\epsilon^3)$-SSE and the $\mathcal{O}(\bar{p}^3)$-HB$\chi$PT
amplitudes, respectively. The data from~\cite{Lund} have been replaced by two 
effective data points, specified in Table~\ref{tab:eff}.
The neutron results are derived 
using the proton values from~\cite{Olmos} as input.
All error bars displayed are only statistical (and added in quadrature).}
\label{tab:effdata}
\end{center}
\end{table}

\section{Attempts to Restore the Thomson Limit}
\label{sec:Thomson1}
In Section~\ref{sec:results}, we demonstrated that our calculation provides a 
good description of the data measured around 68 and 94.2~MeV. However, below a 
certain energy limit, which we found to be of the order of 50-60~MeV, the 
approach
used in this chapter is not valid anymore. This observation is no surprise, as
the power-counting scheme that we apply has been designed for photon energies
of the order of the pion mass. Only for such large energies it is possible to
approximate the nucleon propagator inside the deuteron by $i/\w$, 
whereas the kinetic energy of the nucleon is treated 
perturbatively. Such a perturbative treatment is always possible 
in single-nucleon calculations.
However, the two nucleons bound in the deuteron 
have a non-vanishing momentum relative to each other and therefore cannot be 
treated as static, even if
the energy of the external probe is zero. Therefore, only for large photon 
momenta it is possible to approximate the non-relativistic nucleon propagator
$\frac{i}{\w-p^2/2m_N}$ by $i/\w$, as already explained in 
Section~\ref{sec:theory}.  

\begin{figure}[!htb]
\begin{center} 
\includegraphics*[width=.48\textwidth]{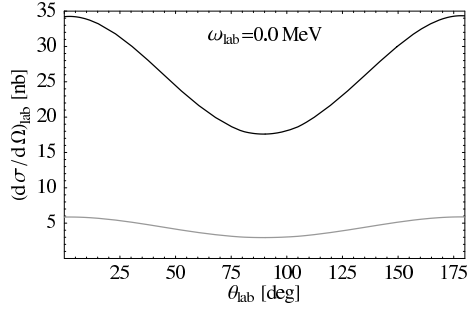}
\parbox{1.\textwidth}{
\caption[Static limit of our deuteron Compton calculation]
{Comparison of the static limit of our deuteron Compton calculation (black)  
to the correct Thomson limit for the deuteron, Eq.~(\ref{eq:Thomson}) (grey).}
\label{fig:static}}
\end{center}
\end{figure}

From these considerations it becomes obvious that~-- unlike e.g. the 
Effective Field Theory with pions integrated out~\cite{Rupak,Ji}~-- 
our calculation cannot 
reach the correct static limit, i.e. the limit of vanishing photon
energy, Eq.~(\ref{eq:Thomson}). There are only two 
contributions to our calculation in this limit: The proton seagull, 
Fig.~\ref{fig:chiPTsingle}(a), and the 
pion-exchange diagrams from Fig.~\ref{fig:chiPTdouble}. The first one 
gives the Thomson limit of the proton, which is larger by a
factor of 4 than the correct deuteron limit, cf. Eqs.~(\ref{eq:Thomson}) and 
(\ref{eq:deuteroncrosssection}). Therefore, one might assume that the 
pion-exchange diagrams partly cancel this contribution, which is, however, not 
the case. In contrast, they even enhance the static limit and render the 
result too large by a factor of 6, a clear signal that gauge invariance is 
violated~\cite{Friar}. This is shown in Fig.~\ref{fig:static} by 
comparing the static cross section of our calculation 
to the correct limit as dictated by gauge invariance, given in 
Eq.~(\ref{eq:Thomson}). 
Although we know that our calculation cannot work for small photon energies, 
this enhancement is surprising. Therefore we undertook some efforts
to~-- at least approximately~-- restore the correct Thomson limit, which we 
describe in the following.

\begin{figure}[!htb]
\begin{center} 
\includegraphics*[width=.17\textwidth]{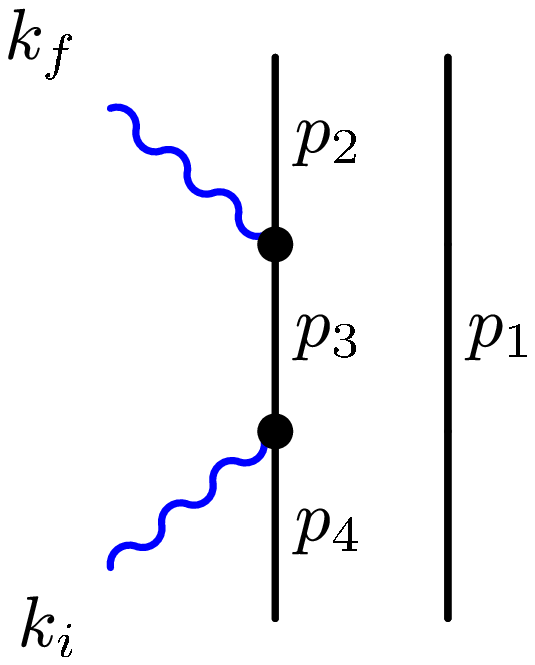}
\qquad
\includegraphics*[width=.17\textwidth]{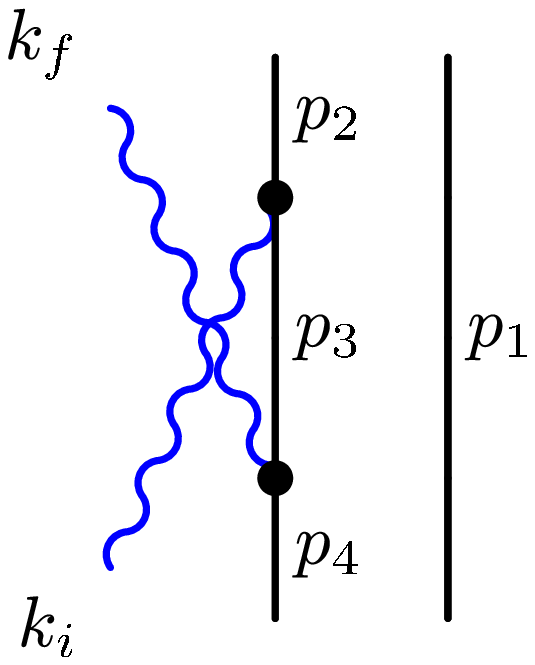}
\parbox{1.\textwidth}{
\caption[Nucleon-pole terms]
{Nucleon-pole terms in the $s$-channel (left) and in the $u$-channel (right).}
\label{fig:polepropagators}}
\end{center}
\end{figure}

So far we used the leading HB$\chi$PT approximation $i/\w$ for the nucleon 
propagator, which 
is, however, not valid for low photon energies, as explained above. Therefore, 
we replace the expression $i/\w$ by the full 
non-relativistic nucleon propagator. We do so in the nucleon-pole diagrams,  
Fig.~\ref{fig:chiPTsingle}(b), which we sketch once again in 
Fig.~\ref{fig:polepropagators}.

In order to calculate the pole diagrams with the full non-relativistic nucleon
propagator, we first write down the nucleon 4-momenta corresponding to 
Fig.~\ref{fig:polepropagators}:
\be
p_1=
\begin{pmatrix}
-\frac{B}{2}+\frac{\w^2}{8m_N}-p_0\\-\vec{p}-\frac{\vec{k}_i}{2}
\end{pmatrix},\;\;\;
p_2=\begin{pmatrix}
-\frac{B}{2}+\frac{\w^2}{8m_N}+p_0\\ \vec{p}+\frac{\vec{k}_i}{2}-\vec{k}_f
\end{pmatrix},\;\;\;
p_4=\begin{pmatrix}
-\frac{B}{2}+\frac{\w^2}{8m_N}+p_0\\ \vec{p}-\frac{\vec{k}_i}{2}
\end{pmatrix},
\nonumber
\ee
\be
p_3^s=
\begin{pmatrix}
\w-\frac{B}{2}+\frac{\w^2}{8m_N}+p_0\\ \vec{p}+\frac{\vec{k}_i}{2}
\end{pmatrix},\;\;\;
p_3^u=\begin{pmatrix}
-\w-\frac{B}{2}+\frac{\w^2}{8m_N}+p_0\\ \vec{p}-\frac{\vec{k}_i}{2}-\vec{k}_f
\end{pmatrix}.
\ee
The photon-nucleon vertex is taken from $\mathcal{L}_{\pi N}^{(2)}$, 
cf.~\cite{BKM}. However, as we are only interested in contributions which do 
not vanish in the static limit, we may restrict ourselves to 
\be
\frac{i\,e}{4\,m_N}\,\left(1+\tau^z\right)\,\epsilon\cdot(p_a+p_b),
\label{eq:photonnucleonvertex}
\ee
whereas we neglect the term
\be
\frac{i\,e}{2\,m_N}\,\left[S\cdot\epsilon,S\cdot k\right]
\left(1+\kappa_s+(1+\kappa_v)\,\tau^z\right).
\ee
After evaluating the $p_0$-integral (we close the contour via the 
four-momentum $p_1$), we find for the two amplitudes
\ba
\label{eq:schannelamplitude}
T^s(\vec{k}_i,\kf;\vec{p})&=-\frac{e^2}{m_N^2}\,
\vec{\epsilon}\,'\cdot(\vec{p}+\vec{k}_i/2)\,\vec{\epsilon}\cdot\vec{p}\,
\frac{1}{\w-B-\frac{\vec{p}\,^2+\vec{p}\cdot\vec{k}_i}{m_N}},\\
T^u(\vec{k}_i,\kf;\vec{p})&=-\frac{e^2}{m_N^2}\,
\vec{\epsilon}\cdot(\vec{p}-\vec{k}_f)\,\vec{\epsilon}\,'\cdot(\vec{p}-\ki/2)\,
\frac{1}{-\w-B-\frac{\w^2+\ki\cdot\kf+2\vec{p}\,^2-2\vec{p}\cdot\vec{k}_f}
{2m_N}}.
\label{eq:uchannelamplitude}
\end{align}
Note that Eqs.~(\ref{eq:schannelamplitude}, \ref{eq:uchannelamplitude}) 
do give non-vanishing contributions in the static 
limit, whereas in the approximation $i/\w$ they cancel each other exactly for 
$\w=|\vec{k}|= 0$. 
 
\begin{figure}[!htb]
\begin{center} 
class 2:\qquad
\parbox[m]{.7\linewidth}{
\includegraphics*[width=.24\linewidth]{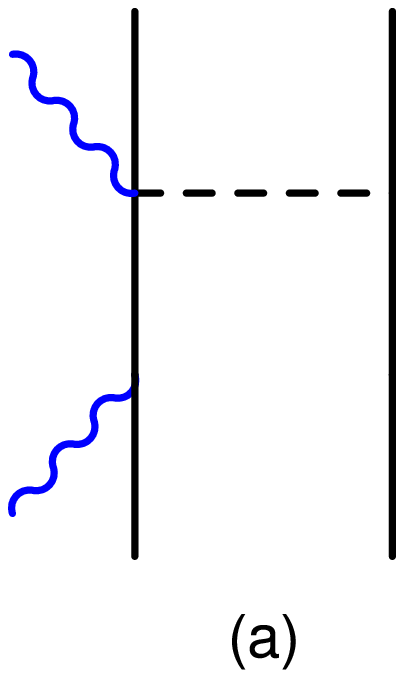}
\includegraphics*[width=.24\linewidth]{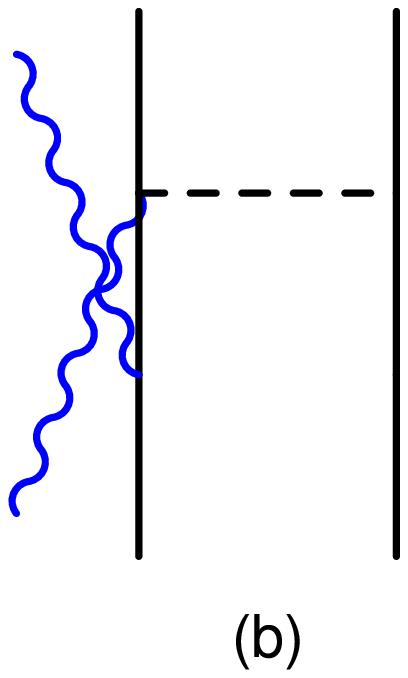}
\includegraphics*[width=.24\linewidth]{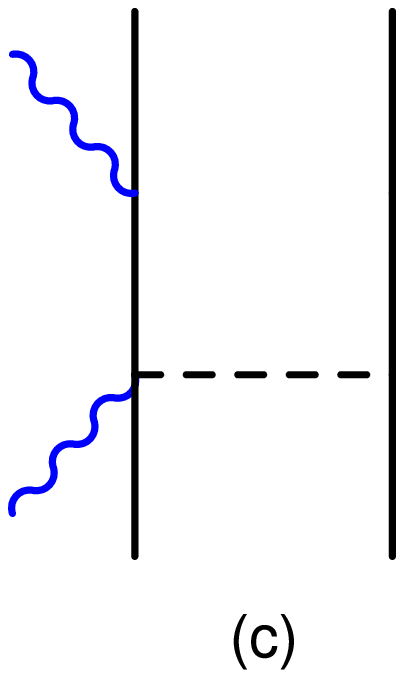}
\includegraphics*[width=.24\linewidth]{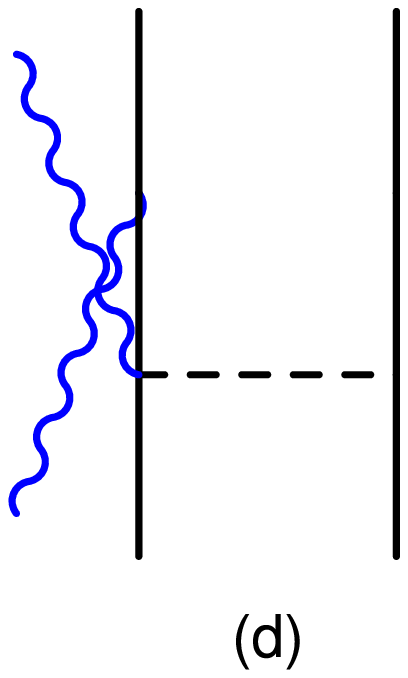}}
$+1\leftrightarrow 2$\\
class 3:\qquad
\parbox[m]{.7\linewidth}{
\includegraphics*[width=.24\linewidth]{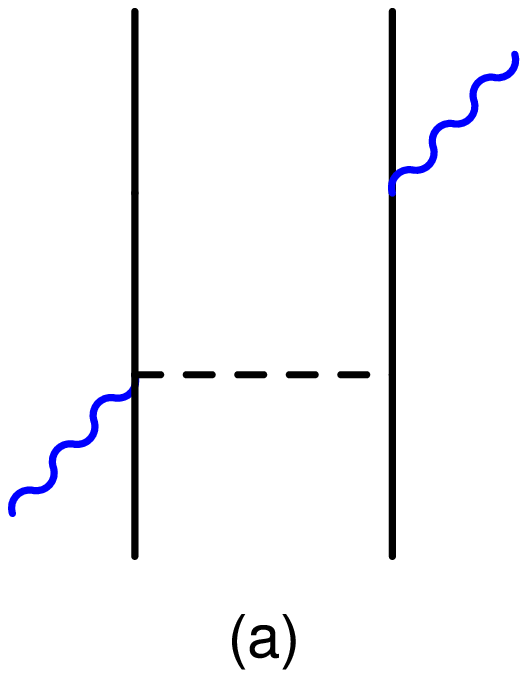}
\includegraphics*[width=.24\linewidth]{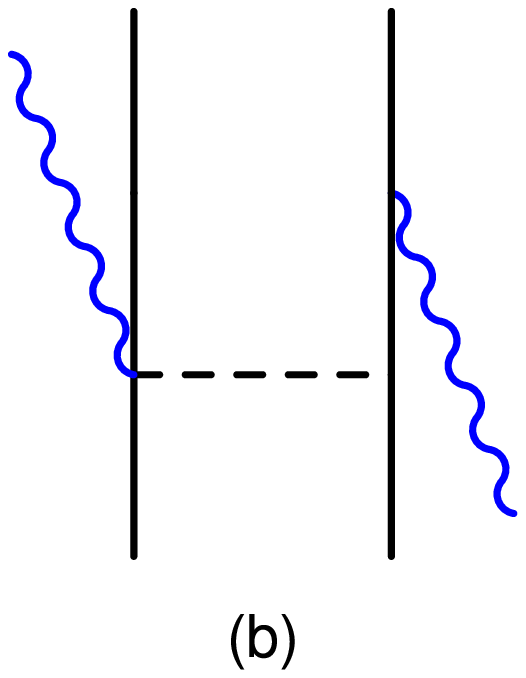}
\includegraphics*[width=.24\linewidth]{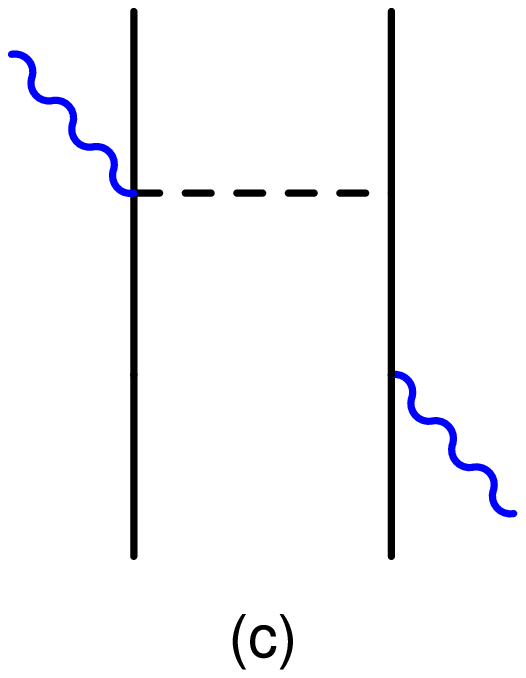}
\includegraphics*[width=.24\linewidth]{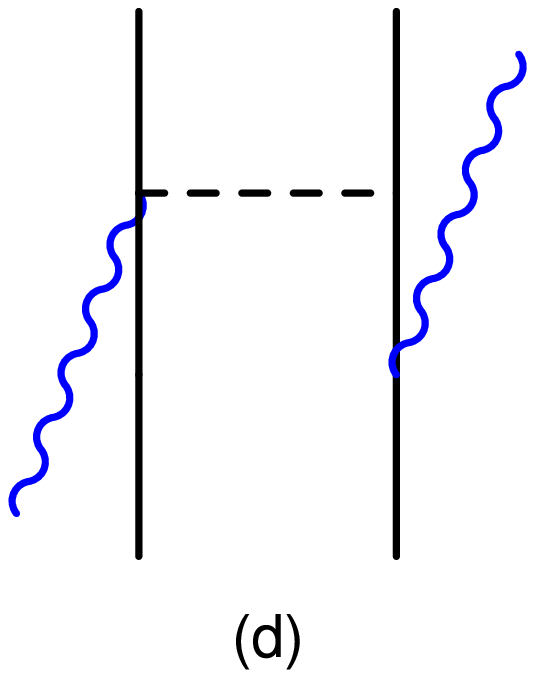}}
$+1\leftrightarrow 2$\\
class 4:\qquad
\parbox[m]{.7\linewidth}{
\includegraphics*[width=.24\linewidth]{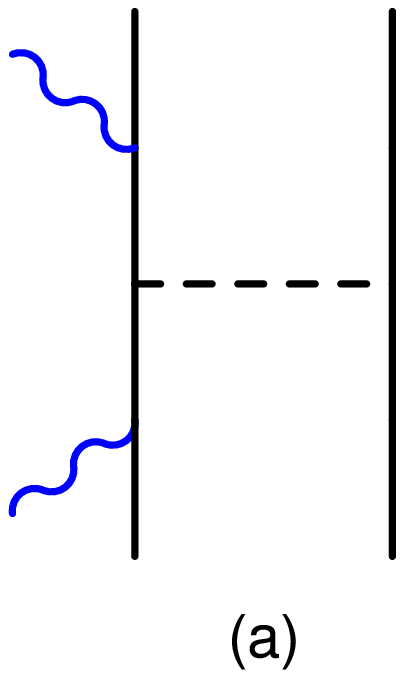}
\includegraphics*[width=.24\linewidth]{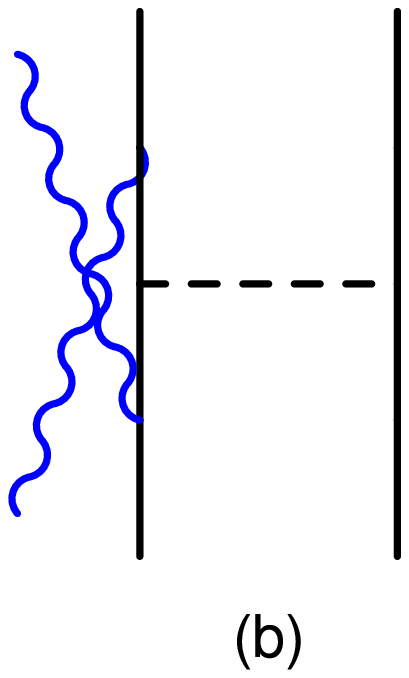}}
$+1\leftrightarrow 2$\\
class 5:\qquad
\parbox[m]{.7\linewidth}{
\includegraphics*[width=.24\linewidth]{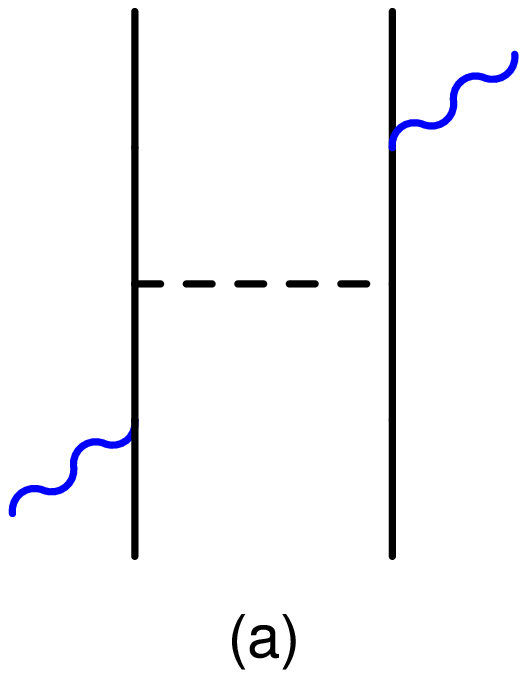}
\includegraphics*[width=.24\linewidth]{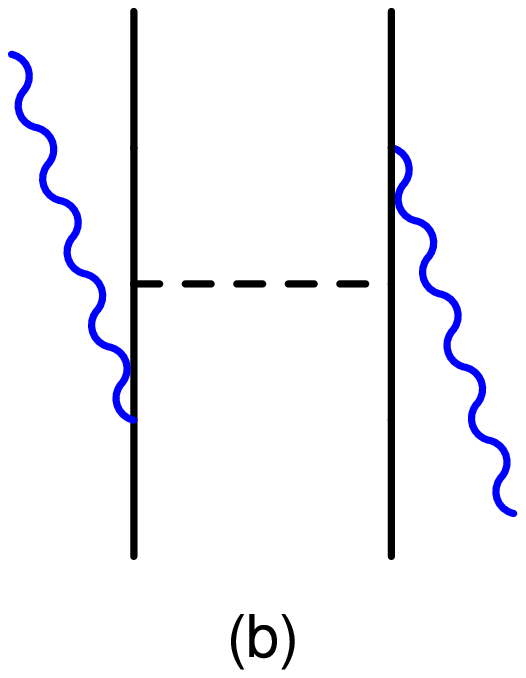}}
$+1\leftrightarrow 2$\\
\vspace{.2cm}
class 6:\qquad
\parbox[m]{.7\linewidth}{
\includegraphics*[width=.24\linewidth]{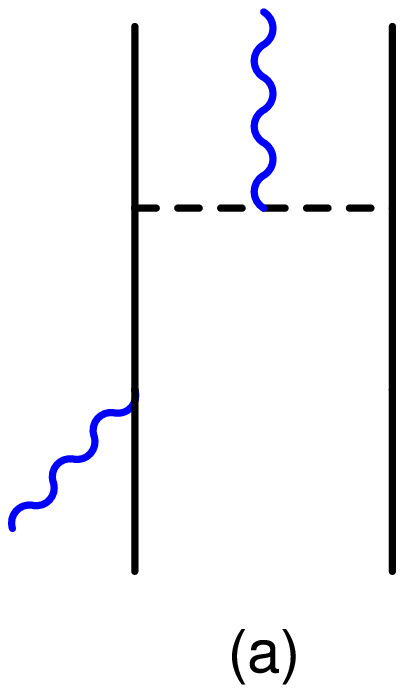}
\includegraphics*[width=.24\linewidth]{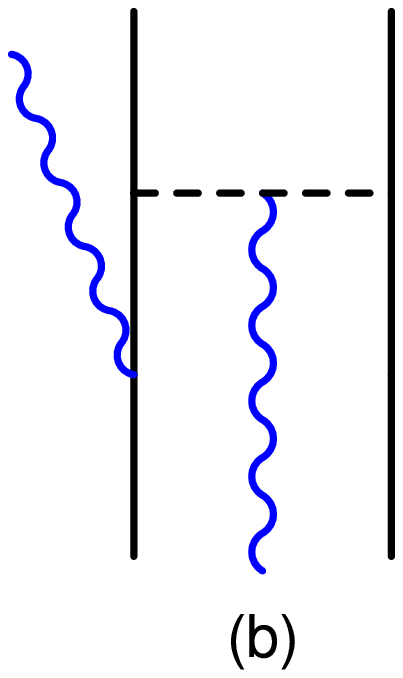}
\includegraphics*[width=.24\linewidth]{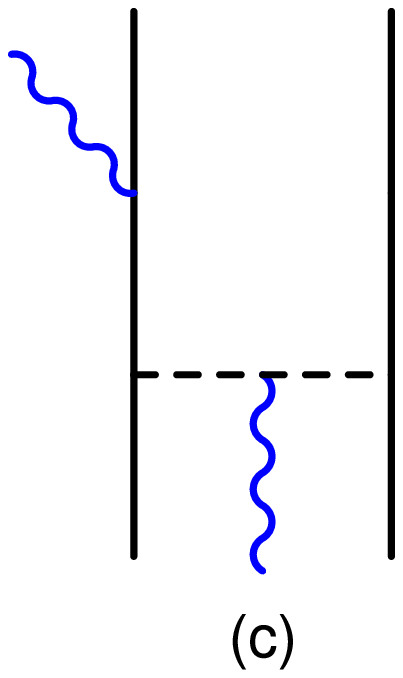}
\includegraphics*[width=.24\linewidth]{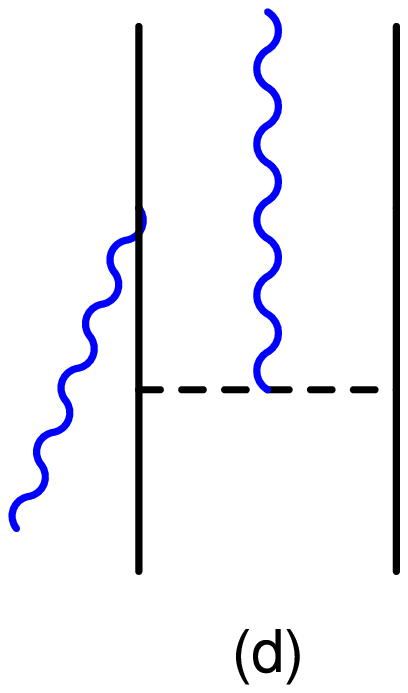}}
$+1\leftrightarrow 2$
\parbox{1.\textwidth}{
\caption[Additional pion-exchange diagrams which appear at 
$\mathcal{O}(\epsilon^3)$  in the low-energy regime]
{Additional pion-exchange diagrams which appear at $\mathcal{O}(\epsilon^3)$ 
in the low-energy regime.}
\label{fig:lowenergydiagrams}}
\end{center}
\end{figure}

So far, we have been counting the nucleon propagator as 
$\calO(\epsilon^{-1})$. However, when we 
look at the propagator of e.g. Eq.~(\ref{eq:schannelamplitude}), we see
that for small photon energies this counting is not justified anymore. For 
$\w\rightarrow 0$, the dominant part in the denominator of the propagator 
is $\vec{p}\,^2/m_N$. Therefore, in the low-energy regime one has to count the 
nucleon propagator as $\calO(\epsilon^{-2})$, as already discussed in 
Section~\ref{sec:theory}. In this counting scheme, however, there are many 
additional pion-exchange diagrams at $\calO(\epsilon^{3})$, which contribute 
in the static limit. 
All these diagrams are 
sketched in Fig.~\ref{fig:lowenergydiagrams}, distributed into several classes
(class 1 corresponds to the diagrams of Fig.~\ref{fig:chiPTdouble}).

\begin{figure}[!htb]
\begin{center} 
\includegraphics*[width=.48\textwidth]{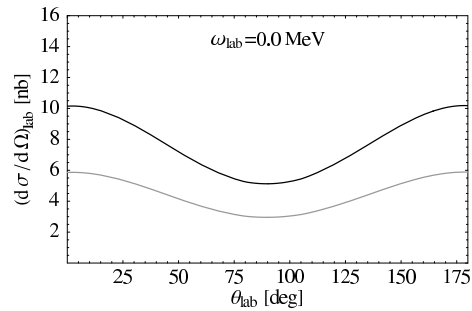}
\parbox{1.\textwidth}{
\caption[Attempt to restore the Thomson limit]
{Attempt to restore the Thomson limit. The grey line marks the correct static 
limit for the deuteron, Eq.~(\ref{eq:Thomson}). The black line is our result 
with the nucleon-pole diagrams calculated with the exact
nucleon propagator and also  including the 
diagrams of Fig.~\ref{fig:lowenergydiagrams}.}
\label{fig:Thomsontrials2}}
\end{center}
\end{figure}

The calculation of the amplitudes corresponding to 
Fig.~\ref{fig:lowenergydiagrams}
is straightforward. Again we only include the part of the photon-nucleon
coupling which does not vanish with the photon energy, 
Eq.~(\ref{eq:photonnucleonvertex}), and of course we use the full nucleon 
propagator. The results are given in 
Appendix~\ref{app:lowenergydiagrams}. Including them in our calculation, 
together with the corrected nucleon-pole terms, leads to the static limit 
shown in Fig.~\ref{fig:Thomsontrials2}. 
We see that these contributions help to 
come closer to the Thomson limit, but still we lack a factor of nearly 2.
In Chapter~\ref{chap:nonperturbative} we show that this discrepancy is 
resolved when we include the full two-nucleon Green's function $G$ 
in the intermediate state, cf. Eq.~(\ref{eq:Ggammagamma}),
i.e. we allow the two nucleons to rescatter.
Therefore we now close our efforts to reach the exact static limit 
and postpone this challenge to Chapter~\ref{chap:nonperturbative} 
(Section~\ref{sec:Thomson2}).

In this chapter we reported a calculation of elastic deuteron Compton 
scattering, using a 
so-called hybrid approach, i.e. we calculated the interaction kernel according
to the rules of perturbative Effective Field Theory in the Small Scale 
Expansion, and folded this kernel with deuteron wave functions derived from 
modern $NN$-potentials. We found good agreement with experimental data at 
photon energies above 60~MeV and thus fitted the Compton cross sections 
to these data in order to determine the isoscalar nucleon polarizabilities. 
The results are in good agreement with the expectation that the isovector 
polarizabilities are small. However, 
there are also certain shortcomings of 
this approach: As it is designed for photon energies of the order of the 
pion mass, our calculation fails to describe the data below 60~MeV and also 
violates the well-known low-energy theorem for deuteron Compton scattering. 
Furthermore, our results show
a stronger sensitivity on the deuteron wave function than we expected.
Therefore we turn now to a partly different approach to deuteron Compton 
scattering, including rescattering of the two nucleons, and we will find that 
all three disadvantages can be cured within this framework. 

\chapter{Non-Perturbative Approach to Deuteron Compton Scattering
\label{chap:nonperturbative} }
\markboth{CHAPTER \ref{chap:nonperturbative}. NON-PERTURBATIVE APPROACH}{}
In this chapter we 
go beyond the strictly perturbative 
approach to deuteron Compton scattering described in the previous chapter, 
in which gauge invariance was not manifest, leading to a violation of 
the low-energy theorem and a surprisingly strong dependence on the 
deuteron wave function.
We do so by use of non-perturbative methods, 
summing over all possible $np$-intermediate states, in combination 
with coupling the photon field to the two nucleons in 
a more complete manner than in Chapter~\ref{chap:perturbative}. 
This part of our 
calculation follows closely the work of J.~Karakowski and 
G.A.~Miller~\cite{Karakowski}\footnote{We note that the sign convention of 
Ref.~\cite{Karakowski} for the deuteron Compton amplitude is different from 
ours. This difference will show up several times in our work. We fix the 
relative sign via the proton seagull diagram, 
Fig.~\ref{fig:chiPTsingle}(a).}. The results of this chapter are contained 
in Ref.~\cite{deuteron2}.

\section{Theoretical Framework \label{sec:theory2}}

\begin{figure}[!htb]
\begin{center} 
\parbox[m]{.158\linewidth}{
\includegraphics*[width=\linewidth]{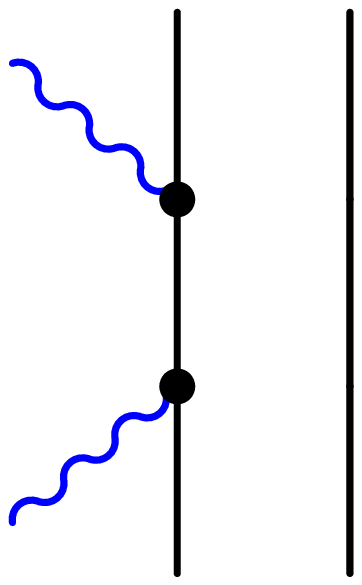}}
\hspace{.5cm}
\parbox[m]{.16\linewidth}{
\includegraphics*[width=\linewidth]{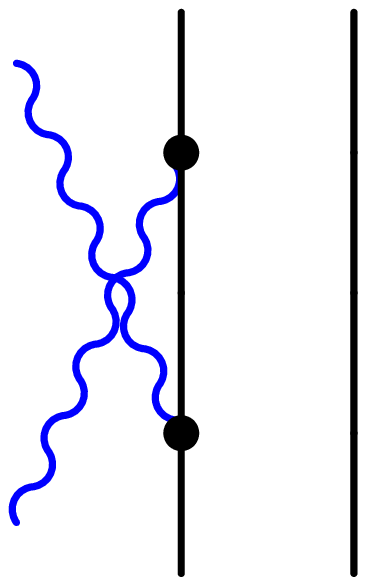}}
$\;\;\;\;\;\longrightarrow\;$
\parbox[m]{.168\linewidth}{
\includegraphics*[width=\linewidth]{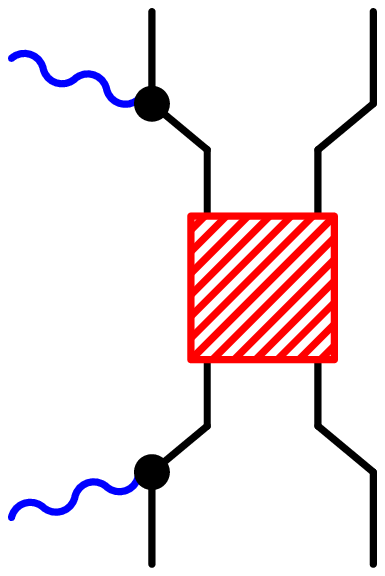}}
\hspace{.5cm}
\parbox[m]{.17\linewidth}{
\includegraphics*[width=\linewidth]{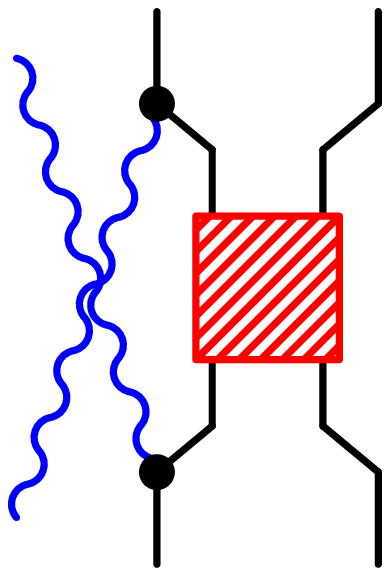}}
\parbox{1.\textwidth}{
\caption[Nucleon-pole diagrams with rescattering]
{Sketch of the main difference between the purely perturbative 
approach to the deuteron Compton kernel used in 
Chapter~\ref{chap:perturbative} and the non-perturbative calculation 
described in this chapter. The square symbolises the 
possible interaction of the two nucleons.}
\label{fig:disp}}
\end{center}
\end{figure}

The main difference to the previous, perturbative expansion of the interaction 
kernel is that we now allow the two nucleons in the 
intermediate state to interact with each other, i.e. to rescatter, whereas such
processes contribute only at higher orders in the power counting applied in 
Chapter~\ref{chap:perturbative}. Diagrammatically, 
this difference is expressed by replacing the nucleon pole diagrams 
(cf. Fig.~\ref{fig:chiPTsingle}(b)) by analogous diagrams with interacting 
nucleons between the two photon vertices. We distinguish between the 
two scenarios~-- with and without rescattering~-- 
in Fig.~\ref{fig:disp} by a square, denoting the $np$-potential.
In other words, we no longer do a perturbative treatment of the two-nucleon 
Green's function $G$ which is contained in the two-nucleon reducible part of 
the interaction kernel $K_\gamma G K_\gamma$, cf. Eq.~(\ref{eq:Ggammagamma}). 
In the next two sections we explain how to calculate the thus modified  
nucleon-pole diagrams.

\subsection{Dominant Diagrams with Intermediate $NN$-Scattering 
\label{sec:dominant}}

The diagrams shown in  Fig.~\ref{fig:disp} are calculated 
using second-order time-ordered perturbation theory in the two-photon 
interaction,
while the intermediate two-nucleon system is treated non-perturbatively.
In general, the scattering amplitude for these processes can be written as
\begin{align}
\Mfi{}&=\sum_C\left
     \{\frac{\mx{d_f,\gamma_f}{\Hint}{C}
             \mx{C}{\Hint}{d_i,\gamma_i}}
    { \w+\frac{\w^2}{2\md}-B-\EC-\frac{\PCsq}{2\mC}}\right.\nonumber\\
      &+\left.
       \frac{\mx{d_f,\gamma_f}{\Hint}{C,\gamma_f,\gamma_i}
             \mx{C,\gamma_f,\gamma_i}{\Hint}{d_i,\gamma_i}}
    {-\w+\frac{\w^2}{2\md}-B-\EC-\frac{\PCsq}{2\mC}}\right\}.
\label{eq:disp}
\end{align}
The amplitude is constructed by writing down (from right to left) the 
deuteron in the initial state, the interaction Hamiltonian which describes 
the coupling of the incoming (outgoing) photon, intermediate states $C$ 
which are summed over, another interaction Hamiltonian and the final-state 
deuteron. In the denominator, the energy of the 
intermediate nucleons appears. Note that~-- as in 
Chapter~\ref{chap:perturbative}~-- we calculate the amplitudes in the 
$\gamma d$-cm frame. Therefore, the incoming and outgoing photons have the 
same energy. The energies appearing in the intermediate state of the 
$s$-channel diagram (the first diagram on both sides of Fig.~\ref{fig:disp}, 
corresponding to the first amplitude in Eq.~(\ref{eq:disp})) are as follows:
\begin{itemize}
\item the photon energy $+\w$
\item the deuteron energy $-B$ with $B\approx 2.225$~MeV denoting the 
deuteron binding energy
\item the excitation energy of the intermediate state $C$, $-\EC$
\item the kinetic energy of the incoming deuteron $+\frac{\w^2}{2\md}$ with 
$m_d=2m_N-B$ the deuteron mass (in the $\gamma d$-cm frame, 
$\vec{P}_i=-\vec{k}_i$ with $\vec{P}_i$ ($\vec{k}_i$) the initial deuteron 
(photon) momentum)
\item the kinetic energy $-\frac{\PCsq}{2\mC}$  of the intermediate 
two-nucleon system (for our numerical evaluations we use $\mC=2m_N$)
\end{itemize}
In the $u$-channel, the same energy denominator appears, except for the 
replacement $\w\rightarrow -\w$. As we calculate in the cm frame of the 
$\gamma d$ system, we have $\frac{\PCsq}{2\mC}=0$ in the $s$-channel diagram, 
whereas $\PC=-\ki-\kf$ in the $u$-channel, i.e. 
$\frac{\PCsq}{2\mC}=\frac{\w^2}{m_C}\,(1+\cos\theta)$. Therefore, an amplitude 
containing $\PCsq$ always corresponds to a $u$-channel diagram throughout this 
work.

The interaction Hamiltonian is given by
\be
\Hint=-\int\vec{J}\ofxi\cdot\vec{A}\ofxi\,d^3\xi.
\label{eq:Hint}
\ee
For the photon field $\vec{A}\ofxi$,
we use the multipole expansion derived in \cite{Karakowski} in analogy to 
Chapter~7 of Ref.~\cite{Rose}, see also   
Appendix~\ref{app:multipoleexpansion}. The result is 
\begin{align}
\label{eq:multipoleexp}
\hat{\epsilon}_\lambda\,\e^{i\vec{k}\cdot\vec{\xi}}&=
\sum_{L=1}^\infty\sum_{M=-L}^L
\wignerd{L}{M}{\lambda}\,i^L\,\sqrt{\frac{2\pi\,(2L+1)}{L\,(L+1)}}\\
&\times\left\{-\frac{i}{\w}\,\nab_\xi\,\psi_L(\w\xi)\,Y_{L\,M}(\hat{\xi})-
i\,\w\,\vec{\xi}\,j_L(\w\xi)\,Y_{L\,M}(\hat{\xi})-
\lambda\,\vec{L}\,Y_{L\,M}(\hat{\xi})\,j_L(\w\xi)\right\},\nonumber
\end{align}
cf. Eq.~(\ref{eq:multipoleexpfinal}). Note that in this chapter we use 
photon polarization-vectors given in the spherical basis, 
Eq.~(\ref{eq:spherpolarizations}), which 
we denote by $\hat{\epsilon}$, whereas so far we used Cartesian polarization
vectors, written as $\vec{\epsilon}$.
The functions $\wignerd{L}{M}{\lambda}$ are part of 
the Wigner $D$-functions, as explained in detail in 
Appendix~\ref{app:multipoleexpansion}, $Y_{L\,M}$ denote the spherical 
harmonics and $\psi_L(\w r)=(1+r\,\frac{d}{dr})\,j_L(\w r)$ with the spherical 
Bessel functions $j_L(z)$ defined in Eq.~(\ref{eq:spherBesseldefinition}).
In the static (long-wavelength) limit only the gradient term in 
Eq.~(\ref{eq:multipoleexp}) survives, as for $\w\rightarrow0$, 
$j_   1(\w r)\rightarrow\frac{1}{3}\,\w r$ and 
$\psi_1(\w r)\rightarrow\frac{2}{3}\,\w r$.
This term will turn out to be the dominant part of the photon field for 
all energies under consideration. Therefore, we define two scalar functions
\begin{align}
\label{eq:phidefinition}
\phi_i(\vec{r})&=-\sum_{L=1}^\infty\sum_{M=-L}^L\delta_{M,\lambda_i}\,
\frac{i^{L+1}}{\w}\,\sqrt{\frac{2\pi\,(2L+1)}{L\,(L+1)}}\,\psi_L(\w r)\,
Y_{L\,M}(\hat{r}),\nonumber\\
\phi_f(\vec{r})&= \sum_{L'=1}^\infty\sum_{M'=-L'}^{L'}(-1)^{L'-\lambda_f}\,
\wignerd{L'}{M'}{-\lambda_f}\,\frac{i^{L'+1}}{\w}\,\sqrt{\frac{2\pi\,(2L'+1)}
{L'\,(L'+1)}}\,\psi_{L'}(\w r)\,Y_{L'\,M'}(\hat{r}),\nonumber\\
\end{align}
which allow us to write
\begin{align}
\label{eq:gradphi}
\eps  \,\e^{ i\vec{k}_i\cdot\vec{\xi}}&\approx\nab\phi_i\ofxi,\nonumber\\
\epspr\,\e^{-i\vec{k}_f\cdot\vec{\xi}}&\approx\nab\phi_f\ofxi,
\end{align}
cf. Eqs.~(\ref{eq:multipoleexpin}) and (\ref{eq:multipoleexpout}). The 
largest contributions to Eq.~(\ref{eq:disp}) are thus the ones where we replace
$\vec{A}\ofxi\rightarrow\nab\phi\ofxi$ in both interaction Hamiltonians. 
Further terms, where the replacement 
$\vec{A}\ofxi\rightarrow\nab\phi\ofxi$ is made only once, are calculated in 
Appendices~\ref{app:subleading} and \ref{app:two-body}. Only a few 
combinations of interactions without the gradient part of $\vec{A}$ give 
sizeable contributions. These are also calculated 
in Appendices~\ref{app:subleading} and \ref{app:two-body}. 

In this section, however, we calculate Eq.~(\ref{eq:disp}) with 
$\Hint\rightarrow-\int\vec{J}\ofxi\cdot\nab\phi\ofxi\,d^3\xi$ simultaneously 
at both vertices, i.e. we restrict ourselves to the terms arising from 
minimal coupling.
In order to simplify the calculation on one hand, and to ensure gauge 
invariance and the correct Thomson limit on the other, we integrate by parts 
and use current conservation:
\ba
-\int\vec{J}\ofxi\cdot\nab\phi\ofxi\,d^3\xi&=
 \int\nab\cdot\vec{J}\ofxi\,\phi\ofxi\,d^3\xi\label{eq:pI}\\
\nab\cdot\vec{J}\ofxi&=-\frac{\partial\rho\ofxi}{\partial t}=
-i\left[H,\rho\ofxi\right]
\label{eq:continuityequation}
\end{align}
The fact that we only need to know the charge density $\rho$ in order to 
calculate the amplitude in the long-wavelength limit is referred to as 
``Siegert's theorem''~\cite{Siegert}.
For $\rho(\vec{\xi})$ one can find in Section~8.3 of Ref.~\cite{Ericson} 
the general separation
\be
\rho\ofxi=\rho^{(0)}\ofxi+\rho^{\mathrm{ex}}(\vec{\xi};\vec{x}_1,\vec{x}_2)
\label{eq:chargedensity}
\ee
with $\rho^{(0)}$ the charge density of the two nucleons and 
$\vec{x}_1,\;\vec{x}_2$ the position of nucleon 1 and 2, respectively.
$\rho^{\mathrm{ex}}(\vec{\xi};\vec{x}_1,\vec{x}_2)$ is the charge 
density due to meson-exchange currents.
The dominant term in Eq.~(\ref{eq:chargedensity}) is 
\be
\rho^{(0)}\ofxi=\sum_{j=n,p}e_j\,\delta(\vec{\xi}-\vec{x}_j)=
e\,\delta(\vec{\xi}-\vec{x}_p),
\label{eq:rhonull}
\ee
which is the only non-vanishing contribution to $\rho\ofxi$ in the static 
limit (``Siegert's hypothesis''~\cite{Siegert}). 
Note that the $\delta$-functions in Eq.~(\ref{eq:rhonull}) indicate
that the two nucleons are treated as pointlike particles, i.e. we do not 
introduce any form factors as e.g. the authors of Ref.~\cite{Lvov}.
We also performed calculations including 
$\rho^{\mathrm{ex}}(\vec{\xi};\vec{x}_1,\vec{x}_2)$. 
From these investigations, which are 
reported in Appendix~\ref{app:rhoex}, we conclude that $\rho^{\mathrm{ex}}$ is 
negligible in the energy range considered. Therefore in the following we are 
only concerned with $\rho^{(0)}\ofxi$. 

\begin{figure}[!htb]
\begin{center}
\includegraphics*[width=.55\linewidth]{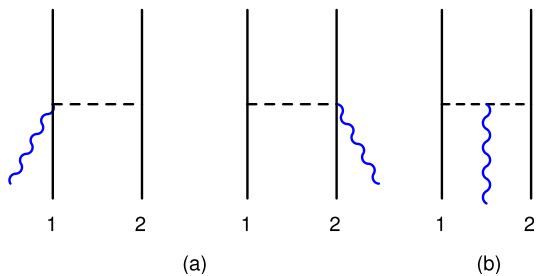}
\caption[Meson-exchange currents]
{One-pion-exchange currents contributing to our calculation: 
the ``Kroll-Ruderman current'' (a) and the ``pion-pole current'' (b).}
\label{fig:mesonexchange}
\end{center}
\end{figure}

\begin{figure}[!htb]
\begin{center} 
\parbox[m]{.168\linewidth}{
\includegraphics*[width=\linewidth]{dispuncrossed.eps}}
\hspace{.5cm}
\parbox[m]{.171\linewidth}{
\includegraphics*[width=\linewidth]{dispcrossed.eps}}
$\;\longrightarrow\;$
\parbox[m]{.098\linewidth}{
\includegraphics*[width=\linewidth]{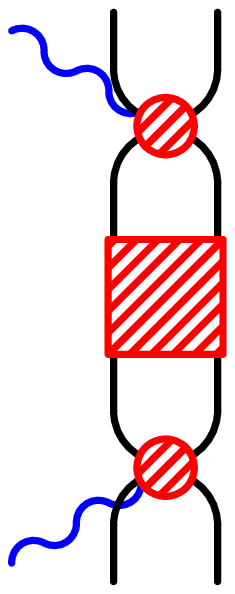}}
\hspace{.5cm}
\parbox[m]{.110\linewidth}{
\includegraphics*[width=\linewidth]{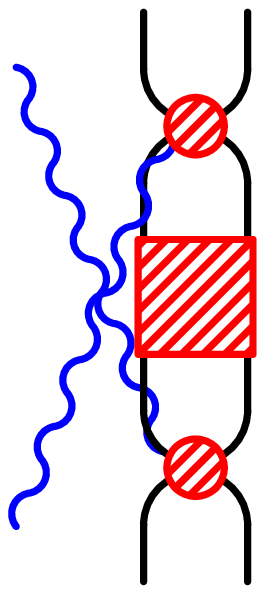}}
\parbox{1.\textwidth}{
\caption[Pole diagrams with possible vertex corrections due to pion exchange]
{Sketch of the diagrams arising from the replacement 
$\vec{A}\rightarrow \nab\phi$ at both vertices simultaneously. The blobs 
denote coupling to the one-body current and possible one-pion exchange, 
the square symbolises the $np$-potential.}
\label{fig:blobs}}
\end{center}
\end{figure}

Albeit it is not obvious, meson-exchange currents (cf. 
Fig.~\ref{fig:mesonexchange}) are also implicitly included in the 
calculation, as by the continuity equation~(\ref{eq:continuityequation})
\be 
\nab\cdot\vec{J}^{\mathrm{ex}}=
-i\left[V^{\mathrm{ex}}\,\vec{\tau}_1\cdot\vec{\tau}_2,\rho^{(0)}\right]
-i\left[H,\rho^{\mathrm{ex}}\right]
\label{eq:implicit}
\ee
with the $np$-potential from one-pion exchange 
$V^{\mathrm{ex}}$~\cite{Ericson}, which is part of the Hamiltonian $H$.
$\vec{\tau}_i$ is the isospin operator of the $i$th nucleon. 
Therefore, the diagrams sketched in Fig.~\ref{fig:disp}, which show only
one-body currents, are to be replaced by Fig.~\ref{fig:blobs}, which includes 
the two-body exchange currents, Fig.~\ref{fig:mesonexchange}.
The $np$-potential is again sketched as a square.

Substituting $\nab\cdot\vec{J}$ by $-i\left[H,\rho^{(0)}\right]$ in 
Eq.~(\ref{eq:pI}), the integral 
over the dummy variable $\vec{\xi}$  can easily be performed to yield
\be
\Hint\approx-\int\vec{J}\ofxi\cdot\nab\phi\ofxi\,d^3\xi=
-i\left[H,e\,\phi(\vec{x}_p)\right],
\label{eq:commutator}
\ee
where $H$ is the full Hamiltonian of the $np$-system
\be
H=\frac{\vec{p}_p^{\,\,2}}{2m_p}+\frac{\vec{p}_n^{\,\,2}}{2m_n}+V
\label{eq:fullHamiltonian}
\ee
with the $np$-potential $V$ including $V^\mathrm{ex}$, cf. 
Eq.~(\ref{eq:implicit}).
With $\vec{p}_p=-i\nab_{x_p}$ and 
$\left.\nab_{x_p}\!\!\!\!\!\!\right.^2\,\phi(\vec{x}_p)\approx 0$ for 
real photons (cf. Eq.~(\ref{eq:gradphi})), 
the part of the commutator containing the proton kinetic energy yields 
\be
H^{\mathrm{int},p}=
-i\left[\frac{\vec{p}_p^{\,\,2}}{2m_p},e\,\phi(\vec{x}_p)\right]
=-\frac{e}{m_p}\,\nab_{x_p}\phi(\vec{x}_p)\cdot\vec{p}_p
\approx-\frac{e}{m_p}\,\vec{A}(\vec{x}_p)\cdot\vec{p}_p.
\ee
This approximate relation becomes exact in the extreme long-wavelength limit, 
where only the gradient part of $\vec{A}$ contributes,
as explained after Eq.~(\ref{eq:multipoleexp}). Therefore, the 
operators responsible for electric dipole $(E1)$ transitions are contained in 
the commutators of $H$ and $\phi$. 
In order to evaluate the commutator 
(\ref{eq:commutator}), we now switch to cm variables, i.e. 
\be
\vec{p}=\frac{\vec{p}_p-\vec{p}_n}{2},\;\;\vec{P}=\vec{p}_p+\vec{p}_n,\;\;
\vec{r}=\vec{x}_p-\vec{x}_n,\;\;\vec{R}=\frac{\vec{x}_p+\vec{x}_n}{2}.
\label{eq:cmvariables}
\ee
Omitting the $np$-potential $V$ for the moment we find
\ba
\Hint_\mathrm{kin}&=-i\left[\frac{\vec{p}_p^{\,\,2}}{2m_p}+
\frac{\vec{p}_n^{\,\,2}}{2m_n},
e\,\phi(\vec{x}_p)\right]=
-i\left[\frac{\vec{p}_p^{\,\,2}}{2m_p},e\,\phi(\vec{x}_p)\right]\nonumber\\
&=-\frac{e}{m_p}\,\nab_{x_p}\phi(\vec{x}_p)\cdot
\left(\vec{p}+\vec{P}/2\right).
\label{eq:pplusPhalf}
\end{align}
The term including the total momentum operator $\vec{P}$ is a recoil correction
and has been calculated in~\cite{Karakowski}. Our evaluation of this term 
showed that its contribution to the cross section is invisibly small.
Therefore we neglect such corrections throughout this work. 

The cm coordinate 
$\vec{R}$ only gives rise to a momentum-conserving $\delta$-function, which 
has to be separated off the scattering amplitude, cf.~\cite{Karakowski}. 
Therefore we may set $\vec{R}=\vec{0}$, i.e. we neglect the cm velocity of the 
two nucleons,
which means that $\vec{x}_p=\frac{\vec{r}}{2}$, $\vec{x}_n=-\frac{\vec{r}}{2}$.
Note that this procedure is consistent with neglecting the recoil corrections 
in Eq.~(\ref{eq:pplusPhalf}). Now we can simplify this equation further:
\ba
-\frac{e}{m_p}\,\nab_{x_p}\phi(\vec{x}_p)\cdot\vec{p}&=
-\frac{e}{m_p}\,\left(\nab_{x_p}-\nab_{x_n}\right)\phi(\vec{x}_p)\cdot\vec{p}
\nonumber\\
&=-i\frac{e}{m_p}\,2\vec{p}\,\phi(\vec{r}/2)\cdot\vec{p}=
-i\left[\frac{\vec{p}^{\,2}}{m_p},e\,\phi(\vec{r}/2)\right]
\label{eq:replacexpxn}
\end{align} 
Defining the ``internal'' Hamiltonian 
\be
H^{np}=\frac{\vec{p}^{\,2}}{m_p}+V,
\label{eq:Hinternal}
\ee
which is identical to Eq.~(\ref{eq:fullHamiltonian}) under the assumption 
$\vec{p}_p=-\vec{p}_n$, we end up with 
\be
\Hint=-i\left[H^{np},e\,\phi(\vec{r}/2)\right].
\label{eq:equivalent}
\ee
This expression is equivalent to Eq.~(\ref{eq:commutator}), neglecting the 
cm motion of the deuteron. Inserting the commutator~(\ref{eq:equivalent}) 
into Eq.~(\ref{eq:disp}) 
and defining $\phiihat=e\,\phi_i(\vec{r}/2)$, $\phifhat=e\,\phi_f(\vec{r}/2)$ 
in analogy to~\cite{Karakowski}
we get (remember, $\frac{\PCsq}{2\mC}=0$ in the $s$-channel diagram)
\ba
\Mfi{\phi\phi}&=-\sum_C\left
     \{\frac{\mx{d_f}{\left[H^{np},\phifhat\right]}{C}
             \mx{C}{\left[H^{np},\phiihat\right]}{d_i}}
    { \w+\frac{\w^2}{2\md}-B-\EC}\right.\nonumber\\
      &+\left.
       \frac{\mx{d_f}{\left[H^{np},\phiihat\right]}{C}
             \mx{C}{\left[H^{np},\phifhat\right]}{d_i}}
    {-\w+\frac{\w^2}{2\md}-B-\EC-\frac{\PCsq}{2\mC}}\right\}.
\end{align}
In order to keep track of the various combinations of interaction 
Hamiltonians we have labelled the double-$\phi$ transition matrix 
'$\phi\phi$'; the photon states have been skipped for brevity. 
Now the commutators are expanded and, as $\ket{d_{i,f}}$, $\ket{C}$ are 
eigenstates of $H^{np}$, we can act with $H^{np}$ on these states. Looking 
only at the $s$-channel term for the moment we find that 
\ba
\Mfi{\phi\phi,s}&=-\sum_C\left
     \{\frac{1}{2}\frac{\mx{d_f}{(-B-\EC)\,\phifhat}{C}
             \mx{C}{\left[H^{np},\phiihat\right]}{d_i}}
    { \w+\frac{\w^2}{2\md}-B-\EC}\right.\nonumber\\
      &+\left.
       \frac{1}{2}\frac{\mx{d_f}{\left[H^{np},\phifhat\right]}{C}
             \mx{C}{(\EC+B)\,\phiihat}{d_i}}
    { \w+\frac{\w^2}{2\md}-B-\EC}\right\}.
\end{align}
Now we add and subtract terms in order to perform some cancellations against 
the denominator. In the resulting expressions without energy denominator, the 
sum over $C$ can be collapsed. In the other terms, the remaining commutator is 
expanded and again some cancellations against the energy denominator are made.
We find 
\ba
\label{eq:Mfiphiphis}
\Mfi{\phi\phi,s}&=\frac{1}{2}\mx{d_f}
{\left[H^{np},\phifhat\right]\,\phiihat}{d_i}-
\frac{1}{2}\mx{d_f}{\phifhat\,\left[H^{np},\phiihat\right]}{d_i}
\\
&-\left(\w+\frac{\w^2}{2\md}\right)\mx{d_f}{\phifhat\,\phiihat}{d_i}+
\left(\w+\frac{\w^2}{2\md}\right)^2
\sum_C\frac{\mx{d_f}{\phifhat}{C}\mx{C}{\phiihat}{d_i}}
{\w+\frac{\w^2}{2\md}-B-\EC}.\nonumber
\end{align}
The $u$-channel amplitude is derived via 
$\w+\frac{\w^2}{2m_d}\rightarrow -\w-\frac{\PCsq}{2\mC}+\frac{\w^2}{2\md}$, 
$\phiihat\leftrightarrow\phifhat$ as 
\ba
\label{eq:Mfiphiphiu}
&\Mfi{\phi\phi,u}=\frac{1}{2}\mx{d_f}
{\left[H^{np},\phiihat\right]\,\phifhat}{d_i}-
\frac{1}{2}\mx{d_f}{\phiihat\,\left[H^{np},\phifhat\right]}{d_i}
\nonumber\\
&+\left(\w+\frac{\PCsq}{2\mC}-\frac{\w^2}{2\md}\right)
\mx{d_f}{\phiihat\,\phifhat}{d_i}\nonumber\\
&+\left(\w+\frac{\PCsq}{2\mC}-\frac{\w^2}{2\md}\right)^2
\sum_C\frac{\mx{d_f}{\phiihat}{C}\mx{C}{\phifhat}{d_i}}
{-\w-\frac{\PCsq}{2\mC}+\frac{\w^2}{2\md}-B-\EC}.
\end{align}
Adding the amplitudes (\ref{eq:Mfiphiphis}) and (\ref{eq:Mfiphiphiu}) we get
($\phiihat\,\phifhat=\phifhat\,\phiihat$)
\ba
\label{eq:Mfiphiphi1added}
\Mfi{\phi\phi 1}&=\left(\w+\frac{\w^2}{2\md}\right)^2
\sum_C\frac{\mx{d_f}{\phifhat}{C}\mx{C}{\phiihat}{d_i}}
{\w+\frac{\w^2}{2\md}-B-\EC},\\
\label{eq:Mfiphiphi2added}
\Mfi{\phi\phi 2}&=\left(\w+\frac{\PCsq}{2\mC}-\frac{\w^2}{2\md}\right)^2
\sum_C\frac{\mx{d_f}{\phiihat}{C}\mx{C}{\phifhat}{d_i}}
{-\w-\frac{\PCsq}{2\mC}+\frac{\w^2}{2\md}-B-\EC},\\
\label{eq:Mfiphiphi3added}
\Mfi{\phi\phi 3}&=\left(\frac{\PCsq}{2\mC}-\frac{\w^2}{\md}\right)
\mx{d_f}{\phifhat\,\phiihat}{d_i},\\
\label{eq:Mfiphiphi4added}
\Mfi{\phi\phi 4}&=\frac{1}{2}
\mx{d_f}{\left[\left[H^{np},\phiihat\right],\phifhat\right]+
         \left[\left[H^{np},\phifhat\right],\phiihat\right]}{d_i}.
\end{align}
These four amplitudes have already been derived in~\cite{Karakowski}, albeit 
in the lab frame.
$\Mfi{\phi\phi 4}$ is the only one which contributes 
in the static limit, since for $\w=0$ also $\PCsq=0$. 
It is responsible for the correct low-energy behaviour of the calculation, as 
will be discussed in great detail in Section~\ref{sec:Thomson2}.

$\Mfi{\phi\phi 1}$ will be calculated first. Defining the shortcut 
$E_0\equiv\w+\frac{\w^2}{2\md}-B$ 
and neglecting prefactors for the moment we can write this amplitude as 
\be
\Mfi{\phi\phi 1}\propto\sum_C\frac{\mx{d_f}{\psi_{L'}\,Y_{L'\,M'}}{C}
\mx{C}{\psi_L\,Y_{L\,M}}{d_i}}{E_0-\EC},
\ee
where we suppressed the sums over $L,M$ and $L',M'$ for brevity, cf. 
Eq.~(\ref{eq:phidefinition}).
Each wave function can be separated into a radial part, denoted by the index 
'$\mathrm{rad}$', and an angular part, denoted by a hat. 
Furthermore, $\ket{C}$ is an 
eigenstate to $H^{np}$ with eigenvalue $\EC$, so we can write 
\be
\Mfi{\phi\phi 1}\propto\sum_{C_\mathrm{rad}\,\hat{C}}
\mx{d_{f\mathrm{rad}}\,\hat{d}_f}{\psi_{L'}\,Y_{L'\,M'}\,\frac{1}
{E_0-H_{\hat{C}}^{np}}}{C_\mathrm{rad}\,\hat{C}}
\mx{C_\mathrm{rad}\,\hat{C}}{\psi_L\,Y_{L\,M}}{d_{i\,\mathrm{rad}}\,\hat{d}_i}.
\ee
$\hat{C}$ is used as a shorthand notation for all angular quantum numbers of 
the intermediate state, i.e. \mbox{$\ket{\hat{C}}=\ket{L_C\,S_C\,J_C\,M_C}$}. 
After acting with the Hamiltonian on the angular state $\ket{\hat{C}}$, which 
will be explained in detail in Eq.~(\ref{eq:HhatCnp}), $H^{np}$ 
depends on the quantum numbers of the intermediate state $\ket{C}$. However, 
the only differential operator in $H^{np}$ is a radial one. Therefore we 
may separate the radial from the angular part of $\ket{C}$. Finally we insert 
two complete sets of radial states $\ket{r}$ and $\ket{r'}$ to get 
\ba
\label{eq:separation}
\Mfi{\phi\phi 1}&\propto\sum_{C_\mathrm{rad}\,\hat{C}}
\int_0^\infty\int_0^\infty r^2 dr\,r'^2 dr'\mx{\hat{d}_f}{Y_{L'\,M'}}{\hat{C}}
\mx{d_{f\mathrm{rad}}}{\psi_{L'}}{r'}\\
&\times\mx{r'}{\frac{1}{E_0-H_{\hat{C}}^{np}}}{C_\mathrm{rad}}
\mxemp{C_\mathrm{rad}}{r}\mx{r}{\psi_L}{d_{i\,\mathrm{rad}}}
\mx{\hat{C}}{Y_{L\,M}}{\hat{d}_i}.\nonumber
\end{align}
Now we replace the deuteron wave function by the position-space
expression given in Eq.~(\ref{eq:deuteronwavefunction}), 
$\Psi_{1m}(\vec{r}\,)=\sum_{l=0,2}\frac{u_l(r)}{r}\mathcal{Y}_m^{l11}(\hat{r})$
with $u_0(r)\equiv u(r)$ and $u_2(r)\equiv w(r)$, see e.g. Section~3.4 
of~\cite{Ericson}. 
The indices $l11$ of the angular wave functions $\mathcal{Y}$ denote 
orbital angular momentum, spin and total angular momentum of the deuteron 
state, $m\in\{-1,0,1\}$ is the projection of the total angular momentum 
of the deuteron onto the quantization axis. Using this notation we can write 
Eq.~(\ref{eq:separation}) as
\ba
\label{eq:Mfiwithint}
\Mfi{\phi\phi 1}&\propto\sum_{\hat{C}}\sum_{l=0,2}\sum_{l'=0,2}\,
\mx{l'\,1\,1\,M_f}{Y_{L'\,M'}}{\hat{C}}\mx{\hat{C}}{Y_{L\,M}}{l\,1\,1\,M_i}\\
&\times\doubleint r dr\,r' dr' u_{l'}(r')\,\psi_{L'}(\frac{\w r'}{2})\,
\mx{r'}{\frac{1}{E_0-H_{\hat{C}}^{np}}}{r}\,
\psi_L(\frac{\w r}{2})\,u_l(r),\nonumber
\end{align}
where we have removed the sum over $C_\mathrm{rad}$. Integrals without limits 
are always integrated from $0$ to infinity throughout this work. 

We now have to evaluate the double integral in Eq.~(\ref{eq:Mfiwithint}), 
including the Green's function 
\be
\green=\mx{r'}{\frac{1}{E_0-H_{\hat{C}}^{np}}}{r}.
\label{eq:greendefinition}
\ee
However, we need to evaluate the integral for arbitrary 
functions of $r$. Therefore we describe how to calculate 
\be
\mathcal{I}_{fi}^{ll'\hat{C}}=
\doubleint r dr\,r' dr'\,u_{l'}(r')\,J_f(r')\,\green\,J_i(r)\,u_l(r).
\label{eq:doubleint}
\ee
We do so in two steps and define
\be
\chi_f^{l'\hat{C}}(r)\equiv\int r' dr'\,u_{l'}(r')\,J_f(r')\,\green.
\ee
Once we have solved this first part of the integral, it is easy to numerically
calculate the remaining integral
\be
\mathcal{I}_{fi}^{ll'\hat{C}}=
\int r dr\,u_l(r)\,J_i(r)\,\chi_f^{l'\hat{C}}(r).
\ee
In order to find the function $\chi_f^{l'\hat{C}}(r)$~-- in the following we 
use the abbreviation $\chi_{\hat{C}}(r)$ for simplicity~-- we first note that 
\be
\left(E_0-H_{\hat{C}}^{np}\right)\,\green=\mxemp{r'}{r}=
\frac{\delta(r'-r)}{r^2}.
\label{eq:Greensequation}
\ee
Eq.~(\ref{eq:Greensequation}) defines the Green's function corresponding to 
Schr\"odinger's equation with a central potential and
the Hamiltonian 
\be
H=\frac{\vec{p}^{\,2}}{2m}+V(r)=-\frac{1}{2m}\,
\left(\frac{1}{r}\frac{d^2}{dr^2}r\right)+\frac{\Lsq}{2m\,r^2}+V(r),
\ee
where $\vec{L}$ denotes the orbital angular momentum operator. 
Note that the neutron-proton potential
contains a tensor part and 
therefore not only depends on the distance $r$ but on the vector $\vec{r}$.
The tensor force mixes e.g. the deuteron $s$- and $d$-states in 
Schr\"odinger's equation. Nevertheless, on the level of the Green's function 
this matrix equation decouples, cf. Section~3.7 of~Ref.~\cite{Ericson}. 
Therefore, we may replace the Hamiltonian in Eq.~(\ref{eq:Greensequation}) by 
\be
H_{\hat{C}}^{np}=\frac{\vec{p}^{\,2}}{m_p}+V_{\hat{C}}(r)=-\frac{1}{m_p}\,
\left(\frac{1}{r}\frac{d^2}{dr^2}r\right)+\frac{L_C\,(L_C+1)}{m_p\,r^2}+
V_{\hat{C}}(r),
\label{eq:HhatCnp}
\ee
where we explicitly show the dependence of the potential on the quantum 
numbers of the interim state $\ket{C}$. The decoupling of 
Eq.~(\ref{eq:Greensequation}) guarantees that only the diagonal terms 
of the tensor force contribute. Therefore, the orbital angular momentum is well
defined, which allows us to replace $\vec{L}^2\rightarrow L_C\,(L_C+1)$ in 
Eq.~(\ref{eq:HhatCnp}). 
Usually we use the AV18 potential as published
in~\cite{AV18}, one of the modern phenomenological high-precision potentials.
An explicit expression for this potential is given in Appendix~\ref{app:AV18}.
In Section~\ref{sec:potentialdep}, however, we compare to results  
with the LO chiral potential, and we demonstrate that our calculation
is rather insensitive to this choice.

Eqs.~(\ref{eq:Greensequation}) and (\ref{eq:HhatCnp}) combine to 
\be
\left[E_0+\frac{1}{m_p}\frac{d^2}{dr^2}-
\frac{L_C\,(L_C+1)}{m_p\,r^2}-V_{\hat{C}}(r)\right]\,r\,\green=
\frac{\delta(r'-r)}{r},
\label{eq:radialoperator}
\ee
which e.g. in the deuteron case reduces to the two differential equations
\be
\left[E_0+\frac{1}{m_p}\frac{d^2}{dr^2}-V_\mathrm{cent}(r)\right]
\,r\,G_{0}(r,r';E_0)=\frac{\delta(r'-r)}{r},
\label{eq:Greens}
\ee
\be
\left[E_0+\frac{1}{m_p}\frac{d^2}{dr^2}-
\frac{6}{m_p\,r^2}-V_\mathrm{cent}(r)+2\,V_\mathrm{ten}(r)\right]
\,r\,G_{2}(r,r';E_0)=\frac{\delta(r'-r)}{r}
\label{eq:Greend}
\ee
with $V_\mathrm{cent}(r)$ the central part and 
$V_\mathrm{ten}(r)$ the tensor part of the potential. 
The indices of the Green's functions in 
Eqs.~(\ref{eq:Greens},~\ref{eq:Greend}) 
reflect the orbital angular momentum state, whereas $J=1,\;S=1$ is not written 
down explicitly. The factor $-2$ in front of $V_\mathrm{ten}(r)$ in 
Eq.~(\ref{eq:Greend}) can be read off Table~\ref{tab:Sij}.

Returning to the calculation of  the double-integral~(\ref{eq:doubleint}), 
we act now with the operator given in square brackets in 
Eq.~(\ref{eq:radialoperator}) on $\chi_{\hat{C}}(r)$.
The integral over $r'$ collapses due to the $\delta$-function and we find 
\be
\left[\frac{d^2}{dr^2}+m_p\,\left(E_0-V_{\hat{C}}(r)\right)-
\frac{L_C\,(L_C+1)}{r^2}\right]\,r\,\chi_{\hat{C}}(r)=m_p\,u_{l'}(r)\,J_f(r).
\label{eq:diffeq}
\ee
This is a second-order differential equation in $r$ with an inhomogeneity, 
which can be interpreted as a source term. Its solutions are real for 
$E_0<0$ and complex for $E_0>0$. The latter case corresponds to $\w>B$, i.e. 
the photon carries enough energy to break up the deuteron into its two 
constituents. Whenever such a new channel opens, an imaginary part starts to 
exist. Another example of an amplitude becoming complex at a 
particle-production threshold is the single-nucleon Compton amplitude for 
$\w\sim m_\pi$, cf.~Section~\ref{sec:polarizabilities1}. There, the imaginary 
part signals that the photon energy suffices to put a pion on-shell; here it 
indicates that there is enough energy to 
split the deuteron into two free nucleons. Like in 
Section~\ref{sec:polarizabilities1}, such an imaginary part  appears 
only in the $s$-channel diagrams, where the incoming photon is absorbed 
before the other one is emitted. In Section~\ref{sec:photodisintegration} we 
will use the imaginary part of the amplitudes to derive total 
deuteron-photodisintegration cross sections via the optical theorem.

For $r\rightarrow\infty,\;$
$u_{l'}(r)J_f(r)\rightarrow 0$ due to $u_{l'}(r)\rightarrow 0$, i.e. 
Eq.~(\ref{eq:diffeq}) reduces to a homogeneous differential equation. 
Furthermore, $V_{\hat{C}}(r)\rightarrow 0$ for 
$r\rightarrow\infty$ because of the finite range of the $NN$-potential, 
see Appendix~\ref{app:AV18}. Therefore, we are for large distances left with
\be
\left[\frac{d^2}{dr^2}+E_0\,m_p-\frac{L_C\,(L_C+1)}{r^2}\right]
\,r\,\chi_{\hat{C}}(r)=0.
\ee 
This equation is known to be solved by a linear combination of the spherical 
Bessel functions of the first and second kind, $j_{L_C}(Qr)$ and $n_{L_C}(Qr)$
with $Q=\sqrt{m_p\,E_0}$, see e.g. Section~17.3 in~\cite{Schwabl}. 
Note that $Q$ can be real
or imaginary\footnote{In order to fix the sign of the 
imaginary solution, one has to add an infinitesimal 
imaginary part to $-B$, i.e. $B\rightarrow B-i\epsilon$. 
The sign of this imaginary part follows from the pole structure of the nucleon 
propagators.}
depending on $E_0$. In our case the boundary condition is that 
$\chi_{\hat{C}}(r)$ must be an outgoing spherical wave for large $r$, cf. e.g.
Section~7.1 of~\cite{Sakurai}. 
Therefore we may write
\be
\lim_{r\rightarrow\infty}\chi_{\hat{C}}(r)\propto h_{L_C}^{(1)}(Qr),
\label{eq:condition}
\ee 
with $h_{L_C}^{(1)}(Qr)$ the spherical Hankel function of the first kind, 
defined as
\be
h_{L_C}^{(1)}(Qr)=j_{L_C}(Qr)+i n_{L_C}(Qr).
\ee
We now have to find solutions for the homogeneous and the inhomogeneous 
differential equation~(\ref{eq:diffeq}). Numerically, this can be easily done. 
As the equation is of second order, we have to specify two initial conditions,
e.g. $\chi_{\hat{C}}(r=0)$ and $\frac{d}{dr}\chi_{\hat{C}}(r)|_{r=0}$.
In order to improve the stability of the numerics routine it is worth 
thinking about proper values for these conditions, which are, however, not to
be confused with the boundary condition~(\ref{eq:condition}), that the 
final solution has to fulfill. For $r\rightarrow0$
the right-hand side of Eq.~(\ref{eq:diffeq}) vanishes due to the wave function
$u_{l'}(r)$. Furthermore, as $V_{\hat{C}}(r)$ is regular at the origin,
cf. Fig.~\ref{fig:AV18plots}, 
$\frac{L_C\,(L_C+1)}{r^2}\gg m_p\,\left(E_0-V_{\hat{C}}(r)\right)$ for $L_C>0$
and sufficiently small $r$. Therefore, for $r\rightarrow 0$ 
Eq.~(\ref{eq:diffeq}) can be approximated by
\be
\left[\frac{d^2}{dr^2}-\frac{L_C\,(L_C+1)}{r^2}\right]\,r\,\chi_{\hat{C}}(r)=0.
\label{eq:diffeqa}
\ee
The general solution of Eq.~(\ref{eq:diffeqa}) is given by 
$r\,\chi_{\hat{C}}(r)=A\,r^{L_C+1}+B\,r^{-L_C}$. As we want 
$\chi_{\hat{C}}(r)$ to be regular at the origin, i.e. 
$r\,\chi_{\hat{C}}(r)=0$ for $r\rightarrow 0$, we find $B=0$ and therefore 
close to the origin $r\,\chi_{\hat{C}}(r)$ is of order $r^{L_C+1}$ plus higher
powers in $r$, which means that the leading contribution to 
$\chi_{\hat{C}}(r)$ is proportional to $r^{L_C}$. Therefore it is advantageous
to choose $\chi_{\hat{C}}(r=0)\neq 0$, 
$\frac{d}{dr}\chi_{\hat{C}}(r)|_{r=0}\neq 0$ for $L_C=0$, 
$\chi_{\hat{C}}(r=0)=    0$, $\frac{d}{dr}\chi_{\hat{C}}(r)|_{r=0}\neq 0$ 
for $L_C=1$ and 
$\chi_{\hat{C}}(r=0)\approx 0$, $\frac{d}{dr}\chi_{\hat{C}}(r)|_{r=0}\approx0$
for higher angular momenta\footnote{The choice $0$ in both cases is 
impossible, as it would give the trivial solution $\chi_{\hat{C}}(r)\equiv0$ 
for the homogeneous equation.}. 

Once we have 
solutions for the homogeneous and the inhomogeneous differential equation, we 
need to find the correct linear combination which satisfies the 
condition~(\ref{eq:condition}). In other words we have to determine the 
coefficient $\lambda$ which fulfills 
\be
\lim_{r\rightarrow\infty}\left\{\chi_{\hat{C}}^\mathrm{in}(r)+
\lambda\,\chi_{\hat{C}}^\mathrm{hom}(r)\right\}\propto h_{L_C}^{(1)}(Qr),
\ee
where $\chi_{\hat{C}}^\mathrm{in}(r)$ ($\chi_{\hat{C}}^\mathrm{hom}(r)$) 
denote the solution to the inhomogeneous (homogeneous) differential equation. 
In the asymptotic limit,  $\chi_{\hat{C}}(r)$ must 
be a linear combination of $j_{L_C}(Qr)$ and $n_{L_C}(Qr)$ or, equivalently, 
of $j_{L_C}(Qr)$ and $h_{L_C}^{(1)}(Qr)$. Therefore we can write the general 
solutions in the following way:
\ba
\label{eq:chihom}
\chi_{\hat{C}}^\mathrm{hom}(r)&=C_{\hat{C}}^\mathrm{hom}(r)\,
\left[j_{L_C}(Qr)+t_{\hat{C}}^\mathrm{hom}(r)\,h_{L_C}^{(1)}(Qr)\right],\\
\chi_{\hat{C}}^\mathrm{in }(r)&=C_{\hat{C}}^\mathrm{in }(r)\,
\left[j_{L_C}(Qr)+t_{\hat{C}}^\mathrm{in }(r)\,h_{L_C}^{(1)}(Qr)\right]
\label{eq:chiin}
\end{align}
with functions $C_{\hat{C}}^\mathrm{in/hom}(r)$, 
$t_{\hat{C}}^\mathrm{in/hom}(r)$ which become the constants 
$C_{\hat{C}}^\mathrm{in/hom}$, $t_{\hat{C}}^\mathrm{in/hom}$ for large $r$. 
With the choice 
$\lambda=-C_{\hat{C}}^\mathrm{in}/C_{\hat{C}}^\mathrm{hom}$ we find
\be
\lim_{r\rightarrow\infty}\left\{\chi_{\hat{C}}^\mathrm{in}(r)+
\lambda\,\chi_{\hat{C}}^\mathrm{hom}(r)\right\}=
C_{\hat{C}}^\mathrm{in}\,
\left(t_{\hat{C}}^\mathrm{in}-t_{\hat{C}}^\mathrm{hom}\right)\,
h_{L_C}^{(1)}(Qr),
\ee
which satisfies the condition~(\ref{eq:condition}). Therefore we need to 
determine the coefficients $C_{\hat{C}}^\mathrm{in}$, 
$C_{\hat{C}}^\mathrm{hom}$. This has to be done in the region where 
$C_{\hat{C}}(r)$, $t_{\hat{C}}(r)$ are constant, i.e. their derivatives 
vanish. In this region
\be
\ln\chi_{\hat{C}}^\mathrm{in/hom}(r)=
\ln C_{\hat{C}}^\mathrm{in/hom}+\ln\left[j_{L_C}(Qr)
+t_{\hat{C}}^\mathrm{in/hom}\,h_{L_C}^{(1)}(Qr)\right].
\ee
Defining $D^\mathrm{in/hom}=\frac{d}{dr}\ln\chi_{\hat{C}}^\mathrm{in/hom}=
\frac{{\chi_{\hat{C}}'}^\mathrm{in/hom}}{\chi_{\hat{C}}^\mathrm{in/hom}}$ 
we find
\be
D^\mathrm{in/hom}=
\frac{d}{dr}\ln\left[j_{L_C}(Qr)+t_{\hat{C}}^\mathrm{in/hom}\,
h_{L_C}^{(1)}(Qr)\right]=
\frac{\frac{d}{dr}j_{L_C}(Qr)+t_{\hat{C}}^\mathrm{in/hom}
\,\frac{d}{dr}h_{L_C}^{(1)}(Qr)}
{j_{L_C}(Qr)+t_{\hat{C}}^\mathrm{in/hom}\,h_{L_C}^{(1)}(Qr)}.
\ee
This equation is easily solved for $t_{\hat{C}}^\mathrm{in/hom}$:
\be
t_{\hat{C}}^\mathrm{in/hom}=\frac{D^\mathrm{in/hom}\,j_{L_C}(Qr)-
\frac{d}{dr}j_{L_C}(Qr)}{\frac{d}{dr}h_{L_C}^{(1)}(Qr)-D^\mathrm{in/hom}\,
h_{L_C}^{(1)}(Qr)}
\ee
Using these results we can solve Eqs.~(\ref{eq:chihom}, \ref{eq:chiin}) for
$C_{\hat{C}}^\mathrm{in/hom}$ and so determine $\lambda$.

Numerically, this is one of the most involved parts of this work. 
Fortunately, a nice 
and valuable cross-check to the routine can be performed. For this we consider
again the double integral to be calculated, Eq.~(\ref{eq:doubleint}). This 
integral is obviously invariant under the interchange $r\leftrightarrow r'$. 
A general feature of Green's functions is 
that they are symmetric under $r\leftrightarrow r'$, i.e. 
$G_{\hat{C}}(r',r;E_0)=\green$~\cite{Feshbach}.
Therefore
\be
\mathcal{I}_{fi}^{ll'\hat{C}}=
\doubleint r' dr'\,r dr\,u_{l'}(r)\,J_f(r)\,\green\,J_i(r')\,u_l(r').
\ee
This expression is identical to $\mathcal{I}_{fi}^{ll'\hat{C}}$, 
Eq.~(\ref{eq:doubleint}), with 
$i\leftrightarrow f$, $l\leftrightarrow l'$, i.e. our results must be 
symmetric under $i\leftrightarrow f$, $l\leftrightarrow l'$. This is a 
non-trivial check, because for $J_f(r)\neq J_i(r)$ completely different 
functions $\chi_f^{l'\hat{C}}(r)$ are generated. Our routine agrees well with 
this symmetry~-- the deviation caused by numerical uncertainties is less than 
1\%. 

Now all tools to calculate $\Mfi{\phi\phi 1,2}$ are prepared. However, as the 
algebraic manipulations are not too complicated, we shift this rather technical
part to Appendix~\ref{app:dominant}. There we also calculate 
$\Mfi{\phi\phi 3}$ and $\Mfi{\phi\phi 4}$.

We turn now to the calculation of those contributions where the replacement 
$\vec{A}\ofxi\rightarrow\nab\phi\ofxi$ (cf. the beginning of this section) is 
made at most once.

\subsection{Subleading Terms \label{sec:subleading}}

So far we only considered contributions arising from minimal coupling of the 
photon field to the two-nucleon system at both vertices. 
In the following we describe how to calculate the amplitudes given in 
Eq.~(\ref{eq:disp}), when the replacement
\be
\Hint=-\int\vec{J}\ofxi\cdot\vec{A}\ofxi\, d^3\xi\rightarrow 
-\int\vec{J}\ofxi\cdot\nab\phi\ofxi\,d^3\xi
\ee
is made only once, again drawing substantially from Ref.~\cite{Karakowski}. 
The term 'subleading' refers to the fact that the resulting 
amplitude is numerically less important than that of 
Section~\ref{sec:dominant}. It is denoted by 
$\Mfi{\phi}$ and follows immediately from Eq.~(\ref{eq:disp}):
\ba
\Mfi{\phi}&=\sum_C\left\{
\frac{\mx{d_f}{\int\vec{J}\ofxi\cdot\nab\phi_f\ofxi\,d^3\xi}{C}
\mx{C}{\int\vec{J}\ofxi\cdot\vec{A}\ofxi\,d^3\xi}{d_i}}{\denoms}\right.
\nonumber\\&+
\frac{\mx{d_f}{\int\vec{J}\ofxi\cdot\vec{A}\ofxi\,d^3\xi}{C}
\mx{C}{\int\vec{J}\ofxi\cdot\nab\phi_i\ofxi\,d^3\xi}{d_i}}{\denoms}
\nonumber\\&+
\frac{\mx{d_f}{\int\vec{J}\ofxi\cdot\nab\phi_i\ofxi\,d^3\xi}{C}
\mx{C}{\int\vec{J}\ofxi\cdot\vec{A}\ofxi\,d^3\xi}{d_i}}{\denomu}
\nonumber\\&+
\left.\frac{\mx{d_f}{\int\vec{J}\ofxi\cdot\vec{A}\ofxi\,d^3\xi}{C}
\mx{C}{\int\vec{J}\ofxi\cdot\nab\phi_f\ofxi\,d^3\xi}{d_i}}{\denomu}\right\}
\end{align}
Now we perform the same steps as described in 
Eqs.~(\ref{eq:commutator}-\ref{eq:Mfiphiphis}), i.e. we first replace 
$\int\vec{J}\ofxi\cdot\nab\phi\ofxi\,d^3\xi$ by 
$i\left[H^{np},e\,\phi(\vec{r}/2)\right]$, then act with $H^{np}$ on $\ket{d}$
and $\ket{C}$, respectively, and finally add and subtract terms in order to 
do some cancellations against the denominator. We find, again 
neglecting recoil terms and the deuteron velocity,
\ba
\label{eq:Mfiphi}
\Mfi{\phi}&=i\sum_C\bigg\{
\mx{d_f}{\phifhat}{C}\mx{C}{\int\vec{J}\ofxi\cdot\vec{A}\ofxi\,d^3\xi}{d_i}
\nonumber\\&-
\bigg(\w+\frac{\w^2}{2m_d}\bigg)\frac{\mx{d_f}{\phifhat}{C}
\mx{C}{\int\vec{J}\ofxi\cdot\vec{A}\ofxi\,d^3\xi}{d_i}}{\denoms}
\nonumber\\&-
\mx{d_f}{\int\vec{J}\ofxi\cdot\vec{A}\ofxi\,d^3\xi}{C}\mx{C}{\phiihat}{d_i}
\nonumber\\&+
\bigg(\w+\frac{\w^2}{2m_d}\bigg)\frac{\mx{d_f}
{\int\vec{J}\ofxi\cdot\vec{A}\ofxi\,d^3\xi}{C}\mx{C}{\phiihat}{d_i}}{\denoms}
\nonumber\\&+
\mx{d_f}{\phiihat}{C}\mx{C}{\int\vec{J}\ofxi\cdot\vec{A}\ofxi\,d^3\xi}{d_i}
\nonumber\\&+
\bigg(\w-\frac{\w^2-\PCsq}{2m_C}\bigg)\frac{\mx{d_f}{\phiihat}{C}
\mx{C}{\int\vec{J}\ofxi\cdot\vec{A}\ofxi\,d^3\xi}{d_i}}{\denomu}
\nonumber\\&-
\mx{d_f}{\int\vec{J}\ofxi\cdot\vec{A}\ofxi\,d^3\xi}{C}\mx{C}{\phifhat}{d_i}
\nonumber\\&-
\bigg(\w-\frac{\w^2-\PCsq}{2m_C}\bigg)\frac{\mx{d_f}
{\int\vec{J}\ofxi\cdot\vec{A}\ofxi\,d^3\xi}{C}\mx{C}{\phifhat}{d_i}}{\denomu}
\bigg\}.
\end{align}
In those terms in which the energy denominator has been cancelled, the sum 
over $C$ may be collapsed. As 
$\hat{\phi}\left(\int\vec{J}\ofxi\cdot\vec{A}\ofxi\,d^3\xi\right)=
\left(\int\vec{J}\ofxi\cdot\vec{A}\ofxi\,d^3\xi\right)\hat{\phi}$, 
these four terms cancel exactly, and only the terms including an energy 
denominator remain.

Now we have to specify the current $\vec{J}\ofxi$ and the relevant parts of 
the photon field $\vec{A}\ofxi$, which are the non-gradient terms in 
Eq.~(\ref{eq:multipoleexp}).
We want to decompose the photon field in its electric and magnetic part. 
Therefore, we write schematically $\vec{A}=\nab\phi+\Aone+\Atwo$ with
\ba
\label{eq:Aone}
\Aone\ofxi&=-\sum_{\vec{k},\lambda=\pm1}\sum_{L=1}^\infty\sum_{M=-L}^L\lambda
\,\sqrt{\frac{2\pi\,(2L+1)}{L\,(L+1)}}\,i^L\,j_L(\w\xi)\,\vec{L}\,
Y_{L\,M}(\hat{\xi})\nonumber\\
&\times\left[a_{\vec{k},\lambda}\,\delta_{M,\lambda}-
a_{\vec{k},\lambda}^\dagger\,(-1)^{L+\lambda}\,\wignerd{L}{M}{-\lambda}\right],
\\
\Atwo\ofxi&=-\sum_{\vec{k},\lambda=\pm1}\sum_{L=1}^\infty\sum_{M=-L}^L
\sqrt{\frac{2\pi\,(2L+1)}{L\,(L+1)}}\,i^{L+1}\,\w\,\vec{\xi}\,j_L(\w\xi)\,
Y_{L\,M}(\hat{\xi})\nonumber\\
&\times\left[a_{\vec{k},\lambda}\,\delta_{M,\lambda}-
a_{\vec{k},\lambda}^\dagger\,(-1)^{L+\lambda}\,\wignerd{L}{M}{-\lambda}\right],
\label{eq:Atwo}
\end{align}
cf. Eqs.~(\ref{eq:multipoleexpin}), (\ref{eq:multipoleexpout}) 
and (\ref{eq:photonfield}). $\Aone$ constitutes the magnetic part of the 
photon field, $\nab\phi+\Atwo$ is the electric field, 
cf. e.g. Chapter~7 of Ref.~\cite{Rose}. 
The operators $a_{\vec{k},\lambda}^\dagger$ ($a_{\vec{k},\lambda}$) create 
(destroy) a photon with momentum $\vec{k}$ and polarization $\lambda$.

Now we specify which currents we include in our calculation. The 
one-body current is considered first. It consists of two parts, which we call 
$\Jsigma$ and $\Jp$, with
\ba
\label{eq:Jsigma}
\Jsigma\ofxi&=\frac{e}{2m_N}\sum_{j=n,p}\left[\nab_\xi\times\mu_j\,
\vec{\sigma}_j\,\delta(\vec{\xi}-\vec{x}_j)\right],\\
\Jp    \ofxi&=\frac{1}{2m_N}\sum_{j=n,p}\left\{
e_j\,\delta(\vec{\xi}-\vec{x}_j),\vec{p}_j\right\},
\label{eq:Jp}
\end{align}
cf. e.g.~\cite{Ericson}, Section~8.2. 
$\mu_j$ is the magnetic moment, $\vec{\sigma}_j$ the 
spin operator and $\vec{p}_j$ the momentum of the $j$th nucleon; 
as in the previous chapter we neglect isospin-breaking 
effects, i.e. we set $m_p\equiv m_n\equiv m_N$. All possible combinations of 
$\Aone,\;\Atwo$ and $\Jsigma,\;\Jp$ have been calculated in \cite{Karakowski}.
We also evaluated all these amplitudes, however we found that only $\Jsigma$ 
gives visible contributions to the deuteron Compton cross sections. 
The reason is that in the amplitudes resulting from the replacement
$\vec{J}\cdot\vec{A}\rightarrow\Jp\cdot\Aone$ in Eq.~(\ref{eq:Mfiphi}),
the leading multipoles of the photon field
$L=L'=1$ are forbidden due to the 
matrix elements involved.
The amplitudes including $\Jp\cdot\Atwo$, on the other
hand, cancel exactly for $L,L'\leq2$ in the $\gamma d$-cm frame. 
Therefore,
we may restrict ourselves to the following combinations: ($\Jsigma,\,\Aone$), 
denoted by $\sigma1$, and ($\Jsigma,\,\Atwo$), denoted by $\sigma2$.

We now calculate $\int\Jsigma\ofxi\cdot\vec{A}^{(1,2)}\ofxi\,d^3\xi$.
We start with the derivation for $\Aone$, writing only the $\xi$-dependent 
terms  for simplicity.
\ba
\int\Jsigma\ofxi\cdot\Aone\ofxi\,d^3\xi&\propto\int\sum_{j=n,p}
\left[\nab_\xi\times\mu_j\,\vec{\sigma}_j\,\delta(\vec{\xi}-\vec{x}_j)\right]\,
j_L(\w \xi)\cdot\vec{L}\,Y_{L\,M}(\hat{\xi})\,d^3\xi\nonumber\\
&=\int\sum_{j=n,p}\left[\varepsilon_{ikl}\,\partial_k\,\mu_j\,\sigma_{j,l}\,
\delta(\vec{\xi}-\vec{x}_j)\right]\,
j_L(\w\xi)\,L_i\,Y_{L\,M}(\hat{\xi})\,d^3\xi\nonumber\\
&=-\int\sum_{j=n,p}\left[\varepsilon_{ikl}\,\mu_j\,\sigma_{j,l}\,
\delta(\vec{\xi}-\vec{x}_j)\right]\,\partial_k\,
j_L(\w\xi)\,L_i\,Y_{L\,M}(\hat{\xi})\,d^3\xi\nonumber\\
&=\int\sum_{j=n,p}\mu_j\,\vec{\sigma}_j\,\delta(\vec{\xi}-\vec{x}_j)\cdot
\nab_\xi\times \left(j_L(\w\xi)\,\vec{L}\,Y_{L\,M}(\hat{\xi})\right)\,d^3\xi,
\end{align}
where one partial integration has been performed. Now we evaluate the integral
and afterwards replace 
$\vec{x}_p\rightarrow \frac{\vec{r}}{2},\;
 \vec{x}_n\rightarrow-\frac{\vec{r}}{2}$, as  in 
Eq.~(\ref{eq:replacexpxn}), yielding
\be
\int\Jsigma\ofxi\cdot\Aone\ofxi\,d^3\xi\propto2\left[\nab_r\times 
\left(j_L(\frac{\w r}{2})\,\vec{L}\,Y_{L\,M}(\hat{r})\right)\right]\cdot
\left(\mu_p\,\vec{\sigma}_p-(-1)^L\,\mu_n\,\vec{\sigma}_n\right),
\label{eq:replacexpxnsigma1}
\ee
where we used $\nab_{\pm r/2}=\pm2\nab_r$ and 
Eq.~(\ref{eq:Yofmr}).
By the help of Eq.~(\ref{eq:LonY}), this becomes
\be
\int\Jsigma\ofxi\cdot\Aone\ofxi\,d^3\xi\propto2\sqrt{L\,(L+1)}
\left[\nab_r\times j_L(\frac{\w r}{2})\,\vsh{L}{L}{M}\right]\cdot
\left(\mu_p\,\vec{\sigma}_p-(-1)^L\,\mu_n\,\vec{\sigma}_n\right).
\ee
Now we can use the curl formula (\ref{eq:curlL}) and the recursion relations 
for spherical Bessel functions, Eq.~(\ref{eq:recursionrelations}),
to write
\ba
\int\Jsigma\ofxi\cdot\Aone\ofxi\,d^3\xi&\propto
\left[i\,\w\,j_{L-1}(\frac{\w r}{2})\,\sqrt{\frac{L\,(L+1)^2}{2L+1}}\,
\vsh{L}{L-1}{M}\right.\\
&-\left.i\,\w\,j_{L+1}(\frac{\w r}{2})\,\sqrt{\frac{L^2\,(L+1)}{2L+1}}\,
\vsh{L}{L+1}{M}\right]\cdot
\left(\mu_p\,\vec{\sigma}_p-(-1)^L\,\mu_n\,\vec{\sigma}_n\right).\nonumber
\end{align}
We found that the numerical importance of the various contributions rapidly
decreases with increasing photon multipolarity $L$. A similar observation has 
been made in Chapters~\ref{chap:spinaveraged} and~\ref{chap:spinpolarized} 
for our multipole expansion of single-nucleon Compton scattering. 
Therefore, the term proportional to 
$\vsh{L}{L+1}{M}$ may be neglected. Defining 
$\vec{S}=\frac{\vec{\sigma}_p+\vec{\sigma}_n}{2}$ and 
$\vec{t}=\frac{\vec{\sigma}_p-\vec{\sigma}_n}{2}$
and including all prefactors, we get the result
\ba
\label{eq:intJsigmaAonescalar}
\int\Jsigma&\ofxi\cdot\Aone\ofxi\,d^3\xi=\nonumber\\
&-\sum_{\vec{k},\lambda=\pm1}\sum_{L=1}^\infty\sum_{M=-L}^L\lambda\,
\sqrt{2\pi\,(L+1)}\,\frac{e\,\w}{2m_N}\,i^{L+1}\,j_{L-1}(\frac{\w r}{2})\,
\vsh{L}{L-1}{M}\\
&\times\left[\left(\mu_p-(-1)^L\,\mu_n\right)\,\vec{S}+
              \left(\mu_p+(-1)^L\,\mu_n\right)\,\vec{t}\,\right]
  \cdot \left[a_{\vec{k},\lambda}\,\delta_{M,\lambda}-
a_{\vec{k},\lambda}^\dagger\,(-1)^{L+\lambda}\,\wignerd{L}{M}{-\lambda}\right].
\nonumber
\end{align}
The scalar products are replaced according to the relation
\be
\vsh{J}{L}{M}\cdot\vec{V}=\left[Y_L\otimes V\right]_{J\,M},
\label{eq:TdotV}
\ee
which holds for any vector (rank 1) operator ($\otimes$ denotes the 
irreducible tensor product). An explicit proof of this identity is given in 
Appendix~\ref{app:subleading}.
We use it to finally rewrite Eq.~(\ref{eq:intJsigmaAonescalar}):
\ba
\int\Jsigma\ofxi\cdot\Aone\ofxi\,d^3\xi&=-\sum_{\vec{k},\lambda=\pm1}
\sum_{L=1}^\infty\sum_{M=-L}^L\lambda\,
\sqrt{2\pi\,(L+1)}\,\frac{e\,\w}{2m_N}\,i^{L+1}\,j_{L-1}(\frac{\w r}{2})
\nonumber\\
&\times
\left\{\left(\mu_p-(-1)^L\,\mu_n\right)\,\left[Y_{L-1}\otimes S\right]_{L\,M}+
      \left(\mu_p+(-1)^L\,\mu_n\right)\,\left[Y_{L-1}\otimes t\right]_{L\,M}\,
\right\}\nonumber\\
&\times\left[a_{\vec{k},\lambda}\,\delta_{M,\lambda}-
a_{\vec{k},\lambda}^\dagger\,(-1)^{L+\lambda}\,\wignerd{L}{M}{-\lambda}\right]
\label{eq:intJsigmaAone}
\end{align}

We turn now to the calculation of $\int\Jsigma\ofxi\cdot\Atwo\ofxi\,d^3\xi$.
Again we restrict ourselves in the derivation to the $\xi$-dependent terms,
finding 
\ba
\int\Jsigma\ofxi\cdot\Atwo\ofxi\,d^3\xi&\propto\int\sum_{j=n,p}
\left[\nab_\xi\times\mu_j\,\vec{\sigma}_j\,\delta(\vec{\xi}-\vec{x}_j)\right]
\cdot\vec{\xi}\,j_L(\w\xi)\,Y_{L\,M}(\hat{\xi})\,d^3\xi\nonumber\\
&=\int\sum_{j=n,p}\mu_j\,\vec{\sigma}_j\,\delta(\vec{\xi}-\vec{x}_j)\cdot
\nab_\xi\times\left(\vec{\xi}\,j_L(\w\xi)\,Y_{L\,M}(\hat{\xi})\right)\,d^3\xi
\nonumber\\
&=\nab_r\times\left(\vec{r}\,j_L(\frac{\w r}{2})\,Y_{L\,M}(\hat{r})\right)\cdot
\left(\mu_p\,\vec{\sigma}_p+(-1)^L\,\mu_n\,\vec{\sigma}_n\right),
\label{eq:intJsigmaAtwo}
\end{align}
where we have performed the same steps as in the derivation of 
Eq.~(\ref{eq:replacexpxnsigma1}). From Eq.~(\ref{eq:RHS1}) we know that 
\be
\vec{r}\,j_L(\frac{\w r}{2})\,Y_{L\,M}(\hat{r})=
\sqrt{L\,(L+1)}\,r\,j_L(\frac{\w r}{2})\,
\left[\frac{\vsh{L}{L-1}{M}}{\sqrt{(L+1)\,(2L+1)}}-
      \frac{\vsh{L}{L+1}{M}}{\sqrt{L\,(2L+1)}}\right],
\label{eq:rjYequal}
\ee
which we plug into Eq.~(\ref{eq:intJsigmaAtwo}). Using the curl formulae 
Eqs.~(\ref{eq:curlLp1}) and (\ref{eq:curlLm1}) we find
\be
\nab_r\times\left(r\,j_L(\frac{\w r}{2})\,
\left[\frac{\vsh{L}{L-1}{M}}{\sqrt{(L+1)\,(2L+1)}}-
      \frac{\vsh{L}{L+1}{M}}{\sqrt{L\,(2L+1)}}\right]\right)=
-i\,j_L(\frac{\w r}{2})\,\vsh{L}{L}{M}.
\label{eq:usingcurl}
\ee
Combining Eqs.~(\ref{eq:intJsigmaAtwo}-\ref{eq:usingcurl}), 
together with the definitions of 
$\vec{S}$ and $\vec{t}$, cf. Eq.~(\ref{eq:intJsigmaAonescalar}), yields
\ba
\int\Jsigma\ofxi\cdot\Atwo\ofxi\,d^3\xi&\propto
-i\,\sqrt{L\,(L+1)}\,j_L(\frac{\w r}{2})\,\vsh{L}{L}{M}\nonumber\\
&\cdot\left[(\mu_p+(-1)^L\,\mu_n)\,\vec{S}
                             +(\mu_p-(-1)^L\,\mu_n)\,\vec{t}\right].
\end{align}
Including all prefactors, we get
\ba
\int\Jsigma\ofxi&\cdot\Atwo\ofxi\,d^3\xi=
-\sum_{\vec{k},\lambda=\pm1}\sum_{L=1}^\infty\sum_{M=-L}^L\sqrt{2\pi\,(2L+1)}
\frac{e\,\w}{2m_N}\,i^L\,j_L(\frac{\w r}{2})\vsh{L}{L}{M}\nonumber\\
&\cdot
\left[(\mu_p+(-1)^L\,\mu_n)\,\vec{S}+(\mu_p-(-1)^L\,\mu_n)\,\vec{t}\,\right]
\cdot
\left[a_{\vec{k},\lambda}\,\delta_{M,\lambda}-a_{\vec{k},\lambda}^\dagger\,
(-1)^{L+\lambda}\,\wignerd{L}{M}{-\lambda}\right]\nonumber\\
&=
-\sum_{\vec{k},\lambda=\pm1}\sum_{L=1}^\infty\sum_{M=-L}^L\sqrt{2\pi\,(2L+1)}\,
\frac{e\,\w}{2m_N}\,i^L\,j_L(\frac{\w r}{2})
   \left\{ (\mu_p+(-1)^L\,\mu_n)\,[Y_L\otimes S]_{L\,M}\right.\nonumber\\
&\left.+(\mu_p-(-1)^L\,\mu_n)\,[Y_L\otimes t]_{L\,M}\right\}\cdot
\left[a_{\vec{k},\lambda}\,\delta_{M,\lambda}-a_{\vec{k},\lambda}^\dagger\,
(-1)^{L+\lambda}\,\wignerd{L}{M}{-\lambda}\right],
\label{eq:intJsigmaAtwofinal}
\end{align}
where we again used Eq.~(\ref{eq:TdotV}).

We are now ready to calculate the amplitudes $\Mfi{\phi\,\sigma 1}$ and 
$\Mfi{\phi\,\sigma 2}$. The results are given in 
Appendix~\ref{app:subleading}, 
together with the amplitudes $\Mfi{\sigma 1\,\sigma 1}$ and 
$\Mfi{\sigma 2\,\sigma 2}$, which do not contain the gradient part of the 
photon field. Nevertheless, these contributions are strong, cf. 
Fig.~\ref{fig:separation}, due to the numerically large factor 
$(\mu_p-\mu_n)^2\approx22$. As the amplitudes with both photons coupling to 
the current $\Jp$, Eq.~(\ref{eq:Jp}), are not supported by this factor, 
these contributions are tiny. Therefore, they are not included
in our work. We also found the mixed amplitudes 
$\Mfi{\sigma 1\,\sigma 2}$, $\Mfi{\sigma 2\,\sigma 1}$ 
negligibly small, which is no surprise, 
as explained in Appendix~\ref{app:subleading}.

So far we (explicitly) only considered one-body currents. However, there are 
also non-negligible contributions from pion-exchange currents, cf. 
Fig.~(\ref{fig:mesonexchange}). The corresponding expressions for the currents 
in coordinate-space representation are
\be
\vec{J}\,^\mathrm{KR}_\mathrm{stat}(\vec{\xi};\vec{x}_1,\vec{x}_2)=
\frac{e\,f^2}{m_\pi^2}\,
\left(\vec{\tau}_1\times\vec{\tau}_2\right)_z\,
\left[
\vec{\sigma}_1\,\delta(\vec{x}_1-\vec{\xi})\,(\vec{\sigma}_2\cdot\hat{r})+
\vec{\sigma}_2\,\delta(\vec{x}_2-\vec{\xi})\,(\vec{\sigma}_1\cdot\hat{r})
\right]\,\frac{\partial}{\partial r}\frac{\e^{-m_\pi r}}{r}
\label{eq:KR}
\ee
for the Kroll-Ruderman (pair) current (Fig.~\ref{fig:mesonexchange}(a)), and
\be
\vec{J}\,^\mathrm{pole}_\mathrm{stat}(\vec{\xi};\vec{x}_1,\vec{x}_2)=-
\frac{e\,f^2}{4\pi}\,
\left(\vec{\tau}_1\times\vec{\tau}_2\right)_z\cdot
\left(\nab_1-\nab_2\right)\,
(\vec{\sigma}_1\cdot\nab_1)\,
(\vec{\sigma}_2\cdot\nab_2)\,
\frac{\e^{-m_\pi|\vec{x}_1-\vec{\xi}|}}{m_\pi\,|\vec{x}_1-\vec{\xi}|}\,
\frac{\e^{-m_\pi|\vec{x}_2-\vec{\xi}|}}{m_\pi\,|\vec{x}_2-\vec{\xi}|}
\label{eq:pole}
\ee
for the so-called pion-pole current (Fig.~\ref{fig:mesonexchange}(b)), cf. 
e.g. \cite{Ericson}, Section~8.3\footnote{We note that our convention for 
the pion-nucleon coupling $f^2$ differs by a factor $4\pi$ 
from that used in \cite{Ericson}.}. $\vec{\tau}_i$ denotes the isospin 
operator of the $i$th nucleon, $\vec{x}_i$ its position. 
The relative vector $\vec{r}$ is defined as $\vec{r}=\vec{x}_1-\vec{x}_2$. 

Our numerical evaluations show that the explicit inclusion of the pole 
current is well negligible in the process and energies under 
consideration. Therefore we are only concerned with the Kroll-Ruderman current,
Eq.~(\ref{eq:KR}). This expression, however, is derived in the limit of 
static nucleons (denoted by the index 'stat' in Eq.~(\ref{eq:KR})), 
i.e. the correction due to the photon energy is neglected. This 
being a rather crude approximation for $\w\sim100$~MeV, which is close to the 
pion mass, we use instead of Eq.~(\ref{eq:KR})
\be
\vec{J}\,^\mathrm{KR}(\vec{\xi};\vec{x}_1,\vec{x}_2)=
\frac{e\,f^2}{m_\pi^2}\,
\left(\vec{\tau}_1\times\vec{\tau}_2\right)_z\,
\left[
\vec{\sigma}_1\,\delta(\vec{x}_1-\vec{\xi})\,(\vec{\sigma}_2\cdot\hat{r})+
\vec{\sigma}_2\,\delta(\vec{x}_2-\vec{\xi})\,(\vec{\sigma}_1\cdot\hat{r})
\right]\,\frac{\partial}{\partial r}f^\mathrm{KR}(r).
\label{eq:KRreal}
\ee
The function $f^\mathrm{KR}(r)$ depends on the 
photon energy and is defined in 
Appendix~\ref{app:two-body}, where we derive Eq.~(\ref{eq:KRreal}).

Now we have another current at hand, which we can use to replace $\vec{J}\ofxi$
in Eq.~(\ref{eq:disp}). However, one has to be careful in order not to 
double-count\footnote{We are grateful to W. Weise for useful comments made on 
this point.} certain contributions; e.g. it is not allowed to combine 
$\Hint=-\int\vec{J}^\mathrm{\,KR}\ofxi\cdot\vec{A}^\mathrm{full}\ofxi\,d^3\xi$
at one vertex with 
$\Hint=-\int\vec{J}\ofxi\cdot\nab\phi\ofxi\,d^3\xi$ at the other, 
because $\vec{J}^\mathrm{\,KR}$ is part of $\vec{J}$.
Therefore, this combination is 
already included~-- at least partly~-- in the dominant 
terms of Section~\ref{sec:dominant}, cf. discussion around 
Eq.~(\ref{eq:implicit}). 

First we note that the Kroll-Ruderman current changes isospin, i.e. 
$\Hint=-\int\vec{J}^\mathrm{\,KR}\ofxi\cdot\vec{A}\ofxi\,d^3\xi$ transforms 
the isospin-0 deuteron into an isospin-1 object. This can easily be seen when 
we act with $\left(\vec{\tau}_1\times\vec{\tau}_2\right)_z$ on the deuteron 
isospin wave function (Eq.~(\ref{eq:isospinwavefunction})):
\be
\left(\vec{\tau}_1\times\vec{\tau}_2\right)_z\,\frac{1}{\sqrt{2}}\,
\ket{p\,n-n\,p}=
\left(\tau_1^x\,\tau_2^y-\tau_1^y\,\tau_2^x\right)\,\frac{1}{\sqrt{2}}\,
\ket{p\,n-n\,p}=-2\,i\,\frac{1}{\sqrt{2}}\,\ket{p\,n+n\,p}
\label{eq:taucrosstauond}
\ee
Therefore we need another isospin-changing interaction at the second vertex. 
Pauli's principle guarantees that the total wave function of the two-nucleon 
system has to be antisymmetric under the exchange of the two constituents, as 
nucleons are fermions. Stated differently, the wave function has to fulfill 
$(-1)^{S+L+T}=-1$, i.e. in order to have $T=1$ we need $S+L$ even. The 
operators at our disposal are $Y_L$ from $\hat{\phi}_{i,f}$ 
(Eq.~(\ref{eq:phidefinition})), $[Y_{L-1}\otimes S]_L$ and 
$[Y_{L-1}\otimes t]_L$ from $\int\Jsigma\cdot\Aone$ 
(Eq.~(\ref{eq:intJsigmaAone})) and $[Y_{L}\otimes S]_L$, $[Y_{L}\otimes t]_L$
from $\int\Jsigma\cdot\Atwo$ (Eq.~(\ref{eq:intJsigmaAtwofinal})). 
$Y_L$ and $[Y_{L'}\otimes S]_L$ are spin-conserving operators, 
$[Y_{L'}\otimes t]_L$ is spin-changing, i.e. this operator 
corresponds to $S_C=0$, cf.
Eqs.~(\ref{eq:mxY}-\ref{eq:mxt}). When we restrict ourselves to $L,\,L'=1$,
which is a reasonable approximation as higher multipoles are strongly 
suppressed, we find $Y_1$~-- remember, the expansion $\phi\ofxi$ starts with 
$L=1$ (Eq.~(\ref{eq:phidefinition})), $[Y_0\otimes S]_1$, $[Y_1\otimes S]_1$, 
 $[Y_0\otimes t]_1$ and $[Y_1\otimes t]_1$. 
The operator  $[Y_0\otimes S]_1$
is isospin-conserving, as it transforms the deuteron into an $S=1$-state with
even orbital angular momentum, cf. Eq.~(\ref{eq:mxL})\footnote{The latter 
claim follows from the fact that the deuteron has even orbital angular 
momentum and from Eq.~(\ref{eq:threejprops5}).}. So is the spin-changing 
operator $[Y_1\otimes t]_1$. Therefore the possible candidates are
$Y_1$, $[Y_1\otimes S]_1$ and $[Y_0\otimes t]_1$.
We found that the numerically  most important operator is $[Y_0\otimes t]_1$. 
The same observation was made in \cite{Karakowski}. Nevertheless, also 
the operator $Y_1$, which stems from $\hat{\phi}_i$, $\hat{\phi}_f$, gives 
non-negligible contributions, whereas we found the amplitudes including 
$[Y_1\otimes S]_1$ invisibly small and therefore these terms are not
written down in our work. However, in the amplitudes including $\hat{\phi}_i$ 
or $\hat{\phi}_f$ at the non-KR vertex, 
one is not allowed to use the full photon field in 
$\Hint=-\int\vec{J}^\mathrm{\,KR}\ofxi\cdot\vec{A}\ofxi\,d^3\xi$, as 
explained after Eq.~(\ref{eq:KRreal}). Therefore we are left with 
$\Mfi{\mathrm{KR}\,\mathrm{full}\,\sigma1}$, 
$\Mfi{\phi\,\mathrm{KR}1}$ and $\Mfi{\phi\,\mathrm{KR}2}$, where 'KR$\,$full'
denotes the integral over the Kroll-Ruderman current, multiplied by the full 
photon field. There is no danger of double-counting in 
$\Mfi{\mathrm{KR}\,\mathrm{full}\,\sigma1}$, 
as we only take into account the operator $[Y_{L-1}\otimes t]_L$. 
This operator, however, changes the deuteron 
spin, cf. Eq.~(\ref{eq:mxt}), whereas the matrix elements arising from 
$\phi_{i,f}$ are spin-conserving, see Eq.~(\ref{eq:mxY}). 
Further contributions, like the one where 
$\vec{J}\ofxi=\vec{J}^\mathrm{\,KR}\ofxi$ at both vertices, turned out to be 
small. Therefore we only have to consider the three combinations above. 
The evaluation of these contributions is given in Appendix~\ref{app:two-body}.

\begin{figure}[!htb]
\begin{center}
\includegraphics*[width=.4\linewidth]
{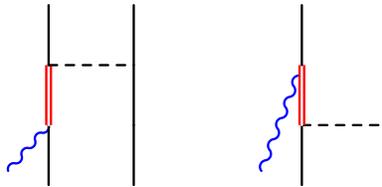}
\caption[Pion-exchange currents with explicit $\Delta$ degrees of freedom]
{Exemplary one-pion-exchange currents with explicit $\Delta(1232)$ degrees of 
freedom.} 
\label{fig:mesonexchangeDelta}
\end{center}
\end{figure}

Two-body currents with explicit $\Delta(1232)$ degrees of freedom, as 
displayed in Fig.~\ref{fig:mesonexchangeDelta}, are suppressed by one order in
$\epsilon$ with respect to the Kroll-Ruderman current, 
due to the 
$\gamma N \Delta$~vertex being part of $\mathcal{L}_{N\Delta}^{(2)}$, 
see Eq.~(\ref{eq:LND2}).
This is in agreement with the findings of \cite{Lvov}, where such 
contributions to elastic deuteron Compton scattering below 100~MeV 
were claimed to be negligibly small (of the order of 2\%). 
A similarly strong influence of the $\Delta$(1232) current 
to the process $np\rightarrow d\gamma$ is reported in Section~8.5 
of~\cite{Ericson}, where 
the contributions from these currents turn out considerably smaller 
than those from pionic exchange currents. Therefore and due to the excellent
agreement of the total deuteron-photodisintegration cross section, extracted
from our elastic Compton amplitude, with experimental data up to $\w=100$~MeV,
cf. Section~\ref{sec:photodisintegration}, we so 
far refrain from including these terms into our calculation. Nevertheless, 
it would be an interesting future task to perform a detailed investigation 
of their size. 

In the last two subsections we prepared all ingredients of our deuteron 
Compton calculation, which are new with respect to 
Chapter~\ref{chap:perturbative}. In the next section we demonstrate that our 
non-perturbative approach fulfills the well-known low-energy theorem, i.e. we 
reach the exact static limit in this approach. In Section~\ref{sec:results2}, 
we present our results at non-zero energies and compare them to those from 
the strictly perturbative approach of Chapter~\ref{chap:perturbative} and to
data.

\section[Low-Energy Limit in the Non-Perturbative Approach]
{Low-Energy Limit in the Non-Perturbative\\Approach \label{sec:Thomson2}}

One of the aims to be achieved with the 
approach to deuteron Compton scattering described in this chapter is to extend
the calculation of Chapter~\ref{chap:perturbative} to lower energies. 
In this section we prove that we have indeed \textit{removed} the limitations
of Chapter~\ref{chap:perturbative} at low energies, i.e. our so-called 
non-perturbative approach to deuteron Compton scattering reaches the correct 
limit of vanishing photon energy.

We remind the reader that the strictly perturbative expansion of the 
interaction kernel up to third order in the SSE counting scheme, used in 
Chapter~\ref{chap:perturbative}, leads to a low-energy cross section that is 
too large by a factor of 6. The amplitude in this limit, which is the 
well-known Thomson term for scattering of an electromagnetic wave from a 
charged particle, cf. Eq.~(\ref{eq:Thomson}), is
\be
A^\mathrm{Thomson}=\frac{Q^2\,e^2}{A\,m_N}\,\eps\cdot\epspr.
\label{eq:Thomson2}
\ee
The overall sign of the amplitude is convention and differs e.g. in 
Ref.~\cite{Karakowski} from this work.
As already discussed in Section~\ref{sec:theory}, Friar showed that 
Eq.~(\ref{eq:Thomson2}) is a consequence of gauge invariance
\cite{Friar}. Therefore the ansatz used in 
Chapter~\ref{chap:perturbative} obviously violates gauge invariance, which is
one of the shortcomings that we cure~-- at least approximately~-- in 
this chapter. 
The violation appears when evaluating the kernel between the deuteron
wave functions, without allowing the two nucleons in the intermediate state 
to interact with each other. The reason is that the deuteron wave function 
implies this interaction, which can be interpreted as the exchange of mesons,
e.g. of pions, between the two nucleons. In order to really achieve gauge 
invariance, it is therefore mandatory to include rescattering of the two 
nucleons on one hand and to couple the photons to these meson-exchange 
currents on the other, cf. Figs.~\ref{fig:mesonexchange} and~\ref{fig:blobs} 
and Refs.~\cite{Lvov, Karakowski}. 
Full gauge invariance will however not be obtained within our calculation, as 
the $np$-potential we use (AV18 \cite{AV18}, cf. Appendix~\ref{app:AV18}), 
contains more than only the one-pion exchange. The short-distance 
part of such a phenomenological potential may be interpreted as the 
exchange of mesons heavier 
than the pion (e.g. the $\omega$- or $\rho$-meson). As we only allow 
for explicit pion-exchange currents, gauge invariance will not be fulfilled 
exactly~\cite{Lvov}.

For a deuteron target, Eq.~(\ref{eq:Thomson2}) reads
\be
A^\mathrm{Thomson}_d=\frac{e^2}{m_d }\,\eps\cdot\epspr\approx
                     \frac{e^2}{2m_N}\,\eps\cdot\epspr.
\ee
This is a non-trivial result because the deuteron mass is involved, whereas
the Thomson seagull for Compton scattering from the proton, 
Fig.~\ref{fig:chiPTsingle}(a), yields the amplitude
\be
A^\mathrm{Thomson}_p=\frac{e^2}{m_p}\,\eps\cdot\epspr,
\label{eq:Thomsonp}
\ee
cf. Eq.~(\ref{eq:poleterms}). The neutron amplitude is zero
in the static limit. Therefore, all other 
contributions to deuteron Compton scattering in the limit $\w\rightarrow 0$ 
have to cancel half of the proton amplitude~(\ref{eq:Thomsonp}).
The only non-vanishing terms in the low-energy limit, 
except for the proton seagull, are two-body diagrams, namely 
the explicit pion-exchange diagrams, 
Fig.~\ref{fig:chiPTdouble}, and the double-commutator term, 
Eq.~(\ref{eq:Mfiphiphi4added}), as explained in Appendix~\ref{app:dominant}.
This double-commutator involves the internal Hamiltonian 
$H^{np}=\frac{\vec{p}\,^2}{m_N}+V$, cf. Eq.~(\ref{eq:Hinternal}),
and therefore can be separated into a kinetic energy and a  
potential part, cf. Appendix~\ref{app:dominant}. Arenh\"ovel~\cite{Arenhoevel}
showed analytically that in the static limit 
the potential energy part, using the one-pion-exchange 
potential, cancels exactly the contributions from explicit pion exchange, 
Fig.~\ref{fig:chiPTdouble}. It is this consistency between the diagrams 
explicitly included in the interaction kernel and the potential used in 
the double-commutator, Eq.~(\ref{eq:Mfiphiphi4added}), which guarantees
the correct low-energy behaviour of our calculation. 
Therefore we only include the one-pion exchange in the 
amplitude~(\ref{eq:Mfiphiphi4added}). Since two-nucleon contact 
terms, which parameterize the exchange of heavier particles than pions, 
are of higher than third order in the SSE scheme, we do not
include such diagrams explicitly, and therefore they are also not included 
in the double-commutator. While this procedure
is strictly speaking not consistent with the $NN$-potential~\cite{AV18} that 
we use for constructing the $np$-Green's function, it is nevertheless a 
legitimate prescription, as we show in Section~\ref{sec:potentialdep} that 
using only the one-pion-exchange potential does not change our results 
significantly.

As the sum of the explicit pion-exchange diagrams from 
Fig.~\ref{fig:chiPTdouble} and the potential energy part of 
Eq.~(\ref{eq:Mfiphiphi4added}) gives no contribution in the low-energy limit, 
it is clear
that the kinetic energy part of the double-commutator has to cancel half of 
the proton seagull amplitude~(\ref{eq:Thomsonp}). This can easily be shown to 
be true:
in the long-wavelength limit, i.e. for $|\vec{k}|\rightarrow 0$, the photon 
field reduces to the polarization vector (dipole approximation), as 
$\hat{\epsilon}_\lambda\,
\e^{i\vec{k}\vec{r}}\rightarrow\hat{\epsilon}_\lambda$. 
As we already know that only the gradient 
part of the photon field survives in the static limit, cf. 
Section~\ref{sec:dominant}, we find
\be
\left.\nab_\xi\phi\ofxi\right|_{\w\rightarrow 0}=\hat{\epsilon}_\lambda.
\ee
Therefore, $\phi\ofxi|_{\w\rightarrow 0}=\hat{\epsilon}_\lambda\cdot\vec{\xi}$
and, using $\hat{\phi}=e\,\phi(\vec{r}/2)$,
\be
\left.\hat{\phi}\right|_{\w\rightarrow 0}=
e\,\frac{\vec{r}}{2}\cdot\hat{\epsilon}_\lambda.
\ee
Now we need to evaluate the double-commutator in this limit, finding 
\ba
\lim_{\w\rightarrow 0}
\left[[\frac{\vec{p}\,^2}{m_N},\phiihat],\phifhat\right]&=
\frac{e^2}{4m_N}\,\left[[\vec{p}\,^2,\vec{r}\cdot\eps],
\vec{r}\cdot\epspr\right]\nonumber\\
&=\frac{e^2}{2m_N}\,\left[(\vec{p}\,(\vec{r}\cdot\eps))\cdot\vec{p},
\vec{r}\cdot\epspr\right]\nonumber\\
&=\frac{-i\,e^2}{2m_N}\,\left[\eps\cdot\vec{p},\vec{r}\cdot\epspr\right]
\nonumber\\
&=-\frac{e^2}{2m_N}\,\eps\cdot\epspr,
\label{eq:doublecommutatorstatic}
\end{align}
where we used $\vec{p}=\frac{\vec{p}_p-\vec{p}_n}{2}=
-i\,\frac{\nab_{x_p}-\nab_{x_n}}{2}$ and $\vec{r}=\vec{x}_p-\vec{x}_n$, cf. 
Eq.~(\ref{eq:cmvariables}). Eq.~(\ref{eq:doublecommutatorstatic}) is 
obviously symmetric under $i\leftrightarrow f$. Therefore the second
double-commutator in Eq.~(\ref{eq:Mfiphiphi4added}) gives the same result.
Adding both commutators cancels the factor $\frac{1}{2}$ in front of the 
matrix element and we see that the result accounts for exactly half of the 
negative Thomson amplitude of the proton. Note that this result is independent
of the deuteron wave function and the $np$-potential chosen.

\begin{figure}[!htb]
\begin{center}
\includegraphics*[width=.48\linewidth]{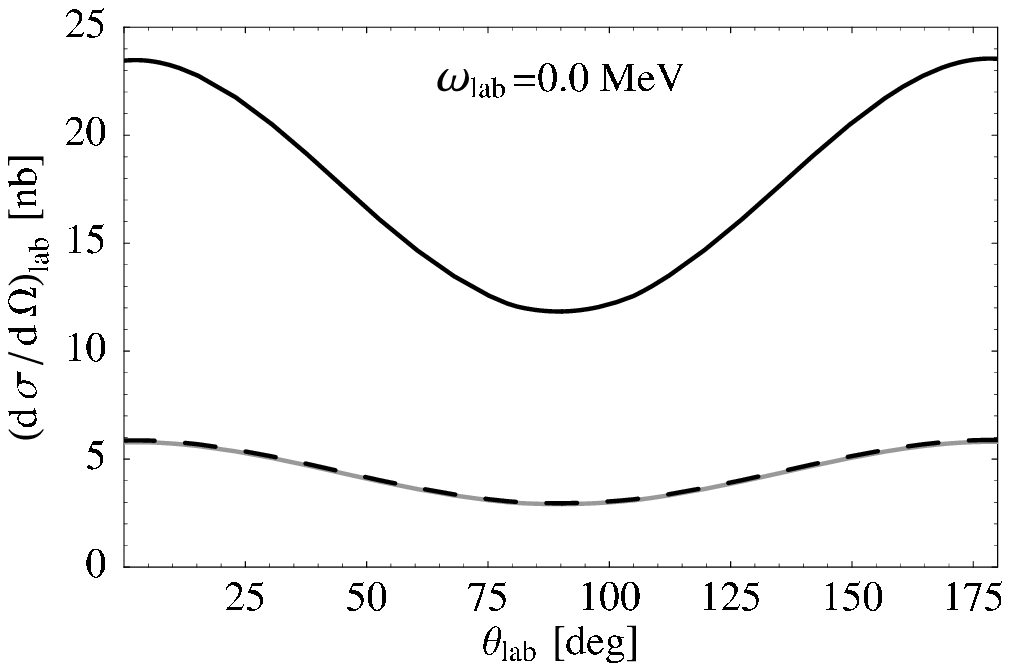}
\hfill
\includegraphics*[width=.48\linewidth]{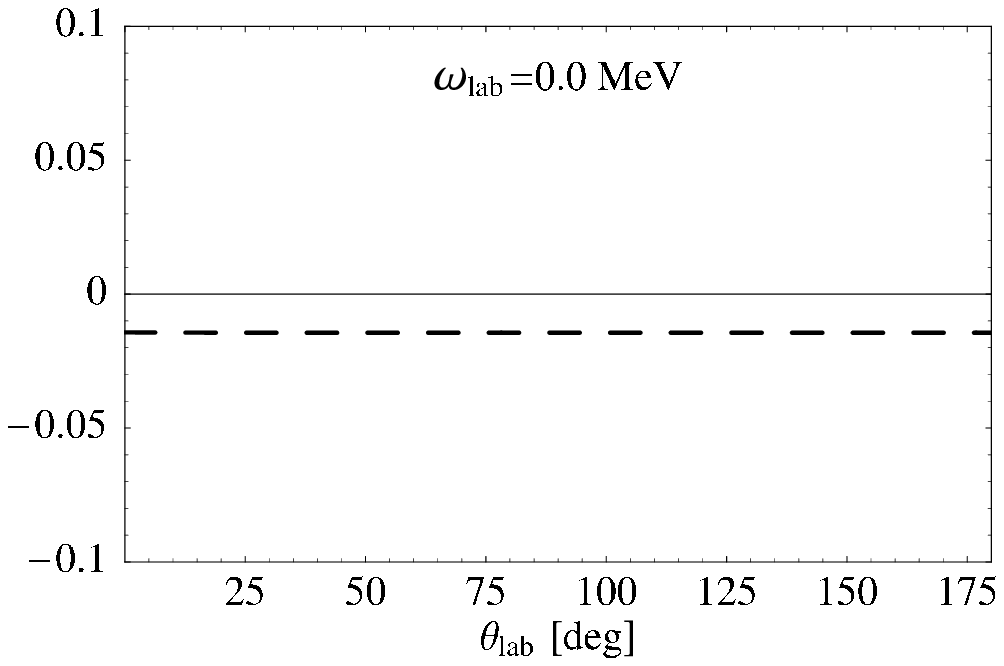}
\caption[Deuteron Compton cross section in the static limit]
{Left panel: Comparison of the proton (black, solid), deuteron (grey, solid) 
and $1/4$ of the proton (black, dashed) Compton cross section in the static 
limit. The function plotted in the right panel is
$\left(           \frac{d\sigma}{d\Omega}\right)_d/
 \left(\frac{1}{4}\frac{d\sigma}{d\Omega}\right)_p-1$.}
\label{fig:Thomson2}
\end{center}
\end{figure}

Our numerical evaluation agrees well with the Thomson limit 
(\ref{eq:Thomson2}), as demonstrated in Fig.~\ref{fig:Thomson2}, where we see
a comparison between the proton Compton cross section, 
$\left(\frac{1}{2}\right)^2=\frac{1}{4}$ of this cross section 
and the deuteron Compton cross section 
at zero photon energy (we remind the reader that 
$\frac{d\sigma}{d\Omega}\propto|\Mfi{}|^2$, 
cf. Eq.~(\ref{eq:deuteroncrosssection})); the latter two curves are nearly 
indistinguishable. In the right panel of Fig.~\ref{fig:Thomson2}, where we 
show the relative error 
$\left(           \frac{d\sigma}{d\Omega}\right)_d/
 \left(\frac{1}{4}\frac{d\sigma}{d\Omega}\right)_p-1$, we see that the 
deviation is less than 2\% and angle-independent. Therefore it can be 
accounted for by a constant factor. 
When we use the AV18 wave function~\cite{AV18}, the curve in the right panel of
Fig.~\ref{fig:Thomson2} turns out of about the same absolute size, but with  
opposite sign. This observation suggests that
the main part of this discrepancy is due to numerical 
uncertainties in the normalization of the wave function within our code. 

In this section we showed that the approach to deuteron Compton 
scattering, that we use in this chapter, fulfills the low-energy theorem and 
therefore guarantees at least approximate gauge invariance of the 
calculation. In the next section we present our results for non-zero photon 
energies, demonstrating that we  have achieved a 
consistent description of $\gamma d$ scattering 
for photon energies ranging from 0~MeV up to $\w\sim100$~MeV.

\section[Predictions for Deuteron Compton Cross Sections]
{Predictions for\\Deuteron Compton Cross Sections  
\label{sec:results2}}

\begin{figure}[!htb]
\begin{center} 
\includegraphics*[width=.48\linewidth]{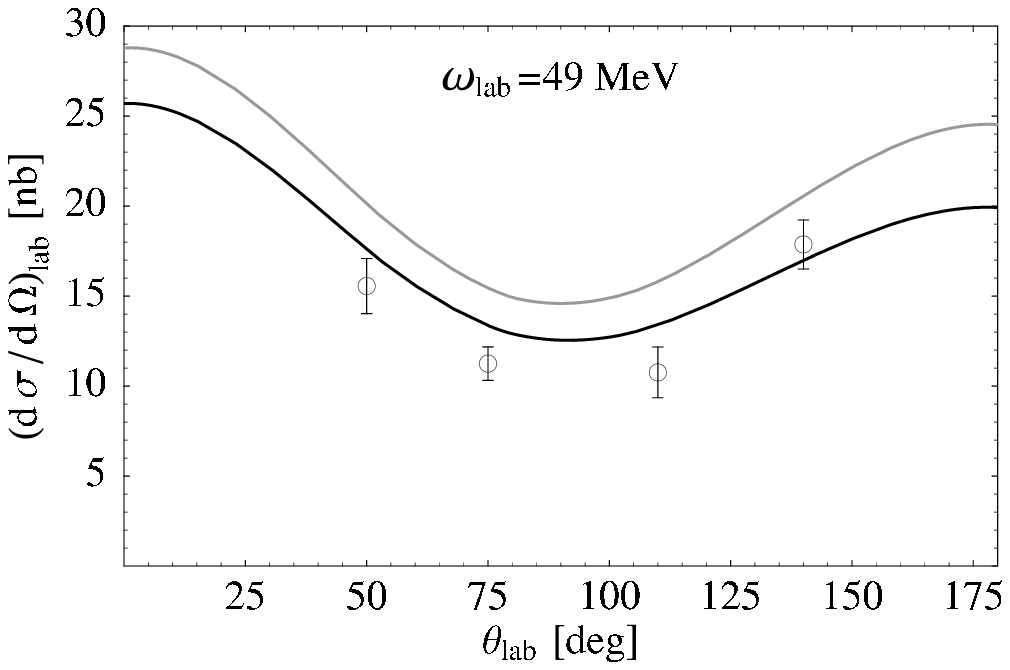}
\hfill
\includegraphics*[width=.48\linewidth]{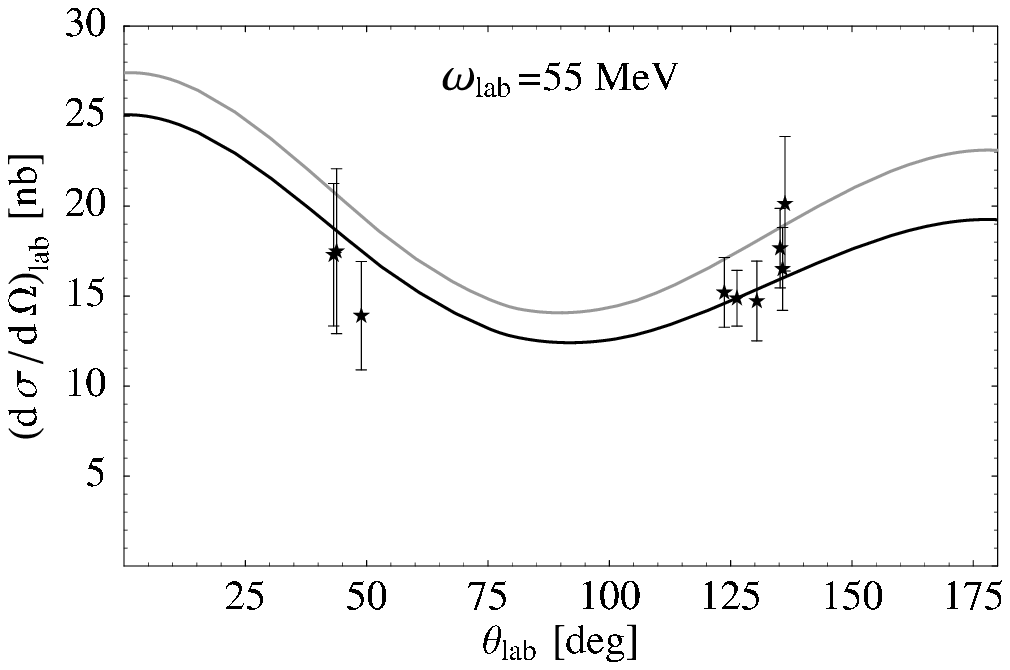}\\
\includegraphics*[width=.48\linewidth]{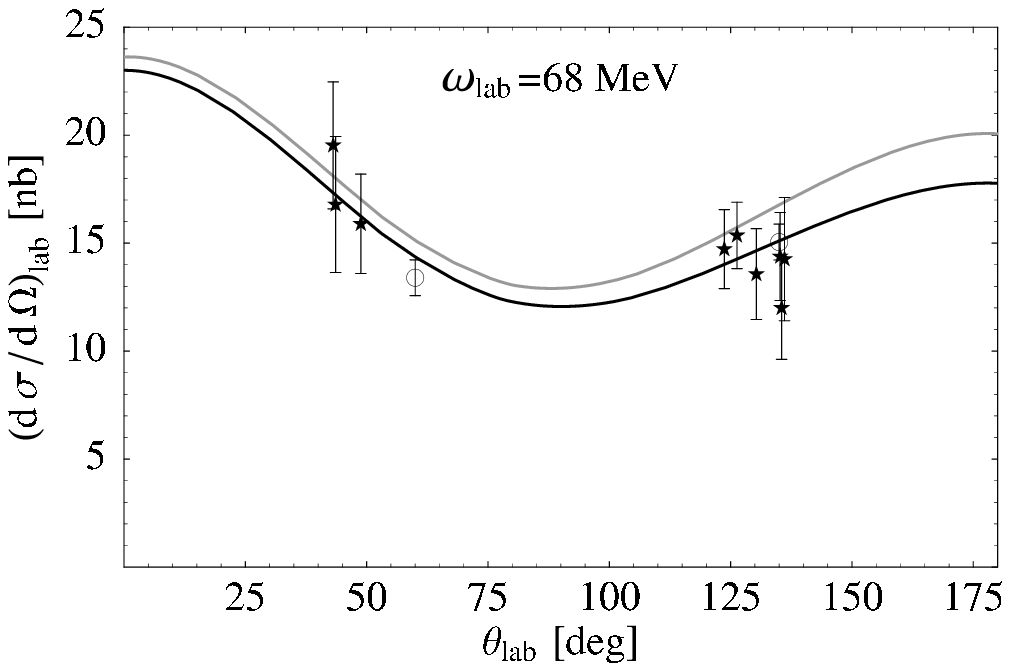}
\hfill
\includegraphics*[width=.48\linewidth]{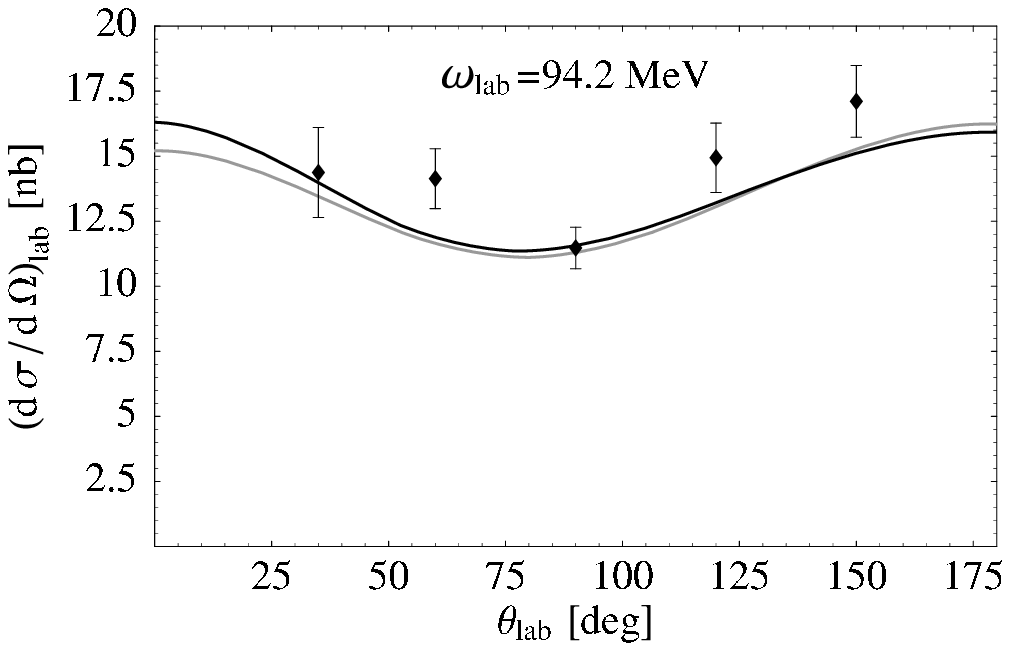}
\parbox{1.\textwidth}{
\caption[Comparison of deuteron Compton results to those from 
Chapter~\ref{chap:perturbative}]
{Comparison of our results from the two different approaches to 
deuteron Compton scattering: The black lines are the results of the 
non-perturbative approach, Chapter~\ref{chap:nonperturbative}, the grey lines 
are the SSE-results of Fig.~\ref{fig:SSEHBplots}. 
The data are from \cite{Lucas}~(circle), 
\cite{Lund}~(star) and \cite{Hornidge}~(diamond).}
\label{fig:comparisonpertnonpert}}
\end{center}
\end{figure}
In Fig.~\ref{fig:comparisonpertnonpert}, we show our parameter-free predictions
for the elastic deuteron Compton cross sections,
achieved with the ``non-perturbative'' calculation of this chapter. These 
results are compared to the predictions from Section~\ref{sec:results} and 
to the data from Illinois~\cite{Lucas}, Lund~\cite{Lund} and 
SAL~\cite{Hornidge}. Obviously we have reached our final goal: 
we have achieved a (chirally) consistent calculation for elastic deuteron 
Compton scattering, which describes all existing data reasonably well and 
also satisfies the low-energy theorem exactly, cf.~Section~\ref{sec:Thomson2}. 
In both calculations presented in Fig.~\ref{fig:comparisonpertnonpert} we use 
for the isoscalar polarizabilities the numbers from our 2-parameter SSE fit 
to proton data, given in Table~\ref{tab:protonfit}, which is justified by the 
fact that the isovector polarizabilities vanish at leading-one-loop order in 
the Small Scale Expansion. 

There are still minor deviations from the experiments, e.g. 
our calculation lies slightly above the three 49~MeV data from 
\cite{Lucas}, which have been measured at 
angles below $120^\circ$. However, this is a feature that our calculation has 
in common with other approaches, which also reach the correct static limit, 
e.g. \cite{Lvov, Karakowski, Rupak}. 
At higher energies the two calculations of Chapters~\ref{chap:perturbative} and
\ref{chap:nonperturbative} approach each other. A heuristic explanation of 
this behaviour, based on the fact that a high-energy photon immediately 
separates the two nucleons from each other, is given in 
Section~\ref{sec:theory}. This is another important 
cross-check as it demonstrates that the power-counting applied in 
Chapter~\ref{chap:perturbative} is indeed suited for calculating deuteron 
Compton  scattering at photon energies of $\w\sim100$~MeV. Consequently, both 
curves in Fig.~\ref{fig:comparisonpertnonpert} describe the 94.2~MeV data from 
\cite{Hornidge} equally well~-- in fact they nearly lie on top of each other.

Because of the many contributions to the curves in 
Fig.~\ref{fig:comparisonpertnonpert} we resume all amplitudes contained in 
our calculation. These are:
\begin{itemize}
\item The single-nucleon Heavy Baryon Chiral Perturbation Theory (HB$\chi$PT) 
contributions from 
Fig.~\ref{fig:chiPTsingle}, except for the nucleon pole diagrams 
(Fig.~\ref{fig:chiPTsingle}(b) and its $u$-channel analog). These two diagrams 
are discussed 
under the final item. As in Chapter~\ref{chap:perturbative}, the pion pole, 
Fig.~\ref{fig:chiPTsingle}(c), gives no contribution because isospin-breaking 
effects like the mass difference between proton and neutron are neglected.
\item The single-nucleon contributions sketched in Fig.~\ref{fig:SSEsingle}, 
which include the explicit $\Delta(1232)$ resonance and
occur at third order in the Small Scale Expansion in addition to 
third-order HB$\chi$PT. For the coupling constants of the two isoscalar 
short-distance operators, Fig.~\ref{fig:SSEsingle}(f), we use the numbers 
derived from our Baldin-constrained fit to the 
proton Compton data, cf. Section~\ref{sec:spin-averaged}.
\item The nine two-body diagrams with both photons coupling to the 
exchanged pion, see Fig.~\ref{fig:chiPTdouble}.
\item All terms with an intermediate two-nucleon state, which replace the 
nucleon-pole diagrams in this chapter, cf. Figs.~\ref{fig:disp} 
and~\ref{fig:blobs}. These include 
the amplitudes $\Mfi{\phi\phi}$ given in 
Eqs.~(\ref{eq:Mfiphiphi1added})-(\ref{eq:Mfiphiphi4added}), which have been 
derived from minimal coupling, i.e. by replacing the photon field in the 
interaction Hamiltonian by the gradient term at both vertices, 
cf. Eqs.~(\ref{eq:Hint}, \ref{eq:multipoleexp}). Our results for these 
amplitudes are given in 
Appendix~\ref{app:dominant}. Further contributions are the amplitudes 
calculated in Appendices~\ref{app:subleading} and \ref{app:two-body}, where 
the above replacement is done at only one or even at none of the photon 
vertices, whereas in the other interaction Hamiltonians we replace the photon
field by $\Aone$ or 
$\Atwo$, respectively, cf. Eqs.~(\ref{eq:Aone}) and (\ref{eq:Atwo}). 
$\Aone$ is the magnetic part of the photon field in our notation, the other 
two parts are of electric nature.
The currents that we use for these interactions are the spin current $\Jsigma$ 
(Eq.~(\ref{eq:Jsigma})) and the Kroll-Ruderman current 
$\vec{J}^\mathrm{KR}$ (Eq.~(\ref{eq:KRreal}), see also 
Fig.~\ref{fig:mesonexchange}(a)), 
which are the only currents that we found to give non-negligible 
contributions. The corresponding amplitudes are called $\Mfi{\phi\,\sigma1}$, 
$\Mfi{\phi\,\sigma2}$, $\Mfi{\sigma1\,\sigma1}$, 
$\Mfi{\sigma2\,\sigma2}$, $\Mfi{\mathrm{KR}\,\sigma1}$, 
$\Mfi{\phi\,\mathrm{KR}1}$ and $\Mfi{\phi\,\mathrm{KR}2}$. 
The indices are '$\phi$' for the vertex arising from minimal coupling, 
'$\sigma 1$' ('$\sigma 2$') for the coupling of $\Aone$ ($\Atwo$) to the spin 
current and analogously for the Kroll-Ruderman current 'KR'. 
For example, $\Mfi{\sigma1\,\sigma1}$ describes the 
coupling of a magnetic photon to the spin current at both vertices. 
This amplitude dominates the deuteron-photodisintegration cross section 
at threshold, cf. Section~\ref{sec:photodisintegration}.
\end{itemize}
It is straightforward to
combine the contributions from Chapter~\ref{chap:perturbative} with those
from this chapter. The only modification necessary
is to replace the Cartesian polarization vectors $\vec{\epsilon}$, used
in Chapter~\ref{chap:perturbative}, by spherical ones, cf. 
Eq.~(\ref{eq:spherpolarizations}). 
Of course, one also has to make sure that the convention adopted for the 
overall sign of the amplitude is the same in both parts. We use the 
proton-seagull diagram (Fig.~\ref{fig:chiPTsingle}(a)) in order to fix the 
sign. The convention we chose is such that the Thomson amplitude is given by 
Eq.~(\ref{eq:Thomson}).

\begin{figure}[!htb]
\begin{center} 
\includegraphics*[width=.48\linewidth]{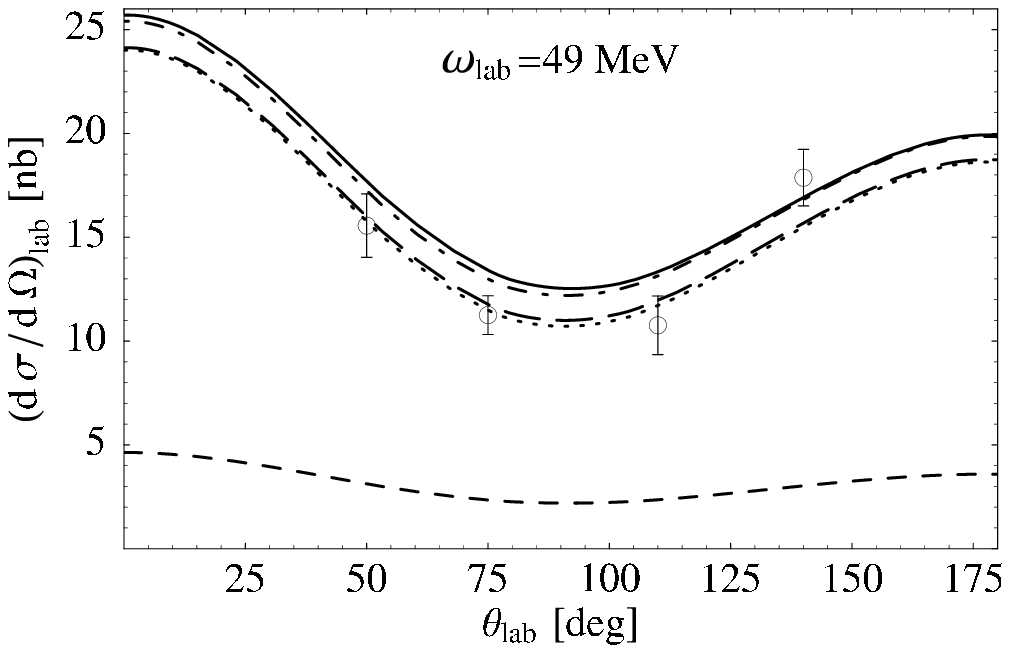}
\hfill
\includegraphics*[width=.48\linewidth]{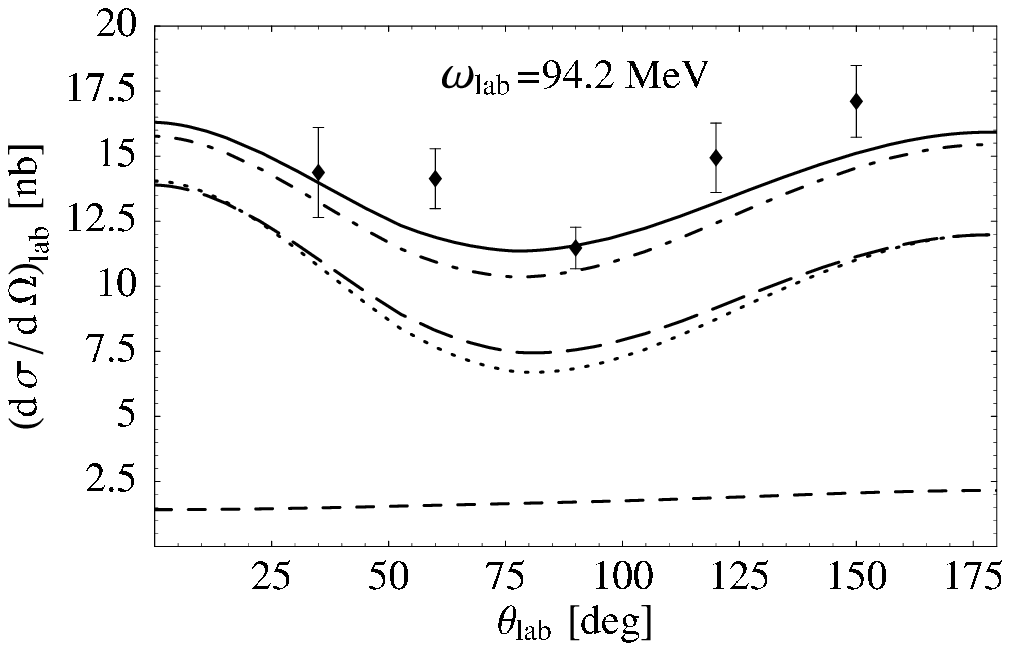}\\
\includegraphics*[width=.48\linewidth]{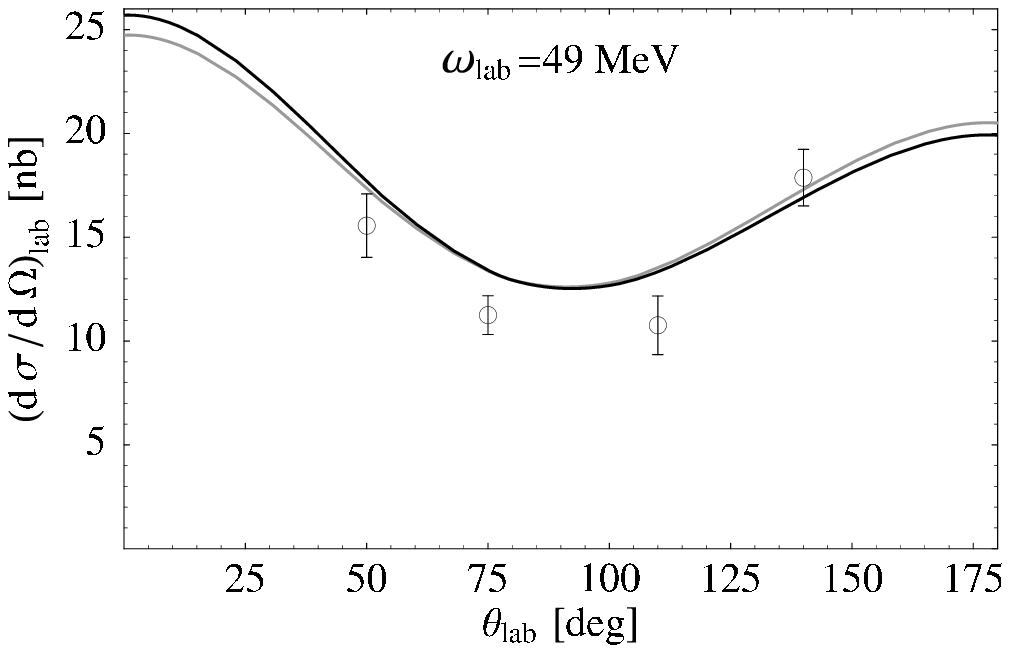}
\hfill
\includegraphics*[width=.48\linewidth]{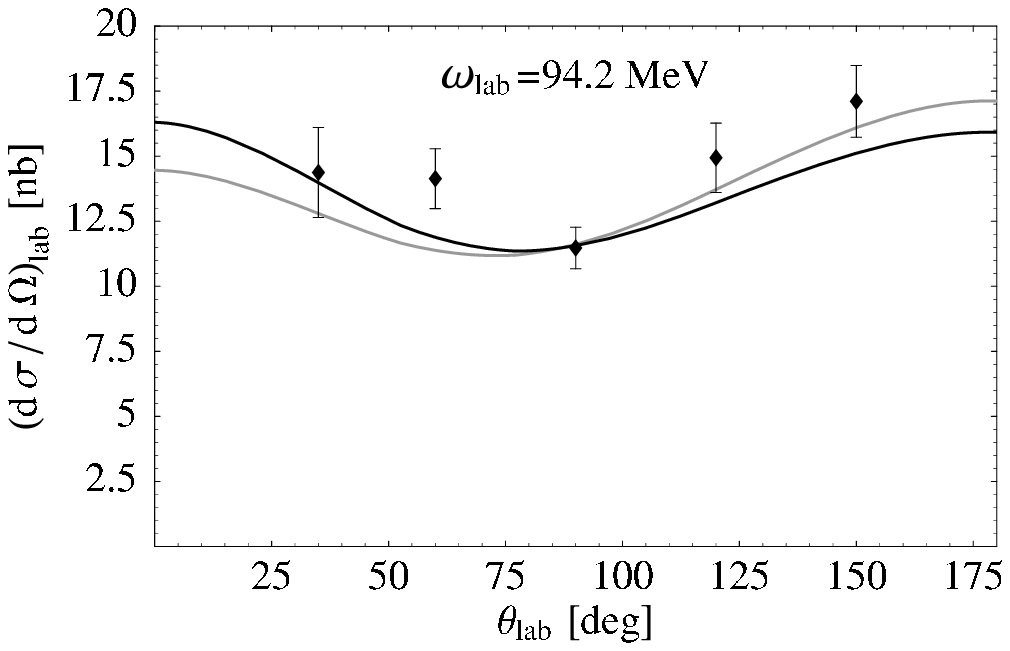}
\parbox{1.\textwidth}{
\caption[Comparison of separate contributions to deuteron Compton 
scattering]
{Comparison of separate contributions to our deuteron Compton-scattering 
results. In the upper panels we compare the full result (solid) to curves with 
several amplitudes subtracted; the subtracted amplitudes are:
$\Mfi{\mathrm{KR}}$ (dotdashed), 
$\Mfi{\mathrm{KR}}+\Mfi{\sigma\sigma}$ (dotted),
$\Mfi{\mathrm{KR}}+\Mfi{\sigma\sigma}+\Mfi{\phi\sigma}$ (longdashed),
$\Mfi{\mathrm{KR}}+\Mfi{\sigma\sigma}+\Mfi{\phi\sigma}+\Mfi{\phi\phi1,2,3}$ 
(shortdashed).
In the lower panels we compare our full result (black) to a curve where 
the amplitudes $\Mfi{\sigma\sigma}$ and $\Mfi{\phi\phi}$ have been replaced 
by their $L=L'=1$-approximations (grey).
The data are from \cite{Lucas}~(circle) and \cite{Hornidge}~(diamond).}
\label{fig:separation}}
\end{center}
\end{figure}

We now discuss the strength of  several contributions separately.
However, there are certain amplitudes which are closely related to each 
other: The kinetic energy part of the double commutator, 
Eq.~(\ref{eq:Mfiphiphi4kefinal}), cancels half 
of the proton seagull in the static limit, cf. Section~\ref{sec:Thomson2}.
The sum of the potential energy part of the commutator 
(\ref{eq:Mfiphiphi4final}) and the nine two-body
contributions from Fig.~\ref{fig:chiPTdouble} is zero
in the limit of vanishing photon energy.  
It stays small
in the whole energy range considered in this work, as already observed in 
Refs.~\cite{Karakowski,ArenhoevelII}.
Therefore we do not separate these contributions from each other. 
Nevertheless, there are a few issues worth investigating in more detail: 
\begin{itemize}
\item[1)]
The prominent role of the amplitudes 
$\Mfi{\phi\phi1,2}$, which include the case of an $E1$-interaction at both 
vertices. 
\item[2)]
The importance of the amplitudes 
$\Mfi{\phi\,\sigma}$ and $\Mfi{\sigma\sigma}$, with $\sigma$ denoting the 
coupling to the spin current. 
\item[3)]
The strength of
the amplitudes with the explicit Kroll-Ruderman current at one vertex, 
$\Mfi{\mathrm{KR}}$. 
\end{itemize}
In the upper two panels of Fig.~\ref{fig:separation}~-- 
we investigate the two extreme energies of 
Fig.~\ref{fig:comparisonpertnonpert}~--
these contributions are successively added to the 
remaining terms:  The single-nucleon amplitudes from 
Figs.~\ref{fig:chiPTsingle} and \ref{fig:SSEsingle} (except for the 
nucleon-pole diagrams Fig.~\ref{fig:chiPTsingle}(b)), the two-nucleon 
diagrams from Fig.~\ref{fig:chiPTdouble} and the double-commutator amplitude 
$\Mfi{\phi\phi4}$, cf. Eq.~(\ref{eq:Mfiphiphi4added}). The amplitude 
$\Mfi{\phi\phi3}$ (Eq.~(\ref{eq:Mfiphiphi3added})) is a small correction and 
has been added to the leading amplitudes $\Mfi{\phi\phi1,2}$. 

Obviously,  the amplitudes $\Mfi{\phi\phi}$ are the dominant ones in 
Fig.~\ref{fig:separation}. 
This observation holds for both energies considered. However,
also the amplitudes $\Mfi{\sigma\sigma}$ give important contributions.
The same pattern occurs in the calculation of total 
deuteron-photodisintegration cross sections, cf. 
Section~\ref{sec:photodisintegration}.
The contributions from $\Mfi{\phi\sigma}$ are nearly negligible. The 
small size of these terms is due to the fact that the 
amplitudes $\Mfi{\phi\sigma1}$ and $\Mfi{\phi\sigma2}$ 
largely cancel each other, 
cf. Appendix~\ref{app:subleading}. The diagrams with one photon 
explicitly coupling to the Kroll-Ruderman current are tiny for low energies,
but give a sizeable correction at 94.2~MeV. Their contribution is stronger in 
our calculation than it appears in Ref.~\cite{Karakowski}. This discrepancy may
be attributed
to the use of a different pion propagator 
in the Kroll-Ruderman current, cf. Appendix~\ref{app:two-body}. 
The difference is that we do not neglect the 
photon energy in the denominator, which is a crude approximation for 
$\w\sim100$~MeV. However, we have to caution 
that comparing cross sections is sometimes misleading, as they are 
non-additive quantities and strongly affected by interference between the 
various amplitudes, cf. Eq.~(\ref{eq:deuteroncrosssection}).

We also estimate the strength of those contributions where the photons 
coupling to the $NN$-rescattering diagrams, Fig.~\ref{fig:disp}, 
have multipolarity $L=2$. In Ref.~\cite{Karakowski}, these next-to-leading 
terms in the multipole expansion of the photon field are claimed to be  
small and therefore have been neglected. However, we slightly disagree from
this statement, as can be seen in the lower two panels of 
Fig.~\ref{fig:separation}. There we show our full results compared
to curves which only include the $L=L'=1$-approximation of the dominant 
amplitudes $\Mfi{\phi\phi1,2,3}$ and $\Mfi{\sigma\sigma}$. For low energies
these corrections are certainly negligible, however in the high-energy
regime of our calculation they are of the order of 10\% in the forward and 
in the backward direction.

Comparing to Ref.~\cite{Karakowski}, we see the main difference of our 
calculation to this work in our more involved description 
of the single-nucleon structure. In \cite{Karakowski} the structure of the 
nucleon is respected only via the static polarizabilities $\bar{\alpha}_{E1}$ 
and $\bar{\beta}_{M1}$, i.e.
via the leading terms of a Taylor expansion of the single-nucleon Compton 
multipoles, cf. Section~\ref{sec:polarizabilities1}. In our work, these
multipoles have been calculated up to third order in the Small Scale 
Expansion, as explained in detail in Chapters~\ref{chap:spinaveraged} and 
\ref{chap:perturbative}, 
and are included with their full energy dependence. Another advantage of 
our approach with respect to \cite{Karakowski} is the treatment of the 
pion propagator in the pion-exchange diagrams of Fig.~\ref{fig:chiPTdouble}.
We calculate these diagrams using the full pion propagator, whereas
the authors of \cite{Karakowski} always make the assumption that the photon 
energy is small compared to the energy of the virtual pion and therefore may 
be neglected. This, however, is no longer a good approximation as soon as
the photon energy comes close to the pion mass. A similar difference  
occurs in the Kroll-Ruderman amplitudes, as discussed above. Finally we do 
not agree with the statement of \cite{Karakowski} that $L=2$-contributions
are negligible for all amplitudes and energies considered, see 
Fig.~\ref{fig:separation}.

\begin{figure}[!htb]
\begin{center} 
\includegraphics*[width=.6\linewidth]{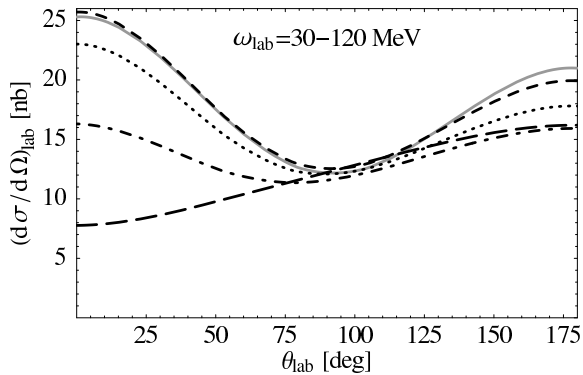}
\parbox{1.\textwidth}{
\caption[Comparison of differential Compton cross sections at various energies]
{Comparison of our results for differential deuteron Compton cross sections 
at various energies: 30~MeV (grey), 49~MeV (shortdashed), 68~MeV (dotted), 
94.2~MeV (dotdashed), 120~MeV (longdashed).}
\label{fig:energies}}
\end{center}
\end{figure}

We believe to have proven, in the last two sections, that our calculation
provides a consistent description of elastic deuteron Compton scattering
below $\w\sim100$~MeV. We can trust that it also gives reasonable 
results for even higher energies, say up to the pion mass. 
There, however, threshold corrections in analogy to Eq.~(\ref{eq:substitut}) 
should not be neglected anymore. Therefore the 120~MeV curve in 
Fig.~\ref{fig:energies}, where we 
show our results at various energies, is only a qualitative 
statement about the behaviour of the differential cross section for 
$\w\rightarrow m_\pi$. We also show our prediction at $\w_\mathrm{lab}=30$~MeV,
which is comparable in magnitude to the 49~MeV curve, but its shape is already
closer to the forward-backward symmetry exhibited by the static cross section, 
cf. Fig.~\ref{fig:Thomson2}. 

In the next section we investigate the sensitivity of our results on the 
deuteron wave function, which is another shortcoming of the approach 
presented in Chapter~\ref{chap:perturbative}, where the observed influence 
turned out unexpectedly strong.

\subsection{Dependence on the Deuteron Wave Function 
\label{sec:wavefunctiondep2}}
\begin{figure}[!htb]
\begin{center}
\includegraphics*[width=.48\linewidth]{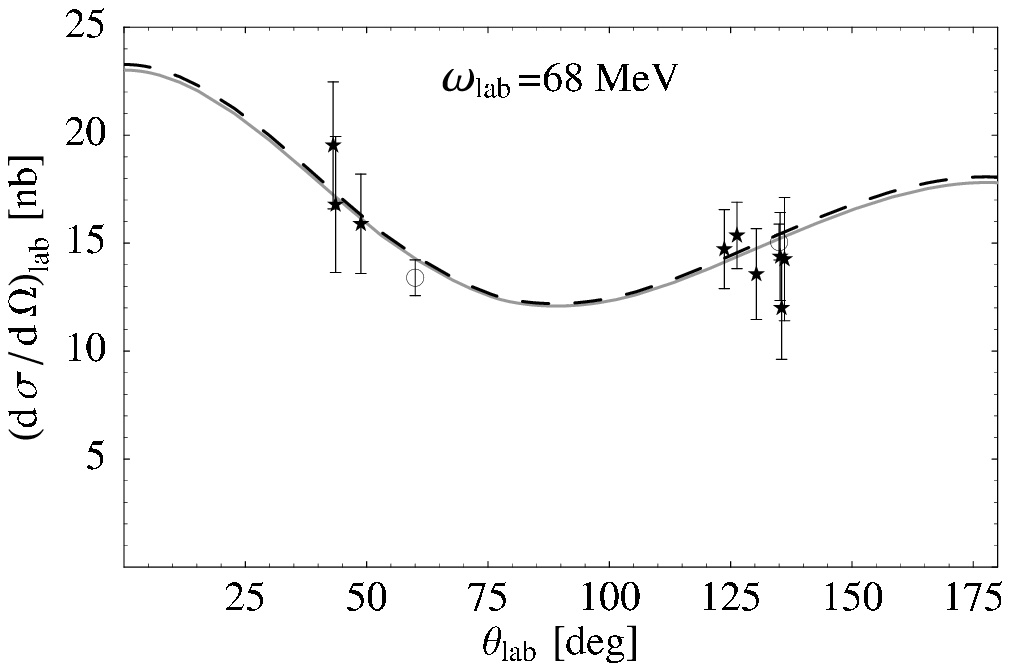}
\hfill
\includegraphics*[width=.48\linewidth]{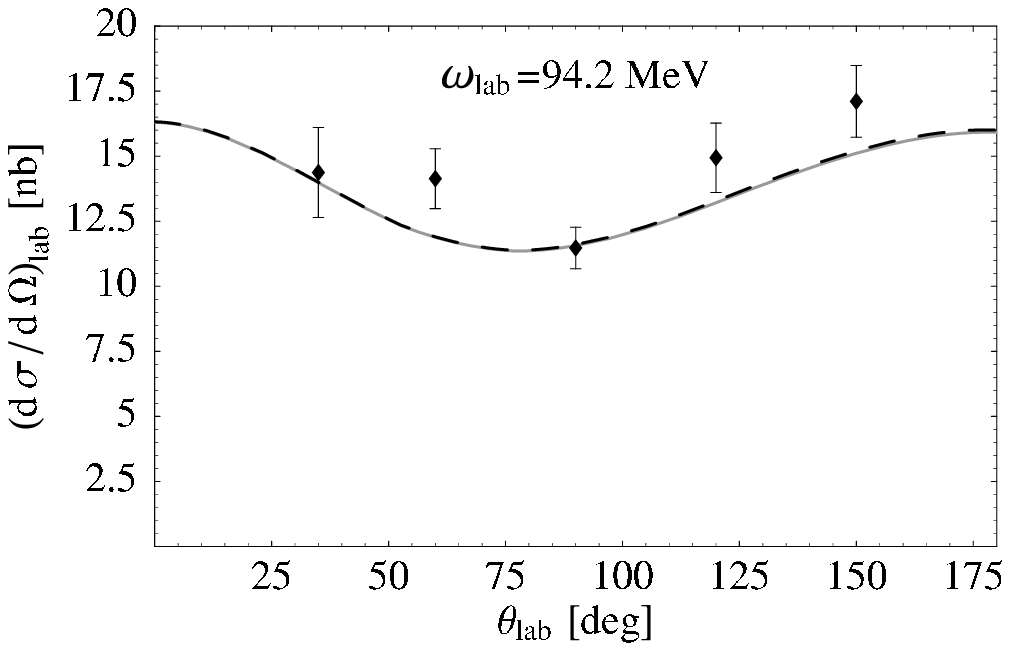}
\includegraphics*[width=.48\linewidth]{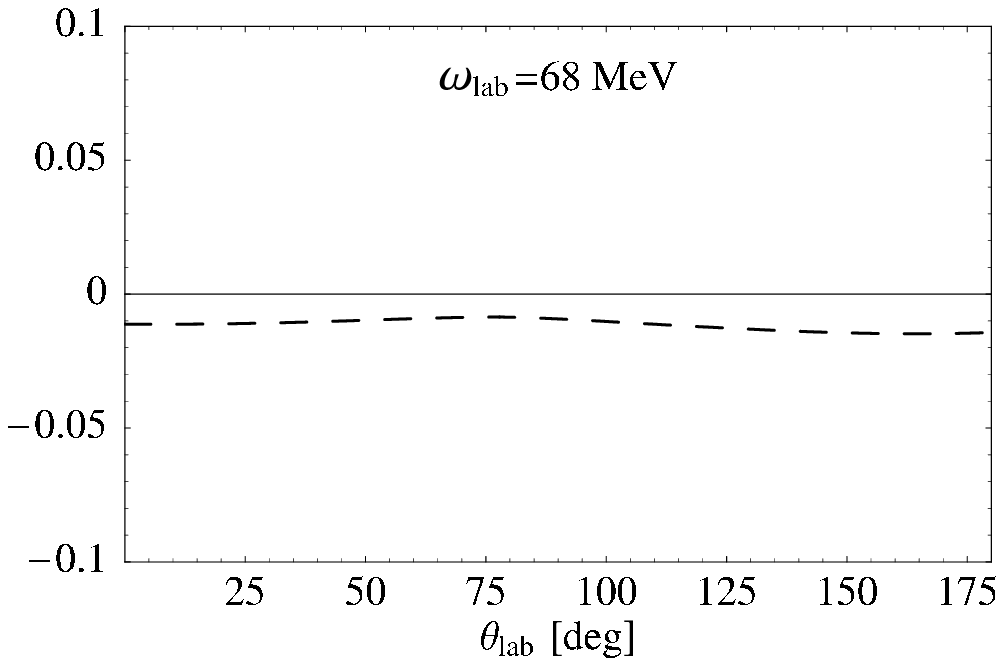}
\hfill
\includegraphics*[width=.48\linewidth]{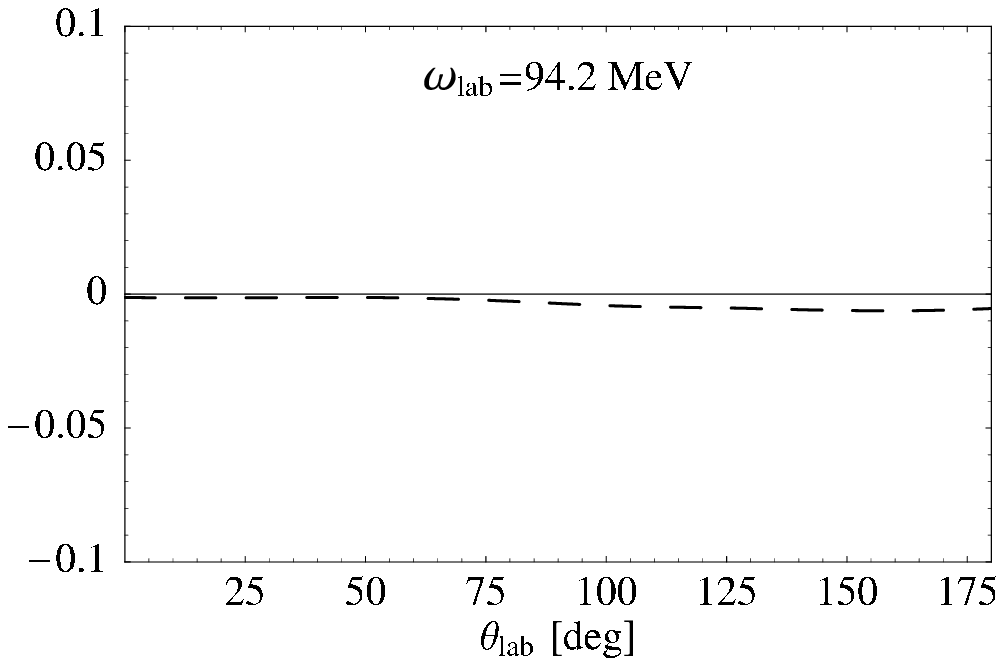}
\caption[Dependence of the deuteron Compton cross sections on the wave 
function]
{Comparison of our deuteron Compton cross-section results for 68 and 94.2~MeV,
using two different wave functions: NNLO $\chi$PT (grey)~\cite{Epelbaum} and 
AV18 (dashed)~\cite{AV18}. In the lower two panels we show 
$\left(\frac{d\sigma}{d\Omega}\right)_\mathrm{NNLO}/
 \left(\frac{d\sigma}{d\Omega}\right)_\mathrm{AV18}-1$.}
\label{fig:wavefunctiondep2}
\end{center}
\end{figure}

As demonstrated in Section~\ref{sec:Thomson2}, our calculation fulfills 
the low-energy theorem, Eq.~(\ref{eq:Thomson2}), independently of the wave 
function chosen. Therefore and because of the nearly energy-independent offset 
between the cross sections calculated with the chiral wave 
function~\cite{Epelbaum} and the AV18-wave function~\cite{AV18}, which we 
observe in Fig.~\ref{fig:wavefdep}, it is not surprising that also at 
non-zero energies the wave-function dependence is reduced with respect to our 
previous approach, where we did not have this low-energy constraint. In fact, 
the remaining dependence is of the order of 1\% and therefore nearly 
invisible, 
cf. Fig.~\ref{fig:wavefunctiondep2}, where we compare our cross sections 
calculated with two of the extreme wave functions of Fig.~\ref{fig:wavefdep}: 
the AV18 \cite{AV18} and the NNLO $\chi$PT \cite{Epelbaum} wave function (the 
same observation holds for other state-of-the-art deuteron wave functions). 
This is another important success of the non-perturbative approach, as it 
demonstrates that our calculation is not sensitive to details of 
short-distance physics anymore. The 10\%-effect observed in 
Fig.~\ref{fig:wavefdep}, however, manifests a much stronger sensitivity to 
short-distance effects of the wave function 
than would be expected from a low-energy Effective Field Theory, as discussed
in Section~\ref{sec:wfdep}.

\subsection{Dependence on the Potential 
\label{sec:potentialdep}}
In this subsection we investigate briefly the sensitivity of our 
calculation to the $np$-potential. Usually we use the 
AV18-potential~\cite{AV18}, 
given in App.~\ref{app:AV18}, which provides an excellent theoretical 
description of the Nijmegen partial-wave analysis. Here we compare 
our results achieved with this modern 'high-precision' potential to those when 
we use the leading-order chiral potential, which includes only the one-pion
exchange and a simple parameterization of short-distance effects via two
point-like, momentum-independent contact operators. This potential is given 
e.g. in~\cite{Rho} as
\ba
V_\mathrm{LO}^{^1\!S_0}(\vec{r})&=-f^2\,v_\Lambda(r)
+\frac{4\pi}{m_N}\,C_0^{^1\!S_0}\,\delta_\Lambda^3(r),\nonumber\\  
V_\mathrm{LO}^d(\vec{r})&=
-f^2\left[v_\Lambda(r)+S_{12}(\hat{r})\,t_\Lambda(r)\right]+
\frac{4\pi}{m_N}\,C_0^d\,\delta_\Lambda^3(r)
\label{eq:VLO}
\end{align} 
for the $^1\!S_0$ and the (deuteron) $^3\!S_1$-$^3\!D_1$ channel, respectively.
$S_{12}(\hat{r})=3\,(\vec{\sigma}_1\cdot\hat{r})\,(\vec{\sigma}_2\cdot\hat{r})-
\vec{\sigma}_1\cdot\vec{\sigma}_2$ is the tensor operator, see also 
Appendix~\ref{app:AV18}. 

The authors of Ref.~\cite{Rho} use a 
Gaussian regulator in order to render the pion-exchange potential finite at the
origin. The resulting central and tensor potential reads
\ba
\label{eq:vLambda}
v_\Lambda(r)&=\frac{1}{2r}
\left[\e^{-m_\pi r}\,
\mathrm{erfc}\left(\frac{-\Lambda r+\frac{m_\pi}{\Lambda}}{\sqrt{2}}\right)
     -\e^{ m_\pi r}\,
\mathrm{erfc}\left(\frac{ \Lambda r+\frac{m_\pi}{\Lambda}}{\sqrt{2}}\right)
\right],\\
\label{eq:tLambda}
t_\Lambda(r)&=\frac{r}{m_\pi^2}\frac{\partial}{\partial r}\frac{1}{r}
\frac{\partial}{\partial r}\,v_\Lambda(r)
\end{align}
with 
$$\mathrm{erfc}(x)=
1-\mathrm{erf}(x)=1-\frac{2}{\sqrt{\pi}}\int_0^xdt\,\e^{-t^2}.$$
The functions given in Eqs.~(\ref{eq:vLambda}, \ref{eq:tLambda}) become 
Yukawa functions for large distances:
\ba
v_\Lambda(r)\bigg|_{r\rightarrow\infty}&=\frac{\e^{-m_\pi r}}{r}\\
t_\Lambda(r)\bigg|_{r\rightarrow\infty}&=
\left(1+\frac{3}{m_\pi r}+\frac{3}{(m_\pi r)^2}\right)\,
\frac{\e^{-m_\pi r}}{r}
\end{align}
$\delta_\Lambda^3(r)$ is the Fourier transform of the regulator,
\be
\delta_\Lambda^3(r)=\int\frac{d^3q}{(2\pi)^3}\,\e^{i\vec{q}\cdot\vec{r}}\,
\e^{-\frac{q^2}{2\Lambda^2}}=\frac{\Lambda^3\,\e^{-\frac{\Lambda^2\,r^2}{2}}}
{(2\pi)^{3/2}}.
\ee

At leading order, there are two free parameters, $C_0^{^1\!S_0}$ and $C_0^d$, 
cf. Eq.~(\ref{eq:VLO}). $C_0^{^1\!S_0}$ 
has been fixed in~\cite{Rho} via the $^1\!S_0$ scattering length, 
$a_0\approx-23.75$~fm, $C_0^d$ at the deuteron binding energy. 
We use these two constants to parameterize any short-distance physics in the
spin-singlet and -triplet channel, respectively.
The results for $C_0^{^1\!S_0}$ and $C_0^d$, achieved in~\cite{Rho} for the 
cutoff-value $\Lambda=600$~MeV, are given in Table~\ref{tab:Rho}.
\begin{table}[!htb] 
\begin{center}
\begin{tabular}{|l|l|}
\hline 
$C_0^{^1\!S_0}$&$C_0^d$\\
\hline
$-0.422$~fm&$0.795$~fm\\
\hline 
\end{tabular}
\end{center}
\caption[Parameters of the LO chiral potential]
{Parameters of the LO chiral potential as determined in~\cite{Rho} for 
$\Lambda=600$~MeV.}
\label{tab:Rho}
\end{table}

\begin{figure}[!htb]
\begin{center}
\includegraphics*[width=.48\linewidth]{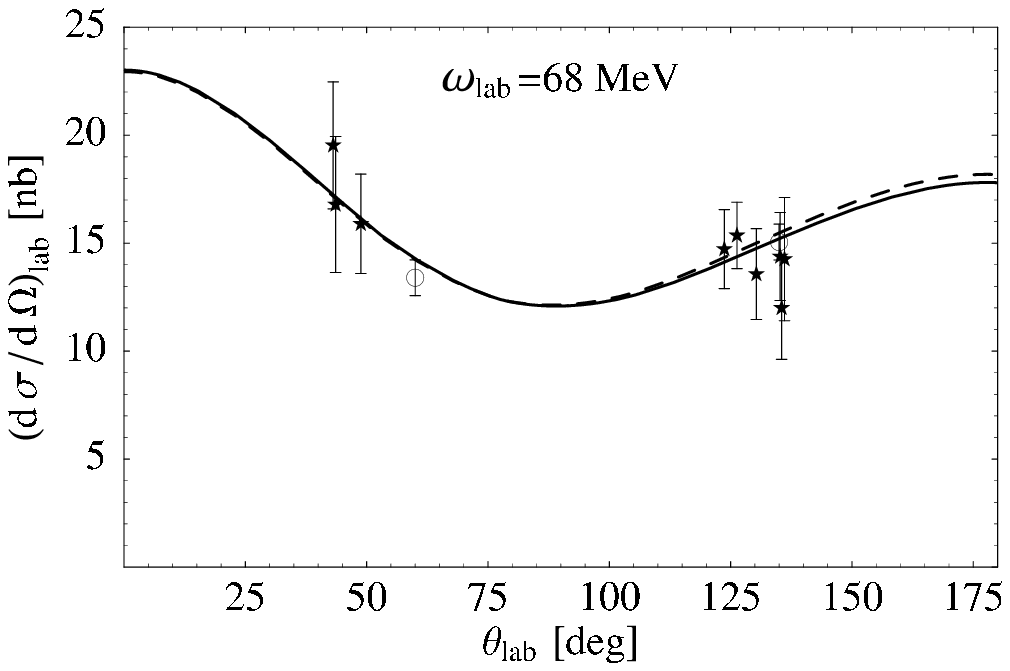}
\hfill
\includegraphics*[width=.48\linewidth]{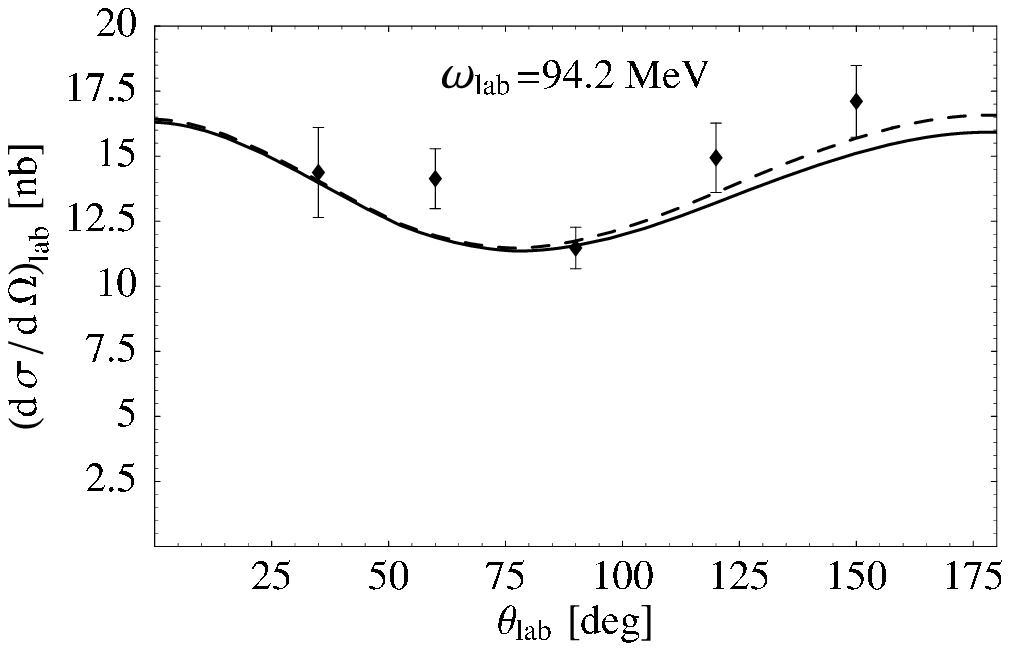}
\includegraphics*[width=.48\linewidth]{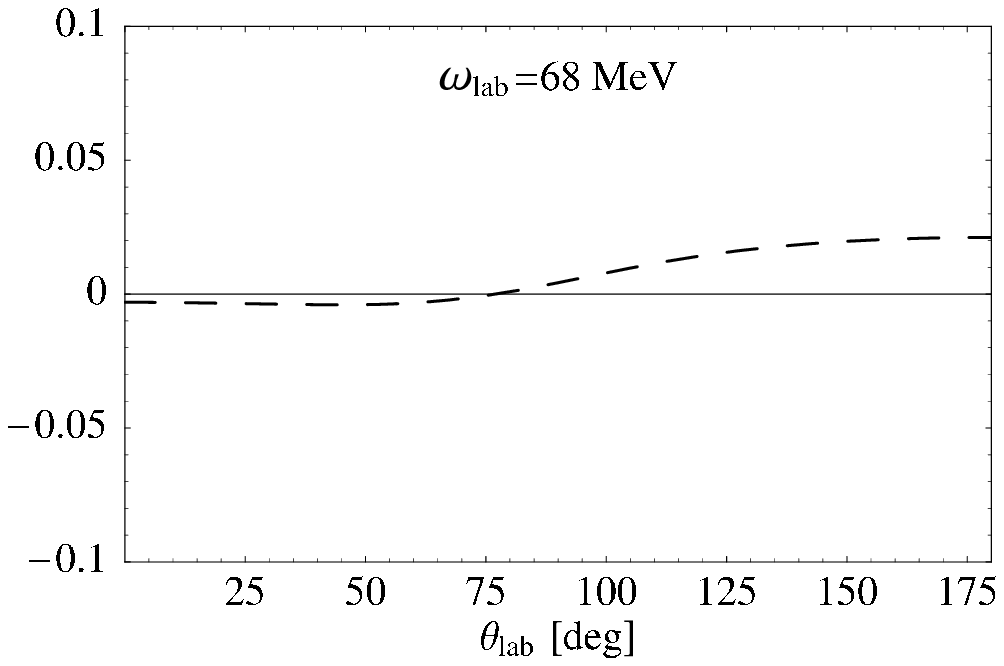}
\hfill
\includegraphics*[width=.48\linewidth]{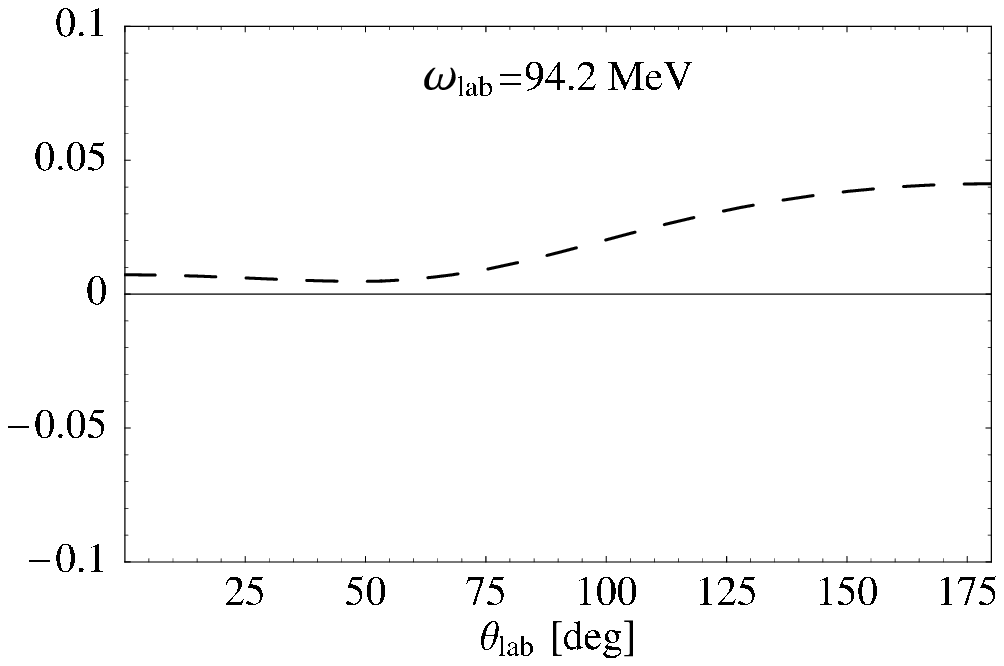}
\caption[Comparison of our result at 68~MeV and 94.2~MeV using two different 
$np$-potentials]
{Upper panels: Comparison of our result at 68~MeV and 94.2~MeV using two 
different $np$-potentials: the AV18-potential~\cite{AV18} (solid) and the LO 
chiral potential~\cite{Rho} (dashed). For both curves the chiral wave 
function~\cite{Epelbaum} has been used.
Lower panels: Corresponding error plots, calculated in analogy to 
Fig.~\ref{fig:Thomson2}.}
\label{fig:LOchiPT}
\end{center}
\end{figure}

Even with this rather crude approximation of the 
neutron-proton interaction we obtain results close to those 
of the AV18-potential, cf. Fig.~\ref{fig:LOchiPT}. 
The observed deviations are small (of the order of $\leq4\%$), even for 
photon energies close to 100~MeV. Obviously 
the one-pion-exchange potential, adequately regulated for $r\rightarrow 0$, 
together with the simplest parameterization of the hard core gives an  
approximation of the potential which is well sufficient for the process 
under consideration. We conclude that we are mainly sensitive to the 
long-range part of the potential.
Nevertheless, there are minor deviations visible 
in Fig.~\ref{fig:LOchiPT}, which justify the application of a more 
sophisticated potential than the leading-order chiral one. 
Not surprisingly, these deviations, which arise due to the poor description
of high-energy dynamics in the LO chiral potential, increase with increasing 
photon energy.

\section[Total Deuteron-Photodisintegration Cross Section]
{Total Deuteron-Photodisintegration\\Cross Section
\label{sec:photodisintegration}}

Besides complying with the low-energy theorem, cf. 
Section~\ref{sec:Thomson2}, another important check on our calculation
is to extract total deuteron-photodisintegration cross 
sections from the Compton amplitude via the optical theorem. This process has 
been studied more extensively than elastic deuteron Compton scattering and 
there is plenty of data below 100~MeV to compare with. In 
\cite{Karakowski}, a comparison to an older calculation of the process 
\cite{Partovi} is given, not only of the sum of all contributions, but also 
of several terms separately, which we use to cross-check our results  for 
certain amplitudes.

The optical theorem in our normalization reads
\be
\sigma^\mathrm{tot}=
\frac{1}{\w}\cdot\frac{1}{6}\,\sum_{i=f}\,\mathrm{Im}[\Mfi{}(\theta=0)],
\ee
i.e. the total cross section is the sum over the imaginary part of all deuteron
Compton amplitudes in the forward direction with identical initial and final 
states ($\lf=\li$, $M_f=M_i$), divided by the photon energy $\w$.
Like in Eq.~(\ref{eq:deuteroncrosssection}), this sum is divided by 6, as 
we have to average over the initial states.

We calculate this cross section in the lab frame, in order to be able to
compare to data and the theoretical works \cite{Partovi, Karakowski}\footnote{
We note that many more authors have been working on this process, 
see e.g. \cite{Levchukdisintegration} or \cite{Arenhoevelreview} and 
references therein.}. The amplitudes given in the appendix have been derived 
in the $\gamma d$-cm frame, but they are easily transformed into the lab frame.
First we note that we only need to sum over the $s$-channel diagrams, as only
they become complex for photon energies above the deuteron binding energy $B$, 
while the $u$-channel amplitudes stay real for all photon energies, cf. 
Section~\ref{sec:dominant}. As the authors of Ref.~\cite{Karakowski} 
calculate in the lab frame, we convert our calculation according to their 
work. In the $s$-channel the only change is 
$\w+\frac{\w^2}{2m_d}\leftrightarrow\w-\frac{\w^2}{2m_d}$, because in the lab 
frame, the deuteron's initial kinetic energy vanishes, whereas the total 
intermediate momentum is $\vec{P}_C=\vec{k}_i$. In the cm frame we have 
$\vec{P}_i=-\vec{k}_i$ and $\vec{P}_C=\vec{0}$ in the $s$-channel.

\begin{figure}[!htb]
\begin{center}
\includegraphics*[width=.6\linewidth]{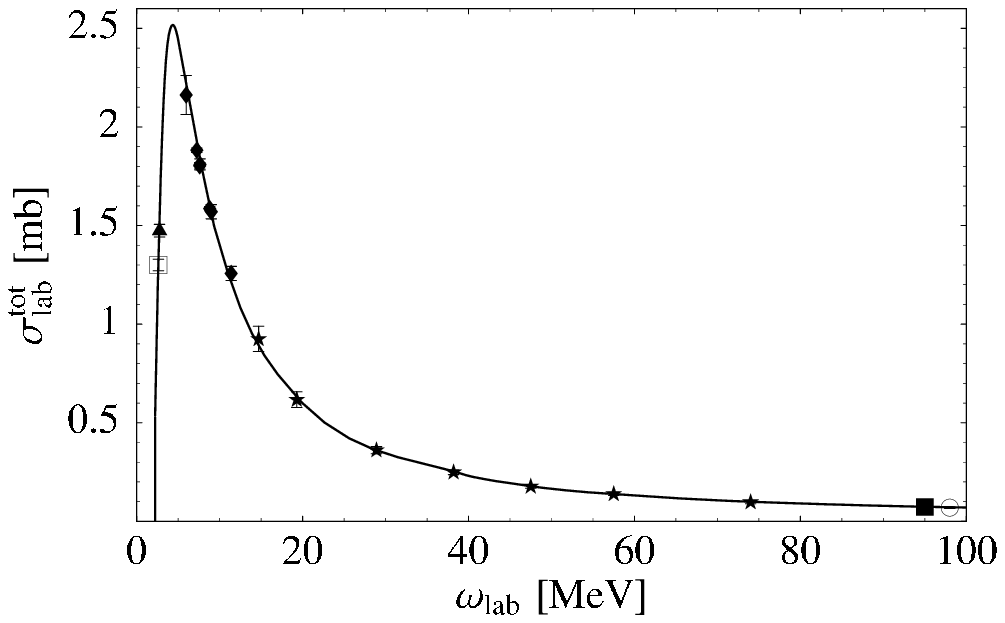}\\
\includegraphics*[width=.48\linewidth]{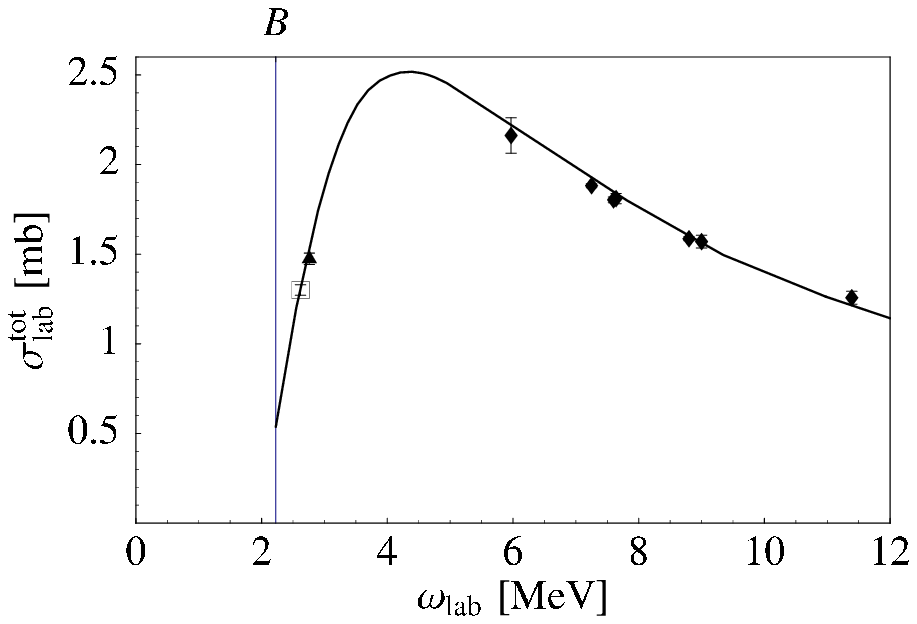}
\hfill
\includegraphics*[width=.48\linewidth]{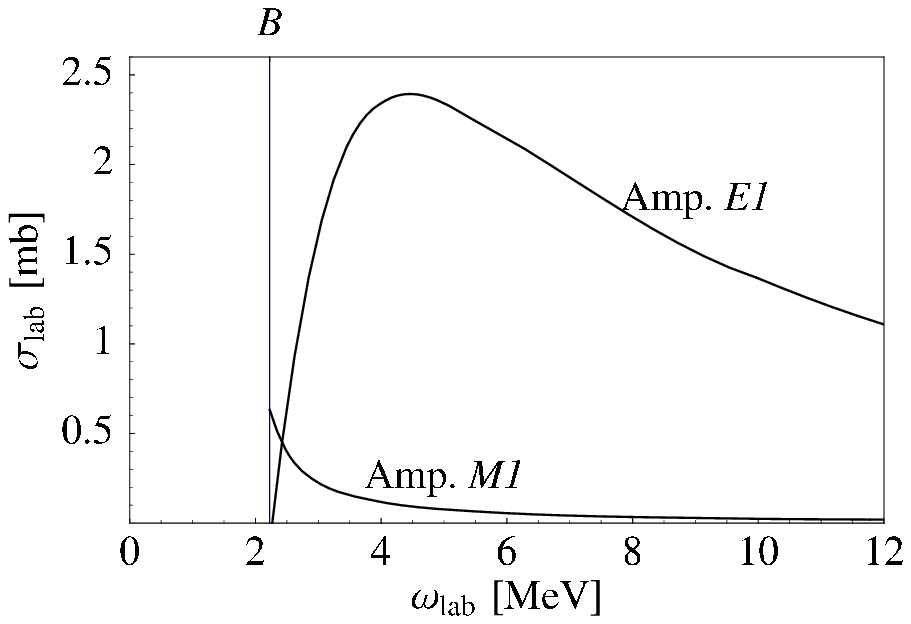}
\caption[Total deuteron-photodisintegration cross section]
{Total deuteron-photodisintegration cross section derived from our deuteron 
Compton amplitudes, 
together with data from \cite{MaB} (open box), \cite{Birenbaum} (diamond), 
\cite{Bernabei} (star), \cite{Meyer} (box), \cite{Sanctis} (circle).
The triangle corresponds to the weighted average of the data measured at 
2.76~MeV \cite{Moreh, SnB, MMC}, as determined in~\cite{Arenhoevelreview}.
'Amp.~$E1$, $M1$' denotes the contributions from the $E1$- and the singlet 
$M1$-transition, respectively. $B$ is the binding energy of the deuteron.}
\label{fig:photodisintegration}
\end{center}
\end{figure}

Our result for the total deuteron-photodisintegration cross section is shown 
in Fig.~\ref{fig:photodisintegration}, together with data from 
[\ref{MAB}-\ref{MMC}], 
which are 
described well by our calculation. In the lower left panel the low-energy 
regime is enlarged, in order to emphasize the non-vanishing value at threshold.
The by far most important contribution at threshold stems from the singlet 
$M1$-transition of $\Mfi{\sigma1\,\sigma1\,a}$, see 
Appendix~\ref{app:subleading}. It corresponds to the operator 
$[Y_{0}\otimes t]_1$, which transforms the deuteron into a (singlet) 
$S_C=0$-state, cf. Eq.~(\ref{eq:mxt}). $M1$ is the shorthand notation for 
the magnetic coupling of a photon with $L=1$. 
Nevertheless, as is well-known, already 1~MeV above threshold the cross 
section is completely dominated by the amplitude $\Mfi{\phi\phi\,1}$, where 
for $L,L'=1$ we have an $E1$-interaction at each vertex, and this dominance 
holds for all higher energies. In \cite{Partovi}, the $E1$-contribution is 
called 'Approximation A', whereas 'Approximation B' is the singlet $M1$ 
amplitude added to this first approximation. We simplify the notation 
and call the $E1$-transition 'Amp.~$E1$', the singlet $M1$-amplitude 
'Amp.~$M1$'. These two (most important) contributions to the total 
photodisintegration cross section are plotted in the lower right panel of 
Fig.~\ref{fig:photodisintegration}, demonstrating the well-known rise 
of 'Amp.~$M1$', as $\w$ approaches the breakup threshold, cf. e.g. 
\cite{Arenhoevelreview, Brown}, whereas 'Amp.~$E1$' is zero for $\w=B$. Note
that Amp.~$E1$ not only consists of $\Mfi{\phi\phi\,1}$ but of all amplitudes 
with an $E1$-interaction at the vertex of the incoming photon. Another example
for such an amplitude is $\Mfi{\phi\,\sigma2\,b}$ 
(Appendix~\ref{app:subleading}).

Strictly speaking there are also contributions from the one-body current
$\Jp\ofxi$, Eq.~(\ref{eq:Jp}). The corresponding amplitudes are given in 
\cite{Karakowski} but are not written
down in this work, as we found that their contributions to the elastic deuteron
Compton cross sections are tiny (of the order of 1\%) 
and so is their effect on the total
disintegration cross section. Nevertheless, in the high-energy regime of our 
calculation, say for $\w\sim 100$~MeV, they do give visible contributions to 
Amp.~$E1$, but these cancel nearly exactly against other terms which also
contain $\Jp$.
Therefore, when we only look at the sum of all amplitudes contributing to 
$\sigma^\mathrm{tot}$, we may well neglect the current $\Jp$, cf. 
Table~\ref{tab:photodisintegration}. However, in this table we  
compare our results for Amp.~$E1$,$\;$Amp.~$M1$ with those of 
\cite{Partovi, Karakowski}, where $\Jp$ was included.
Therefore we give two numbers for Amp.~$E1$ and $\sigma^\mathrm{tot}$, 
the one in bracket including $\Jp$, the other one not including 
this current. 
We report contributions from $\Jp$ 
only for reasons of cross-checking our calculation, and
we find that the deviations from \cite{Partovi, Karakowski} may well be 
attributed to the use of different wave functions and potentials. This also 
holds for $\sigma^\mathrm{tot}$, which does, however, not include the two-body 
(Kroll-Ruderman) current diagrams of Appendix~\ref{app:two-body}, as explicit
meson-exchange currents have been neglected in \cite{Partovi}. 

\begin{table}[!htb] 
\begin{center}
\begin{tabular}{|c|c|c|c|}
\hline 
 &\cite{Partovi}&\cite{Karakowski} & this work  \\
\hline
\hline
 & \multicolumn{3}{|c|}{20~MeV} \\ 
\hline 
Amp.~$E1$~$[\mu b]$&579.1&583.3&580.0 $\;$ (583.5)\\
\hline 
Amp.~$M1$~$[\mu b]$&10.1&9.9&9.4\\
\hline 
$\sigma^\mathrm{tot}$~$[\mu b]$&588.2&591.2&594.4 $\;$ (594.7)\\
\hline 
\hline 
 & \multicolumn{3}{|c|}{80~MeV} \\ 
\hline 
Amp.~$E1$~$[\mu b]$&77.2&80.5&75.8 $\;$ (79.1)\\
\hline 
Amp.~$M1$~$[\mu b]$&6.4&5.3&5.9\\
\hline 
$\sigma^\mathrm{tot}$~$[\mu b]$&87.4&86.4&87.4 $\;$ (88.2)\\
\hline 
\hline 
 & \multicolumn{3}{|c|}{140~MeV} \\ 
\hline 
Amp.~$E1$~$[\mu b]$&34.0&34.6&33.2 $\;$ (36.1)\\
\hline 
Amp.~$M1$~$[\mu b]$&5.7&4.0&4.7\\
\hline 
$\sigma^\mathrm{tot}$~$[\mu b]$&44.5&39.5&44.5 $\;$ (44.9)\\
\hline 
\end{tabular}
\end{center}
\caption[Comparison of the dominant contributions to the total 
deuteron-photodisintegration cross section]
{Comparison of our results for the two dominant amplitudes contributing to 
the total deuteron-photodisintegration cross section with former works at 
three different energies. The total cross section is also compared, 
however excluding diagrams with exlicit pion exchange. Amp.~$E1$ 
denotes the contribution due to an $E1$-,
Amp.~$M1$ due to a singlet $M1$-transition. The numbers in brackets include 
contributions from $\Jp\ofxi$.}
\label{tab:photodisintegration}
\end{table}

We also compare our results with predictions for the strengths of 
electric and magnetic transitions close to threshold from the 
\textit{Effective Range Expansion}~\cite{Bethe1, Bethe2}, given by 
\be
\sigma_\mathrm{ER}^\mathrm{el}=\frac{2}{3}\,\frac{e^2}{\gamma^2}\,
\frac{(\frac{\w}{B}-1)^{3/2}}{(\frac{\w}{B})^3\,(1-\gamma\,r_t)},
\label{eq:EREel}
\ee
\be
\sigma_\mathrm{ER}^\mathrm{mag}=\frac{1}{6}\,\frac{e^2}{m_N^2}\,
(\mu_p-\mu_n)^2\,\frac{k\,\gamma}{k^2+\gamma^2}\,\frac{(1-\gamma\,a_s+
\frac{1}{4}\,a_s\,(r_s+r_t)\,\gamma^2-\frac{1}{4}\,a_s\,(r_s-r_t)\,k^2)^2}
{(1+k^2\,a_s^2)\,(1-\gamma\,r_t)}
\label{eq:EREmag}
\ee
with $\gamma=\sqrt{m_N\,B}$.
The final-state relative momentum is 
$k=|\vec{p}_p-\vec{p}_n|/2=\sqrt{m_N\,(\w-B)}$, and
for   the singlet scattering length $a_s$ and the singlet  
(triplet) effective range $r_s$ ($r_t$) we use 
$a_s=-23.749$~fm, $r_s=2.81$~fm, $r_t=1.76$~fm given in \cite{AV18}.
The explicit form of Eqs.~(\ref{eq:EREel},~\ref{eq:EREmag}) is 
adopted from \cite{Arenhoevelreview}.

In order to determine which amplitudes correspond to electric and magnetic 
transitions, we recall that the gradient part 
of the photon field, cf. Eq.~(\ref{eq:multipoleexpfinal}), as well as 
$\Atwo$, given in Eq.~(\ref{eq:Atwo}), are of electric nature. $\Aone$ 
(Eq.~(\ref{eq:Aone})) constitutes the magnetic part of $\vec{A}$.
Therefore, except for two-body-current contributions, 
$\sigma^\mathrm{el}$ is made up by the amplitudes
$\Mfi{\phi\phi\,1}$, $\Mfi{\phi\,\sigma1,2\,b}$, $\Mfi{\phi\,\sigma2\,a}$ and 
$\Mfi{\sigma2\,\sigma2\,a}$, cf. Appendices~\ref{app:dominant} 
and~\ref{app:subleading}, whereas  
$\sigma^\mathrm{mag}$ consists of the amplitudes  
$\Mfi{\phi\,\sigma1\,a}$ and $\Mfi{\sigma1\,\sigma1\,a}$, given in 
Appendix~\ref{app:subleading}.

The only non-negligible contributions to the total disintegration cross 
section at low energies including the 
Kroll-Ruderman current are the amplitudes $\Mfi{\mathrm{KR}\,\sigma1\,a,b}$, 
cf. Appendix~\ref{app:two-body}. $\Mfi{\mathrm{KR}\,\sigma1\,b}$ contributes 
to $\sigma^\mathrm{mag}$, but as we use the full photon field for 
$H^{\mathrm{int\,KR}}$, a unique assignment of $\Mfi{\mathrm{KR}\,\sigma1\,a}$
is not possible. However, as the amplitudes $\Mfi{\mathrm{KR}\,\sigma1}$ have 
a singlet intermediate state, the magnetic part of the photon 
field dominates because flipping the spin is a typical magnetic effect, 
and therefore we assign both relevant KR~amplitudes to $\sigma^\mathrm{mag}$.

\begin{figure}[!htb]
\begin{center}
\includegraphics*[width=.32\linewidth]{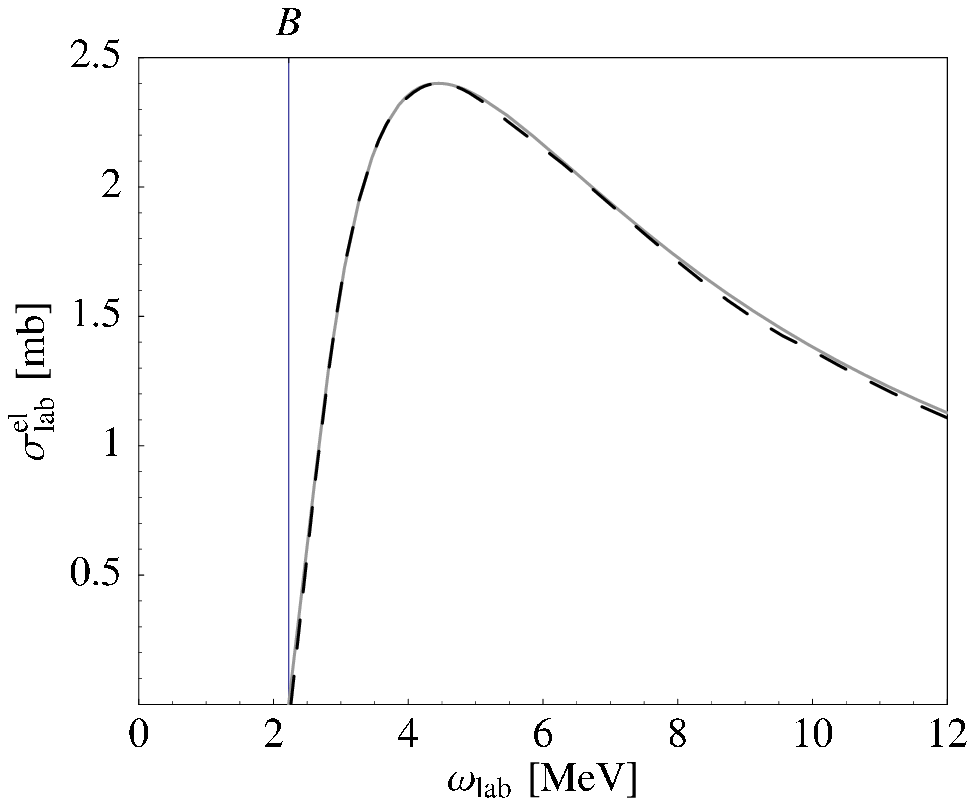}
\includegraphics*[width=.32\linewidth]{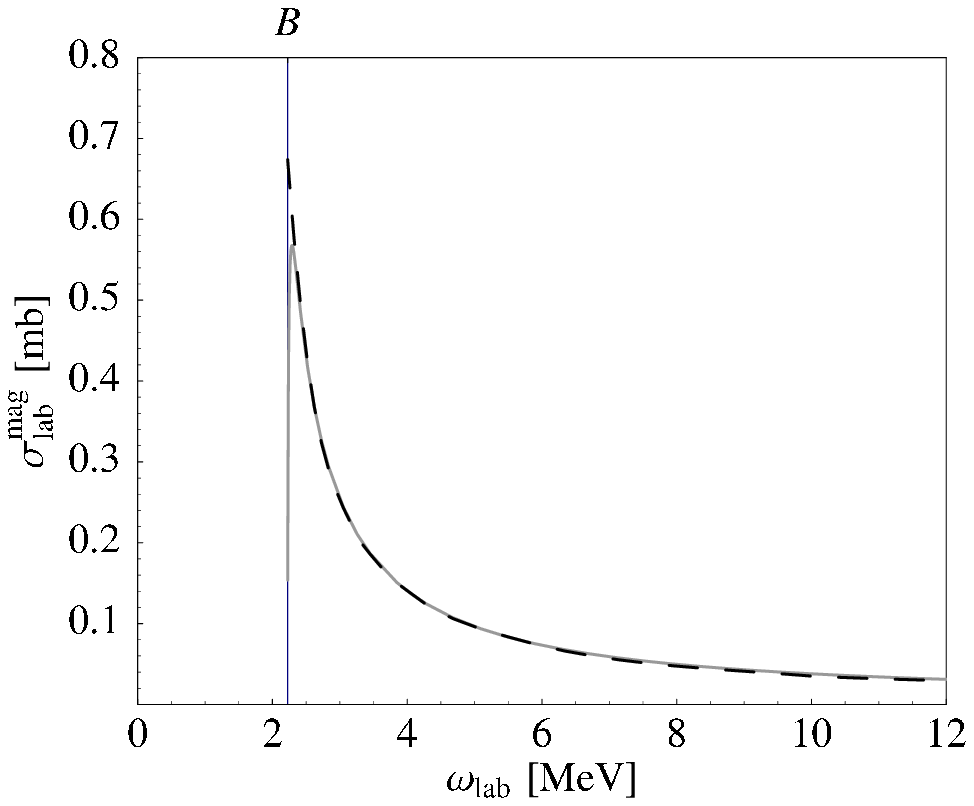}
\includegraphics*[width=.32\linewidth]{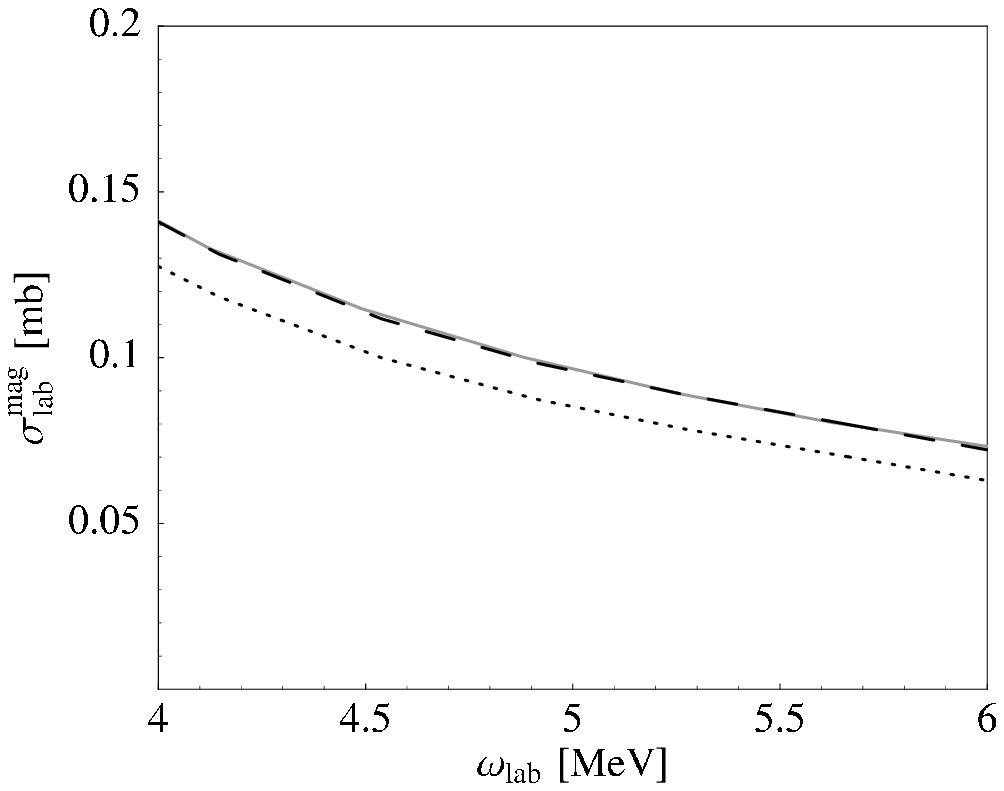}
\caption[Contributions of electric and magnetic transitions to the total 
deuteron-photodisintegration cross section]
{Comparison of our results (dashed) for the contributions of electric (left 
panel) and magnetic (middle and right panel) transitions to the total 
deuteron-photodisintegration cross section with predictions from the 
\textit{Effective Range Expansion} (grey). The dotted curve in the right panel 
does not include the Kroll-Ruderman diagrams.} 
\label{fig:photodisintegration2}
\end{center}
\end{figure}

In Fig.~\ref{fig:photodisintegration2} we compare our results with 
Eqs.~(\ref{eq:EREel}, \ref{eq:EREmag}), finding excellent agreement between 
both approaches. We also demonstrate~-- in the right panel~-- 
the non-negligible size of the KR~diagrams.

The cross-check described in this section, together with the exact 
reproduction of the low-energy theorem, cf. Section~\ref{sec:Thomson2}, gives 
a strong hint that the numerically most important amplitudes have been 
calculated correctly. Therefore, and due to the good agreement of our 
calculation with the elastic deuteron Compton data, which we observed in 
Section~\ref{sec:results2}, we now fit the isoscalar nucleon 
polarizabilities to all existing elastic $\gamma d$ data
in the next section.

\section{Fits of the Isoscalar Polarizabilities \label{sec:fits2}}

Our results for the elastic deuteron Compton cross sections 
obtained with the non-perturbative approach give a good 
description of all existing data, cf. 
Section~\ref{sec:results2}. Therefore, as in Section~\ref{sec:fits1}, we use 
our deuteron Compton cross sections to fit
the static isoscalar nucleon polarizabilities $\bar{\alpha}_{E1}^s$ and 
$\bar{\beta}_{M1}^s$ to these 
data. This time, however, we may use \textit{all} data for the fits, whereas 
in Section~\ref{sec:fits1} we had to restrict ourselves to the experiments 
performed around 68 and 94.2~MeV. 

\begin{figure}[!htb]
\begin{center}
\includegraphics*[width=.4\linewidth]{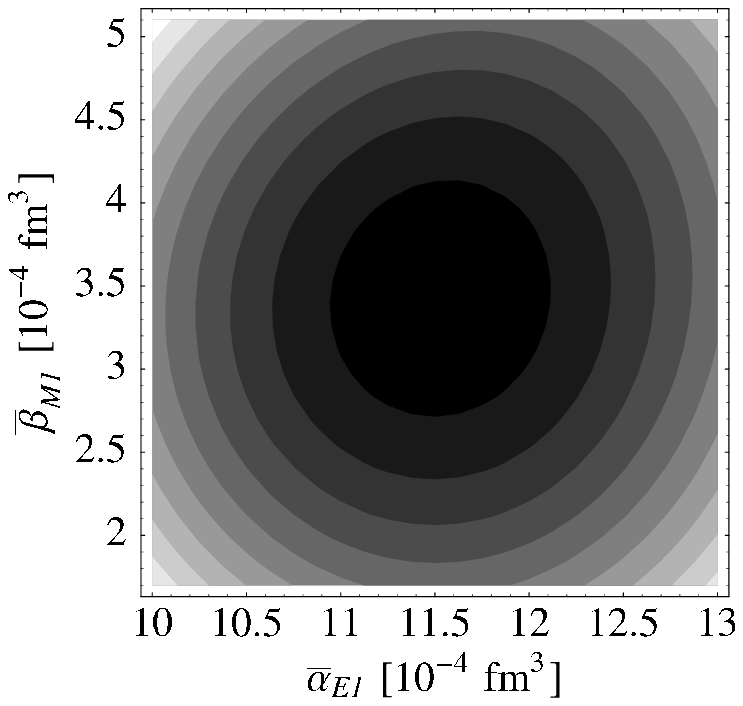}
\hspace{.1\linewidth}
\includegraphics*[width=.4\linewidth]{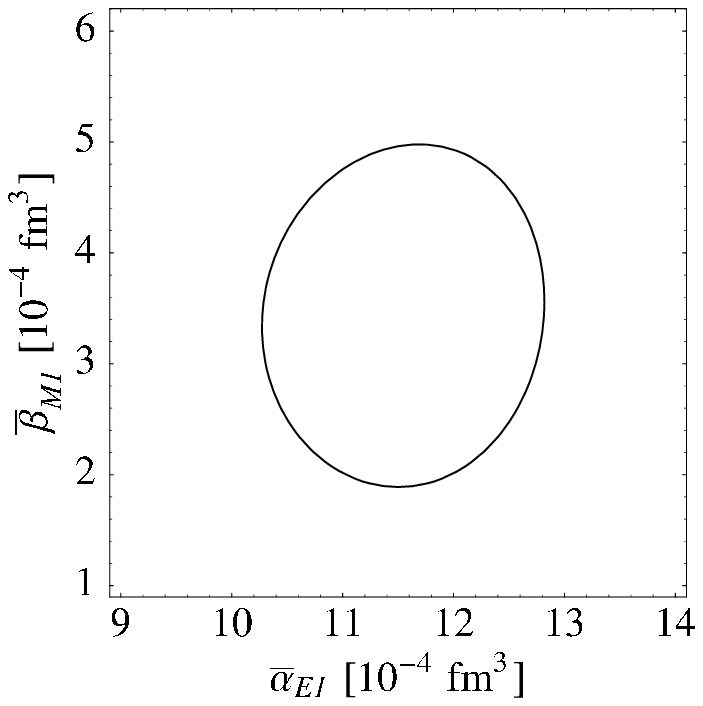}
\caption[Contour plot of the $\chi^2$ achieved in our global 2-parameter fit 
and the 70\%~confidence ellipse]
{Contour plot of the $\chi^2$ achieved in our global 2-parameter
fit for $\bar{\alpha}_{E1}^s$ and $\bar{\beta}_{M1}^s$ 
(left panel) and the 70\%~confidence ellipse (right panel).}
\label{fig:chisquareplot}
\end{center}
\end{figure}

The fitting procedure used here is the same as in Section~\ref{sec:fits1}, i.e.
we do a least-$\chi^2$ fit, cf. 
Eq.~(\ref{eq:chisquared}), using the chiral NNLO wave 
function~\cite{Epelbaum}. Our results for the isoscalar polarizabilities 
from the global fit to all data read
\ba
\left.\phantom{\PCsq}\bar{\alpha}_{E1}^s\right|_\mathrm{global}&=
(11.5\pm1.4\,(\mathrm{stat}))\cdot10^{-4}\;\fm^3,\nonumber\\
\left.\phantom{\PCsq}\bar{\beta}_{M1}^s \right|_\mathrm{global}&=
( 3.4\pm1.6\,(\mathrm{stat}))\cdot10^{-4}\;\fm^3.
\label{eq:globalfit}
\end{align}
We only give the statistical error as we neglect further uncertainties, e.g. 
the error induced by the dependence on the deuteron wave function. 
This error may well be set to zero, due to the tiny wave-function dependence 
observed in Fig.~\ref{fig:wavefunctiondep2}, whereas the wave function 
introduces a sizeable uncertainty in Chapter~\ref{chap:perturbative}, see
Fig.~\ref{fig:wavefdep} and Tables~\ref{tab:fulldata} and \ref{tab:effdata}.
Theoretical errors from higher orders are also neglected, 
albeit we are aware that they 
may be comparable in size with our statistical error: in 
Eq.~(\ref{eq:higherorder}), they were estimated to be 
$|\bar{\alpha}_\mathrm{NLO}|\sim|\bar{\beta}_\mathrm{NLO}|\sim
1\cdot 10^{-4}\; \fm^3 $ from
na\"ive dimensional analysis.
The corresponding $\chi^2$ per degree of freedom is dramatically reduced
with respect to Table~\ref{tab:fulldata}: 
\be
\left.\frac{\chi^2}{d.o.f.}\right|_\mathrm{global}=0.98
\ee
with 27 degrees of freedom (4 data points from \cite{Lucas} at 49~MeV, 9 from 
\cite{Lund} at 55~MeV, 2 from \cite{Lucas} and 9 from \cite{Lund} around 
68~MeV and 5 from \cite{Hornidge} around 94.2~MeV, along with two fit 
parameters). In Fig.~\ref{fig:chisquareplot} we give a contour plot of 
the achieved $\chi^2$, together with the ellipse corresponding to the 
70\%~confidence level.

\begin{figure}[!htb]
\begin{center}
\includegraphics*[width=.48\linewidth]{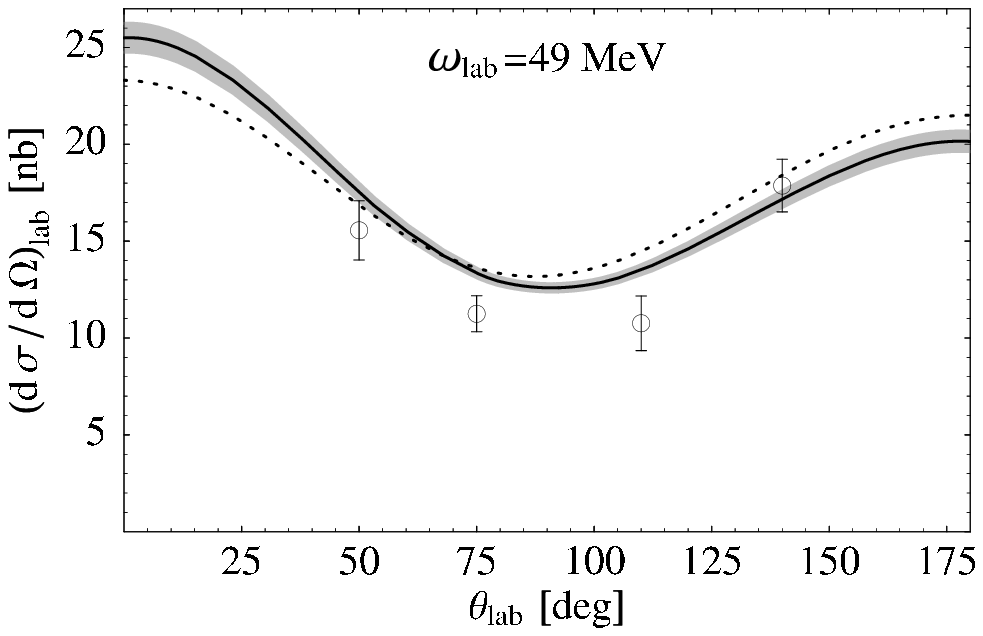}
\hfill
\includegraphics*[width=.48\linewidth]{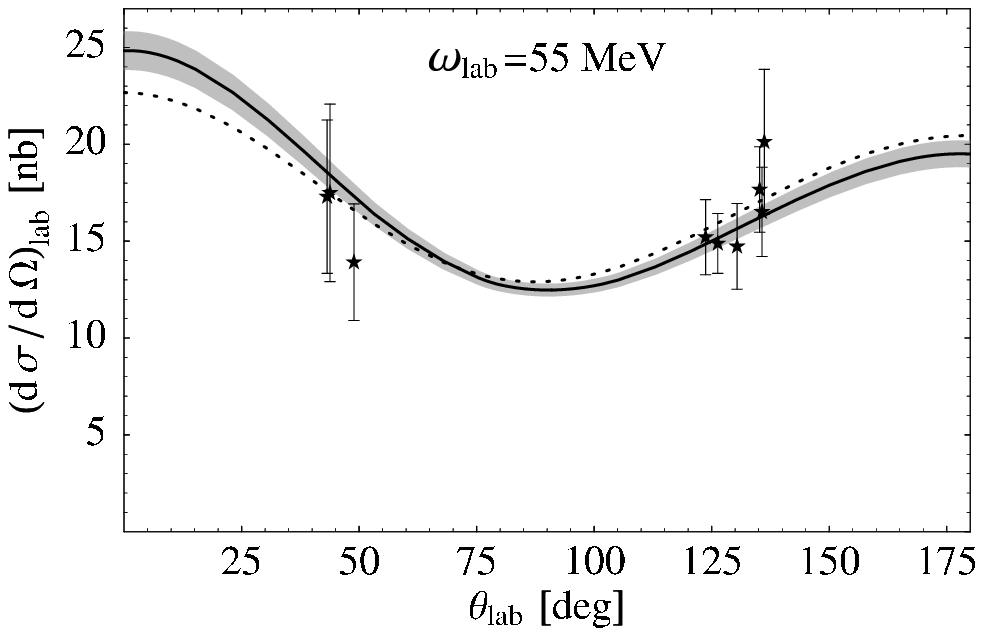}
\\
\includegraphics*[width=.48\linewidth]{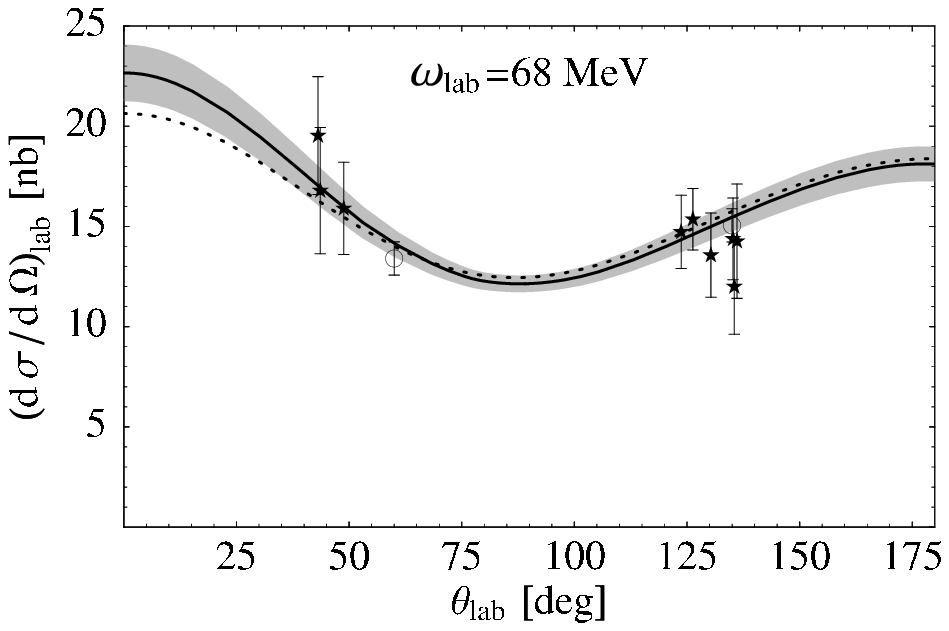}
\hfill
\includegraphics*[width=.48\linewidth]{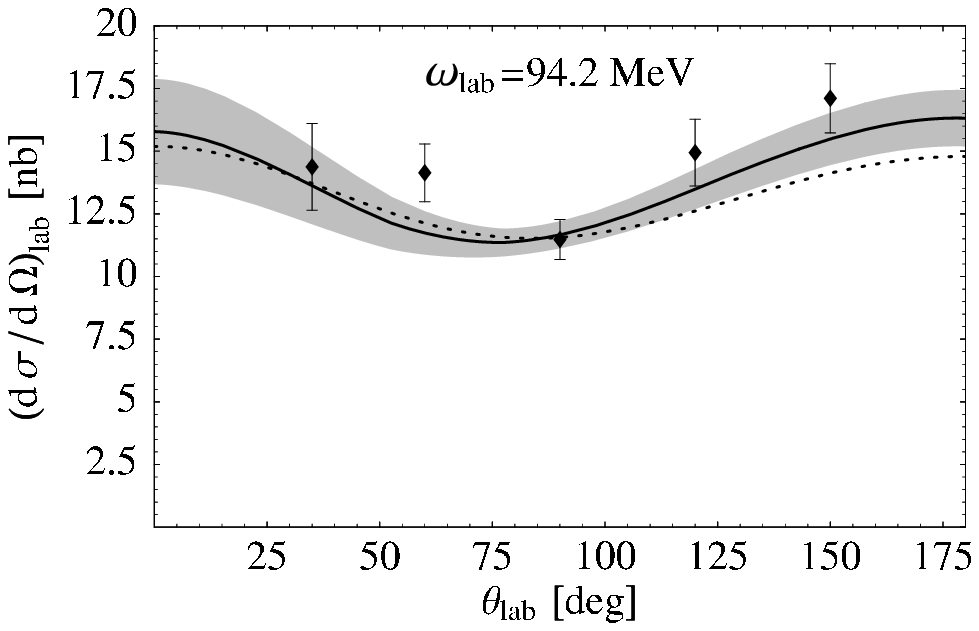}
\\
\includegraphics*[width=.48\linewidth]{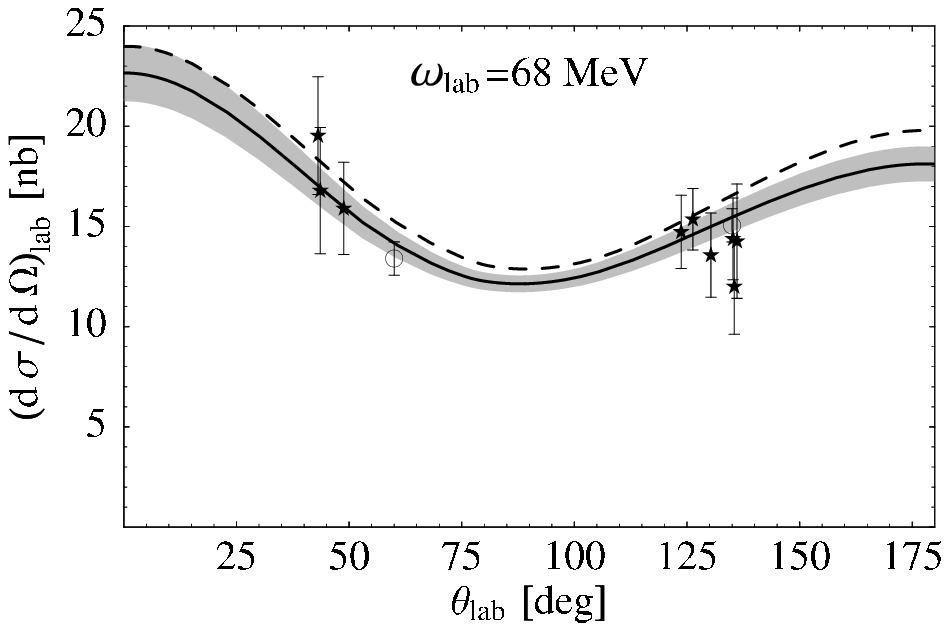}
\hfill
\includegraphics*[width=.48\linewidth]{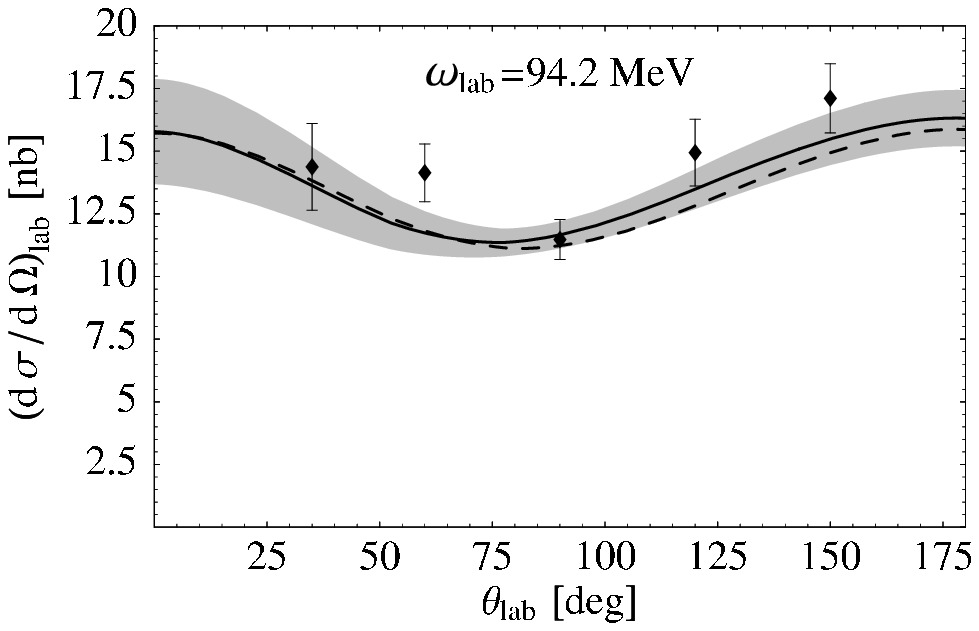}
\caption[Fit of $\bar{\alpha}_{E1}^s$ and $\bar{\beta}_{M1}^s$ to all existing 
elastic $\gamma d$ data]
{Results from a global fit of $\bar{\alpha}_{E1}^s$ and $\bar{\beta}_{M1}^s$ 
to all existing elastic $\gamma d$ data (solid). The grey bands are derived 
from our statistical errors. The dotted line represents ``fit IV'', one of the 
$\mathcal{O}(q ^4)$-HB$\chi$PT fits from Ref.~\cite{McGPhil}, 
with central values $\bar{\alpha}_{E1}^s=11.5$, $\bar{\beta}_{M1}^s=0.3$. For 
the $\mathcal{O}(q ^4)$ calculation the NLO chiral wave function of 
Ref.~\cite{NLO} has been used, whereas our results were derived with the 
NNLO-version of this wave function~\cite{Epelbaum}. In the lower two panels 
we compare to our 2-parameter-fit results from Section~\ref{sec:fits1}, 
using the chiral wave function~\cite{Epelbaum} (dashed).}
\label{fig:2parameterfit2}
\end{center}
\end{figure}

The corresponding plots, together with the (statistical) error 
bands, are given in Fig.~\ref{fig:2parameterfit2}. Our predictions, using our
results for $\bar{\alpha}_{E1}^p$ and  $\bar{\beta} _{M1}^p$  
for the proton \textit{and} the neutron polarizabilities,
describe the data already well, see Fig.~\ref{fig:comparisonpertnonpert}. 
It is therefore no surprise that also the fitted curves are in 
good agreement with experiment. Like in Section~\ref{sec:fits1} we compare our
fit results to ``fit IV'' from Ref.~\cite{McGPhil}, which is the 
$\mathcal{O}(q ^4)$ HB$\chi$PT fit to all data with central values 
$\bar{\alpha}_{E1}^s=11.5$, $\bar{\beta}_{M1}^s=0.3$. The only sizeable 
deviations are again observed at 94.2~MeV in the backward 
direction, due to the $\Delta$-resonance diagram, Fig.~\ref{fig:SSEsingle}(a),
which is not included in the calculation of Ref.~\cite{McGPhil}, as explained 
in detail in Section~\ref{sec:fits1}. In 
the lower two panels of Fig.~\ref{fig:2parameterfit2}, we also compare to our 
2-parameter fit from Section~\ref{sec:unbiased}, which was performed within 
the strictly perturbative approach, using the chiral
 wave function~\cite{Epelbaum} and the effective data set, cf. 
Table~\ref{tab:eff}. Here we observe a constant offset at 68~MeV, 
similarly to Fig.~\ref{fig:comparisonpertnonpert}, whereas at 94.2~MeV the two
curves are quite close to each other.

The value of our global fit for $\bar{\alpha}_{E1}^s$ is slightly smaller, the 
one for $\bar{\beta}_{M1}^s$ slightly larger than the fit results of  
Eq.~(\ref{eq:finalunbiased}). Nevertheless, both extractions agree 
well with each other within their error bars, and there is also very good 
agreement of Eq.~(\ref{eq:globalfit}) with the values quoted in 
\cite{Kossert} and those recommended in~\cite{Schumacher}. 
Furthermore, we find that the numbers given in Eq.~(\ref{eq:globalfit}) add up
nearly exactly to the isoscalar Baldin sum rule, 
$\bar{\alpha}_{E1}^s+\bar{\beta}_{M1}^s=(14.5\pm0.6)\cdot10^{-4}\;\fm^3$, cf.
Eq.~(\ref{eq:Baldin}). Therefore, in order to reduce the statistical error, 
we repeat our global fit, using the central sum-rule value as an additional 
fit constraint like in Section~\ref{sec:fits1}. The results are 
\ba
\left.\phantom{\PCsq}\bar{\alpha}_{E1}^s\right|_\mathrm{global\;Baldin}&=
(11.3\pm0.7\,(\mathrm{stat})\pm0.6\,(\mathrm{Baldin}))\cdot10^{-4}\;\fm^3,
\nonumber\\
\left.\phantom{\PCsq}\bar{\beta}_{M1}^s \right|_\mathrm{global\;Baldin}&=
( 3.2\mp0.7\,(\mathrm{stat})\pm0.6\,(\mathrm{Baldin}))\cdot10^{-4}\;\fm^3
\label{eq:globalfitBaldin}
\end{align}
with $\chi^{2}/d.o.f.=0.95$.
Of course the central values of Eq.~(\ref{eq:globalfitBaldin}) are very 
similar to the ones of Eq.~(\ref{eq:globalfit}), due to 
the nearly perfect agreement of the 2-parameter-fit result
with the sum-rule value. However, the statistical error is reduced by 
about 50\%.

\begin{figure}[!htb]
\begin{center}
\includegraphics*[width=.48\linewidth]{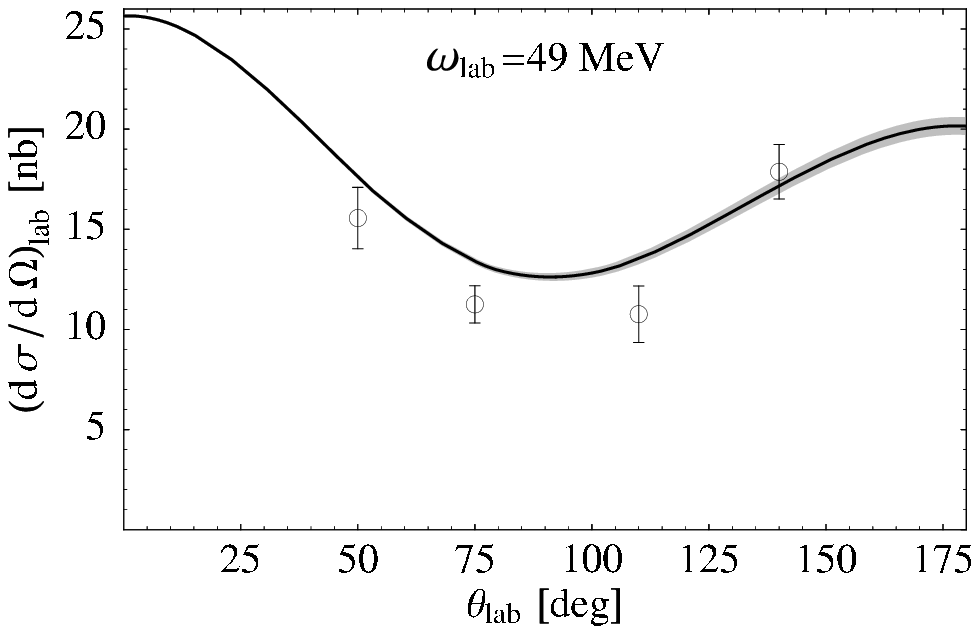}
\hfill
\includegraphics*[width=.48\linewidth]{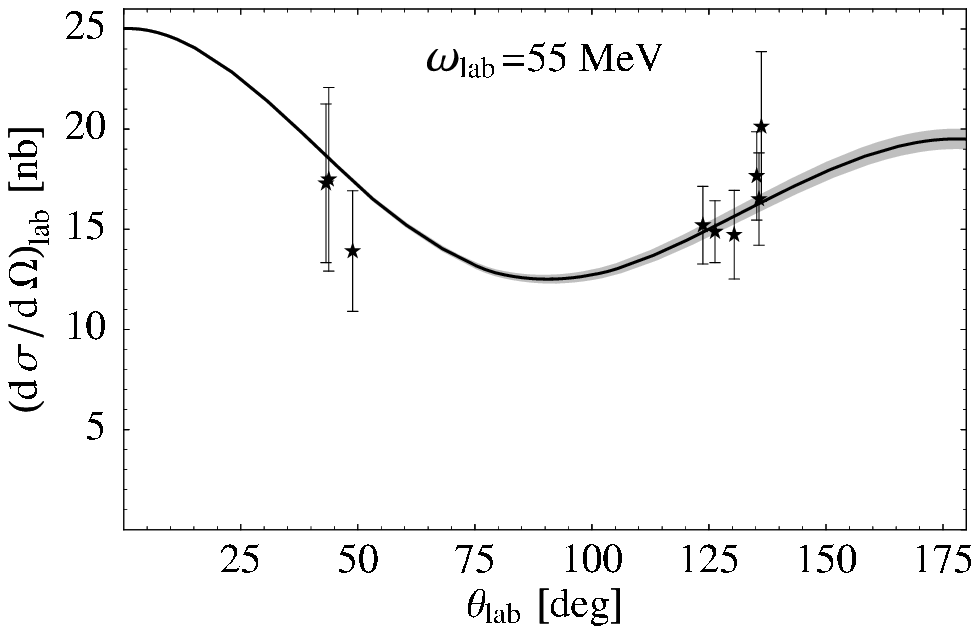}
\\
\includegraphics*[width=.48\linewidth]{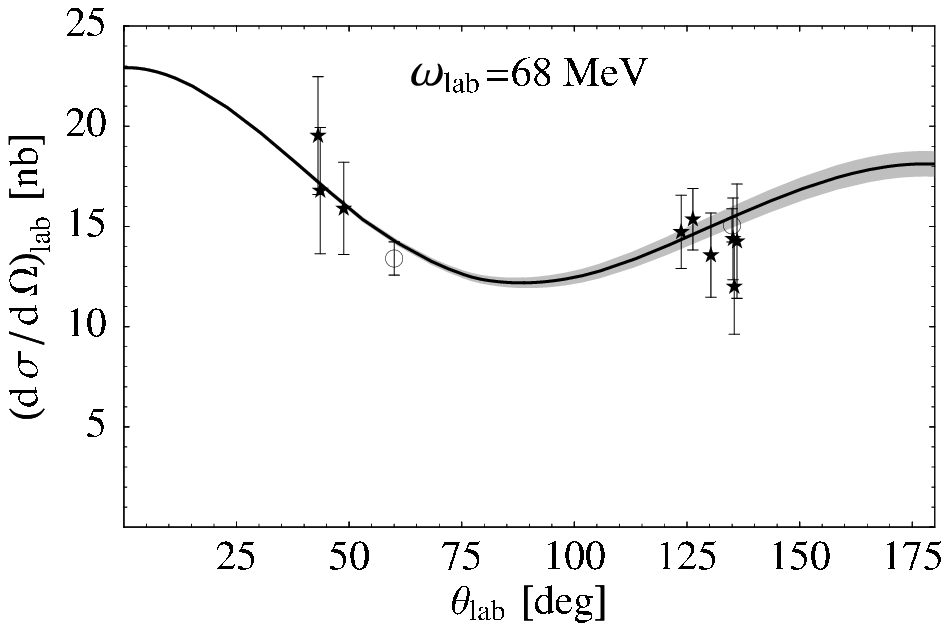}
\hfill
\includegraphics*[width=.48\linewidth]{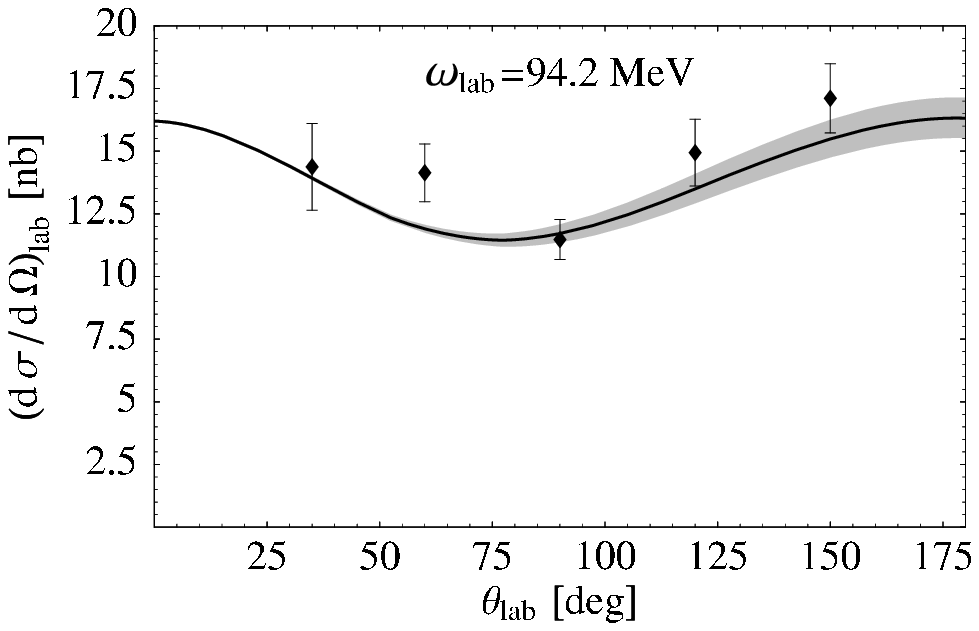}
\caption[Fit of $\bar{\alpha}_{E1}^s$ to all existing elastic $\gamma d$ data]
{Results from a global fit of $\bar{\alpha}_{E1}^s$ to all existing 
elastic $\gamma d$ data, using the chiral wave function~\cite{Epelbaum}. 
$\bar{\beta}_{M1}^s$ is fixed via the Baldin sum rule 
$\bar{\alpha}_{E1}^s+\bar{\beta}_{M1}^s=14.5\cdot10^{-4}\;\fm^3$. 
The grey bands are derived from our statistical errors.}
\label{fig:2parameterfitBaldin2}
\end{center}
\end{figure}

The plots arising from the global, Baldin-constrained fit, together with the 
corresponding error bars 
are shown in Fig.~\ref{fig:2parameterfitBaldin2}.
The central curves are nearly indistinguishable from 
the ones of Fig.~\ref{fig:2parameterfit2}.

Combining Eqs.~(\ref{eq:globalfit}) or (\ref{eq:globalfitBaldin}), 
respectively, with the (Baldin-constrained) 
2-parameter-fit results of Table~\ref{tab:protonfit}, 
we calculate the values for the neutron polarizabilities as 
\ba
\left.\phantom{\PCsq}\bar{\alpha}_{E1}^n\right|_\mathrm{global}&=
(12.0\pm2.0\,(\mathrm{stat})\pm0.4\,(\text{Baldin}))\cdot10^{-4}\;\fm^3,
\nonumber\\
\left.\phantom{\PCsq}\bar{\beta}_{M1}^n \right|_\mathrm{global}&=
( 4.0\pm2.1\,(\mathrm{stat})\pm0.4\,(\text{Baldin}))\cdot10^{-4}\;\fm^3
\label{eq:globalfitneutron}
\end{align}
for the 2-parameter fit and 
\ba
\left.\phantom{\PCsq}\bar{\alpha}_{E1}^n\right|_\mathrm{global\;Baldin}&=
(11.6\pm1.5\,(\mathrm{stat})\pm0.6\,(\mathrm{Baldin}))\cdot10^{-4}\;\fm^3,
\nonumber\\
\left.\phantom{\PCsq}\bar{\beta}_{M1}^n \right|_\mathrm{global\;Baldin}&=
( 3.6\mp1.5\,(\mathrm{stat})\pm0.6\,(\mathrm{Baldin}))\cdot10^{-4}\;\fm^3
\label{eq:globalfitBaldinneutron}
\end{align}
for the fit of the isoscalar polarizabilities 
including the Baldin constraint. We regard these values to be 
the most reliable ones of this work, as the isoscalar 
polarizabilities, from which they are derived, have been determined by 
fitting our deuteron Compton calculation, which 
fulfills the low-energy theorem, to all existing elastic deuteron 
Compton-scattering data. That means there is no restriction on either the 
energy, as in Section~\ref{sec:fits1}, or on the angle, like 
in \cite{McGPhil}. From these values we deduce that the neutron is paramagnetic
and that the isovector polarizabilities are considerably smaller than the  
isoscalar ones. In other words, our analysis shows that within the precision of
the published data, elastic Compton scattering from the proton and the 
deuteron is in agreement with 
\ba
\bar{\alpha}_{E1}^p&\approx\bar{\alpha}_{E1}^n,\nonumber\\
\bar{\beta}_{ M1}^p&\approx\bar{\beta}_{ M1}^n.
\end{align}
These findings agree with those of Section~\ref{sec:fits1} and of
Refs.~\cite{Schumacher, Kossert}.

\chapter{Conclusion \label{chap:conclusion} }
In this work, Compton scattering from the single nucleon and the deuteron has 
been studied theoretically. The framework that we choose for our investigations
is a Chiral Effective Field Theory based on Heavy Baryon Chiral Perturbation 
Theory, extended for the $\Delta(1232)$ resonance as an explicit degree of 
freedom. 
We also treat important non-perturbative aspects of the $NN$-system in 
Chapter~\ref{chap:nonperturbative} on the deuteron.

One of the central aims of our studies is to extract both the proton and the 
neutron polarizabilities from experiments. For the proton polarizabilities 
one can rely on the wealth of elastic 
proton Compton-scattering data, see e.g. \cite{Olmos,Hallin,Fed91,Mac95}.
A single-neutron target does not exist, however. 
Therefore, in order to extract the neutron polarizabilities one 
depends on other methods, such as quasi-free Compton scattering from the
neutron bound in the deuteron, or elastic deuteron Compton scattering. In 
this work we have chosen the latter process for our extraction of the neutron 
polarizabilities. Strictly speaking, from such experiment one cannot 
directly determine the neutron polarizabilities, because the deuteron, i.e. the
bound state of a proton and a neutron, is an isoscalar target. Therefore we 
use the experiments performed at Illinois, Lund and Saskatoon 
\cite{Lucas,Lund,Hornidge} to extract the isoscalar polarizabilities, i.e. the 
average over proton and neutron. These numbers may then be combined with the 
proton values in order to derive $\bar{\alpha}_{E1}^n$ and 
$\bar{\beta}_{M1}^n$. 

Our study of elastic proton Compton scattering is based on a multipole 
analysis of this process: a systematic expansion of
the proton Compton cross sections in the multipole order (dipole, quadrupole, 
...) by projecting the Compton amplitude on the various multipoles. 
We are not only interested in cross sections, but also in the 
multipole amplitudes themselves which we combine to so-called 
\textit{dynamical polarizabilities}, first introduced in Ref.~\cite{GH1}. 
These quantities turn out to be a useful tool for studying the internal 
nucleonic degrees of freedom
in their response to the external electromagnetic field. We find in
Chapter~\ref{chap:spinaveraged} that in the energy range where we expect our 
calculation to be valid, contributions to the spin-averaged Compton cross 
sections from $l\geq2$ are negligible. This
observation also holds in spin-polarized quantities, as demonstrated by our 
calculation of several asymmetries in Chapter~\ref{chap:spinpolarized}. 

Our results for the dynamical polarizabilities are compared to a 
Dispersion-Relation Analysis \cite{HGHP}. Both frameworks agree well with each
other in
most multipole channels~-- albeit 
the upper energy limit of our calculation is lower than that of the 
Dispersion-Relation approach,
because up to third order in SSE, which is the order chosen in 
this work, the $\Delta(1232)$ is treated as a stable particle. Therefore, our 
calculation is certainly invalid above 170-200~MeV, where the finite width of 
the $\Delta$ resonance cannot be neglected anymore. Predictions from Dispersion
Theory are at least reliable up to the two-pion threshold.

Not all of our results for the dynamical polarizabilities are 
predictions. In the original formulation of the Small Scale Expansion there is
no free parameter up to third order. Nevertheless, in our procedure of fitting
the static dipole polarizabilities $\bar{\alpha}_{E1}$ and $\bar{\beta}_{M1}$ 
to proton Compton data, we include two energy-independent short-distance 
operators into our calculation, which provide the desired two free parameters.
The energy dependence of the (dynamical) polarizabilities is therefore still 
predicted. 

Our fit results for $\bar{\alpha}_{E1}^p$ and $\bar{\beta}_{M1}^p$ are in good 
agreement with the Baldin sum rule for the proton. Therefore we use this
sum-rule value in order to further reduce the number of free parameters. The
proton polarizabilities thus achieved, 
\begin{align}
\bar{\alpha}_{E1}^p&=(11.04\pm1.36\,(\text{stat})\pm0.4\,(\text{Baldin}))
\cdot10^{-4}\;\text{fm}^3,\nonumber\\
\bar{\beta}_{M1}^p &=( 2.76\mp1.36\,(\text{stat})\pm0.4\,(\text{Baldin}))
\cdot10^{-4}\;\text{fm}^3,
\label{eq:alphabetaconclusion}
\end{align}
agree well with the analysis from \cite{Olmos} within statistical 
error bars. Systematic uncertainties from higher orders are not included in 
Eq.~(\ref{eq:alphabetaconclusion}). They have been estimated in 
Section~\ref{sec:staticpolas} as
$|\bar{\alpha}_\mathrm{NLO}|\sim|\bar{\beta}_\mathrm{NLO}|\sim
1\cdot 10^{-4}\; \fm^3 $ from na\"ive dimensional analysis and are suppressed
also in the following equations.

From the good agreement between our results for the Compton 
multipoles and the Dispersion-Relation Analysis, and from the 
fact that our fits describe the low-energy proton Compton data well over the 
whole range of scattering angles, we conclude that all relevant nucleonic 
degrees of freedom are included in our calculation. Comparison with 
third-order HB$\chi$PT demonstrates that it is advantageous and in fact 
necessary to include the explicit
$\Delta(1232)$ resonance in a leading-one-loop-order calculation. Otherwise
one misses the data in the backward direction and fails to reproduce the shape
of the resonant multipole channels, such as $\beta_{M1}(\w)$.

In Chapter~\ref{chap:spinaveraged} we demonstrate 
that an $l=1$-approximation
of the multipole expansion suffices to describe spin-averaged Compton cross
sections. In  Chapter~\ref{chap:spinpolarized} the same feature is confirmed 
for spin-polarized observables, namely for various asymmetries using 
circularly and linearly polarized photons. This observation suggests that it 
is possible to directly determine the six dipole polarizabilities from 
experiment~-- apart from the two spin-independent ones, $\alpha_{E1}(\w)$ and 
$\beta_{M1}(\w)$, there are four spin polarizabilities at dipole order. 
This suggestion is
further confirmed as we observe a non-negligible influence of the spin 
polarizabilities on the spin-averaged Compton cross sections. We show that 
this dependence suffices to determine~-- at least qualitatively~-- two of 
the four spin polarizabilities, but it is obvious that spin-averaged 
experiments alone are not enough to extract all six dipole polarizabilities. 
Therefore we investigate which configurations are especially well suited in 
order to access the spin polarizabilities experimentally. We also 
propose a model-independent way to extract the dynamical spin 
polarizabilities from a combination of spin-averaged and polarized 
experimental data. 

We investigate not only proton observables but also the corresponding 
quantities for the neutron. 
Direct Compton experiments on the neutron are however not possible. Therefore,
in the second main part of this work we focus on the theoretical 
description of elastic Compton scattering from the deuteron, which we use to 
extract the isoscalar analogues to Eq.~(\ref{eq:alphabetaconclusion}) from 
experiments. 

Our deuteron Compton calculation is performed 
in two steps. First, we calculate the $\gamma d$ interaction kernel
strictly following the power-counting rules of the Small Scale Expansion.
This first attempt provides a good description of the experimental data above
60~MeV, but it breaks down for lower energies. The two main reasons for this 
breakdown are the applied power-counting scheme and the approximation of the 
nucleon propagator used in this scheme. Both are only valid in the energy 
regime $\w\gg20$~MeV. Nevertheless, 
encouraged by the good agreement at higher energies, we fit the
isoscalar polarizabilities $\bar{\alpha}_{E1}^s$ and $\bar{\beta}_{M1}^s$ to 
the data from \cite{Lucas,Lund,Hornidge}, which have been measured
around 68~MeV and 94.2~MeV, yielding results in good agreement with the 
quasi-elastic experiment from Ref.~\cite{Kossert} and with the numbers 
recommended in Ref.~\cite{Schumacher}, which both suggest rather small 
isovector polarizabilities. Motivated by the statistical imbalance between 
experimental data around 94.2~MeV and 68~MeV, we reduce in a second fit the 
nine data points given in~\cite{Lund} at $\sim$68~MeV to only two points by 
rebinning, in order to obtain an equal weighting between the two 
energy sets. The results from this fit, which we consider as the more 
reliable of the two, 
confirm our findings of small values for $\bar{\beta}_{M1}^s$, also implying 
small isovector components. As we observe a relatively 
strong dependence of this first approach on the deuteron wave function,
we fit twice, using the two extreme wave functions, which are the chiral NNLO 
wave function~\cite{Epelbaum} and the Nijm93 wave function~\cite{Nijm}.
Averaging over the results of our two 2-parameter SSE fits to the reduced set 
of data results in the isoscalar polarizabilities
\begin{align}
\bar{\alpha}_{E1}^s&=(12.8\pm1.4\,(\mathrm{stat})\pm1.1\,(\mathrm{wf}))
            \cdot 10^{-4}\;\mathrm{fm}^3\,,\nonumber\\
\bar{\beta}_{M1}^s &=( 2.1\pm1.7\,(\mathrm{stat})\pm0.1\,(\mathrm{wf}))
            \cdot 10^{-4}\;\mathrm{fm}^3\,.
\label{eq:alphasbetasconclusionunbiased}
\end{align}
The systematic error due to the wave-function dependence (wf) is 
estimated to be half of the difference between the results obtained with the 
extreme wave functions. 
As the numbers presented in Eq.~(\ref{eq:alphasbetasconclusionunbiased})
are in very good agreement with the isoscalar Baldin sum rule, we also use the 
central sum-rule value as an additional fit constraint, obtaining
\begin{align}
\bar{\alpha}_{E1}^s&=(12.6\pm0.8\,(\mathrm{stat})\pm0.7\,(\mathrm{wf})
                 \pm0.6\,(\mathrm{Baldin}))
            \cdot 10^{-4}\;\mathrm{fm}^3\,,\nonumber\\
\bar{\beta}_{M1}^s &=( 1.9\mp0.8\,(\mathrm{stat})\mp0.7\,(\mathrm{wf})
                 \pm0.6\,(\mathrm{Baldin}))
            \cdot 10^{-4}\;\mathrm{fm}^3\,.
\label{eq:alphasbetasconclusionunbiasedBaldin}
\end{align}

Combining the numbers of Eq.~(\ref{eq:alphasbetasconclusionunbiased}) with our 
results for the proton polarizabilities, given in 
Eq.~(\ref{eq:alphabetaconclusion}), we obtain a consistent Effective Field 
Theory determination of the neutron polarizabilities with a precision 
comparable to~\cite{Kossert}:
\begin{align}
\bar{\alpha}_{E1}^n&=(14.6\pm2.0\,(\mathrm{stat})\pm1.1\,(\mathrm{wf})
\pm0.4\,(\text{Baldin}))\cdot 10^{-4}\;\mathrm{fm}^3\,\nonumber\\
\bar{\beta}_{M1}^n &=( 1.4\pm2.2\,(\mathrm{stat})\pm0.1\,(\mathrm{wf})
\pm0.4\,(\text{Baldin}))\cdot 10^{-4}\;\mathrm{fm}^3\,
\label{eq:conclusionneutron}
\end{align} 
The isoscalar input of Eq.~(\ref{eq:conclusionneutron}) does not include the 
Baldin-sum-rule 
constraint, whereas the one-parameter fit using the Baldin sum rule gives
\begin{align}
\bar{\alpha}_{E1}^n&=(14.2\pm1.6\,(\mathrm{stat})\pm0.7\,(\mathrm{wf})
         \pm0.6\,(\mathrm{Baldin}))\cdot 10^{-4}\;\mathrm{fm}^3\,,\nonumber\\
\bar{\beta}_{M1}^n &=( 1.0\mp1.6\,(\mathrm{stat})\mp0.7\,(\mathrm{wf})
         \pm0.6\,(\mathrm{Baldin}))\cdot 10^{-4}\;\mathrm{fm}^3\,.
\label{eq:conclusionneutronBaldin}
\end{align} 
Eqs.~(\ref{eq:conclusionneutron}) and~(\ref{eq:conclusionneutronBaldin}) prove
that small isovector nucleon polarizabilities are not in contradiction with 
elastic deuteron Compton-scattering data. 
This finding is in good agreement with~\cite{Kossert}, where quasi-elastic 
Compton scattering off the proton and neutron was measured. 

Furthermore we use the $\mathcal{O}(p^3)$-HB$\chi$PT amplitudes for 
analogous fits, finding similar values for $\bar{\alpha}_{E}$ but larger ones 
for $\bar{\beta}_{M}$, which is not surprising, as the dynamics of the 
resonant Compton multipoles is not well captured in third-order HB$\chi$PT. 
Therefore, the static value becomes large, since it must correct for the 
missing $\Delta$ resonance, leading $\mathcal{O}(p^3)$ HB$\chi$PT to a 
disagreement with the single-nucleon 
Compton multipoles as observed in Chapter~\ref{chap:spinaveraged}.
Obviously, $\gamma d$ scattering alone is not sufficient
to investigate the relevant low-energy degrees of freedom in nuclear Compton 
scattering, but one has to combine information from $\gamma d$ and $\gamma p$
scattering and analyze both in the same framework.

The results for the isoscalar polarizabilities, extracted in our first 
chapter on deuteron Compton scattering, along with the good description of 
the high-energy data is certainly a first success. However, 
as already explained, there are severe shortcomings of this way of calculating
$\gamma d$ scattering: the unexpectedly strong sensitivity on the deuteron 
wave function and, even more importantly, the fact that the calculation 
completely fails in the low-energy regime, which is a clear 
indication that gauge invariance is violated \cite{Friar}.  
We report on several attempts 
to restore gauge invariance and the Thomson limit by 
inclusion of additional diagrams and the full non-relativistic nucleon 
propagator. Although  we are able to improve on the static limit~-- we reduce 
the factor of 6 that our previous ``power-counting'' calculation 
is off to less than 2~-- we cannot 
restore it exactly. Therefore, in Chapter~\ref{chap:nonperturbative} we 
turn to a refined approach to deuteron Compton scattering, following
closely Ref.~\cite{Karakowski}. This calculation is based on second-order 
perturbation theory in the interaction of the photon with the two-nucleon 
system,  with
summation over all possible intermediate two-nucleon states. For the photon 
coupling we make use of Siegert's theorem~\cite{Siegert}, which is well-known
to guarantee the exact static limit~\cite{Karakowski}. 

Besides the compliance with the low-energy theorem, this approach 
provides another valuable cross-check of our calculation: the 
extraction of the total deuteron-photodisintegration cross section from the
Compton amplitude via the optical theorem. There is a wealth of experimental 
data on this process, see e.g.
[\ref{MAB}-\ref{MMC}], and we demonstrate in 
Section~\ref{sec:photodisintegration} that our calculation agrees well with 
these data. Nevertheless, our primary goal is to have a consistent description
of elastic deuteron Compton scattering in the whole range from 0~MeV up to 
$\w\sim 100$~MeV or even up to the pion mass. In Section~\ref{sec:results2},
we show that we have largely achieved this aim. We are able to improve 
the low-energy regime of our calculation, i.e. in this second approach we 
obtain a good description of the data published below 60~MeV. This improvement
is of course connected to the correct static limit. Other calculations, 
reaching this limit, are also able to describe the low-energy data
well, see e.g.~\cite{Lvov, Karakowski, Rupak}. However, unlike most of 
these calculations, we achieve good agreement also with the high-energy data, 
i.e. we have resolved the so-called 'SAL-puzzle'.
In fact, at 94.2~MeV  both calculations of ours are very close to each other. 
In this energy regime the work of Ref.~\cite{Karakowski} fails, the theory of 
\cite{Rupak} is inapplicable and even the 
authors of \cite{Lvov} have problems to describe the data in the backward 
direction, at least without introducing unrealistically large isovector 
polarizabilities. The main difference between their approach 
and ours is in the dynamics of the resonant multipoles, 
which is well captured in our calculation due to the inclusion of the 
$\Delta$-resonance diagram. 

Having achieved a good description of all elastic deuteron Compton-scattering 
data enables us to perform a global fit of the isoscalar 
polarizabilities to all existing data points, published in 
\cite{Lucas,Lund,Hornidge}. Our 2-parameter-fit results are
\ba
\left.\phantom{\PCsq}\bar{\alpha}_{E1}^s\right|_\mathrm{global}&=
(11.5\pm1.4\,(\mathrm{stat}))\cdot10^{-4}\;\fm^3,\nonumber\\
\left.\phantom{\PCsq}\bar{\beta}_{M1}^s \right|_\mathrm{global}&=
( 3.4\pm1.6\,(\mathrm{stat}))\cdot10^{-4}\;\fm^3.
\label{eq:globalfitconclusion}
\end{align}
We only need to give the statistical error in this case since we have 
demonstrated in Section~\ref{sec:wavefunctiondep2} that the wave-function 
dependence of our ``non-perturbative'' approach is tiny. This source of 
uncertainty discussed in  Chapter~\ref{chap:perturbative} has therefore also 
been removed. We notice that 
the numbers of Eq.~(\ref{eq:globalfitconclusion}) are very close to
our results for the proton (Eq.~(\ref{eq:alphabetaconclusion})), which leaves 
little space for large isovector polarizabilities. Furthermore they are 
consistent within the error bars with our previous extraction, 
Eq.~(\ref{eq:alphasbetasconclusionunbiased}), and agree 
extraordinarily well with the isoscalar Baldin-sum-rule value, cf. 
Eq.~(\ref{eq:Baldin}). Therefore, in order to reduce the statistical error,  
we repeat our fits including this additional constraint, achieving 
\ba
\left.\phantom{\PCsq}\bar{\alpha}_{E1}^s\right|_\mathrm{global\;Baldin}&=
(11.3\pm0.7\,(\mathrm{stat})\pm0.6\,(\mathrm{Baldin}))\cdot10^{-4}\;\fm^3,
\nonumber\\
\left.\phantom{\PCsq}\bar{\beta}_{M1}^s \right|_\mathrm{global\;Baldin}&=
( 3.2\mp0.7\,(\mathrm{stat})\pm0.6\,(\mathrm{Baldin}))\cdot10^{-4}\;\fm^3.
\label{eq:globalfitBaldinconclusion}
\end{align}
Combining Eqs.~(\ref{eq:globalfitconclusion}) or 
(\ref{eq:globalfitBaldinconclusion}), 
respectively, with the Baldin-constrained proton results of 
Eq.~(\ref{eq:alphabetaconclusion}), we find the neutron polarizabilities 
\ba
\left.\phantom{\PCsq}\bar{\alpha}_{E1}^n\right|_\mathrm{global}&=
(12.0\pm2.0\,(\mathrm{stat})\,\pm0.4\,(\text{Baldin}))\cdot10^{-4}\;\fm^3,
\nonumber\\
\left.\phantom{\PCsq}\bar{\beta}_{M1}^n \right|_\mathrm{global}&=
( 4.0\pm2.1\,(\mathrm{stat})\,\pm0.4\,(\text{Baldin}))\cdot10^{-4}\;\fm^3
\label{eq:globalfitneutronconclusion}
\end{align}
for the 2-parameter fit and 
\ba
\left.\phantom{\PCsq}\bar{\alpha}_{E1}^n\right|_\mathrm{global\;Baldin}&=
(11.6\pm1.5\,(\mathrm{stat})\pm0.6\,(\mathrm{Baldin}))\cdot10^{-4}\;\fm^3,
\nonumber\\
\left.\phantom{\PCsq}\bar{\beta}_{M1}^n \right|_\mathrm{global\;Baldin}&=
( 3.6\mp1.5\,(\mathrm{stat})\pm0.6\,(\mathrm{Baldin}))\cdot10^{-4}\;\fm^3
\label{eq:globalfitBaldinneutronconclusion}
\end{align}
for the fit including the Baldin constraint also in the extraction of the 
isoscalar polarizabilities. We consider the values given in
Eqs.~(\ref{eq:alphabetaconclusion}) 
and~(\ref{eq:globalfitBaldinneutronconclusion}) to be most reliable because 
of the fact that our second approach to deuteron Compton scattering 
fulfills the low-energy theorem and
enables us to  include all experimental data into our fit of the isoscalar 
polarizabilities. From these results we deduce that the magnetic response
of the neutron is comparable to that of the proton and that both nucleons 
are paramagnetic. We also conclude that the proton and neutron 
polarizabilities are identical within the precision of our analysis.
In both points our two deuteron Compton calculations  
agree with each other.

Nevertheless, we strongly advocate  enlarging the deuteron Compton data base. 
If further experiments, as planned at TUNL/HI$\gamma$S or at MAXlab, provide 
additional data below the pion mass, an improved global fit with increased 
statistics would be possible, which would reduce the 
statistical error in our determination of the neutron polarizabilities.

\appendix
\chapter{Numerical Values of Physical Constants  \label{app:parameters} }
\markboth{APPENDIX \ref{app:parameters}. 
NUMERICAL VALUES OF PHYSICAL CONSTANTS}
         {APPENDIX \ref{app:parameters}. 
NUMERICAL VALUES OF PHYSICAL CONSTANTS}
In Table~\ref{tab:const} we list the numerical values  that we use 
for all parameters, except for those which are determined within this work:
$b_1$, $g_{1}$, $g_{2}$, cf. Section~\ref{sec:protonfits}.
The numbers are from~\cite{AV18, Rho, PDG}.

\begin{table}[!htb] 
\begin{center}
\begin{tabular}{|c|c|c|}
\hline 
Parameter         & Value         & Comment \\
\hline 
$m_\pi$           & $139.6$~MeV   & charged pion mass \\
$m_{\pi^0}$       & $135.0$~MeV   & neutral pion mass \\
$m_N$             & $938.9$~MeV   & isoscalar nucleon mass \\
$m_C$             & $1877.8$~MeV  & twice the isoscalar nucleon mass \\
$f^2$             & $0.075$       & pion-nucleon coupling constant \\
$g_A$             & $1.267$       & axial coupling constant \\
$f_\pi$           & $92.4$~MeV    & pion-decay constant \\
$\alpha$          & $1/137$       & QED fine-structure constant \\
$\mu_p$           & $2.795 $      & magnetic moment of the proton \\
$\mu_n$           & $-1.913$      & magnetic moment of the neutron \\
\hline
$\Delta_0$        & $271.1$~MeV   & $N\Delta$ mass splitting \\
$g_{\pi N\Delta}$ & $1.125$       & $\pi N\Delta$ coupling constant \\
\hline
$m_d$             & $1875.58$~MeV & deuteron mass \\
$B$               & $2.2246$~MeV  & deuteron binding energy\\
\hline
\end{tabular}
\end{center}
\caption[$\chi$EFT parameters determined independently of this work]
{$\chi$EFT parameters determined independently of this work. 
Magnetic moments are given in nuclear magnetons.}
\label{tab:const}
\end{table}
\noindent
The pion-nucleon coupling constant $f^2$ is at leading order connected to 
$g_A$, $m_\pi$ and $f_\pi$ via the Goldberger-Treiman 
relation~\cite{Goldberger}:
\be
f^2\approx\left(\frac{g_A\,m_\pi}{2\,f_\pi}\right)^2\cdot\frac{1}{4\pi}=0.073.
\ee

\chapter{Pole Contributions to Nucleon Compton Scattering  
\label{app:poleterms} }
\markboth
{APPENDIX \ref{app:poleterms}. POLE CONTRIBUTIONS}
{APPENDIX \ref{app:poleterms}. POLE CONTRIBUTIONS}
\begin{figure}[!htb]
\begin{center} 
\includegraphics*[width=.5\textwidth]{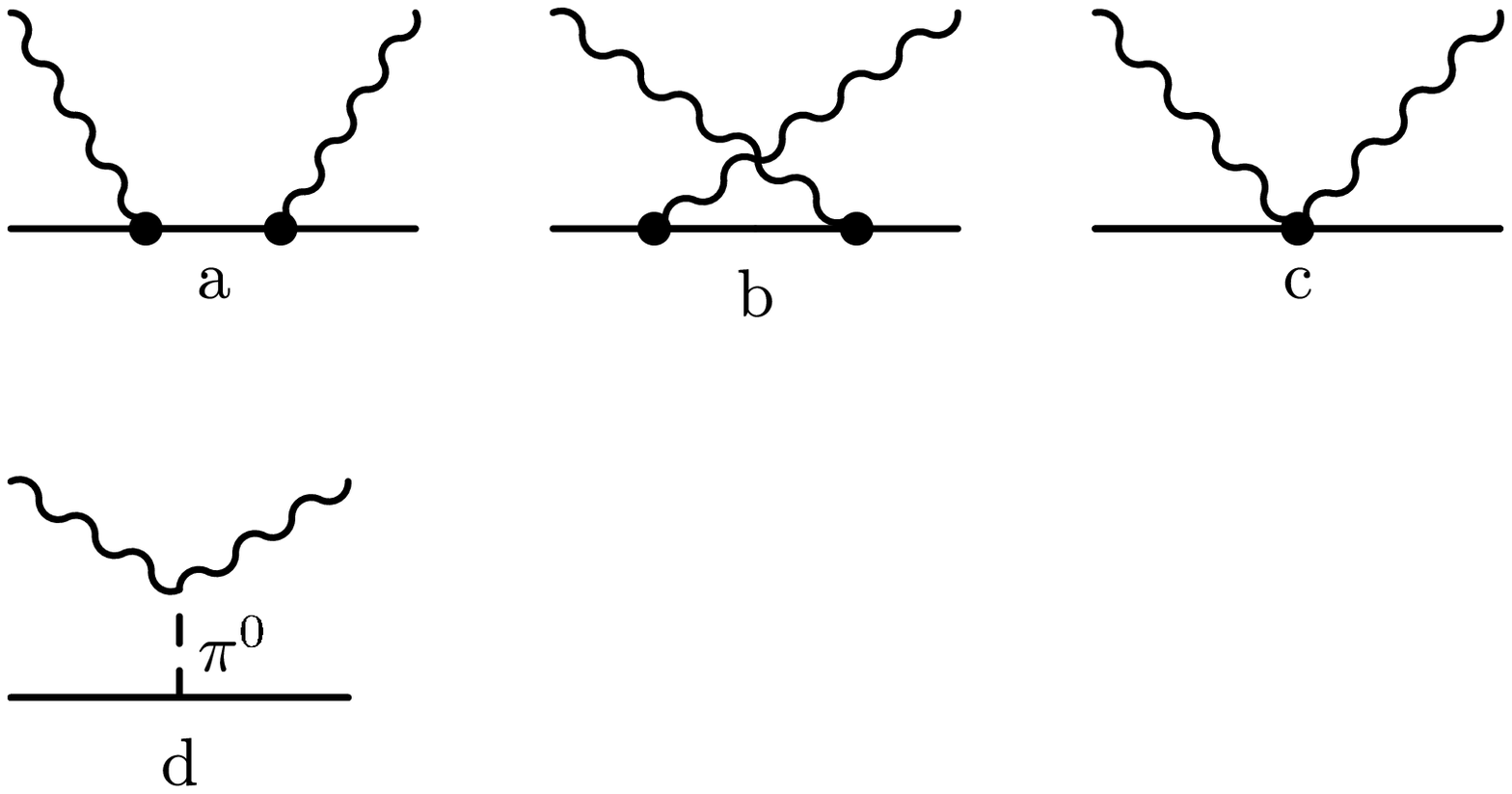}
\hspace{.3cm}
\includegraphics*[width=.12\textwidth]{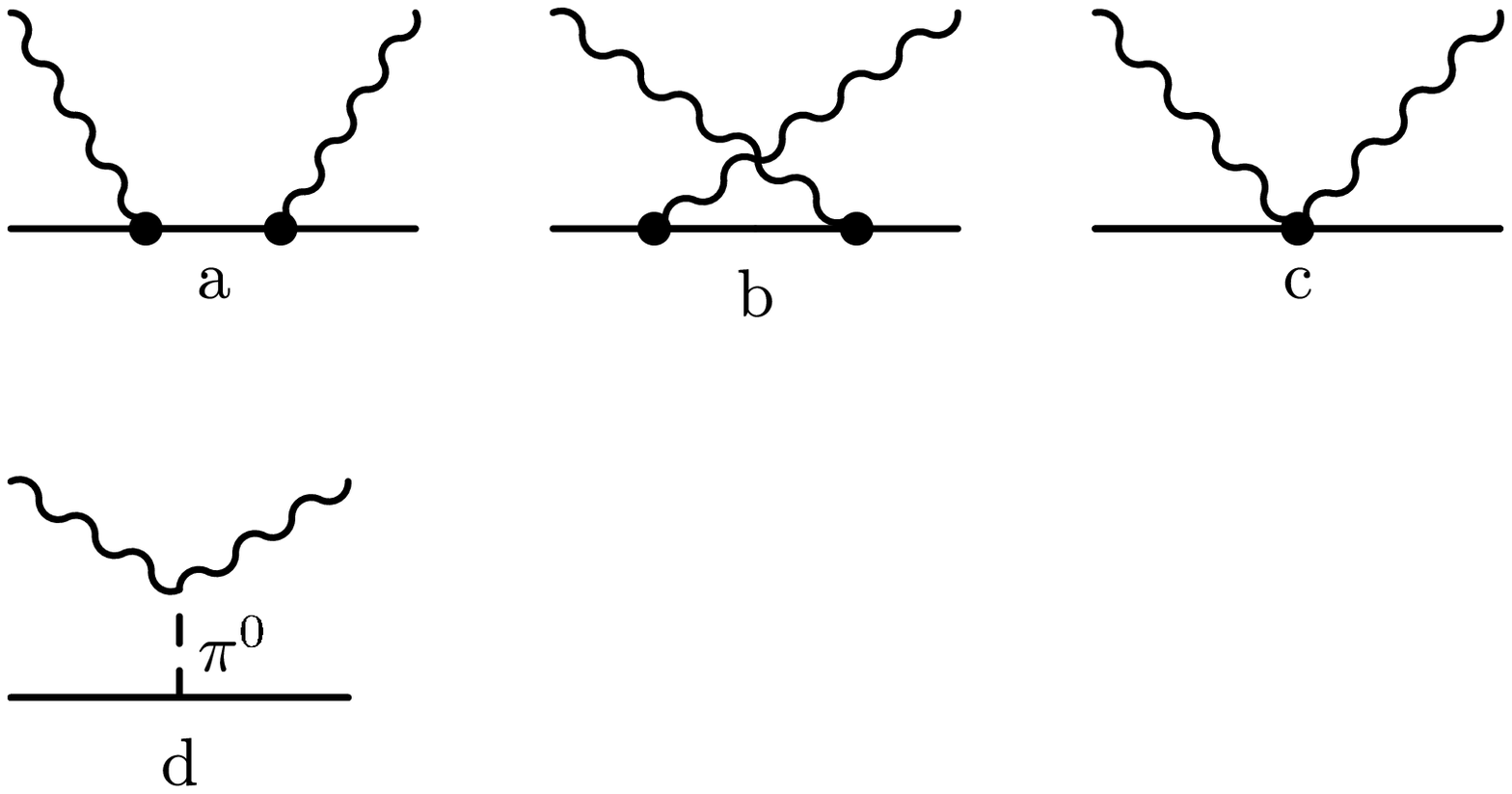}
\caption[Pole contributions to nucleon Compton scattering]
{Pole contributions to nucleon Compton scattering in the $s$-channel 
(a), $u$-channel (b) and $t$-channel (d). The proton seagull (c) occurs in the 
non-relativistic reduction of diagrams (a),~(b).}
\label{fig:app:pole}
\end{center}
\end{figure}
\noindent
In this appendix we explicitly write down the non-structure or pole 
amplitudes of nucleon Compton scattering in the cm frame up to third order in 
the Small Scale Expansion\footnote{We note that these amplitudes are identical
in third-order HB$\chi$PT and SSE.}, corresponding to Fig.~\ref{fig:app:pole}.
These are (with the charge $Q=1$~$(0)$ for a proton (neutron) target)
\begin{align}
A_1^{\mathrm{pole}}(\w,\theta)&=-\frac{Q\,e^2}{m_N}
+\mathcal{O}(\epsilon^4),  \nonumber\\
A_2^{\mathrm{pole}}(\w,\theta)&= \frac{Q^2\,e^2\,\w}{m_N^2}
+\mathcal{O}(\epsilon^4),  \nonumber\\
A_3^{\mathrm{pole}}(\w,\theta)&=
\frac{e^2\,\w\,\left(Q\,(1+2\kappa)-(Q+\kappa)^2\,\cos\theta\right)}
{2\,m_N^2}\nonumber\\
&-(2\,Q-1)\,\frac{e^2\,g_A}{4\,\pi^2\,f_\pi^2}\,
\frac{\w^3\,(1-\cos\theta)}{m_{\pi^0}^2+2\, \w^2\,(1-\cos\theta)}
+\mathcal{O}(\epsilon^4),\nonumber\\
A_4^{\mathrm{pole}}(\w,\theta)&=-\frac{e^2\,\w\,(Q+\kappa)^2} {2\,m_N^2}
+\mathcal{O}(\epsilon^4),\nonumber\\
A_5^{\mathrm{pole}}(\w,\theta)&=\frac{e^2\,\w\,(Q+\kappa)^2} {2\,m_N^2}
-(2\,Q-1)\,\frac{e^2\,g_A}{8\,\pi^2\,f_\pi^2}\,\frac{\w^3}{m_{\pi^0}^2+2\,
\w^2\,(1-\cos\theta)}+\mathcal{O}(\epsilon^4),  \nonumber\\
A_6^{\mathrm{pole}}(\w,\theta)&=-\frac{e^2\,\w\,Q\,(Q+\kappa)}{2\,m_N^2}
+(2\,Q-1)\,\frac{e^2\,g_A}{8\,\pi^2\,f_\pi^2}\,\frac{\w^3}{m_{\pi^0}^2+2\,
\w^2\,(1-\cos\theta)} +\mathcal{O}(\epsilon^4).
\label{eq:poleterms}
\end{align}
For the numerical values of the various parameters and their meaning, cf.
Appendix~\ref{app:parameters} 
($\kappa=\mu-Q$).
The proton seagull, Fig.~\ref{fig:app:pole}(c),
contributes to $A_1^{\mathrm{pole}}$ and  $A_3^{\mathrm{pole}}$. The 
contribution to $A_1^{\mathrm{pole}}$ stems from the two-photon vertex from
$\mathcal{L}_{N\pi}^{(2)}$. The seagull term in $A_3^{\mathrm{pole}}$, which 
is the one proportional to $Q\,(1+2\kappa)$, is the leading relativistic 
correction from spin-orbit coupling 
and part of $\mathcal{L}_{N\pi}^{(3)}$.
The terms depending on the axial coupling constant $g_A$ are the contributions 
from the pion pole, Fig.~\ref{fig:app:pole}(d).  All other terms correspond to 
the nucleon-pole diagrams, Figs.~\ref{fig:app:pole}(a) and (b).

\chapter{Projection Formulae
\label{app:projection} }
\markboth{APPENDIX \ref{app:projection}. PROJECTION FORMULAE}
         {APPENDIX \ref{app:projection}. PROJECTION FORMULAE}
The connection between the Compton structure amplitudes
$\bar{A}_i(\omega,z),\;i=1,\ldots, 6$, given in Appendix~B of 
Ref.~\cite{HGHP}, and the cm Compton multipoles 
$f_{XX'}^{l\pm}(\omega),\;X,X'=E,M$, introduced in 
Section~\ref{sec:amplitudestomultipoles}, has been derived in \cite{DA}.  
It reads:
\begin{align*}
f_{EE}^{1+}(\w)&=\int\limits_{-1}^{1}\frac{m_N}{16\cdot 4\pi\,W}
\bigg[ \bar{A}_3(\w,z)\,\left(-3+z^2\right)+4\bar{A}_6(\w,z)\,
\left(-1+z^2\right)\\
&+\left(2\bar{A}_2(\w,z)+\bar{A}_4(\w,z)+2\bar{A}_5(\w,z)\right)
\,z\,\left(-1+z^2\right)+2\bar{A}_1(\w,z)\,\left(1+z^2\right)\bigg]dz\\
f_{EE}^{1-}(\w)&=\int\limits_{-1}^{1}\frac{m_N}{ 8\cdot 4\pi\,W}
\bigg[-\bar{A}_3(\w,z)\,\left(-3+z^2\right)-4\bar{A}_6(\w,z)\,
\left(-1+z^2\right)\\
&-\left(-\bar{A}_2(\w,z)+\bar{A}_4(\w,z)+2\bar{A}_5(\w,z)\right)
\,z\,\left(-1+z^2\right)+ \bar{A}_1(\w,z)\,\left(1+z^2\right)\bigg]dz\\
f_{MM}^{1+}(\w)&=\int\limits_{-1}^{1}\frac{m_N}{16\cdot 4\pi\,W}
\bigg[2\bar{A}_2(\w,z)\,\left(-1+z^2\right)\\
&+\bar{A}_4(\w,z)\,\left(-1+z^2\right)+2\left(\bar{A}_5(\w,z)\,
\left( 1-z^2\right)+\bar{A}_1(\w,z)\,2z-\bar{A}_3(\w,z)\,z\right)\bigg]dz\\
f_{MM}^{1-}(\w)&=\int\limits_{-1}^{1}\frac{m_N}{ 8\cdot 4\pi\,W}
\bigg[\bar{A}_4(\w,z)\,\left(1-z^2\right)\\
&+\bar{A}_2(\w,z)\,\left(-1+z^2\right)+2\left(\bar{A}_5(\w,z)
\left(-1+z^2\right)+\bar{A}_1(\w,z)\, z+\bar{A}_3(\w,z)\,z\right)\bigg]dz\\
f_{EE}^{2+}(\w)&=\int\limits_{-1}^{1}\frac{m_N}{72\cdot 4\pi\,W}
\bigg[\bar{A}_4(\w,z)\,\left(-1-3z^2+4z^4\right)
+\bar{A}_2(\w,z)\,\left(3-9z^2+6z^4\right)+2\left(\bar{A}_5(\w,z)\right.\\
&\times\left.\left(-1-3z^2+4z^4\right)+\bar{A}_1(\w,z)\,3z^3
+\bar{A}_3(\w,z)\,\left(2z^3-3z\right)+\bar{A}_6(\w,z)\,
\left(6z^3-6z\right)\right)\bigg]dz\\
f_{EE}^{2-}(\w)&=\int\limits_{-1}^{1}\frac{m_N}{48\cdot 4\pi\,W}
\bigg[\bar{A}_4(\w,z)\,\left( 1+3z^2-4z^4\right)
+\bar{A}_2(\w,z)\,\left(2-6z^2+4z^4\right)+2\left(\bar{A}_5(\w,z)\right.\\
&\times\left.\left( 1+3z^2-4z^4\right)+\bar{A}_1(\w,z)\,2z^3
+\bar{A}_3(\w,z)\,\left(3z-2z^3\right)+\bar{A}_6(\w,z)\,
\left(6z-6z^3\right)\right)\bigg]dz
\end{align*}

\begin{align}
f_{MM}^{2+}(\w)&=\int\limits_{-1}^{1}\frac{m_N}{72\cdot 4\pi\,W}
\bigg[\bar{A}_3(\w,z)\,\left( 1-3z^2\right)\nonumber\\
&+\left(3\bar{A}_2(\w,z)+5\bar{A}_4(\w,z)-2\bar{A}_5(\w,z)\right)
\,z\,\left(-1+z^2\right)+\bar{A}_1(\w,z)\,\left(-3+9z^2\right)\bigg]dz
\nonumber\\
f_{MM}^{2-}(\w)&=\int\limits_{-1}^{1}\frac{m_N}{48\cdot 4\pi\,W}
\bigg[\bar{A}_3(\w,z)\,\left(-1+3z^2\right)\nonumber\\
&+\left(2\bar{A}_2(\w,z)-5\bar{A}_4(\w,z)+2\bar{A}_5(\w,z)\right)
\,z\,\left(-1+z^2\right)+\bar{A}_1(\w,z)\,\left(-2+6z^2\right)\bigg]dz
\nonumber\\
f_{EM}^{1+}(\w)&=\int\limits_{-1}^{1}\frac{m_N}{16\cdot 4\pi\,W}
\bigg[\bar{A}_3(\w,z)\,\left(1-3z^2\right)\nonumber\\
&-2\bar{A}_6(\w,z)\,\left(-1+z^2\right)-\left(\bar{A}_4(\w,z)
+4\bar{A}_5(\w,z)\right)\,z\,\left(-1+z^2\right)\bigg]dz\nonumber\\
f_{ME}^{1+}(\w)&=\int\limits_{-1}^{1}\frac{m_N}{16\cdot 4\pi\,W}
\bigg[\bar{A}_4(\w,z)\,\left(1- z^2\right)
-2z\,\left(\bar{A}_3(\w,z)+\bar{A}_6(\w,z)\,\left(1-z^2\right)\right)
\bigg]dz
\end{align}

\chapter{The Deuteron Wave Function  
\label{app:wavefunction} }
\begin{figure}[!htb]
\begin{center} 
\includegraphics*[width=.48\textwidth]{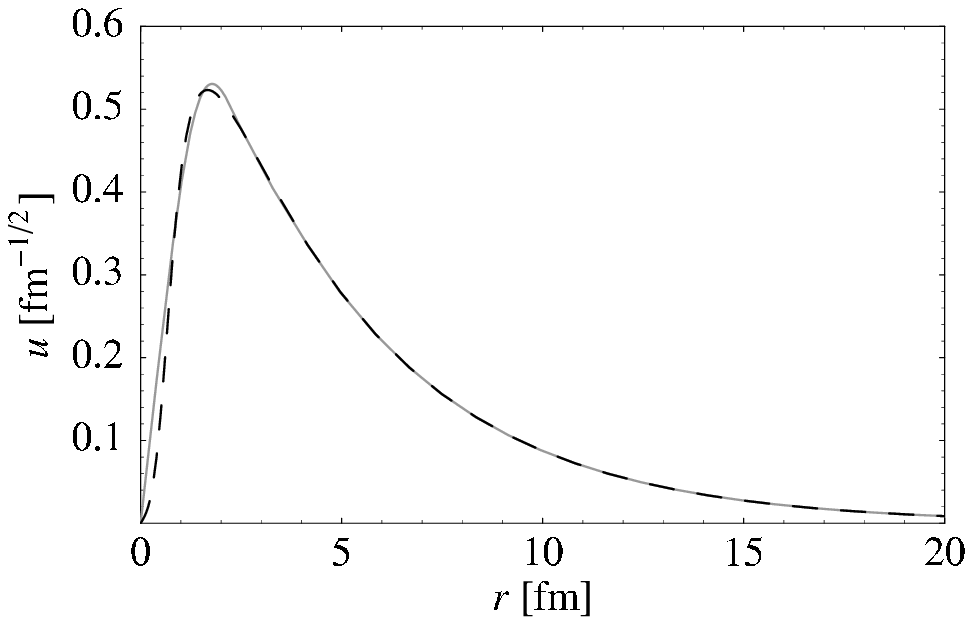}
\hfill
\includegraphics*[width=.48\textwidth]{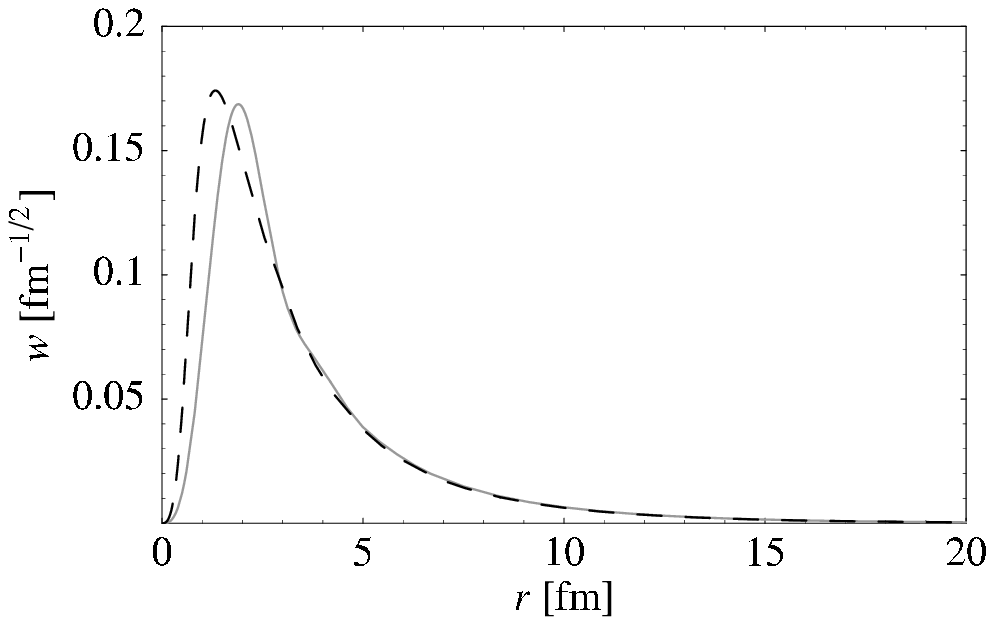}\\
\includegraphics*[width=.48\textwidth]{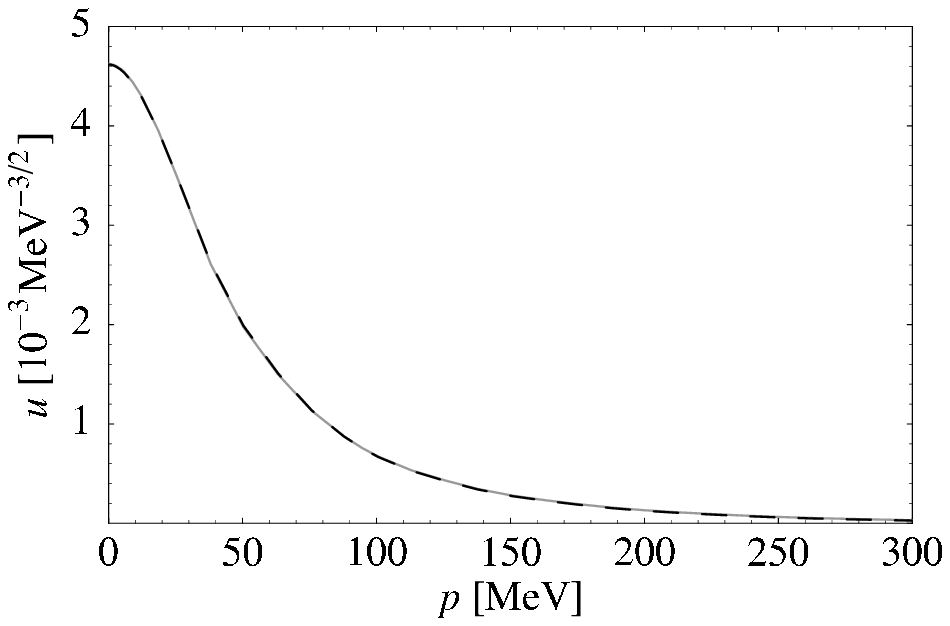}
\hfill
\includegraphics*[width=.48\textwidth]{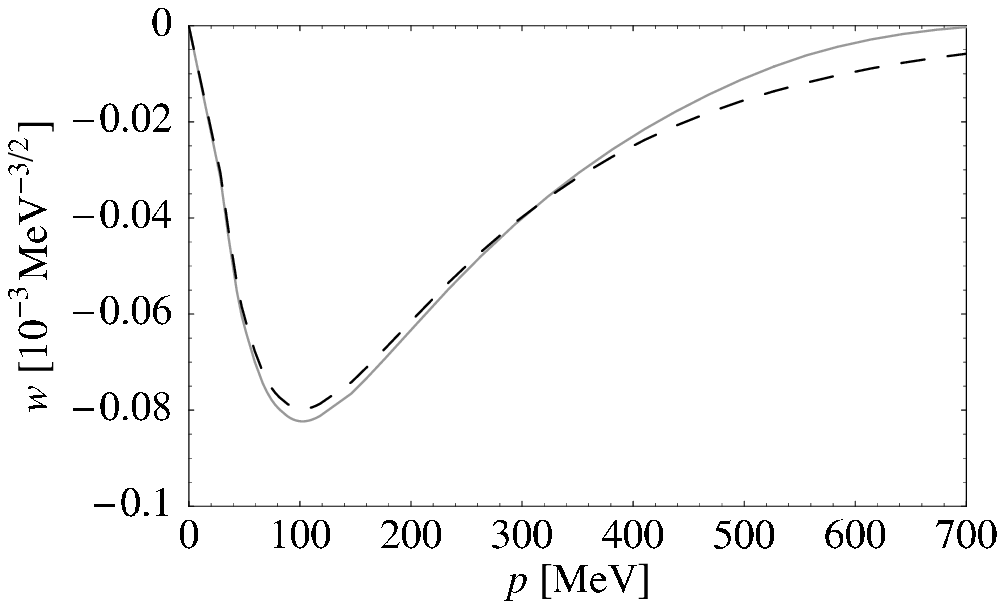}
\caption[Position- and momentum-space representation of two modern deuteron 
wave functions]
{The radial wave functions generated from the NNLO chiral potential with 
cut-off $\Lambda=650$~MeV \cite{Epelbaum} (grey) 
and the AV18-potential \cite{AV18} (dashed). In the upper row
we compare the position-space wave functions, in the lower the representation 
in momentum space. The first column corresponds to the $s$-state wave function 
$u$, the second to $w$, the radial wave function of the $d$-state.}
\label{fig:wavefunctions}
\end{center}
\end{figure}

\markboth{APPENDIX \ref{app:wavefunction}. THE DEUTERON WAVE FUNCTION}
         {APPENDIX \ref{app:wavefunction}. THE DEUTERON WAVE FUNCTION}
For the deuteron Compton calculation of Chapters~\ref{chap:perturbative} and
\ref{chap:nonperturbative} we need an explicit expression for the deuteron 
wave function (total angular momentum $j=1$). 
This can in momentum space be written as, see e.g. Section~3.4 
in~\cite{Ericson}, 
\be
\Psi_{1m}(\vec{p}\,)=u(p)\,\mathcal{Y}_m^{011}(\hat{p})+
                w(p)\,\mathcal{Y}_m^{211}(\hat{p}),
\label{eq:deuteronwavefunctionp}
\ee
where the radial wave functions $u(p),\;w(p)$ fulfill the normalization 
condition 
\be
\int_0^\infty dp\,p^2\,\left(u(p)^2+w(p)^2\right)=1.
\ee
The indices of the angular wave functions $\mathcal{Y}$ are $l11$ for 
orbital angular momentum, spin and total angular momentum of the deuteron 
state. $m\in\{-1,0,1\}$ denotes the projection of the total angular momentum 
of the deuteron onto the quantization axis.
We know from experiment
that the deuteron is composed of an $s$-wave ($l=0$) and a $d$-wave state 
($l=2$), cf. Eq.~(\ref{eq:deuteronwavefunctionp})~-- 
however, the $s$-wave part is by far the dominant one, 
see Fig.~\ref{fig:wavefunctions}. 

In position space the wave function is usually written as \cite{Ericson} 
\be
\Psi_{1m}(\vec{r}\,)=\frac{u(r)}{r}\,\mathcal{Y}_m^{011}(\hat{r})+
                \frac{w(r)}{r}\,\mathcal{Y}_m^{211}(\hat{r}),
\label{eq:deuteronwavefunctionrliterature}
\ee
with the normalization
\be
\int_0^\infty dr\,\left(u(r)^2+w(r)^2\right)=1.
\ee
However, as we want to write down a sum over the two orbital angular momentum 
states we use the notation
\be
\Psi_{1m}(\vec{r}\,)=\sum_{l=0,2}\frac{u_l(r)}{r}\mathcal{Y}_m^{l11}(\hat{r}),
\label{eq:deuteronwavefunction}
\ee
with $u_0(r)\equiv u(r)$ and $u_2(r)\equiv w(r)$.

The angular wave functions are built of the spherical harmonics, multiplied
with the corresponding Clebsch-Gordan coefficients 
$\Clebsch{j_1}{m_1}{j_2}{m_2}{j_3}{m_3}$, cf. 
Eq.~(\ref{eq:CGdefinition}) \cite{Ericson}. 
In order to determine these coefficients, we first write down 
the non-vanishing Clebsch-Gordan coefficients for combining the spins of the 
two nucleons to the total spin $S=1$: 
\be
\Clebsch{\frac{1}{2}}{-\frac{1}{2}}{\frac{1}{2}}{-\frac{1}{2}}{1}{-1}=
1\;\;\;\;\;\;\;
\Clebsch{\frac{1}{2}}{-\frac{1}{2}}{\frac{1}{2}}{ \frac{1}{2}}{1}{ 0}=
\Clebsch{\frac{1}{2}}{ \frac{1}{2}}{\frac{1}{2}}{-\frac{1}{2}}{1}{ 0}=
\frac{1}{\sqrt{2}}\;\;\;\;\;\;\;
\Clebsch{\frac{1}{2}}{ \frac{1}{2}}{\frac{1}{2}}{ \frac{1}{2}}{1}{ 1}=1
\ee
In the $s$-wave state ($l=0$), we have $m_l=0$ and therefore $m_j=m_s$. 
The corresponding Clebsch-Gordan coefficients 
are
\be
\Clebsch{0}{0}{1}{m_s}{1}{m_j}=\delta_{m_s,m_j},
\ee
always assuming that the projections $m$ only take on physical values, i.e.
$m_L\in\{-L,-L+1,..,L\}$.

We also need the $d$-wave coefficients, which project the states $l=2,m_l$, 
$s=1,m_s$ onto $j=1,m_j$. 
The relation $m_l+m_s=m_j$ guarantees that there are 
only nine non-vanishing Clebsch-Gordan coefficients:
\be
\parbox{3cm}{
\begin{align}
\Clebsch{2}{-2}{1}{ 1}{1}{-1}&=\sqrt{\frac{3}{5}}\nonumber\\
\Clebsch{2}{-1}{1}{ 1}{1}{ 0}&=\sqrt{\frac{3}{10}}\nonumber\\
\Clebsch{2}{ 0}{1}{ 1}{1}{ 1}&=\frac{1}{\sqrt{10}}\nonumber
\end{align}}\;\;\;\;\;\;
\parbox{3cm}{
\begin{align}
\Clebsch{2}{-1}{1}{ 0}{1}{-1}&=-\sqrt{\frac{3}{10}}\nonumber\\
\Clebsch{2}{ 0}{1}{ 0}{1}{ 0}&=-\sqrt{\frac{2}{5}}\nonumber\\
\Clebsch{2}{ 1}{1}{ 0}{1}{ 1}&=-\sqrt{\frac{3}{10}}\nonumber
\end{align}}\;\;\;\;\;\;
\parbox{3cm}{
\begin{align}
\Clebsch{2}{ 0}{1}{-1}{1}{-1}&=\frac{1}{\sqrt{10}}\nonumber\\
\Clebsch{2}{ 1}{1}{-1}{1}{ 0}&=\sqrt{\frac{3}{10}}\nonumber\\
\Clebsch{2}{ 2}{1}{-1}{1}{ 1}&=\sqrt{\frac{3}{5}} \nonumber
\end{align}}
\ee

Now we construct the deuteron wave function, e.g. in position 
space. The notation is $\Psi_{m_s^1 m_s^2\,m_j}$ with 
$m_s^1,\,m_s^2\in\{\uparrow,\downarrow\}$ the spin projections of nucleon~1 
and~2, respectively, and $m_j$ the projection of the total angular momentum of 
the deuteron. Note that we skip the spinors of the nucleons for brevity.
\begin{align}
\Psi_{\downarrow\downarrow\,-1}(\vec{r})&=\frac{u(r)}{r}\,Y_{0\,0}(\hat{r})+
\frac{w(r)}{r}\,Y_{2\,0}(\hat{r})\cdot\Clebsch{2}{0}{1}{-1}{1}{-1}\nonumber\\
\Psi_{\uparrow\downarrow\,-1}(\vec{r})&=
\Psi_{\downarrow\uparrow\,-1}(\vec{r}) = \frac{w(r)}{r}\,Y_{2\,-1}(\hat{r})
\cdot\Clebsch{2}{-1}{1}{0}{1}{-1}\cdot\frac{1}{\sqrt{2}}\nonumber\\
\Psi_{\uparrow\uparrow\,-1}(\vec{r})&=\frac{w(r)}{r}\,Y_{2\,-2}(\hat{r})
\cdot\Clebsch{2}{-2}{1}{1}{1}{-1}\nonumber\\
\Psi_{\downarrow\downarrow\, 0}(\vec{r})&=
\frac{w(r)}{r}\,Y_{2\,1}(\hat{r})\cdot\Clebsch{2}{1}{1}{-1}{1}{0}\nonumber\\
\Psi_{\uparrow\downarrow\, 0}(\vec{r})&=
\Psi_{\downarrow\uparrow\, 0}(\vec{r}) = 
\left[\frac{u(r)}{r}\,Y_{0\,0}(\hat{r})+\frac{w(r)}{r}\,Y_{2\,0}(\hat{r})
\cdot\Clebsch{2}{0}{1}{0}{1}{0}\right]\cdot\frac{1}{\sqrt{2}}\nonumber\\
\Psi_{\uparrow\uparrow\,0}(\vec{r})&=\frac{w(r)}{r}\,Y_{2\,-1}(\hat{r})
\cdot\Clebsch{2}{-1}{1}{1}{1}{0}\nonumber\\
\Psi_{\downarrow\downarrow\, 1}(\vec{r})&=
\frac{w(r)}{r}\,Y_{2\,2}(\hat{r})\cdot\Clebsch{2}{2}{1}{-1}{1}{1}\nonumber\\
\Psi_{\uparrow\downarrow\,1}(\vec{r})&=
\Psi_{\downarrow\uparrow\,1}(\vec{r}) = \frac{w(r)}{r}\,Y_{2\,1}(\hat{r})
\cdot\Clebsch{2}{1}{1}{0}{1}{1}\cdot\frac{1}{\sqrt{2}}\nonumber\\
\Psi_{\uparrow\uparrow\, 1}(\vec{r})&=\frac{u(r)}{r}\,Y_{0\,0}(\hat{r})+
\frac{w(r)}{r}\,Y_{2\,0}(\hat{r})\cdot\Clebsch{2}{0}{1}{1}{1}{1}
\label{eq:Psi}
\end{align}

Finally, in Fig.~\ref{fig:wavefunctions} we show the radial functions $u(r)$ 
($u(p)$) and $w(r)$ ($w(p)$)
for two typical wave functions that we use, namely the NNLO chiral wave 
function with cutoff $\Lambda=650$~MeV~\cite{Epelbaum} and the AV18-wave 
function~\cite{AV18}. Significant differences are only visible in the 
$d$-state. At large distances (small momenta), 
both wave functions are dominated by one-pion exchange and therefore
lie nearly exactly on top of each other.

\chapter{Additional Pion-Exchange Diagrams
\label{app:lowenergydiagrams} }
\markboth
{APPENDIX \ref{app:lowenergydiagrams}. ADDITIONAL PION-EXCHANGE DIAGRAMS}
{APPENDIX \ref{app:lowenergydiagrams}. ADDITIONAL PION-EXCHANGE DIAGRAMS}
This appendix, where we give our results for the diagrams shown in 
Fig.~\ref{fig:lowenergydiagrams}, refers exclusively to 
Section~\ref{sec:Thomson1}. These pion-exchange diagrams appear at third-order
SSE in addition to those of Fig.~\ref{fig:chiPTdouble}, when we count 
the nucleon propagator as $\calO(\epsilon^{-2})$, which is the correct power 
in the low-energy regime, cf. Section~\ref{sec:Thomson1}. As we are only 
interested in contributions to the static limit, we restrict ourselves to the 
photon-nucleon coupling given in Eq.~(\ref{eq:photonnucleonvertex}). The 
full non-relativistic nucleon propagator is included, like in 
Eqs.~(\ref{eq:schannelamplitude}, \ref{eq:uchannelamplitude}).
The notation employed can be read off Fig.~\ref{fig:lowenergydiagrams}. 
Due to the pion exchange, all diagrams calculated in this appendix 
contribute to $T^{\gamma NN}(\ki,\kf;\pvec,\ppvec)$, cf. Eq.~(\ref{eq:Mfi}).

Using the abbreviations
\ba
\vec{p}_{\sss{ppp}}&=\pvec+\ppvec+\frac{\ki+\kf}{2},\nonumber\\
\vec{p}_{\sss{mpp}}&=\pvec-\ppvec+\frac{\ki+\kf}{2},\nonumber\\
\vec{p}_{\sss{mpm}}&=\pvec-\ppvec+\frac{\ki-\kf}{2},\nonumber\\
\vec{p}_{\sss{mmp}}&=\pvec-\ppvec-\frac{\ki+\kf}{2}
\end{align}
we find:
\ba
T^{(2a)}&=\frac{e^2\,g_A^2}{4\,f_\pi^2\,m_N}\,
\frac{\epspvec\cdot\sigone\,
\vec{p}_{\sss{mpm}}\cdot\sigtwo\,\epsvec\cdot\pvec}
{\bigg[-\mpi^2-\vec{p}_{\sss{mpm}}^2\bigg]
\bigg[\w-B-\frac{\pvec^2+\pvec\cdot\ki}{m_N}\bigg]}
\nonumber\\
T^{(2b)}&=\frac{e^2\,g_A^2}{4\,f_\pi^2\,m_N}\,
\frac{\epsvec\cdot\sigone\,
\vec{p}_{\sss{mpm}}\cdot\sigtwo\,
\epspvec\cdot\left(\pvec-\frac{\ki}{2}\right)}
{\bigg[-\mpi^2-\vec{p}_{\sss{mpm}}^2\bigg]
\bigg[-\w-B-\frac{2\pvec^2+\w^2+\ki\cdot\kf-2\pvec\cdot\kf}{2m_N}\bigg]}
\nonumber\\
T^{(2c)}&=-\frac{e^2\,g_A^2}{4\,f_\pi^2\,m_N}\,
\frac{\epsvec\cdot\sigtwo\,
\vec{p}_{\sss{mpm}}\cdot\sigone\,\epspvec\cdot\ppvec}
{\bigg[-\mpi^2-\vec{p}_{\sss{mpm}}^2\bigg]
\bigg[\w-B-\frac{\ppvec^2+\ppvec\cdot\kf}{m_N}\bigg]}
\nonumber\\
T^{(2d)}&=-\frac{e^2\,g_A^2}{4\,f_\pi^2\,m_N}\,
\frac{\epspvec\cdot\sigone\,
\vec{p}_{\sss{mpm}}\cdot\sigtwo\,
\epsvec\cdot\left(\ppvec-\frac{\kf}{2}\right)}
{\bigg[-\mpi^2-\vec{p}_{\sss{mpm}}^2\bigg]
\bigg[-\w-B-\frac{2\ppvec^2+\w^2+\ki\cdot\kf-2\ppvec\cdot\ki}{2m_N}\bigg]}
\end{align}
\ba
T^{(3a)}&=-\frac{e^2\,g_A^2}{4\,f_\pi^2\,m_N}\,
\frac{\epsvec\cdot\sigtwo\,
\vec{p}_{\sss{mmp}}\cdot\sigone\,\epspvec\cdot\ppvec}
{\bigg[\w^2-\mpi^2-\vec{p}_{\sss{mmp}}^2\bigg]
\bigg[\w-B-\frac{\ppvec^2+\ppvec\cdot\kf}{m_N}\bigg]}
\nonumber\\
T^{(3b)}&=-\frac{e^2\,g_A^2}{4\,f_\pi^2\,m_N}\,
\frac{\epspvec\cdot\sigtwo\,
\vec{p}_{\sss{mpp}}\cdot\sigone\,
\epsvec\cdot\left(\ppvec-\frac{\kf}{2}\right)}
{\bigg[\w^2-\mpi^2-\vec{p}_{\sss{mpp}}^2\bigg]
\bigg[-\w-B-\frac{2\ppvec^2+\w^2+\ki\cdot\kf-2\ppvec\cdot\ki}{2m_N}\bigg]}
\nonumber\\
T^{(3c)}&=\frac{e^2\,g_A^2}{4\,f_\pi^2\,m_N}\,
\frac{\epspvec\cdot\sigtwo\,
\vec{p}_{\sss{mpp}}\cdot\sigone\,\epsvec\cdot\pvec}
{\bigg[\w^2-\mpi^2-\vec{p}_{\sss{mpp}}^2\bigg]
\bigg[\w-B-\frac{\pvec^2+\pvec\cdot\ki}{m_N}\bigg]}
\nonumber\\
T^{(3d)}&=\frac{e^2\,g_A^2}{4\,f_\pi^2\,m_N}\,
\frac{\epsvec\cdot\sigtwo\,
\vec{p}_{\sss{mmp}}\cdot\sigone\,
\epspvec\cdot\left(\pvec-\frac{\ki}{2}\right)}
{\bigg[\w^2-\mpi^2-\vec{p}_{\sss{mmp}}^2\bigg]
\bigg[-\w-B-\frac{2\pvec^2+\w^2+\ki\cdot\kf-2\pvec\cdot\kf}{2m_N}\bigg]}
\end{align}
\ba
T^{(4a)}&=\frac{e^2\,g_A^2}{8\,f_\pi^2\,m_N^2}\,
\frac{\vec{p}_{\sss{mpm}}\cdot\sigone\,
\vec{p}_{\sss{mpm}}\cdot\sigtwo\,
\epsvec\cdot\pvec\,\epspvec\cdot\ppvec}
{\bigg[-\mpi^2-\vec{p}_{\sss{mpm}}^2\bigg]
\bigg[\w-B-\frac{\pvec^2+\pvec\cdot\ki}{m_N}\bigg]
\bigg[\w-B-\frac{\ppvec^2+\ppvec\cdot\kf}{m_N}\bigg]}
\nonumber\\
T^{(4b)}&=\frac{e^2\,g_A^2}{8\,f_\pi^2\,m_N^2}\,
\frac{\vec{p}_{\sss{mpm}}\cdot\sigone\,
\vec{p}_{\sss{mpm}}\cdot\sigtwo\,
\epsvec\cdot\left(\ppvec-\frac{\kf}{2}\right)\,
\epspvec\cdot\left(\pvec-\frac{\ki}{2}\right)}
{\bigg[-\mpi^2-\vec{p}_{\sss{mpm}}^2\bigg]
\bigg[-\w-B-\frac{2\ppvec^2+\w^2+\ki\cdot\kf-2\ppvec\cdot\ki}{2m_N}\bigg]}
\nonumber\\
&\times\frac{1}
{\bigg[-\w-B-\frac{2\pvec^2 +\w^2+\ki\cdot\kf-2\pvec \cdot\kf}{2m_N}\bigg]}
\end{align}
\ba
T^{(5a)}&=\frac{e^2\,g_A^2}{4\,f_\pi^2\,m_N^2}\,
\frac{\vec{p}_{\sss{ppp}}\cdot\sigone\,
\vec{p}_{\sss{ppp}}\cdot\sigtwo\,
\epsvec\cdot\pvec\,\epspvec\cdot\ppvec}
{\bigg[\w^2-\mpi^2-\vec{p}_{\sss{ppp}}^2\bigg]
\bigg[\w-B-\frac{\pvec^2+\pvec\cdot\ki}{m_N}\bigg]
\bigg[\w-B-\frac{\ppvec^2+\ppvec\cdot\kf}{m_N}\bigg]}
\nonumber\\
T^{(5b)}&=-\frac{e^2\,g_A^2}{4\,f_\pi^2\,m_N^2}\,
\frac{\vec{p}_{\sss{mmp}}\cdot\sigone\,
\vec{p}_{\sss{mmp}}\cdot\sigtwo\,
\epsvec \cdot\left(\ppvec+\frac{\kf}{2}\right)\,
\epspvec\cdot\left(\pvec -\frac{\ki}{2}\right)}
{\bigg[\w^2-\mpi^2-\vec{p}_{\sss{mmp}}^2\bigg]
\bigg[-\w-B-\frac{2\ppvec^2+\w^2+\ki\cdot\kf+2\ppvec\cdot\ki}{2m_N}\bigg]}
\nonumber\\
&\times\frac{1}
{\bigg[-\w-B-\frac{2\pvec^2 +\w^2+\ki\cdot\kf-2\pvec \cdot\kf}{2m_N}\bigg]}
\end{align}
\ba
T^{(6a)}&=\frac{e^2\,g_A^2}{2\,f_\pi^2\,m_N}\,
\frac{\vec{p}_{\sss{mpp}}\cdot\sigone\,
\vec{p}_{\sss{mpm}}\cdot\sigtwo\,
\epsvec\cdot\pvec\,\epspvec\cdot\left(\pvec-\ppvec+\frac{\ki}{2}\right)}
{\bigg[\w^2-\mpi^2-\vec{p}_{\sss{mpp}}^2\bigg]
\bigg[-\mpi^2-\vec{p}_{\sss{mpm}}^2\bigg]
\bigg[\w-B-\frac{\pvec^2+\pvec\cdot\ki}{m_N}\bigg]}
\nonumber\\
T^{(6b)}&=\frac{e^2\,g_A^2}{2\,f_\pi^2\,m_N}\,
\frac{\vec{p}_{\sss{mmp}}\cdot\sigone\,
\vec{p}_{\sss{mpm}}\cdot\sigtwo\,
\epsvec\cdot\left(\pvec-\ppvec-\frac{\kf}{2}\right)\,
\epspvec\cdot\left(\pvec-\frac{\ki}{2}\right)}
{\bigg[\w^2-\mpi^2-\vec{p}_{\sss{mmp}}^2\bigg]
\bigg[-\mpi^2-\vec{p}_{\sss{mpm}}^2\bigg]
\bigg[-\w-B-\frac{2\pvec^2+\w^2+\ki\cdot\kf-2\pvec\cdot\kf}{2m_N}\bigg]}
\nonumber\\
T^{(6c)}&=-\frac{e^2\,g_A^2}{2\,f_\pi^2\,m_N}\,
\frac{\vec{p}_{\sss{mmp}}\cdot\sigone\,
\vec{p}_{\sss{mpm}}\cdot\sigtwo\,
\epsvec\cdot\left(\pvec-\ppvec-\frac{\kf}{2}\right)\,\epspvec\cdot\ppvec}
{\bigg[\w^2-\mpi^2-\vec{p}_{\sss{mmp}}^2\bigg]
\bigg[-\mpi^2-\vec{p}_{\sss{mpm}}^2\bigg]
\bigg[\w-B-\frac{\ppvec^2+\ppvec\cdot\kf}{m_N}\bigg]}
\nonumber\\
T^{(6d)}&=-\frac{e^2\,g_A^2}{2\,f_\pi^2\,m_N}\,
\frac{\vec{p}_{\sss{mpp}}\cdot\sigone\,
\vec{p}_{\sss{mpm}}\cdot\sigtwo\,
\epsvec\cdot\left(\ppvec-\frac{\kf}{2}\right)\,
\epspvec\cdot\left(\pvec-\ppvec+\frac{\ki}{2}\right)}
{\bigg[\w^2-\mpi^2-\vec{p}_{\sss{mpp}}^2\bigg]
\bigg[-\mpi^2-\vec{p}_{\sss{mpm}}^2\bigg]
\bigg[-\w-B-\frac{2\ppvec^2+\w^2+\ki\cdot\kf-2\ppvec\cdot\ki}{2m_N}\bigg]}
\end{align}
Exchange of the nucleons ($+(1\leftrightarrow2)$, cf. 
Fig.~\ref{fig:lowenergydiagrams}) has been skipped for brevity.

\chapter{Multipole Expansion of the Photon Field 
\label{app:multipoleexpansion} }
\markboth{APPENDIX \ref{app:multipoleexpansion}. MULTIPOLE EXPANSION}
         {APPENDIX \ref{app:multipoleexpansion}. MULTIPOLE EXPANSION}
The multipole expansion of $\vec{A}$ that we use for the 
non-perturbative approach to deuteron Compton scattering, described in 
Chapter~\ref{chap:nonperturbative}, has been derived in~\cite{Karakowski} 
in analogy to Chapter~7 of Ref.~\cite{Rose}. 
Nevertheless, it is worthwhile to repeat the derivation here.

We start with the expansion for 
$\hat{\epsilon}_\lambda\,\e^{i\vec{k}\cdot\vec{r}}$ with 
$\vec{k}\parallel\vec{e}_z$. For 
real photons, $\hat{\epsilon}_\lambda\perp \vec{k}$, therefore 
$\hat{\epsilon}_\lambda\equiv\hat{r}_\lambda$, the unit vector in the 
spherical basis, with $\lambda=\pm1$. The spherical basis
together with its scalar product is explained in App.~\ref{app:formulae}, 
Eqs.~(\ref{eq:sphericalbasis})-(\ref{eq:scalarproductspherical}).

Using the well-known expansion of the exponential function, cf. 
e.g.~\cite{Jackson}, we find
\ba
\hat{\epsilon}_\lambda\,\e^{i\vec{k}\cdot\vec{r}}|_{\hat{k}=\hat{z}}=
\hat{r}       _\lambda\,\e^{i\vec{k}\cdot\vec{r}}|_{\hat{k}=\hat{z}}&=
\hat{r}       _\lambda\sum_{l=0}^\infty\sum_{m=-l}^l\,4\pi\,i^l\,j_l(\w r)\,
Y_{l\,m}^\ast(\hat{k})\,Y_{l\,m}(\hat{r})\nonumber\\&=
\hat{r}       _\lambda\sum_{l=0}^\infty i^l\,\sqrt{4\pi\,(2l+1)}\,j_l(\w r)\,
Y_{l\,0}(\hat{r}),
\label{eq:expansionstart}
\end{align}
where we have used 
Eq.~(\ref{eq:Yofez}). $j_l(z)$ are 
the spherical Bessel functions of the first kind, 
see Eq.~(\ref{eq:spherBesseldefinition}).

The vector spherical harmonic $\vsh{J}{l}{M}$ is defined as
\be
\vsh{J}{l}{M}=\sum_{\nu=-1,0,1} 
(-1)^{1-l-M}\,\sqrt{2J+1}\,\threej{l}{1}{J}{M+\nu}{-\nu}{-M}\,
Y_{l\,M+\nu}\,\hat{r}_{-\nu},
\label{eq:vsh}
\ee
cf. Eq.~(\ref{eq:vshdefinition}). Eq.~(\ref{eq:vsh}) can be inverted to
\be
Y_{l\,m}\,\hat{r}_{-\lambda}=\sum_{J=0}^\infty\sum_{M=-J}^J(-1)^{-l+M+1}\,
\sqrt{2J+1}\,\threej{l}{1}{J}{m}{-\lambda}{-M}\,\vsh{J}{l}{M}.
\label{eq:Yr}
\ee
This is easily proven when we use the definition of the Clebsch-Gordan 
coefficients, Eq.~(\ref{eq:CGdefinition}), to write
\be
Y_{l\,m}\,\hat{r}_{-\lambda}=\sum_{J=0}^\infty\sum_{M=-J}^J
\Clebsch{l}{m}{1}{-\lambda}{J}{M}\,\vsh{J}{l}{M}
\label{eq:YrClebsch}
\ee
and
\be
\vsh{J}{l}{M}=\sum_{\nu=-1,0,1}\Clebsch{l}{M+\nu}{1}{-\nu}{J}{M}\,
Y_{l\,M+v}\,\hat{r}_{-\nu}=
\sum_{\nu,\tilde{m}}\Clebsch{l}{\tilde{m}}{1}{-\nu}{J}{M}\,
Y_{l\,\tilde{m}}\,\hat{r}_{-\nu},
\label{eq:vshClebsch}
\ee
where we have used $\tilde{m}-\nu-M=0$, cf. 
Eq.~(\ref{eq:projections}).
Inserting (\ref{eq:vshClebsch}) into (\ref{eq:YrClebsch}) and making use of 
the unitarity of the Clebsch-Gordan coefficients~(\ref{eq:threejunitarity}),
\be
\sum_{J,M}\Clebsch{l}{m}{1}{-\lambda}{J}{M}\,
\Clebsch{l}{\tilde{m}}{1}{-\nu}{J}{M}=
\delta_{\lambda,\nu}\,\delta_{m,\tilde{m}},
\ee
Eq.~(\ref{eq:Yr}) follows immediately and can be used to write 
Eq.~(\ref{eq:expansionstart}) as
\be
\hat{\epsilon}_\lambda\,\e^{i\vec{k}\cdot\vec{r}}|_{\hat{k}=\hat{z}}=
\sum_{l=0}^\infty\sum_{J=|l-1|}^{l+1}(-1)^{-l+\lambda+1}\,i^l\,
\sqrt{4\pi\,(2l+1)\,(2J+1)}\,\threej{l}{1}{J}{0}{\lambda}{-\lambda}\,
j_l(\w r)\,\vsh{J}{l}{\lambda}.
\ee 
The possible values for $J$ are constrained by the triangular condition of the 
Clebsch-Gordan coefficients, Eq.~(\ref{eq:triangularcondition}). This 
condition guarantees that for $l=0$ only $J=1$ gives a non-vanishing 
contribution. Therefore
\be
\hat{\epsilon}_\lambda\,\e^{i\vec{k}\cdot\vec{r}}|_{\hat{k}=\hat{z}}=
\sum_{l=0}^\infty\sum_{J=l-1}^{l+1}(-1)^{-l}\,i^l\,
\sqrt{4\pi\,(2l+1)\,(2J+1)}\,\threej{l}{1}{J}{0}{\lambda}{-\lambda}\,
j_l(\w r)\,\vsh{J}{l}{\lambda},
\ee
where we have used $(-1)^{\lambda+1}=1$ as $\lambda=\pm1$. Explicitly writing
out the sum over $J$ yields
\ba
\hat{\epsilon}_\lambda\,&\e^{i\vec{k}\cdot\vec{r}}|_{\hat{k}=\hat{z}}
=\sum_{l=0}^\infty(-1)^{-l}\,i^l\,
\sqrt{4\pi\,(2l+1)}\,j_l(\w r)\,\left\{
 \sqrt{2l-1}\,\threej{l}{1}{l-1}{0}{\lambda}{-\lambda}\,\vsh{l-1}{l}{\lambda}
\right.
\nonumber\\
&+\left.
\sqrt{2l+1}\,\threej{l}{1}{l}  {0}{\lambda}{-\lambda}\,\vsh{l}  {l}{\lambda}
+\sqrt{2l+3}\,\threej{l}{1}{l+1}{0}{\lambda}{-\lambda}\,\vsh{l+1}{l}{\lambda}
\right\}.
\label{eq:Jexplicitly}
\end{align}
The 3-$j$~symbols in Eq.~(\ref{eq:Jexplicitly}) are given by 
\ba
\threej{l}{1}{l-1}{0}{\lambda}{-\lambda}&=
 \frac{(-1)^{-l}\,\sqrt{l-1}}{\sqrt{2\,(2l-1)\,(2l+1)}},\nonumber\\
\threej{l}{1}{l}  {0}{\lambda}{-\lambda}&=
-\lambda\,\frac{(-1)^{-l}}{\sqrt{2\,(2l+1)}},\nonumber\\
\threej{l}{1}{l+1}{0}{\lambda}{-\lambda}&=
 \frac{(-1)^{l}\,\sqrt{l+2}}{\sqrt{2\,(2l+1)\,(2l+3)}},\nonumber
\end{align}
so we end up with ($(-1)^{2\,l}=1$)
\be
\hat{\epsilon}_\lambda\,\e^{i\vec{k}\cdot\vec{r}}|_{\hat{k}=\hat{z}}=
\sum_{l=0}^\infty i^l\,\sqrt{2\pi}\,j_l(\w r)\,\left\{
\sqrt{l-1}\,\vsh{l-1}{l}{\lambda}-\lambda\,\sqrt{2\,l+1}\,\vsh{l}{l}{\lambda}+
\sqrt{l+2}\,\vsh{l+1}{l}{\lambda}\right\}.
\label{eq:TTT}
\ee
For $l=0$, the first two terms vanish due to the 3-$j$~symbol in the definition
of $\vsh{J}{l}{M}$, cf. Eq.~(\ref{eq:vsh}), for $l=1$, the first term vanishes 
due to $\sqrt{l-1}$. Therefore, we can rearrange Eq.~(\ref{eq:TTT}) such that 
all of the vector spherical harmonics have the same total angular momentum, 
which we call $L$. The result is
\ba
\hat{\epsilon}_\lambda\,\e^{i\vec{k}\cdot\vec{r}}|_{\hat{k}=\hat{z}}&=
\sum_{L=1}^\infty i^L\,
\sqrt{2\pi\,(2L+1)}\,\left\{
i\,\sqrt{\frac{L}{2L+1}}\,j_{L+1}(\w r)\,\vsh{L}{L+1}{\lambda}\right.
\nonumber\\
&-\left.\lambda\,j_L(\w r)\,\vsh{L}{L}{\lambda}
-i\sqrt{\frac{L+1}{2L+1}}\,j_{L-1}(\w r)\,\vsh{L}{L-1}{\lambda}
\right\}\!.
\label{eq:E8}
\end{align}
Now we want to generalize this expression to arbitrary directions of $\vec{k}$.
This can be achieved by acting with the rotation operator on the 
angle-dependent quantities on the right-hand side of Eq.~(\ref{eq:E8}), which 
are the vector spherical harmonics. The matrix elements of the rotation 
operator are
\be
\mx{j\,m}{\hat{D}(\alpha,\beta,\gamma)}{j'\,m'}=
\delta_{j,j'}\,D_{m,m'}^j(\alpha,\beta,\gamma)
\ee
with the Wigner $D$-functions $D_{m,m'}^j(\alpha,\beta,\gamma)$. These 
can be written as
\be
D_{m,m'}^j(\alpha,\beta,\gamma)=
\e^{-i\,m\,\alpha}\,d_{m,m'}^j(\beta)\,\e^{-i\,m\,\gamma}.
\ee
As we only have to take into account one Euler angle, namely the scattering 
angle $\theta$, we may choose $\alpha=0,\;\,\beta=\theta,\;\,\gamma=0$.
Using the well-known relation (see e.g. \cite{Edmonds}, Chapter~4)
\be
\hat{D}(\alpha,\beta,\gamma)\,u(j\,m)=
\sum_{m'}u(j\,m')\,D_{m',m}^j(\alpha,\beta,\gamma),
\ee
which in our case reduces to
\be
\hat{D}(0,\theta,0)\,u(j\,m)=
\sum_{m'}u(j\,m')\,\wignerd{j}{m'}{m},
\ee
we find
\ba
\hat{\epsilon}_\lambda\,\e^{i\vec{k}\cdot\vec{r}}&=
\sum_{L=1}^\infty \sum_{M=-L}^L i^L\,\sqrt{2\pi\,(2L+1)}\,\left\{
i\,\sqrt{\frac{L}{2L+1}}\,j_{L+1}(\w r)\,\vsh{L}{L+1}{M}\right.
\nonumber\\
&-\left.\lambda\,j_L(\w r)\,\vsh{L}{L}{M}
-i\sqrt{\frac{L+1}{2L+1}}\,j_{L-1}(\w r)\,\vsh{L}{L-1}{M}
\right\}\,\wignerd{L}{M}{\lambda}.
\label{eq:E8arbitrary}
\end{align}
The second term can be simplified via \cite{Edmonds}
\be
\vsh{L}{L}{M}=\frac{1}{\sqrt{L\,(L+1)}}\,\vec{L}\,Y_{L\,M}(\hat{r})
\label{eq:LonY}
\ee
with the orbital angular momentum operator $\vec{L}$.
In order to simplify the other two terms we show that 
\ba
\label{eq:othertwoterms}
&\sqrt{\frac{L+1}{2L+1}}\,j_{L-1}(\w r)\,
\vsh{L}{L-1}{M}-\sqrt{\frac{L}{2L+1}}\,j_{L+1}(\w r)\,\vsh{L}{L+1}{M}
\nonumber\\
=&\frac{1}{\sqrt{L\,(L+1)}}
\left\{\w\,\vec{r}\,j_L(\w r)\,Y_{L\,M}(\hat{r})+
\frac{1}{\w}\,\nab_r\left(1+r\frac{d}{dr}\right)\,j_L{\w r}\,
Y_{L\,M}(\hat{r})\right\}.
\end{align}
First we decompose $\vec{r}$ into spherical components via 
Eqs.~(\ref{eq:spherdecomposition}) and~(\ref{eq:usefulidentity}):
\be
\vec{r}=\sum_{\nu=-1,0,1}r\,(-1)^\nu\,r_\nu\,\hat{r}_{-\nu}=
\sum_{\nu=-1,0,1}r\,(-1)^\nu\,\sqrt{\frac{4\pi}{3}}\,Y_{1\,\nu}(\hat{r})\,
\hat{r}_{-\nu}
\ee
Using this relation we can write the first term on the right-hand side ($RHS1$)
of Eq.~(\ref{eq:othertwoterms}) as
\be
RHS1=\sum_{\nu=-1,0,1}\frac{r\,\w}{\sqrt{L\,(L+1)}}\,j_L(\w r)\,(-1)^\nu\,
\sqrt{\frac{4\pi}{3}}\,Y_{1\,\nu}(\hat{r})\,\hat{r}_{-\nu}\,Y_{L\,M}(\hat{r}).
\ee
Now, using Eq.~(\ref{eq:additiontheorem}), the spherical harmonics are 
combined to
\ba
RHS1&=\sum_{\nu=-1,0,1}\sum_{\tilde{J}=|L-1|}^{L+1}
\sum_{\tilde{M}=-\tilde{J}}^{\tilde{J}}\frac{r\,\w}{\sqrt{L\,(L+1)}}\,
j_L(\w r)\,(-1)^{\nu+\tilde{M}}\,
\sqrt{(2L+1)\,(2\tilde{J}+1)}\nonumber\\
&\times\threej{1}{L}{\tilde{J}}{0}{0}{0}\,
\threej{1}{L}{\tilde{J}}{\nu}{M}{-\tilde{M}}\,
Y_{\tilde{J}\,\tilde{M}}(\hat{r})\,\hat{r}_{-\nu}.
\end{align}
The spherical harmonic can be replaced by a vector spherical harmonic, making 
use of Eq.~(\ref{eq:Yr}):
\ba
RHS1&=\sum_{\nu=-1,0,1}\sum_{\tilde{J}=|L-1|}^{L+1}
\sum_{J'=|\tilde{J}-1|}^{\tilde{J}+1}\sum_{\tilde{M},M'}
\frac{r\,\w}{\sqrt{L\,(L+1)}}\,
j_L(\w r)\,(-1)^{\nu+\tilde{M}+\tilde{J}-1+M'}\,
\sqrt{(2L+1)}\nonumber\\
&\times\sqrt{(2\tilde{J}+1)\,(2J'+1)}
\threej{1}{L}{\tilde{J}}{0}{0}{0}\,
\threej{1}{L}{\tilde{J}}{\nu}{M}{-\tilde{M}}\,
\threej{\tilde{J}}{1}{J'}{\tilde{M}}{-\nu}{-M'}\,\vsh{J'}{\tilde{J}}{M'}
\label{eq:spherharmreplaced}
\end{align}
The 3-$j$~symbol is symmetric under even permutations of rows and 
$$\threej{\tilde{J}}{1}{J'}{ \tilde{M}}{-\nu}{-M'}=(-1)^{\tilde{J}+1+J'}\,
 \threej{\tilde{J}}{1}{J'}{-\tilde{M}}{ \nu}{ M'},$$ 
cf. Eq.~(\ref{eq:threejprops3}). Furthermore 
$(-1)^{\nu+M'+\tilde{M}}=(-1)^{\nu+M'-\tilde{M}}$, as we are only concerned
with integer quantum numbers throughout this work, and 
$(-1)^{\nu+M'-\tilde{M}}=1$ due to the 3-$j$~symbols in 
Eq.~(\ref{eq:spherharmreplaced}).
Therefore we find
\ba
RHS1&=\sum_{\nu=-1,0,1}\sum_{\tilde{J}=|L-1|}^{L+1}
\sum_{J'=|\tilde{J}-1|}^{\tilde{J}+1}\frac{r\,\w}{\sqrt{L\,(L+1)}}\,
j_L(\w r)\,(-1)^{J'}\,\sqrt{(2L+1)\,(2\tilde{J}+1)\,(2J'+1)}\nonumber\\
&\times
\threej{1}{L}{\tilde{J}}{0}{0}{0}\,\vsh{J'}{\tilde{J}}{M'}\,\sum_{\tilde{M},M'}
\threej{\tilde{J}}{1}{L}{-\tilde{M}}{\nu}{M}\,
\threej{\tilde{J}}{1}{J'}{-\tilde{M}}{\nu}{M'}.
\end{align}
By use of the unitarity of the 3-$j$~symbols, 
Eq.~(\ref{eq:threejunitarity}), this becomes
\be
RHS1=\sum_{\tilde{J}=|L-1|}^{L+1}r\,\w\,(-1)^L\,j_L(\w r)\,
\sqrt{\frac{2\tilde{J}+1}{L\,(L+1)}}\,
\threej{1}{L}{\tilde{J}}{0}{0}{0}\,\vsh{L}{\tilde{J}}{M}.
\ee
The 3-$j$~symbol takes on the values 
$\frac{(-1)^{1-L}\,\sqrt{L+1}}{\sqrt{(2L+1)\,(2L+3)}}$ for $\tilde{J}=L+1$,
$0$ for $\tilde{J}=L$ and 
$\frac{(-1)^{L}\,\sqrt{L}}{\sqrt{(2L-1)\,(2L+1)}}$ for $\tilde{J}=L-1$. 
Therefore we get
\be
RHS1=r\,\w\,j_L(\w r)\,\left[\frac{\vsh{L}{L-1}{M}}{\sqrt{(L+1)\,(2L+1)}}
-\frac{\vsh{L}{L+1}{M}}{\sqrt{L\,(2L+1)}}
\right].
\label{eq:RHS1}
\ee
Now we have to write the second part on the right-hand side of 
Eq.~(\ref{eq:othertwoterms}) in terms of vector spherical harmonics. This can 
be achieved using the gradient formula, Eq.~(\ref{eq:gradientformula}), which 
gives
\ba
\nab_r&\left(1+r\frac{d}{dr}\right)j_L(\w r)\,Y_{L\,M}(\hat{r})=
\nonumber\\
&+\sqrt{\frac{  L}{2L+1}}\,\left((L+3)\,\frac{d}{dr}+r\frac{d^2}{dr^2}+
\frac{L+1}{r}\right)\,j_L(\w r)\,\vsh{L}{L-1}{M}\nonumber\\
&-\sqrt{\frac{L+1}{2L+1}}\,\left((2-L)\,\frac{d}{dr}+r\frac{d^2}{dr^2}-
\frac{L  }{r}\right)\,j_L(\w r)\,\vsh{L}{L+1}{M}.
\label{eq:gradientformulaonjY}
\end{align}
Using the recursion relations for spherical Bessel functions, 
Eq.~(\ref{eq:recursionrelations}), we find
\ba
\frac{1}{r}\,j_L(\w r)&=
\frac{\w}{2L+1}\,\left[j_{L-1}(\w r)+j_{L+1}(\w r)\right],\nonumber\\
\frac{d}{dr}j_L(\w r)&=\frac{\w}{2L+1}\,
\left[L\,j_{L-1}(\w r)-(L+1)\,j_{L+1}(\w r)\right],\nonumber\\
r\frac{d^2}{dr^2}j_L(\w r)&=\frac{r\,\w}{2L+1}\,
\left[\frac{L\,(L-1)}{r}\,j_{L-1}(\w r)
-\w\,(2L+1)\,j_L(\w r)+
\frac{(L+1)\,(L+2)}{r}\,j_{L+1}(\w r)\right].\nonumber
\end{align}
Inserting these three expressions into Eq.~(\ref{eq:gradientformulaonjY}), 
which is used to replace the corresponding term in 
Eq.~(\ref{eq:othertwoterms}), we end up with
\ba
RHS2&=\frac{\vsh{L}{L+1}{M}}{\sqrt{L\,(2L+1)}}\,
\left(r\,\w\,j_L(\w r)-L\,j_{L+1}(\w r)\right)\nonumber\\
&+\frac{\vsh{L}{L-1}{M}}{\sqrt{(L+1)\,(2L+1)}}\,
\left((L+1)\,j_{L-1}(\w r)-r\,\w\,j_L(\w r)\right).
\label{eq:RHS2}
\end{align}
Adding Eqs.~(\ref{eq:RHS1}) and (\ref{eq:RHS2}) we immediately see the identity
(\ref{eq:othertwoterms}). Combining 
Eqs.~(\ref{eq:E8arbitrary}),~(\ref{eq:LonY}) and~(\ref{eq:othertwoterms}), we 
get the final result for the multipole expansion:
\ba
\label{eq:multipoleexpfinal}
\hat{\epsilon}_\lambda\,\e^{i\vec{k}\cdot\vec{r}}&=
\sum_{L=1}^\infty \sum_{M=-L}^L \wignerd{L}{M}{\lambda}\,i^L\,
\sqrt{\frac{2\pi\,(2L+1)}{L\,(L+1)}}\\
&\times\left\{-\frac{i}{\w}\,\nab_r\left(1+r\frac{d}{dr}\right)\,j_L(\w r)\,
Y_{L\,M}(\hat{r})-i\,\w\,\vec{r}\,j_L(\w r)\,Y_{L\,M}(\hat{r})-\lambda\,
\vec{L}\,Y_{L\,M}(\hat{r})\,j_L(\w r)\right\}\nonumber
\end{align}
The first two terms correspond to electric, the third to magnetic photons, 
cf. e.g.~\cite{Rose}, Chapter~7.

In our calculation the incoming photon always moves along the $z$-direction, 
i.e. $\theta_i=0$. However, $d_{M,\lambda}^L(\theta=0)=\delta_{M,\lambda}$, 
cf. e.g.~\cite{VMK}. 
Therefore we get for the incoming photon
\ba
\label{eq:multipoleexpin}
\hat{\epsilon}_{\lambda_i}\,\e^{i\vec{k}_i\cdot\vec{r}}&=
-\sum_{L=1}^\infty \sum_{M=-L}^L \delta_{M,\lambda_i}\,i^L\,
\sqrt{\frac{2\pi\,(2L+1)}{L\,(L+1)}}\\
&\times\left\{\frac{i}{\w}\,\nab_r\left(1+r\frac{d}{dr}\right)\,j_L(\w r)\,
Y_{L\,M}(\hat{r})+\lambda_i\,\vec{L}\,Y_{L\,M}(\hat{r})\,j_L(\w r)+
i\,\w\,\vec{r}\,j_L(\w r)\,Y_{L\,M}(\hat{r})\right\}.\nonumber
\end{align}
Considering the case of an outgoing photon, the expansion would start with
\ba
\hat{\epsilon}_\lambda^\ast\,\e^{-i\vec{k}\cdot\vec{r}}|_{\hat{k}=\hat{z}}&=
\hat{r}       _\lambda^\ast\sum_{l=0}^\infty\sum_{m=-l}^l\,4\pi\,(-1)^l\,i^l\,
j_l(\w r)\,Y_{l\,m}^\ast(\hat{k})\,Y_{l\,m}(\hat{r})\\&=
(-1)^\lambda\,\hat{r}_{-\lambda}\sum_{l=0}^\infty\sum_{m=-l}^l\,4\pi\,(-1)^l\,
i^l\,j_l(\w r)\,Y_{l\,m}^\ast(\hat{k})\,Y_{l\,m}(\hat{r}).\nonumber
\end{align}
The equivalent to Eq.~(\ref{eq:E8}) therefore is
\ba
\hat{\epsilon}_\lambda^\ast\,\e^{-i\vec{k}\cdot\vec{r}}|_{\hat{k}=\hat{z}}&=
\sum_{L=1}^\infty i^L\,\sqrt{2\pi\,(2L+1)}\,(-1)^{L+\lambda}\,\left\{
-i\,\sqrt{\frac{L}{2L+1}}\,j_{L+1}(\w r)\,\vsh{L}{L+1}{-\lambda}\right.
\nonumber\\
&+\left.\lambda\,j_L(\w r)\,\vsh{L}{L}{-\lambda}
+i\sqrt{\frac{L+1}{2L+1}}\,j_{L-1}(\w r)\,\vsh{L}{L-1}{-\lambda}\right\}.
\end{align}
The signs follow from $\lambda\rightarrow -\lambda$ and 
$(-1)^{L+1}=(-1)^{L-1}=-(-1)^L$. Finally, $\wignerd{L}{M}{-\lambda}$ appears 
instead of $\wignerd{L}{M}{\lambda}$ in the pendant to 
Eq.~(\ref{eq:E8arbitrary}). Therefore we end up with
\ba
\label{eq:multipoleexpout}
\hat{\epsilon}_{\lambda_f}^\ast\,&\e^{-i\vec{k}_f\cdot\vec{r}}=
\sum_{L'=1}^\infty \sum_{M'=-L'}^{L'} \wignerd{L'}{M'}{-\lambda_f}\,i^{L'}\,
\sqrt{\frac{2\pi\,(2L'+1)}{L'\,(L'+1)}}\,(-1)^{L'+\lambda_f}\\
&\times\left\{\frac{i}{\w}\,\nab_r\left(1+r\frac{d}{dr}\right)\,j_{L'}(\w r)\,
Y_{L'\,M'}(\hat{r})+\lambda_f\,\vec{L}\,Y_{L'\,M'}(\hat{r})\,j_{L'}(\w r)+
i\,\w\,\vec{r}\,j_{L'}(\w r)\,Y_{L'\,M'}(\hat{r})\right\},\nonumber
\end{align}
where we use primed quantities in order to distinguish between outgoing and 
incoming photons, as we will throughout the whole work.

\chapter{Calculation of the Dominant Terms with $NN$-Rescattering 
\label{app:dominant} }
\markboth{APPENDIX \ref{app:dominant}. DOMINANT TERMS WITH $NN$-RESCATTERING}
         {APPENDIX \ref{app:dominant}. DOMINANT TERMS WITH $NN$-RESCATTERING}
We now evaluate Eqs.~(\ref{eq:Mfiphiphi1added}-\ref{eq:Mfiphiphi4added}), 
which are the amplitudes arising from 
the replacement $\vec{A}\rightarrow\nab\phi$ at both vertices. 
The results are already given (in the lab frame) in Ref.~\cite{Karakowski}.
We start with 
the two amplitudes including the intermediate state $\ket{C}$ and an energy 
denominator, $\Mfi{\phi\phi1}$ and $\Mfi{\phi\phi2}$.

We write out $\phiihat$ and $\phifhat$ according to 
Eq.~(\ref{eq:phidefinition})~-- remember, $\hat{\phi}_{i,f}$ was defined as 
$e\,\phi_{i,f}(\vec{r}/2)$~-- separate the radial from the angular wave 
function of the intermediate state $\ket{C}$ and insert two complete sets of 
radial states $\ket{r}$, $\ket{r'}$, as explained in Eq.~(\ref{eq:separation}).
We find
\ba
\Mfi{\phi\phi1}&=\left(\w+\frac{\w^2}{2m_d}\right)^2\sum_{\hat{C}}\doubleint
r^2dr\,r'^2dr'\,\bra{d_f}e\,\sum_{L'=1}^{\infty}\sum_{M'=-L'}^{L'}(-1)^{L'-\lf}
\,\wignerd{L'}{M'}{-\lf}\nonumber\\
&\times\frac{i^{L'+1}}{\w}\,
\sqrt{\frac{2\pi\,(2L'+1)}{L'\,(L'+1)}}\,\psi_{L'}(\frac{\w r'}{2})\,Y_{L'\,M'}
\ket{\hat{C}\,r'}\mx{r'}{\frac{1}{E_0-H_{\hat{C}}^{np}}}{r}\nonumber\\
&\times\mx{r\,\hat{C}}{-e\,\sum_{L=1}^{\infty}\sum_{M=-L}^{L}\delta_{M,\li}\,
\frac{i^{L+1}}{\w}\sqrt{\frac{2\pi\,(2L+1)}{L\,(L+1)}}\,\psi_L(\frac{\w r}{2})
\,Y_{L\,M}}{d_i}.
\label{eq:Mfiphiphi1first}
\end{align}
$E_0$ is defined as $E_0=\w+\frac{\w^2}{2m_d}-B$, cf. 
Sect.~\ref{sec:dominant}. We now use the form 
(\ref{eq:deuteronwavefunction}) for the deuteron wave 
functions and introduce the Green's function $\green$ according to 
Eq.~(\ref{eq:greendefinition}).
\ba
\Mfi{\phi\phi1}&=\sum_{L=1}^{\infty}\sum_{M=-L}^{L}
\sum_{L'=1}^{\infty}\sum_{M'=-L'}^{L'}\sum_{L_C,S_C,J_C,M_C}\sum_{l=0,2}
\sum_{l'=0,2}\left(1+\frac{\w}{2m_d}\right)^2\,2\pi\,e^2\,(-1)^{L'-\lf}
\nonumber\\
&\times i^{L+L'}\,\delta_{M,\li}\,\wignerd{L'}{M'}{-\lf}\,
\sqrt{\frac{(2L+1)\,(2L'+1)}{L\,(L+1)\,L'\,(L'+1)}}\nonumber\\
&\times\doubleint rdr\,r'dr'\,u_l(r)\,\psi_L(\frac{\w r}{2})\,\green\,
\psi_{L'}(\frac{\w r'}{2})\,u_{l'}(r')\\
&\times\mx{l'\,1\,1\,M_f}{Y_{L'\,M'}}{L_C\,S_C\,J_C\,M_C}\,
\mx{L_C\,S_C\,J_C\,M_C}{Y_{L\,M}}{l\,1\,1\,M_i}\nonumber
\end{align}
Now the Wigner-Eckart theorem (\ref{eq:WE}) is applied to both matrix 
elements. 
The resulting reduced matrix elements are given in Eq.~(\ref{eq:mxY}) and 
guarantee that $S_C=1$. Further we know that $M=\li$. Therefore we read off 
Eq.~(\ref{eq:projections}) $M_C=M_i+\li$ and $M'=M_f-M_i-\li$, which removes 
the formal sums over $M,\;M',\;M_C$. The possible values for $J_C$ are 
determined by the triangular condition (\ref{eq:triangularcondition}), which 
the 3-$j$~symbol has to fulfill. $L_C$ of course takes on the values 
$|J_C-S_C|,\cdots,J_C+S_C$. Therefore we find the final result
\ba
\label{eq:Mfiphiphi1final}
\Mfi{\phi\phi1}&=\sum_{L=1}^{\infty}
\sum_{L'=1}^{\infty}\sum_{J_C=|L-1|}^{L+1}\sum_{L_C=|J_C-1|}^{J_C+1}
\sum_{l,l'}\left(1+\frac{\w}{2m_d}\right)^2\,2\pi\,e^2\,
(-1)^{L'-\lf+1-M_f+J_C-M_i-\li}
\nonumber\\
&\times i^{L+L'}\,\wignerd{L'}{M_f-M_i-\li}{-\lf}\,
\sqrt{\frac{(2L+1)\,(2L'+1)}{L\,(L+1)\,L'\,(L'+1)}}\,
\threej{J_C}{L}{1}{-M_i-\li}{\li}{M_i}\nonumber\\
&\times\doubleint rdr\,r'dr'\,u_l(r)\,\psi_L(\frac{\w r}{2})\,\green\,
\psi_{L'}(\frac{\w r'}{2})\,u_{l'}(r')\\
&\times\threej{1}{L'}{J_C}{-M_f}{M_f-M_i-\li}{M_i+\li}\,
\mxred{l'\,1\,1}{Y_{L'}}{L_C\,1\,J_C}\,
\mxred{L_C\,1\,J_C}{Y_{L}}{l\,1\,1},\nonumber
\end{align}
where we used the shortcut $\sum_{l,l'}=\sum_{l=0,2}\sum_{l'=0,2}$ as we
always will throughout the whole work. Due to 
$\psi_1(\frac{\w r}{2})\rightarrow\frac{1}{3}\,\w r$ for $\w\rightarrow 0$,
we see that the amplitude $\Mfi{\phi\phi1}$~-- as well as $\Mfi{\phi\phi2}$ 
and $\Mfi{\phi\phi3}$~-- cannot contribute in the static limit.

The evaluation of $\Mfi{\phi\phi2}$ is quite similar. Therefore we only give
the final result:
\ba
\label{eq:Mfiphiphi2final}
\Mfi{\phi\phi2}&=\sum_{L=1}^{\infty}
\sum_{L'=1}^{\infty}\sum_{J_C=|L-1|}^{L+1}\sum_{L_C=|J_C-1|}^{J_C+1}
\sum_{l,l'}\left(\w-\frac{\w^2}{2m_d}+\frac{\PCsq}{2m_C}\right)^2\,
\frac{2\pi\,e^2}{\w^2}\,
(-1)^{L'-\lf+1+J_C-\li}
\nonumber\\
&\times i^{L+L'}\,\wignerd{L'}{M_f-M_i-\li}{-\lf}\,
\sqrt{\frac{(2L+1)\,(2L'+1)}{L\,(L+1)\,L'\,(L'+1)}}\,
\threej{1}{L}{J_C}{-M_f}{\li}{M_f-\li}\nonumber\\
&\times\doubleint rdr\,r'dr'\,u_{l'}(r)\,\psi_L(\frac{\w r}{2})\,\greenpr\,
\psi_{L'}(\frac{\w r'}{2})\,u_{l}(r')\\
&\times\threej{J_C}{L'}{1}{-M_f+\li}{M_f-M_i-\li}{M_i}\,
\mxred{l'\,1\,1}{Y_{L}}{L_C\,1\,J_C}\,
\mxred{L_C\,1\,J_C}{Y_{L'}}{l\,1\,1}\nonumber
\end{align}
We remind the reader that we use $m_C=2m_N$.
Note that in the $u$-channel diagrams, the Green's function depends on 
$E_0'=-\w-\frac{\PCsq}{2m_C}+\frac{\w^2}{2m_d}-B$, rather than on $E_0$ and 
$u_l$ ($u_{l'}$) depends on $r'$ ($r$), as we insert the complete set of 
states $\ket{r'}$ ($\ket{r}$) at the vertex of the outgoing (incoming) photon.

How to evaluate the double integrals over $r$, $r'$ in 
Eqs.~(\ref{eq:Mfiphiphi1final}) and~(\ref{eq:Mfiphiphi2final}) is described in 
Sect.~\ref{sec:dominant}. However, in the $u$-channel diagrams we do an 
approximation of the energy denominator as we want to avoid the scattering 
angle $\theta$ entering the denominator via $\PCsq$. The reason is that 
the numerical effort in solving the double integral is rather high, so 
we prefer to solve it only once for each energy. The momentum of the 
intermediate two-nucleon state in the $u$-channel, calculated
in the $\gamma d$-cm frame is $\PC=-\vec{k}_i-\vec{k}_f$. Therefore 
$\PCsq=(\vec{k}_i+\vec{k}_f)^2=2\,\w^2\,(1+\cos\theta)$. Now we expand the 
denominator of Eq.~(\ref{eq:Mfiphiphi2added}) for $\frac{\PCsq}{2m_C}\ll\w$
and $m_C\approx m_d$:
\be
\frac{1}{-\w+\frac{\w^2}{2m_d}-B-E_C-\frac{\PCsq}{2m_C}}\approx
\frac{1}{-\w-\frac{\w^2}{2m_d}-B-E_C}+
\frac{\frac{\w^2}{m_d}\,\cos\theta}
{\left(-\w-\frac{\w^2}{2m_d}-B-E_C\right)^2}
\label{eq:denomexp}
\ee
When we assume that the photon energy is the dominant quantity in the 
denominator, which should be justified as for low energies the  
diagrams including an energy denominator become negligible, we can further 
simplify Eq.~(\ref{eq:denomexp}):
\be
\frac{1}{-\w+\frac{\w^2}{2m_d}-B-E_C-\frac{\PCsq}{2m_C}}\approx
\frac{1-\frac{\w}{m_d}\,\cos\theta}
{-\w-\frac{\w^2}{2m_d}-B-E_C}
\ee
Therefore, in this approximation we just have to evaluate the double integral
for $\theta=\frac{\pi}{2}$ and multiply the result with 
$\left(1-\frac{\w}{m_d}\,\cos\theta\right)$. Numerical checks, that we 
performed for all energies at which we calculate the $\gamma d$ cross sections 
and for $\theta=5^\circ,\;180^\circ$ exhibit that the error on the integrals 
introduced by this approximation is well below 2\% in all considered cases.

Evaluation of $\Mfi{\phi\phi3}$ (Eq.~(\ref{eq:Mfiphiphi3added})) is 
straightforward. It is suppressed by $\frac{\w}{m_d}$ and therefore 
a small correction, as the photon energy which appears in
the prefactors of Eqs.~(\ref{eq:Mfiphiphi1added}, \ref{eq:Mfiphiphi2added}) 
drops out. Inserting the explicit expressions for $\phiihat,\;\phifhat$ we get
\ba
\Mfi{\phi\phi3}&=\sum_{L=1}^{\infty}\sum_{M=-L}^{L}
\sum_{L'=1}^{\infty}\sum_{M'=-L'}^{L'}\sum_{l,l'}
\left(\frac{\PCsq}{2m_C}-\frac{\w^2}{m_d}\right)\,
\frac{2\pi\,e^2}{\w^2}\,(-1)^{L'-\lf}
\nonumber\\
&\times i^{L+L'}\,\delta_{M,\li}\,\wignerd{L'}{M'}{-\lf}\,
\sqrt{\frac{(2L+1)\,(2L'+1)}{L\,(L+1)\,L'\,(L'+1)}}\\
&\times\int dr\,u_l(r)\,\psi_L(\frac{\w r}{2})\,
\psi_{L'}(\frac{\w r}{2})\,u_{l'}(r)
\,\mx{l'\,1\,1\,M_f}{Y_{L\,M}\,Y_{L'\,M'}}{l\,1\,1\,M_i}.\nonumber
\end{align}
We only need one complete set of states $\ket{r}$ due to the missing 
intermediate state. The spherical harmonics are combined using 
Eq.~(\ref{eq:additiontheorem}). We find
\ba
\Mfi{\phi\phi3}&=\sum_{L=1}^{\infty}\sum_{M=-L}^{L}
\sum_{L'=1}^{\infty}\sum_{M'=-L'}^{L'}\sum_{\tilde{L}=|L-L'|}^{L+L'}\sum_{l,l'}
\left(\frac{\PCsq}{2m_C}-\frac{\w^2}{m_d}\right)\,
\frac{\sqrt{\pi}\,e^2}{\w^2}\,
(-1)^{L'-\lf+M+M'}
\nonumber\\
&\times i^{L+L'}\,\delta_{M,\li}\,\wignerd{L'}{M'}{-\lf}\,(2L+1)\,(2L'+1)\,
\sqrt{\frac{2\tilde{L}+1}{L\,(L+1)\,L'\,(L'+1)}}\nonumber\\
&\times
\threej{L}{L'}{\tilde{L}}{0}{0}{0}\,\threej{L}{L'}{\tilde{L}}{M}{M'}{-M-M'}
\nonumber\\
&\times\int dr\,u_l(r)\,\psi_L(\frac{\w r}{2})\,
\psi_{L'}(\frac{\w r}{2})\,u_{l'}(r)
\,\mx{l'\,1\,1\,M_f}{Y_{\tilde{L}\,M+M'}}{l\,1\,1\,M_i}.
\end{align}
By use of the Wigner-Eckart theorem (\ref{eq:WE}) we obtain the final result
\ba
\Mfi{\phi\phi3}&=\sum_{L=1}^{\infty}
\sum_{L'=1}^{\infty}\sum_{\tilde{L}=|L-L'|}^{L+L'}\sum_{l,l'}
\left(\frac{\PCsq}{2m_C}-\frac{\w^2}{m_d}\right)\,
\frac{\sqrt{\pi}\,e^2}{\w^2}\,
(-1)^{L'-\lf+1-M_i}
\nonumber\\
&\times i^{L+L'}\,\wignerd{L'}{M_f-M_i-\li}{-\lf}\,(2L+1)\,(2L'+1)\,
\sqrt{\frac{2\tilde{L}+1}{L\,(L+1)\,L'\,(L'+1)}}\nonumber\\
&\times
\threej{L}{L'}{\tilde{L}}{0}{0}{0}\,
\threej{L}{L'}{\tilde{L}}{\li}{M_f-M_i-\li}{-M_f+M_i}\,
\threej{1}{\tilde{L}}{1}{-M_f}{M_f-M_i}{M_i}\nonumber\\
&\times\int dr\,u_l(r)\,\psi_L(\frac{\w r}{2})\,
\psi_{L'}(\frac{\w r}{2})\,u_{l'}(r)
\,\mxred{l'\,1\,1}{Y_{\tilde{L}}}{l\,1\,1}.
\end{align}
Now we turn to the amplitude containing the double commutators, 
Eq.~(\ref{eq:Mfiphiphi4added}). We evaluate this contribution by splitting
the Hamiltonian $H^{np}$ into a kinetic and a potential energy part. First we 
discuss the kinetic energy part, denoted by 'kE', i.e. 
$H^{np}\rightarrow \frac{\vec{p}^2}{m_N}$
in Eq.~(\ref{eq:Mfiphiphi4added}). The momentum operator $\vec{p}$ is
\be
\vec{p}=\frac{\vec{p}_p-\vec{p}_n}{2}=\frac{-i\,\nab_{x_p}+i\,\nab_{x_n}}{2}
=-i\,\nab_r.
\ee
The only $\vec{r}$-dependent quantities in $\phiihat,\;\phifhat$ are
$\psi_L\,Y_{L\,M}$ and $\psi_{L'}\,Y_{L'\,M'}$, respectively. Therefore we 
only consider the commutators 
$-\frac{1}{m_N}\,\left[[\nabsq,\psi_L\,Y_{L\,M}],
\psi_{L'}\,Y_{L'\,M'}\right]$ and 
$-\frac{1}{m_N}\,\left[[\nabsq,\psi_{L'}\,Y_{L'\,M'}],
\psi_{L}\,Y_{L\,M}\right]$.
Evaluating the first double commutator yields
\be
-\frac{1}{m_N}\,\left[[\nabsq,\psi_L\,Y_{L\,M}],
\psi_{L'}\,Y_{L'\,M'}\right]=-\frac{2}{m_N}
\,\left[\nab(\psi_L\,Y_{L\,M})\right]
\cdot\left[\nab(\psi_{L'}\,Y_{L'\,M'})\right].
\label{eq:firstcommutator}
\ee
The second double commutator in Eq.~(\ref{eq:Mfiphiphi4added}) obviously 
gives the same contribution, as Eq.~(\ref{eq:firstcommutator}) is symmetric 
under $L\leftrightarrow L'$, $M\leftrightarrow M'$. Now we use 
$(\nab f)\cdot(\nab g)=
\frac{1}{2}\left(\nabsq (f\,g)-f\,\nabsq g-g\,\nabsq f\right)$
to rewrite Eq.~(\ref{eq:firstcommutator}) as 
\ba
\label{eq:firstcommutatorrewritten}
-\frac{1}{m_N}\,\left[[\nabsq,\psi_L\,Y_{L\,M}],
\psi_{L'}\,Y_{L'\,M'}\right]&=-\frac{1}{m_N}
\,\left[\nabsq\left(\psi_L\,Y_{L\,M}\,\psi_{L'}\,Y_{L'\,M'}\right)\right.\\
&-\left.
\psi_L\,Y_{L\,M}\,\nabsq\left(\psi_{L'}\,Y_{L'\,M'}\right)-
\psi_{L'}\,Y_{L'\,M'}\,\nabsq\left(\psi_{L}\,Y_{L\,M}\right)\right].\nonumber
\end{align}
The two spherical harmonics in the first term on the right hand side
of Eq.~(\ref{eq:firstcommutatorrewritten}) combine to one.
Therefore we only need to evaluate the structure 
$\nabsq\left(f(r)\, Y_{L\,M}(\hat{r})\right)$. We do so by use of the 
gradient formula (\ref{eq:gradientformula}) and 
Eqs.~(\ref{eq:divergenceLp1}, \ref{eq:divergenceLm1}).
\ba
\nabsq\left(f(r)\,Y_{L\,M}(\hat{r})\right)&=\nab\,\left\{
-\sqrt{\frac{L+1}{2L+1}}\,\left(
\frac{\partial}{\partial r}-\frac{L}{r}\right)f(r)\,\vsh{L}{L+1}{M}\right.
\nonumber\\
&+\left.\sqrt{\frac{L}{2L+1}}\,
\left(\frac{\partial}{\partial r}+\frac{L+1}{r}\right)f(r)\,
\vsh{L}{L-1}{M}\right\}\nonumber\\
&=\frac{L+1}{2L+1}\left(\frac{\partial}{\partial r}+\frac{L+2}{r}\right)
\left(\frac{\partial}{\partial r}-\frac{L}{r}\right)f(r)\,Y_{L\,M}(\hat{r})
\nonumber\\
&+\frac{L}{2L+1}\left(\frac{\partial}{\partial r}-\frac{L-1}{r}\right)
\left(\frac{\partial}{\partial r}+\frac{L+1}{r}\right)f(r)\,Y_{L\,M}(\hat{r})
\nonumber\\
&=\left(\frac{\partial^2}{\partial r^2}+
\frac{2}{r}\,\frac{\partial}{\partial r}-\frac{L\,(L+1)}{r^2}\right)
\,f(r)\,Y_{L\,M}(\hat{r})\nonumber\\
&=Y_{L\,M}(\hat{r})\,\left(\frac{1}{r}\,\frac{\partial^2}{\partial r^2}\,r
-\frac{L\,(L+1)}{r^2}\right)f(r)
\end{align}
By the help of Eq.~(\ref{eq:additiontheorem}) we find
\ba
&-\frac{1}{m_N}\,\left[[\nabsq,\psi_L\,Y_{L\,M}],
\psi_{L'}\,Y_{L'\,M'}\right]=-\frac{1}{m_N}\sum_J (-1)^{M+M'}\,
\sqrt{\frac{(2L+1)\,(2L'+1)\,(2J+1)}{4\pi}}\nonumber\\
&\times\threej{L}{L'}{J}{0}{0}{0}\,
\threej{L}{L'}{J}{M}{M'}{-M-M'}\,Y_{J\,M+M'}\,
\left\{\left(\frac{1}{r}\,\frac{\partial^2}{\partial r^2}\,r-
\frac{J\,(J+1)}{r^2}\right)\,\psi_L\,\psi_{L'}\right.\nonumber\\
&-\left.
\psi_L\,\left(\frac{1}{r}\,\frac{\partial^2}{\partial r^2}\,r-
\frac{L'\,(L'+1)}{r^2}\right)\,\psi_{L'}-
\psi_{L'}\,\left(\frac{1}{r}\,\frac{\partial^2}{\partial r^2}\,r-
\frac{L\,(L+1)}{r^2}\right)\,\psi_{L}\right\}.
\end{align}
Including all prefactors and inserting one complete set of radial states
$\ket{r}$, we get for $\Mfi{\phi\phi4\,\mathrm{kE}}$ (remember, the second 
double commutator gives exactly the same contribution):
\ba
\Mfi{\phi\phi4\,\mathrm{kE}}&=-\sum_{L=1}^{\infty}\sum_{M=-L}^{L}
\sum_{L'=1}^{\infty}\sum_{M'=-L'}^{L'}\sum_{J=|L-L'|}^{L+L'}\sum_{l,l'}
\frac{\sqrt{\pi}\,e^2}{\w^2\,m_N}\,(-1)^{L'-\lf+M+M'}
\,i^{L+L'}\,(2L+1)\nonumber\\
&\times(2L'+1)\,
\sqrt{\frac{2J+1}{L\,(L+1)\,L'\,(L'+1)}}\,\threej{L}{L'}{J}{0}{0}{0}
\,\threej{L}{L'}{J}{M}{M'}{-M-M'}\nonumber\\
&\times\delta_{M,\li}\,\wignerd{L'}{M'}{-\lf}\,
\mx{l'\,1\,1\,M_f}{Y_{J\,M+M'}}{l\,1\,1\,M_i}\nonumber\\
&\times\int dr\,u_l(r)\,u_{l'}(r)\,\left\{
\left(\frac{1}{r}\,\frac{\partial^2}{\partial r^2}\,r-
\frac{J\,(J+1)}{r^2}\right)\,\psi_L(\frac{\w r}{2})\,
\psi_{L'}(\frac{\w r}{2})-\psi_L(\frac{\w r}{2})\,\left(\frac{1}{r}\,
\frac{\partial^2}{\partial r^2}\,r\right.\right.\nonumber\\
&-\left.\left.
\frac{L'\,(L'+1)}{r^2}\right)\,\psi_{L'}(\frac{\w r}{2})-
\psi_{L'}(\frac{\w r}{2})\,\left(\frac{1}{r}\,
\frac{\partial^2}{\partial r^2}\,r-
\frac{L\,(L+1)}{r^2}\right)\,\psi_{L}(\frac{\w r}{2})
\right\}.
\end{align}
The derivatives can be written in a more compact form and after using the 
Wigner-Eckart theorem (\ref{eq:WE}) the formal sums over $M,\;M'$ are removed 
to yield the result
\ba
\Mfi{\phi\phi4\,\mathrm{kE}}&=\sum_{L=1}^{\infty}
\sum_{L'=1}^{\infty}\sum_{J=|L-L'|}^{L+L'}\sum_{l,l'}
\frac{\sqrt{\pi}\,e^2}{\w^2\,m_N}\,(-1)^{L'-\lf-M_i}
\,i^{L+L'}\,(2L+1)\,(2L'+1)\nonumber\\
&\times
\sqrt{\frac{2J+1}{L\,(L+1)\,L'\,(L'+1)}}\,\threej{L}{L'}{J}{0}{0}{0}
\,\threej{L}{L'}{J}{\li}{M_f-M_i-\li}{-M_f+M_i}\nonumber\\
&\times\threej{1}{J}{1}{-M_f}{M_f-M_i}{M_i}\,
\wignerd{L'}{M_f-M_i-\li}{-\lf}\,
\mxred{l'\,1\,1}{Y_{J}}{l\,1\,1}\nonumber\\
&\times\int dr\,u_l(r)\,u_{l'}(r)\,\left\{
2\,\left(\frac{\partial}{\partial r}\,\psi_L   (\frac{\w r}{2})\right)
   \left(\frac{\partial}{\partial r}\,\psi_{L'}(\frac{\w r}{2})\right)
\right.\nonumber\\
&+\left.
\frac{L\,(L+1)+L'\,(L'+1)-J\,(J+1)}{r^2}\,
\psi_{L}(\frac{\w r}{2})\psi_{L'}(\frac{\w r}{2})\right\}.
\label{eq:Mfiphiphi4kefinal}
\end{align}

The final task of this appendix is to calculate Eq.~(\ref{eq:Mfiphiphi4added})
with the potential energy part of $H^{np}$ inserted into the commutators.
It was shown in Ref.~\cite{Friar} that the correct low-energy limit is a direct
consequence from demanding gauge invariance of the calculation. 
Therefore, in order to fulfill the low-energy theorem,
it is necessary to be consistent between the explicitly included 
exchange particles and the potential used in the double commutator, which is 
the only part of the two-nucleon reducible amplitude that survives in the 
static limit.
In fact it was proven in Ref.~\cite{Arenhoevel}, that the pion-exchange 
diagrams of Fig.~\ref{fig:chiPTdouble} cancel exactly in the static limit 
against the double commutator including the one-pion-exchange potential 
$V^\mathrm{OPE}$.
The short-distance (hard-core) part of the two-nucleon potential may be 
interpreted as the exchange of mesons heavier than the pion
($\rho,\;\omega,...$), see e.g. the CD-Bonn potential~\cite{Bonn}.
As pointed out in Ref.~\cite{Lvov}, in order to achieve full
consistency one would also have to allow for the explicit exchange of such 
particles. Nevertheless, below the pion-production threshold our approximation
to only include the one-pion exchange explicitly is certainly sufficient, and, 
consequently, we only include $V^\mathrm{OPE}$ in the double commutator. 
The validity of this procedure is further confirmed in 
Section~\ref{sec:potentialdep}, where we demonstrate that our calculation is 
nearly insensitive to features of the $np$-potential which go beyond the
one-pion exchange.

The only operators in this potential, which do not commute with 
$\phi_{i,f}(\vec{x}_p)$, are the isospin operators,
because also $\phi(\vec{x}_p)$ depends on the isospin.
In the AV18-notation \cite{AV18}, $V^\mathrm{OPE}$ reads,
cf. Eq.~(\ref{eq:VpiAV18}),
\ba
V_\pi^{AV18}(\vec{r})&=
-f^2\,v_\pi^{AV18}(m_{\pi^0})+(-1)^{T+1}\,2\,f^2\,v_\pi^{AV18}(m_{\pi^\pm})
\nonumber\\
&\approx\left(2\,(-1)^{T+1}-1\right)\,f^2\,v_\pi^{AV18}(m_\pi)\nonumber\\
&=f^2\,v_\pi^{AV18}(m_\pi)\cdot
\left\{\parbox{.2\linewidth}{$-3\qquad T=0$\\$\phantom{-}1\qquad T=1$,}\right.
\label{eq:VpiAV18AV18}
\end{align}
where we neglect the mass difference between charged and neutral pions. The 
isospin dependence can be written as 
\be
V_\pi^{AV18}(\vec{r})=f^2\,v_\pi^{AV18}(m_\pi)\,\taudottau,
\label{eq:Vpilong}
\ee
because
\ba
\frac{1}{2}\,\mx{p\,n-n\,p}{\taudottau}{p\,n-n\,p}&=-3,\nonumber\\
\frac{1}{2}\,\mx{p\,n+n\,p}{\taudottau}{p\,n+n\,p}&=1.
\end{align}
Here we used the isospin-0 (deuteron) wave function 
$\frac{1}{\sqrt{2}}\ket{p\,n-n\,p}$ and the isospin-1 wave function 
$\frac{1}{\sqrt{2}}\ket{p\,n+n\,p}$, respectively. The potential 
$v_\pi^{AV18}(m_\pi)$
is given as, cf. \cite{AV18} and Eq.~(\ref{eq:vpiAV18}),
\be
v_\pi^{AV18}(m_\pi)=\frac{1}{3}\,m_\pi\,
\left[Y^{AV18}(r)\,\vec{\sigma}_1\cdot\vec{\sigma}_2+
T^{AV18}(r)\,S_{12}\right],
\ee
where we set $m_{\pi^\pm}=m_{\pi^0}=m_\pi$, like in 
Eq.~(\ref{eq:VpiAV18AV18}). The functions 
$Y^{AV18}(r)$, $T^{AV18}(r)$ are (see Eqs.~(\ref{eq:YmAV18}, \ref{eq:TmAV18}))
\ba
Y^{AV18}(r)&=\frac{\e^{-m_\pi r}}{m_\pi r}\,\left(1-\e^{-c r^2}\right),
\nonumber\\
T^{AV18}(r)&=\left(1+\frac{3}{m_\pi r}+\frac{3}{(m_\pi r)^2}\right)\,
\frac{\e^{-m_\pi r}}{m_\pi r}\,\left(1-\e^{-c r^2}\right)^2,
\end{align}
with the cutoff parameter $c$ assigned the value $c=2.1$~$\fm^{-2}$ in 
\cite{AV18}, and the operator $S_{12}$ is defined via
\be
S_{1 2}=3\,\left(\vec{\sigma}_1\,\cdot\hat{r}\right)\,
\left(\vec{\sigma}_2\,\cdot\hat{r}\right)-\vec{\sigma}_1\cdot\vec{\sigma}_2.
\ee
However, it is known \cite{Arenhoevel,Karakowski} that the one-pion-exchange 
potential\footnote{We omit the term proportional to $\delta(\vec{r})$,
wich gives no contribution due to the vanishing deuteron wave function 
at the origin.}
\be
V_\pi(\vec{r})=f^2\,m_\pi\,\taudottau\,
\left[S_{12}\,\left(\frac{1}{3}+\frac{1}{m_\pi r}+\frac{1}{(m_\pi r)^2}\right)
\frac{\e^{-m_\pi r}}{m_\pi r}+\vec{\sigma}_1\cdot\vec{\sigma}_2\,
\frac{\e^{-m_\pi r}}{3\,m_\pi r}\right],
\label{eq:Vpishort}
\ee
cf. e.g. \cite{Ericson}, Section~3.2, together with the pion-exchange diagrams
of Fig.~\ref{fig:chiPTdouble} generates the correct Thomson limit, cf. 
Section~\ref{sec:Thomson2}. Therefore we remove the cutoff factors 
$\left(1-\e^{-c r^2}\right)$, $\left(1-\e^{-c r^2}\right)^2$, 
 which are introduced in 
\cite{AV18} in order to have a finite potential at the origin, and rather use
the potential (\ref{eq:Vpishort}) instead of (\ref{eq:Vpilong}). 
We mark the difference between the thus defined functions and the ones 
used in \cite{AV18} by skipping the index '$AV18$'.
We are aware that this procedure is strictly speaking not consistent, as we 
are using $V_\pi^{AV18}(\vec{r})$ in all other parts of this work, but our 
aim is to achieve the exact static limit on the one hand and to use a modern
$np$-potential as far as possible on the other.

In order to evaluate the double commutators 
$\left[[V_\pi,\phiihat],\phifhat\right]$ and 
$\left[[V_\pi,\phifhat],\phiihat\right]$, we explicitly write the 
isospin dependence of $\hat{\phi}_{i,f}$ (cf. Eq.~(\ref{eq:rhonull})):
\be
\hat{\phi}_{i,f}=e\,\phi_{i,f}(\vec{x}_p)=
\sum_{j=n,p}e_j\,\phi_{i,f}(\vec{x}_j)=
\sum_{j=n,p}\frac{1}{2}\left(1+\tau_j^z\right)\,e\,\phi_{i,f}(\vec{x}_j)
\label{eq:isospindep}
\ee
Therefore, we have
\ba
\left[[\taudottau,\phiihat],\phifhat\right]&=\frac{e^2}{4}\sum_{l,m=n,p}
\left[[\taudottau,(1+\tau_l^z)],(1+\tau_m^z)\right]\,
\phi_i(\vec{x}_l)\,\phi_f(\vec{x}_m)\nonumber\\
&=\frac{e^2}{4}\sum_{l,m=n,p}
\left[[\tau_p^i\,\tau_n^i,\tau_l^z],\tau_m^z\right]\,
\phi_i(\vec{x}_l)\,\phi_f(\vec{x}_m)\nonumber\\
&=\frac{e^2}{4}\left\{\left[[\tau_p^i\,\tau_n^i,\tau_p^z],\tau_p^z\right]\,
\phi_i(\vec{x}_p)\,\phi_f(\vec{x}_p)
+\left[[\tau_p^i\,\tau_n^i,\tau_p^z],\tau_n^z\right]\,
\phi_i(\vec{x}_p)\,\phi_f(\vec{x}_n)\right.\nonumber\\
&+\left.\left[[\tau_p^i\,\tau_n^i,\tau_n^z],\tau_p^z\right]\,
\phi_i(\vec{x}_n)\,\phi_f(\vec{x}_p)
+\left[[\tau_p^i\,\tau_n^i,\tau_n^z],\tau_n^z\right]\,
\phi_i(\vec{x}_n)\,\phi_f(\vec{x}_n)\right\}.
\end{align}
Using the commutator relation $[\tau_i,\tau_j]=2\,i\,\varepsilon_{ijk}\,\tau_k$
we find
\ba
\left[[\taudottau,\phiihat],\phifhat\right]&=
e^2\left\{(\vec{\tau}_1\cdot\vec{\tau}_2-\tau_1^z\,\tau_2^z)\,
\phi_i(\vec{x}_p)\,\phi_f(\vec{x}_p)-
(\vec{\tau}_1\cdot\vec{\tau}_2-\tau_1^z\,\tau_2^z)\,
\phi_i(\vec{x}_p)\,\phi_f(\vec{x}_n)\right.\nonumber\\
&-\left.(\vec{\tau}_1\cdot\vec{\tau}_2-\tau_1^z\,\tau_2^z)\,
\phi_i(\vec{x}_n)\,\phi_f(\vec{x}_p)+
(\vec{\tau}_1\cdot\vec{\tau}_2-\tau_1^z\,\tau_2^z)\,
\phi_i(\vec{x}_n)\,\phi_f(\vec{x}_n)\right\}.
\end{align}
From this equation it is obvious that 
$\left[[\taudottau,\phifhat],\phiihat\right]=
 \left[[\taudottau,\phiihat],\phifhat\right]$. Evaluating the isospin operator
$(\vec{\tau}_1\cdot\vec{\tau}_2-\tau_1^z\,\tau_2^z)$ between two deuteron 
($T=0$) wave functions gives
\be
\frac{1}{2}\mx{p\,n-n\,p}{\vec{\tau}_1\cdot\vec{\tau}_2-\tau_1^z\,\tau_2^z}
{p\,n-n\,p}=-2.
\ee
Therefore we can rewrite the potential energy part of 
Eq.~(\ref{eq:Mfiphiphi4added}) as
\ba
\Mfi{\phi\phi4\,\mathrm{pot}}&=\bra{d_f} -2\,e^2\,f^2\,v_\pi(m_\pi)\,
[\phi_i(\vec{x}_p)\,\phi_f(\vec{x}_p)-
 \phi_i(\vec{x}_p)\,\phi_f(\vec{x}_n)\nonumber\\
&-
 \phi_i(\vec{x}_n)\,\phi_f(\vec{x}_p)+
 \phi_i(\vec{x}_n)\,\phi_f(\vec{x}_n)]\ket{d_i}.
\end{align}
We replace $\vec{x}_p\rightarrow \frac{\vec{r}}{2}$, 
$\vec{x}_n\rightarrow -\frac{\vec{r}}{2}$ as in Eq.~(\ref{eq:replacexpxn}).
Inserting one complete set of radial states and the explicit expressions for 
$\phi_i(\pm\vec{r}/2)$, $\phi_f(\pm\vec{r}/2)$, cf. 
Eq.~(\ref{eq:phidefinition}), we find
\ba
\label{eq:Mfiphiphi4first}
\Mfi{\phi\phi4\,\mathrm{pot}}&=\sum_{L=1}^\infty\sum_{M=-L}^L
\sum_{L'=1}^\infty\sum_{M'=-L'}^{L'}\sum_{l,l'}
\frac{4\pi\,e^2\,f^2\,m_\pi}{3\,\w^2}\,(-1)^{1+L'-\lf}\,i^{L+L'}\,
\delta_{M,\li}\,\wignerd{L'}{M'}{-\lf}\nonumber\\
&\times\sqrt{\frac{(2L+1)\,(2L'+1)}{L\,(L+1)\,L'\,(L'+1)}}\,
\bigg\{\int dr\,u_l(r)\,u_{l'}(r)\,
\psi_L(\frac{\w r}{2})\,\psi_{L'}(\frac{\w r}{2})\,Y(r)\nonumber\\
&\times
\mx{l'\,1\,1\,M_f}
{\vec{\sigma}_1\cdot\vec{\sigma}_2\,Y_{L\,M}(\hat{r})\,Y_{L'\,M'}(\hat{r})}
{l\,1\,1\,M_i}\nonumber\\
&+\int dr\,u_l(r)\,u_{l'}(r)\,
\psi_L(\frac{\w r}{2})\,\psi_{L'}(\frac{\w r}{2})\,T(r)\\
&\times
\mx{l'\,1\,1\,M_f}
{S_{12}\,Y_{L\,M}(\hat{r})\,Y_{L'\,M'}(\hat{r})}{l\,1\,1\,M_i}\bigg\}
\left(1-(-1)^{L'}-(-1)^L+(-1)^{L+L'}\right)\nonumber.
\end{align}
Here we used Eq.~(\ref{eq:Yofmr}), i.e. 
$Y_{L\,M}(-\hat{r})=(-1)^L\,Y_{L\,M}(\hat{r})$. As explained before, 
$T(r),\;Y(r)$ are identical to 
$T^{AV18}(r),\;Y^{AV18}(r)$ with the cutoff functions removed.
We rewrite Eq.~(\ref{eq:Mfiphiphi4first}), separating 
the two operators $\vec{\sigma}_1\cdot\vec{\sigma}_2$ and 
$\left(\vec{\sigma}_1\,\cdot\hat{r}\right)\,
\left(\vec{\sigma}_2\,\cdot\hat{r}\right)$ from each other. 
The sum over $M$ may already be removed.
\ba
\Mfi{\phi\phi4\,\mathrm{pot}}&=\sum_{L=1}^\infty
\sum_{L'=1}^\infty\sum_{M'=-L'}^{L'}\sum_{l,l'}\sum_{i}
\frac{4\pi\,e^2\,f^2\,m_\pi}{3\,\w^2}\,(-1)^{1+L'-\lf+i}\,i^{L+L'}\,
\wignerd{L'}{M'}{-\lf}\nonumber\\
&\times\sqrt{\frac{(2L+1)\,(2L'+1)}{L\,(L+1)\,L'\,(L'+1)}}\,
\bigg\{\int dr\,u_l(r)\,u_{l'}(r)\,
\psi_L(\frac{\w r}{2})\,\psi_{L'}(\frac{\w r}{2})\,(Y(r)-T(r))
\nonumber\\
&\times
\mx{l'\,1\,1\,M_f}
{\sigma_{2\,i}\,\sigma_{1\,-i}\,Y_{L\,\li}\,Y_{L'\,M'}}{l\,1\,1\,M_i}
\nonumber\\
&+\sum_j 3\,(-1)^j\,\int dr\,u_l(r)\,u_{l'}(r)\,
\psi_L(\frac{\w r}{2})\,\psi_{L'}(\frac{\w r}{2})\,T(r)\nonumber\\
&\times
\mx{l'\,1\,1\,M_f}
{\sigma_{2\,j}\,\sigma_{1\,i}\,r_{-j}\,r_{-i}\,Y_{L\,\li}\,Y_{L'\,M'}}
{l\,1\,1\,M_i}\bigg\}\nonumber\\
&\times\left(1-(-1)^{L'}-(-1)^L+(-1)^{L+L'}\right)
\end{align}
In this step we also expanded the scalar products into spherical components, 
according to Eq.~(\ref{eq:scalarproductspherical}). $\sum_{i}$
is a shortcut for $\sum_{i=-1,0,1}$, which we will use 
throughout this work.
Now we replace $r_{-j},\;r_{-i}$ according to Eq.~(\ref{eq:usefulidentity})
and afterwards combine $Y_{1\,-j}\,Y_{1\,-i}$ and $Y_{L\,\li}\,Y_{L'\,M'}$ in 
the second matrix element, using Eq.~(\ref{eq:additiontheorem}). We get
\ba
\label{eq:Mfiphiphi4second}
\Mfi{\phi\phi4\,\mathrm{pot}}&=\sum_{L=1}^\infty
\sum_{L'=1}^\infty\sum_{M'=-L'}^{L'}\sum_{l,l'}\sum_{i}
\frac{4\pi\,e^2\,f^2\,m_\pi}{3\,\w^2}\,(-1)^{1+L'-\lf+i}\,i^{L+L'}\,
\wignerd{L'}{M'}{-\lf}\nonumber\\
&\times\sqrt{\frac{(2L+1)\,(2L'+1)}{L\,(L+1)\,L'\,(L'+1)}}\,
\bigg\{\int dr\,u_l(r)\,u_{l'}(r)\,
\psi_L(\frac{\w r}{2})\,\psi_{L'}(\frac{\w r}{2})\,(Y(r)-T(r))
\nonumber\\
&\times
\mx{l'\,1\,1\,M_f}
{\sigma_{2\,i}\,\sigma_{1\,-i}\,Y_{L\,\li}\,Y_{L'\,M'}}{l\,1\,1\,M_i}
\nonumber\\
&+\sum_j\sum_{J=0,2}\sum_{\tilde{J}=|L-L'|}^{L+L'}(-1)^{-i+\li+M'}\,
3\,\sqrt{(2J+1)\,(2L+1)\,(2L'+1)\,(2\tilde{J}+1)}\nonumber\\
&\times
\threej{1}{1}{J}{0}{0}{0}\,\threej{1}{1}{J}{-j}{-i}{i+j}\,
\threej{L}{L'}{\tilde{J}}{0}{0}{0}\threej{L}{L'}{\tilde{J}}{\li}{M'}{-\li-M'}
\nonumber\\
&\times
\int dr\,u_l(r)\,u_{l'}(r)\,
\psi_L(\frac{\w r}{2})\,\psi_{L'}(\frac{\w r}{2})\,T(r)\\
&\times
\mx{l'\,1\,1\,M_f}
{\sigma_{2\,j}\,\sigma_{1\,i}\,Y_{J\,-i-j}\,Y_{\tilde{J}\,\li+M'}}
{l\,1\,1\,M_i}\bigg\}\left(1-(-1)^{L'}-(-1)^L+(-1)^{L+L'}\right),\nonumber
\end{align}
where the possible values for $J,\;\tilde{J}$ are given by the 3-$j$~symbols, 
cf. Eqs.~(\ref{eq:triangularcondition}) and (\ref{eq:threejprops5}). 
Therefore we need to evaluate the matrix-element structure
$\mx{l'\,1\,1\,M_f}{\sigma_{2\,j}\,\sigma_{1\,i}\,Y_{L\,M}\,Y_{L'\,M'}}
{l\,1\,1\,M_i}$. However, as we will be faced with further combinations of 
spin and orbital 
angular momenta in the initial and final state 
in App.~\ref{app:two-body}, we derive the 
element 
\mbox{$\mx{L_1\,S_1\,J_1\,M_1}
{\sigma_{2\,j}\,\sigma_{1\,i}\,Y_{l\,m}\,Y_{l'\,m'}}
{L_2\,S_2\,J_2\,M_2}$.}

First we write $\sigma_{2\,j}\,\sigma_{1\,i}$ as sum over tensor products, 
using Eq.~(\ref{eq:tensorproductinversion}) and the fact that the vector 
operator $\vec{\sigma}$ is of rank~1.
\ba
\sigma_{2\,j}\,\sigma_{1\,i}&=\sum_{S'=0}^2\sum_{M'=-S'}^{S'}(-1)^{M'}\,
\sqrt{2S'+1}\,\threej{1}{1}{S'}{j}{i}{-M'}\,(\sigma_2\otimes\sigma_1)_{S'\,M'}
\nonumber\\
&=\sum_{S'=0}^2(-i)^{i+j}\,\sqrt{2S'+1}\,\threej{1}{1}{S'}{j}{i}{-i-j}\,
(\sigma_2\otimes\sigma_1)_{S'\,i+j}
\end{align}
Combining the two spherical harmonics via Eq.~(\ref{eq:additiontheorem}) and 
making use of Eq.~(\ref{eq:tensorproductinversion}) once more we find
\ba
\sigma_{2\,j}\,&\sigma_{1\,i}\,Y_{l\,m}\,Y_{l'\,m'}=
\sum_{S'=0}^2\sum_{{L''}=|l-l'|}^{l+l'}\sum_{\tilde{J}=|{L''}-S'|}^{{L''}+S'}
(-1)^{S'-{L''}}\,\threej{l}{l'}{{L''}}{0}{0}{0}\\
&\times\threej{l}{l'}{{L''}}{m}{m'}{-m-m'}
\threej{1}{1}{S'}{j}{i}{-i-j}\,
\threej{S'}{{L''}}{\tilde{J}}{i+j}{m+m'}{-i-j-m-m'}\nonumber\\
&\times
\sqrt{\frac{(2l+1)\,(2l'+1)\,(2{L''}+1)\,(2S'+1)\,(2\tilde{J}+1)}{4\pi}}\,
\left[(\sigma_2\otimes\sigma_1)_{S'}
\otimes Y_{L''}\right]_{\tilde{J}\,i+j+m+m'}.
\nonumber
\end{align}
Now we use the Wigner-Eckart theorem.
\ba
&\mx{L_1\,S_1\,J_1\,M_1}{\sigma_{2\,j}\,\sigma_{1\,i}\,Y_{l\,m}\,Y_{l'\,m'}}
{L_2\,S_2\,J_2\,M_2}\nonumber\\
&\qquad=\sum_{S'}\sum_{{L''}}\sum_{\tilde{J}}(-1)^{S'-{L''}+J_1-M_1}\,
\threej{l}{l'}{{L''}}{0}{0}{0}\,\threej{l}{l'}{{L''}}{m}{m'}{-m-m'}\nonumber\\
&\qquad\times\threej{1}{1}{S'}{j}{i}{-i-j}\,
\threej{S'}{{L''}}{\tilde{J}}{i+j}{m+m'}{-i-j-m-m'}\nonumber\\
&\qquad\times\threej{J_1}{\tilde{J}}{J_2}{-M_1}{i+j+m+m'}{M_2}\,
\sqrt{\frac{(2l+1)\,(2l'+1)\,(2{L''}+1)\,(2S'+1)\,(2\tilde{J}+1)}{4\pi}}
\nonumber\\
&\qquad\times\mxred{L_1\,S_1\,J_1}
{\left[(\sigma_2\otimes\sigma_1)_{S'}\otimes Y_{L''}\right]_{\tilde{J}}}
{L_2\,S_2\,J_2}
\end{align}
We separate 
this matrix element into spin and orbital angular momentum parts, according to
Eqs.~(\ref{eq:uncoupling1}) and~(\ref{eq:uncoupling2}).
\ba
&\mx{L_1\,S_1\,J_1\,M_1}{\sigma_{2\,j}\,\sigma_{1\,i}\,Y_{l\,m}\,Y_{l'\,m'}}
{L_2\,S_2\,J_2\,M_2}\nonumber\\
&\qquad=\sum_{S'}\sum_{{L''}}\sum_{\tilde{J}}(-1)^{S'-{L''}+J_1-M_1}\,
\sqrt{\frac{(2l+1)\,(2l'+1)\,(2{L''}+1)\,(2S'+1)\,(2J_1+1)\,
(2J_2+1)}{4\pi}}\nonumber\\
&\qquad\times(2\tilde{J}+1)\,\threej{l}{l'}{{L''}}{0}{0}{0}\,
\threej{l}{l'}{{L''}}{m}{m'}{-m-m'}\,
\threej{1}{1}{S'}{j}{i}{-i-j}\nonumber\\
&\qquad\times\threej{S'}{{L''}}{\tilde{J}}{i+j}{m+m'}{-i-j-m-m'}\,
\threej{J_1}{\tilde{J}}{J_2}{-M_1}{i+j+m+m'}{M_2}\nonumber\\
&\qquad\times
\ninej{S_1}{L_1}{J_1}{S_2}{L_2}{J_2}{S'}{{L''}}{\tilde{J}}\,
\mxred{S_1}{(\sigma_2\otimes\sigma_1)_{S'}}{S_2}\,
\mxred{L_1}{Y_{L''}}{L_2}\nonumber\\
&\qquad=\sum_{S'}\sum_{{L''}}\sum_{\tilde{J}}(-1)^{-{L''}+J_1-M_1+S_1+S_2}\,
(2\tilde{J}+1)\,(2S'+1)\,
\threej{J_1}{\tilde{J}}{J_2}{-M_1}{i+j+m+m'}{M_2}\nonumber\\
&\qquad\times
\threej{1}{1}{S'}{j}{i}{-i-j}\,
\threej{l}{l'}{{L''}}{m}{m'}{-m-m'}\,
\threej{S'}{{L''}}{\tilde{J}}{i+j}{m+m'}{-i-j-m-m'}\nonumber\\
&\qquad\times
\threej{l}{l'}{{L''}}{0}{0}{0}
\ninej{S_1}{L_1}{J_1}{S_2}{L_2}{J_2}{S'}{{L''}}{\tilde{J}}\,
\sqrt{\frac{(2l+1)\,(2l'+1)\,(2{L''}+1)\,(2J_1+1)\,(2J_2+1)}{4\pi}}\nonumber\\
&\qquad\times
\mxred{L_1}{Y_{L''}}{L_2}\,\sum_s\sixj{1}{1}{S'}{S_2}{S_1}{s}\,
\mxred{S_1}{S-t}{s}\,\mxred{s}{S+t}{S_2}
\label{eq:master}
\end{align}
In the last line we defined $\vec{S}=\frac{\vec{\sigma}_1+\vec{\sigma}_2}{2}$
and $\vec{t}=\frac{\vec{\sigma}_1-\vec{\sigma}_2}{2}$ in analogy to 
Eq.~(\ref{eq:intJsigmaAonescalar}).
The reduced matrix elements for $S$ and $t$ are given in 
App.~\ref{app:formulae}, Eqs.~(\ref{eq:mxSshort}, \ref{eq:mxtshort}), and we 
find $S$ spin-conserving, $t$ spin-changing, which we also see in 
Eqs.~(\ref{eq:mxL}, \ref{eq:mxt}).
For evaluating Eq.~(\ref{eq:Mfiphiphi4second}) we need a spin-conserving 
operator; thus only two of the four combinations 
$\mxred{S_1}{S-t}{s}\,\mxred{s}{S+t}{S_2}$ survive:
\be
\mxred{S_1}{S-t}{s}\,\mxred{s}{S+t}{S_2}\rightarrow
\mxred{S_1}{S}{s}\,\mxred{s}{S}{S_2}-
\mxred{S_1}{t}{s}\,\mxred{s}{t}{S_2}
\ee
These products of matrix elements can be evaluated according to 
Eqs.~(\ref{eq:mxSshort},\ref{eq:mxtshort}), however we rather keep the 
compact notation of Eq.~(\ref{eq:Mfiphiphi4second}) for the final result.
Nevertheless, in order to simplify the last bracket in 
Eq.~(\ref{eq:Mfiphiphi4second}), we show that $L+L'$ has to be an even number.
First we note that $J$ is even in Eq.~(\ref{eq:Mfiphiphi4second}). The index
$L''$, which is summed over in Eq.~(\ref{eq:master}), has to be even due to 
$\mxred{l'}{Y_{L''}}{l}$ with even $l,\,l'$, cf. Eqs.~(\ref{eq:mxYshort}) and
(\ref{eq:threejprops5}). Therefore also $J+\tilde{J}$ is an even number, which
follows from Eq.~(\ref{eq:master}). With $J$ even also $\tilde{J}$ has to be 
even and the 3-$j$~symbol $\threej{L}{L'}{\tilde{J}}{0}{0}{0}$ guarantees
$L+L'$ even. This is also true for the first matrix element in 
Eq.~(\ref{eq:Mfiphiphi4second}), as again the index $L''$ in 
Eq.~(\ref{eq:master}) is even and therefore $L+L'$ has to be even, i.e. 
$(-1)^{L+L'}=1$ and $(-1)^{L'}=(-1)^L$. Finally we read $M'=M_f-M_i-\li$ off
Eq.~(\ref{eq:master}), which enables us to remove the sum over $M'$ and write 
down the final result:
\ba
\label{eq:Mfiphiphi4final}
\Mfi{\phi\phi4\,\mathrm{pot}}&=\sum_{L=1}^\infty
\sum_{L'=1}^\infty\sum_{l,l'}\sum_{i}
\frac{4\pi\,e^2\,f^2\,m_\pi}{3\,\w^2}\,(-1)^{1+L'-\lf+i}\,i^{L+L'}\,
\wignerd{L'}{M_f-M_i-\li}{-\lf}\nonumber\\
&\times\sqrt{\frac{(2L+1)\,(2L'+1)}{L\,(L+1)\,L'\,(L'+1)}}\,
\bigg\{\int dr\,u_l(r)\,u_{l'}(r)\,
\psi_L(\frac{\w r}{2})\,\psi_{L'}(\frac{\w r}{2})\,(Y(r)-T(r))
\nonumber\\
&\times\mx{l'\,1\,1\,M_f}
{\sigma_{2\,i}\,\sigma_{1\,-i}\,Y_{L\,\li}\,Y_{L'\,M_f-M_i-\li}}
{l\,1\,1\,M_i}\nonumber\\
&+\sum_j\sum_{J=0,2}\sum_{\tilde{J}=|L-L'|}^{L+L'}(-1)^{-i+M_f-M_i}\,
3\,\sqrt{(2J+1)\,(2L+1)\,(2L'+1)\,(2\tilde{J}+1)}\nonumber\\
&\times
\threej{1}{1}{J}{0}{0}{0}\,\threej{1}{1}{J}{-j}{-i}{i+j}\,
\threej{L}{L'}{\tilde{J}}{0}{0}{0}\nonumber\\
&\times
\threej{L}{L'}{\tilde{J}}{\li}{M_f-M_i-\li}{-M_f+M_i}\,
\int dr\,u_l(r)\,u_{l'}(r)\,
\psi_L(\frac{\w r}{2})\,\psi_{L'}(\frac{\w r}{2})\,T(r)\nonumber\\
&\times
\mx{l'\,1\,1\,M_f}
{\sigma_{2\,j}\,\sigma_{1\,i}\,Y_{J\,-i-j}\,Y_{\tilde{J}\,M_f-M_i}}
{l\,1\,1\,M_i}\bigg\}\,2\left(1-(-1)^{L}\right)
\end{align}
We never show results beyond $L,L'=2$, which we found to be a sufficiently 
good approximation in all amplitudes. Therefore we may restrict ourselves 
to $L=L'=1$, as the $L=2$-contribution vanishes and our result demands 
$L+L'$ even, i.e. $L=1,\; L'=2$ is forbidden.

Now all amplitudes with the photon field replaced by $\nab\phi$ at both 
vertices have been evaluated.
In the next appendix we calculate the amplitudes where 
this replacement is made only once. We also give the results for the 
non-negligible  contributions obtained from 
$\vec{A}\rightarrow\Aone,\,\Atwo$ at both vertices.

\chapter{Calculation of further Terms with $NN$-Rescattering 
\label{app:subleading} }
\markboth{APPENDIX \ref{app:subleading}. FURTHER TERMS WITH $NN$-RESCATTERING}
         {APPENDIX \ref{app:subleading}. FURTHER TERMS WITH $NN$-RESCATTERING}
This appendix contains the evaluation of the amplitudes given in 
Eq.~(\ref{eq:Mfiphi}), with $\vec{J}\ofxi$ replaced by $\Jsigma\ofxi$, cf. 
Eq.~(\ref{eq:Jsigma}). We also calculated the amplitudes including 
$\Jp\ofxi$ (Eq.~(\ref{eq:Jp})), but found these contributions invisibly small.
Therefore we abstain from writing down those results. The amplitudes 
(\ref{eq:Mfiphi}) were derived by replacing $\vec{A}\rightarrow\nab\phi$ at 
one interaction vertex, but there are also non-negligible contributions which 
do not include the $\nab\phi$-term. These are calculated at the end of this 
appendix. All following amplitudes can also be found in 
Ref.~\cite{Karakowski}, albeit in the lab frame.

Before we turn to evaluating Eq.~(\ref{eq:Mfiphi}), we prove the 
relation $\vsh{J}{L}{M}\cdot\vec{V}=\left[Y_L\otimes V\right]_{J\,M}$, 
cf. Eq.~(\ref{eq:TdotV}), making use of
Eqs.~(\ref{eq:tensorproductdef}) and (\ref{eq:vshdefinition}). 
First we decompose the scalar product into spherical components, cf.
Eq.~(\ref{eq:scalarproductspherical}):
\be
\vec{T}_{J\,L\,M}\cdot\vec{V}=
\sum_{\nu=-1,0,1}(-1)^\nu\,T_{J\,L\,M\,\,\nu}\,V_{1\,-\nu}
\ee
Now we replace $T_{J\,L\,M\,\,\nu}$ according to 
Eqs.~(\ref{eq:spherdecomposition})
and (\ref{eq:vshdefinition}) by 
\be
T_{J\,L\,M\,\,\nu}=\vec{T}_{J\,L\,M}\cdot\hat{r}_\nu=(-1)^\nu\,(-1)^{-L-M+1}\,
\sqrt{2J+1}\,\threej{L}{1}{J}{M+\nu}{-\nu}{-M}\,Y_{L\,M+\nu}.
\label{eq:T_JLMnu}
\ee
Therefore,
\ba
\vec{T}_{J\,L\,M}\cdot\vec{V}&=
\sum_{\nu=-1,0,1}(-1)^{-L-M+1}\,
\sqrt{2J+1}\,\threej{L}{1}{J}{M+\nu}{-\nu}{-M}\,Y_{L\,M+\nu}\,V_{1\,-\nu}
\nonumber\\
&=\sum_{\nu=-1,0,1}(-1)^{-L-M+1}\,
\sqrt{2J+1}\,\threej{L}{1}{J}{M-\nu}{ \nu}{-M}\,Y_{L\,M-\nu}\,V_{1\, \nu},
\nonumber\\
&
\label{eq:TdotV2}
\end{align}
where we replaced $\nu\rightarrow -\nu$ which does not change the sum 
over $\nu$. On the other hand, Eq.~(\ref{eq:tensorproductdef}) gives
\ba
\left[Y_L\otimes V\right]_{J\,M}&=\sum_{\mu,\nu}(-1)^{-L+1-M}\,\sqrt{2J+1}\,
\threej{L}{1}{J}{\mu}{\nu}{-M}\,Y_{L\,\mu}\,V_{1\,\nu}\\
&=\sum_\nu(-1)^{-L+1-M}\,\sqrt{2J+1}\,
\threej{L}{1}{J}{M-\nu}{\nu}{-M}\,Y_{L\,M-\nu}\,V_{1\,\nu},\nonumber
\end{align}
which is identical to Eq.~(\ref{eq:TdotV2}). 

Eq.~(\ref{eq:Mfiphi}) consists of eight amplitudes, however we found that four 
of them cancel exactly. Only the amplitudes including an energy 
denominator remain. These are
\ba
\Mfi{\phi\,\sigma\,a}&=
-i\,\left(\w+\frac{\w^2}{2m_d}\right)\,
\sum_C\frac{\mx{d_f}{\phifhat}{C}
\mx{C}{\int\Jsigma\ofxi\cdot\vec{A}\ofxi\,d^3\xi}{d_i}}{\denoms},\nonumber\\
\Mfi{\phi\,\sigma\,b}&=
i\,\left(\w+\frac{\w^2}{2m_d}\right)\,
\sum_C\frac{\mx{d_f}{\int\Jsigma\ofxi\cdot\vec{A}\ofxi\,d^3\xi}{C}
\mx{C}{\phiihat}{d_i}}{\denoms},\nonumber\\
\Mfi{\phi\,\sigma\,c}&=
i\,\left(\w-\frac{\w^2}{2m_d}+\frac{\PCsq}{2m_C}\right)\,
\sum_C\frac{\mx{d_f}{\phiihat}{C}
\mx{C}{\int\Jsigma\ofxi\cdot\vec{A}\ofxi\,d^3\xi}{d_i}}{\denomu},\nonumber\\
\Mfi{\phi\,\sigma\,d}&=
-i\,\left(\w-\frac{\w^2}{2m_d}+\frac{\PCsq}{2m_C}\right)\,
\sum_C\frac{\mx{d_f}{\int\Jsigma\ofxi\cdot\vec{A}\ofxi\,d^3\xi}{C}
\mx{C}{\phifhat}{d_i}}{\denomu}.
\label{eq:phisigmadiagrams}
\end{align}
The photon field $\vec{A}\ofxi$ is replaced either by $\Aone\ofxi$ 
or $\Atwo\ofxi$ (Eqs.~(\ref{eq:Aone}, \ref{eq:Atwo})). We start 
with $\Aone\ofxi$ and use the expression derived for 
$\int\Jsigma\ofxi\cdot\Aone\ofxi\,d^3\xi$ in Eq.~(\ref{eq:intJsigmaAone}).
The corresponding amplitudes are denoted by $\Mfi{\phi\,\sigma1}$, whereas 
we use the notation $\Mfi{\phi\,\sigma2}$ when we replace 
$\vec{A}\ofxi\rightarrow\Atwo\ofxi$ later on.

In order to calculate $\Mfi{\phi\,\sigma1}$, we insert two complete sets of 
radial states $\ket{r}$, $\ket{r'}$, as we did in 
Eq.~(\ref{eq:Mfiphiphi1first}). From the explicit expressions for $\phifhat$ 
and $\int\Jsigma\ofxi\cdot\Aone\ofxi\,d^3\xi$, Eqs.~(\ref{eq:phidefinition}) 
and (\ref{eq:intJsigmaAone}), we get
\ba
\Mfi{\phi\,\sigma1\,a}&=
-i\,\left(\w+\frac{\w^2}{2m_d}\right)\sum_{\hat{C}}\doubleint
r^2dr\,r'^2dr'\,\bra{d_f}e\,\sum_{L'=1}^{\infty}\sum_{M'=-L'}^{L'}(-1)^{L'-\lf}
\nonumber\\
&\times\wignerd{L'}{M'}{-\lf}\,
\frac{i^{L'+1}}{\w}\,
\sqrt{\frac{2\pi\,(2L'+1)}{L'\,(L'+1)}}\,\psi_{L'}(\frac{\w r'}{2})\,Y_{L'\,M'}
\ket{\hat{C}\,r'}\nonumber\\
&\times\mx{r'}{\frac{1}{E_0-H_{\hat{C}}^{np}}}{r}\,
\bra{r\,\hat{C}}-\sum_{L=1}^{\infty}\sum_{M=-L}^{L}\li\,
\sqrt{2\pi\,(L+1)}\,\frac{e\,\w}{2\,m_N}\,i^{L+1}\nonumber\\
&\times j_{L-1}(\frac{\w r}{2})\,
\left(\mu_p-(-1)^L\,\mu_n\right)\,
[Y_{L-1}\otimes S]_{L\,M}\,\delta_{M,\li}\ket{d_i}.
\label{eq:Mfiphisigma1first}
\end{align}
The term proportional to $[Y_{L-1}\otimes t]_{L\,M}$ in 
Eq.~(\ref{eq:intJsigmaAone}) changes the spin of the two-nucleon state, cf. 
Eq.~(\ref{eq:mxt}). Therefore it gives no contribution as $Y_{L'\,M'}$ demands
$S_C=S_d=1$, as can be seen from Eq.~(\ref{eq:mxY}), and has been dropped.

We rewrite Eq.~(\ref{eq:Mfiphisigma1first}) as
\ba
\Mfi{\phi\,\sigma1\,a}&=\sum_{L=1}^{\infty}
\sum_{L'=1}^{\infty}\sum_{M'=-L'}^{L'}\sum_{L_C,J_C,M_C}\sum_{l,l'}
\frac{\pi\,e^2\,\li}{m_N}\,\left(\w+\frac{\w^2}{2m_d}\right)\,(-1)^{L'-\lf+1}
\nonumber\\
&\times i^{L+L'+1}\,\wignerd{L'}{M'}{-\lf}\,
\sqrt{\frac{(L+1)\,(2L'+1)}{L'\,(L'+1)}}\,
\left(\mu_p-(-1)^L\,\mu_n\right)\nonumber\\
&\times\doubleint rdr\,r'dr'\,u_l(r)\,j_{L-1}(\frac{\w r}{2})\,\green\,
\psi_{L'}(\frac{\w r'}{2})\,u_{l'}(r')\\
&\times\mx{l'\,1\,1\,M_f}{Y_{L'\,M'}}{L_C\,1\,J_C\,M_C}\,
\mx{L_C\,1\,J_C\,M_C}{[Y_{L-1}\otimes S]_{L\,M}}{l\,1\,1\,M_i}.\nonumber
\end{align}
Applying the Wigner-Eckart theorem (\ref{eq:WE}) twice and removing the 
formal sums over $M'$ and $M_C$, we get the final result
\ba
\label{eq:Mfiphisigma1final}
\Mfi{\phi\,\sigma1\,a}&=\sum_{L=1}^{\infty}
\sum_{L'=1}^{\infty}\sum_{J_C=|L-1|}^{L+1}\sum_{L_C=|J_C-1|}^{J_C+1}\sum_{l,l'}
\frac{\pi\,e^2\,\li}{m_N}\,\left(\w+\frac{\w^2}{2m_d}\right)\,
(-1)^{L'-\lf-\li-M_f-M_i+J_C}
\nonumber\\
&\times i^{L+L'+1}\,\wignerd{L'}{M_f-M_i-\li}{-\lf}\,
\sqrt{\frac{(L+1)\,(2L'+1)}{L'\,(L'+1)}}\,
\left(\mu_p-(-1)^L\,\mu_n\right)\nonumber\\
&\times\doubleint rdr\,r'dr'\,u_l(r)\,j_{L-1}(\frac{\w r}{2})\,\green\,
\psi_{L'}(\frac{\w r'}{2})\,u_{l'}(r')\nonumber\\
&\times\threej{1}{L'}{J_C}{-M_f}{M_f-M_i-\li}{M_i+\li}
\,\threej{J_C}{L}{1}{-M_i-\li}{\li}{M_i}\nonumber\\
&\times\mxred{l'\,1\,1}{Y_{L'}}{L_C\,1\,J_C}\,
\mxred{L_C\,1\,J_C}{[Y_{L-1}\otimes S]_{L}}{l\,1\,1}.
\end{align}
The values possible for $J_C$ are determined by the second (or, as well, 
first) 3-$j$~symbol, and $L_C$ takes on the values 
$|J_C-S_C|,\cdots,J_C+S_C$. The reduced matrix elements can be found in 
Eqs.~(\ref{eq:mxY}, \ref{eq:mxL}).

$\Mfi{\phi\,\sigma1\,b,c,d}$ are calculated quite similarly. Therefore we 
only give the results.
\ba
\label{eq:Mfiphisigma1bfinal}
\Mfi{\phi\,\sigma1\,b}&=\sum_{L=1}^{\infty}
\sum_{L'=1}^{\infty}\sum_{J_C=|L-1|}^{L+1}\sum_{L_C=|J_C-1|}^{J_C+1}\sum_{l,l'}
\frac{\pi\,e^2\,\lf}{m_N}\,\left(\w+\frac{\w^2}{2m_d}\right)\,
(-1)^{L'-\lf-\li-M_f-M_i+J_C+1}
\nonumber\\
&\times i^{L+L'+1}\,\wignerd{L'}{M_f-M_i-\li}{-\lf}\,
\sqrt{\frac{(2L+1)\,(L'+1)}{L\,(L+1)}}\,
\left(\mu_p-(-1)^{L'}\,\mu_n\right)\nonumber\\
&\times\doubleint rdr\,r'dr'\,u_l(r)\,\psi_{L}(\frac{\w r}{2})\,\green\,
j_{L'-1}(\frac{\w r'}{2})\,u_{l'}(r')\nonumber\\
&\times\threej{1}{L'}{J_C}{-M_f}{M_f-M_i-\li}{M_i+\li}
\,\threej{J_C}{L}{1}{-M_i-\li}{\li}{M_i}\nonumber\\
&\times\mxred{l'\,1\,1}{[Y_{L'-1}\otimes S]_{L'}}{L_C\,1\,J_C}\,
\mxred{L_C\,1\,J_C}{Y_{L}}{l\,1\,1}
\end{align}
\ba
\label{eq:Mfiphisigma1cfinal}
\Mfi{\phi\,\sigma1\,c}&=\sum_{L=1}^{\infty}
\sum_{L'=1}^{\infty}\sum_{J_C=|L-1|}^{L+1}\sum_{L_C=|J_C-1|}^{J_C+1}\sum_{l,l'}
\frac{\pi\,e^2\,\lf}{m_N}\,
\left(\w-\frac{\w^2}{2m_d}+\frac{\PCsq}{2m_C}\right)\,(-1)^{L'-\lf-\li+J_C+1}
\nonumber\\
&\times i^{L+L'+1}\,\wignerd{L'}{M_f-M_i-\li}{-\lf}\,
\sqrt{\frac{(2L+1)\,(L'+1)}{L\,(L+1)}}\,
\left(\mu_p-(-1)^{L'}\,\mu_n\right)\nonumber\\
&\times\doubleint rdr\,r'dr'\,u_{l'}(r)\,\psi_{L}(\frac{\w r}{2})\,\greenpr\,
j_{L'-1}(\frac{\w r'}{2})\,u_{l}(r')\nonumber\\
&\times\threej{1}{L}{J_C}{-M_f}{\li}{M_f-\li}
\,\threej{J_C}{L'}{1}{-M_f+\li}{M_f-M_i-\li}{M_i}\nonumber\\
&\times\mxred{l'\,1\,1}{Y_{L}}{L_C\,1\,J_C}\,
\mxred{L_C\,1\,J_C}{[Y_{L'-1}\otimes S]_{L'}}{l\,1\,1}
\end{align}
\ba
\label{eq:Mfiphisigma1dfinal}
\Mfi{\phi\,\sigma1\,d}&=\sum_{L=1}^{\infty}
\sum_{L'=1}^{\infty}\sum_{J_C=|L-1|}^{L+1}\sum_{L_C=|J_C-1|}^{J_C+1}\sum_{l,l'}
\frac{\pi\,e^2\,\li}{m_N}\,
\left(\w-\frac{\w^2}{2m_d}+\frac{\PCsq}{2m_C}\right)\,
(-1)^{L'-\lf-\li+J_C}
\nonumber\\
&\times i^{L+L'+1}\,\wignerd{L'}{M_f-M_i-\li}{-\lf}\,
\sqrt{\frac{(L+1)\,(2L'+1)}{L'\,(L'+1)}}\,
\left(\mu_p-(-1)^{L}\,\mu_n\right)\nonumber\\
&\times\doubleint rdr\,r'dr'\,u_{l'}(r)\,j_{L-1}(\frac{\w r}{2})\,\greenpr\,
\psi_{L'}(\frac{\w r'}{2})\,u_{l}(r')\nonumber\\
&\times\threej{1}{L}{J_C}{-M_f}{\li}{M_f-\li}
\,\threej{J_C}{L'}{1}{-M_f+\li}{M_f-M_i-\li}{M_i}\nonumber\\
&\times\mxred{l'\,1\,1}{[Y_{L-1}\otimes S]_{L}}{L_C\,1\,J_C}\,
\mxred{L_C\,1\,J_C}{Y_{L'}}{l\,1\,1}
\end{align}
Now we calculate the amplitudes $\Mfi{\phi\,\sigma2}$ which we get from 
replacing $\vec{A}\ofxi$ by $\Atwo\ofxi$ in Eq.~(\ref{eq:phisigmadiagrams}).
As there are no further difficulties in the calculation, once we have the 
formulae for $\hat{\phi}_{i,f}$ (Eq.~(\ref{eq:phidefinition})) and for
$\int\Jsigma\ofxi\cdot\Atwo\ofxi\,d^3\xi$ 
(Eq.~(\ref{eq:intJsigmaAtwofinal})) 
at hand, we can immediately write down the results.
\ba
\label{eq:Mfiphisigma2final}
\Mfi{\phi\,\sigma2\,a}&=\sum_{L=1}^{\infty}
\sum_{L'=1}^{\infty}\sum_{J_C=|L-1|}^{L+1}\sum_{L_C=|J_C-1|}^{J_C+1}\sum_{l,l'}
\frac{\pi\,e^2}{m_N}\,\left(\w+\frac{\w^2}{2m_d}\right)\,
(-1)^{L'-\lf-\li-M_f-M_i+J_C}
\nonumber\\
&\times i^{L+L'}\,\wignerd{L'}{M_f-M_i-\li}{-\lf}\,
\sqrt{\frac{(2L+1)\,(2L'+1)}{L'\,(L'+1)}}\,
\left(\mu_p+(-1)^L\,\mu_n\right)\nonumber\\
&\times\doubleint rdr\,r'dr'\,u_l(r)\,j_{L}(\frac{\w r}{2})\,\green\,
\psi_{L'}(\frac{\w r'}{2})\,u_{l'}(r')\nonumber\\
&\times\threej{1}{L'}{J_C}{-M_f}{M_f-M_i-\li}{M_i+\li}
\,\threej{J_C}{L}{1}{-M_i-\li}{\li}{M_i}\nonumber\\
&\times\mxred{l'\,1\,1}{Y_{L'}}{L_C\,1\,J_C}\,
\mxred{L_C\,1\,J_C}{[Y_{L}\otimes S]_{L}}{l\,1\,1}
\end{align}
\ba
\label{eq:Mfiphisigma2bfinal}
\Mfi{\phi\,\sigma2\,b}&=\sum_{L=1}^{\infty}
\sum_{L'=1}^{\infty}\sum_{J_C=|L-1|}^{L+1}\sum_{L_C=|J_C-1|}^{J_C+1}\sum_{l,l'}
\frac{\pi\,e^2}{m_N}\,\left(\w+\frac{\w^2}{2m_d}\right)\,
(-1)^{L'-\lf-\li-M_f-M_i+J_C+1}
\nonumber\\
&\times i^{L+L'}\,\wignerd{L'}{M_f-M_i-\li}{-\lf}\,
\sqrt{\frac{(2L+1)\,(2L'+1)}{L\,(L+1)}}\,
\left(\mu_p+(-1)^{L'}\,\mu_n\right)\nonumber\\
&\times\doubleint rdr\,r'dr'\,u_l(r)\,\psi_{L}(\frac{\w r}{2})\,\green\,
j_{L'}(\frac{\w r'}{2})\,u_{l'}(r')\nonumber\\
&\times\threej{1}{L'}{J_C}{-M_f}{M_f-M_i-\li}{M_i+\li}
\,\threej{J_C}{L}{1}{-M_i-\li}{\li}{M_i}\nonumber\\
&\times\mxred{l'\,1\,1}{[Y_{L'}\otimes S]_{L'}}{L_C\,1\,J_C}\,
\mxred{L_C\,1\,J_C}{Y_{L}}{l\,1\,1}
\end{align}
\ba
\label{eq:Mfiphisigma2cfinal}
\Mfi{\phi\,\sigma2\,c}&=\sum_{L=1}^{\infty}
\sum_{L'=1}^{\infty}\sum_{J_C=|L-1|}^{L+1}\sum_{L_C=|J_C-1|}^{J_C+1}\sum_{l,l'}
\frac{\pi\,e^2}{m_N}\,\left(\w-\frac{\w^2}{2m_d}+\frac{\PCsq}{2m_C}\right)\,
(-1)^{L'-\lf-\li+J_C+1}
\nonumber\\
&\times i^{L+L'}\,\wignerd{L'}{M_f-M_i-\li}{-\lf}\,
\sqrt{\frac{(2L+1)\,(2L'+1)}{L\,(L+1)}}\,
\left(\mu_p+(-1)^{L'}\,\mu_n\right)\nonumber\\
&\times\doubleint rdr\,r'dr'\,u_{l'}(r)\,\psi_{L}(\frac{\w r}{2})\,\greenpr\,
j_{L'}(\frac{\w r'}{2})\,u_{l}(r')\nonumber\\
&\times\threej{1}{L}{J_C}{-M_f}{\li}{M_f-\li}
\,\threej{J_C}{L'}{1}{-M_f+\li}{M_f-M_i-\li}{M_i}\nonumber\\
&\times\mxred{l'\,1\,1}{Y_{L}}{L_C\,1\,J_C}\,
\mxred{L_C\,1\,J_C}{[Y_{L'}\otimes S]_{L'}}{l\,1\,1}
\end{align}
\ba
\label{eq:Mfiphisigma2dfinal}
\Mfi{\phi\,\sigma2\,d}&=\sum_{L=1}^{\infty}
\sum_{L'=1}^{\infty}\sum_{J_C=|L-1|}^{L+1}\sum_{L_C=|J_C-1|}^{J_C+1}\sum_{l,l'}
\frac{\pi\,e^2}{m_N}\,\left(\w-\frac{\w^2}{2m_d}+\frac{\PCsq}{2m_C}\right)\,
(-1)^{L'-\lf-\li+J_C}
\nonumber\\
&\times i^{L+L'}\,\wignerd{L'}{M_f-M_i-\li}{-\lf}\,
\sqrt{\frac{(2L+1)\,(2L'+1)}{L'\,(L'+1)}}\,
\left(\mu_p+(-1)^{L}\,\mu_n\right)\nonumber\\
&\times\doubleint rdr\,r'dr'\,u_{l'}(r)\,j_{L}(\frac{\w r}{2})\,\greenpr\,
\psi_{L'}(\frac{\w r'}{2})\,u_{l}(r')\nonumber\\
&\times\threej{1}{L}{J_C}{-M_f}{\li}{M_f-\li}
\,\threej{J_C}{L'}{1}{-M_f+\li}{M_f-M_i-\li}{M_i}\nonumber\\
&\times\mxred{l'\,1\,1}{[Y_{L}\otimes S]_{L}}{L_C\,1\,J_C}\,
\mxred{L_C\,1\,J_C}{Y_{L'}}{l\,1\,1}
\end{align} 
The amplitudes $\Mfi{\phi\,\sigma1}$ ($\Mfi{\phi\,\sigma2}$) \textit{alone}
are numerically large. Their size is a correction of about 5\%-10\% at
94.2~MeV. However, they partly cancel each other. Therefore the net effect of
the contributions $\Mfi{\phi\,\sigma1}$, $\Mfi{\phi\,\sigma2}$ is well below 
5\% for all angles at 94.2~MeV, see Fig.~\ref{fig:separation}. 

Numerically more important than the amplitudes calculated so far in this 
appendix are those, where we replace $\vec{J}\ofxi\cdot\vec{A}\ofxi$ by
$\Jsigma\ofxi\cdot\vec{A}^{(1,2)}\ofxi$ at both vertices. 
However, we found that the mixed 
contributions, i.e. $\Hint=-\int\Jsigma\ofxi\cdot\Aone\ofxi\,d^3\xi$ at one 
vertex and $\Hint=-\int\Jsigma\ofxi\cdot\Atwo\ofxi\,d^3\xi$ at the other
are negligibly small. The reason for this suppression is the product of 
matrix elements occuring in these amplitudes, which forbids 
the leading contribution $L=L'=1$. It is given by 
Eq.~(\ref{eq:mxL}) with $L'=0$ and $L'=1$, respectively. However, it follows 
from Eq.~(\ref{eq:threejprops5}) and the fact that $l,\,l'$ are even numbers 
that $L_C$ must be even, odd for $L'=0,\;1$. Therefore, this 
product vanishes for the leading contribution and
we only need to calculate the following four amplitudes:
\ba
\Mfi{\sigma1\,\sigma1\,a}&=
\sum_C\frac{\mx{d_f}{\int\Jsigma\ofxi\cdot\Aone\ofxi\,d^3\xi}{C}
\mx{C}{\int\Jsigma\ofxi\cdot\Aone\ofxi\,d^3\xi}{d_i}}{\denoms}\nonumber\\
\Mfi{\sigma1\,\sigma1\,b}&=
\sum_C\frac{\mx{d_f}{\int\Jsigma\ofxi\cdot\Aone\ofxi\,d^3\xi}{C}
\mx{C}{\int\Jsigma\ofxi\cdot\Aone\ofxi\,d^3\xi}{d_i}}{\denomu}\nonumber\\
\Mfi{\sigma2\,\sigma2\,a}&=
\sum_C\frac{\mx{d_f}{\int\Jsigma\ofxi\cdot\Atwo\ofxi\,d^3\xi}{C}
\mx{C}{\int\Jsigma\ofxi\cdot\Atwo\ofxi\,d^3\xi}{d_i}}{\denoms}\nonumber\\
\Mfi{\sigma2\,\sigma2\,b}&=
\sum_C\frac{\mx{d_f}{\int\Jsigma\ofxi\cdot\Atwo\ofxi\,d^3\xi}{C}
\mx{C}{\int\Jsigma\ofxi\cdot\Atwo\ofxi\,d^3\xi}{d_i}}{\denomu}
\label{eq:sigmasigmadiagrams}
\end{align}
The evaluation of these amplitudes is straightforward, using 
Eqs.~(\ref{eq:intJsigmaAone}) and (\ref{eq:intJsigmaAtwofinal}). The main 
difference to $\Mfi{\phi\,\sigma1}$ and $\Mfi{\phi\,\sigma2}$ is the fact that
now the intermediate state may have total spin $S_C=1$ \textit{or} $S_C=0$.
In fact, the singlet ($S_C=0$) part of 
$\Mfi{\sigma1\,\sigma1\,a}$, which is the amplitude with two $M1$-interactions,
is the dominant contribution to the total deuteron-photodisintegration cross 
section at threshold, cf. Section~\ref{sec:photodisintegration} and 
Fig.~\ref{fig:photodisintegration}. 
Among the amplitudes~(\ref{eq:sigmasigmadiagrams}),
it is also the singlet transition of $\Mfi{\sigma1\,\sigma1}$ which gives
the most important contribution to the deuteron Compton cross sections, whereas
the triplet amplitude and $\Mfi{\sigma2\,\sigma2}$ are only minor corrections.
The results are:
\ba
\label{eq:Mfisigma1sigma1afinal}
\Mfi{\sigma1\,\sigma1\,a}&=\sum_{L=1}^{\infty}
\sum_{L'=1}^{\infty}\sum_{J_C=|L-1|}^{L+1}\sum_{S_C=0,1}
\sum_{L_C=|J_C-S_C|}^{J_C+S_C}\sum_{l,l'}
\frac{\pi\,e^2\,\w^2}{2\,m_N^2}\,(-1)^{L'-\lf-\li-M_f-M_i+J_C+1}
\nonumber\\
&\times i^{L+L'}\,\li\,\lf\,\wignerd{L'}{M_f-M_i-\li}{-\lf}\,
\sqrt{(L+1)\,(L'+1)}\nonumber\\
&\times\threej{1}{L'}{J_C}{-M_f}{M_f-M_i-\li}{M_i+\li}\,
\threej{J_C}{L}{1}{-M_i-\li}{\li}{M_i}\nonumber\\
&\times
\bigg\{(\mu_p-(-1)^{L'}\,\mu_n)\,(\mu_p-(-1)^{L}\,\mu_n)\,
\mxred{l'\,1\,1}{[Y_{L'-1}\otimes S]_{L'}}{L_C\,S_C\,J_C}\nonumber\\
&\times\mxred{L_C\,S_C\,J_C}{[Y_{L-1}\otimes S]_{L}}{l\,1\,1}
+(\mu_p+(-1)^{L'}\,\mu_n)\,(\mu_p+(-1)^{L}\,\mu_n)
\nonumber\\
&\times\mxred{l'\,1\,1}{[Y_{L'-1}\otimes t]_{L'}}{L_C\,S_C\,J_C}\,
\mxred{L_C\,S_C\,J_C}{[Y_{L-1}\otimes t]_{L}}{l\,1\,1}\bigg\}
\nonumber\\
&\times\doubleint rdr\,r'dr'\,u_l(r)\,j_{L-1}(\frac{\w r}{2})\,\green\,
j_{L'-1}(\frac{\w r'}{2})\,u_{l'}(r')
\end{align}
\ba
\label{eq:Mfisigma1sigma1bfinal}
\Mfi{\sigma1\,\sigma1\,b}&=\sum_{L=1}^{\infty}
\sum_{L'=1}^{\infty}\sum_{J_C=|L-1|}^{L+1}\sum_{S_C=0,1}
\sum_{L_C=|J_C-S_C|}^{J_C+S_C}\sum_{l,l'}
\frac{\pi\,e^2\,\w^2}{2\,m_N^2}\,(-1)^{L'-\lf-\li+J_C+1}
\nonumber\\
&\times i^{L+L'}\,\li\,\lf\,\wignerd{L'}{M_f-M_i-\li}{-\lf}\,
\sqrt{(L+1)\,(L'+1)}\nonumber\\
&\times\threej{1}{L}{J_C}{-M_f}{\li}{M_f-\li}\,
\threej{J_C}{L'}{1}{-M_f+\li}{M_f-M_i-\li}{M_i}\nonumber\\
&\times
\bigg\{(\mu_p-(-1)^{L}\,\mu_n)\,(\mu_p-(-1)^{L'}\,\mu_n)\,
\mxred{l'\,1\,1}{[Y_{L-1}\otimes S]_{L}}{L_C\,S_C\,J_C}\nonumber\\
&\times\mxred{L_C\,S_C\,J_C}{[Y_{L'-1}\otimes S]_{L'}}{l\,1\,1}
+(\mu_p+(-1)^{L}\,\mu_n)\,(\mu_p+(-1)^{L'}\,\mu_n)
\nonumber\\
&\times\mxred{l'\,1\,1}{[Y_{L-1}\otimes t]_{L}}{L_C\,S_C\,J_C}\,
\mxred{L_C\,S_C\,J_C}{[Y_{L'-1}\otimes t]_{L'}}{l\,1\,1}\bigg\}
\nonumber\\
&\times\doubleint rdr\,r'dr'\,u_{l'}(r)\,j_{L-1}(\frac{\w r}{2})\,\greenpr\,
j_{L'-1}(\frac{\w r'}{2})\,u_{l}(r')
\end{align}
\ba
\label{eq:Mfisigma2sigma2final}
\Mfi{\sigma2\,\sigma2\,a}&=\sum_{L=1}^{\infty}
\sum_{L'=1}^{\infty}\sum_{J_C=|L-1|}^{L+1}\sum_{S_C=0,1}
\sum_{L_C=|J_C-S_C|}^{J_C+S_C}\sum_{l,l'}
\frac{\pi\,e^2\,\w^2}{2\,m_N^2}\,(-1)^{L'-M_f-M_i+J_C-\li-\lf}
\nonumber\\
&\times i^{L+L'}\,\wignerd{L'}{M_f-M_i-\li}{-\lf}\,
\sqrt{(2L+1)\,(2L'+1)}\nonumber\\
&\times\threej{1}{L'}{J_C}{-M_f}{M_f-M_i-\li}{M_i+\li}\,
\threej{J_C}{L}{1}{-M_i-\li}{\li}{M_i}\nonumber\\
&\times
\bigg\{(\mu_p+(-1)^{L'}\,\mu_n)\,(\mu_p+(-1)^{L}\,\mu_n)\,
\mxred{l'\,1\,1}{[Y_{L'}\otimes S]_{L'}}{L_C\,S_C\,J_C}\nonumber\\
&\times\mxred{L_C\,S_C\,J_C}{[Y_{L}\otimes S]_{L}}{l\,1\,1}
+(\mu_p-(-1)^{L'}\,\mu_n)\,(\mu_p-(-1)^{L}\,\mu_n)
\nonumber\\
&\times\mxred{l'\,1\,1}{[Y_{L'}\otimes t]_{L'}}{L_C\,S_C\,J_C}\,
\mxred{L_C\,S_C\,J_C}{[Y_{L}\otimes t]_{L}}{l\,1\,1}\bigg\}
\nonumber\\
&\times\doubleint rdr\,r'dr'\,u_l(r)\,j_{L}(\frac{\w r}{2})\,\green\,
j_{L'}(\frac{\w r'}{2})\,u_{l'}(r')
\end{align}
\ba
\label{eq:Mfisigma2sigma2bfinal}
\Mfi{\sigma2\,\sigma2\,b}&=\sum_{L=1}^{\infty}
\sum_{L'=1}^{\infty}\sum_{J_C=|L-1|}^{L+1}\sum_{S_C=0,1}
\sum_{L_C=|J_C-S_C|}^{J_C+S_C}\sum_{l,l'}
\frac{\pi\,e^2\,\w^2}{2\,m_N^2}\,(-1)^{L'-\lf-\li+J_C}
\nonumber\\
&\times i^{L+L'}\,\wignerd{L'}{M_f-M_i-\li}{-\lf}\,
\sqrt{(2L+1)\,(2L'+1)}\nonumber\\
&\times\threej{1}{L}{J_C}{-M_f}{\li}{M_f-\li}\,
\threej{J_C}{L'}{1}{-M_f+\li}{M_f-M_i-\li}{M_i}\nonumber\\
&\times
\bigg\{(\mu_p+(-1)^{L}\,\mu_n)\,(\mu_p+(-1)^{L'}\,\mu_n)\,
\mxred{l'\,1\,1}{[Y_{L}\otimes S]_{L}}{L_C\,S_C\,J_C}\nonumber\\
&\times\mxred{L_C\,S_C\,J_C}{[Y_{L'}\otimes S]_{L'}}{l\,1\,1}
+(\mu_p-(-1)^{L}\,\mu_n)\,(\mu_p-(-1)^{L'}\,\mu_n)
\nonumber\\
&\times\mxred{l'\,1\,1}{[Y_{L}\otimes t]_{L}}{L_C\,S_C\,J_C}\,
\mxred{L_C\,S_C\,J_C}{[Y_{L'}\otimes t]_{L'}}{l\,1\,1}\bigg\}
\nonumber\\
&\times\doubleint rdr\,r'dr'\,u_{l'}(r)\,j_{L}(\frac{\w r}{2})\,\greenpr\,
j_{L'}(\frac{\w r'}{2})\,u_{l}(r')
\end{align}
The reduced matrix elements are given in Eqs.~(\ref{eq:mxL}, \ref{eq:mxt}).
The operator 
$[Y_{l}\otimes t]_{l}$ corresponds to $S_C=0$ (singlet), 
$[Y_{l}\otimes S]_{l}$ to             $S_C=1$ (triplet).

In this appendix we calculated diagrams where only 
one-body currents are explicitly involved. In the next one we allow for 
explicit coupling of the photon to two-body, i.e. to meson-exchange currents. 

\chapter{Diagrams with Explicit Two-Body Currents \label{app:two-body} }
\markboth{APPENDIX \ref{app:two-body}. EXPLICIT TWO-BODY CURRENTS}
         {APPENDIX \ref{app:two-body}. EXPLICIT TWO-BODY CURRENTS}
In this appendix we calculate diagrams explicitly including the two-body 
currents shown in Fig.~\ref{fig:mesonexchange}. However, it turned out from 
our calculation that the only non-negligible of these currents is the 
Kroll-Ruderman (pair) current, Fig.~\ref{fig:mesonexchange}(a). Therefore the 
pion-pole current, Fig.~\ref{fig:mesonexchange}(b), is only included 
via charge conservation, cf. Eq.~(\ref{eq:implicit}).

First we derive the Kroll-Ruderman current, including corrections due to the 
photon energy. We then calculate several contributions with one of the photons
coupling to this current. Diagrams with explicit pion exchange at both 
vertices were found to be small and are not considered in this work. 

\begin{figure}[!htb]
\begin{center}
\includegraphics*[width=.6\linewidth]{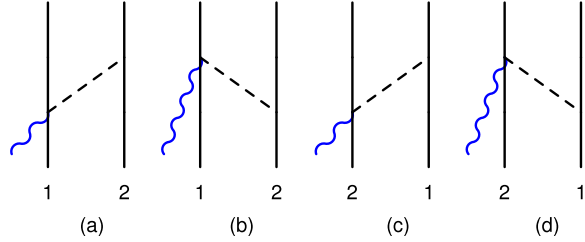}\\
\caption[Time orderings for Kroll-Ruderman current]
{Possible time-orderings for an incoming photon coupling to the 
Kroll-Ruderman current.}
\label{fig:KRtimeorderings}
\end{center}
\end{figure}

In order to calculate the interaction Hamiltonian, from which we derive the 
Kroll-Ruderman current, we first sketch the relevant time-ordered diagrams 
in Fig.~\ref{fig:KRtimeorderings}.

Now we need to define the Hamiltonians for the $\gamma\pi N$ and the $\pi N N$
coupling. These are given in Eqs.~(\ref{eq:HpiNN}, \ref{eq:HgammapiN}) and 
can be used, together with Eqs.~(\ref{eq:pionfield}) and 
(\ref{eq:taupm}-\ref{eq:taucrossphi}) 
to write down the amplitude for the process Fig.~\ref{fig:KRtimeorderings}(a):
\ba
\label{eq:twobody1}
\Mfi{\mathrm{KR}\,a}&=\sum_B\sum_\pm\int\frac{d^3q}{(2\pi)^3}\,
\frac{2\pi}{E_\pi}\,\mx{C}{-\frac{f}{m_\pi}\,(\vec{\sigma}_2\cdot\nab_2)\,
\frac{1}{\sqrt{2}}\,\tau_2^\pm\,\e^{i\vec{q}\cdot\vec{x}_2}}{B}\\
&\times\frac{1}{\w+\frac{\w^2}{2m_d}-B-E_B-E_\pi}\,
\mx{B}{\frac{i\,e\,f}{m_\pi}\,
(\vec{\sigma}_1\cdot\vec{A}(\vec{x}_1))\,
\frac{1}{\sqrt{2}}\,(\mp\tau_1^\mp)\,\e^{-i\vec{q}\cdot\vec{x}_1}}{d_i}
\nonumber
\end{align}
$E_B$ is the energy of the two-nucleon state $\ket{B}$ while the pion is in 
flight. In order to derive the expression given for the Kroll-Ruderman current
in Eq.~(\ref{eq:KR}) one would have to make the assumption that 
$E_\pi=\sqrt{m_\pi^2+\vec{q}^{\,2}}$ is the by far dominant term in the energy 
denominator of Eq.~(\ref{eq:twobody1}). However, for $\w\sim 100$~MeV, this is
a rather crude approximation. Therefore we keep the photon energy in the 
denominator and do the approximation 
$\w+\frac{\w^2}{2m_d}-B-E_B-E_\pi\approx\w-E_\pi$. Nevertheless, we first 
approximate the denominator by the negative pion energy $-E_\pi$, i.e. we set 
$\w=0$, to check whether we get the correct result for 
$\vec{J}_\mathrm{stat}^\mathrm{\,KR}$.

Neglecting the energy of the intermediate state $\ket{B}$ in the denominator, 
one can collapse the sum over $B$ and we find
\be
M_{fi\,\mathrm{stat}}^{\mathrm{KR}\,a}=\frac{\pi\,e\,f^2}{i\,m_\pi^2}\,
\mx{C}{(\tau_2^+\,\tau_1^- - \tau_2^-\,\tau_1^+ )\,\int
\frac{d^3q}{(2\pi)^3}\,(\vec{\sigma}_2\cdot\nab_2)\,
\frac{\e^{i\vec{q}\cdot(\vec{x}_2-\vec{x}_1)}}{m_\pi^2+\vec{q}^{\,2}}\,
(\vec{\sigma}_1\cdot\vec{A}(\vec{x}_1))}{d_i}.
\label{eq:twobody1collapsed}
\ee
From our definition for $\tau_\pm$ (Eq.~(\ref{eq:taupm})) follows 
$\tau_2^+\,\tau_1^- - \tau_2^-\,\tau_1^+ =2\,i\,\taucrosstau$.
Furthermore we define, as in Section~8.3 of~\cite{Ericson}, 
$\vec{x}_1-\vec{x}_2=\vec{r}$.
Therefore $\nab_2=\nab_{x_2-x_1}=\nab_{-r}=-\nab_r$. The integral over $d^3q$
can easily be evaluated, yielding
\be
\int\frac{d^3q}{(2\pi)^3}\,\frac{\e^{-i\vec{q}\cdot\vec{r}}}
{m_\pi^2+\vec{q}^{\,2}}=
\frac{\e^{-m_\pi r}}{4\pi\,r}.
\ee
Therefore we find
\ba
M_{fi\,\mathrm{stat}}^{\mathrm{KR}\,a}&=-\frac{e\,f^2}{2\,m_\pi^2}\,
\mx{C}{\taucrosstau\,(\vec{\sigma}_2\cdot\nab_r)\,
\frac{\e^{-m_\pi r}}{r}\,
(\vec{\sigma}_1\cdot\vec{A}(\vec{x}_1))}{d_i}\nonumber\\
&=-\frac{e\,f^2}{2\,m_\pi^2}\,
\mx{C}{\taucrosstau\,(\vec{\sigma}_2\cdot\hat{r})\,
\frac{\partial}{\partial r}\frac{\e^{-m_\pi r}}{r}\,
(\vec{\sigma}_1\cdot\vec{A}(\vec{x}_1))}{d_i}\!.
\label{eq:MfiKRstata}
\end{align}
The calculation of diagram~\ref{fig:KRtimeorderings}(b), where the 
pion is destroyed rather than created at the $\gamma\pi N$ vertex, is very 
similar to the derivation of Eq.~(\ref{eq:MfiKRstata}). The amplitude is 
\ba
\label{eq:twobody2}
\Mfi{\mathrm{KR}\,b}&=\sum_B\sum_\pm\int\frac{d^3q}{(2\pi)^3}\,
\frac{2\pi}{E_\pi}\,\mx{C}{\frac{i\,e\,f}{m_\pi}\,
(\vec{\sigma}_1\cdot\vec{A}(\vec{x}_1))\,
\frac{1}{\sqrt{2}}\,(\mp\tau_1^\mp)\,\e^{i\vec{q}\cdot\vec{x}_1}}{B}\nonumber\\
&\times\frac{1}{-B-E_B-E_\pi}\,
\mx{B}{-\frac{f}{m_\pi}\,(\vec{\sigma}_2\cdot\nab_2)\,
\frac{1}{\sqrt{2}}\,\tau_2^\pm\,\e^{-i\vec{q}\cdot\vec{x}_2}}{d_i},
\end{align}
so the only differences are that $\nab_2\rightarrow -i\,\vec{q}$ rather than
$i\,\vec{q}$ and the interchange of $\vec{\tau}_1,\;\vec{\tau}_2$, which 
gives an additional sign. These two signs cancel and therefore 
$\Mfi{\mathrm{KR}\,b}=\Mfi{\mathrm{KR}\,a}$. Diagrams 
\ref{fig:KRtimeorderings}(c),~(d) are identical to diagrams (a),~(b) 
with the nucleons exchanged. This exchange leads to 
$\taucrosstau\rightarrow-\taucrosstau$ and $\hat{r}\rightarrow -\hat{r}$ in 
Eq.~(\ref{eq:MfiKRstata}), so we get for diagram \ref{fig:KRtimeorderings}(c) 
(which is, of course, identical to diagram \ref{fig:KRtimeorderings}(d))
\be
M_{fi\,\mathrm{stat}}^{\mathrm{KR}\,c}=-\frac{e\,f^2}{2\,m_\pi^2}\,
\mx{C}{\taucrosstau\,(\vec{\sigma}_1\cdot\hat{r})\,
\frac{\partial}{\partial r}\frac{\e^{-m_\pi r}}{r}\,
(\vec{\sigma}_2\cdot\vec{A}(\vec{x}_2))}{d_i}.
\ee
Adding the four amplitudes \ref{fig:KRtimeorderings}(a)-(d), we find for the 
corresponding interaction Hamiltonian, by removing the initial and final state,
\be
H^{\mathrm{int}\,\mathrm{KR}}_\mathrm{stat}=-\frac{e\,f^2}{m_\pi^2}\,
\taucrosstau\,\left[(\vec{\sigma}_2\cdot\hat{r})\,
(\vec{\sigma}_1\cdot\vec{A}(\vec{x}_1))+(\vec{\sigma}_1\cdot\hat{r})\,
(\vec{\sigma}_2\cdot\vec{A}(\vec{x}_2))\right]\,
\frac{\partial}{\partial r}\frac{\e^{-m_\pi r}}{r},
\label{eq:Hintstat}
\ee
which is identical to 
$-\int\vec{J}_\mathrm{stat}^\mathrm{\,KR}(\vec{\xi};\vec{x}_1,\vec{x}_2)\cdot
\vec{A}\ofxi$ with 
$\vec{J}_\mathrm{stat}^\mathrm{\,KR}(\vec{\xi};\vec{x}_1,\vec{x}_2)$ given in 
Eq.~(\ref{eq:KR}).

Now we discuss how this result changes when we retain the photon 
energy in the denominator. Again we start with diagram~(a) of 
Fig.~\ref{fig:KRtimeorderings} and find instead of 
Eq.~(\ref{eq:twobody1collapsed})
\ba
M_{fi}^{\mathrm{KR}\,a}&=-\frac{\pi\,e\,f^2}{i\,m_\pi^2}\,
\bra{C}(\tau_2^+\,\tau_1^- - \tau_2^-\,\tau_1^+ )\\
&\times\int
\frac{d^3q}{(2\pi)^3}\,(\vec{\sigma}_2\cdot\nab_2)\,
\frac{\e^{i\vec{q}\cdot(\vec{x}_2-\vec{x}_1)}}
{\sqrt{m_\pi^2+\vec{q}^{\,2}}\,(\w-\sqrt{m_\pi^2+\vec{q}^{\,2}})}\,
(\vec{\sigma}_1\cdot\vec{A}(\vec{x}_1))\ket{d_i}.\nonumber
\end{align}
In order to perform the integral, we do an expansion of the integrand in $\w$.
This expansion is possible to any order, and it converges quite fast. We found 
that our results achieved with an expansion up to $\calO(\w^7)$ only deviate 
by about $0.1\%$ from analogous results using the expansion up to 
$\calO(\w^6)$. Therefore the error introduced by cutting off the expansion
at $\calO(\w^7)$ is certainly negligibly small. The resulting interaction
Hamiltonian is
\ba
H^{\mathrm{int}\,\mathrm{KR}\,a}&\approx
-\frac{e\,f^2}{m_\pi^2}\,\taucrosstau\,
(\vec{\sigma}_2\cdot\hat{r})\,(\vec{\sigma}_1\cdot\vec{A}(\vec{x}_1))\,
\frac{\partial}{\partial r}\left\{\frac{\e^{-m_\pi r}}{2\,r}+
\frac{K_0(m_\pi r)}{\pi}\,\w\right.\nonumber\\
&+\frac{\e^{-m_\pi r}}{4\,m_\pi}\,\w^2+
\frac{r\,K_1(m_\pi r)}{3\,m_\pi\,\pi}\,\w^3+
\frac{\e^{-m_\pi r}\,(1+m_\pi r)}{16\,m_\pi^3}\,\w^4+
\frac{r^2\,K_2(m_\pi r)}{15\,m_\pi^2\,\pi}\,\w^5\nonumber\\
&+\left.\frac{\e^{-m_\pi r}\,(3+3\,m_\pi r+m_\pi^2\,r^2)}{96\,m_\pi^5}\,\w^6+
\frac{r^3\,K_3(m_\pi r)}{105\,m_\pi^3\,\pi}\,\w^7\right\},
\end{align}
$K_i(z)$ being the modified Bessel functions of the second kind. 

The photon energy $\w$ does not appear in the 
denominator of 
diagram~\ref{fig:KRtimeorderings}(b). Therefore we find the same interaction
Hamiltonian as in the static case. Exchange of the nucleons is treated as 
before, leaving us with
\be
H^{\mathrm{int}\,\mathrm{KR}}_s=-\frac{e\,f^2}{m_\pi^2}\,
\taucrosstau\,\left[(\vec{\sigma}_2\cdot\hat{r})\,
(\vec{\sigma}_1\cdot\vec{A}(\vec{x}_1))+(\vec{\sigma}_1\cdot\hat{r})\,
(\vec{\sigma}_2\cdot\vec{A}(\vec{x}_2))\right]\,
\frac{\partial}{\partial r}f^\mathrm{KR}_s(r),
\label{eq:HintKRs}
\ee
with 
\ba
f^\mathrm{KR}_s(r)&=\frac{\e^{-m_\pi r}}{r}+
\frac{K_0(m_\pi r)}{\pi}\,\w+
\frac{\e^{-m_\pi r}}{4\,m_\pi}\,\w^2\nonumber\\
&+
\frac{r\,K_1(m_\pi r)}{3\,m_\pi\,\pi}\,\w^3+
\frac{\e^{-m_\pi r}\,(1+m_\pi r)}{16\,m_\pi^3}\,\w^4+
\frac{r^2\,K_2(m_\pi r)}{15\,m_\pi^2\,\pi}\,\w^5\nonumber\\
&+\frac{\e^{-m_\pi r}\,(3+3\,m_\pi r+m_\pi^2\,r^2)}{96\,m_\pi^5}\,\w^6+
\frac{r^3\,K_3(m_\pi r)}{105\,m_\pi^3\,\pi}\,\w^7.
\label{eq:fsKR}
\end{align}
The index $s$ signals that Eq.~(\ref{eq:HintKRs}) is only valid for 
$s$-channel diagrams. In the $u$-channel, the both time orderings correspond 
to the (approximate) energy denominators $\frac{1}{-E_\pi}$ and 
$\frac{1}{-\w-E_\pi}$, respectively. The net effect is an alternating sign in 
$f^\mathrm{KR}(r)$, say
\ba
f^\mathrm{KR}_u(r)&=\frac{\e^{-m_\pi r}}{r}-
\frac{K_0(m_\pi r)}{\pi}\,\w+
\frac{\e^{-m_\pi r}}{4\,m_\pi}\,\w^2\nonumber\\
&-
\frac{r\,K_1(m_\pi r)}{3\,m_\pi\,\pi}\,\w^3+
\frac{\e^{-m_\pi r}\,(1+m_\pi r)}{16\,m_\pi^3}\,\w^4-
\frac{r^2\,K_2(m_\pi r)}{15\,m_\pi^2\,\pi}\,\w^5\nonumber\\
&+\frac{\e^{-m_\pi r}\,(3+3\,m_\pi r+m_\pi^2\,r^2)}{96\,m_\pi^5}\,\w^6-
\frac{r^3\,K_3(m_\pi r)}{105\,m_\pi^3\,\pi}\,\w^7.
\label{eq:fuKR}
\end{align}
We note that the amplitudes which we calculate in this appendix deviate 
by about 25\%, depending on whether we use the static interaction Hamiltonian 
(\ref{eq:Hintstat}) or a corrected one, including 
Eq.~(\ref{eq:fsKR}) or~(\ref{eq:fuKR}), respectively.

\begin{figure}[!htb]
\begin{center}
\includegraphics*[width=.6\linewidth]{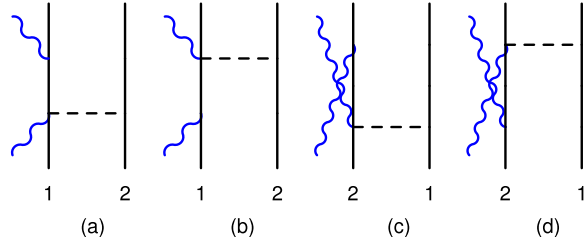}\\
\caption[Diagrams with explicit Kroll-Ruderman current]
{Diagrams with one photon coupling to the Kroll-Ruderman current.}
\label{fig:KRdiagrams}
\end{center}
\end{figure}

We are now able to use the Kroll-Ruderman interaction Hamiltonian to calculate
the diagrams displayed in Fig.~\ref{fig:KRdiagrams}.
As explained in Sect.~\ref{sec:subleading}, there are only a few combinations 
to be taken 
into account. These are $\Mfi{\mathrm{KR}\,\mathrm{full}\,\sigma 1}$,
$\Mfi{\phi\,\mathrm{KR}1}$ and $\Mfi{\phi\,\mathrm{KR}2}$. The indices 
'full', '1', '2' denote the coupling of the full photon field $\vec{A}$ 
or of $\Aone,\;\Atwo$, respectively. First we compute the contributions 
$\Mfi{\mathrm{KR}\,\mathrm{full}\,\sigma 1}$, where we skip the index 'full'
for brevity. These four amplitudes have already been calculated 
in~\cite{Karakowski}, however using the static Kroll-Ruderman 
Hamiltonian~(\ref{eq:Hintstat}).
They read
\ba
\Mfi{\mathrm{KR}\,\sigma 1\,a}&=
-\sum_C\frac{\mx{d_f}{\int\Jsigma\ofxi\cdot\Aone\ofxi\,d^3\xi}{C}
\mx{C}{H^{\mathrm{int}\,\mathrm{KR}}_s}{d_i}}{\denoms},\nonumber\\
\Mfi{\mathrm{KR}\,\sigma 1\,b}&=
-\sum_C\frac{\mx{d_f}{H^{\mathrm{int}\,\mathrm{KR}}_s}{C}
\mx{C}{\int\Jsigma\ofxi\cdot\Aone\ofxi\,d^3\xi}{d_i}}{\denoms},\nonumber\\
\Mfi{\mathrm{KR}\,\sigma 1\,c}&=
-\sum_C\frac{\mx{d_f}{\int\Jsigma\ofxi\cdot\Aone\ofxi\,d^3\xi}{C}
\mx{C}{H^{\mathrm{int}\,\mathrm{KR}}_u}{d_i}}{\denomu},\nonumber\\
\Mfi{\mathrm{KR}\,\sigma 1\,d}&=
-\sum_C\frac{\mx{d_f}{H^{\mathrm{int}\,\mathrm{KR}}_u}{C}
\mx{C}{\int\Jsigma\ofxi\cdot\Aone\ofxi\,d^3\xi}{d_i}}{\denomu},
\label{eq:KRsigma1diagrams}
\end{align}
which follows immediately from Eq.~(\ref{eq:disp}). 
As explained in Section~\ref{sec:subleading}, 
we only consider that part of 
$\int\Jsigma\ofxi\cdot\Aone\ofxi\,d^3\xi$, which contains the operator  
$[Y_{L-1}\otimes t]_{L}$, cf. Eq.~(\ref{eq:intJsigmaAone}). Plugging this 
expression and the Kroll-Ruderman-Hamiltonian into the first amplitude of 
Eq.~(\ref{eq:KRsigma1diagrams}) we find
\ba
\mathcal{M}&_{fi}^{\mathrm{KR}\,\sigma 1\,a}=
-\sum_C\bra{d_f}-\sum_{L'=1}^\infty
\sum_{M'=-L'}^{L'}\lambda_f\,\sqrt{2\pi\,(L'+1)}\,\frac{e\,\w}{2m_d}\,i^{L'+1}
\,j_{L'-1}(\frac{\w r}{2})\nonumber\\
&\times\left(\mu_p+(-1)^{L'}\,\mu_n\right)\,
[Y_{L'-1}\otimes t]_{L'\,M'}\,\left(-(-1)^{L'+\lambda_f}\,
\wignerd{L'}{M'}{-\lambda_f}\right)\ket{C}\frac{1}{\denoms}\nonumber\\
&\times\mx{C}{-\frac{e\,f^2}{m_\pi^2}\,\taucrosstau\,
\left\{(\vec{\sigma}_1\cdot\eps)\,(\vec{\sigma}_2\cdot\hat{r})\,
\e^{i\vec{k}_i\vec{x}_1}+
(\vec{\sigma}_2\cdot\eps)\,(\vec{\sigma}_1\cdot\hat{r})\,
\e^{i\vec{k}_i\vec{x}_2}\right\}\,
\frac{\partial}{\partial r}f^\mathrm{KR}_s(r)}{d_i}.
\end{align}
We decompose the exponentials into multipoles according to 
Eq.~(\ref{eq:expansionstart}), replacing 
$\vec{x}_1\rightarrow \frac{\vec{r}}{2}$, 
$\vec{x}_2\rightarrow-\frac{\vec{r}}{2}$ as usual, and use
\be
\frac{1}{\sqrt{2}}\mx{p\,n+n\,p}{\taucrosstau\,\frac{1}{\sqrt{2}}}{p\,n-n\,p}=
-2\,i,
\ee
cf. Eq.~(\ref{eq:taucrosstauond}). Therefore,
\ba
\mathcal{M}&_{fi}^{\mathrm{KR}\,\sigma 1\,a}=
-\sum_{\hat{C}}\sum_{L'=1}^\infty\sum_{M'=-L'}^{L'}
\sum_{L=0}^\infty\sum_{M=-L}^{L}\sum_{l,l'}\frac{4\pi\,\lambda_f\,e^2\,f^2\,\w}
{m_N\,m_\pi^2}\,i^{L+L'}\,(-1)^{L'+\lambda_f}\,Y_{L\,M}^\ast(\hat{k}_i)
\nonumber\\
&\times\left(\mu_p+(-1)^{L'}\,\mu_n\right)\,\wignerd{L'}{M'}{-\lambda_f}
\sqrt{2\pi\,(L'+1)}\,\mxbc{[Y_{L'-1}\otimes t]_{L'\,M'}}\nonumber\\
&\times\mxca{\left\{(\vec{\sigma}_1\cdot\eps)\,(\vec{\sigma}_2\cdot\hat{r})+
(-1)^L(\vec{\sigma}_2\cdot\eps)\,(\vec{\sigma}_1\cdot\hat{r})\right\}\,
Y_{L\,M}(\hat{r})}\nonumber\\
&\times
\doubleint r\,dr\,r'\,dr'\,u_{l'}(r')\,j_{L'-1}(\frac{\w r'}{2})\,\green
\,j_L(\frac{\w r}{2})\,u_l(r)\,\frac{\partial}{\partial r}f_s^\mathrm{KR}(r).
\end{align}
We now apply the Wigner-Eckart theorem (\ref{eq:WE}), the identity 
$Y_{L\,M}^\ast(\hat{k}_i)=\frac{1}{2}\,\sqrt{\frac{2L+1}{\pi}}\,\delta_{M,0}$ 
and decompose the scalar products into spherical components, using 
$r_j=\sqrt{\frac{4\pi}{3}}\,Y_{1\,j}(\hat{r})$. We further put $S_C=0$
due to Eq.~(\ref{eq:mxt}).
\ba
\mathcal{M}&_{fi}^{\mathrm{KR}\,\sigma 1\,a}=
\sum_{\hat{C}}\sum_{L'=1}^\infty\sum_{M'=-L'}^{L'}
\sum_{L=0}^\infty\sum_{l,l'}\sum_{i,j}\frac{4\pi\,\lambda_f\,e^2\,f^2\,\w}
{m_N\,m_\pi^2}\,\sqrt{\frac{\pi}{3}}\,i^{L+L'}\,(-1)^{L'+\lambda_f+1-M_f+i+j}
\nonumber\\
&\times\left(\mu_p+(-1)^{L'}\,\mu_n\right)\,(\eps)_{-i}\,
\wignerd{L'}{M'}{-\lambda_f}\sqrt{2\,(L'+1)\,(2L+1)}\nonumber\\
&\times
\doubleint r\,dr\,r'\,dr'\,u_{l'}(r')\,j_{L'-1}(\frac{\w r'}{2})\,\green\,
j_L(\frac{\w r}{2})\,u_l(r)\,\frac{\partial}{\partial r}f_s^\mathrm{KR}(r)
\nonumber\\
&\times\threej{1}{L'}{J_C}{-M_f}{M'}{M_C}\,
\mxred{l'\,1\,1}{[Y_{L'-1}\otimes t]_{L'}}{J_C\,0\,J_C}\nonumber\\
&\times\mx{J_C\,0\,J_C\,M_C}
{\sigma_{2\,j}\,\sigma_{1\,i}\,Y_{1\,-j}\,Y_{L\,0}+
(-1)^L\,\sigma_{1\,j}\,\sigma_{2\,i}\,Y_{1\,-j}\,Y_{L\,0}}{l\,1\,1\,M_i}
\label{eq:MfiKRsigma1a}
\end{align}
The second matrix element in Eq.~(\ref{eq:MfiKRsigma1a}) is evaluated according
to Eq.~(\ref{eq:master}). From 
$\mxred{S_1}{S-t}{s}\,\mxred{s}{S+t}{S_2}$ and the definitions of $\vec{S}$, 
$\vec{t}$, cf. Eq.~(\ref{eq:master}), we see that the product 
$\sigma_{2\,j}\,\sigma_{1\,i}$ is 
antisymmetric under the interchange $1\leftrightarrow 2$ in the spin-changing,
symmetric in the spin-conserving case. In Eq.~(\ref{eq:MfiKRsigma1a}) we have
$S_1=0$ and $S_2=1$. Therefore only two of the four combinations are possible:
\be
\mxred{S_1}{S-t}{s}\,\mxred{s}{S+t}{S_2}\rightarrow
\mxred{S_1}{S}{s}\,\mxred{s}{t}{S_2}-
\mxred{S_1}{t}{s}\,\mxred{s}{S}{S_2}
\ee
From Eqs.~(\ref{eq:mxSshort}, \ref{eq:mxtshort}) we find 
that the only non-vanishing product of spin matrix elements is
\be
\mxred{0}{t}{s}\,\mxred{s}{S}{1}=-3\sqrt{2}\,\delta_{s,1}.
\ee
The 9-$j$~symbol guarantees that $S'=1$, cf. Eq.~(\ref{eq:ninejzero}).
Therefore we can easily evaluate the sum over $s$ and find
\be
\sum_s\sixj{1}{1}{1}{1}{0}{s}\,\mxred{0}{S-t}{s}\,\mxred{s}{S+t}{1}=-\sqrt{2}.
\ee
Nevertheless, we keep the compact notation of Eq.~(\ref{eq:MfiKRsigma1a}) to 
write down the final result:
\ba
\mathcal{M}&_{fi}^{\mathrm{KR}\,\sigma 1\,a}=
\sum_{L=0}^\infty\sum_{L'=1}^\infty
\sum_{J_C=|L'-1|}^{L'+1}\sum_{l,l'}\sum_{i,j}
\frac{4\pi\,\lambda_f\,e^2\,f^2\,\w}
{m_N\,m_\pi^2}\,\sqrt{\frac{\pi}{3}}\,i^{L+L'}\,(-1)^{L'+\lambda_f+1-M_f+i+j}
\nonumber\\
&\times\left(\mu_p+(-1)^{L'}\,\mu_n\right)\,(\eps)_{-i}\,
\wignerd{L'}{M_f-M_i-i}{-\lambda_f}\,\sqrt{2\,(L'+1)\,(2L+1)}\nonumber\\
&\times
\doubleint r\,dr\,r'\,dr'\,u_{l'}(r')\,j_{L'-1}(\frac{\w r'}{2})\,\green\,
j_L(\frac{\w r}{2})\,u_l(r)\,\frac{\partial}{\partial r}f_s^\mathrm{KR}(r)
\nonumber\\
&\times\threej{1}{L'}{J_C}{-M_f}{M_f-M_i-i}{M_i+i}\,
\mxred{l'\,1\,1}{[Y_{L'-1}\otimes t]_{L'}}{J_C\,0\,J_C}\nonumber\\
&\times\mx{J_C\,0\,J_C\,M_i+i}
{\sigma_{2\,j}\,\sigma_{1\,i}\,Y_{1\,-j}\,Y_{L\,0}}{l\,1\,1\,M_i}\,
\left(1-(-1)^L\right)
\label{eq:MfiKRsigma1aresult}
\end{align}
Obviously the leading orbital angular momenta are $L=L'=1$. As 
$|\mu_p-\mu_n|\gg|\mu_p+\mu_n|$, one may restrict oneself to this contribution.

The evaluation of $\Mfi{\mathrm{KR}\,\sigma 1\,b\text{-}d}$ is quite similar 
to the derivation of Eq.~(\ref{eq:MfiKRsigma1aresult}). Therefore we only 
give the results:
\ba
\mathcal{M}&_{fi}^{\mathrm{KR}\,\sigma 1\,b}=
\sum_{L=1}^\infty\sum_{L'=0}^\infty
\sum_{J_C=|L-1|}^{L+1}\sum_{l,l'}\sum_{i,j}
\frac{8\pi\,\lambda_i\,e^2\,f^2\,\w}
{m_N\,m_\pi^2}\,\sqrt{\frac{\pi}{3}}\,i^{L+L'}\,(-1)^{J_C-M_i-\lambda_i+i+j}
\nonumber\\
&\times\left(\mu_p+(-1)^{L}\,\mu_n\right)\,(\epspr)_{-i}\,
Y_{L'\,M_f-M_i-\lambda_i-i}^\ast(\hat{k}_f)\,\sqrt{2\pi\,(L+1)}\nonumber\\
&\times
\doubleint r\,dr\,r'\,dr'\,u_{l}(r)\,j_{L-1}(\frac{\w r}{2})\,\green\,
j_{L'}(\frac{\w r'}{2})\,u_{l'}(r')\,\frac{\partial}{\partial r'}f_s^
\mathrm{KR}(r')\nonumber\\
&\times\mx{l'\,1\,1\,M_f}
{\sigma_{2\,j}\,\sigma_{1\,i}\,Y_{1\,-j}\,Y_{L'\,M_f-M_i-\lambda_i-i}}
{J_C\,0\,J_C\,M_i+\lambda_i}\nonumber\\
&\times
\mxred{J_C\,0\,J_C}{[Y_{L-1}\otimes t]_{L}}{l\,1\,1}\,
\threej{J_C}{L}{1}{-M_i-\lambda_i}{\lambda_i}{M_i}\,\left((-1)^{L'}-1\right)
\end{align}
\ba
\mathcal{M}&_{fi}^{\mathrm{KR}\,\sigma 1\,c}=
\sum_{L=1}^\infty\sum_{L'=0}^\infty
\sum_{J_C=|L-1|}^{L+1}\sum_{l,l'}\sum_{i,j}
\frac{8\pi\,\lambda_i\,e^2\,f^2\,\w}
{m_N\,m_\pi^2}\,\sqrt{\frac{\pi}{3}}\,i^{L+L'}\,(-1)^{-M_f+i+j}
\nonumber\\
&\times\left(\mu_p+(-1)^{L}\,\mu_n\right)\,(\epspr)_{-i}\,
Y_{L'\,M_f-M_i-\lambda_i-i}^\ast(\hat{k}_f)\,\sqrt{2\pi\,(L+1)}\nonumber\\
&\times
\doubleint r\,dr\,r'\,dr'\,u_{l'}(r)\,j_{L-1}(\frac{\w r}{2})\,\greenpr\,
j_{L'}(\frac{\w r'}{2})\,u_l(r')\,
\frac{\partial}{\partial r'}f_u^\mathrm{KR}(r')
\nonumber\\
&\times\threej{1}{L}{J_C}{-M_f}{\lambda_i}{M_f-\lambda_i}\,
\mxred{l'\,1\,1}{[Y_{L-1}\otimes t]_{L}}{J_C\,0\,J_C}\\
&\times\mx{J_C\,0\,J_C\,M_f-\lambda_i}
{\sigma_{2\,j}\,\sigma_{1\,i}\,Y_{1\,-j}\,Y_{L'\,M_f-M_i-\li-i}}
{l\,1\,1\,M_i}\,\left((-1)^{L'}-1\right)\nonumber
\end{align}
\ba
\mathcal{M}&_{fi}^{\mathrm{KR}\,\sigma 1\,d}=
\sum_{L=0}^\infty\sum_{L'=1}^\infty
\sum_{J_C=|L'-1|}^{L'+1}\sum_{l,l'}\sum_{i,j}
\frac{4\pi\,\lambda_f\,e^2\,f^2\,\w}
{m_N\,m_\pi^2}\,\sqrt{\frac{\pi}{3}}\,i^{L+L'}\,(-1)^{L'-\lf+1+j+J_C-M_f}
\nonumber\\
&\times\left(\mu_p+(-1)^{L'}\,\mu_n\right)\,(\eps)_{-i}\,
\wignerd{L'}{M_f-M_i-i}{-\lf}\,\sqrt{2\,(L'+1)\,(2L+1)}\nonumber\\
&\times
\doubleint r\,dr\,r'\,dr'\,u_{l}(r')\,j_{L'-1}(\frac{\w r'}{2})\,\greenpr\,
j_{L}(\frac{\w r}{2})\,u_{l'}(r)\,\frac{\partial}{\partial r}
f_u^\mathrm{KR}(r)\nonumber\\
&\times\mx{l'\,1\,1\,M_f}
{\sigma_{2\,j}\,\sigma_{1\,i}\,Y_{1\,-j}\,Y_{L\,0}}
{J_C\,0\,J_C\,M_f-i}\\
&\times
\mxred{J_C\,0\,J_C}{[Y_{L'-1}\otimes t]_{L'}}{l\,1\,1}\,
\threej{J_C}{L'}{1}{-M_f+i}{M_f-M_i-i}{M_i}\,\left(1-(-1)^{L}\right)\nonumber
\end{align}
In $\Mfi{\mathrm{KR}\,\sigma 1\,b,d}$ we used
\be
\frac{1}{2}\mx{p\,n-n\,p}{\taucrosstau}{p\,n+n\,p}=2\,i.
\ee

We turn now to the calculation of $\Mfi{\phi\,\mathrm{KR}1}$ and 
$\Mfi{\phi\,\mathrm{KR}2}$, going beyond Ref.~\cite{Karakowski}. 
First we demonstrate the 
isospin-changing character of $\phi(\vec{x}_p)$. Therefore, we explicitly 
write the isospin-dependence of $\phi$, as we did in Eq.~(\ref{eq:isospindep}) 
and evaluate this operator between two isospin wave functions with $T=0$ and
$T=1$, respectively.
\ba
&\frac{1}{2}\,\mx{p\,n-n\,p}
{\sum_{j=1,2}\frac{1}{2}\,(1+\tau_z^j)\,\phi(\vec{x}_j)}{p\,n+n\,p}\nonumber\\
=&\frac{1}{4}\,\mx{p\,n-n\,p}
{\tau_z^1\,\phi(\vec{x}_1)+\tau_z^2\,\phi(\vec{x}_2)}{p\,n+n\,p}\nonumber\\
=&\frac{1}{2}\,(\phi(\vec{x}_1)-\phi(\vec{x}_2))=
   \frac{1}{2}\,(\phi(\vec{r}/2)-\phi(-\vec{r}/2))
\label{eq:isospinchangingcharacter}
\end{align}
The last step reflects our usual choice 
$\vec{x}_1=\frac{\vec{r}}{2},\;\vec{x}_2=-\frac{\vec{r}}{2}$. 
The only angular operator contained in $\phi$ is $Y_L$. As we need an 
isospin-changing matrix element, $L$ has to be odd, which follows from 
Eq.~(\ref{eq:mxL}), see also Section~\ref{sec:subleading}.
Due to this observation and the definition of $\phi(\vec{r})$, 
Eq.~(\ref{eq:phidefinition}), which 
implies $\phi_{L\,\mathrm{odd}}(-\vec{r}/2)=
-\phi_{L\,\mathrm{odd}}(\vec{r}/2)$, we may replace 
$\frac{1}{2}\,(\phi(\vec{r}/2)-\phi(-\vec{r}/2))\rightarrow\phi(\vec{r}/2)$ 
in the following. An explicit proof of our claim $L$ odd is given after 
Eq.~(\ref{eq:MfiphiKR1aresult}).

We start with the calculation of $\Mfi{\phi\,\mathrm{KR}1}$. The four 
amplitudes follow immediately from Eq.~(\ref{eq:Mfiphi}):
\ba
\Mfi{\phi\,\mathrm{KR}1\,a}&=
i\,\left(\w+\frac{\w^2}{2m_d}\right)\,
\sum_C\frac{\mx{d_f}{\phifhat}{C}
\mx{C}{H^{\mathrm{int}\,\mathrm{KR}1}_s}{d_i}}{\denoms}\nonumber\\
\Mfi{\phi\,\mathrm{KR}1\,b}&=
-i\,\left(\w+\frac{\w^2}{2m_d}\right)\,
\sum_C\frac{\mx{d_f}{H^{\mathrm{int}\,\mathrm{KR}1}_s}{C}
\mx{C}{\phiihat}{d_i}}{\denoms}\nonumber\\
\Mfi{\phi\,\mathrm{KR}1\,c}&=
-i\,\left(\w-\frac{\w^2}{2m_d}+\frac{\PCsq}{2m_C}\right)\,
\sum_C\frac{\mx{d_f}{\phiihat}{C}
\mx{C}{H^{\mathrm{int}\,\mathrm{KR}1}_u}{d_i}}{\denomu}\nonumber\\
\Mfi{\phi\,\mathrm{KR}1\,d}&=
i\,\left(\w-\frac{\w^2}{2m_d}+\frac{\PCsq}{2m_C}\right)\,
\sum_C\frac{\mx{d_f}{H^{\mathrm{int}\,\mathrm{KR}1}_u}{C}
\mx{C}{\phifhat}{d_i}}{\denomu}
\label{eq:phiKR1diagrams}
\end{align}
Now we have to specify $H^{\mathrm{int}\,\mathrm{KR}1}$, which is achieved 
by replacing $\vec{A}$ by $\Aone$ in Eq.~(\ref{eq:HintKRs}).
We find, using Eq.~(\ref{eq:LonY}), 
\ba
H&^{\mathrm{int}\,\mathrm{KR}1}_{s,u}=\sum_{\vec{k},\lambda}
\sum_{L=1}^{\infty}\sum_{M=-L}^L\frac{e\,f^2}{m_\pi^2}\,\taucrosstau\,\lambda\,
\sqrt{2\pi\,(2L+1)}\,i^L\,j_L(\frac{\w r}{2})\,
\left[a_{\vec{k},\lambda}\,\delta_{M,\lambda}\right.\nonumber\\
&-
\left.a_{\vec{k},\lambda}^\dagger\,(-1)^{L+\lambda}\,\wignerd{L}{M}{-\lambda}
\right]\,\left\{
(\vec{\sigma}_1\cdot\vec{T}_{L\,L\,M}(\vec{x}_1))\,
(\vec{\sigma}_2\cdot\hat{r})+
(\vec{\sigma}_2\cdot\vec{T}_{L\,L\,M}(\vec{x}_2))\,
(\vec{\sigma}_1\cdot\hat{r})\right\}\,\frac{\partial}{\partial r}
f_{s,u}^\mathrm{KR}(r).
\end{align}
The scalar products are expanded into spherical components, according to 
Eq.~(\ref{eq:T_JLMnu}).
\ba
H^{\mathrm{int}\,\mathrm{KR}1}_{s,u}&=-\sum_{\vec{k},\lambda}
\sum_{L=1}^{\infty}\sum_{M=-L}^L\sum_{i,j}\frac{2\pi\,e\,f^2}{m_\pi^2}\,
\taucrosstau\,(-1)^{j-L-M}\,i^L\,\sqrt{\frac{2}{3}}\,\lambda\,(2L+1)\,
j_L(\frac{\w r}{2})\nonumber\\
&\times
\left[a_{\vec{k},\lambda}\,\delta_{M,\lambda}
-a_{\vec{k},\lambda}^\dagger\,(-1)^{L+\lambda}\,\wignerd{L}{M}{-\lambda}
\right]\,\threej{L}{1}{L}{M-i}{i}{-M}\nonumber\\
&\times
\sigma_{2\,j}\,\sigma_{1\,i}\,Y_{1\,-j}\,
Y_{L\,M-i}\,\left(1+(-1)^L\right)\,\frac{\partial}{\partial r}
f_{s,u}^\mathrm{KR}(r)
\label{eq:HintKR1}
\end{align}
Here we used the fact that the matrix element arising from $\phi$ does not 
change the spin, i.e. $S_C=1$ in the following amplitudes, and the symmetry of
$\sigma_{2\,j}\,\sigma_{1\,i}$  under $1\leftrightarrow2$
for spin-conserving matrix elements, as observed in Eq.~(\ref{eq:master}). 
Therefore, the 
Hamiltonian (\ref{eq:HintKR1}) only holds for spin-conserving matrix elements.

Now the evaluation of the amplitudes (\ref{eq:phiKR1diagrams}) is 
straightforward and very similar to the calculation of 
$\Mfi{\mathrm{KR}\,\sigma1}$. Therefore we only give the results.
\ba
\mathcal{M}&_{fi}^{\phi\,\mathrm{KR}1\,a}=
\sum_{L=1}^\infty\sum_{L'=1}^\infty
\sum_{J_C=|L'-1|}^{L'+1}\sum_{L_C=|J_C-1|}^{J_C+1}\sum_{l,l'}\sum_{i,j}
\frac{8\pi\,e^2\,f^2}
{m_\pi^2}\,\left(1+\frac{\w}{2m_d}\right)\nonumber\\
&\times
(-1)^{L'-L-\lf-\li+j-M_f}\,i^{L+L'+1}\,\li\,
\wignerd{L'}{M_f-M_i-\li}{-\lambda_f}\,(2L+1)\,
\sqrt{\frac{\pi\,(2L'+1)}{3\,L'\,(L'+1)}}\nonumber\\
&\times
\doubleint r\,dr\,r'\,dr'\,u_{l'}(r')\,\psi_{L'}(\frac{\w r'}{2})\,\green\,
j_L(\frac{\w r}{2})\,u_l(r)\,\frac{\partial}{\partial r}f_s^\mathrm{KR}(r)
\nonumber\\
&\times
\threej{L}{1}{L}{\li-i}{i}{-\li}
\threej{1}{L'}{J_C}{-M_f}{M_f-M_i-\li}{M_i+\li}\,
\mxred{l'\,1\,1}{Y_{L'}}{L_C\,1\,J_C}\nonumber\\
&\times\mx{L_C\,1\,J_C\,M_i+\li}
{\sigma_{2\,j}\,\sigma_{1\,i}\,Y_{1\,-j}\,Y_{L\,\li-i}}{l\,1\,1\,M_i}\,
\left(1+(-1)^L\right)
\label{eq:MfiphiKR1aresult}
\end{align}
Before we write down the remaining amplitudes we prove explicitly our claim 
in the context of 
Eq.~(\ref{eq:isospinchangingcharacter}) that $L'$ has to be an odd number.
The last bracket of Eq.~(\ref{eq:MfiphiKR1aresult}) demands $L$ even. From 
Eq.~(\ref{eq:threejprops5}) we know
that 
the index $L''$, which is summed over in Eq.~(\ref{eq:master}), has to be odd 
and from $\mxred{L_C}{Y_{L''}}{l}$ 
follows that $L_C$ has to be an odd number too, cf.
Eq.~(\ref{eq:mxYshort}), because $l$ is even. That is exactly what we need, 
as $L_C+S_C+T_C$ has to be odd (cf. Section~\ref{sec:subleading}) and 
$S_C=T_C=1$. Eq.~(\ref{eq:mxY}) then tells us 
that in fact $L'$, the orbital angular momentum number of $\phi$, has to be 
an odd number.
This also holds in the 
following three amplitudes, albeit we don't show it explicitly. These are
\ba
\mathcal{M}&_{fi}^{\phi\,\mathrm{KR}1\,b}=
\sum_{L=1}^\infty\sum_{L'=1}^\infty
\sum_{J_C=|L-1|}^{L+1}\sum_{L_C=|J_C-1|}^{J_C+1}\sum_{l,l'}\sum_{i,j}
\frac{8\pi\,e^2\,f^2}{m_\pi^2}\,\left(1+\frac{\w}{2m_d}\right)\nonumber\\
&\times
(-1)^{1-\lf+j-M_f+J_C}\,i^{L+L'+1}\,\lf\,
\wignerd{L'}{M_f-M_i-\li}{-\lambda_f}\,(2L'+1)\,
\sqrt{\frac{\pi\,(2L+1)}{3\,L\,(L+1)}}\nonumber\\
&\times
\doubleint r\,dr\,r'\,dr'\,u_{l}(r)\,\psi_{L}(\frac{\w r}{2})\,\green\,
j_{L'}(\frac{\w r'}{2})\,u_{l'}(r')\,\frac{\partial}{\partial r'}
f_s^\mathrm{KR}(r')\\
&\times
\threej{L'}{1}{L'}{M_f-M_i-\li-i}{i}{-M_f+M_i+\li}
\threej{J_C}{L}{1}{-M_i-\li}{\li}{M_i}\,
\left(1+(-1)^{L'}\right)\nonumber\\
&\times\mx{l'\,1\,1\,M_f}
{\sigma_{2\,j}\,\sigma_{1\,i}\,Y_{1\,-j}\,Y_{L'\,M_f-M_i-\li-i}}
{L_C\,1\,J_C\,M_i+\li}\,
\mxred{L_C\,1\,J_C}{Y_{L}}{l\,1\,1},\nonumber
\end{align}
\ba
\mathcal{M}&_{fi}^{\phi\,\mathrm{KR}1\,c}=
\sum_{L=1}^\infty\sum_{L'=1}^\infty
\sum_{J_C=|L-1|}^{L+1}\sum_{L_C=|J_C-1|}^{J_C+1}\sum_{l,l'}\sum_{i,j}
\frac{8\pi\,e^2\,f^2}{m_\pi^2\,\w}\,
\left(\w-\frac{\w^2}{2m_d}+\frac{\PCsq}{2m_C}\right)\nonumber\\
&\times
(-1)^{1-\lf+\li+j+M_i}\,i^{L+L'+1}\,\lf\,
\wignerd{L'}{M_f-M_i-\li}{-\lambda_f}\,(2L'+1)\,
\sqrt{\frac{\pi\,(2L+1)}{3\,L\,(L+1)}}\nonumber\\
&\times
\doubleint r\,dr\,r'\,dr'\,u_{l'}(r)\,\psi_{L}(\frac{\w r}{2})\,\greenpr\,
j_{L'}(\frac{\w r'}{2})\,u_l(r')\,\frac{\partial}{\partial r'}
f_u^\mathrm{KR}(r')\\
&\times
\threej{L'}{1}{L'}{M_f-M_i-\li-i}{i}{-M_f+M_i+\li}
\threej{1}{L}{J_C}{-M_f}{\li}{M_f-\li}\,\left(1+(-1)^{L'}\right)
\nonumber\\
&\times
\mxred{l'\,1\,1}{Y_{L}}{L_C\,1\,J_C}\,
\mx{L_C\,1\,J_C\,M_f-\li}
{\sigma_{2\,j}\,\sigma_{1\,i}\,Y_{1\,-j}\,Y_{L'\,M_f-M_i-\li-i}}{l\,1\,1\,M_i},
\nonumber
\end{align}
\ba
\mathcal{M}&_{fi}^{\phi\,\mathrm{KR}1\,d}=
\sum_{L=1}^\infty\sum_{L'=1}^\infty
\sum_{J_C=|L'-1|}^{L'+1}\sum_{L_C=|J_C-1|}^{J_C+1}\sum_{l,l'}\sum_{i,j}
\frac{8\pi\,e^2\,f^2}{m_\pi^2\,\w}\,
\left(\w-\frac{\w^2}{2m_d}+\frac{\PCsq}{2m_C}\right)\nonumber\\
&\times
(-1)^{j-L+L'-\lf+J_C-M_f}\,i^{L+L'+1}\,\li\,
\wignerd{L'}{M_f-M_i-\li}{-\lf}\,(2L+1)\,
\sqrt{\frac{\pi\,(2L'+1)}{3\,L'\,(L'+1)}}\nonumber\\
&\times
\doubleint r\,dr\,r'\,dr'\,u_{l}(r')\,\psi_{L'}(\frac{\w r'}{2})\,\greenpr\,
j_{L}(\frac{\w r}{2})\,u_{l'}(r)\,\frac{\partial}{\partial r}
f_u^\mathrm{KR}(r)\\
&\times
\threej{L}{1}{L}{\li-i}{i}{-\li}
\threej{J_C}{L'}{1}{-M_f+\li}{M_f-M_i-\li}{M_i}\,
\left(1+(-1)^{L}\right)\nonumber\\
&\times\mx{l'\,1\,1\,M_f}
{\sigma_{2\,j}\,\sigma_{1\,i}\,Y_{1\,-j}\,Y_{L\,\li-i}}
{L_C\,1\,J_C\,M_f-\li}\,
\mxred{L_C\,1\,J_C}{Y_{L'}}{l\,1\,1}.\nonumber
\end{align}
Finally, we need $H^{\mathrm{int}\,\mathrm{KR}2}_{s,u}$ in order to calculate 
the amplitudes $\Mfi{\phi\,\mathrm{KR}2}$. We replace $\vec{A}$ by $\Atwo$ in 
Eq.~(\ref{eq:HintKRs}) and find
\ba
H&^{\mathrm{int}\,\mathrm{KR}2}_{s,u}=\sum_{\vec{k},\lambda}
\sum_{L=1}^{\infty}\sum_{M=-L}^L\frac{e\,f^2\,\w}{m_\pi^2}\,\taucrosstau\,
\sqrt{\frac{2\pi\,(2L+1)}{L\,(L+1)}}\,i^{L+1}\,j_L(\frac{\w r}{2})\,
\left[a_{\vec{k},\lambda}\,\delta_{M,\lambda}\right.\nonumber\\
&-
\left.a_{\vec{k},\lambda}^\dagger\,(-1)^{L+\lambda}\,\wignerd{L}{M}{-\lambda}
\right]\left\{
(\vec{\sigma}_1\cdot\vec{x}_{1})\,
(\vec{\sigma}_2\cdot\hat{r})\,Y_{L\,M}(\vec{x}_1)+
(\vec{\sigma}_2\cdot\vec{x}_{2})\,
(\vec{\sigma}_1\cdot\hat{r})\,Y_{L\,M}(\vec{x}_2)\right\}\,
\frac{\partial}{\partial r}f_{s,u}^\mathrm{KR}(r).
\end{align}
We substitute $\vec{x}_1\rightarrow\frac{\vec{r}}{2}=\frac{r}{2}\cdot\hat{r}$,
$\vec{x}_2\rightarrow-\frac{r}{2}\cdot\hat{r}$ and combine the two 
spherical harmonics resulting from 
$(\vec{\sigma}_1\cdot\hat{r})\,(\vec{\sigma}_2\cdot\hat{r})$ via
Eq.~(\ref{eq:additiontheorem}). The result is, again assuming spin 
conservation,
\ba
H^{\mathrm{int}\,\mathrm{KR}2}_{s,u}&=\sum_{\vec{k},\lambda}
\sum_{L=1}^{\infty}\sum_{M=-L}^L\sum_{J=0,2}\sum_{i,j}
\frac{2\pi\,e\,f^2\,\w}{m_\pi^2}\,\taucrosstau\,i^{L+1}\,
\sqrt{\frac{(2L+1)\,(2J+1)}{2\,L\,(L+1)}}\,j_L(\frac{\w r}{2})\nonumber\\
&\times
\left[a_{\vec{k},\lambda}\,\delta_{M,\lambda}
-a_{\vec{k},\lambda}^\dagger\,(-1)^{L+\lambda}\,\wignerd{L}{M}{-\lambda}
\right]\,\threej{1}{1}{J}{0}{0}{0}\,\threej{1}{1}{J}{-i}{-j}{i+j}\nonumber\\
&\times
\sigma_{2\,j}\,\sigma_{1\,i}
\,Y_{J\,-i-j}\,Y_{L\,M}\,\left(1-(-1)^L\right)\,
r\frac{\partial}{\partial r}f_{s,u}^\mathrm{KR}(r).
\end{align}
The amplitudes are identical to Eq.~(\ref{eq:phiKR1diagrams}), except for 
$H^{\mathrm{int}\,\mathrm{KR}1}\rightarrow H^{\mathrm{int}\,\mathrm{KR}2}$.
Therefore we can immediately write down the results:
\ba
M&_{fi}^{\phi\,\mathrm{KR}2\,a}=
\sum_{L=1}^\infty\sum_{L'=1}^\infty
\sum_{J_C=|L'-1|}^{L'+1}\sum_{L_C=|J_C-1|}^{J_C+1}\sum_{J=0,2}\sum_{l,l'}
\sum_{i,j}\frac{4\pi\,e^2\,f^2}{m_\pi^2}\,
\left(\w+\frac{\w^2}{2m_d}\right)\nonumber\\
&\times
(-1)^{L'-\lf-M_f}\,i^{L+L'}\,
\wignerd{L'}{M_f-M_i-\li}{-\lf}\,
\sqrt{\frac{\pi\,(2L+1)\,(2L'+1)\,(2J+1)}{L\,(L+1)\,L'\,(L'+1)}}\nonumber\\
&\times
\doubleint r\,dr\,r'\,dr'\,u_{l'}(r')\,\psi_{L'}(\frac{\w r'}{2})\,\green\,
j_L(\frac{\w r}{2})\,u_l(r)\,r\frac{\partial}{\partial r}f_s^\mathrm{KR}(r)
\nonumber\\
&\times\threej{1}{L'}{J_C}{-M_f}{M_f-M_i-\li}{M_i+\li}
      \threej{1}{1}{J}{0}{0}{0}\, \threej{1}{1}{J}{-i}{-j}{i+j}\,
\left(1-(-1)^L\right)\nonumber\\
&\times\mxred{l'\,1\,1}{Y_{L'}}{L_C\,1\,J_C}\,\mx{L_C\,1\,J_C\,M_i+\li}
{\sigma_{2\,j}\,\sigma_{1\,i}\,Y_{J\,-i-j}\,Y_{L\,\li}}{l\,1\,1\,M_i}\,
\label{eq:MfiphiKR2aresult}
\end{align}
Obviously, $L$ has to be an odd number. $J$ is even, which means $L_C$ has to 
be odd, as well as $L'$. Therefore our prediction of odd angular quantum 
numbers for $\phi$ still holds. 

The remaining three amplitudes are:
\ba
M&_{fi}^{\phi\,\mathrm{KR}2\,b}=
\sum_{L=1}^\infty\sum_{L'=1}^\infty
\sum_{J_C=|L-1|}^{L+1}\sum_{L_C=|J_C-1|}^{J_C+1}\sum_{J=0,2}\sum_{l,l'}
\sum_{i,j}\frac{4\pi\,e^2\,f^2}{m_\pi^2}\,
\left(\w+\frac{\w^2}{2m_d}\right)\nonumber\\
&\times
(-1)^{1+L'-\lf-\li+J_C-M_i}\,i^{L+L'}\,
\wignerd{L'}{M_f-M_i-\li}{-\lf}\,
\sqrt{\frac{\pi\,(2L+1)\,(2L'+1)\,(2J+1)}{L\,(L+1)\,L'\,(L'+1)}}\nonumber\\
&\times
\doubleint r\,dr\,r'\,dr'\,u_{l}(r)\,\psi_{L}(\frac{\w r}{2})\,\green\,
j_{L'}(\frac{\w r'}{2})\,u_{l'}(r')\,r'\frac{\partial}{\partial r'}
f_s^\mathrm{KR}(r')
\nonumber\\
&\times\threej{J_C}{L}{1}{-M_i-\li}{\li}{M_i}
      \threej{1}{1}{J}{0}{0}{0}\, \threej{1}{1}{J}{-i}{-j}{i+j}\,
\left(1-(-1)^{L'}\right)\nonumber\\
&\times\mx{l'\,1\,1\,M_f}
{\sigma_{2\,j}\,\sigma_{1\,i}\,Y_{J\,-i-j}\,Y_{L'\,M_f-M_i-\li}}
{L_C\,1\,J_C\,M_i+\li}\,
\mxred{L_C\,1\,J_C}{Y_{L}}{l\,1\,1}
\end{align}
\ba
M&_{fi}^{\phi\,\mathrm{KR}2\,c}=
\sum_{L=1}^\infty\sum_{L'=1}^\infty
\sum_{J_C=|L-1|}^{L+1}\sum_{L_C=|J_C-1|}^{J_C+1}\sum_{J=0,2}\sum_{l,l'}
\sum_{i,j}\frac{4\pi\,e^2\,f^2}{m_\pi^2}\,
\left(\w-\frac{\w^2}{2m_d}+\frac{\PCsq}{2m_C}\right)\nonumber\\
&\times
(-1)^{1+L'-\lf-M_f}\,i^{L+L'}\,
\wignerd{L'}{M_f-M_i-\li}{-\lf}\,
\sqrt{\frac{\pi\,(2L+1)\,(2L'+1)\,(2J+1)}{L\,(L+1)\,L'\,(L'+1)}}\nonumber\\
&\times
\doubleint r\,dr\,r'\,dr'\,u_{l'}(r)\,\psi_{L}(\frac{\w r}{2})\,\greenpr\,
j_{L'}(\frac{\w r'}{2})\,u_l(r')\,r'\frac{\partial}{\partial r'}
f_u^\mathrm{KR}(r')
\nonumber\\
&\times\threej{1}{L}{J_C}{-M_f}{\li}{M_f-\li}
      \threej{1}{1}{J}{0}{0}{0}\, \threej{1}{1}{J}{-i}{-j}{i+j}\,
\left(1-(-1)^{L'}\right)\nonumber\\
&\times\mxred{l'\,1\,1}{Y_{L}}{L_C\,1\,J_C}\,\mx{L_C\,1\,J_C\,M_f-\li}
{\sigma_{2\,j}\,\sigma_{1\,i}\,Y_{J\,-i-j}\,Y_{L'\,M_f-M_i-\li}}
{l\,1\,1\,M_i}
\end{align}
\ba
M&_{fi}^{\phi\,\mathrm{KR}2\,d}=
\sum_{L=1}^\infty\sum_{L'=1}^\infty
\sum_{J_C=|L'-1|}^{L'+1}\sum_{L_C=|J_C-1|}^{J_C+1}\sum_{J=0,2}\sum_{l,l'}
\sum_{i,j}\frac{4\pi\,e^2\,f^2}{m_\pi^2}\,
\left(\w-\frac{\w^2}{2m_d}+\frac{\PCsq}{2m_C}\right)\nonumber\\
&\times
(-1)^{L'-\lf+\li+J_C-M_f}\,i^{L+L'}\,
\wignerd{L'}{M_f-M_i-\li}{-\lf}\,
\sqrt{\frac{\pi\,(2L+1)\,(2L'+1)\,(2J+1)}{L\,(L+1)\,L'\,(L'+1)}}\nonumber\\
&\times
\doubleint r\,dr\,r'\,dr'\,u_{l}(r')\,\psi_{L'}(\frac{\w r'}{2})\,\greenpr\,
j_{L}(\frac{\w r}{2})\,u_{l'}(r)\,r\frac{\partial}{\partial r}
f_u^\mathrm{KR}(r)
\nonumber\\
&\times\threej{J_C}{L'}{1}{-M_f+\li}{M_f-M_i-\li}{M_i}
      \threej{1}{1}{J}{0}{0}{0}\, \threej{1}{1}{J}{-i}{-j}{i+j}\,
\left(1-(-1)^{L}\right)\nonumber\\
&\times\mx{l'\,1\,1\,M_f}
{\sigma_{2\,j}\,\sigma_{1\,i}\,Y_{J\,-i-j}\,Y_{L\,\li}}
{L_C\,1\,J_C\,M_f-\li}\,
\mxred{L_C\,1\,J_C}{Y_{L'}}{l\,1\,1}
\end{align}

\chapter{Corrections to the Charge Density\label{app:rhoex} }
\markboth{APPENDIX \ref{app:rhoex}. CORRECTIONS TO THE CHARGE DENSITY}
         {APPENDIX \ref{app:rhoex}. CORRECTIONS TO THE CHARGE DENSITY}
In this appendix we give an estimate which shows that we do not need 
to take into account corrections to the charge density due to meson exchange,
cf. Sect.~\ref{sec:dominant}. As we found that only the Kroll-Ruderman current
(Fig.~\ref{fig:mesonexchange}) gives non-negligible contributions, we assume 
that we may also neglect the charge density corresponding to the pion-pole 
current. Nevertheless we briefly demonstrate the small size of this term.

\begin{figure}[!htb]
\begin{center}
\includegraphics*[width=.2\linewidth]{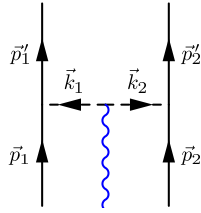}
\caption[Momenta in pion-pole current]
{Momenta in the coupling of a photon to the pion-pole current.}
\label{fig:pionpolelabel}
\end{center}
\end{figure}
The pion-pole current is proportional to $\vec{k}_1-\vec{k}_2$, with 
$\vec{k}_1,\;\vec{k}_2$ the momenta transferred to nucleon~1 and~2, see 
Fig.~\ref{fig:pionpolelabel}.
Therefore, $\rho_\mathrm{ex}^\mathrm{pion-pole}\propto 
k_1^0-k_2^0=p_1'^0-p_1^0-(p_2'^0-p_2^0)=
p_1'^0-p_2'^0-(p_1^0-p_2^0)$ with $p_i^0=\sqrt{m_N^2+\vec{p}_i^{\,2}}\approx
m_N+\frac{\vec{p}_i^{\,2}}{2m_N}$. In the limit of a static deuteron, which 
is assumed throughout the whole work, except for energy denominators, the 
kinetic energies of the two nucleons are equal and $k_1^0-k_2^0$ 
vanishes. We therefore conclude that we only have to take care of 
$\rho_\mathrm{ex}^\mathrm{KR}$, which is the charge density corresponding to 
the Kroll-Ruderman current.

First we derive an explicit expression for $\rho_\mathrm{ex}^\mathrm{KR}$.
The $\gamma\pi N$-vertex is proportional to $\epsilon\cdot S$ with the 
Pauli-Lubanski spin vector defined as, see e.g. \cite{BKM},
\be
S^\mu=\frac{i}{2}\,\gamma_5\,\sigma^{\mu\nu}\,v_{\nu},
\ee
where $\sigma^{\mu\nu}=\frac{i}{2}\,\left[\gamma^\mu,\gamma^\nu\right]$.
We use $\gamma_5=\matrixtwobytwo{0}{1}{1}{0}$ and 
$\sigma^{0i}=i\,\matrixtwobytwo{0}{\sigma^i}{\sigma^i}{0}$, cf. e.g. 
\cite{Itzykson}. The four-velocity $v^\nu$ is \cite{Landau}
\be
v^\nu=\begin{pmatrix}
\frac{1}{\sqrt{1-v^2/c^2}}\\
\frac{\vec{v}}{c\,\sqrt{1-v^2/c^2}}
\end{pmatrix},
\ee
which in the static limit reduces to
\be
v_\nu^\mathrm{stat}=\begin{pmatrix}
1\\
0\\
0\\
0
\end{pmatrix}.
\ee
In this limit we find for the spin vector
\be
S^\mu=\frac{i}{2}\,\gamma_5\,\sigma^{\mu 0}=\frac{1}{2}\,
\begin{pmatrix}
0\\
\matrixtwobytwo{\vec{\sigma}}{0}{0}{\vec{\sigma}}
\end{pmatrix},
\ee
which means that in the non-relativistic reduction we may replace 
$\vec{S}\rightarrow \frac{1}{2}\,\vec{\sigma}$, i.e. 
$\vec{J}^\mathrm{\,KR}\propto\vec{\sigma}$, in agreement with 
Eq.~(\ref{eq:KR}). Our task is therefore to calculate the leading 
contribution to $S^0$. Approximating $\frac{v^2}{c^2}\approx 0$ and returning
to units with $c=1$, we find
\be
S^0=\frac{i}{2}\,\gamma_5\,\sigma^{0\nu}\,v_\nu=
-\frac{i}{2}\,\gamma_5\,\sigma^{0i}\,v_i\approx\frac{1}{2}\,
\matrixtwobytwo{\vec{\sigma}\cdot\vec{v}}{0}{0}{\vec{\sigma}\cdot\vec{v}}.
\ee 
Therefore we have to replace 
\be
S^0\rightarrow\frac{1}{2}\,\vec{\sigma}\cdot\vec{v}=
\frac{1}{2}\,\vec{\sigma}\cdot\frac{\vec{p}}{m_N},
\ee
which leads us to \cite{Ericson}
\be
\rho_\mathrm{ex}^\mathrm{KR}(\vec{k}_1,\vec{k}_2)=-i\,e\,
\left(\frac{4\pi\,f^2}{m_N\,m_\pi^2}\right)\,\taucrosstau\,
\left\{\frac{(\vec{\sigma}_1\cdot\vec{p}_1)\,(\vec{\sigma}_2\cdot\vec{k}_2)}
{\left.\vec{k}_2\!\!\!\right.^2+m_\pi^2}
-      \frac{(\vec{\sigma}_2\cdot\vec{p}_2)\,(\vec{\sigma}_1\cdot\vec{k}_1)}
{\left.\vec{k}_1\!\!\!\right.^2+m_\pi^2}\right\},
\ee
or, in position space, 
\ba
\rho_\mathrm{ex}^\mathrm{KR}(\vec{\xi};\vec{x}_1,\vec{x}_2)&=
\frac{e\,f^2}{m_N\,m_\pi^2}\,\taucrosstau\,
\left\{\delta(\vec{x}_1-\vec{\xi})\,(\vec{\sigma}_2\cdot\hat{r})\,
(\vec{\sigma}_1\cdot\vec{p}_1)\right.\nonumber\\
&+\left.\delta(\vec{x}_2-\vec{\xi})\,(\vec{\sigma}_1\cdot\hat{r})\,
(\vec{\sigma}_2\cdot\vec{p}_2)\right\}\,\frac{\partial}{\partial r}
\frac{\e^{-m_\pi r}}{r}.
\end{align}
We neglect the correction to $\frac{\e^{-m_\pi r}}{r}$ described in 
App.~\ref{app:two-body}, as we only want to give an estimate for the size of 
corrections due to the meson-exchange charge density.

Now we have to think about how to include $\rho^\mathrm{ex}$ into the 
calculation. In the long-wavelength limit, the photon cannot resolve the 
deuteron, i.e. the deuteron appears as a charged, pointlike, static object to 
the photon. Therefore also the nucleons appear to be static in this limit
and Siegert's hypothesis \cite{Siegert} 
guarantees that the $\gamma d$ 
scattering amplitude is described by $\rho^0(\vec{\xi})$ only, whereas 
meson-exchange effects cannot be resolved. Thus $\rho^\mathrm{ex}$ does
not contribute in the static limit. However this does not need to be the case
for non-zero photon energies. In the high-energy regime of our calculation, say
in the order of $\w\sim 100$~MeV, we found the amplitudes containing the 
structure $\mx{d_f}{\hat{\phi}}{C}\mx{C}{\hat{\phi}}{d_i}$ 
to be the dominant terms arising from
$\nab\cdot\vec{J}=-i\,[H,\rho]$ (cf. Sect.~\ref{sec:dominant} and 
App.~\ref{app:dominant}). Therefore we assume that for a numerical estimate of 
the importance of $\rho^\mathrm{ex}$ we may replace 
$\nab\cdot\vec{J}\rightarrow -i\,[H,\rho^0]$ at one interaction vertex and 
$\nab\cdot\vec{J}\rightarrow -i\,[H,\rho^\mathrm{ex}]$ at the other. Repeating
the steps described in Eqs.~(\ref{eq:commutator}-\ref{eq:Mfiphiphiu}) we find
\ba
\Mfi{\rho^\mathrm{ex}\,a}&=
\left(\w+\frac{\w^2}{2m_d}\right)^2\,
\sum_C\frac{\mx{d_f}{\phifhat}{C}
\mx{C}{\int\rho^\mathrm{ex}(\vec{\xi};\vec{x}_1,\vec{x}_2)\,
\phi_i\ofxi\,d^3\xi}{d_i}}{\denoms},\nonumber\\
\Mfi{\rho^\mathrm{ex}\,b}&=
\left(\w+\frac{\w^2}{2m_d}\right)^2\,
\sum_C\frac{\mx{d_f}{\int\rho^\mathrm{ex}(\vec{\xi};\vec{x}_1,\vec{x}_2)\,
\phi_f\ofxi\,d^3\xi}{C}
\mx{C}{\phiihat}{d_i}}{\denoms},\nonumber\\
\Mfi{\rho^\mathrm{ex}\,c}&=
\left(\w-\frac{\w^2}{2m_d}+\frac{\PCsq}{2m_C}\right)^2\,
\sum_C\frac{\mx{d_f}{\phiihat}{C}
\mx{C}{\int\rho^\mathrm{ex}(\vec{\xi};\vec{x}_1,\vec{x}_2)\,
\phi_f\ofxi\,d^3\xi}{d_i}}{\denomu},\nonumber\\
\Mfi{\rho^\mathrm{ex}\,d}&=
\left(\w-\frac{\w^2}{2m_d}+\frac{\PCsq}{2m_C}\right)^2\,
\sum_C\frac{\mx{d_f}{\int\rho^\mathrm{ex}(\vec{\xi};\vec{x}_1,\vec{x}_2)\,
\phi_i\ofxi\,d^3\xi}{C}
\mx{C}{\phifhat}{d_i}}{\denomu}.
\label{eq:rhoexdiagrams}
\end{align}
The index 'KR' has been skipped for convenience as we are only concerned with 
the Kroll-Ruderman charge density. So we need to calculate 
$\int\rho^\mathrm{ex}(\vec{\xi};\vec{x}_1,\vec{x}_2)\,\phi_{i,f}\ofxi\,d^3\xi$,
which is trivial due to the $\delta$-functions in $\rho^\mathrm{ex}$.
\ba
\label{eq:intrhoexphi}
\int\rho^\mathrm{ex}(\vec{\xi};\vec{x}_1,\vec{x}_2)\,\phi_{i,f}\ofxi\,d^3\xi&=
\frac{e\,f^2}{m_N\,m_\pi^2}\,\taucrosstau\,
\left(\frac{\partial}{\partial r}\frac{\e^{-m_\pi r}}{r}\right)\\
&\times
\left\{\phi_{i,f}(\vec{x}_1)\,(\vec{\sigma}_2\cdot\hat{r})\,
(\vec{\sigma}_1\cdot\vec{p}_1)+
       \phi_{i,f}(\vec{x}_2)\,(\vec{\sigma}_1\cdot\hat{r})\,
(\vec{\sigma}_2\cdot\vec{p}_2)\right\}\nonumber
\end{align} 
We now replace 
$\vec{x}_1\rightarrow \frac{\vec{r}}{2},\;
 \vec{x}_2\rightarrow-\frac{\vec{r}}{2},\;
 \vec{p}_1\rightarrow-i\,\nab_1=-i\,\nab_r$ and 
$\vec{p}_2\rightarrow-i\,\nab_2= i\,\nab_r$ with $\nab_r$ acting on the 
deuteron wave function. For our estimate we restrict ourselves to the 
$s$-wave part of this wave function, cf. Eq.~(\ref{eq:deuteronwavefunction}),
which is by far ($\sim95$\%) dominating over the $d$-wave function. 
The reason is that in 
this approximation $\nab_r$ reduces to $\hat{r}\frac{\partial}{\partial r}$, as
the $s$-wave function is spherically symmetric. We therefore may rewrite 
Eq.~(\ref{eq:intrhoexphi}) as
\ba
\int\rho^\mathrm{ex}(\vec{\xi};\vec{x}_1,\vec{x}_2)\,\phi_{i,f}\ofxi\,d^3\xi&=
-\frac{i\,e\,f^2}{m_N\,m_\pi^2}\,\taucrosstau\,
\left(\frac{\partial}{\partial r}\frac{\e^{-m_\pi r}}{r}\right)\\
&\times
(\vec{\sigma}_1\cdot\hat{r})\,(\vec{\sigma}_2\cdot\hat{r})\,
\left(\phi_{i,f}(\vec{r}/2)-\phi_{i,f}(-\vec{r}/2)\right)\,
\frac{\partial}{\partial r}.\nonumber
\end{align} 
Under these assumptions it is easy to evaluate the amplitudes 
(\ref{eq:rhoexdiagrams}). However, we found them to be about a 
factor of 4 smaller than e.g. the amplitudes $\Mfi{\phi\,\mathrm{KR}2}$, which
are small corrections themselves~-- their contributions account for less than 
5\% of the differential cross sections 
for all energies and angles under consideration.
Therefore we claim that
contributions due to $\rho^\mathrm{ex}$ may well be neglected in our 
calculation and thus abstain from showing the final results for the amplitudes 
(\ref{eq:rhoexdiagrams}).

\chapter{The AV18 Neutron-Proton Potential\label{app:AV18} }
\begin{figure}[!htb]
\begin{center} 
\includegraphics*[width=.48\textwidth]{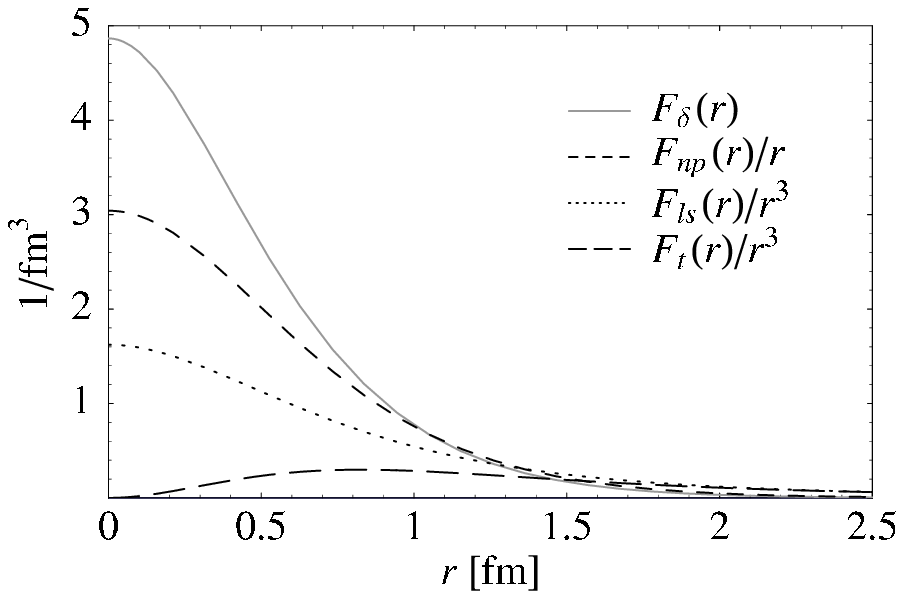}
\hfill
\includegraphics*[width=.48\textwidth]{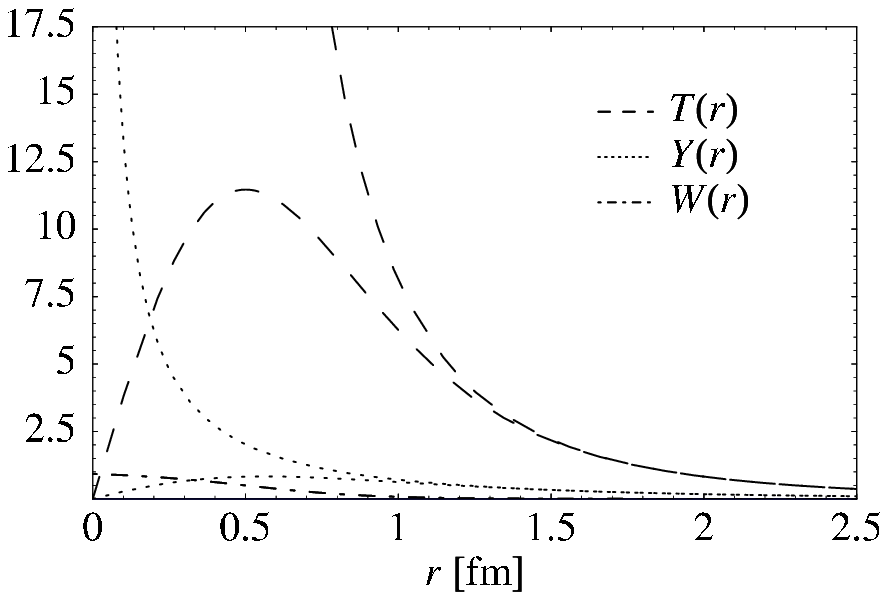}
\caption[Various functions used in the AV18-potential]
{Various functions used in the AV18-potential. Those which appear in 
Eq.~(\ref{eq:Vmm}) are divided by the corresponding
power of $r$. In the divergent functions 
on the right-hand side the cutoff functions, cf. 
Eqs.~(\ref{eq:YmAV18}, \ref{eq:TmAV18}), have been removed.} 
\label{fig:AV18functions}
\end{center}
\end{figure}

\begin{figure}[!htb]
\begin{center} 
\includegraphics[width=.32\linewidth]{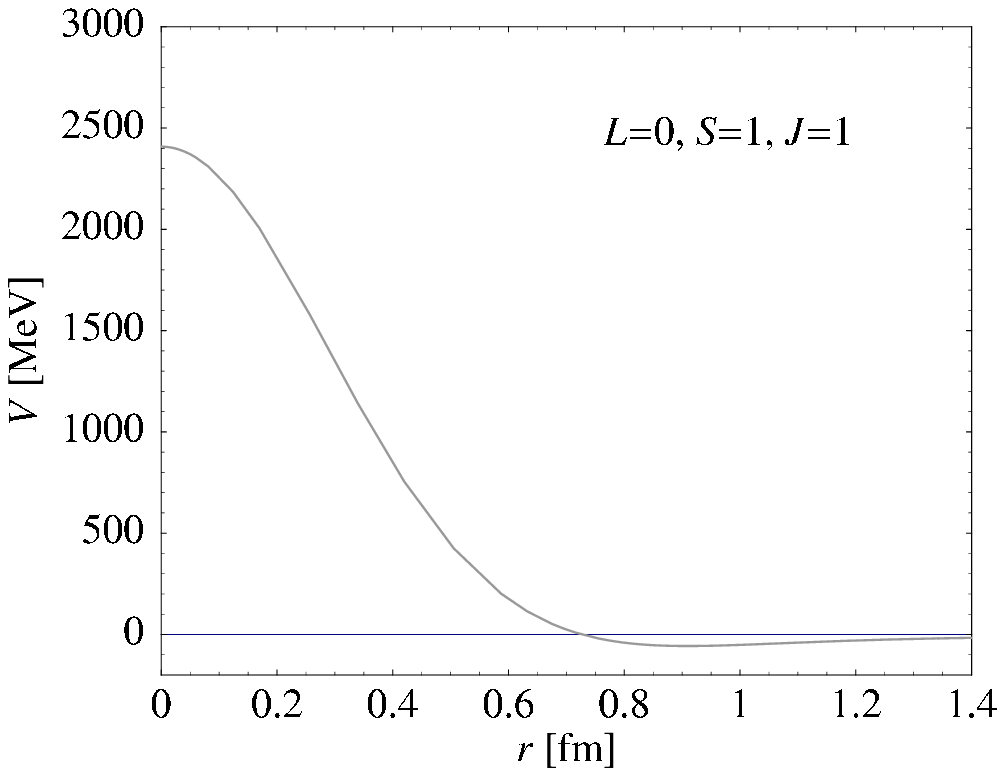}
\includegraphics[width=.32\linewidth]{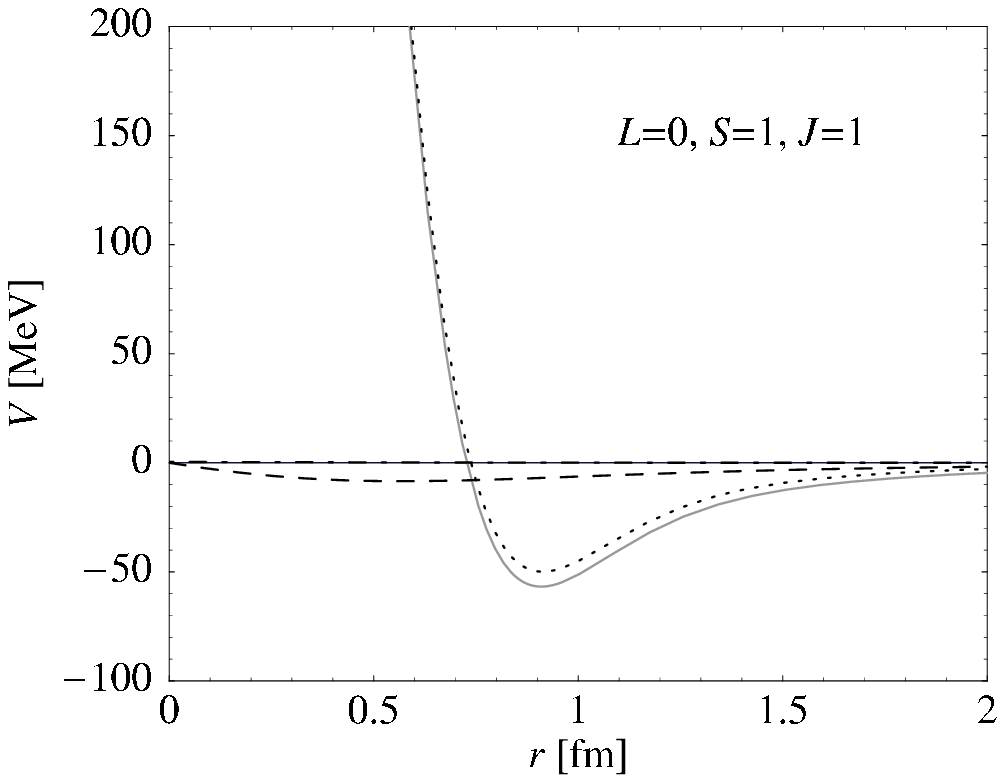}
\includegraphics[width=.32\linewidth]{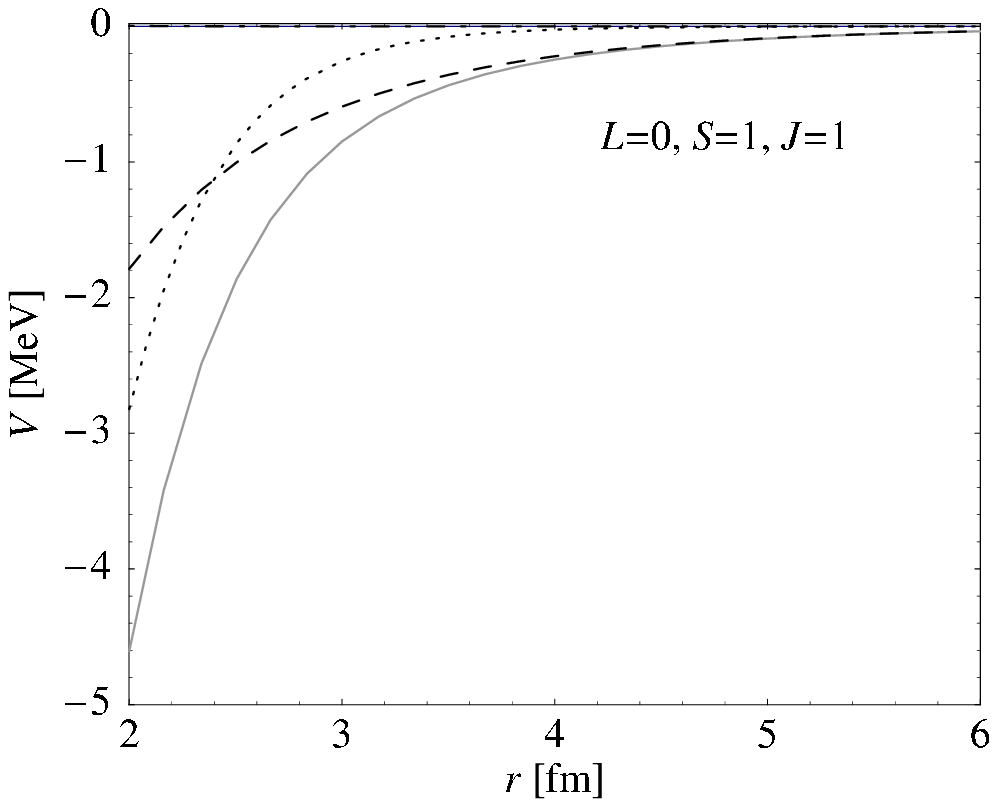}
\\
\includegraphics[width=.32\linewidth]{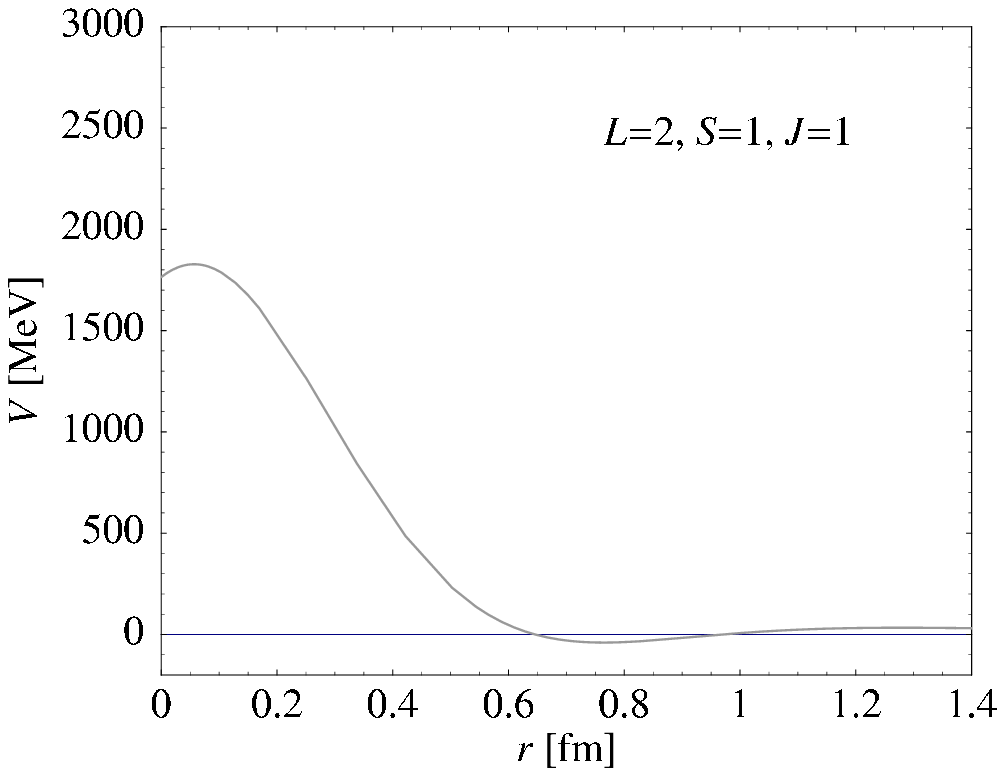}
\includegraphics[width=.32\linewidth]{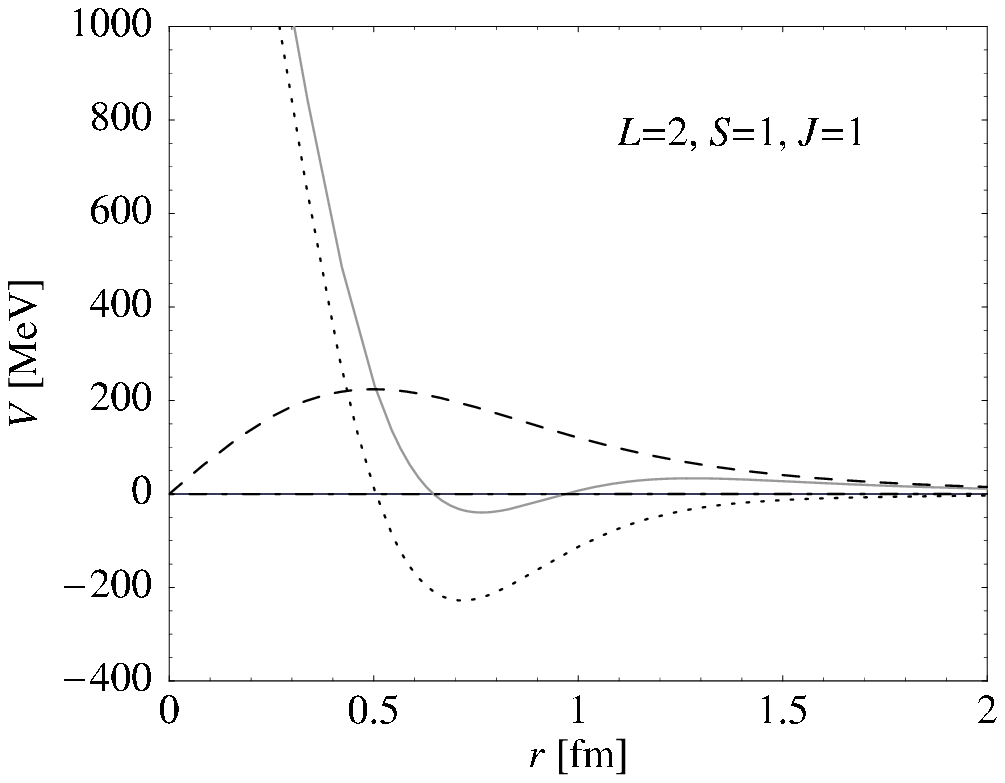}
\includegraphics[width=.32\linewidth]{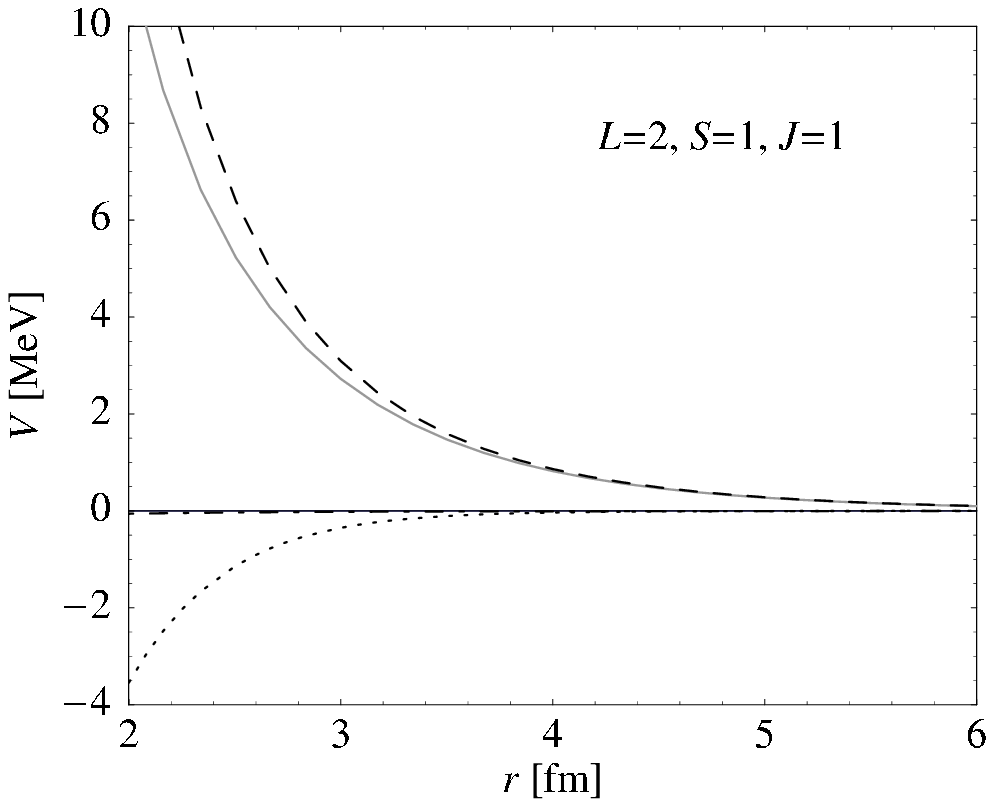}
\\
\includegraphics[width=.32\linewidth]{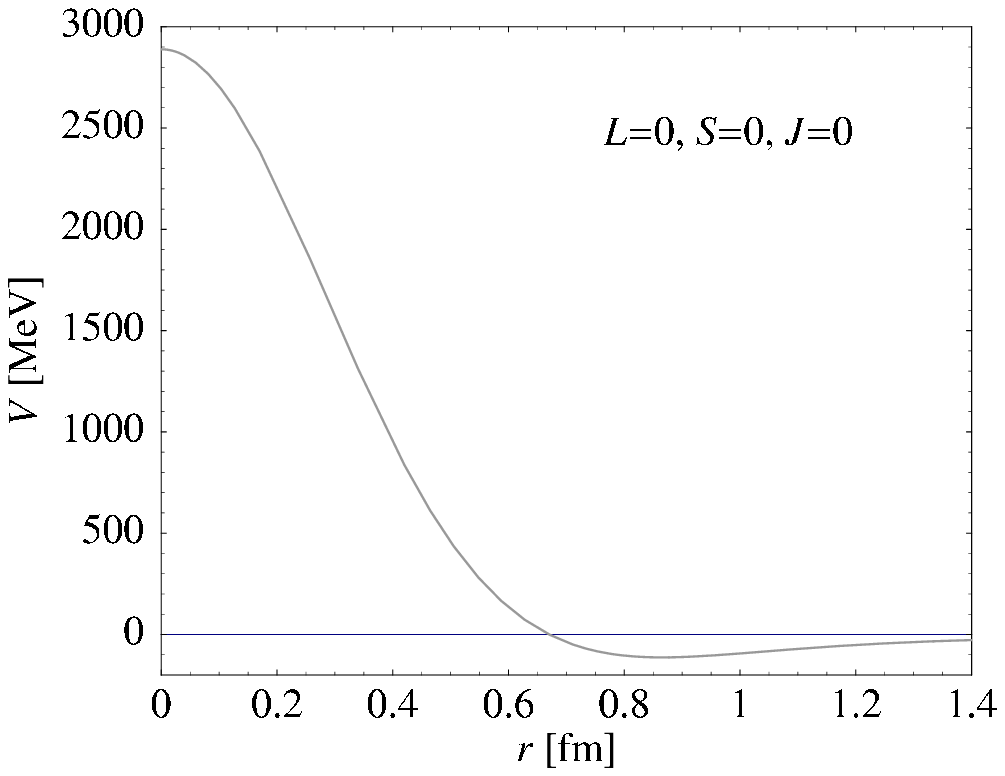}
\includegraphics[width=.32\linewidth]{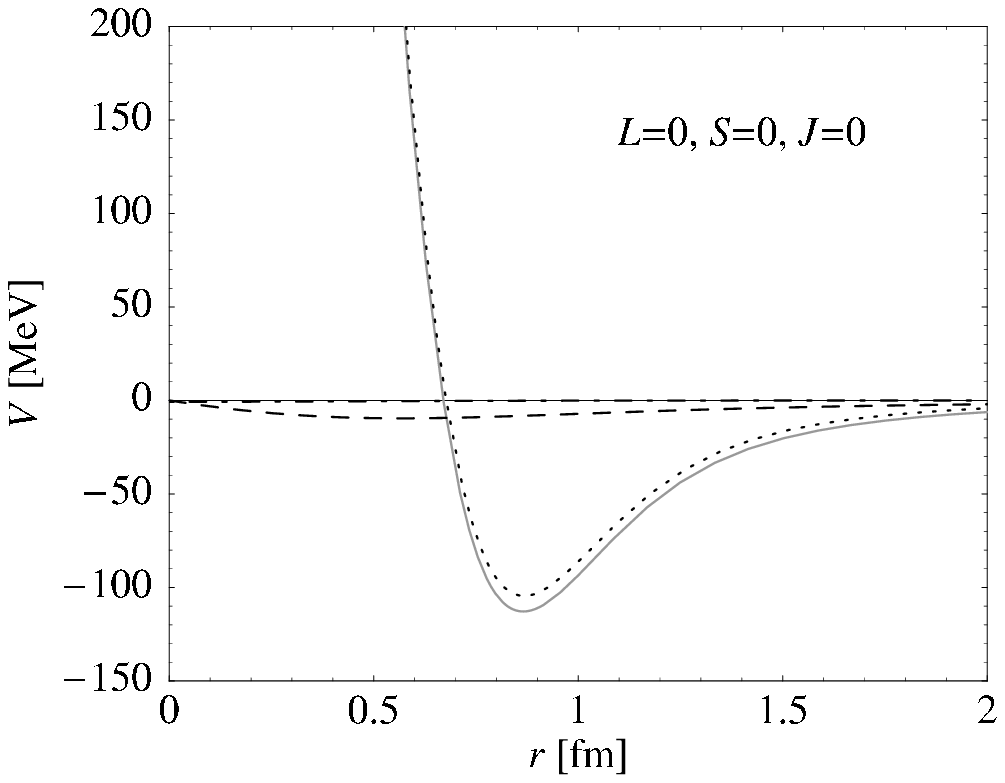}
\includegraphics[width=.32\linewidth]{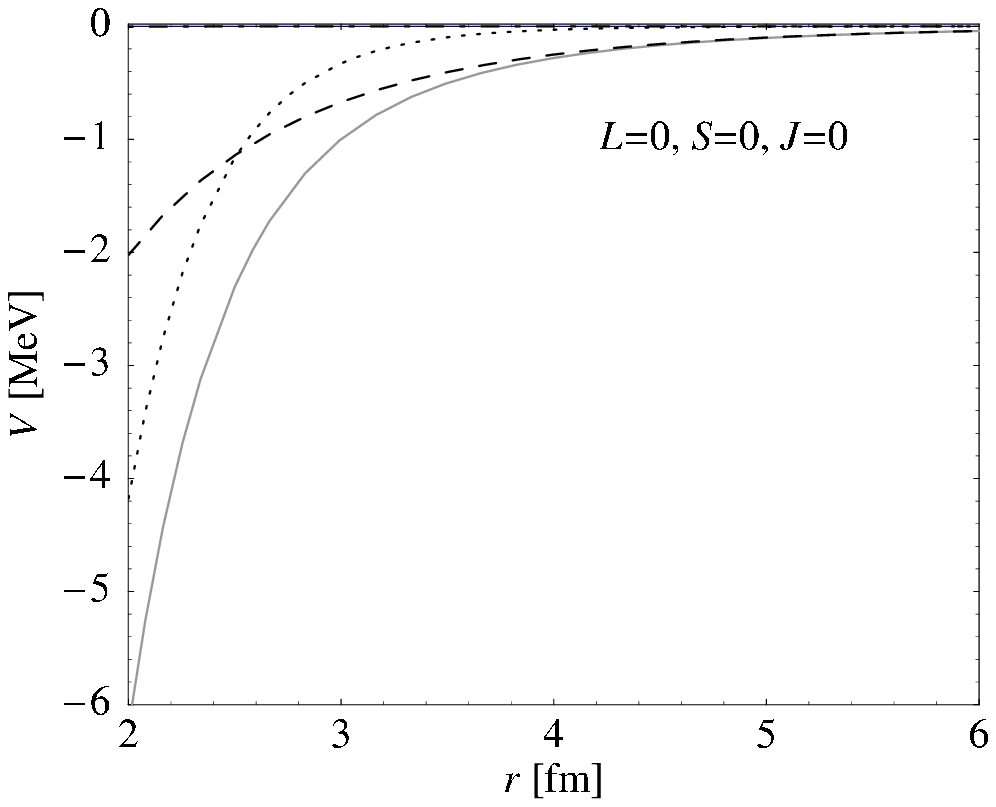}
\\
\includegraphics[width=.32\linewidth]{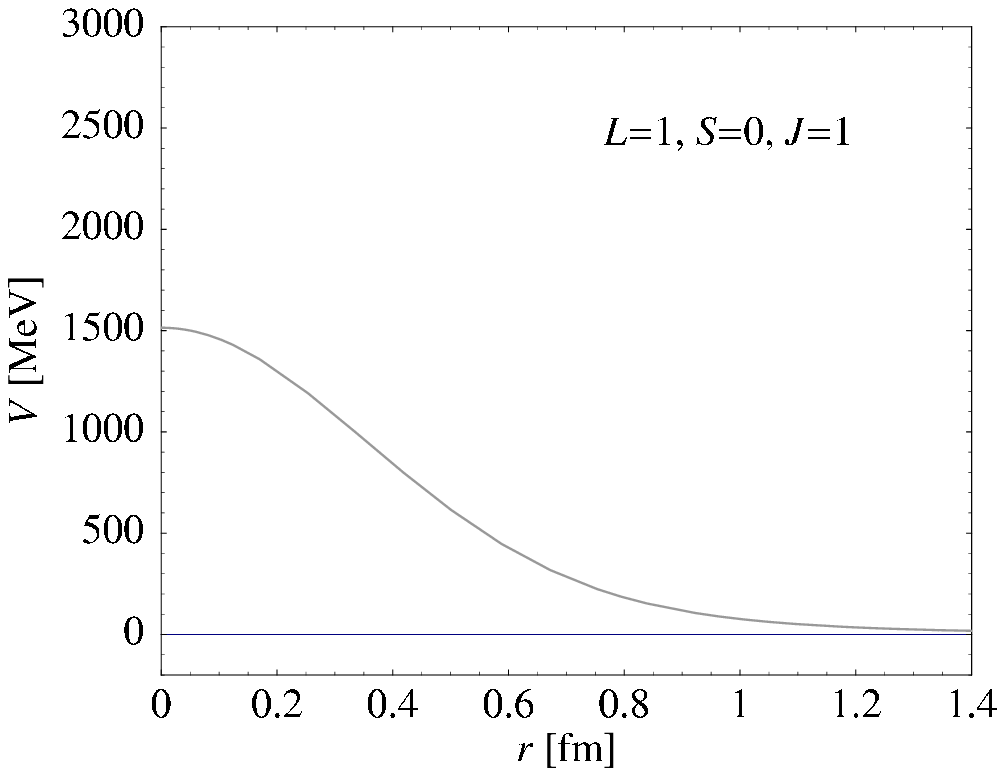}
\includegraphics[width=.32\linewidth]{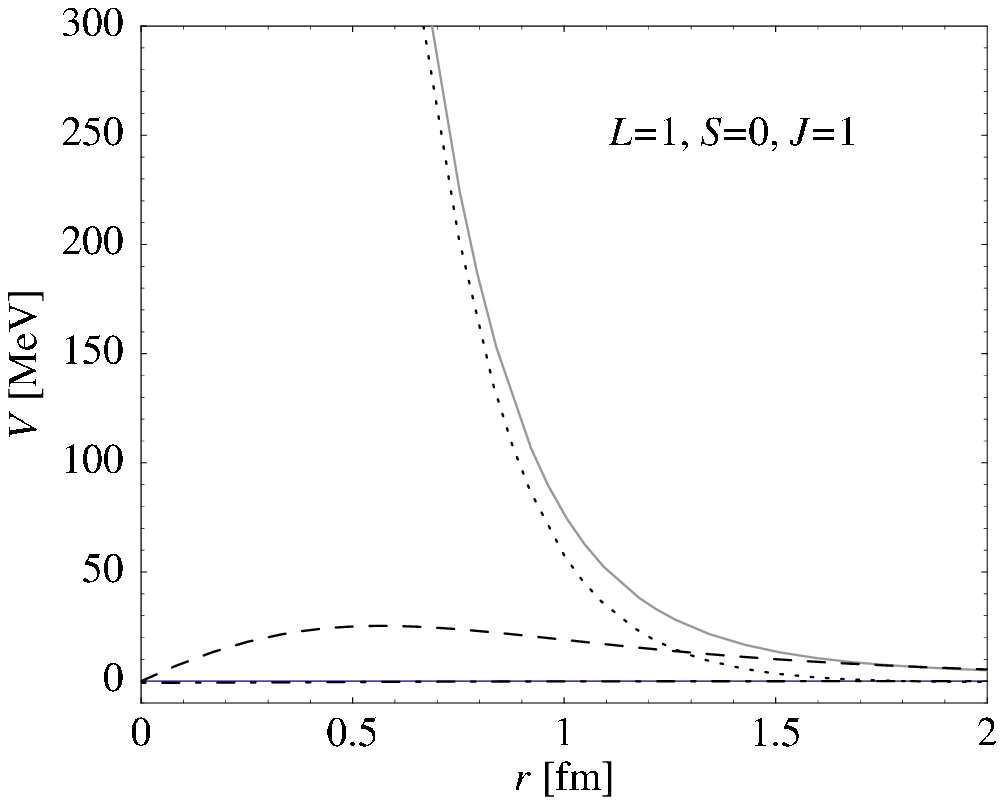}
\includegraphics[width=.32\linewidth]{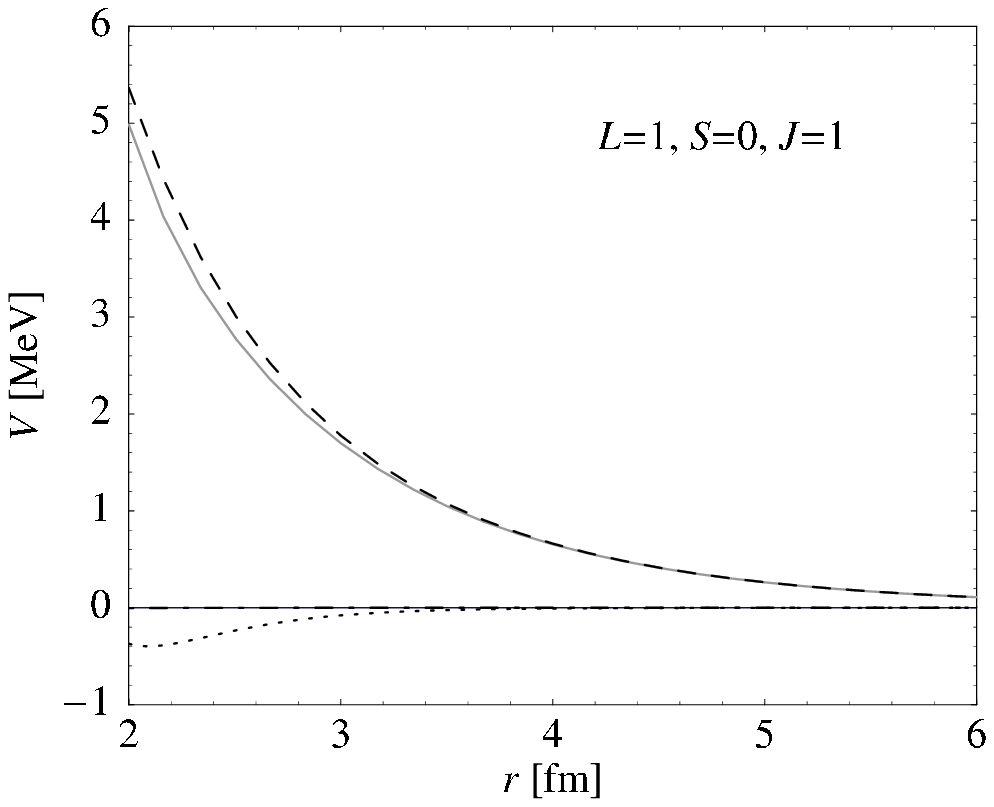}
\caption[Four examples of the AV18-potential]
{The AV18-potential for four different combinations of $L,\;S,\;J$. 
In the left panels we show the full potential, in the middle column we compare 
the full potential (solid) with the three contributions $em$ (dotdashed), 
$\pi$ (dashed) and $sd$ (dotted) at short distances, in the right panel we give
this comparison for larger distances.} 
\label{fig:AV18plots}
\end{center}
\end{figure}

\markboth{APPENDIX \ref{app:AV18}. THE AV18 NEUTRON-PROTON POTENTIAL}
         {APPENDIX \ref{app:AV18}. THE AV18 NEUTRON-PROTON POTENTIAL}
Here we sketch the neutron-proton potential that we use in our 
Schr\"odinger-like differential equation, Eq.~(\ref{eq:diffeq}). It is 
the so-called AV18-potential as published in \cite{AV18}. This potential 
consists of an electromagnetic part ($em$), a one-pion-exchange part ($\pi$), 
and an intermediate- and short-distance part ($sd$):
\be
V(np)=V^{em}(np)+V^\pi(np)+V^{sd}(np)
\label{eq:AV18first}
\ee
The electromagnetic part is very small for the neutron-proton system, and 
therefore nearly invisible in
Fig.~\ref{fig:AV18plots}. Nevertheless, we include it for completeness  
although we do not expect any significant influence on our cross sections.
It is composed of a Coulomb term, arising from the 
neutron charge distribution, and of the interaction between the magnetic 
moments of the two nucleons,
\be
V^{em}(np)=V_C(np)+V_{mm}(np)
\ee
with 
\be
V_C(np)=\alpha\,\beta_n\,\frac{F_{np}(r)}{r}.
\ee
$\alpha$ is the fine-structure constant and $\beta_n$ the slope of the 
neutron electric form factor. The authors of \cite{AV18} use 
$\beta_n=0.0189\;\fm^2$, which is the value determined experimentally in 
\cite{Krohn}. The function $F_{np}(r)$ is defined as
\be
F_{np}(r)=b^2\,\left(15\,br+15\,(br)^2+6\,(br)^3+(br)^4\right)\,
\frac{\e^{-br}}{384}
\ee
with $b=4.27\;\fm^{-1}$  
and is shown in Fig.~\ref{fig:AV18functions}, together
with several other functions contained in the AV18-potential.

The magnetic-moment interaction is given by
\ba
V_{mm}(np)&=-\frac{\alpha}{4\,m_n\,m_p}\,\mu_n\,\mu_p\,
\left[\frac{2}{3}\,F_\delta(r)\,\vec{\sigma}_i\cdot\vec{\sigma}_j+
\frac{F_t(r)}{r^3}\,S_{ij}\right]\nonumber\\
&-\frac{\alpha}{2\,m_n\,m_r}\,\mu_n\,\frac{F_{ls}(r)}{r^3}\,
\left(\vec{L}\cdot\vec{S}+\vec{L}\cdot\vec{t}\right).
\label{eq:Vmm}
\end{align}
$\mu_n,\;\mu_p$ are the magnetic moments of the neutron and proton, 
respectively, $m_n,\;m_p$ their masses. $m_r$ is the reduced nucleon mass
$m_r=\frac{m_n\,m_p}{m_n+m_p}$. 
The spin operators $S_{ij},\;\vec{S}$ and $\vec{t}$ are
\ba
S_{ij}&=3\,(\vec{\sigma}_i\cdot\hat{r})\,(\vec{\sigma}_j\cdot\hat{r})-
\vec{\sigma}_i\cdot\vec{\sigma}_j,\nonumber\\
\vec{S}&=\frac{1}{2}\,(\vec{\sigma}_i+\vec{\sigma}_j),\nonumber\\
\vec{t}&=\frac{1}{2}\,(\vec{\sigma}_i-\vec{\sigma}_j).
\end{align}
However, the operator $\vec{t}$ is designated to be very small, cf. 
\cite{AV18}, and therefore is neglected in our calculation, as also the other 
contributions to $V^{em}(np)$ are tiny, see 
Fig.~\ref{fig:AV18plots}. $\vec{L}=\vec{r}\times\vec{p}$ is the 
orbital angular momentum operator and the functions $F_\delta(r),\;F_t(r)$ and
$F_{ls}(r)$~-- for their derivation cf.~\cite{AV18} and references therein~-- 
are given by 
\ba
F_{\delta}(r)&=b^3\,\left(\frac{1}{16}+\frac{1}{16}\,br+\frac{1}{48}\,(br)^2
\right)\,\e^{-br},\nonumber\\
F_t(r)      &=1-\left(1+br+\frac{1}{2}\,(br)^2+\frac{1}{6}\,(br)^3+
\frac{1}{24}\,(br)^4+\frac{1}{144}\,(br)^5\right)\,\e^{-br},\nonumber\\
F_{ls}(r)  &=1-\left(1+br+\frac{1}{2}\,(br)^2+\frac{7}{48}\,(br)^3+
\frac{1}{48}\,(br)^4\right)\,\e^{-br}
\end{align}
and are shown in Fig.~\ref{fig:AV18functions}, divided by the same power
of $r$ as they appear in Eq.~(\ref{eq:Vmm}). 

The tensor operator $S_{ij}$
only contributes in the triplet state ($S=1$) \cite{Ericson}. The values of 
its matrix elements $\mx{L\,1\,J}{S_{ij}}{L'\,1\,J}$ are given in 
Table~\ref{tab:Sij}, cf. Appendix~10 of~\cite{Ericson}\footnote{We note that 
there is a typo in \cite{Ericson} for $L=L'=J+1$, namely an erroneous factor 
$J$.}, for all possible orbital angular momentum 
numbers $L,\,L'$ at given total angular momentum $J$.

\begin{table}[!htb] 
\begin{center}
\begin{tabular}{|c||c c c|}
\hline 
 &  & $L'$ &  \\
\hline
$L$  & $J+1$ &$J$ & $J-1$ \\
\hline 
\hline 
$J+1$&{\Large$-\frac{2\,(J+2)}{2J+1}$}&0&
{\Large$\frac{6\,\sqrt{J\,(J+1)}}{2J+1}$}\\
&&&\\
$J$  &0&2&0\\
&&&\\
$J-1$&{\Large$\frac{6\,\sqrt{J\,(J+1)}}{2J+1}$}&0&
{\Large$-\frac{2\,(J-1)}{2J+1}$}\\
\hline
\end{tabular}
\end{center}
\caption[Matrix elements of the tensor operator]
{Matrix elements of the tensor operator 
$\mx{L\,1\,J}{S_{ij}}{L'\,1\,J}$.}
\label{tab:Sij}
\end{table}

The eigenvalues of the spin-orbit operator $\vec{L}\cdot\vec{S}$, which 
obviously only contributes in the triplet case due to the total spin 
$\vec{S}$, are given in Table~\ref{tab:LS}, those of 
$\vec{\sigma}_i\cdot\vec{\sigma}_j$ are 
\be
\vec{\sigma}_i\cdot\vec{\sigma}_j\,\ket{S}=2\,\left[S\,(S+1)-3/2\right]\,
\ket{S}.
\ee

\begin{table}[!htb] 
\begin{center}
\begin{tabular}{|c c c|}
\hline 
 & $L$ &  \\
\hline
$J+1$ &$J$ & $J-1$ \\
\hline 
\hline 
$-(L+1)$&$-1$&$L$\\
\hline
\end{tabular}
\end{center}
\caption[Matrix elements of the spin-orbit operator]
{Matrix elements of the spin-orbit operator 
$\mx{L\,1\,J}{\vec{L}\cdot\vec{S}}{L\,1\,J}$.}
\label{tab:LS}
\end{table}

The next component of the potential to be discussed is the one-pion-exchange 
potential, which is the dominant part at large distances, cf. 
Fig.~\ref{fig:AV18plots}. It reads
\be
V^\pi(np)=-f^2\,v_\pi(m_{\pi^0})+(-1)^{T+1}\,2\,f^2\,v_\pi(m_{\pi^\pm}),
\label{eq:VpiAV18}
\ee
where $T$ is the total isospin of the $np$-system, $f$ the pion-nucleon 
coupling ($f^2=0.075$), cf. Section~\ref{sec:dominant} and 
Appendix~\ref{app:parameters}, and
\be
v_\pi(m)=\left(\frac{m}{m_{\pi^\pm}}\right)^2\,\frac{1}{3}\,m\,
\left[Y_m^{AV18}(r)\,\vec{\sigma}_i\cdot\vec{\sigma}_j+T_m^{AV18}(r)\,S_{ij}
\right].
\label{eq:vpiAV18}
\ee
We mark the Yukawa and tensor functions $Y_m(r),\;T_m(r)$ by the index 
$AV18$, as we use slightly different functions $Y(r),\;T(r)$ in 
App.~\ref{app:dominant}. The AV18-functions are defined as
\ba
\label{eq:YmAV18}
Y_m^{AV18}(r)&=\frac{\e^{-mr}}{mr}\,\left(1-\e^{-cr^2}\right),\\
T_m^{AV18}(r)&=\left(1+\frac{3}{mr}+\frac{3}{(mr)^2}\right)\,
\frac{\e^{-mr}}{mr}\,\left(1-\e^{-cr^2}\right)^2
\label{eq:TmAV18}
\end{align}
and are plotted in the right panel of Fig.~\ref{fig:AV18functions}, where we 
set $m\equiv\mu\equiv\frac{1}{3}\,(m_{\pi^0}+2\,m_{\pi^\pm})$, which is the 
average over the pion masses. 
The cutoff parameter $c$ 
is assigned the value $c=2.1\;\fm^{-2}$ 
in \cite{AV18}. The cutoff functions $\left(1-\e^{-cr^2}\right)$, 
$\left(1-\e^{-cr^2}\right)^2$ are introduced in \cite{AV18} in order to avoid 
the $\frac{1}{r^n}$-divergency for $r\rightarrow 0$. However, in 
App.~\ref{app:dominant} we remove the cutoff functions for calculating one 
certain amplitude. The thus truncated functions are also 
displayed in Fig.~\ref{fig:AV18functions}.

The remaining part of the potential is the so-called intermediate- and 
short-distance phenomenological potential ($sd$), which is expressed as a sum 
of central~($c$), $\vec{L}^2$~($l2$), tensor~($t$), spin-orbit~($ls$) and 
quadratic spin-orbit~($ls2$) terms. The parameters of the short-distance 
potential depend on the total spin $S$ and isospin $T$ of the system.
\be
V_{ST}^{sd}(np)=v_{ST}^c(r)+v_{ST}^{l2}(r)\,\Lsq+v_{ST}^t(r)\,S_{ij}+
v_{ST}^{ls}\,\vec{L}\cdot\vec{S}+v_{ST}^{ls2}(r)\,
\left(\vec{L}\cdot\vec{S}\right)^2
\label{eq:Vsd}
\ee
This part of the potential is the dominant one for short distances, see 
Fig.~\ref{fig:AV18plots}. Each of the functions $v^i_{ST}(r)$ in 
Eq.~(\ref{eq:Vsd}) has the general form
\be
v^i_{ST}(r)=I^i_{ST}\,\left(T_\mu^{AV18}(r)\right)^2+
\left[P_{ST}^i+\mu r\,Q_{ST}^i+(\mu r)^2\,R_{ST}^i\right]\,W(r),
\label{eq:ViST}
\ee
where $\mu$ is the averaged pion mass as before. $T_\mu^{AV18}(r)$ is defined 
in Eq.~(\ref{eq:TmAV18}), the Woods-Saxon function 
\be
W(r)=\frac{1}{1+\e^{(r-r_0)/a}}
\ee
is plotted in 
Fig.~\ref{fig:AV18functions}; it dampens out the square bracket in 
Eq.~(\ref{eq:ViST}) for large distances and therefore ensures the short-range 
character of $V^{sd}$.
The parameters of the Woods-Saxon function chosen in \cite{AV18} 
are $r_0=0.5\;\fm$ and $a=0.2\;\fm$. 
The spin- and isospin-dependent parameters $I,\;P,\;Q,\;R$ are listed in 
\cite{AV18}. The eigenvalues of $\Lsq$ are $L\,(L+1)$ and those of 
$(\vec{L}\cdot\vec{S})^2$ are the squares of the numbers given in 
Table~\ref{tab:LS}.

The only remaining subtlety is the question, which combinations of $S,\;T$ 
and $L$ are allowed by Pauli's principle. This principle guarantees that 
the total wave function of the two-nucleon system has to be antisymmetric
under exchange of the two nucleons due to their fermionic nature, i.e. the 
wave function has to fulfill
$(-1)^{S+T+L}=-1$. Therefore, for $S=1,\;T=0$, which is the deuteron case, 
and for $S=0,\;T=1$ we find $L$ even, for $S=0,\;T=0$ and $S=1,\;T=1$ we 
have $L$ odd. 

Four examples of the 
resulting potentials for the various combinations of $L,\;S,\;J$ are shown in 
the first column of Fig.~\ref{fig:AV18plots}, 
together with a comparison of the full potential with  
the three parts separated from each other, cf. Eq.~(\ref{eq:AV18first}). 
This comparison is done at short distances (middle column) and at larger 
distances (right column). At short distances the contribution $sd$ (dotted) 
dominates, whereas above about $4\;\fm$ the potential is made up nearly 
entirely by the one-pion-exchange (dashed) potential. The 
electromagnetic part (dotdashed) is negligible in all cases 
considered. It only gives sizeable contributions when we look at the 
proton-proton rather than the proton-neutron interaction \cite{AV18}. 
The first row in Fig.~\ref{fig:AV18plots} corresponds to the deuteron 
$s$-state, the second to the deuteron $d$-state, which are both triplet 
($S=1$) states. The lower two rows show singlet states.

\chapter{Formulae \label{app:formulae} }
\markboth{APPENDIX \ref{app:formulae}. FORMULAE}
         {APPENDIX \ref{app:formulae}. FORMULAE}
Photon field $\vec{A}(\vec{x})$ and  (charged) pion field $\phi(\vec{x})$, 
normalized according to Ref.~\cite{Karakowski}:
\ba
\label{eq:photonfield}
\vec{A}(\vec{x})&=\sum_{\vec{k},\lambda}
\left(a_{\vec{k}}\,\hat{\epsilon}_\lambda\,\e^{i\vec{k}\cdot\vec{x}}+
      a_{\vec{k}}^\dagger\,\hat{\epsilon}_\lambda^\ast\,
\e^{-i\vec{k}\cdot\vec{x}}
\right)\\
\label{eq:pionfield}
\phi_\pm(\vec{x})&=\sum_{\vec{q}}\sqrt{\frac{2\pi}{E_\pi}}
\left(a_{\mp,\vec{q}}\,\e^{i\vec{q}\cdot\vec{x}}+
      a_{\pm,\vec{q}}^\dagger\,\e^{-i\vec{q}\cdot\vec{x}}
\right)
\end{align}
$a_\pm^\dagger$ ($a_\pm$) creates (annihilates) a $\pi^\pm$,
$E_\pi=\sqrt{\vec{q}^{\,2}+m_\pi^2}$.
\\ 
\\
Hamiltonians for $\pi N$ and $\gamma\pi N$ coupling, cf. e.g.~\cite{Ericson}, 
Section~8.2:
\ba
\label{eq:HpiNN}
H^{\pi N N}&=-\frac{f}{m_\pi}\,(\vec{\sigma}\cdot\nab)\,
(\vec{\tau}\cdot\vec{\phi}(\vec{x}))\\
\label{eq:HgammapiN}
H^{\gamma\pi N}&=\frac{e\,f}{m_\pi}\,(\vec{\sigma}\cdot\vec{A}(\vec{x}))\,
(\vec{\tau}\times\vec{\phi}(\vec{x}))_z
\end{align}
Definitions:

\be
\parbox{4cm}{
\ba
\phi_\pm&= \frac{1}{\sqrt{2}}\,(\phi_x\pm i\,\phi_y) \nonumber\\
&\nonumber\\
\Rightarrow\qquad\phi_x  &= \frac{1}{\sqrt{2}}\,(\phi_+ +\phi_-) \nonumber\\
\phi_y  &=-\frac{i}{\sqrt{2}}\,(\phi_+ -\phi_-) \nonumber
\end{align}}
\qquad
\parbox{4cm}{
\ba
\label{eq:taupm}
\tau_\pm&= \tau_x\pm i\,\tau_y\phantom{\frac{1}{\sqrt{2}}}\nonumber\\
&\nonumber\\
\tau_x  &= \frac{1}{2}\,(\tau_+ +\tau_-)\phantom{\frac{1}{\sqrt{2}}}\nonumber\\
\tau_y  &=-\frac{i}{2}\,(\tau_+ -\tau_-)\phantom{\frac{1}{\sqrt{2}}} \nonumber
\end{align}}
\ee
\ba
\Rightarrow\qquad\vec{\tau}\cdot\vec{\phi}&=\tau_x\,\phi_x+\tau_y\,\phi_y=
\frac{1}{\sqrt{2}}\,(\tau_+\,\phi_-+\tau_-\,\phi_+)\\
(\vec{\tau}\times\vec{\phi})_z&=\tau_x\,\phi_y-\tau_y\,\phi_x=
\frac{i}{\sqrt{2}}\,(\tau_+\,\phi_--\tau_-\,\phi_+)
\label{eq:taucrossphi}
\end{align}

\vspace{.2cm}
\noindent
The following formulae can all be found in textbooks on the quantum theory
of angular momentum, e.g. \cite{VMK,Edmonds,Rose}.
\\
\\
The spherical basis:
\ba
\label{eq:sphericalbasis}
\hat{r}_{+1}&=-\frac{1}{\sqrt{2}}\,(\vec{e}_x+i\,\vec{e}_y)\nonumber\\
\hat{r}_{0} &=\vec{e}_z\nonumber\\
\hat{r}_{-1}&= \frac{1}{\sqrt{2}}\,(\vec{e}_x-i\,\vec{e}_y)\\
&\nonumber\\
\Rightarrow \hat{r}_i\cdot\hat{r}_j&=(-1)^i\,\delta_{i,-j}
\end{align}
Polarization vectors in the spherical basis:

\be
\parbox{1.5cm}{
\ba
\hat{\epsilon}_{+1}&=\hat{r}_{+1}=
-\frac{1}{\sqrt{2}}\,
\begin{pmatrix}
1\\
i\\
0
\end{pmatrix}
\nonumber\\
\hat{\epsilon}_{-1}&=\hat{r}_{-1}= 
\frac{1}{\sqrt{2}}\,
\begin{pmatrix}
1\\
-i\\
0
\end{pmatrix}
\nonumber
\end{align}}
\qquad\qquad
\parbox{1.5cm}{
\ba
\label{eq:spherpolarizations}
\hat{\epsilon}_{+1}'^\ast&=
\frac{1}{\sqrt{2}}\, 
\begin{pmatrix}
-\cos\theta\\
i\\
\sin\theta
\end{pmatrix}
\nonumber\\
\hat{\epsilon}_{-1}'^\ast&=
\frac{1}{\sqrt{2}}\,
\begin{pmatrix}
\cos\theta\\
i\\
-\sin\theta
\end{pmatrix}\nonumber
\end{align}}
\ee
Expansion of a vector $\vec{V}$ into spherical components:
\be
\vec{V}=\sum_{\nu=-1,0,1}(-1)^\nu\,V_\nu\,\hat{r}_{-\nu}\;;\qquad\qquad
V_\nu=\vec{V}\cdot\hat{r}_\nu
\label{eq:spherdecomposition}
\ee
Scalar product in spherical coordinates:
\be
\vec{U}\cdot\vec{V}=\sum_{j=-1,0,1}(-1)^j\,U_j\,V_{-j}
\label{eq:scalarproductspherical}
\ee
Useful identity for $r_\nu$:
\be
r_\nu=\hat{r}\cdot\hat{r}_\nu=\sqrt{\frac{4\pi}{3}}\,Y_{1\,\nu}(\hat{r})
\label{eq:usefulidentity}
\ee
\vspace{.2cm}

\noindent
Clebsch-Gordan coefficients:
\be
C_{j_1\,m_1\;j_2\,m_2}^{j_3\,m_3}=(-1)^{j_1-j_2+m_3}\,\sqrt{2j_3+1}\,
\threej{j_1}{j_2}{j_3}{m_1}{m_2}{-m_3}
\label{eq:CGdefinition}
\ee
$\threej{j_1}{j_2}{j_3}{m_1}{m_2}{-m_3}$ is the Wigner 3-$j$~symbol and the 
angular momenta $j_1,\;j_2,\;j_3$ have to fulfill the triangular condition
\be
j_1\leq j_2+j_3,\qquad j_2\leq j_1+j_3,\qquad j_3\leq j_1+j_2.
\label{eq:triangularcondition}
\ee
The projections on the quantization axis have to add up to zero:
\be
m_1+m_2-m_3=0
\label{eq:projections}
\ee
Further properties of the 3-$j$~symbol:\\
The 3-$j$~symbol is symmetric under even permutations of rows, 
\be
\threej{j_1}{j_2}{j_3}{m_1}{m_2}{m_3}=\threej{j_2}{j_3}{j_1}{m_2}{m_3}{m_1}=
\threej{j_3}{j_1}{j_2}{m_3}{m_1}{m_2}.
\label{eq:threejprops1}
\ee
Under odd permutations it behaves like
\be
\threej{j_1}{j_2}{j_3}{m_1}{m_2}{m_3}=(-1)^{j_1+j_2+j_3}\,
\threej{j_2}{j_1}{j_3}{m_2}{m_1}{m_3}.
\label{eq:threejprops2}
\ee
Further:
\ba
\label{eq:threejprops3}
\threej{j_1}{j_2}{j_3}{m_1}{m_2}{m_3}&=(-1)^{j_1+j_2+j_3}\,
\threej{j_1}{j_2}{j_3}{-m_1}{-m_2}{-m_3}\\
\label{eq:threejprops4}
\Rightarrow\qquad
\threej{j_1}{j_2}{j_3}{0}{0}{0}&=(-1)^{j_1+j_2+j_3}\,
\threej{j_1}{j_2}{j_3}{0}{0}{0}\\
\label{eq:threejprops5}
\Rightarrow\qquad
\threej{j_1}{j_2}{j_3}{0}{0}{0}&=0\;\,\mathrm{for}\;j_1+j_2+j_3\;
\mathrm{odd}
\end{align}
Unitarity of the Clebsch-Gordan-coefficients:
\ba
\sum_{j,m}C_{j_1\,m_1'\;j_2\,m_2'}^{j\,m}\,C_{j_1\,m_1\;j_2\,m_2}^{j\,m}&=
\delta_{m_1',m_1}\,\delta_{m_2',m_2}\nonumber\\
\sum_{m_1,m_2}C_{j_1\,m_1\;j_2\,m_2}^{j'\,m'}\,C_{j_1\,m_1\;j_2\,m_2}^{j\,m}&=
\delta_{j',j}\,\delta_{m',m}
\label{eq:threejunitarity}
\end{align}
\vspace{.2cm}

\noindent
Properties of the 6-$j$ and 9-$j$~symbols:\\
In a non-vanishing 6-$j$~symbol $\sixj{j_1}{j_2}{j_3}{l_1}{l_2}{l_3}$ 
the four triads $(j_1\,j_2\,j_3),$ $(l_1\,l_2\,j_3),$ $(j_1\,l_2\,l_3),$
$(l_1\,j_2\,l_3)$ fulfill the triangular condition 
Eq.~(\ref{eq:triangularcondition}).\\
The 9-$j$~symbol can be written in terms of the 6-$j$~symbols \cite{Rotenberg}:
\ba
\label{eq:ninejdefinition}
\ninej{j_{11}}{j_{12}}{j_{13}}{j_{21}}{j_{22}}{j_{23}}{j_{31}}{j_{32}}{j_{33}}
&=\sum_j(-1)^{2j}\,(2j+1)\\
&\times\sixj{j_{11}}{j_{21}}{j_{31}}{j_{32}}{j_{33}}{j}\,
\sixj{j_{12}}{j_{22}}{j_{32}}{j_{21}}{j}{j_{23}}\,
\sixj{j_{13}}{j_{23}}{j_{33}}{j}{j_{11}}{j_{12}}\nonumber
\end{align}
The upper limit for $j$ is determined by the triangular condition on the triads
of the various 6-$j$~symbols:
\be
j\leq \mathrm{Min}[j_{11}+j_{33},\,j_{21}+j_{32},\,j_{12}+j_{23}]
\ee
We calculate 9-$j$~symbols via Eq.~(\ref{eq:ninejdefinition}), 3-$j$ and 
6-$j$~symbols are standard functions in programs like $Mathematica$.\\
The 9-$j$~symbol is symmetric under even permutations of rows or columns.
Any 9-$j$~symbol with one null entry can be simplified using
\be
\ninej{L_f}{S_f}{J_f}{L_i}{S_i}{J_i}{L}{0}{L'}=
\delta_{S_f,S_i}\,\delta_{L,L'}\,\frac{(-1)^{L_f+S_f+J_i+L'}}
{\sqrt{(2S_i+1)\,(2L+1)}}\,\sixj{L_i}{J_i}{S_i}{J_f}{L_f}{L}.
\label{eq:ninejzero}
\ee
\vspace{.2cm}

\noindent
Product of two spherical harmonics:
\ba
Y_{j\,m}\,Y_{j'\,m'}&=\sum_{J}(-1)^{m+m'}
\sqrt{\frac{(2j+1)\,(2j'+1)\,(2J+1)}{4\pi}}\nonumber\\
&\times
\threej{j}{j'}{J}{0}{0}{0}\,\threej{j}{j'}{J}{m}{m'}{-m-m'}\,Y_{J\,m+m'}
\label{eq:additiontheorem}
\end{align}
Some properties of the spherical harmonics:
\ba
\label{eq:Yofmr}
Y_{l\,m}(-\hat{r}) &=(-1)^l\,Y_{l\,m}(\hat{r})\\
\label{eq:Yofez}
Y_{l\,m}(\vec{e}_z)&=\delta_{m,0}\sqrt{\frac{2l+1}{4\pi}}
\end{align}
Definition of the spherical Bessel functions (we use the same convention as 
e.g. Ref.~\cite{Ericson}): 
\be
j_l(z)=\sqrt{\frac{\pi}{2z}}J_{l+\frac{1}{2}}(z)
\label{eq:spherBesseldefinition}
\ee
$J_l(z)$ is the Bessel function of the first kind.\\
Recursion relations for the spherical Bessel functions:
\ba
\frac{2l+1}{x}\,j_l(x)&=j_{l-1}(x)+j_{l+1}(x)\nonumber\\
(2l+1)\,j_l'(x)&=l\,j_{l-1}(x)-(l+1)\,j_{l+1}(x)
\label{eq:recursionrelations}
\end{align}
Wigner-Eckart theorem:
\be
\mx{J_f\,M_f}{T_{J\,M}}{J_i\,M_i}=(-1)^{J_f-M_f}\,
\threej{J_f}{J}{J_i}{-M_f}{M}{M_i}\,\mxred{J_f}{T_J}{J_i}
\label{eq:WE}
\ee
The matrix element on the right-hand side is called ``reduced matrix element''.
\\
\\
Tensor product ($\otimes$) definition:
\be
[T_{k_1}\otimes T_{k_2}]_{k\,m}=\sum_{m_1,m_2}\sqrt{2k+1}\,(-1)^{-k_1+k_2-m}\,
\threej{k_1}{k_2}{k}{m_1}{m_2}{-m}\,T_{k_1\,m_1}\,T_{k_2\,m_2}
\label{eq:tensorproductdef}
\ee
Inversion:
\be
T_{k_1\,m_1}\,T_{k_2\,m_2}=\sum_{k,m}(-1)^{k_1-k_2+m}\,\sqrt{2k+1}\,
\threej{k_1}{k_2}{k}{m_1}{m_2}{-m}\,[T_{k_1}\otimes T_{k_2}]_{k\,m}
\label{eq:tensorproductinversion}
\ee
Reduced matrix elements:
\ba
\label{eq:mxYshort}
\mxred{l'}{Y_L}{l}&=(-1)^{l'}\,\sqrt{\frac{(2l'+1)\,(2L+1)\,(2l+1)}{4\pi}}\,
\threej{l'}{L}{l}{0}{0}{0}\\
\label{eq:mxSshort}
\mxred{S_f}{S}{S_i}&=\sqrt{6}\,\delta_{S_i,1}\,\delta_{S_f,1}\\
\label{eq:mxtshort}
\mxred{S_f}{t}{S_i}&=\sqrt{3}\,(\delta_{S_i,0}\,\delta_{S_f,1}-
                                 \delta_{S_i,1}\,\delta_{S_f,0})
\end{align}
\ba
\label{eq:mxY}
\mxred{L_f\,S_f\,J_f}{Y_L}{L_i\,S_i\,J_i}&=(2L+1)\,
\sqrt{\frac{(2L_i+1)\,(2L_f+1)\,(2J_i+1)\,(2J_f+1)\,(2S_f+1)}{4\pi}}\nonumber\\
&\times(-1)^{L_f}\,\delta_{S_f,S_i}\,
\ninej{L_f}{S_f}{J_f}{L_i}{S_i}{J_i}{L}{0}{L}\,\threej{L_f}{L}{L_i}{0}{0}{0}
\end{align}
\ba
&
\mxred{L_f\,S_f\,J_f}{[Y_{L'}\otimes S]_L}{L_i\,S_i\,J_i}=
\sqrt{\frac{3\,(2L_i+1)\,(2L_f+1)\,(2J_i+1)\,(2J_f+1)}{2\pi}}\nonumber\\
&\qquad\times\sqrt{(2L+1)\,(2L'+1)}\,(-1)^{L_f}\,\delta_{S_f,S_i}\,
\delta_{S_i,1}\,
\ninej{L_f}{S_f}{J_f}{L_i}{S_i}{J_i}{L'}{1}{L}\,\threej{L_f}{L'}{L_i}{0}{0}{0}
\label{eq:mxL}
\end{align}
\ba
&
\mxred{L_f\,S_f\,J_f}{[Y_{L'}\otimes t]_L}{L_i\,S_i\,J_i}=
\sqrt{\frac{3\,(2L_i+1)\,(2L_f+1)\,(2J_i+1)\,(2J_f+1)}{16\pi}}\nonumber\\
&\qquad\times\sqrt{(2L+1)\,(2L'+1)}\,(-1)^{L_f}\,[(-1)^{S_i}-(-1)^{S_f}]\,
\ninej{L_f}{S_f}{J_f}{L_i}{S_i}{J_i}{L'}{1}{L}\,\threej{L_f}{L'}{L_i}{0}{0}{0}
\label{eq:mxt}
\end{align}
$S$ and $t$ are related to 
$\vec{S}=\frac{\vec{\sigma}_1+\vec{\sigma}_2}{2}$, 
$\vec{t}=\frac{\vec{\sigma}_1-\vec{\sigma}_2}{2}$; 
$\vec{L}$ is the angular momentum operator.\\
\\
Formulae for uncoupling tensor products: 
\ba
\label{eq:uncoupling1}
\mxred{j_1\,j_2\,j}{[T_{k_1}\otimes T_{k_2}]_{k}}{j_1'\,j_2'\,j'}&=
\sqrt{(2j+1)\,(2j'+1)\,(2k+1)}\\
&\times
\ninej{j_1}{j_2}{j}{j_1'}{j_2'}{j'}{k_1}{k_2}{k}\,\mxred{j_1}{T_{k_1}}{j_1'}\,
\mxred{j_2}{T_{k_2}}{j_2'}\nonumber\\
\label{eq:uncoupling2}
\mxred{j}{[T_{k_1}\otimes T_{k_2}]_{k}}{j'}&=
(-1)^{k+j+j'}\,\sqrt{2k+1}\\
&\times
\sum_{J}\sixj{k_1}{k_2}{k}{j'}{j}{J}\,\mxred{j}{T_{k_1}}{J}\,
\mxred{J}{T_{k_2}}{j'}\nonumber
\end{align}
Definition of the vector spherical harmonic:
\be
\vsh{J}{L}{M}=\sum_{\nu=-1,0,1}(-1)^{-L-M+1}\,\sqrt{2J+1}\,
\threej{L}{1}{J}{M+\nu}{-\nu}{-M}\,\hat{r}_{-\nu}\,Y_{L\,M+\nu}(\hat{r})
\label{eq:vshdefinition}
\ee
The gradient formula:
\ba
\label{eq:gradientformula}
\nab\,f(r)\,Y_{L\,M}(\hat{r})&=-\sqrt{\frac{L+1}{2L+1}}\,
\left(\frac{\partial}{\partial r}-\frac{L}{r}\right)\,f(r)\,\vsh{L}{L+1}{M}
\nonumber\\
&+\sqrt{\frac{L}{2L+1}}\,
\left(\frac{\partial}{\partial r}+\frac{L+1}{r}\right)\,f(r)\,
\vsh{L}{L-1}{M}
\end{align}
Curl version of the gradient formula:
\ba
\label{eq:curlL}
\nab\times\left[f(r)\,\vsh{L}{L}{M}\right]&=
i\,\left(\frac{\partial}{\partial r}-\frac{L}{r}\right)\,f(r)\,
\sqrt{\frac{L}{2L+1}}\,\vsh{L}{L+1}{M}\\
&+
i\,\left(\frac{\partial}{\partial r}+\frac{L+1}{r}\right)\,f(r)\,
\sqrt{\frac{L+1}{2L+1}}\,\vsh{L}{L-1}{M}\nonumber\\
\label{eq:curlLp1}
\nab\times\left[f(r)\,\vsh{L}{L+1}{M}\right]&=
i\,\left(\frac{\partial}{\partial r}+\frac{L+2}{r}\right)\,f(r)\,
\sqrt{\frac{L}{2L+1}}\,\vsh{L}{L}{M}\\
\label{eq:curlLm1}
\nab\times\left[f(r)\,\vsh{L}{L-1}{M}\right]&=
i\,\left(\frac{\partial}{\partial r}-\frac{L-1}{r}\right)\,f(r)\,
\sqrt{\frac{L+1}{2L+1}}\,\vsh{L}{L}{M}
\end{align}
Divergence version of the gradient formula:
\ba
\label{eq:divergenceL}
\nab\cdot\left[f(r)\,\vsh{L}{L}{M}\right]&=0\\
\label{eq:divergenceLp1}
\nab\cdot\left[f(r)\,\vsh{L}{L+1}{M}\right]&=-\sqrt{\frac{L+1}{2L+1}}\,
\left(\frac{\partial}{\partial r}+\frac{L+2}{r}\right)\,f(r)\,Y_{L\,M}(\hat{r})
\\
\label{eq:divergenceLm1}
\nab\cdot\left[f(r)\,\vsh{L}{L-1}{M}\right]&= \sqrt{\frac{L  }{2L+1}}\,
\left(\frac{\partial}{\partial r}-\frac{L-1}{r}\right)\,f(r)\,Y_{L\,M}(\hat{r})
\end{align}

\newpage
\thispagestyle{empty}
\phantom{hello}

\chapter*{Acknowledgments}

\noindent
There are many people who supported me~-- in one way or the other~--
in the development of this thesis. Especially I would like to thank
\begin{itemize}
\item Prof. Dr. W. Weise for giving me the opportunity to graduate at his chair
and for numerous valuable discussions and comments.
\item my advisors Dr. habil. H.W. Grie{\ss}hammer and Dr. T.R. Hemmert for all
their patience and help.
\item my second assessor Prof.~Dr.~H.~Friedrich and chairman 
      Prof.~Dr.~R.~Kr\"ucken. 
\item my parents for their support in all non-physical matters.
\item Estelle for the nice coffee breaks and her friendship in general.
\item D.R. Phillips for his kind invitation to the Ohio University, Athens,
      and for providing us with his deuteron Compton code.
\item many physicists for interesting and helpful discussions, especially
D.R.~Phillips, N.~Kaiser, B.~Pasquini, H.~Gao and A.~Nogga.
\item all members of T39 for the nice working climate, especially  
my room mate and friend M.~Procura.
\end{itemize}
\noindent
This work was supported in part by the Bundesministerium f\"ur Forschung und 
Technologie, and by the Deutsche Forschungsgemeinschaft under contracts 
GR1887/2-1 and GR1887/2-2.

\end{document}